%% file: _0_Tesis.tex
\documentclass[10pt,a4paper]{book}

\newcommand{\imprimir}{noimprimir}          
\newcommand{\contodos}{sintodos}            
\newcommand{\conquotes}{true}

\newcommand{\margenizquierda}{3.3cm}   
\newcommand{\margenderecha}{2.8cm}       
\newcommand{\margensuperior}{4cm}      
\newcommand{\margeninferior}{3.5cm}    

\newcommand{\titulo}{Towards general relativity through parametrized theories}
\newcommand{\tema}{GR}

\input{_1_preambulo}

\makeindex

\begin{document}

\input{_2_portada}

\include{1_introduccion}

\pagestyle{fancy}    \mainmatter
\fancyhf{}
\fancyhead[LO]{\nouppercase \rightmark}   
\fancyhead[LE,RO]{\thepage}               
\fancyhead[RE]{\leftmark}                 
\renewcommand{\headrulewidth}{0.2pt}

\setlength{\headheight}{12.2pt}
\include{2_mathematical_background}      
\include{3_string_masses}  

\include{4_parametrized_theories}            
\include{5_parametrized_EM}

\include{6_parametrized_scalar}        %
\include{7_parametrized_MCS}        %
\include{8_GR}           
\include{9_conclusions}           
\include{A_more_mathematics}

\backmatter

\fancypagestyle{plain}{%
	\fancyhf{} 
	\renewcommand{\headrulewidth}{0pt} 
}

\titleformat{\chapter}[hang]
{\gdef\chapterlabel{}\normalfont\sffamily\Huge\bfseries\scshape}
{\gdef\chapterlabel{\thechapter\ }}
{0pt}
{\begin{tikzpicture}[remember picture,overlay]
	\node[yshift=-2.5cm] at (current page.north west)
	{\begin{tikzpicture}[remember picture, overlay]
		\fill[gray!20] (0,0) rectangle (\paperwidth,3cm);
		\node[anchor=east,
		xshift=.9\paperwidth,
		rectangle,
		rounded corners=19pt,
		inner sep=8pt,
		fill=MidnightBlue]
		{\rule{0ex}{1.9ex}\color{white}\hspace*{.15ex} #1\phantom{\fontsize{30}{50}\selectfont g}\hspace*{-.45ex}\mbox{}};
		\end{tikzpicture}
	};
\end{tikzpicture}{\vspace*{-15ex}\normalfont\normalsize\starwars[Admiral Ackbar]{It's a trap!\phantom{.-}}{Return of the Jedi\phantom{.-}}\mbox{}\newline\vspace*{35ex}}}

\bibliographystyle{bio}      

\thispagestyle{empty}
\small
\bibliography{bibliografia.bib}

\normalsize

\newcommand{\sepuno}{\\[-2.2ex]\hspace*{70ex}}
\newcommand{\sepdos}{\\[-2.2ex]\hspace*{70ex}}
\newcommand{\nonumeres}[1]{{ }}
\index{A@\textbf{\huge{A\sepuno}\thispagestyle{empty}}|nonumeres}       \index{B@\textbf{\huge{B\sepdos}}|nonumeres}
\index{C@\textbf{\huge{C\sepuno}}|nonumeres}       \index{D@\textbf{\huge{D\sepuno}}|nonumeres}
\index{E@\textbf{\huge{E\sepuno}}|nonumeres}       \index{F@\textbf{\huge{F\sepdos}}|nonumeres}
\index{G@\textbf{\huge{G\sepdos}}|nonumeres}       \index{H@\textbf{\huge{H\sepdos}}|nonumeres}
\index{I@\textbf{\huge{I\sepdos}}|nonumeres}       
\index{K@\textbf{\huge{K\sepdos}}|nonumeres}
\index{L@\textbf{\huge{L\sepdos}}|nonumeres}
\index{M@\textbf{\huge{M\sepdos}}|nonumeres}       \index{N@\textbf{\huge{N\sepuno}}|nonumeres}
\index{O@\textbf{\huge{O\sepdos}}|nonumeres}
\index{P@\textbf{\huge{P\sepuno}}|nonumeres}       
\index{Q@\textbf{\huge{Q\sepdos}}|nonumeres}
\index{R@\textbf{\huge{R\sepuno}}|nonumeres} 
\index{S@\textbf{\huge{S\sepuno}}|nonumeres}
\index{T@\textbf{\huge{T\sepuno}}|nonumeres}       
\index{V@\textbf{\huge{V\sepuno}}|nonumeres}       \index{W@\textbf{\huge{W\sepdos}}|nonumeres}

\cleardoublepage  
\phantomsection

\titleformat{\chapter}[hang]
{\gdef\chapterlabel{}\normalfont\sffamily\Huge\bfseries\scshape}
{\gdef\chapterlabel{\thechapter\ }}
{0pt}
{\begin{tikzpicture}[remember picture,overlay]
	\node[yshift=-2.5cm] at (current page.north west)
	{\begin{tikzpicture}[remember picture, overlay]
		\fill[gray!20] (0,0) rectangle (\paperwidth,3cm);
		\node[anchor=east,
		xshift=.9\paperwidth,
		rectangle,
		rounded corners=19pt,
		inner sep=8pt,
		fill=MidnightBlue]
		{\rule{0ex}{1.9ex}\color{white}\hspace*{.15ex} #1\phantom{\fontsize{30}{50}\selectfont g}\hspace*{-.45ex}\mbox{}};
		\end{tikzpicture}
	};
\end{tikzpicture}{\vspace*{-20ex}\normalfont\normalsize\starwars[Obi-Wan Kenobi]{These aren’t the droids you’re looking for.\phantom{.-}}{A New Hope\phantom{.-}}\mbox{}\newline\vspace*{45ex}}}

\printindex\phantomsection
\cleardoublepage

 \pagestyle{empty}

\ifjuanquotes
 \mbox{}\vspace*{40ex}

\begin{center}
	\begin{minipage}{.7\linewidth}
		\starwars[Obi-Wan Kenobi]{Remember\ldots the Force will be with you, always.}{A New Hope}
	\end{minipage}
\end{center}

\else
\vspace*{10ex}
\centerline{\huge \today \qquad - \qquad \currenttime}
\fi

\ifjuanimprimirconportada
\else

\cleardoublepage
\mbox{}\newpage

\renewcommand{\thepage}{Last}
\begin{tikzpicture}[remember picture,overlay]
\node at (current page.center) {\includegraphics[width=\pdfpagewidth,height=\pdfpageheight]{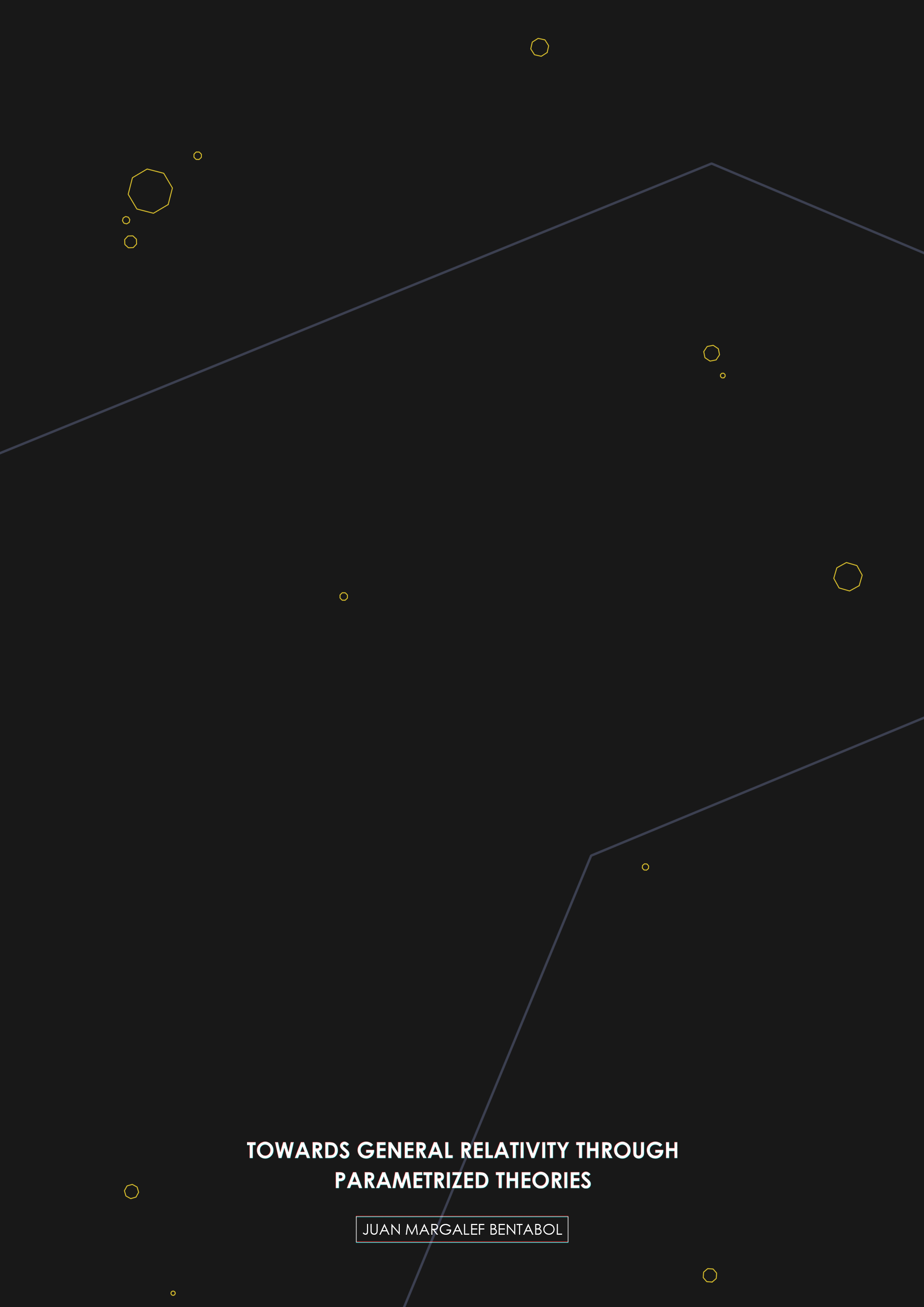}};
\end{tikzpicture}
\fi

\end{document}

%% file: _1_preambulo.tex
\usepackage{amsmath}                    
\usepackage[bbgreekl]{mathbbol}

\usepackage{amsfonts,amssymb,amsthm}    
\DeclareSymbolFontAlphabet{\mathbbm}{bbold}
\DeclareSymbolFontAlphabet{\mathbb}{AMSb}%
\usepackage{tensor,mathrsfs}
\usepackage[english]{babel}             
\usepackage[breaklinks=true]{hyperref}  
\usepackage{cite}   
\usepackage[utf8]{inputenc} 
\usepackage[T1]{fontenc}
\usepackage{geometry} 				
\usepackage{smartref}               \addtoreflist{chapter}   
\usepackage{lmodern}
\usepackage{enumerate}
\usepackage{bm}
\usepackage{etex}
\usepackage{doi}

\usepackage[margin=6ex,centerlast,textfont=small,labelfont={bf,small}]{caption}

\usepackage{array}
\usepackage{appendix}

\usepackage{fancyhdr}                   
\usepackage{ifthen}
\usepackage{makeidx}                    
\usepackage[nottoc]{tocbibind}          
\usepackage[dvipsnames]{xcolor}            

\usepackage[all]{xy}                    
\usepackage{fancybox}
\usepackage{verbatim}
\usepackage[explicit]{titlesec}
\usepackage{sectsty}   \allsectionsfont{\normalfont\sffamily\bfseries}
\usepackage{soul}
\usepackage{dialogue}
\usepackage{float}

\usepackage{wrapfig}
\usepackage{cutwin}
\usepackage[export]{adjustbox}
\usepackage{tikz}
\usepackage[pdf,usenames,dvipsnames]{pstricks}
\usepackage{datetime}
\usepackage{tikz-cd}
\usetikzlibrary{arrows}
\usetikzlibrary{calc}
\usetikzlibrary{positioning}
\usepackage{xifthen}%
\usepackage{xpatch}
\makeatletter
\AtBeginDocument{\xpatchcmd{\@thm}{\thm@headpunct{.}}{\thm@headpunct{}}{}{}}
\makeatother

\usepackage{braket}
\usepackage{upgreek}
\usepackage{linegoal}

\usepackage{filecontents}

\usepackage{mathtools}

\graphicspath{{./Images/}}

\newcommand{\tikzmark}[1]{\tikz[overlay,remember picture] \node (#1) {};}
\tikzset{square arrow/.style={to path={-- ++(0,-.25) -| (\tikztotarget)}}}

\renewcommand{\thechapter}{\arabic{chapter}} 

\sectionfont{\normalfont\fontsize{16}{15}\sffamily\bfseries}
\subsectionfont{\normalfont\fontsize{13}{15}\sffamily\bfseries}
\subsubsectionfont{\normalfont\fontsize{12}{15}\sffamily\bfseries}

\newcommand{\barrita}{}
\newcommand*\chapterlabel{}
\titleformat{\chapter}[hang]
            {\gdef\chapterlabel{}\normalfont\sffamily\Huge\bfseries\scshape}
            {\gdef\chapterlabel{\thechapter\ }}
            {0pt}
            {\begin{tikzpicture}[remember picture,overlay]
			  \node[yshift=-2.5cm] at (current page.north west)
			  {\begin{tikzpicture}[remember picture, overlay]
				 \fill[gray!20] (0,0) rectangle (\paperwidth,3cm);
				 \node[anchor=east,
					   xshift=.9\paperwidth,
					   rectangle,
					   rounded corners=19pt,
					   inner sep=8pt,
					   fill=MidnightBlue]
		{\rule{0ex}{1.9ex}\color{white}\hspace*{.15ex} #1\phantom{\fontsize{30}{50}\selectfont g}\hspace*{-.45ex}\mbox{}};
				\end{tikzpicture}
				};
			\end{tikzpicture}}
			
\titlespacing*{\chapter}{0pt}{60ex}{-50ex}

\titlespacing{\section}{0pt}{3ex+\parskip}{.5ex-\parskip}
\titlespacing{\subsection}{0pt}{3ex}{0.5ex-\parskip}
\titlespacing{\subsubsection}{0pt}{1.5ex}{-\parskip}

\newlength{\separacion}
\newcommand{\separ}{\vspace*{2ex}}
\newcommand{\separprevia}{\vspace*{.9ex}}
\newcommand{\separpost}{\vspace*{2.5ex}}
\newif\ifjuannoimprimir
\newif\ifjuancontodos
\newif\ifjuanquotes
\newif\ifjuanimprimirconportada

\ifthenelse{\equal{\imprimir}{noimprimir}}{\juannoimprimirtrue}{\juannoimprimirfalse}%
\ifthenelse{\equal{\contodos}{sintodos}}{\juancontodosfalse}{\juancontodostrue}%
\ifthenelse{\equal{\conquotes}{false}}{\juanquotesfalse}{\juanquotestrue}%

\ifjuancontodos
\usepackage[
   spanish,
   colorinlistoftodos,
   textsize=small,
   linecolor=blue,
   bordercolor=blue,
   backgroundcolor=brightturquoise,
   textwidth=2.8cm]{todonotes}   
\else
\usepackage[disable]{todonotes}
\fi

\ifjuannoimprimir
  \newcommand{\corurl}{BrickRed}  \newcommand{\corcite}{red}
  \newcommand{\corlink}{blue}    \newcommand{\corfile}{black}
\else
  \newcommand{\corurl}{black}    \newcommand{\corcite}{black}
  \newcommand{\corlink}{black}   \newcommand{\corfile}{black}
\fi

\hypersetup{
	linktocpage,
	colorlinks,
	urlcolor=\corurl,	
	citecolor=\corcite,
	linkcolor=\corlink,
	filecolor=\corfile,
	pdfnewwindow,
	pdftitle={\titulo},
	pdfauthor={Juan Margalef Bentabol},
	pdfsubject={\tema}}

\allowdisplaybreaks  

\definecolor{brightturquoise}{rgb}{0.03, 0.91, 0.87}

\newenvironment{remarks}{\goodbreak%
\begin{remarkss}\mbox{}\\[-4ex]
	\begin{enumerate}\setcounter{enumi}{\value{theorem}}}
	{\setcounter{theorem}{\value{enumi}}
	\end{enumerate}
\end{remarkss}}

\newenvironment{definitions}{\goodbreak%
\begin{definitionss}\mbox{}\\[-4ex]
\begin{enumerate}\setcounter{enumi}{\value{theorem}}}
{\setcounter{theorem}{\value{enumi}}
\end{enumerate}
\end{definitionss}}

\newenvironment{properties}{\goodbreak%
	\begin{propertiess}\mbox{}\\[-4ex]
		\begin{enumerate}\setcounter{enumi}{\value{theorem}}}
		{\setcounter{theorem}{\value{enumi}}
		\end{enumerate}
\end{propertiess}}

\newcommand{\deriv}[3][{}]{\frac{\mathrm{d}^{#1}#2}{\mathrm{d}#3^{#1}}}
\newcommand{\parcial}[3][{}]{\frac{\partial^{#1}#2}{\partial #3^{#1}}}
\newcommand{\vecc}[3][{}]{\peqsub{#2}{{\textsf{#3}_{#1}}}}

\newcommand{\updown}[3]{\overset{#1}{\underset{#2}{#3}}}   
\newcommand{\myrightleftarrows}[1]{\mathrel{\substack{\xrightarrow{#1} \\[-.6ex] \xleftarrow{#1}}}}

\newcommand{\Cinf}[1]{\ifthenelse{\equal{#1}{}}{\mathcal{C}^\infty}{\mathcal{C}^\infty \hspace{-0.25ex}(#1)}}
\newcommand{\parc}[2][]{\frac{\partial #1}{\partial #2}}

\makeatletter
\newcommand{\raisemath}[1]{\mathpalette{\raisem@th{#1}}}
\newcommand{\raisem@th}[3]{\raisebox{#1}{$#2#3$}}
\makeatother
\newcommand{\peqsub}[2]{#1_{\raisemath{.2pt}{\hspace*{-0.1ex}\scriptscriptstyle #2}\hspace*{-0.2ex}}}
\newcommand{\peqsubfino}[3]{#1_{\raisemath{#3}{\hspace*{-0.1ex}\scriptscriptstyle #2}\hspace*{-0.2ex}}}

\newcommand{\longlongrightarrow}[1]{\xrightarrow{\hspace*{#1ex}}}
\newcommand{\longlongmapsto}[1]{\xmapsto{\hspace*{#1ex}}}

\newcommand{\falta}{green}         

\newcommand{\todoin}[2][cyan]{\todo[inline,color=#1]{#2}}

\newcommand{\refconchap}[1]{\hyperref[#1]{\textcolor{\corlink}{\chapterref{#1}.}\ref*{#1}}}   
\newcommand{\refconchapp}[1]{\chapterref{#1}.\ref*{#1}}   
\newcommand{\eqrefconchap}[1]{(\hyperref[#1]{\textcolor{\corlink}{\chapterref{#1}.}\ref*{#1}})}   

\makeatletter
\DeclareRobustCommand{\notheader}[1]{\ifx\label\@gobble\else#1\fi}
\makeatother

\newcommand{\N}{\mathbb{N}}     
\newcommand{\R}{\mathbb{R}}     
\newcommand{\C}{\mathbb{C}}     
\renewcommand{\S}{\mathbb{S}}   
\newcommand{\K}{\mathbb{K}}     

\def\QED{{\boldmath$\rule{0.5em}{0.5em}$}}                                
\def\markatright#1{\leavevmode\unskip\nobreak\quad\hspace*{\fill}{#1}}    
\def\qed{\markatright{\QED}}                                              

\DeclareMathOperator{\cotan}{cotan}

\newtheorem{theorem}{Theorem}[chapter]

\makeatletter
\let\c@equation\c@theorem
\makeatother

\newtheorem{definition}[theorem]{Definition}       \newtheorem*{definitionss}{Definitions}
                   \newtheorem{proposition}[theorem]{Proposition}
              \newtheorem{remark}[theorem]{Remark}
                  
\newtheorem{lemma}[theorem]{Lemma}
\newtheorem*{lemma*}{Lemma}                        \newtheorem{corollary}[theorem]{Corollary}

\renewenvironment{proof}[1][Proof]{\textbf{#1.} }{\qed\\}     

\theoremstyle{definition}

\newtheorem*{remarkss}{Remarks}  
\newtheorem*{propertiess}{Properties}

\newcommand{\domDelta}{H^2_\mu[0,\ell]}
\newcommand{\domRobin}{H^1_{\mu,R}[0,\ell]}

\newcommand{\trassub}{\mbox{}\vspace*{-3.5ex}}
\newcommand{\starwars}[3][]{%
	\ifthenelse{\isempty{#1}}%
	  {\ifjuanquotes\mbox{}\vspace*{-17ex}
	  	
	  	\begin{quote}%
	  	  \raggedleft{#2}\vspace*{1ex}%
	  	  
	  	  \raggedleft--- \emph{#3}%
  	   \end{quote}\fi}
      {\ifjuanquotes\mbox{}\vspace*{-17ex}
      	
      	\begin{quote}%
  	     \raggedleft{#2}\vspace*{1ex}%
  	     
  	     \raggedleft--- #1, \emph{#3}%
        \end{quote}\fi}
	}

\newcommand{\starwarssection}[3][]{%
	\ifthenelse{\isempty{#1}}%
	{\ifjuanquotes\mbox{}\vspace*{-6.5ex}
		
		\begin{quote}%
			\raggedleft{#2}\vspace*{1ex}%
			
			\raggedleft--- \emph{#3}%
		\end{quote}\vspace*{1.5ex}\fi}
	{\ifjuanquotes\mbox{}\vspace*{-6.5ex}
		
		\begin{quote}%
			\raggedleft{#2}\vspace*{1ex}%
			
			\raggedleft--- #1, \emph{#3}%
		\end{quote}\vspace*{1.5ex}\fi}
}

\newlength{\figwidth}%
\newlength{\figwidthleft}%

\makeatletter
\DeclareFontFamily{OMX}{MnSymbolE}{}
\DeclareSymbolFont{MnLargeSymbols}{OMX}{MnSymbolE}{m}{n}
\SetSymbolFont{MnLargeSymbols}{bold}{OMX}{MnSymbolE}{b}{n}
\DeclareFontShape{OMX}{MnSymbolE}{m}{n}{
	<-6>  MnSymbolE5
	<6-7>  MnSymbolE6
	<7-8>  MnSymbolE7
	<8-9>  MnSymbolE8
	<9-10> MnSymbolE9
	<10-12> MnSymbolE10
	<12->   MnSymbolE12
}{}
\DeclareFontShape{OMX}{MnSymbolE}{b}{n}{
	<-6>  MnSymbolE-Bold5
	<6-7>  MnSymbolE-Bold6
	<7-8>  MnSymbolE-Bold7
	<8-9>  MnSymbolE-Bold8
	<9-10> MnSymbolE-Bold9
	<10-12> MnSymbolE-Bold10
	<12->   MnSymbolE-Bold12
}{}

\let\llangle\@undefined
\let\rrangle\@undefined
\DeclareMathDelimiter{\llangle}{\mathopen}%
{MnLargeSymbols}{'164}{MnLargeSymbols}{'164}
\DeclareMathDelimiter{\rrangle}{\mathclose}%
{MnLargeSymbols}{'171}{MnLargeSymbols}{'171}
\makeatother

\newcommand{\smallcirc}{\mathbin{\text{\raisebox{0.185ex}{\scalebox{0.7}{$\circ$}}}}}
\newcommand{\ccdot}{\hspace*{-0.25ex}\cdot{}\hspace*{-0.25ex}}

\newcommand{\riemann}{Riem}

\newcommand{\coma}{\hspace*{0.4ex},\hspace*{0.1ex}}

%% file: _2_portada.tex
    \newcommand{\imagenportada}{portada}

    \pagestyle{empty}
    
    \ifjuanimprimirconportada
    \else
    \frontmatter\renewcommand{\thepage}{Cover}
    \pdfbookmark{Cover}{cover}
    \begin{tikzpicture}[remember picture,overlay]
    \node at (current page.center) {\includegraphics[width=\pdfpagewidth,height=\pdfpageheight]{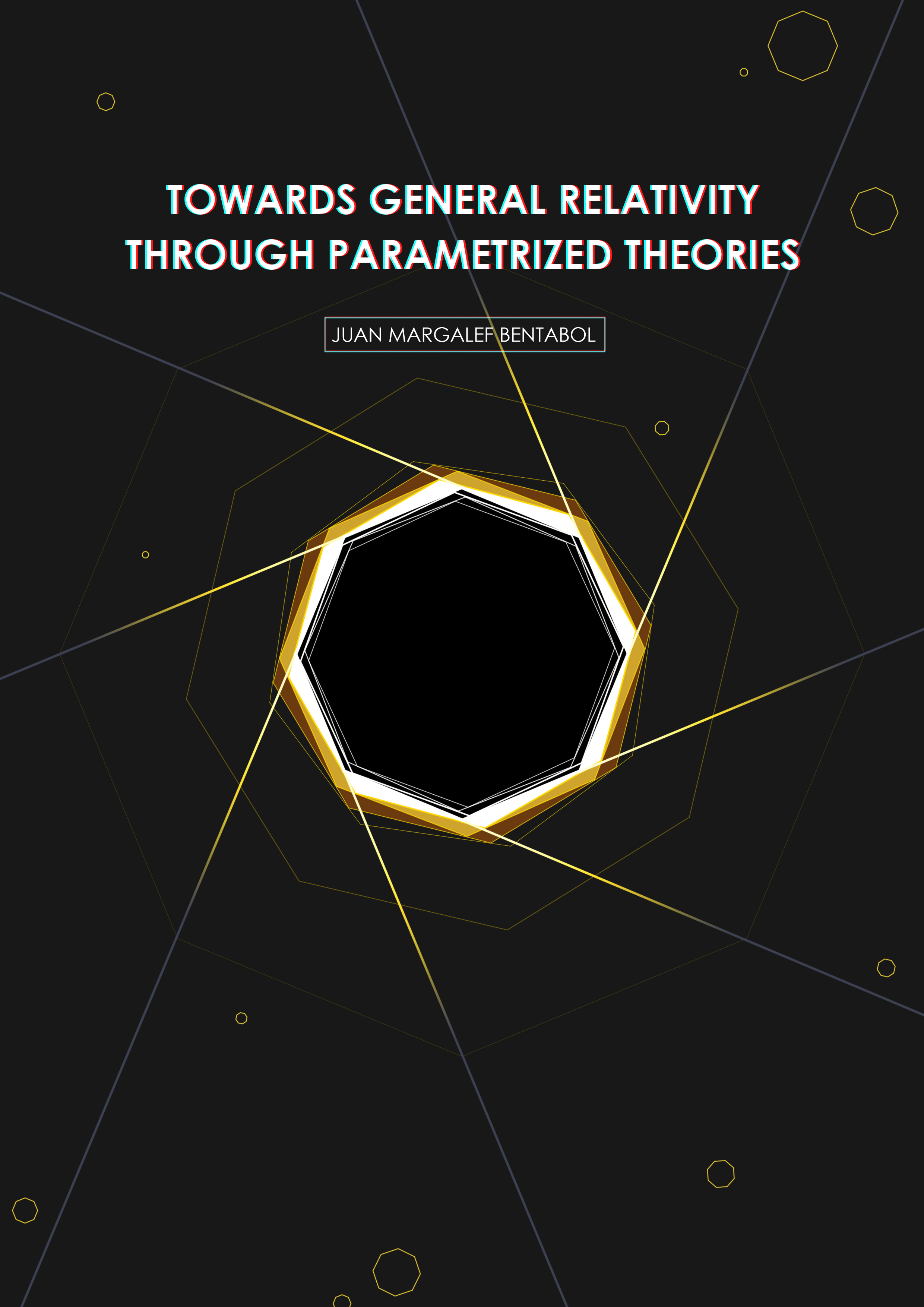}};
    \end{tikzpicture}
    
    \newpage 
    \renewcommand{\thepage}{-}
    
    \cleardoublepage
    \fi
    
    \renewcommand{\thepage}{Full title}
    \begin{tikzpicture}[remember picture,overlay]
    \node at (current page.center) {\includegraphics[width=\pdfpagewidth,height=\pdfpageheight]{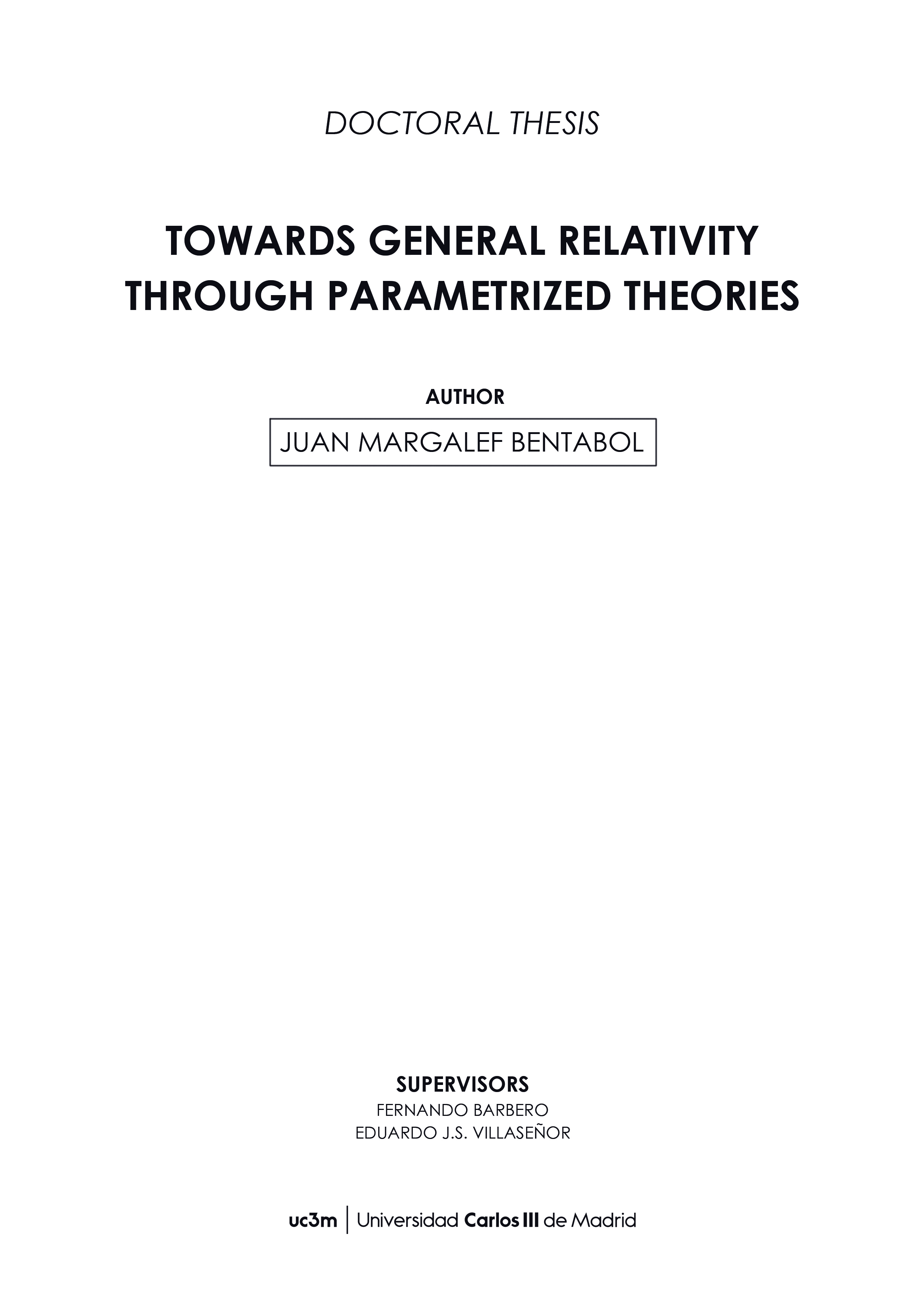}};
    \end{tikzpicture}
    
        \restoregeometry

    \setcounter{page}{0}
    \renewcommand{\thepage}{\Roman{page}}
    
\ifjuannoimprimir
\else    
\cleardoublepage
     \pagestyle{empty}
    \newgeometry{inner=2cm,outer=2cm,top=2.9cm,bottom=2.8cm}
\cleardoublepage
\fi

    \ifjuanquotes
    \cleardoublepage
    \mbox{}\vspace*{40ex}
    
    \begin{center}
    	\begin{minipage}{.6\linewidth}
    		\starwars[Master Yoda]{Do, or do not. There is no try.}{The Empire Strikes Back}
    	\end{minipage}
    \end{center}
    
    \cleardoublepage\fi

    \pagestyle{plain}
    \newgeometry{inner=3.5cm,outer=3.1cm,top=4cm,bottom=3cm}
    \cleardoublepage\tableofcontents\pdfbookmark{Table of contents}{toc}%
    \cleardoublepage
    \restoregeometry
    \listoftodos

%% file: 1_introduccion.tex
    \chapter{Acknowledgments}
    
    \starwars[Darth Vader]{Impressive. Most impressive. Obi-Wan has taught you well.}{The Empire Strikes Back}

     \vspace*{2ex}
     Picture me at 10 years old being asked about my parent's jobs. My mom's was easy: she was a mathematician, teaching in a high-school, I got it. But my dad's, that's another story. He was a scientific researcher in mathematics at the research council of Spain. Needless to say, I had no clue whatsoever of what it meant. Nonetheless, I knew that answer by heart. Now, so many years later, I have started to grasp its meaning which makes this moment, the submission of my thesis dissertation, even more special to me. Becoming, as my dad, a doctor in mathematics fills me with pride and, above all, gratitude. Indeed, I am in debt to so many people that have helped me so much during this path, that I am afraid of forgetting someone. If I do, please forgive me and also understand that I have had to restrict my acknowledgments to those who have contributed to my academic success. Otherwise, believe me when I say that I would not have space enough to thank all the incredible friends I have one by one. Instead, accept this generic but heartful: thank you!\separ
     
     I would like to begin with my undergraduate period. First my awesome overseas friends, Lucas and Victor, \emph{el boludo}. Although I met you so many years ago at the International Chemistry Olympiad, you are still among my best friends no matter the distance. We have had an amazing time in so many countries that it is difficult to keep track but, for sure, we will keep adding places to the list. Then I had the pleasure to share time and laughs at the mathematical faculty of the Universidad Complutense de Madrid with Milena, Alf, Eloy, and (years later) María. Lovely people with whom, luckily, I still share time and laughs. I also want to thank my friends from the dark side: the physics faculty. I delved deep into the physics realm with my great friends Izan and Marcio, thanks to you I enjoyed every class and every lab session even when we did them in a crazy rush. Finally, my Erasmus at the Université Libre de Bruxelles was a milestone in my life. Not only did I learn advanced physics and French, but I also made many awesome friends. Céline, you helped me with physics, French, and party time so much and so often that I cannot even count. Hélène, Ilenia, and Ilaria, you are beautiful souls and we should have hanged out more. My beloved Piera, the best \emph{coinquilina} ever, always ready to leave your \emph{archi} stuff to have a drink with me regardless of the hour. And finally Malvina, my Italian \emph{sorela}, you made my \emph{séjour} the coolest I could have ever imagined. So much memories, laughs, parties, laughs, funny fights, and laughs. I am certain that there will be many more to come. \emph{Grazie per tutto ma belle Malvina}.\separ
     
     After the Erasmus, thanks to a PhD fellowship from ``La Caixa'' Foundation, I got into the UC3M where I have had lots of fun with great people. A special mention to Jacobo, Miguel, Filippo, and Carlos, with whom I shared incredible moments and lots of weddings. I do not know how many weddings are still to come, but I wish we will have many more amazing moments together. By the way, Amanda and Mariana, although you are not in the university, feel yourselves included in the previous wish because you are the best. Also Andrés, Emanuel, Rafa, Rocío, Alex, and my cosmonauts peers Sergei and Carolina, you all have done a great difference in my PhD. I am really grateful for having shared my daily life, organized seminars, and hanged out with all of you.\separ
     
     During my PhD I had the opportunity to do two research stays of three months each. The first one was at the Technical University of Vienna under the supervision of Martin Bauer. I am extraordinarily thankful to you for your kindness, for all the scientific talks we had and, of course, for showing me great places in Vienna. I also met at the TU my beloved María, who arrived shortly before me, and whom I miss so much. I miss your questions, eating everyday with you, lecturing you to be more positive, and enjoying Vienna together. Vienna was also the place where I met who is now, without any doubt, one of my best friends: Nico. We had a crazy time together talking, traveling, drinking, partying, and sending awkward audios to random people. You also introduced me to all your friends: Vir, Pablo, Bea, my great friend Manu\ldots Thanks to all of you I had a stay which was, not only scientifically challenging and exciting, but also worth to be lived.\separ
     
     My second research stay was at the University of Erlangen-Nürnberg in Germany, under the supervision of Hanno Sahlmann. I am deeply grateful to you for all your help and for being so nice. It was a pleasure to be there with you and the rest of the Quantum Gravity group who welcomed me so warmly. I want to specially thank David, one of the funniest persons I have ever met, who totally made the difference. You showed me around, invited me to pub quizzes, took me to \emph{Der Berch} and helped me with my German. \emph{Vielen vielen Dank!}\separ
     
     I must admit that I have not been a typical PhD student because I cannot remember any negative moment related to this work. It is true that I am an annoyingly positive person, but most certainly, this is thanks to the guide and help of my supervisors Eduardo and Fernando. You are, and will be, my mentors and my friends. Since the very beginning you treated me like your scientific peer, taking into account my ideas and comments even though I was so scientifically behind you. I cannot express enough my gratitude for all these years, for the scientific and non-scientific talks, and for all the great times we have shared. You have always been available, read my drafts, answered my questions and concerns and, in short, worked with me. I could not have had better supervisors and I deeply thank you for everything, specially during the last weeks before the due date. I take also this opportunity to thank Andrea Chapela, Hanno Sahlmann, José Luis Jaramillo, Juan Carlos Marrero, and Miguel Sánchez for the careful reading of this thesis and your helpful comments.\separ
     
	Although sometimes it doesn't seem so, there is life outside the Academia and I have many people to thank for their support. First to Alberto and Isma, it is overwhelming that after so many years we are still best friends despite the distance. Also to Helena for our French parties. To Marina, my emergency philosopher, one day we will meet in Berlin. And, of course, to Luu. I learned so much from you and I went always beyond my limits thanks to you. You are so inspiring and encouraging that I owe you a huge part of my academic and personal success. \emph{Obrigado mesmo por tudo!}\separ
     
     A deep thank you is due to my LFI ``shadow cabinet'', the council of wise people, my awesome friends. Pablo, Nico, Sabela, and Violeta, I am so lucky to have you by my side, specially to put out so many fires. Obviously to Efrén, one of the persons I most admire. To  Domin, Miguel, Héctor (thanks for most beautiful cover ever!) and, \emph{malgré tout}, Cris who cured me the stone disease when I needed it most. I am also in debt to the historical institution of \emph{La Residencia de Estudiantes}. I am privileged to have lived there and I kindly thank all the staff and friends I have met there. To name a few among so many amazing people: Miguel ---living and working together, who'd have said it would turn out so well---, my neighbor Andrea with our messages across the wall at ungodly hours, Ana for being always there (except all the months you have been away), Mariano for your \emph{jeje}'s and the awesome pictures for this thesis, Álvaro my wingman in \emph{Trivulgando}, Miguel Alirangues for always trying to lecture me to be a better self, Alba the sweetest and most random girl, and Raquel for our conversations at our tea parties. I am extremely thankful to all of you. The last incorporation to this special list is my incredible FameLab family. We will popularize science wherever it takes. Finally, to the incredible Vicky and her wingman Buri. Thanks for taking me out from the social retreat which is writing a thesis. You made me really happy and I enjoyed so much our reckless journey that I beg it is not the last one. Luckily, we'll always have Rome!\separ
     
     Let me finish with the warmest thank to all my family as I owe them everything. To my lovely cousins, aunts, uncles, and my grandparents, who would have been very proud. To Belén, Txiki, Arantxa, Keko, and Ondina. To Carla, no matter how much we fought in the past, I love to have you in my life and visit you around the world. To Berta, for your visits, concern, care, love, and for always being there for me. Finally, to my parents to whom I dedicate this thesis. You have supported me, believed in me, and encouraged me beyond my wildest imagination. You two were, are, and will be an inspiration. I will cherish all the teachings I have learned from you. I probably do not tell you this enough but for sure you know: \emph{¡os quiero mucho y gracias de todo corazón!}
     
     \cleardoublepage

     \mbox{}\vspace*{40ex}
     
     \centerline{\includegraphics[width=.75\linewidth]{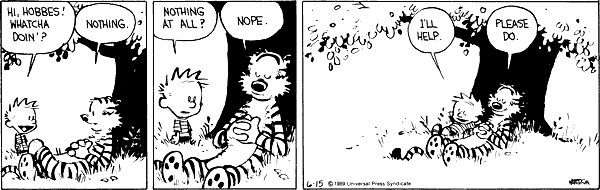}}

    \chapter{Introduction}

        \section*{A bit of history}\addcontentsline{toc}{section}{A bit of history}   
           
        \starwarssection[Master Yoda]{When 900 years old, you reach\ldots\ Look as good, you will not.}{The Empire Strikes Back}

		The satisfactory quantization of general relativity (GR) is one of the most important open problems in theoretical physics. Its resolution is crucial to understand the physics hidden at high energies and small scales such as the behavior of black holes or the Big Bang itself.\separ
        
        The fundamental feature that distinguishes general relativity from the rest of field theories is the absence of a background geometrical structure, like the Minkowski metric in the usual field theories, which suggests that new methods are necessary to tackle this problem. It was probably Einstein the first who, through heuristic reasonings, pointed out that the new theory of relativity had to be modified by quantum effects. He did so as early as 1916 in his first paper about gravitational radiation \cite{einstein1916naherungsweise}. Later on, several physicists like Oskar Klein, Rosenfeld, Fierz, or Pauli arrived at the same conclusion. However, contrary to Einstein who moved on to believe in the necessity of a new brand unifying theory, they mostly considered that similar arguments as the ones applied for electromagnetism would suffice. In fact Rosenfeld wrote the first papers on quantum gravity by applying some Pauli's quantization ideas to linearized GR \cite{rosenfeld1930quantelung}. Soon after that, the relation of this theory with the linear spin-two quantum field theory was discovered. It is now believed that Matvei Pretrovich Bronstein, a young Russian physicist, was the first person to realize that the quantization of the (non-linear) gravitational field required a special treatment due to its unique features \cite{Bronstein}. His ideas, nonetheless, were doomed to oblivion as he was arrested and executed in 1938 during the Great Purge in the Soviet Union. It seems that the French physicist Jacques Salomon was the only one outside the Soviet Union to acknowledge and develop his ideas, although sadly he was executed in 1942 by the Nazis during the German occupation of France. For further details and references I strongly recommend to the reader the excellent historical reviews of Stachel \cite{stachel1999early} and Rovelli \cite{rovelli2002notes}. In fact the latter contains a brief history of quantum gravity up to the beginning of the 21st century, of which I will mention some of its landmarks in the following.\separ
        
        During the late 40's and early 50's Bergmann and his collaborators started to study the phase space quantization of non-linear field theories and the observables in GR \cite{bergmann1949non,bergmann1949non2}. At the same time, Dirac developed his procedure to deal with general constrained Hamiltonian systems \cite{dirac1950generalized,dirac1964lectures}. However, its application to the Hamiltonian formulation of GR was quite unclear. At this point Gupta introduced some of the elements necessary for the perturbative quantization of GR \cite{gupta1952sn}. In particular, he pinpointed out the necessity to have a background metric.  Years later, in the 70's, t'Hooft and Veltman proved that such approach was unsuccessful due to the non-renormazibility of the resulting theory \cite{t1974one}. In the meantime, several physicists like Feynman, DeWitt, Wheeler, or Penrose worked to solve this problem, already considered at that point as a Herculean task. Probably, the most important result found in the context of GR at that moment was due to Arnowit, Deser, and Misner who, following the program started by Dirac, obtained the Hamiltonian formulation of GR using the so-called ADM variables \cite{arnowitt2008republication}.\separ
        
        The following milestone of this very brief history was the discovery by Hawking of black hole radiation \cite{hawking1974black}. He used some techniques developed in the context of QFT in curved space-times to show that a spherical black hole with mass $M$ emits thermal radiation at a temperature
        \[T=\frac{\hbar c^3}{8\pi \peqsub{k}{B}\peqsub{G}{N}M}\]
        This was in agreement with some earlier observations by Bekenstein who found a formal analogy between thermodynamics and black holes \cite{bekenstein1972black,bekenstein1973black,bekenstein1974generalized}. Hawking proved that, in fact, this correspondence was also physical. The formalization of black hole thermodynamics implies the existence of some kind of entropy $S$ (the Bekenstein-Hawking entropy), which is related to the area $A$ according to
        \[S=\frac{A}{4\peqsub{\ell}{P}^2}\]
        This result, that has been derived in several alternative ways, is now used as a consistency condition for candidate quantum gravity theories. The first of them, proposed in the 70's and 80's, were supergravity \cite{freedman1976progress}, higher derivative gravity \cite{stelle1977renormalization}, and the connection formulation of GR \cite{ashtekar1987new}. From those seminal ideas two important candidates emerged: string theory and loop quantum gravity (LQG). Both theories were able to derive, almost at the same time in the late 90's, the Bekestein-Hawking radiation law \cite{ashtekar1998quantum,strominger1996microscopic}. As a final remark of this quick historical review, I would like to mention the quantization of $2+1$ gravity carried out in 1988 by Witten \cite{witten19882}.

\section*{Main goal}\addcontentsline{toc}{section}{Main goal}

\starwarssection[Han Solo]{Never tell me the odds.}{The Empire Strikes Back}

The aim of this thesis can be neatly summarized in the following sentence
\begin{quote}\centering%
	\begin{minipage}{.72\linewidth}\centering
		\emph{Boundaries, GNH, and parametrized theories.\\It takes three to tango.}
	\end{minipage}
\end{quote}
Let us proceed to explain each term and why we can take advantage of their interrelation.\separ

$\blacktriangleright$ Boundaries\separprevia

The world is full of boundaries. They do matter. For instance, the objects that we use everyday are limited by their ``edges''. Also, in the case of sound and electromagnetic waves, which are essential for our lives, their behavior depends strongly on the shape of the walls that they encounter. Then, it is natural to include boundaries in our physical models to faithfully represent our reality. However, boundaries are tricky. If we are located within the bulk of an object, when we arrive at the boundary there is a \textbf{sudden} change in dimensionality as it is reduced by $1$. This is probably the origin of many of the difficulties that boundaries pose but, on the other hand, their presence leads to more fruitful and interesting theories. The best example is probably the general theory of relativity, where boundaries are essential to understand black holes or infinity. In particular, it seems mandatory to have a clean description of GR with boundaries in order to give a proper explanation of black hole entropy.\separ

This is precisely one of our main goals: to understand the role of boundaries in some field theories in order to apply what we learn to GR.\separpost

$\blacktriangleright$ GNH algorithm\separprevia

The Dirac's ``algorithm'' allows us to deal with singular mechanical system in which some constrains (functional relations between dynamical variables) must be preserved by its evolution. Despite its success in dealing with finite dimensional systems, its application to field theories is not as clean, specially in the presence of boundaries.\separ

The GNH algorithm was developed to generalize and simplify Dirac's algorithm, and it does so by relying on geometric methods. This in turn provides a clearer understanding of the procedure. Besides, boundaries can be included without any conceptual change in the algorithm, although in the case of field theories some care must be exercised due to functional analytic issues.\separ

All things considered, we have that another important goal of this thesis is the following: to obtain the dynamics of several field theories through the GNH algorithm.\separpost

$\blacktriangleright$ Parametrized theories\separprevia

The absence of background geometric objects in GR makes this theory invariant under diffeomorphisms. This group is infinite dimensional, which renders this field theory much more complicated than the usual ones. It is then desirable to have some simpler dynamical models which are also diff-invariant and use them to understand, for instance, the application of the GNH algorithm or the role of boundaries in the theory. This is exactly what it is achieved through parametrization, a procedure that introduces diff-invariance into any theory involving background geometric objects.\separ

Regarding this issue, the main goal is: to understand the role of the diff-invariance in theories simpler than GR and its relation with other gauge symmetries they may possess.\separpost

$\blacktriangleright$ Interplay of the three\separprevia

Those three elements together conform the main thread of this thesis. We will study in chapter \ref{Chapter - Scalar fields coupled to point masses} some field theories with boundaries using the GNH algorithm which will serve, in turn, to underline the importance of some functional analytic subtleties that must be taken into account in more complicated models. Then, we will proceed in chapter \ref{Chapter - Parametrized theories} to introduce parametrized theories and develop the simplest case: parametrized classical mechanics. It will be useful as a warm-up for the study of parametrized electromagnetism with boundaries in chapter \ref{Chapter - Parametrized EM}, the revisiting of the parametrized scalar field in chapter \ref{Chapter - parametrized scalar} (with a detailed description of the behavior at the boundary of this simpler theory), and the study of the parametrized Maxwell-Chern-Simons theory with boundaries in chapter~\ref{Chapter - parametrized MCS}.\separ

Finally, to tie up the thesis, in chapter \ref{Chapter - General relativity} we proceed with the theory that served as a beacon since the beginning: the general theory of relativity. We apply what we have learned with the aforementioned theories to the Hamiltonian formulation of GR to derive, in a geometric and easy way, the ADM formulation. We also look at another interesting model ---unimodular gravity--- that we will briefly explain.

\section*{State of the art}\addcontentsline{toc}{section}{State of the art}    
\starwarssection[Darth Vader]{No, I am your father!}{The Empire Strikes Back}

Let us end this introduction by summarizing the state of the art of the research about the three aforementioned nuclear concepts.\separ

$\blacktriangleright$ Boundaries\separprevia

Boundaries are nowadays some sort of trending topic in physics. Many interesting results have been obtained in many contexts like condensed matter (topological insulators), quantum computation, and of course general relativity. Focusing on the topic of this thesis, we found some contributions that are worth mentioning.\separ

First, the Hamiltonian formulation of the parametrized scalar field in bounded spatial regions has been discussed in \cite{andrade2011can}. The authors of that paper encountered some difficulties when Robin boundary conditions were imposed and left it as an open question the existence of the the canonical formalism in such case. It is interesting to mention that, with the proper geometric approach, we answered affirmatively this question in \cite{margalef2015quantization}, as we will explain in chapter \ref{Chapter - Scalar fields coupled to point masses}.

Boundaries are of capital importance in GR and have got a lot of attention for decades. They can show very pathological behaviors even at the topological level \cite{margalef2014topology} and its role in the study of black holes is crucial. Of particular interest is the notion of \textbf{isolated horizons}\index{Isolated horizon}, a quasi-local concept that seems more suited than the event horizon (a global concept) for the Hamiltonian formulation of black holes. For more details about this topic, we recommend the reader the review of Ashtekar and Krishnan \cite{ashtekar2004isolated}. From the properties of isolated horizons, several ideas emerged in the context of LQG that made it possible, among other things, to perform a rigorous state counting to compute the black hole entropy \cite{villasenor2009computation,agullo2009combinatorics}. In fact, although not related to the contents of this thesis, we have recently contributed to this area in \cite{margalef2018distribution}, where we study the spectrum of the area operator in LQG providing a very accurate description of the distribution of its eigenvalues.\separpost

$\blacktriangleright$ GNH algorithm\separprevia

The GNH method has been extensively studied by mathematicians in areas related to symplectic geometry, mechanics, or Hamiltonian reduction. Nonetheless, they are mostly interested in the technical details and extensions, and not so much in its application to physically relevant examples. Physicists, on the other hand, seem to be comfortable with the good old Dirac algorithm although we know that it can crash badly in some interesting examples. Probably the first extensive use of the GNH algorithm for non-trivial physical example in field theories was \cite{Eduardofernando2014}, where a careful study of the scalar and electromagnetic field in the presence of boundaries is carried out. In fact this paper was the origin of my thesis and has lead to several publications \cite{margalef2015quantization,margalef2016hamiltonian,margalef2017functional,margalefboundary}, where the GNH algorithm plays, thanks to its clear geometric interpretation, a central role in the understanding of the problems at hand.\separpost

$\blacktriangleright$ Parametrized theories\separprevia

Introduced by Dirac in the 50's, parametrized theories regained relevance with some works of Kucha\v{r}, Isham, Hájí\v{c}ek, and Torre \cite{kucha1976geometry,hajivcek1996symplectic,isham1985representations2,torre1992covariant} in the boundaryless case. Also, we have already mentioned the work \cite{andrade2011can}, where the parametrized scalar field with different boundary conditions was considered. There has been also some interest in the the context of LQG to explore the role of diffeomorphisms \cite{laddha2010polymer,laddha2011hamiltonian}.\separ

Parametrized field theories are of great importance because they are diff-invariant on one hand, and can be solved in typical examples, on the other. An interesting problem that crops up in this setting is to understand the interplay between diff-invariance and other ``more standard'' gauge symmetries. Several authors have worked on this topic \cite{kuchar1987canonical,rosenbaum2008space,torre1992covariant} with inconclusive results. As we have been able to show, it is actually possible to completely understand how diff-invariance and ordinary gauge symmetries interact by studying the parametrized EM field \cite{margalef2016hamiltonianEM}. To this end we have relied once again on the geometric GNH algoritm as we will explain in chapter \ref{Chapter - Parametrized EM}.


%% file: 2_mathematical_background.tex
\titleformat{\chapter}[hang]
{\gdef\chapterlabel{}\normalfont\sffamily\Huge\bfseries\scshape}
{\gdef\chapterlabel{\thechapter\ }}
{0pt}
{\begin{tikzpicture}[remember picture,overlay]
	\node[yshift=-2.5cm] at (current page.north west)
	{\begin{tikzpicture}[remember picture, overlay]
		\fill[gray!20] (0,0) rectangle (\paperwidth,3cm);
		\node[anchor=east,
		xshift=.9\paperwidth,
		rectangle,
		rounded corners=19pt,
		inner sep=8pt,
		fill=MidnightBlue]
		{\rule{0ex}{1.9ex}\color{white}\hspace*{.18ex} \chapterlabel- #1\phantom{\fontsize{30}{50}\selectfont g}\hspace*{-.45ex}\mbox{}};
		\end{tikzpicture}
	};
\end{tikzpicture}}
\chapter{Mathematical background}\thispagestyle{empty}\label{Chapter - Mathematical background}
	\ifjuanquotes\mbox{}\vspace*{-14ex}
	
	\hfill\begin{minipage}{.78\textwidth}
	\emph{\begin{dialogue}\normalfont
			\speak{Luke} But I need your help. I've come back to complete the training.
			\speak{Yoda} No more training, do you require. Already know you that which you need.
			\speak{Luke} Then I am a Jedi.
			\speak{Yoda} Not yet.
		\end{dialogue}}
	\raggedleft--- \emph{Return of the Jedi}\end{minipage}\fi

\mbox{}

  \section{Introduction}
    \emph{Never underestimate the joy people derive from hearing something they already know}. This sentence, attributed to Enrico Fermi, is one of the two reasons to include this chapter. The other reason, less Machiavellian, is to gather all the basic but important results that will be used through this work and, in the meantime, fix the notation. Much of what has been included in this first chapter might be well known and can be skipped by the advanced readers. Nonetheless, we warn them that some topics are well worth reading as they are not usually covered at a introductory level. In particular, the mathematical description of the spaces of embeddings, the GNH algorithm, and the Fock construction.\separ
    
    Whenever necessary, we will use the abstract index notation\footnote{No coordinates were \st{harmed} used in the making of this thesis (except at few places that will be mentioned).} introduced by Penrose \cite{penrose1984spinors}, although in this chapter we will try to stick to the no-index mathematical convention. Some useful technical results, that might not be so well known, are included in appendix \ref{appendix}. We also gather in section \refconchap{appendix section useful identities} of the appendix all the formulas appearing in this chapter (and some more) together with the abstract index  version of such formulas.

  \section{Differential geometry}
  \subsection{A bit of history}
  
The first traces of the use of geometry date from around 3000BC when some civilizations, such as the Harappans and the Babylonians, made several empirical discoveries concerning angles, lengths, areas, and volumes. Their motivation was most certainly for practical purposes in astronomy, construction, or agriculture. There is also evidence that, fifteen hundred years before Pythagoras was born, the Babylonians and the Egyptians developed already the Pythagoras' theorem. But it was not until Thales and the Pythagoreans, around 500BC, that geometry became an abstract and axiomatic science, which allowed it to flourish in an spectacular way.\separ

The next step forward in the history of geometry happens at the beginning of the 17th century, when Descartes and Fermat developed analytic geometry using coordinates and equations in geometry for the first time. It is nowadays considered as a necessary step in the creation of differential calculus by Newton and Leibniz almost at the end of the century. This brand new tool was further developed by many mathematician like Euler, the Bernoulli, or Agnesi. Despite the huge advances that geometry was undergoing, there was also a problem that had been bugging geometers for centuries: deciding whether the Euclid's fifth axiom, the parallel postulate, could be derived from the other ones. It was not until the 18th century that the first mathematicians succeeded in going beyond the classical geometry established by the Greeks. Gauss, Bolyai, and Lobachevsky studied independently the Euclid's geometry without the fifth postulate. Such studies gave rise to the first non-Euclidean geometries: hyperbolic and elliptic geometry.\separ

Gauss was most likely the first one to think abstractly about geometric spaces when he began to study the curvature of surfaces at each point. He provided the definition of the so-called \textbf{Gaussian curvature} in terms of the curvature of all curves over the surface and proved one of the most important theorems in geometry: the \textbf{Theorema Egregium}. It states that the Gaussian curvature, a magnitude defined using the ambient space, is actually independent of it. This was the starting shot to the study of  intrinsic geometry. The first mathematician that probably thought about intrinsic geometric spaces of general dimension was Lagrange, which might seem a bit surprising because he proudly boasted about his un-geometric point of view in his important \emph{Méchanique analitique} written in 1788:
\begin{quote}\centering
\begin{minipage}{.97\linewidth}\emph{On ne trouvera point de Figures dans cet ouvrage. Les méthodes que j'y expose ne demandent ni construction, ni raisonnemens géométriques ou méchaniques, mais seulement des opérations algébriques, assujetties à une marche régulière et uniforme.}
\end{minipage}
\end{quote}
Nonetheless, Lagrange realized that the degrees of freedom of a mechanical system can be thought of as an abstract space. This was probably the first step towards the concept of manifold. But it was Riemann, a student of Gauss, who finally extend the ideas of non-Euclidean geometry for a general dimension. In the late 19th century and early 20th century a lot of great mathematicians --like Hilbert, Lie, Poincaré, Weyl, Cartan (father and son)-- contributed to give differential geometry the form that we use and love nowadays, and that is presented here.

  \subsection{Crash course on vector bundles}
    We begin our mathematical tour with this section, where we introduce briefly the most relevant aspects of the geometry of vector bundles skipping the mathematical details. For further discussion see \cite{ONeill,gockeler1989differential} and, specially for infinite dimensional manifolds, we recommend \cite{marsden1974applications,abraham2012manifolds}.
      
    \subsubsection*{Vector bundle}\trassub
    
\begin{figure}[h!]
      \centering\includegraphics[width=0.75\linewidth,clip,trim=70ex 70ex 70ex 65ex]{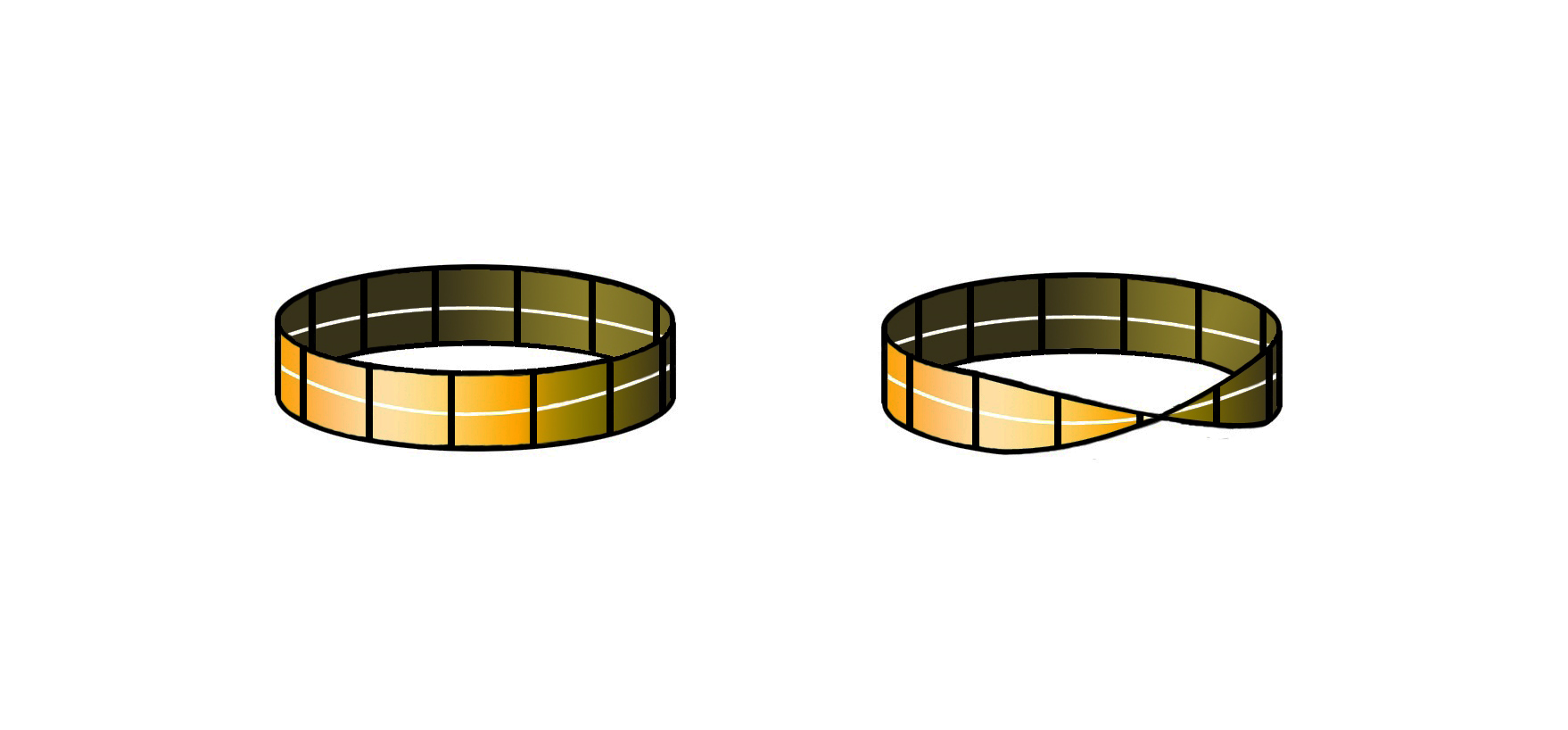}
      \caption{On the left we have a cylinder $C=\S^1\times\R$ that can be understood as a fiber $\R$ over each point of $\S^1$. On the right we have the M\"obius band $M$ formed by fibers $\R$ over each point of the circle but twisted in such a way that, globally, $M$ is not a product manifold.}\label{Mathematical background - image - vector bundles}
      \end{figure}

      Vector bundles\index{Bundle!Vector bundle} are, in a way, a generalization of product manifolds. To see that, consider the manifold $Q=M\times N$ which can be understood as the disjoint union
      \[Q=\bigsqcup_{m\in M}N\]
      where we placed the manifold $N$ as fibers over each point of $M$. The topology and differential structure are such that the fibers are all straight up. On the other hand, the idea of a fiber bundle is that such structures (more technically, the transition functions) are defined in such a way that the fibers might be twisted up like figure \ref{Mathematical background - image - vector bundles} shows for the M\"obius band. We focus our attention to the case of vector bundles i.e.\ when the fibers are vector spaces.      
  
      A vector bundle consists of a total manifold $E$, a base manifold $B$, and a projection $\pi:E\to B$ such that the fibers attached to each point  $\pi^{-1}(\{b\})$ are vector spaces. The topology and differential structure are defined in such a way that, locally, $E$ is a product manifold
      \[E=\bigsqcup_{b \in B}E_b\qquad/\qquad \pi^{-1}(U)\cong U\times F\qquad U\subset B\text{ small open set}\]
      where we denote $E_b=\pi^{-1}(\{b\})\cong F$ the fiber at $b\in B$. It is also customary to write the vector bundle as $E\overset{\pi}{\to}B$.  Notice that, for finite dimensional manifolds, $\dim(E)=\dim(B)+\dim(F)$.\separ

      One of the most important objects related to the vector bundles are the sections\index{Section of a bundle}, which roughly speaking are maps picking smoothly one point of each fiber $E_q$ i.e.\ $s(q)\in E_q$ for every $q\in B$.
      
      \begin{wrapfigure}{r}{0.24\textwidth}
      	\vspace{-.5ex}
      	\centering\includegraphics[clip,trim=6ex 5ex 10ex 10ex,width=1\linewidth]{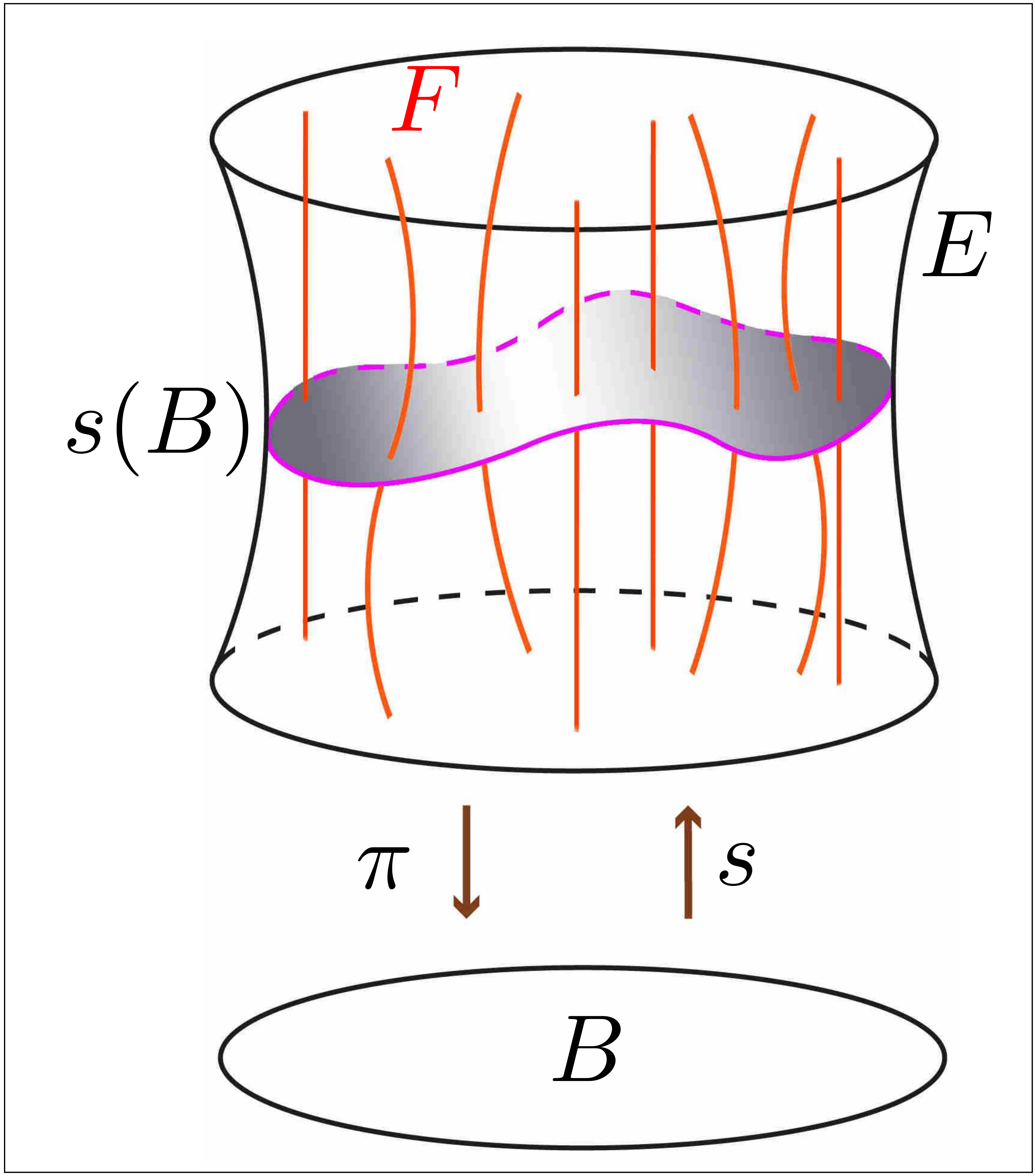}
      \vspace*{-8ex}
      \end{wrapfigure}
  
  \mbox{}\vspace*{-2ex}
      
      \begin{definition}\mbox{}\\
      	A \textbf{section of a vector bundle {\boldmath$E\overset{\pi}{\rightarrow} B$}}\index{Section} is a smooth map $s:B\rightarrow E$ such that $\pi\smallcirc s=\mathrm{Id}_B$. We denote $\Gamma(E)$ the set of sections of the vector bundle.
      \end{definition}  
      There always exists a privileged section, namely the zero section $s_0$, that maps every point $q\in B$ to the zero of the corresponding vector space $E_q$ i.e.\ $s_0(q)=0\in E_q$.
      \begin{remark}\mbox{}\\
      	$\Gamma(E)$ is a $\mathcal{C}^\infty(B)$-module i.e.\ if $s_1,s_2\in\Gamma(E)$ and $f\in\mathcal{C}^\infty(B)$ then $s'=fs_1+s_2\in\Gamma(E)$ where $s'(q)=f(q)s_1(q)+s_2(q)\in E_q$. In particular it is an $\R$-vector space.
      \end{remark}
      We now introduce the most relevant types of vector bundles. We assume that the mathematical construction of the dual\index{Bundle!Dual bundle}, direct sum\index{Bundle!Direct sum}, and tensor product bundles\index{Bundle!Tensor product bundle} is known by the reader.
      
    \subsubsection*{Pullback bundle}\trassub
    
       We have seen how a vector bundle $E\overset{\pi}{\rightarrow}B$ is formed by placing fibers at every point $b\in B$. Now, if we have a map $f:D\to B$ we could pullback to every $d\in D$ the fiber that corresponds to $f(d)\in B$.
      
      \begin{definition}\label{def pullback bundle}\mbox{}\\
      	Let $E\overset{\pi}{\rightarrow}B$ be a vector bundle and $f:D\rightarrow B$ a continuous map, we define the \textbf{pullback bundle}\index{Bundle!Pullback bundle}
      	\[f^*\hspace*{-.2ex}E=\Big\{(d,e)\in D\times E \ /\ f(d)=\pi(e)\Big\}\]
      	equipped with the subspace topology and the projection over the first factor $\pi_1:f^*\!E\rightarrow D$.
      \end{definition}
      
      Notice in particular that the following diagram is commutative
      
      \centerline{\xymatrix{f^*\hspace*{-.2ex}E \ar[r]^{\pi_2} \ar[d]_{\pi_1} & E \ar[d]^\pi\\   D \ar[r]_f & B }}
      
      and that, indeed, with this definition we have placed over $D$ the fibers of $B$ in the sense that $(f^*\hspace*{-.2ex}E)_d\cong E_{f(d)}$ for every $d\in D$.

    \subsubsection*{Trivial line bundle and smooth maps}\trassub
    
      We consider from now on that $M$ is a differentiable $n$-manifold, although most of the concepts and results are valid (with due care and sometimes with restrictions) for infinite dimensional manifolds \cite{marsden1974applications,abraham2012manifolds}.\separ
      
      The \textbf{trivial line bundle}\index{Bundle!Trivial line bundle} is defined simply as $M\times\R\overset{\pi_1}{\longrightarrow}M$.
      
      \begin{definitions}
      	\item A \textbf{smooth function}\index{Smooth function}\index{Section!Smooth function} of $M$ is a section of the trivial line bundle $f\in\Gamma(M\times\R)$.
      	\item We denote $\Cinf{M}=\Gamma(M\times\R)$ the set of all smooth maps $M$. Usually we forget about the base $M$ in the image and consider $g:M\to\R$ by $g=\pi_2\smallcirc f$ instead.
      \end{definitions}
      
    \subsubsection*{Tangent bundle and vector fields}\trassub
    
      The \textbf{tangent bundle}\index{Bundle!Tangent bundle}\index{Tangent bundle} $TM\overset{\pi}{\rightarrow}M$ of $M$ is the disjoint union of the tangent spaces of $M$
      \[TM=\bigsqcup_{p\in M}T_pM=\bigsqcup_{p\in M}\Big\{(p,v_p)\ /\ p\in M\ \ v_p\in T_pM\Big\}\]
      
      \begin{definitions}
      	\item A \textbf{vector field}\index{Section!Vector field}\index{Vector field} of $M$ is a section of the tangent bundle $V\in\Gamma(TM)$ i.e.\ a smooth map $V:M\rightarrow TM$ such that $\pi\smallcirc V=\mathrm{Id}_M$. In particular $V_p\in T_pM$ for every $p\in M$.
      	\item We denote $\mathfrak{X}(M)=\Gamma(TM)$ the set of all vector fields of $M$.
      \end{definitions}
      \centerline{\includegraphics[width=0.42\linewidth,clip,trim=5ex 20ex 5ex 15ex]{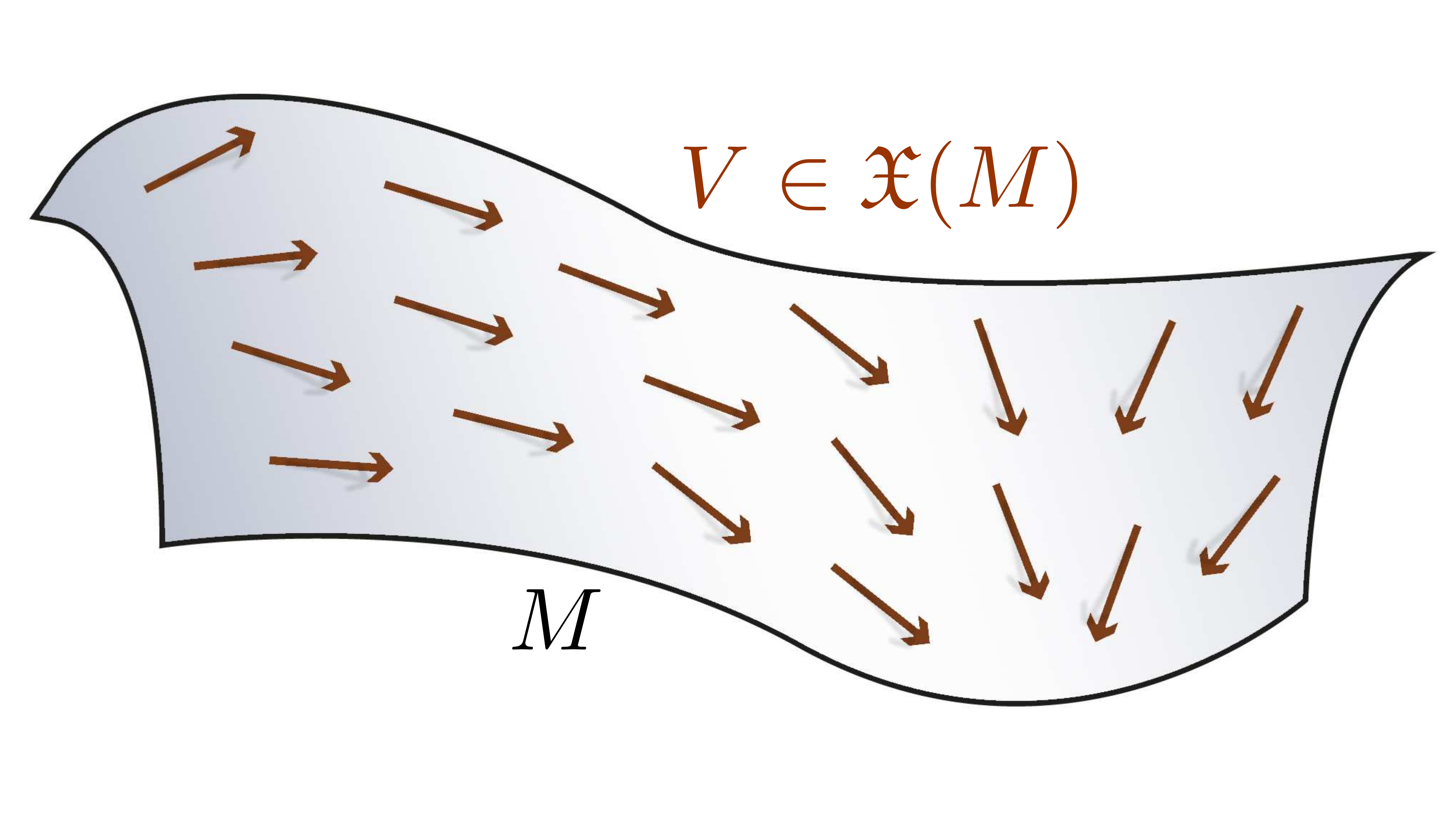}}
      
      A vector field defines an ODE over $M$ whose solution is given by a flow\index{Flow of a vector field}\index{Vector field!Flow} $\phi^V:D\subset\R\times M\to M$. Conversely, given a flow $\phi$ we have an associated vector field given by $V^\phi=\left.\partial_t\right|_{t=0}\phi$.\separ
      
      The flow\index{Vector field!Flow}\index{Flow of a vector field} of a vector field $V$ provides the definition of the directional derivative\index{Directional derivative} of a map
      \[f\smallcirc\phi^V_p:\R\to\R\qquad\qquad V(f)(p)=V_p(f)=\left.\frac{\mathrm{d}}{\mathrm{d}t}\right|_{t=0}f\left(\phi^V_t(p)\right)\]
      which allows, in particular, to consider a vector field as a derivation $V:\Cinf{M}\to\Cinf{M}$.\separ
      
      Finally, notice that in general a vector field cannot be pushed forward nor pulled back. However, if $F:M\rightarrow N$ is a diffeomorphism, then $W=F_*V\in\mathfrak{X}(N)$ where $F_*$ denotes the differential of $F$ i.e.\ $(F_*V)_p=\mathrm{d}_pF(V_p)$. In the following, we will use both notations indistinctively depending on the context.
      
    \subsubsection*{Cotangent bundle and \texorpdfstring{\boldmath$1$}{1}-form fields}\trassub
    
      The \textbf{cotangent bundle}\index{Bundle!Cotangent bundle}\index{Cotangent bundle} $T^*\!M\overset{\pi}{\rightarrow}M$ of $M$ is the disjoint union of the cotangent spaces of $M$
      \[T^*\!M=\bigsqcup_{p\in M}T^*_p\!M=\bigsqcup_{p\in M}\Big\{(p,\alpha_p)\ /\ p\in M\ \ \alpha_p\in T^*_p\!M\Big\}\]
   
      \begin{definitions}
      	\item A \textbf{{\boldmath$1$}-form field}\index{Form field!1-form field}\index{Section!1-form field} of $M$ is a section of the cotangent bundle $\alpha\in\Gamma(T^*\!M)$ i.e.\ a smooth map $\alpha:M\rightarrow T^*\!M$ such that $\pi\smallcirc \alpha=\mathrm{Id}_M$. In particular $\alpha_p\in T^*_p\!M$ for every $p\in M$.
      	\item We denote $\Omega^1(M)$ the set of all $1$-form fields of $M$.
      \end{definitions}
      
      \begin{remarks}
      	\item $T^*\!M$ is the dual bundle\index{Bundle!Dual bundle} of $TM$. Furthermore, $\Omega^1(M)$ can be understood as the ``dual'' of $\mathfrak{X}(M)$ in the sense that $\alpha\in\Omega^1(M)$ is also the map $\alpha:\mathfrak{X}(M)\to\Cinf{M}$ given by \label{remark intro: Omega(M) dual X(M)}\[\alpha(V)(p)=\alpha_p(V_p)\qquad \qquad \alpha_p:T_p^*\!M\to\R\]
      	\item Of course, we have then that $\mathfrak{X}(M)$ is also ``dual'' to $\Omega^1(M)$ because $W\in\mathfrak{X}(M)$ might be considered as the map $W:\Omega^1(M)\to\Cinf{M}$ given by $W(\beta)(p)=\beta_p(V_p)$.\label{remark intro: X(M) dual Omega(M)}
      \end{remarks}
  The prototypical example of a $1$-form field is the differential\index{Differential of a map} of a smooth map $f\in\Cinf{M}$. We denote it as $\mathrm{d}f\in\Omega^1(M)$ and it acts over a vector field $V\in\mathfrak{X}(M)$ as $\mathrm{d}f(V):=V(f)$.\separ
  
  	  It is important to notice that, unlike vector fields, a $1$-form field $\alpha\in\Omega^1(N)$ can always be pulled back through a smooth map $f:M\to N$. Indeed $f^*\alpha\in\Omega^1(M)$ is given by $(f^*\alpha)(V)=\alpha(f_*V)$ for every $V\in\mathfrak{X}(M)$ even though $f_*V$ might not define a vector field (pointwise it defines a vector).
      
    \subsubsection*{Tensor bundle and tensor fields}\trassub
    
    So far we have defined vector fields and $1$-form fields. In order to allow for more general tensor fields, we have to generalize the tangent and cotangent bundle via tensorization. The $(r,s)$-tensor bundle\index{Bundle!Tensor bundle} of a differentiable manifold $M$ is defined as the following tensor bundle
    \[\mathcal{T}^{r,s}M=(TM)^{\otimes r}\bigotimes (T^*\!M)^{\otimes s}\]
    The fiber $\mathcal{T}^{r,s}_pM$ is generated by $v_1\otimes\ldots\otimes v_r\otimes\alpha_1\otimes\ldots\otimes\alpha_s\in(T_pM)^{\otimes r}\bigotimes (T_p^*\!M)^{\otimes s}$.
      
      \begin{definitions}
      	\item An \textbf{{\boldmath$(r,s)$}-tensor field}\index{Tensor field} of $M$ is a section $T\in\Gamma(\mathcal{T}^{r,s}M)$ i.e.\ a smooth map $T:M\rightarrow \mathcal{T}^{r,s}M$ such that $\pi\smallcirc T=\mathrm{Id}_M$.
      	\item We denote $\mathfrak{T}^{r,s}(M)$ the set of all $(r,s)$-tensor fields of $M$.
      \end{definitions}      
      
      This vector bundle includes the three previous ones because $\mathcal{T}^{0,0}M=M\times\R$, $\mathfrak{T}^{0,0}(M)=\Cinf{M}$, $\mathcal{T}^{1,0}M=TM$, $\mathfrak{T}^{1,0}(M)=\mathfrak{X}(M)$, $\mathcal{T}^{0,1}M=T^*\!M$ and $\mathfrak{T}^{0,1}(M)=\Omega^1(M)$.\separ
      
      A tensor field $T$\index{Tensor field}\index{Section!Tensor field} is a field over $M$ assigning an $(r,s)$-tensor $T_p\in\mathcal{T}^{r,s}_pM$ at each $p\in M$ but it can also be understood in several equivalent ways (see remarks \refconchap{remark intro: Omega(M) dual X(M)} and \refconchap{remark intro: X(M) dual Omega(M)}):
  	  \begin{itemize}
  	  	\item  $T:\mathfrak{X}(M)\times\overset{(s)}{\cdots}\times\mathfrak{X}(M)\longrightarrow\mathfrak{X}(M)\otimes\overset{(r)}{\cdots}\otimes\mathfrak{X}(M)$
  	  	\item  $T:\Omega^1(M)\times\overset{(r)}{\cdots}\times\Omega^1(M)\longrightarrow\Omega^1(M)\otimes\overset{(s)}{\cdots}\otimes\Omega^1(M)$
  	  	\item $T:\Omega^1(M)\times\overset{(r)}{\cdots}\times\Omega^1(M)\times\mathfrak{X}(M)\times\overset{(s)}{\cdots}\times\mathfrak{X}(M)\longrightarrow\Cinf{M}$ which is $\Cinf{M}-$multilinear.
  	  \end{itemize}
      
    \subsubsection*{Wedge bundle and \texorpdfstring{\boldmath$n$}{n}-form fields}\trassub
    
      We now define the \textbf{wedge bundle}\index{Bundle!Wedge bundle}\index{Wedge bundle} as the $s$-th exterior power of the cotangent bundle
      \[\Lambda^s(T^*\! M)=\bigsqcup_{p\in M}\Lambda^s(T_p^* M)\]
      The fiber $\Lambda^s(T_p^*M)$ is formed by linear combinations of $\alpha_1\wedge\ldots\wedge\alpha_s\in\Lambda^s(T^*_p\!M)$ where
      \begin{equation}\alpha_1\wedge\dots\wedge \alpha_s:=\frac{1}{s!}\sum_{\sigma\in S_p}(-1)^\sigma\alpha_{\sigma(1)}\otimes\dots\otimes\alpha_{\sigma(s)}
      \end{equation}
      \begin{definitions}
      	\item An \textbf{{\boldmath$s$}-form field}\index{Form field!k-form field}\index{Section!k-form field} of $M$ is a section $\alpha\in\Gamma(\Lambda^s(T^*\! M))$ i.e.\ a smooth map $\alpha:M\rightarrow \Lambda^s(T^*\!M)$ such that $\pi\smallcirc \alpha=\mathrm{Id}_M$.
      	\item We denote $\Omega^s(M)$ the set of all $s$-form fields of $M$. 
      	\item If $\alpha\in\Omega^s(M)$ we say that $|\alpha|:=s$ is the \textbf{degree}\index{k-form field!Degree} of $\alpha$.
      	\item A top dimensional form which is nowhere vanishing is known as \textbf{volume form}\index{Volume form}\index{k-form field!Volume form}. We denote
      	\[\mathrm{Vol}(M)=\Big\{\omega\in\Omega^n(M)\ \ /\ \ \omega_p\neq 0 \text{ for every }p\in M\Big\}\] 
      \end{definitions}
      \begin{remark}\label{Mathematical background - remark - alpha=fw}\mbox{}\\
  	   The modulo $\Omega^s(M)$ has $\binom{n}{s}$ local generators,  in particular $\Omega^n(M)$ has only $1$. This implies that if $\omega$ is a volume form and $\alpha\in\Omega^n(M)$, then $\alpha=f\omega$ for some $f\in\Cinf{M}$.
  	  \end{remark}
      
      An $s$-form field $\alpha:M\to\Lambda^s(T^*\!M)$ can be understood as an antisymmetric multiliniear map
      \[\alpha:\mathfrak{X}(M)\times\overset{(s)}{\cdots}\times\mathfrak{X}(M)\longrightarrow\Cinf{M}\]
      Being antisymmetric implies, in particular, that $\alpha_p$ is zero if it is evaluated over $k$ vectors that do not span a $k$-dimensional subspace (as they would be linearly dependent). Actually, an $s$-form at $p\in M$ gives a volume reference over the $s$-vector spaces of $T_pM$. Of course, a volume form assigns a volume reference at each point of the manifold itself.\separ

      Finally, we have the wedge product\index{Wedge product} $\wedge:\Omega^s(M)\times\Omega^r(M)\to\Omega^{s+r}(M)$, which is bilinear and satisfies
      \begin{align}
        &\alpha\wedge \beta=(-1)^{|\alpha||\beta|}\beta\wedge \alpha\\
        &f^*(\alpha\wedge \beta)=(f^*\alpha)\wedge(f^*\beta)\qquad f:M\to N 
      \end{align}
      
  \subsection{Maps over vector bundles}
  
  Once we have defined the arena where our geometric objects live, we proceed with the description of several maps that we can apply to such objects.
  
    \subsubsection*{Contraction}\trassub
    
      We have seen that $\Omega^1(M)$ is dual to $\mathfrak{X}(M)$, so we can define the natural pairing between a $1$-form field and a vector field.
      
      \begin{lemma}\mbox{}\\
      	There exists a unique map $C:\mathfrak{T}^{1,1}(M)\rightarrow\mathfrak{T}^{0,0}(M)=\Cinf{M}$ such that $C(X\otimes\alpha)=\alpha(X)$ for every $X\in\mathfrak{X}(M)$ and for every $\alpha\in\Omega^1(M)$.
      \end{lemma}
      
      	We define the \textbf{{\boldmath$(a,b)$}-contraction} as the map\index{Contraction} $C^a_b:\mathfrak{T}^{r,s}(M)\longrightarrow\mathfrak{T}^{r-1,s-1}(M)$ given by
      	\[(C^a_bT)(\alpha_1,\ldots,\alpha_{r-1},X_1,\ldots,X_{s-1})=C\Big(T(\alpha_1,\ldots,\overset{\left.a\right)}{\cdot{\tikzmark{a}}},\ldots,\alpha_{r-1},X_1,\ldots,\overset{\left.b\right)}{\cdot{\tikzmark{b}}},\ldots,X_{s-1})\Big)\tikz[overlay,remember picture]{\draw[-,red,square arrow] (a.south) to (b.south);}\]
      \mbox{}\vspace*{-1.2ex}
      
      Notice that for the contraction, no metric is needed, but we can only contract one covariant index with one contravariant. The metric precisely allows us to contract two covariant or two contravariant indices by raising or lowering one of them (see section \ref{Subsection - Semiriemannian geometry}), and then contracting. 
      
    \subsubsection*{Lie Derivative}\trassub
    
      We have seen that a vector field $V$ defines a flow $\phi^V:D\subset\R\times M \to M$ over $M$. It is then natural to think about how objects defined over the manifold vary when we move along with the flow. Physically, it is like navigating a river and study how some physical quantities, like the temperature or the wind, vary (equivalently, we might consider that the objects are dragged by the river while we stand at a fixed position). A quantity like that is modeled by a tensorial field $R\in\mathfrak{T}^{r,s}(M)$ and, in order to compare it with the initial value $R_p\in T^{r,s}_pM$, we need to pull it back to the initial fiber. This suggests the definition of the \textbf{Lie derivative}\index{Lie derivative} along the vector field $V\in\mathfrak{X}(M)$ as
      \begin{equation}\label{def eq Lie deriv}\peqsub{\mathcal{L}}{V}R=\left.\frac{\mathrm{d}}{\mathrm{d}t}\right|_{t=0}(\phi^V_{-t})^*R\qquad\equiv\qquad(\peqsub{\mathcal{L}}{V}R)_p=\lim_{t\rightarrow0}\frac{(\phi^V_{-t})_{\phi_p^V(t)}^*R_{\phi_p^V(t)}-R_{p}}{t}\in\mathcal{T}^{r,s}_pM
      \end{equation}
      Notice that the Lie derivative $\peqsub{\mathcal{L}}{V}:\mathfrak{T}^{r,s}(M)\to\mathfrak{T}^{r,s}(M)$ is a derivation of degree $0$, in particular
      \begin{align*}
        \peqsub{\mathcal{L}}{V}(T\otimes R)=(\peqsub{\mathcal{L}}{V}T)\otimes R+T\otimes(\peqsub{\mathcal{L}}{V}R)
      \end{align*}
      
      Notice that for smooth maps $\peqsub{\mathcal{L}}{V}f=V(f)$. Meanwhile, the action of the Lie derivative over vector fields, also known as the \textbf{Lie bracket}\index{Lie bracket}, has a very nice visual interpretation. Namely, $[V,W]:=\peqsub{\mathcal{L}}{V}W$ measures the difference between the path $\phi^V_\varepsilon\smallcirc\phi^W_\varepsilon$ and the path $\phi^W_\varepsilon\smallcirc\phi^V_\varepsilon$ in the limit $\varepsilon\to0$. It is well known that $(\mathfrak{X}(M),[\,,\hspace*{.1ex}])$ forms a Lie algebra\index{Lie algebra}.\label{Lie alegbra with Lie bracket}
      
      \begin{figure}[H]
      	\centering\includegraphics[width=1\linewidth]{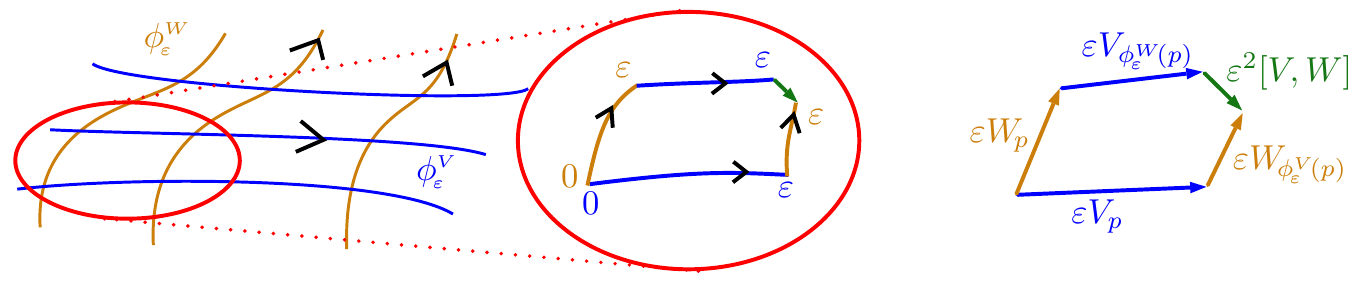}
      	\caption{On the left we represent the flow of two vector fields $V,W$. The zoom-in shows that the flows do not commute at order $\varepsilon$. On the right we have the infinitesimal version (we have included the $\varepsilon$, which is usually omitted, to emphasize the order of the lengths).}\label{Mathematical background - image - corchete de lie flujos}
      \end{figure}
      
      Finally, notice that $\peqsub{\mathcal{L}}{V}\omega\in\Omega^n(M)$ for a given volume form $\omega$. Now remark \ref{Mathematical background - remark - alpha=fw} tells that there exists a function, denoted $\mathrm{div}_\omega V$ and called \textbf{divergence of \bm{$V$}}\index{Divergence}, such that $\peqsub{\mathcal{L}}{V}\hspace*{0.2ex}\omega=\mathrm{div}_\omega V\cdot{}\omega$. The divergence measures the infinitesimal change of the volume $\omega$ in the direction of the vector field.\label{Mathematical background - definition - divergence}

    \subsubsection*{Interior and exterior derivative}\trassub
    
      Given a vector field $V\in\mathfrak{X}(M)$ we define the \textbf{interior derivative with {\boldmath$V$}}\index{Interior derivative} as the derivation $\peqsub{\imath}{V}:\Omega^k(M)\to\Omega^{k-1}(M)$ of degree $-1$ given by
      \begin{equation}
        \peqsub{\imath}{V}\beta(W_1,\ldots,W_{k-1})=\beta(V,W_1,\ldots,W_{k-1})
      \end{equation}
      Notice that $(\peqsub{\imath}{V})^2=0$. Furthermore, as it is a derivation we have
      \begin{equation}
        \peqsub{\imath}{V}(\alpha\wedge\beta)=(\peqsub{\imath}{V}\alpha)\wedge\beta+(-1)^{|\alpha|}\alpha\wedge(\peqsub{\imath}{V}\beta)
      \end{equation}
      Let us now define the \textbf{exterior derivative}\index{Exterior derivative}, a derivation of degree $1$,  $\mathrm{d}_k:\Omega^k(M)\to\Omega^{k+1}(M)$. In order to do that let us first consider the boundary operator $\partial_{k+1}$ that takes a $(k+1)$-submanifold $N$ of the $n$-manifold $M$, and returns the $k$-manifold that corresponds to the topological boundary $\partial N\subset M$. Now, the differential $\mathrm{d}$ can be thought as the ``dual'' of $\partial$ in the sense that
      \[\mathrm{d}_k\beta(N)=\beta(\partial_{k+1} N)\qquad\quad\equiv\quad\qquad \mathrm{d}\beta\left(\includegraphics[height=5.6ex,valign=c]{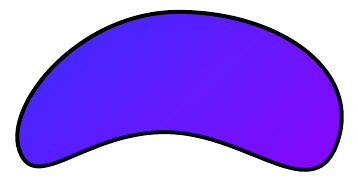}\right)=\beta\left(\includegraphics[height=5.6ex,valign=c]{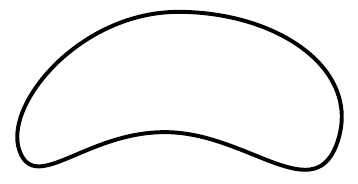}\right)\]
      With this sketchy definition we see already two important properties: $\mathrm{d}^2=0$ and $\mathrm{d}_n=0$. The latter is obvious because there is no $(n+1)$-submanifold of the $n$-manifold $M$. The former follows from the fact that the boundary of a manifold has no boundary, therefore $\partial^2=0$.\separ
      
      The above equation is somehow a ``macroscopic'' one and we need an infinitesimal definition because the forms, as we have defined them previously, act on vectors and not on manifolds. To do so, we have to intuitively replace the $(k+1)$-submanifold $N$ with an ``infinitesimal version'' around each point. This can be achieved by considering $k+1$ small vectors of order $\varepsilon$ and extending using the flow of the rest to build a $(k+1)$-dimensional submanifold $P$. Then $\partial P$ is formed by the $(k+1)$ faces formed by the $k$ vectors, the opposite ones and the faces required to ``close'' the manifold (see the right diagram in figure \ref{Mathematical background - image - corchete de lie flujos}). Let us see how it works for $k=1$ at $p\in M$.
      \begin{align*}
        \mathrm{d}\beta(V,W)&=\frac{1}{\varepsilon^2}\mathrm{d}\beta(\varepsilon V,\varepsilon W)\approx\frac{1}{\varepsilon^2} \mathrm{d}\beta\left(\includegraphics[clip,trim=2.25ex 2ex 2ex 2.5ex,height=14ex,valign=c]{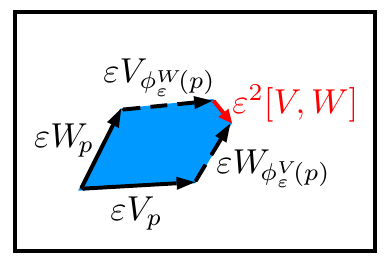}\right)=\frac{1}{\varepsilon}\beta\left( \includegraphics[clip,trim=3.6ex 2.4ex 1.7ex 2.75ex,height=12ex,valign=c]{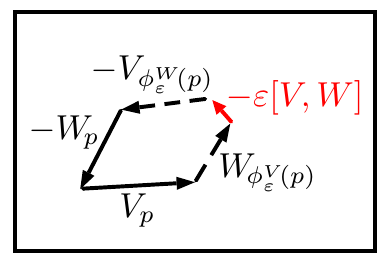}\right)=\\
        &=\frac{1}{\varepsilon}\beta\left( \includegraphics[clip,trim=4.2ex 1.5ex 10ex 1.5ex,height=14ex,valign=c]{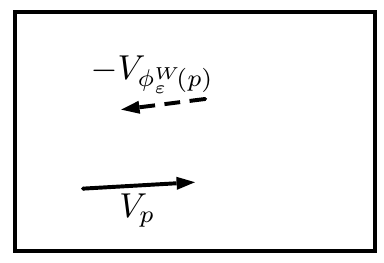}\right)+\frac{1}{\varepsilon}\beta\left( \includegraphics[clip,trim=1.5ex 2.4ex 4.6ex 5.2ex,height=9.5ex,valign=c]{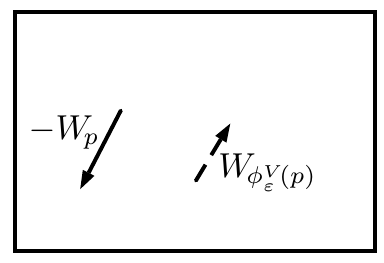}\right)-\beta\Big([V,W]\Big)=\\
        &=-\frac{1}{\varepsilon}\Big[\beta_{\phi_\varepsilon^W\!(p)}\big(V_{\phi_\varepsilon^W\!(p)}\big)-\beta_p\big(V_p\big)\Big]+\frac{1}{\varepsilon}\Big[\beta_{\!\phi_\varepsilon^V\!(p)}\big(W_{\!\phi_\varepsilon^V\!(p)}\big)-\beta_p\big(W_{\!p}\big)\Big]-\beta\Big([V,W]\Big)\approx\\[.8ex]
        &\approx\lim_{\varepsilon\to 0}\frac{[\beta\big(W)]_{\phi_\varepsilon^V\!(p)}-[\beta\big(W)]_p}{\varepsilon}-\lim_{\varepsilon\to 0}\frac{[\beta\big(V)]_{\phi_\varepsilon^W\!(p)}-[\beta\big(V)]_p}{\varepsilon}-\beta\Big([V,W]\Big)=\\[.8ex]
        &=V\Big(\beta(W)\Big)-W\Big(\beta(V)\Big)-\beta\Big([V,W]\Big)
      \end{align*} 
      We could reproduce this computation for any $k$-form noticing that, in order to close higher dimensional manifolds, more faces are needed. Nonetheless, only the Lie brackets will be of order $\varepsilon^2$ while the rest would be of higher order thus, in the limit, go to zero. This leads to the formula
      \begin{align}\label{equation d_k}\begin{split}
        \mathrm{d}\beta(V_1,\ldots,V_{k+1})&=\sum_{j=1}^{k+1}(-1)^{j-1}V_j\Big(\beta(V_1,\ldots,\widehat{V}_j,\ldots,V_{k+1})\Big)+\\
        &\phantom{=}\ \ +\sum_{i<j}(-1)^{i+j}\beta\Big([V_i,V_j],V_1,\ldots,\widehat{V}_i,\ldots,\widehat{V}_j,\ldots,V_{k+1}\Big)
      \end{split}\end{align}
      \begin{properties}
      	\item $\mathrm{d}(\alpha\wedge\beta)=(\mathrm{d}\alpha)\wedge\beta+(-1)^{|\alpha|}\alpha\wedge(\mathrm{d}\beta)$
      	\item Acting over forms we have $\peqsub{\mathcal{L}}{V}=\mathrm{d}\peqsub{\imath}{V}+\peqsub{\imath}{V}\mathrm{d}$, known as Cartan's magic formula.
      	\item Also, over forms, we have $\peqsub{\mathcal{L}}{fV}\beta=f\peqsub{\mathcal{L}}{V}\beta+\mathrm{d}f\wedge\peqsub{\imath}{V}\beta$ for $f\in\Cinf{M}$.
      	\item $\mathrm{d}$ commutes with pullbacks and the Lie derivative i.e.\ $\mathrm{d}f^*=f^*\mathrm{d}$ and $\peqsub{\mathcal{L}}{V}\mathrm{d}=\mathrm{d}\peqsub{\mathcal{L}}{V}$.
      \end{properties}
      
    \subsubsection*{Integration}\index{Integration}\trassub
    
      We have already mentioned how an $n$-form $\omega\in\Omega^n(M)$ over the $n$-dimensional manifold $M$ defines a reference volume over each tangent space $T_pM$. We would like to measure the total volume of $M$ with respect to $\omega$
      \[\omega(M)\qquad\text{ or equivalently }\qquad\int_M\omega\]
      The intuitive idea of the procedure is simple. First split $M$ into tiny pieces, each of which can be approximated by $n$-cubes $C_\alpha$ built using small vectors. Evaluate $\omega_p(C_\alpha)\in\R$ and then add up all together. This is an approximation which can be refined by taking smaller pieces (hence more of them). In the limit, we obtain the desired integral.
      
      \centerline{\includegraphics[clip,trim=10ex 18ex 8ex 16ex,height=0.139\linewidth]{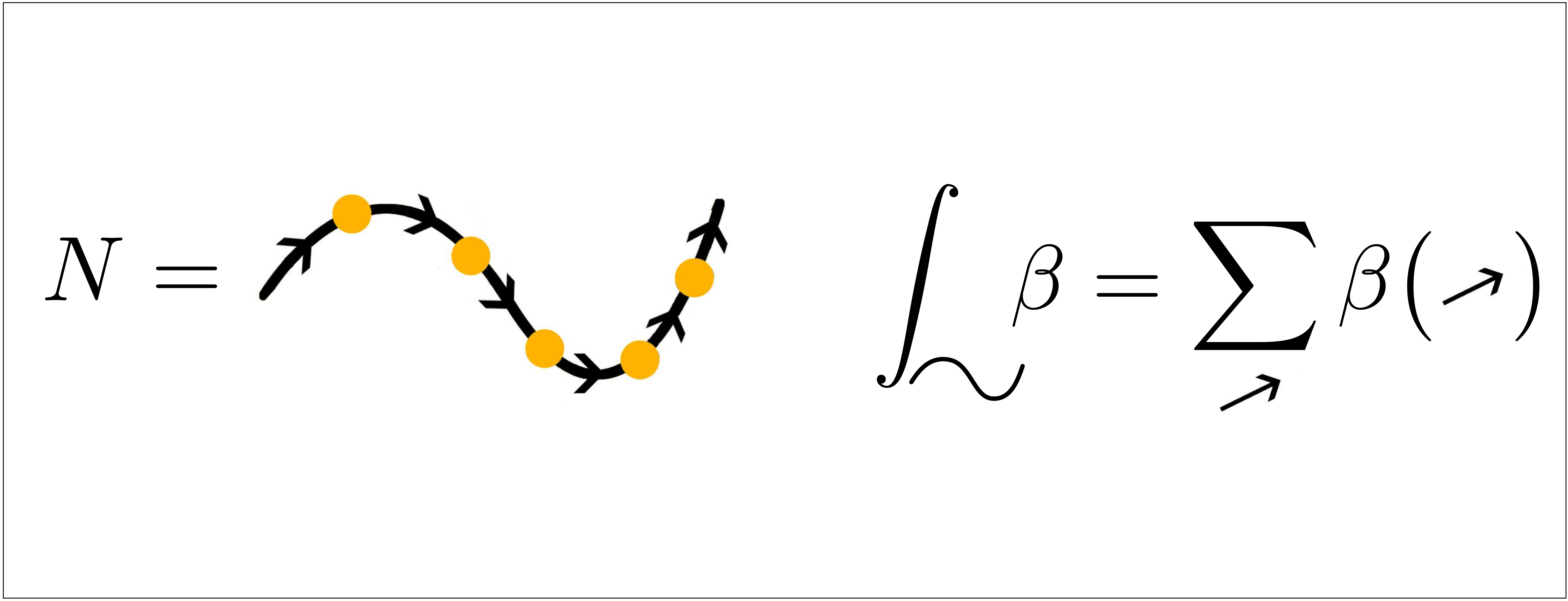}\hfill\includegraphics[clip,trim=10ex 18ex 7ex 16ex,height=0.139\linewidth]{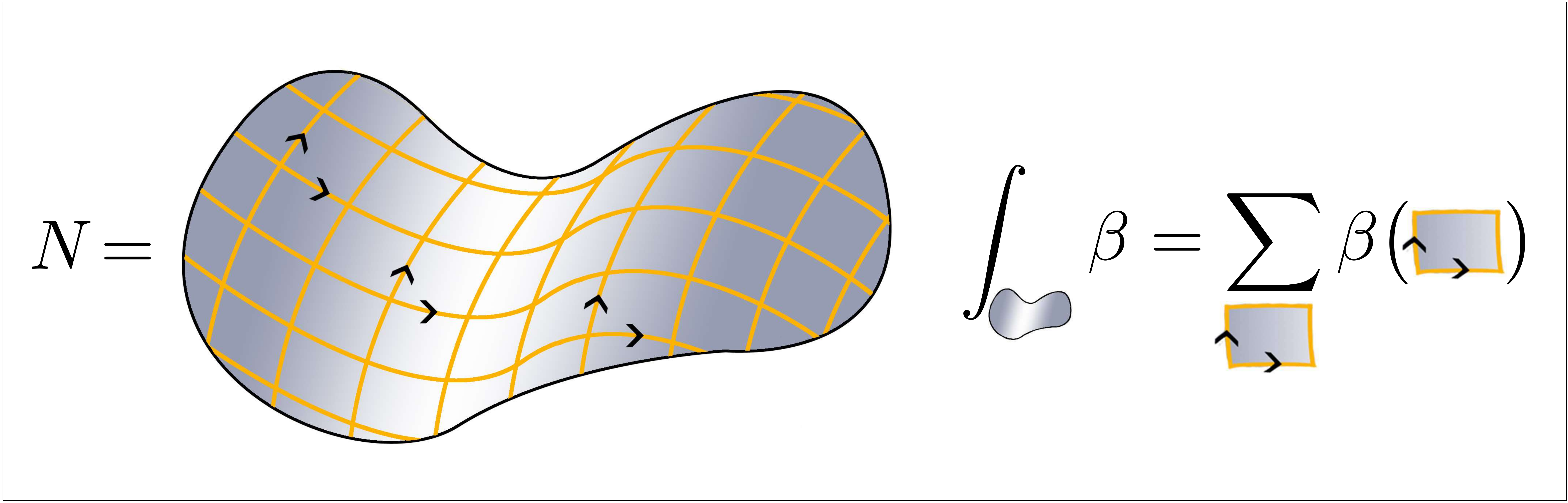}}
      
      Notice that for $\omega_p(C_\alpha)$ to be well defined, an orientation has to be fixed i.e.\ a non-vanishing $n$-form. Then we can take all the $C_\alpha$ to be spanned by a positive basis.\separ
      
      To integrate over a manifold it is enough to know how to do it locally. Let us consider $M=\R^n$ and $\omega\in\Omega^n(\R^n)$. As $\mathrm{d}x_1\wedge\cdots\wedge \mathrm{d}x_n\in\mathrm{Vol}(\R^n)$ we know from remark \ref{Mathematical background - remark - alpha=fw} that there exists some $f\in\Cinf{M}$ such that $\omega=f\mathrm{d}x_1\wedge\cdots\wedge \mathrm{d}x_n$. We thus define
      \[\int_M\omega:=\int_{\R^n}f(x_1,\ldots,x_n)\mathrm{d}x_1\cdots \mathrm{d}x_n\]
      where we use the standard Lebesgue integral of $\R^n$. Now, if we consider a general oriented manifold $M$ and a chart $\varphi:U\subset\R^n\to M$, we would like to have
      
      \vspace{1ex}
      
      \centerline{\includegraphics[width=0.99\linewidth]{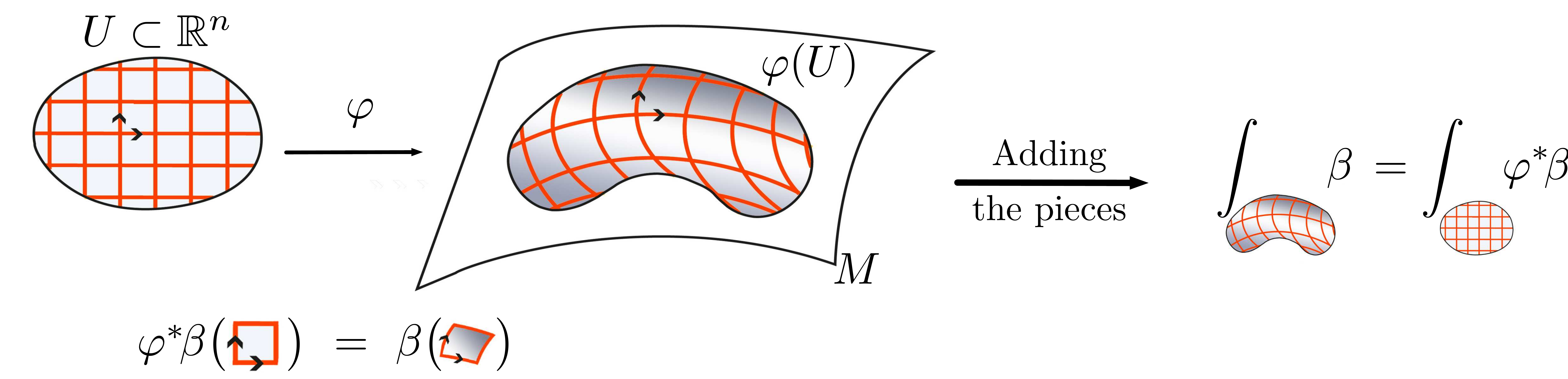}}
      
      \vspace{1.5ex}
      
      The final definition of the integral relies on the existence of partitions of unity which allows us to formalize the aforementioned sketchy procedure (in order to avoid convergence issues, the $n$-forms are taken with compact support).
      
    \subsubsection*{Stokes' theorem}\trassub
    
      We saw when we defined the differential $\mathrm{d}$ that, given $\beta\in\Omega^s(M)$, we have locally the ``formula''
      
      \vspace*{.5ex}
      
      \centerline{\includegraphics[width=0.22\linewidth]{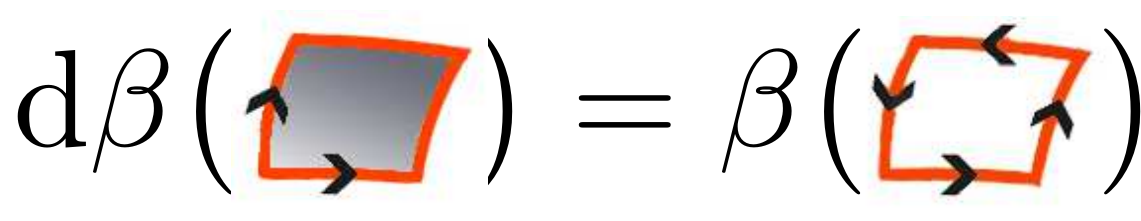}}
      
      If we now consider a manifold to be split into small pieces, the common boundaries cancel out and we obtain a global formula
      
      \centerline{\includegraphics[clip,trim=20ex 20ex 10ex 10ex,width=.99\linewidth]{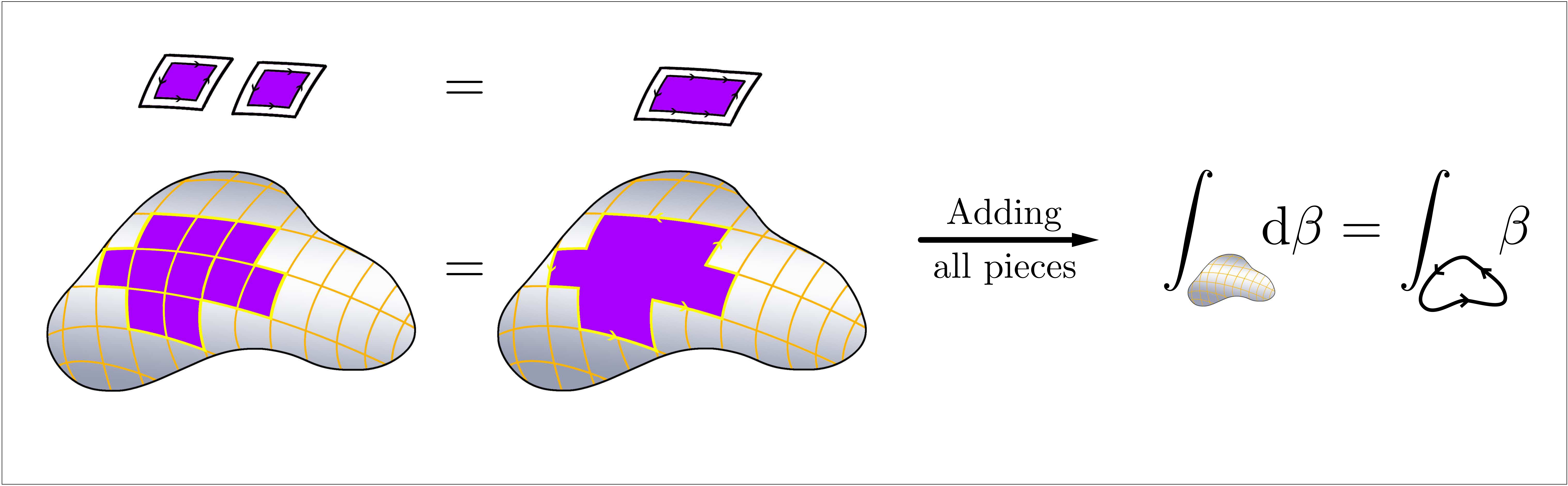}}
      
      This idea is formalized by the Stokes' theorem\index{Stokes' theorem}.
      \begin{theorem}\label{Mathematical background - theorem - stokes}\mbox{}\\
       	Given an oriented $n$-dimensional manifold $M$ and $\alpha\in\Omega^{n-1}(M)$ with compact support, then
       	\[\int_M \mathrm{d}\alpha=\int_{\partial M}\peqsubfino{\jmath}{\partial}{-0.2ex}^*\alpha\]
       	where $\peqsubfino{\jmath}{\partial}{-0.2ex}:\partial M\hookrightarrow M$ and $\partial M$ (possibly empty) is endowed with the induced orientation of $M$.
      \end{theorem}      
         
    \subsubsection*{Connection on a vector bundle}\trassub
    
      Over each fiber of a vector bundle we can add vectors or multiply them by scalars. However, there is no way to compare objects corresponding to two different fibers. To do so we need to prescribe how to ``parallel'' transport objects from a fiber along a curve $\gamma:I\to B$ joining the base points of the two fibers.  One might think that the Lie derivative could do the trick but it does not because to compute $\peqsub{\mathcal{L}}{V}T$ we need $V$ to be defined over a small neighborhood of $\gamma$. This is illustrated, for instance, by the fact that
      \[\left.
      \partial_x\right|_{\{y=0\}}=\left.(y+1)\partial_x\right|_{\{y=0\}}\qquad\text{ but }\qquad\left.\peqsub{{\mathcal{L}_{\,\partial}}}{\!x}\partial_y\right|_{\{y=0\}}=0\neq -1=\left.\peqsub{{\mathcal{L}_{\,(y+1)\partial}}}{\!x}\partial_y\right|_{\{y=0\}}\]
      
      \begin{wrapfigure}{r}{4.5cm}
      	\centering
      	\vspace*{-3.5ex}
      	\includegraphics[width=1\linewidth]{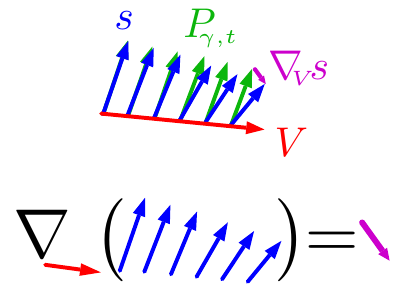}
      \end{wrapfigure}
      Given a vector bundle $\pi:E\to M$, let us consider that we have defined a parallel transport\index{Parallel transport} along any curve $\gamma$, i.e.\ an isomorphism $P_{\gamma,t}:E_{\gamma(0)}\to E_{\gamma(t)}$. Then, it is easy to define a directional derivative\index{Directional derivative} $\peqsub{\nabla}{\!V}s$ of a section $s\in\Gamma(E)$ in the direction $V\in\mathfrak{X}(M)$ as the infinitesimal parallel transportation
      \[(\peqsub{\nabla}{\!V} s)_p=\lim_{t\rightarrow0}\frac{(P_{\gamma,t})^{-1}(s_{\gamma(t)})-s_p}{t}\]
      where $\gamma$ is any curve with $\gamma(0)=p$ and $\dot{\gamma}(0)=V_p$. This map, given a direction and a section along such direction, gives its deviation with respect to the parallel transport.\separ
      
      Conversely, given a directional derivative we can define a parallel transport by integration. It is thus equivalent and it turns out to be much more convenient to work with the former.\vspace*{.8ex}
   
      \begin{definition}\mbox{}\\
        A \textbf{connection on {\boldmath$E\overset{\pi}{\rightarrow}M$}}\index{Connection} is an $\R$-bilinear map $\nabla:\mathfrak{X}(M)\times\Gamma(E)\rightarrow\Gamma(E)$ such that
      	\[\peqsub{\nabla}{\!fV}s=f\peqsub{\nabla}{\!V}s\qquad\qquad\peqsub{\nabla}{\!V}(fs)=V(f)s+f\peqsub{\nabla}{\!V}s\]
      	for every $f\in\Cinf{M}$, $V\in\mathfrak{X}(M)$ and every $s\in\Gamma(E)$.
      \end{definition}
      
      Sometimes it is more useful to consider the ``dual'' map $\nabla:\Gamma(E)\rightarrow\Omega^1(M)\otimes\Gamma(E)$ which is usually called the \textbf{exterior covariant derivative}\index{Exterior covariant derivative} \cite{binz2011geometry,gockeler1989differential} and denoted as
      \[\begin{array}{cccc}
        \peqsub{\mathrm{d}}{\nabla}: & \Omega^0(E):=\Gamma(E) & \longlongrightarrow{6} & \Omega^1(E):=\Omega^1(M)\otimes \Gamma(E)\\[1ex]
                           & s               &  \longlongmapsto{6}  & \left(\rule{0ex}{4ex}\begin{array}{cccc}\peqsub{\mathrm{d}}{\nabla}(s):&\mathfrak{X}(M)&\longrightarrow&\Gamma(E)\\[.6ex] &V&\longmapsto&\peqsub{\mathrm{d}}{\nabla}(s)(V)=\peqsub{\nabla}{\!V}s\end{array}\right)
          \end{array}\]
      where $\Omega^1(E)$ is the set of $1$-forms with values at $E$, meaning that instead of functions (as in $\alpha=3x\mathrm{d}x+y^2\mathrm{d}y$) the coefficients are sections of $E$. Analogously to equation \eqref{equation d_k}, we extend its definition to all degrees $\peqsub{\mathrm{d}}{\nabla}:\Omega^k(E)\to\Omega^{k+1}(E)$. It is easy to see that the Leibniz rule holds
      \[\peqsub{\mathrm{d}}{\nabla}(\alpha\otimes s)=\mathrm{d}\alpha\otimes s+(-1)^{|\alpha|}\alpha\wedge\peqsub{\mathrm{d}}{\nabla}s\in\Omega^{k+1}(E)\qquad \alpha\in\Omega^k(M),\ s\in\Gamma(E)\]
      In the case of the trivial bundle with the trivial connection, we recover the usual exterior derivative.\vspace*{.8ex}      
      
      \begin{remarks}
      	\item There exist a local basis of sections $\mathcal{B}=\{e_1\ldots e_n\}$ and a $\mathcal{B}$-dependent matrix of $1$-forms $A^{\nabla}=(\tensor{A}{_j^k})_{j,k}\in \mathfrak{gl}_n(\Omega^1(M))$ such that
      	\[ \peqsub{\mathrm{d}}{\nabla}e_j=\tensor{A}{_j^k}\otimes e_k\qquad\text{ where }\qquad\tensor{A}{_j^k}=\sum_{i=1}^n\tensor{A}{_i_j^k}\mathrm{d}x^i\in\Omega^1(M)\]
      	The coefficients $\tensor{A}{_i_j^k}$ are known as the \textbf{symbols of the connection}\index{Symbols of the connection}.
        \item Computing $\peqsub{\mathrm{d}}{\nabla}(s^ae_a)$ with the Lebniz rule and the previous formula we see that, locally for all degrees, $\peqsub{\mathrm{d}}{\nabla}=\mathrm{d}+A^\nabla\wedge$.
        \item We can extend $\peqsub{\mathrm{d}}{\nabla}$ for more general tensor fields. For instance, considering $\Gamma(E)\otimes\Gamma(E)^*$ and computing $\peqsub{\mathrm{d}}{\nabla}(\tensor{s}{^a_b}e_a\otimes e^b)$, we get that locally $\peqsub{\mathrm{d}}{\nabla}=\mathrm{d}+A^\nabla\!\wedge-\wedge A^\nabla=\mathrm{d}+[A^\nabla,A^\nabla]$.\label{Mathematical background - remark - d_nabla local R}
      \end{remarks}

    \subsubsection*{Curvature of a connection}\trassub
    
      We know that $\mathrm{d}^2=0$ but this is not necessarily true for $\peqsub{\mathrm{d}}{\nabla}$. Indeed, locally we have for $s\in\Gamma(E)$
      \begin{align*}
        \peqsub{\mathrm{d}}{\nabla}^2s&=(\mathrm{d}+A^\nabla\!\wedge)(\mathrm{d}+A^\nabla\!\wedge)s=(\mathrm{d}+A^\nabla)(\mathrm{d}s+A^\nabla\!\wedge s)=\\
        &=0+(\mathrm{d}A^\nabla)\wedge s+(-1)^{|A|}A^\nabla\!\wedge \mathrm{d}s+A^\nabla\!\wedge\mathrm{d}s+A^\nabla\!\wedge A^\nabla\!\wedge s=\\
        &=(\mathrm{d}A^\nabla+A^\nabla\!\wedge A^\nabla)\wedge s=:F^\nabla\!\!\wedge s
      \end{align*}
      \setlength{\figwidth}{0.4\linewidth}
      \setlength{\figwidthleft}{\dimexpr \linewidth-\figwidth}
      \null\smash{\raisebox{-\dimexpr 7.8\baselineskip+\parskip\relax}%
      	{\includegraphics[width=.355\linewidth]{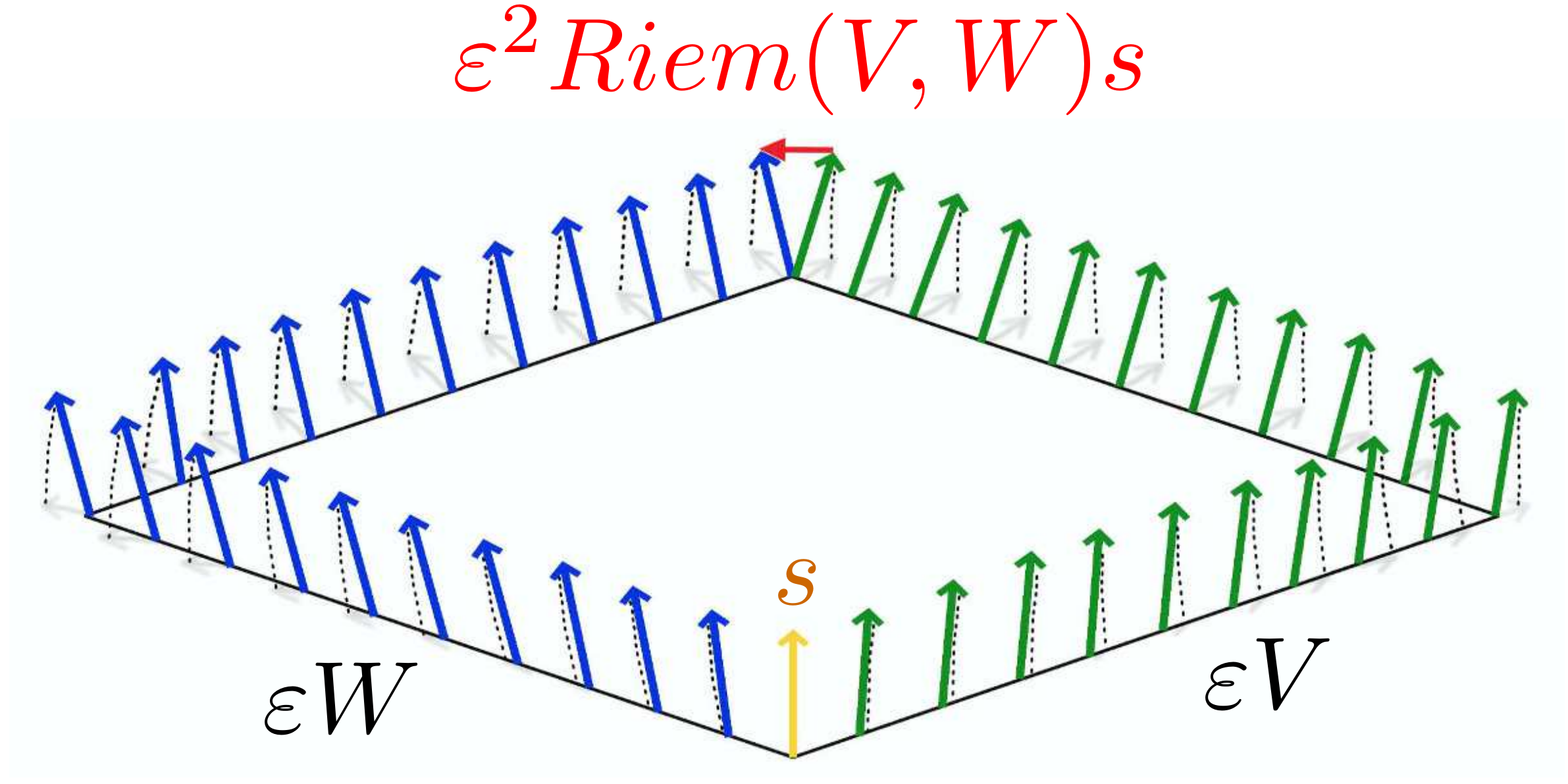}}}\strut\hfill \\[-\dimexpr 2\baselineskip+2\parskip\relax]
      
      \parshape=7 
      0pt \linewidth 
      0pt \linewidth 
      \figwidth \figwidthleft
      \figwidth \figwidthleft
      \figwidth \figwidthleft
      \figwidth \figwidthleft   
      0pt \linewidth 
      where we define $F^\nabla\in\Omega^2(M)\otimes \Gamma(E)\otimes\Gamma(E)^*$ which can be considered as a matrix of two forms. It is called the \textbf{curvature of the connection}\index{Curvature!Riemann}\index{Connection!Curvature} and measures the deviation of the parallel transport around a small square. Notice that on the figure of the left the parallelogram might not close, but this introduces negligible errors of order $\varepsilon^3$. Taking now into account \refconchap{Mathematical background - remark - d_nabla local R} we can obtain the (second) Bianchi identity\index{Bianchi identity!Second Bianchi identity}
      \begin{equation}\label{Mathematical background - equation - bianchi equation}
      \hspace*{22ex}\peqsub{\mathrm{d}}{\nabla}F^\nabla=0
      \end{equation}

      The analog of \eqref{equation d_k} tells that $\peqsub{\mathrm{d}}{\nabla}^2$ is precisely the curvature $\riemann^{\!\nabla}:\mathfrak{X}(M)\times\mathfrak{X}(M)\times\Gamma(E)\rightarrow \Gamma(E)$
      \begin{align*}
      	 \peqsub{\mathrm{d}}{\nabla}^2s(V,W)&=\peqsub{\mathrm{d}}{\nabla}(\peqsub{\mathrm{d}}{\nabla}s)(V,W)=\peqsub{\nabla}{\!V}(\peqsub{\mathrm{d}}{\nabla}s(W))-\peqsub{\nabla}{\!W}(\peqsub{\mathrm{d}}{\nabla}s(V))-\peqsub{\mathrm{d}}{\nabla}s[V,W]=\\
      	 &=\peqsub{\nabla}{\!V}\peqsub{\nabla}{\!W}s-\peqsub{\nabla}{\!W}\peqsub{\nabla}{\!V}s-\peqsub{\nabla}{\![V,W]}s=:\riemann^{\!\nabla}(V,W)s
      \end{align*}
      If $\riemann^{\!\nabla}\equiv0$ then we say that the connection $\nabla$ is \textbf{flat}\index{Connection!Flat}. Alternatively, the curvature can be considered as the obstruction for the vanishing of $\peqsub{\mathrm{d}}{\nabla}^2$.
      
    \subsubsection*{Ricci curvature of a connection}\trassub
    
      Let us consider now the vector bundle $E=TM$. We have that the Riemann tensor $\riemann^{\!\nabla}$ is a $(3,1)$-tensor field and it can be proved that it encodes all the curvature of the manifold. It is easier, and physically very relevant as we will see, to study its trace to obtain what is known as the \textbf{Ricci curvature tensor}\index{Ricci curvature}\index{Curvature!Ricci} $\mathrm{Ric}^{\!\nabla}=C^1_2(\riemann^{\!\nabla})\in\mathfrak{T}^{0,2}(M)$. Equivalently we can write
      \[\mathrm{Ric}^{\nabla}(V,W)=\mathrm{Tr}\Big(Y\mapsto \riemann^{\!\nabla}(V,Y)W\Big)\qquad\quad\equiv\quad\qquad \mathrm{Ric}_{ab}=\tensor{Riem}{^c_a_c_b}\]
      The Ricci curvature is a symmetric tensor field.
      
    \subsubsection*{Torsion of a connection}\trassub  
    
\setlength{\figwidth}{0.65\linewidth}
\null\hfill\smash{\raisebox{-\dimexpr 7.5\baselineskip+\parskip\relax}%
	{\includegraphics[width=.32\linewidth]{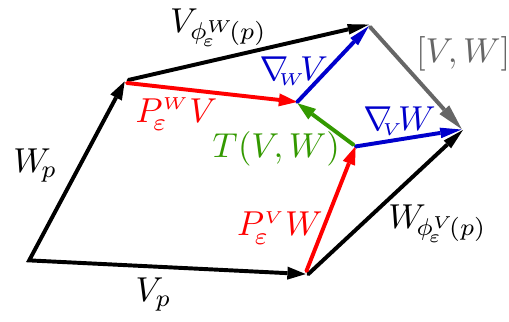}} \,}\strut \\[-\dimexpr 2.95\baselineskip+2\parskip\relax]

\parshape=10 
0pt \linewidth 0pt \linewidth 
0pt \figwidth 0pt \figwidth 0pt \figwidth 0pt \figwidth 0pt \figwidth 0pt \figwidth 0pt \figwidth   
0pt \linewidth 
     	Let us keep working with the vector bundle $E=TM$. We have seen that the Lie bracket measures the gap formed when we try to build up a parallelogram out of two vectors and Lie derivatives (see figure \ref{Mathematical background - image - corchete de lie flujos}). We could, instead, build up the parallelogram using the parallel transport. In that case the gap is measured by the \textbf{torsion}\index{Torsion}\index{Connection!Torsion}. To get a useful expression we notice that, given a vector $V\in T_pM$ we can extend it in the $W\in T_pM$ direction to a vector field by pushforward $(\phi^W_\varepsilon)_*V$. The difference between such vector field and the parallel transported one is precisely the covariant derivative. We have now two parallelograms with two gaps: one measured by the Lie bracket and the other by the torsion. Hence we obtain
      \[T^\nabla(V,W)=\peqsub{\nabla}{\!V}W-\peqsub{\nabla}{\!W}V-[V,W]\qquad\overset{\eqref{equation d_k}}{\equiv}\qquad T^\nabla=\peqsub{\mathrm{d}}{\nabla}\peqsub{\mathrm{Id}}{TM}\]

    \subsubsection*{Geodesics}\trassub
               
      Given a smooth curve $\gamma:I\rightarrow M$ and the tangent bundle $(TM\overset{\pi}{\rightarrow}M,\nabla)$, we have that the pull-back bundle through $\gamma$
      \[\gamma^*(TM)=\Big\{(t,v_q)\in I\times TM\ \ /\ \ \gamma(t)=\pi(v_q)\in M\Big\}=\Big\{(t,v_{\gamma(t)})\in I\times TM\Big\}\]  
      can be endowed with the induced connection
      \[\nabla:\mathfrak{X}(I)\times\Gamma(\gamma^*TM)\longrightarrow \Gamma(\gamma^*TM)=\Big\{V:I\rightarrow TM\ \ /\ \ V(t)\in T_{\gamma(t)}M\Big\}\]
      so we only have one direction to differentiate: the one given by $\dot{\gamma}(t)$.
      
      \begin{definitions}
      	\item We say that $s\in\Gamma(\gamma^*TM)$ is \textbf{parallel along {\boldmath$\gamma$}} if $\peqsub{\nabla}{\!\dot\gamma}s=0$ for every $t\in I$.
      	\item\label{def chap geodesic} A \textbf{geodesic}\index{Geodesic} is a curve $\alpha:I\rightarrow M$ such that $\dot{\alpha}\in\Gamma(\alpha^*(TM))$ is parallel to itself.
      \end{definitions}
      Visually, if we follow a geodesic what we are doing is walking always at the same pace in the direction that our nose is pointing without ever turning our head.
     
    \subsection{Semi-riemannian geometry}\label{Subsection - Semiriemannian geometry}
    \subsubsection*{Introduction}\trassub
    
      Given a smooth manifold $M$, we have at our disposal several tools (differential forms, tensor field, Lie derivative\ldots) and, besides, we can add the additional structure given by the covariant derivative that defines parallel transport. Nonetheless, we cannot measure lengths, areas or angles. For that matter we need yet another structure. Namely, a metric. We will briefly describe the essential elements needed in the following. For further details see \cite{ONeill,wald2010general,hawking1973large}
      \begin{definitions}
    	\item A \textbf{semi-riemannian manifold}\index{Semi-Riemannian manifold} $(M,g)$ is a smooth manifold $M$ equipped with a \textbf{metric}\index{Metric} $g\in\mathfrak{T}^2_0(M)$ i.e.\ a symmetric, non-degenerate and of constant index $(0,2)$-tensor field.
    	\item  We define also the inverse metric $g^{-1}:\Omega^1(M)\times\Omega^1(M)\to\Cinf{M}$ by
    	\[g\Big(g^{-1}(\alpha,\cdot{}\ ),V\Big)=\alpha(V)\qquad\quad\equiv\quad\qquad g_{ab}g^{bc}=\tensor*{\delta}{_a^c}\]
    	\item We can extend the metric to any tensorial bundle $g:\mathfrak{T}^{r,s}(M)\times\mathfrak{T}^{r,s}(M)\to\Cinf{M}$
    	\[\peqsubfino{\langle T,R\rangle}{\!g}{-0.2ex}=g\big(T,R\big)=\frac{1}{(r+s)!}g_{a_1c_1}\ \overset{(s)}{\cdots}\ g_{a_sc_s}g^{b_1d_1}\ \overset{(r)}{\cdots}\ g^{b_rd_r}\tensor{T}{^{a_1}^\cdots^{a_s}_{b_1}_\cdots_{b_s}}\tensor{R}{^{c_1}^\cdots^{c_s}_{d_1}_\cdots_{d_s}}\]
    	\item A \textbf{space-time}\index{Space-time} is a semi-Riemannian manifold of signature $(1,n-1)$. In that context a non zero vector $v\in T_pM$ is \textbf{time-like}\index{Time-like} if $g_p(v,v)<0$, \textbf{space-like}\index{Space-like} if $g_p(v,v)>0$ and \textbf{light-like}\index{Light-like} if $g_p(v,v)=0$.
    	\item $F:(M,\peqsub{g}{M})\to(N,\peqsub{g}{N})$ is an \textbf{isometry}\index{Isometry} if $F$ is a diffeomorphism such that $F^*\peqsub{g}{N}=\peqsub{g}{M}$.
      \end{definitions}
 
  	  Notice that now we have two structures, the metric and the parallel transport, and we want them to be compatible in the sense that the parallel transport respects the geometry (lengths and angles). Besides, the extremal points of the functional
  	  \[S(\gamma)=\int_I ds\,\sqrt{\left|\peqsub{g}{\gamma(s)}\Big(\dot{\gamma}(s),\dot{\gamma}(s)\Big)\right|}\qquad\qquad \gamma:I\to M\]
  	  are the \textbf{metric geodesics}\index{Geodesic}, and we want them to coincide with the geodesics previously defined. All these can be achieved because there exists a unique connection which is torsion-free (parallelograms built by parallel transport are closed) and $g$-compatible (parallel transport is an isometry).
  	  
  	  \begin{theorem}\mbox{}\\
  	  	 For a semi-Riemannian manifold $(M,g)$ there exists a unique connection  $\nabla$, called the \textbf{Levi-Civita connection}\index{Levi-Civita connection}\index{Connection!Levi-Civita}, such that $\peqsub{\mathrm{d}}{\nabla}g=0$  and $T^\nabla=0$.
  	  \end{theorem}
    
      From now on, we will always work with the Levi-Civita (LC) connection. It is important to mention that any manifold admits a Riemannian metric, however there are strong restrictions for a Lorentz metric to exist: a manifold $M$ admits a Lorentzian metric if and only if there exits a non-vanishing vector field $V\in\mathfrak{X}(M)$. For compact manifolds this is in fact equivalent to having zero Euler characteristic $\chi(M)=0$ (see for example \cite{ONeill}).
  
  	\subsubsection*{Killing vector fields}\trassub
  	
  	  A vector field $V$ whose flow preserves the metric structure, $\peqsub{\mathcal{L}}{V}g=0$, is called \textbf{Killing vector field}\index{Killing vector field}\index{Vector field!Killing}. They are very important as they are associated with the symmetries of the metric (it does not change when dragged in such directions). For generic metrics they do not exist.\separ
  	  
  	  A space-time is said to be \textbf{stationary}\index{Space-time!Stationary} if it admits a time-like Killing vector field, while we said that it is \textbf{static}\index{Space-time!Static} if it admits an irrotational time-like Killing vector field.
  	  
  	\subsubsection*{Hodge operator}\trassub
  	
  	  We have mentioned that now we are able to measure lengths, angles, areas, and volumes. One might argue that the volume forms allowed already to determine volumes, however, there was no canonical way of doing that. Now, if $(M,g)$ is an $n$-dimensional oriented semi-Riemannian manifold, we can define the \textbf{metric volume form}\index{Metric volume form}\index{Volume form!Metric} as the unique volume form $\peqsub{\mathrm{vol}}{g}\in\mathrm{Vol}(M)$ of norm $1$, i.e.\ $\langle\peqsub{\mathrm{vol}}{g},\peqsub{\mathrm{vol}}{g}\rangle_g=(-1)^s$ ($s$ is the signature of $g$), among the volume forms defined by the orientation.\separ  	  
  	  	 
  	  The metric volume form allows us to define the isomorphism $\peqsub{\star}{g}:\Omega^{k}(M)\to\Omega^{n-k}(M)$, known as the \textbf{Hodge star operator}\index{Hodge star operator}, where $\peqsub{\star}{g\,}\beta$ is the ``complement'' of $\beta$ to form a volume metric (normalized by its norm). More specifically
  	  \[\alpha\wedge\peqsub{\star}{g\,}\beta=\peqsubfino{\langle\alpha,\beta\rangle}{\!g}{-0.2ex}\,\peqsub{\mathrm{vol}}{g}\ \quad \alpha\in\Omega^k(M)\qquad\quad\equiv\quad\qquad(\peqsub{\star}{g\,}\beta)_{a_{k+1}\cdots a_n}=\frac{1}{k!}\beta_{a_1\cdots a_k}\mathrm{vol}^{a_1\cdots a_k}_{\phantom{a_1\cdots a_k}a_{k+1}\cdots a_n}\]
  	 
      \begin{properties}
      	\item $\star_{n-k}\star_k=(-1)^{k(n-k)+s}\,\mathrm{Id}$  where $s$ is the signature of $g$.
      	\item $\star 1=\peqsub{\mathrm{vol}}{g}$ and $\star\peqsub{\mathrm{vol}}{g}=(-1)^s$.
      	\item $\displaystyle(\alpha,\beta)=\int_M\alpha\wedge\peqsub{\star}{g\,}\beta=\int_M\peqsubfino{\langle\alpha,\beta\rangle}{\!g\,}{-0.2ex}\peqsub{\mathrm{vol}}{g}$ is an scalar product.\label{Mathematical background - property - (a,b)=int <a,b>vol}
      \end{properties}
      We now define the \textbf{codifferential}\index{Codifferential}\index{Differential!Codifferential} $\delta:\Omega^k(M)\to\Omega^{k-1}(M)$ as the map
      \begin{equation}\label{Mathematical background - equation - codifferential divergence}\delta_k=(-1)^{n(k+1)+1+s}\star_{n-k+1} \mathrm{d}_{n-k}\star_k\qquad\equiv\qquad (\delta\alpha)_{a_1\cdots a_{k-1}}=-\nabla^c\alpha_{ca_1\cdots a_{k-1}}\end{equation}
      As $\mathrm{d}^2=0$ and $\star\star=\pm\mathrm{Id}$, we see that $\delta^2=0$. Besides, if $M$ is compact and without boundary, $\delta$ is the adjoint operator of $\mathrm{d}$ with respect to the scalar product $(\,,\,\!)$.
      
      \subsubsection*{Densities}\label{Mathematical background - subsection - densities}\trassub
      
      Later on, we will allow the metric to vary. It is then useful to fix some reference volume $\peqsub{\mathrm{vol}}{M}$ that does not depend on the metric. According to remark \ref{Mathematical background - remark - alpha=fw} there exists a map, that we denote $\sqrt{|g|}$, such that $\peqsub{\mathrm{vol}}{g}=\sqrt{|g|}\peqsub{\mathrm{vol}}{M}$. It is important to realize that this map depends on the metric and, therefore, behaves in a different way than a standard map under certain operations that depend on the metric. For instance, from the definition of the divergence on page \pageref{Mathematical background - definition - divergence} we obtain that \[\peqsub{\mathcal{L}}{V}\sqrt{|g|}=\mathrm{div}_gV\sqrt{|g|}\]
      This can be better understood if we realize that, in a loose sense, the map under consideration is not just from $\Sigma$ to $\R$ but something like $\sqrt{|\cdot{}|}:\mathrm{Vol}(M)\times\Sigma\to \R$. Thus, somehow we have to consider not only the smooth map but also the volume form as a pair $(\sqrt{|g|},\peqsub{\mathrm{vol}}{g})$ which leads to the following formalization.\separ
      
      We define an $\bm{(r,s)}$\textbf{-density field}\index{Density}\index{Section!Density} (of weight $1$) as a section of the tensor bundle $\mathcal{T}^{r,s}(M)\otimes\mathrm{Vol}(M)$ which is equivalent to an $(r,s+n)$-tensor field such that its last $n$ contravariant (abstract) indexes are antisymmetric. Notice that if we fix some auxiliary volume form $\peqsub{\mathrm{vol}}{M}$, then any tensor field $T\in\mathfrak{T}^{r,s}(M)$ gives raise to an $(r,s)$-density field by considering $T\otimes\peqsub{\mathrm{vol}}{g}=\sqrt{|g|}\,T\otimes\peqsub{\mathrm{vol}}{M}$. As the volume form is fixed, we can think that the density field is just $\sqrt{|g|}T$. This is specially useful when using the scalar product
      \[(\alpha,\beta)=\int_M\peqsubfino{\langle\alpha,\beta\rangle}{\!g\,}{-0.2ex}\peqsub{\mathrm{vol}}{g}=\int_M\peqsubfino{\left\langle\alpha,\sqrt{|g|}\beta\right\rangle}{\!\!g}{0ex}\peqsub{\mathrm{vol}}{M}\]

    \subsubsection*{Musical isomorphisms}\trassub
    
  	  Being non-degenerate, the metric establishes the so-called \textbf{musical isomorphisms}\index{Musical isomorphism} between vector fields and $1$-form fields (which can be extended naturally to isomorphisms between $\mathfrak{T}^{r,s}(M)$ and $\mathfrak{T}^{r+k,s-k}(M)$ whenever this makes sense).
  	  \[\mathfrak{X}(M) \updown{\peqsub{\flat}{g}}{\peqsub{\sharp}{g}}{\myrightleftarrows{\rule{1cm}{0cm}}}\ \Omega^1(M)\]
  	  Given some $V\in\mathfrak{X}(M)$, the $1$-form field $\peqsub{\flat}{g}V$ is given by $(\peqsub{\flat}{g}V)(W)=g(V,W)$ while the vector field $\peqsub{\sharp}{g}\beta$, for some $\beta\in\Omega^1(M)$, is defined by the equation $\beta(W)=g(\peqsub{\sharp}{g}\beta,W)$. In abstract index notation they are just denoted
  	  \begin{equation}\label{Mathematical background - equation - musical ishomorphisms}
  	  V_a:=(\peqsub{{\flat}}{g}V)_a=g_{ab}V^b\qquad\qquad\qquad \beta^a:=(\peqsub{\sharp}{g}\beta)^a=g^{ab}\beta_b
  	  \end{equation}
      From such formulas it is clear why they are also known as \textbf{lowering and raising indices}\index{Lower index}\index{Raise index}. Besides, it is clear that now we can contract any pair of indices.
      
    \subsubsection*{Curvature revisited}\trassub
    
    Let us revisit the curvature associated with the LC connection. First we state the most important result: a semi-Riemannian manifold $(M,g)$ is locally flat if and only if $\riemann=0$. In particular it implies that $\riemann$ encodes all the local curvature\index{Curvature!Riemann}. We have also that the \textbf{Bianchi identity}\index{Bianchi identity} \eqref{Mathematical background - equation - bianchi equation} reads now
    \begin{equation}
    0=\peqsub{\mathrm{d}}{\nabla}\riemann(V,W,U)=\peqsub{\nabla}{V} \riemann(W,U)+\peqsub{\nabla}{W}\riemann(U,V)+\peqsub{\nabla}{U}\riemann(V,W)
    \end{equation}
    
    \setlength{\figwidth}{0.3\linewidth}\definecolor{verdecito}{HTML}{009926}
    \setlength{\figwidthleft}{\dimexpr \linewidth-\figwidth}
    \null\smash{\raisebox{-\dimexpr 5.5\baselineskip+\parskip\relax}%
    	{\includegraphics[width=.28\linewidth]{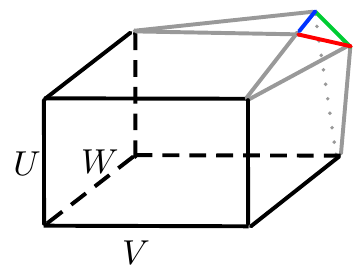}}}\strut\hfill \\[-\dimexpr 4\baselineskip+2\parskip\relax]
    
    \parshape=1 
    \figwidth \figwidthleft    
    Using that the torsion is zero, we get the \textbf{Bianchi symmetry}\index{Bianchi symmetry}\index{Bianchi identity!First Bianchi identity}
    \begin{equation}
    \textcolor{verdecito}{\riemann(V,W)U}+\textcolor{red}{\riemann(U,V)W}+\textcolor{blue}{\riemann(W,U)V}=0
    \end{equation}
    which has a very nice geometrical interpretation as the image on the left shows.\separ
    
\parshape=2 
\figwidth \figwidthleft
0pt \linewidth
    It is interesting to mention that the curvature provides yet another geometric concept which is of vital importance for general relativity: the \textbf{geodesic deviation}\index{Geodesic!Deviation}. Indeed $\riemann$ measures how some initially parallel geodesics might fail to remain so.\separ
    
    Finally, it is worth mentioning that the Ricci\index{Curvature!Ricci} tensor field has now, in the presence of a metric, a clear physical interpretation: it describes how the volume of a small ball of particles changes -in all directions- when all of them follow geodesics.
    
    \subsubsection*{Scalar curvature}\trassub
    
      The contraction of the Ricci tensor (using the musical isomorphisms) leads to the definition of the \textbf{scalar curvature}\index{Scalar curvature}\index{Curvature!Scalar}
      \[R(g)=C\left(\mathrm{Ric}^{\nabla_g}\right)=g^{ab}\tensor{\mathrm{Ric}}{_a_b}\]
      For a Riemannian metric, $R(g)$ measures the deviation of the volume/area of a metric ball $B(p,\varepsilon)$ with respect to the volume/area of the Euclidean ball $B_{\R^n}(0,\varepsilon)$ 
      \[\frac{\mathrm{vol}_g\Big(B_\varepsilon(p)\subset M\Big)}{\peqsub{\mathrm{vol}}{\R^n}\Big(B_\varepsilon(0)\subset \R^n\Big)}=1-\frac{R}{6(n+2)}\varepsilon^2+O(\varepsilon^4)\quad\qquad\frac{\mathrm{area}_g\Big(\partial B_\varepsilon(p)\subset M\Big)}{\peqsub{\mathrm{area}}{\R^n}\Big(\partial B_\varepsilon(0)\subset \R^n\Big)}=1-\frac{R}{6n}\varepsilon^2+O(\varepsilon^4)\]
      In the semi-riemannian case the interpretation is far from being direct \cite{ehrlich2000some}.

	\subsubsection*{Semi-riemannian submanifolds}\trassub
	
			
	  Let $\jmath:\overline{M}\hookrightarrow M$ be an embedding i.e.\ a smooth map such that $\jmath_*$ is injective and the induced map $\jmath:\overline{M}\to\jmath(\overline{M})$ is a homeomorphism. It is well known that its image $\jmath(\overline{M})\subset M$ is then a submanifold. It is customary to identify it with $\overline{M}$ itself, but we will avoid that to prevent some common confusions. Let us now sketch how $\overline{M}$ inherits some of the structures of $M$ and the differences between studying the submanifold from within ($\overline{M}\equiv$ intrinsically) or from the outside ($\jmath(\overline{M})\equiv$ extrinsically).\separ
	  
	  Notice that if $(M,g)$ is semi-riemannian, then $(\overline{M},\overline{g}=\jmath^*g)$ might not be because $\overline{g}$ can be degenerate at some point. In the Lorentzian case we say that $(\overline{M},\overline{g})$ is \textbf{space-like}\index{Metric!Space-like}\index{Space-time!Space-like} if $\overline{g}$ is a Riemannian metric, \textbf{time-like}\index{Metric!Space-like}\index{Space-time!Time-like} if it is a Lorentzian metric and \textbf{light-like}\index{Metric!Light-like}\index{Space-time!Light-like} if the induced metric is degenerate everywhere. It is of great interest in general relativity to study the latter (also known as \textbf{null submanifolds}\index{Space-time!Null submanifold}\index{Null submanifold}) but we will assume from now on that $\overline{g}$ is non-degenerate. This allows us, in particular, to decompose $TM=\jmath_*(T\overline{M})\oplus T^{\scriptscriptstyle\perp}\!\jmath(\overline{M})$ where we define the \textbf{normal subbundle}\index{Bundle!Normal subbundle} as
	  \[T^{\scriptscriptstyle\perp}\!\jmath(\overline{M})=\bigsqcup_{p\in \overline{M}}T^{\scriptscriptstyle\perp}_{\jmath(p)}\jmath(\overline{M})\subset\jmath^*TM\quad\text{ where }\quad T^{\scriptscriptstyle\perp}_{\jmath(p)}\jmath(\overline{M})=\Big\{v\in T_{\jmath(p)}M\ /\ g_{\jmath(p)}\!\left(v,\jmath_*T_p\overline{M}\right)=0\Big\}\]
	  \vspace*{-1.5ex}
	  
	  We denote by $\mathfrak{X}^{\scriptscriptstyle\perp}_\jmath\hspace*{-0.2ex}(M)=\Gamma(T^{\scriptscriptstyle\perp}\overline{M})\subset\Gamma(\jmath^*TM)$ the vector fields over $\jmath(\overline{M})$ that are normal to the submanifold. Likewise, we denote by $\mathfrak{X}^{\scriptscriptstyle\top}_\jmath\!(M)\subset\Gamma(\jmath^*TM)$ the tangent ones i.e.\ if $V\in\mathfrak{X}^{\scriptscriptstyle\top}_\jmath\!(M)$ then $V_p\in j_* T_p\overline{M}\subset T_{\jmath(p)}M$.\separ
	  
	  The fact that $\overline{g}$ is non-degenerate allows us also to consider the associated LC connection
	  \[\overline{\nabla}:\mathfrak{X}(\overline{M})\times\mathfrak{X}(\overline{M})\to\mathfrak{X}(\overline{M})\]
	  together with the intrinsic curvatures (Riemann, Ricci and scalar). We have also the induced LC connection (with the associated extrinsic curvatures) over $\jmath(\overline{M})$ defined as the unique connection over $\jmath$
	  \[\nabla^{(\jmath)}:\mathfrak{X}(\overline{M})\times\Gamma(\jmath^*TM)\to\Gamma(\jmath^*TM)\qquad\text{with}\qquad\peqsub{\nabla^{(\jmath)}}{V}(W\smallcirc\jmath)=\nabla_{\jmath_*\!V}W\]
	  We will in general omit the superscript $(\jmath)$ because the direction of the derivative already tells us which one we are referring to. Actually, if we use abstract index notation, it becomes even more transparent. For instance using Latin indexes for $\overline{M}$ and Greek ones for $M$ we obtain
	  \begin{equation}\label{Mathematical background - equation - nabla_a=tau nabla_alpha}
	  \nabla_b\equiv\nabla^{(\jmath)}_b=(\jmath_*)^\beta_b\nabla_\beta
	  \end{equation}
	  
	  Notice that this is a generalization of what we did for geodesics. Indeed, $V\in\Gamma(\jmath^*TM)$ is a map $V:\overline{M}\to TM$ such that $\pi\smallcirc V=\jmath$. Thus it only makes sense to compute the variation when we move along a direction in $\overline{M}$. One of the most important results about semi-riemannian submanifolds is that if we project $\nabla$ over $\overline{M}$ we obtain $\overline{\nabla}$. Although true in general, it is simpler and enough for our purposes to focus on hypersurfaces.
	  	  
	  \subsubsection*{Semi-riemannian hypersurfaces}\trassub
	  
	   A \textbf{hypersurface}\index{Hypersurface} is a submanifold of codimension $1$. This implies that, at least locally $\overline{U}\subset\overline{M}$, there exists a unique (up to a sign) unitary vector field $\vec{n}\in\mathfrak{X}^{\scriptscriptstyle\perp}_\jmath(\jmath(\overline{U}))$ which is perpendicular to $\jmath(\overline{U})\subset M$. Being unitary $g(\vec{n},\vec{n})=\varepsilon=\pm1$ implies $\peqsub{\nabla}{V}\vec{n}\perp\vec{n}$ for every $V\in\mathfrak{X}(\overline{M})$. In particular, the variation of the normal vector field along $\overline{M}$ is tangent to $\jmath(\overline{M})$ \[\peqsub{\nabla}{V}\vec{n}\in\mathfrak{X}^{\scriptscriptstyle\top}_\jmath\!(M)\subset\Gamma(\jmath^*TM)\]
	   This allows us to define the \textbf{Weingarten map}\index{Weingarten map} $\mathcal{W}:\mathfrak{X}(\overline{M})\to\mathfrak{X}(\overline{M})$ given by
	  \begin{equation}\label{Mathematical background - equation - Weingarten}
	  \jmath_*\big(\mathcal{W}(V)\big)=-\peqsub{\nabla}{V}\vec{n}\qquad\quad\equiv\quad\qquad (\jmath_*)_b^\beta\tensor{\mathcal{W}}{_a^b}=-\nabla_an^\beta
	  \end{equation}
	  which accounts for the extrinsic curvature\index{Curvature!Extrinsic} in the sense that it measures how the hypersurface bends by describing how the normal direction changes in the ambient space.
	  	  The metrically equivalent map $K\in\mathfrak{T}^{0,2}(\overline{M})$ defined by
	  	  \begin{equation}
	  	  K(V,U)=\overline{g}(\mathcal{W}V,U)\qquad\equiv\qquad K_{ab}=\overline{g}_{ac}\tensor{\mathcal{W}}{_b^c}
	  	  \end{equation}
	  	  is known as the \textbf{second fundamental form of {\boldmath$\vec{n}$}}\index{Second fundamental form} which is symmetric (because $W$ is self-adjoint).

	  \begin{lemma}[Gauss' lemma]\label{Mathematical background - theorem - Gauss lemma}\mbox{}\\
	  	Given $V,W\in\mathfrak{X}(\overline{M})$ we have the decomposition
	  	\[\begin{array}{ccccccccc}
	  	\peqsub{\nabla}{V}(\jmath_*W)=&\hspace*{-1.5ex}\jmath_*\peqsub{\overline{\nabla}}{V}W&\hspace*{-1ex}+&\hspace*{-1.5ex}\varepsilon K(V,W)\vec{n}&\quad\equiv\qquad&
	  	\ \ \nabla_a\big((\jmath_*)^\beta_bW^b\big)=&\hspace*{-1.5ex}(\jmath_*)^\beta_b\,\overline{\nabla}_aW^b&\hspace*{-1ex}+&\hspace*{-1.5ex}\varepsilon K_{ab}W^bn^\beta\\[.5ex]
	  	&\hspace*{-2.5ex}\text{tangent} &&\hspace*{-2.5ex}\text{normal} &&
	  	&\hspace*{-1.5ex}\text{tangent} &&\hspace*{-2.5ex}\text{normal}\end{array}
	  	\]
	  	\vspace*{-3.5ex}
	  \end{lemma}
	  This important result shows, as we mentioned on the previous section, that the tangent part of $\nabla$ over $\jmath(M)$ is essentially $\overline{\nabla}$.  Notice in particular that the geodesics of $\overline{M}$ would be geodesics of $M$ if and only if $K=0$. This makes sense as geodesics over a submanifold can only have normal acceleration so if we extrinsically bend the submanifold, the geodesics will change and will no longer coincide with the ones of $M$.
	  
	  \centerline{\includegraphics[clip,trim=30ex 90ex 5ex 70ex,width=0.8\linewidth]{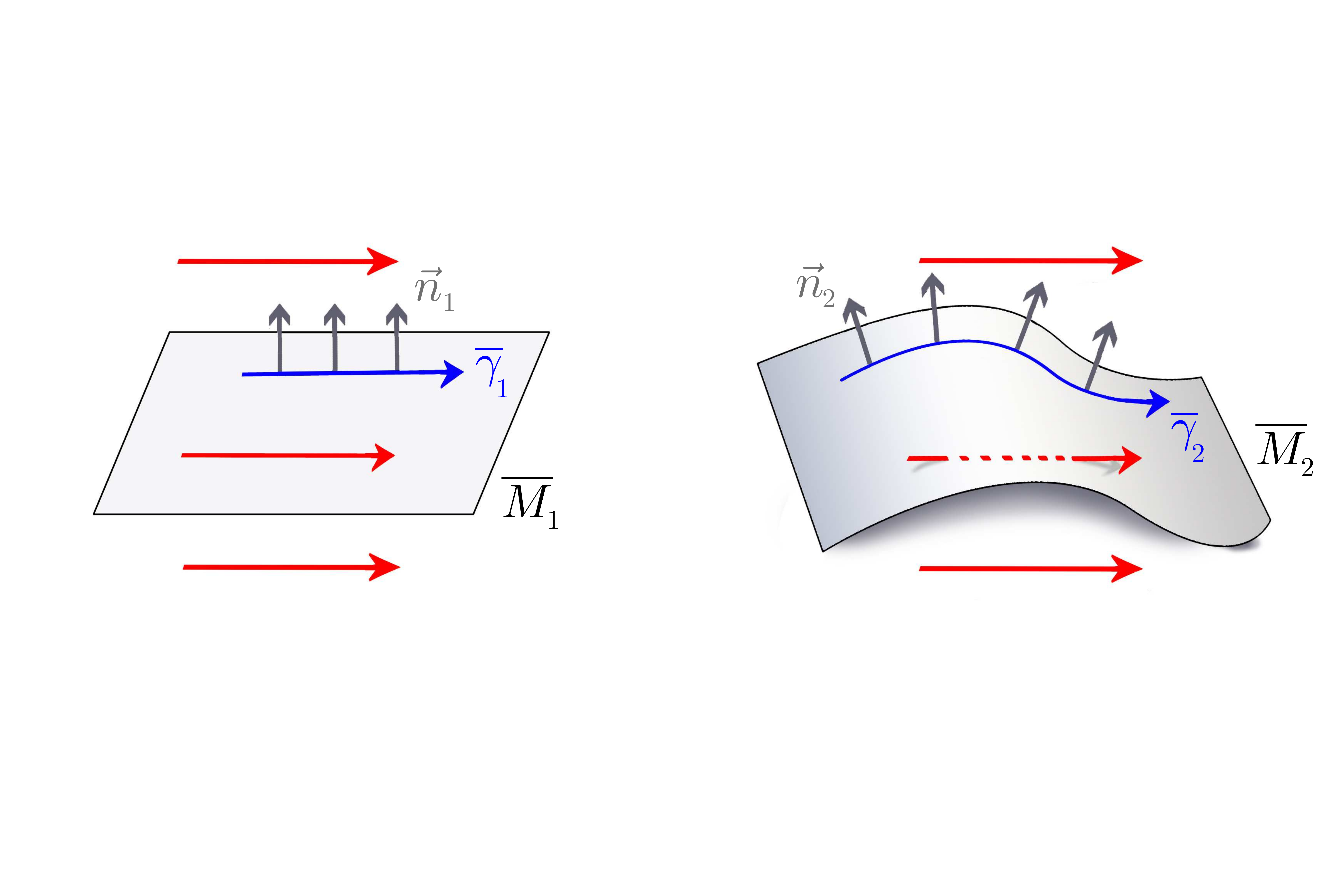}}

	  We have already compared the LC connections $\nabla$ and $\overline{\nabla}$, let us now compare the Riemann curvatures $Riem$ and $\overline{Riem}$ associated with them.
	  
	  \begin{theorem}\label{Mathematical background - theorem - Gauss R bar(R)}\mbox{}\\
	  	Given $V,W,X,Y\in\mathfrak{X}(\overline{M})$ we have
	  	\[g\Big(Riem(\jmath_*V,\jmath_*W)\jmath_*X,\jmath_*Y\Big)=\overline{g}\Big(\overline{Riem}(V,W)X,Y\Big)-\varepsilon\det\begin{pmatrix}K(V,X)&K(W,X)\\K(V,Y)&K(W,Y)\end{pmatrix}\]
	  \end{theorem}
	  
	  Taking twice the trace in the previous formula (see lemma \ref{appendix - lemma - escalar de curvatura}) leads to  
	  \begin{equation}
	  R(g)=\overline{R}(\overline{g})+\varepsilon \mathrm{Tr}_{\overline{g}}(K)^2-2\varepsilon \langle K,K\rangle_{\overline{g}}-2\varepsilon\mathrm{div}\big(\vec{n}\,\mathrm{div}\,\vec{n}-\nabla_{\vec{n}}\vec{n}\big)
	  \end{equation}
	  This formula is fundamental to develop the Hamiltonian formulation of general relativity.\separ
	  
	  Finally, notice that a very important hypersurface is the boundary $\partial M$ of a manifold $M$. We can now reinterpret the Stokes' theorem in terms of the (globally defined as $M$ is oriented) outer vector normal field to the boundary $\vec{\nu}$. For that consider $W\in\mathfrak{X}(M)$ and $\alpha=\flat_gW\in\Omega^1(M)$, then
	  \begin{align}
	  \begin{split}
	  	\mathrm{div}W\peqsub{\mathrm{vol}}{g}&\updown{\eqref{Appendix - equation - star 1=vol}}{\eqref{Appendix - definition - delta=-nabla}}{=}-\delta\alpha\cdot{}\peqsub{\star}{g}1=-\peqsub{\star}{g}\delta\alpha\overset{\eqref{appendix property star delta=d star}}{=}\\
	  	&=\mathrm{d}\peqsub{\star}{g}\alpha\overset{\eqref{appendix property star alpha=i_alpha vol}}{=}\mathrm{d}\peqsub{\imath}{W}\peqsub{\mathrm{vol}}{g}\overset{\dagger}{=}\mathrm{d}\peqsub{\imath}{W}(\nu\wedge\peqsub{{\mathrm{vol}_g}}{\!\partial})=\\
	  	&=\mathrm{d}\Big(\nu(W)\peqsub{{\mathrm{vol}_g}}{\!\partial}-\nu\wedge\peqsub{\imath}{W}\peqsub{{\mathrm{vol}_g}}{\!\partial}\Big)
	  \end{split}
	  \end{align}
	  where $\peqsubfino{g}{\partial}{-.4ex}=\peqsub{\jmath}{\partial}^*g$ is the induced metric over the boundary and $\nu$ the $1$-form field metrically equivalent to $\vec{\nu}$. Besides, we have used in the $\dagger$ equality that over the boundary $\peqsub{\mathrm{vol}}{g}=\nu\wedge\peqsub{{\mathrm{vol}_g}}{\!\partial}$ (see lemma \ref{Appendix - lemma - vol=nu wedge vol}). Finally, applying the Stokes' theorem, taking into account that $\peqsub{\jmath}{\partial}^*\nu=0$, leads to
	  \begin{equation}
	    \int_M\mathrm{div}W\peqsub{\mathrm{vol}}{g}=\int_{\partial M}\nu(W)\peqsubfino{{\mathrm{vol}_g}}{\!\partial}{-0.3ex}
	  \end{equation}
	  
  \subsection{Physics over space-times}\label{Mathematical background - subection - pysics over space-time}
  
  In this section we introduce the Einstein equations and the space-times that are of interest for the Hamiltonian formulation of General Relativity.
  
  \subsubsection*{Einstein equations}\trassub
  
  We have reviewed in the previous section several objects we can define once we have a metric $g$ in our manifold $M$. However, we have not talked yet about how to endow $M$ with a particular metric so that $(M,g)$ models a system of physical interest. For that reason let us take a small detour to sketch briefly the physical motivations behind the Einstein equations. We refer the reader to \cite{misner2017gravitation,weinberg2014gravitation,ONeill,wald2010general,hawking1973large,margalef2012evolution,margalef2013evolution,margalef2014topology} for more details as well as discussions about related topics like cosmology and the role of topology in general relativity.\newpage
  
  $\blacktriangleright$ Force vs curvature\separprevia
  
  \begin{wrapfigure}{r}{4.6cm}
  	\centering
  	\includegraphics[width=1\linewidth]{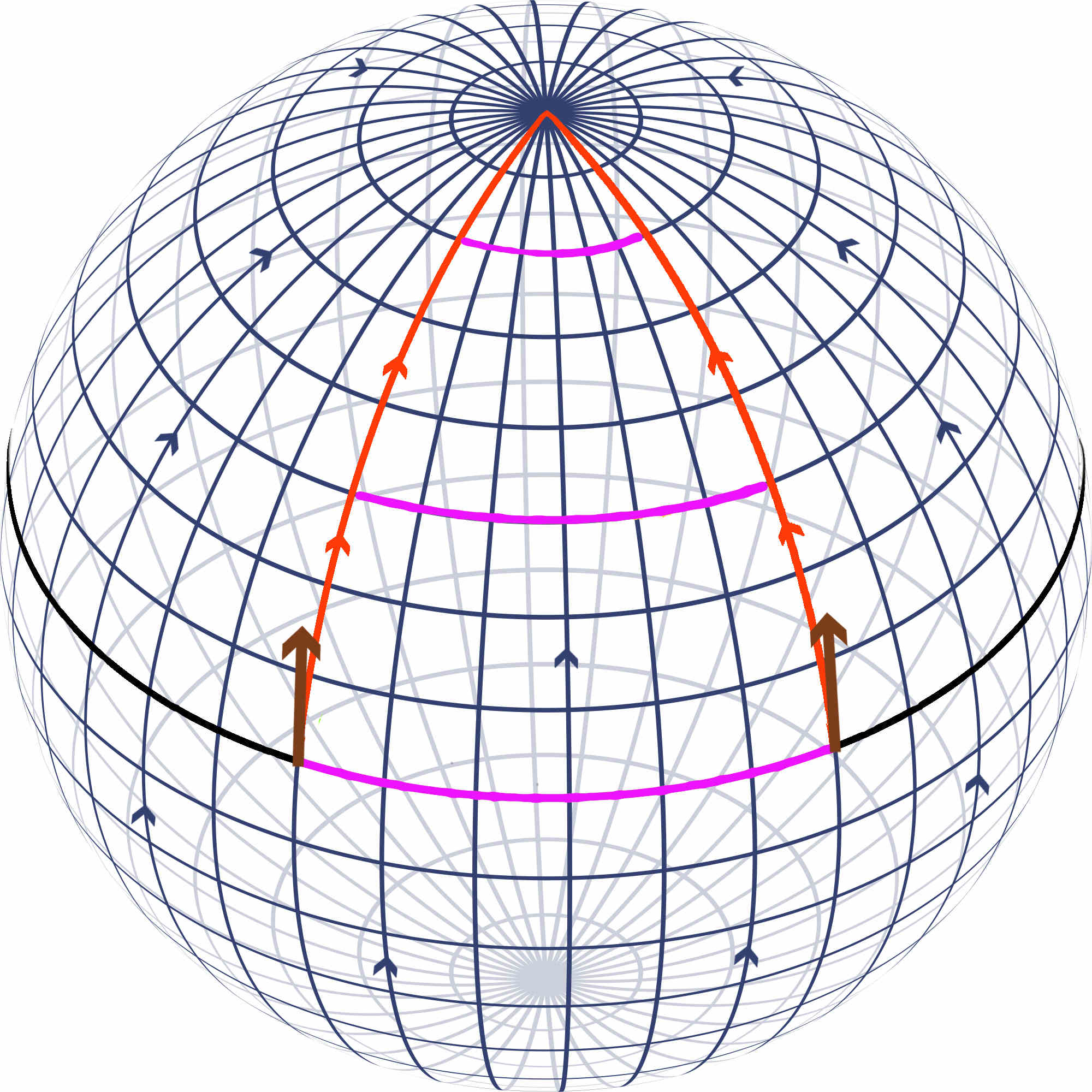}
  	\vspace*{-5ex}
  	
  \end{wrapfigure}
  Newton considered gravity as a force field exerted by body masses that affects all the objects in range. Meanwhile, Einstein showed that it was much more accurate/convenient to consider it as a manifestation of the space-time curvature\index{Curvature!Space-time}. It is useful to consider the following analogy. Assume that we have a huge non-rotating sphere with a source of water emanating at the south pole. The water is constantly flowing towards the north pole, where there is a sink, following geodesics (great circles joining the poles). Notice in particular that the flow of water is normal to the parallels. Now, if two observers let themselves go along the flow at the equator, they both start moving in parallel directions. However, it is clear that as time goes by, they will get closer and closer until they meet at the north pole. As the sphere is big enough, they might think that they are in a plane and that there is an attractive force between them instead of considering that they live in a sphere. That is somehow what happens with gravity. The flow of water is analogous to the flow of time, and the shape of the space-time can make that the geodesics converge or diverge as in the previous example.\separpost
  
  $\blacktriangleright$ Newtonian limit\separprevia
  
  First, we recall that in the Newtonian theory we have a gravitational potential $\phi$ which is determined by the distribution of mass $\rho$ via the Poisson's equation
  \[\Delta\phi=4\pi \peqsub{G}{\!N}\,\rho\]
  where $\peqsub{G}{\!N}$ is the gravitational constant. This potential allows to write down the Newtonian equation $\vec{g}=-\nabla\phi$ to obtain the gravitational field $\vec{g}$. Secondly, it can be proved that in the Newtonian limit the ``slow'' geodesics of the metric $g=\eta+h$, with $\eta$ the Minkowski metric and $h$ a ``small perturbation'', are given by
  \[\vec{a}=\frac{\mathrm{d}^2\vec{x}}{\mathrm{d}\tau^2}\approx\frac{c^2}{2}\nabla g_{00}\] 
  As the acceleration $\vec{a}$ is given, in the Newton's theory, by the gravitational field $\vec{g}$, we obtain the approximation
  \[\peqsubfino{g}{00}{-0.2ex}\approx-1-\frac{c^2}{2}\phi\]
  where the integration constant is chosen using the fact that far from the source $\phi\to0$ and $g\to\eta$. In particular, we see that we can rewrite the Poisson's equation as
  \begin{equation}\label{Mathematical background - equation - approx newtoniana}
  \Delta\peqsubfino{g}{00}{-0.2ex}\approx-\frac{8\pi\peqsub{G}{\!N}}{c^2}\rho\approx-\frac{8\pi\peqsub{G}{\!N}}{c^4}\peqsubfino{T}{00}{-0.1ex}
  \end{equation}
  where $T$ is the \textbf{energy-momentum tensor}\index{Energy-momentum tensor}, a symmetric $(0,2)$-tensor field encoding the density and flux of energy-and-momentum in a space-time. In the Minkowskian case $T_{00}=c^2\rho$ (in the weak gravity case it is only an approximation) is the energy density, $T_{0i}$ is the linear momentum density, $T_{ii}$  the normal stresses and $T_{ij}$ the shear stresses. The previous equation, heuristically obtained, suggests that the metric is directly related to the energy-momentum tensor.\separpost

  $\blacktriangleright$ Einstein's guesses\separprevia
  
  The \emph{Ansatz} is that the metric has to satisfy an equation, linking the geometry with its sources, of the form
  \[\mathcal{G}(g)=\frac{8\pi\peqsub{G}{\!N}}{c^4}\, T\]
  where $\mathcal{G}(g)$, being proportional to $T$, is a symmetric $(0,2)$-tensor field which depends on $g$ and its derivatives. A dimensional study suggests that we neglect terms of order higher than three, which is reinforced by the fact that \eqref{Mathematical background - equation - approx newtoniana} is of order two. This leads to
  \[\mathcal{G}(g)=\alpha_0+\alpha_1+\alpha_2\]
  where $\alpha_0$ depends linearly on $g$, $\alpha_1$ on the derivatives of $g$ and $\alpha_2$ contain terms that are either linear in the second derivatives of $g$ or quadratic in the first derivatives. It can be proved that $\alpha_0=\Lambda g$ for some constant $\Lambda$, $\alpha_1=0$ (we can always find some coordinates such that the first derivatives vanish) and
  \[\alpha_2=\beta \textrm{Ric}+\gamma R\, g\]
  for some constant $\beta,\gamma\in\R$. Now, imposing the \textbf{local conservation of energy} $\mathrm{div}T=0$ we have
  \[0=\frac{8\pi\peqsub{G}{\!N}}{c^4}\,\mathrm{div}T=\mathrm{div}\Big(\Lambda g+\beta\mathrm{Ric}+\gamma R\,g\Big)=\beta\mathrm{div}\mathrm{Ric}+\gamma\nabla R\overset{\eqref{Appendix - equation - div Ric=grad R}}{=}\left(\frac{\beta}{2}+\gamma\right)\nabla R\]
  As we want it to be true for a general metric, we take $\gamma=-\beta/2$. It is not hard to see that $\beta=1$ once we realize that $\mathcal{G}(g)_{00}\approx \beta\Delta g_{00}$ in the weak field approximation. Thus we have obtained the Einstein's equation\index{Einstein's equation} with cosmological constant\index{Cosmological constant} $\Lambda$
  \[\mathrm{Ric}(g)-\frac{R}{2}g+\Lambda g=\frac{8\pi\peqsub{G}{\!N}}{c^4}T\]
  The first two terms $G=\mathrm{Ric}(g)-\frac{R}{2}g$ are known as the \textbf{Einstein's tensor}\index{Einsten's tensor}. Taking the trace in the previous equation we can write the scalar curvature in terms of the trace of $T$ and obtain the equivalent equation
  \[\mathrm{Ric}-\frac{2}{n-2}\Lambda g=\frac{8\pi\peqsub{G}{\!N}}{c^4}\left(T-\frac{\mathrm{Tr}(T)}{n-2}g\right)\]
  Notice that if there is no matter at all and no cosmological constant, then the Ricci curvature is zero but the Riemann curvature might be non-zero (as in the case of the Schwarzschild solution).

  \subsubsection*{Globally hyperbolic space-times}\trassub

    Let us consider a space-time $(M,g)$, our goal is to recover the whole metric information from our knowledge of the metric ``now'' and the evolution given by Einstein equations. For that matter we need first to restrict ourselves to the space-times that admit such evolution.\separ
    
    A \textbf{Cauchy hypersurface}\index{Cauchy hypersurface} is a smooth space-like hypersurface of $M$ such that any inextensible causal-like curve intersects it exactly once. It thus determines the future and past uniquely and can be identified, in a loose sense, as an instant of time. A space-time is said to be \textbf{globally hyperbolic}\index{Globally hyperbolic} if it admits a Cauchy hypersurface. In that case it is possible to define well posed initial value problems.
    
    \begin{proposition}\mbox{}\\
    	If $(M,g)$ is a globally hyperbolic space-time, then there exists a diffeomorphism
    	\[\varphi:\R\times\Sigma\longrightarrow M\]
    	Moreover, $\varphi(\{t\}\times\Sigma)\subset M$ is a Cauchy hypersurface and so $M$ can be foliated by Cauchy hypersurfaces.
    \end{proposition}
\centerline{\includegraphics[width=.7\linewidth]{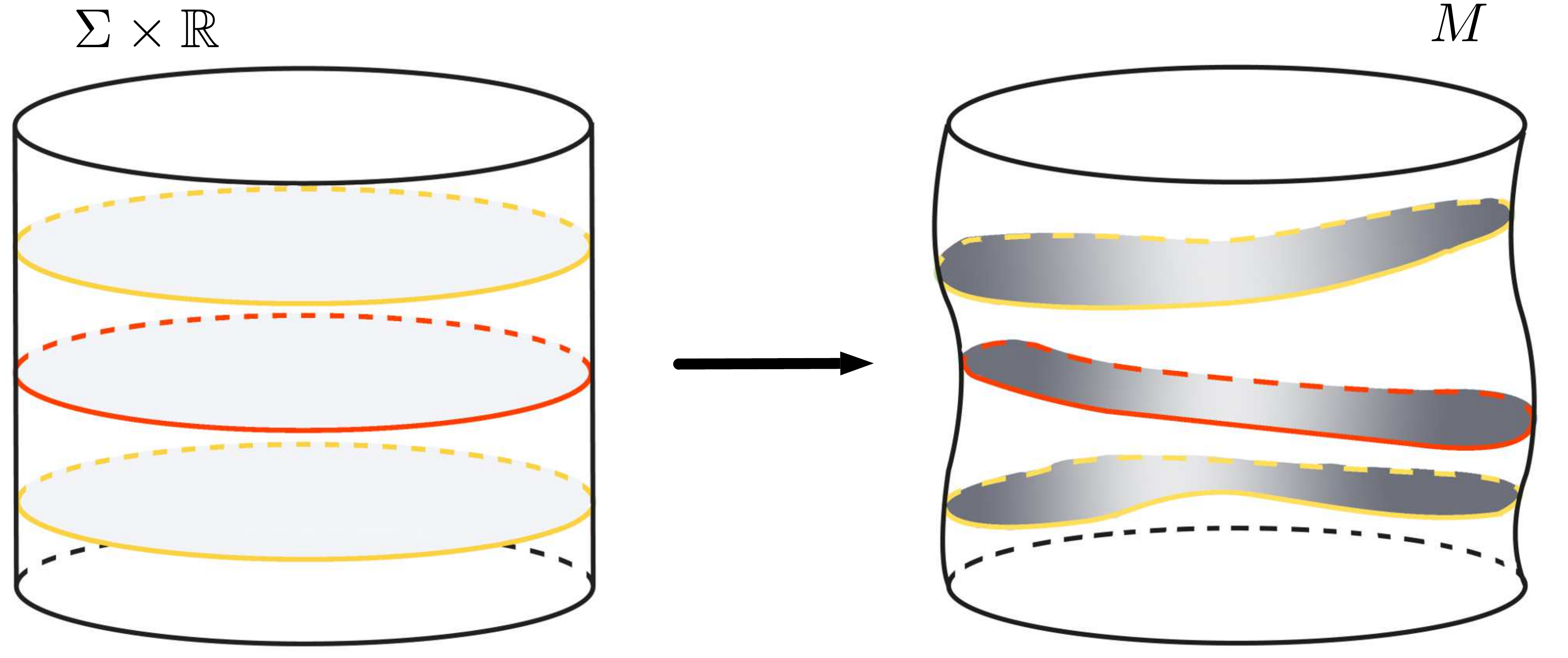}}
    
    The previous result, which actually holds for a more general definition of Cauchy hypersurface \cite{geroch1970domain,bernal2003smooth,sanchez2004causal}, is of vital importance and provides a way to carry out the $n+1$ decomposition, first step towards the Hamiltonian formulation.
    
    \subsubsection*{\texorpdfstring{$\bm{n+1}$}{n+1} decomposition over \texorpdfstring{$\bm{M=I\times\Sigma}$}{M=IxE}}\trassub
    
    Let $M=I\times\Sigma$ with no metric so far. The fact that $TM=TI\oplus T\Sigma$ will allow us to break many geometrical objects into the ``time'' part and the ``spatial'' part.\separ
    
    First notice that we have the projection $t:I\times\Sigma\to I$ which, as it is a smooth map $t\in\Omega^0(M)$, allows us to define $\mathrm{d}t\in\Omega^1(M)$. Now we choose $\partial_t\in\mathfrak{X}(M)$ to be some transversal vector field to the foliation such that $\mathrm{d}t(\partial_t)=1$ everywhere\footnote{Equivalently we can fix a diffeomorphism $Z_0$ such that $Z_0(\{t\}\times\Sigma)$ are the leaves of the foliation and then take $\partial_t$ as the tangent vector of the curves $Z_0(\sigma,t)$ for every $\sigma\in\Sigma$. Picking another $Z_0$ leads to another choice of $\partial_t$.}. With these ingredients we define the $(1,1)$-tensor field $\Pi=\mathrm{Id}-\mathrm{d}t\otimes\partial_t$ which is a ``projector over $\Sigma$''. For instance a vector field $Y\in\mathfrak{X}(I\times\Sigma)$ and a $1$-form field $\alpha\in\Omega^1(I\times\Sigma)$ can be written as
    \[Y=\mathrm{d}t(Y)\partial_t+\Pi_*(Y)\qquad\qquad\alpha=\alpha(\partial_t)\mathrm{d}t+\Pi^*(\alpha)\]
    We have used for clarity the notation $\Pi_*:\mathfrak{X}(M)\to\mathfrak{X}(M)$ and $\Pi^*:\Omega^1(M)\to\Omega^1(M)$ for the maps equivalent to $\Pi:\mathfrak{X}(M)\times\Omega^1(M)\to\Cinf{M}$.\separ
  
   We consider now a Lorentzian metric $g$ over the manifold $M$ such that the leaves $\{t\}\times\Sigma$ are space-like submanifolds. Breaking the vector fields $X,Y\in\mathfrak{X}(M)$ in the expression $g(X,Y)$ leads to
  \[g=\varepsilon \Lambda\, \mathrm{d}t\otimes \mathrm{d}t+\mathrm{d}t\otimes N+N\otimes \mathrm{d}t+\widetilde{\gamma}\qquad \text{ where}\qquad \begin{array}{l}
  \Lambda:=\varepsilon g(\partial_t,\partial_t)\\[0.7ex]
  N(\,\cdot\,):=g(\partial_t,\Pi_*\,\cdot\,)\\[.6ex]
  \widetilde{\gamma}(\,\cdot\,,\cdot\,):=g(\Pi_*\,\cdot\,,\Pi_*\,\cdot\,)
  \end{array}\]
  
  and $\varepsilon=-1$ but we keep it in order to allow for a straightforward extension of our
  results to the Riemannian case. Analogously, the inverse metric can be decomposed as
  \[g^{-1}= \varepsilon\frac{\partial_t-\vec{N}}{\mathbf{N}}\otimes\frac{\partial_t-\vec{N}}{\mathbf{N}}+\zeta\qquad \text{ where}\qquad \begin{array}{l}
  \mathbf{N}^2:=\varepsilon /g^{-1}(\mathrm{d}t,\mathrm{d}t)\\[0.7ex]
  \vec{N}(\,\cdot\,):=-\varepsilon\mathbf{N}^2\,g^{-1}(\mathrm{d}t,\Pi^*\,\cdot\,)\\[0.7ex]
  \zeta(\,\cdot\,,\cdot\,):=g^{-1}(\Pi^*\,\cdot\,,\Pi^*\,\cdot\,)-\dfrac{\varepsilon}{\mathbf{N}^2}\vec{N}\otimes\vec{N}
  \end{array}\]
  Using the explicit expression $\Pi_a^b=\delta_a^b-(\mathrm{d}t)_a\partial_t^b$ it is easy to derive the main properties of the objects involved
  \[\begin{array}{llllll}
   \Pi_*\partial_t=0\quad\ &\Pi_*\vec{N}=\vec{N}\quad\ &N(\partial_t)=0\quad\ &\widetilde{\gamma}(\partial_t,\cdot{})=0\quad\ &\widetilde{\gamma}(\vec{N},\cdot{})=N\ &\varepsilon\Lambda=\varepsilon\mathbf{N}^2+N(\vec{N})\\%
   \Pi^*(\mathrm{d}t)=0\quad&\Pi^*N=N\quad&\mathrm{d}t(\vec{N})=0\quad&\zeta(\mathrm{d}t,\cdot\,)=0\quad&\zeta(N,\cdot{})=\vec{N}\quad&\widetilde{\gamma}_{ab}\zeta^{bc}=\Pi_a^c\\
  \end{array}\]
   Notice that both triples $(\Lambda,N,\widetilde{\gamma})$ and $(\mathbf{N},\vec{N},\zeta)$ do contain the same information. It is important to realize that the previous definitions are made in such a way that they are related by $\zeta$ (raising index) and $\widetilde{\gamma}$ (lowering index) and not by $g$. We have done so because, in the end, we want to pullback everything to a single leaf where $\widetilde{\gamma}$ is a metric whose inverse is $\zeta$. Actually, it is clear from the previous expressions that the metric $g$ can be recovered from the data $(\Lambda,N,\widetilde{\gamma})$ or, equivalently, from $(\mathbf{N},\vec{N},\zeta)$.\separ

   We have seen in the previous section that the normal vector field $\vec{n}\in\mathfrak{X}(M)$ to a hypersurface is essential to understand its extrinsic geometry. Imposing the conditions $g(\vec{n},\vec{n})=\varepsilon$ and $g(\vec{n},\Pi_*)=~0$ it is easy to obtain the vector field and its metrically equivalent $1$-form field
   \begin{equation}\label{Mathematical background - equation - n=varepsilon N dt}
   	\vec{n}=\frac{\partial_t-\vec{N}}{\mathbf{N}}\in\mathfrak{X}(M)\qquad\qquad n=\varepsilon\mathbf{N}\mathrm{d}t\in\Omega^1(M)
   \end{equation}
   Recall that $\partial_t$ was arbitrarily chosen among the transversal vector fields such that $\mathrm{d}t(\partial_t)=1$. We see that $\vec{N}\in\mathfrak{X}(M)$, which is tangent to the foliation and called the \textbf{shift}\index{Shift}, compensates the deviation of $\partial_t$ from the orthogonal direction. Meanwhile $\mathbf{N}=\varepsilon g(\vec{n},\partial_t)\in\Cinf{M}$, known as the \textbf{lapse}\index{Lapse}, is a normalization factor.\separ
   
   Now we can decompose $g^{-1}$ with the help of the normal vector field as
   \begin{equation}\label{Mathematical background - equation - inverse metric n}
   g^{-1}=\varepsilon\vec{n}\otimes\vec{n}+\zeta
   \end{equation}
   Analogously, a vector field $V\in\mathfrak{X}(M)$ and a $k$-form field $\beta\in\Omega^k(M)$ can be decomposed as
  \begin{align}
   &V=V^{\scriptscriptstyle\perp}\vec{n}+\vec{V}^{\scriptscriptstyle\top}&&\hspace*{-4ex}\text{where}\qquad\qquad\begin{array}{|l}
  V^{\scriptscriptstyle\perp}=\varepsilon g(V,\vec{n})\in\Cinf{M}\\[1ex] \vec{V}^{\scriptscriptstyle\top}=V-V^{\scriptscriptstyle\perp}\vec{n}\in\mathfrak{X}(M)\end{array}\label{Mathematical background - equation - descomposicion campo vectorial}\\[2.2ex]
  &\beta=n\wedge\peqsub{\beta}{\perp}+\beta^{\scriptscriptstyle\top}&&\hspace*{-4ex}\text{where}\qquad\qquad\begin{array}{|l}
  \peqsub{\beta}{\perp}=\varepsilon \imath_{\vec{n}}\beta\in\Omega^{k-1}(M)\\[1ex] \beta^{\scriptscriptstyle\top}=\beta-\varepsilon n\wedge\peqsub{\beta}{\perp}\in\Omega^{k}(M)\end{array}\label{Mathematical background - equation - descomposicion campo k-formas}
  \end{align}
  Clearly $g(\vec{V}^{\scriptscriptstyle\top},\vec{n})=0$ so indeed $\vec{V}^{\scriptscriptstyle\top}$ is tangent to the foliation. Analogously $\imath_{\vec{n}}\beta^{\scriptscriptstyle\top}=0$. It is interesting to note that the map
  \begin{equation}\label{Mathematical background - equation - d^top}
  \mathrm{d}^{\!\top}:=\mathrm{d}-\varepsilon n\wedge \imath_{\vec{n}}\mathrm{d}
  \end{equation}
  is defined in such a way that, when pulled back to a single leave, it is the exterior differential map (the normal part vanishes).

	  \subsubsection*{Towards parametrization}\trassub
	  
	  In chapters \ref{Chapter - Parametrized theories} and \ref{Chapter - Parametrized EM} we will introduce some parametrized theories that include diffeomorphisms as dynamical variables. For their study we will need to know how the previous decomposition varies when we perform a diffeomorphism $Z:M\to(M,g)$ because it changes the foliation. Notice that, in order to preserve the space-time structure, we need to consider the submanifold
	  \[\mathcal{D}=\Big\{Z\in\mathrm{Diff}(M)\ \ /\ \ Z\big(\{t\}\times\Sigma\big)\text{ is space-like and }Z_*\partial_t\text{ future directed}\Big\}\subset\mathrm{Diff}(M)\]
	  Given a vector field $V=\mathrm{d}t(V)\partial_t+\Pi_*(V)\in\mathfrak{X}(M)$ we can push it forward and from linearity we have $Z_*V=\mathrm{d}t(V)\dot{Z}+\overline{Z}_*V\in\mathfrak{X}(M)$ where we have defined, for reasons that will be clear in the next section
	  \[\dot{Z}:=Z_*\partial_t\qquad\qquad\overline{Z}_*:=Z_*\Pi_*\]
	  Given a metric $g$ on $M$ we consider the pullback metric $\peqsubfino{g}{Z}{-0.2ex}:=Z^*\!g$ and compute the triples $(\peqsub{\Lambda}{Z},\peqsub{N}{Z},\peqsubfino{\gamma}{Z}{-0.2ex})$ in terms of $g$ and $Z$
	  \begin{align*}
	  &\peqsub{\Lambda}{Z}:=\varepsilon (g\smallcirc Z)\!\left(\dot{Z},\dot{Z}\right)&
	  &\peqsub{N}{Z}(\,\cdot\,):=(g\smallcirc Z)\!\left(\dot{Z},\overline{Z}_*\,\cdot\,\right)&
	  &\peqsubfino{\gamma}{Z}{-0.2ex}(\,\cdot\,,\cdot\,):=(g\smallcirc Z)\!\left(\overline{Z}_*\,\cdot\,,\overline{Z}_*\,\cdot\,\right)
	  \end{align*}
	  We could now define the objects $(\peqsub{\mathbf{N}}{Z},\peqsub{\vec{N}}{Z},\peqsubfino{\zeta}{Z}{-0.2ex})$ in an analogous way but, as the inverse of $\peqsub{g}{Z}$ is $(Z^{-1})_*g^{-1}$, the expressions are a bit cumbersome to handle and it is better to define them in terms of the equivalent data $(\peqsub{\Lambda}{Z},\peqsub{N}{Z},\peqsubfino{\gamma}{Z}{-0.2ex})$ by using the properties of the previous section. First, we define the pseudo-inverse of $\peqsubfino{\gamma}{Z}{-0.2ex}$ by $\peqsubfino{\zeta}{Z}{-0.2ex}\cdot{}\peqsubfino{\gamma}{Z}{-0.2ex}=\Pi$ and $\peqsubfino{\zeta}{Z}{-0.2ex}(\mathrm{d}t,\cdot{}\,)=0$, which allows us to define the lapse and the shift
	  \[\varepsilon\peqsub{\mathbf{N}}{Z}^2=g(\dot{Z},\dot{Z})-\peqsubfino{\zeta}{Z}{-0.2ex}(\peqsub{N}{Z},\peqsub{N}{Z})\qquad\qquad\peqsub{\vec{N}}{Z}(\,\cdot\,)=\peqsubfino{\zeta}{Z}{-0.2ex}(\peqsub{N}{Z},\cdot{}\,)\]
	  We have the future-directed, unitary vector field which is $g$-normal to the foliation $Z(\{t\}\times\Sigma)$
	  \[\peqsub{\vec{n}}{Z}=Z_*\left(\frac{\partial_t-\peqsub{\vec{N}}{Z}}{\peqsub{\mathbf{N}}{Z}}\right)\qquad\equiv\qquad \peqsub{\vec{n}}{Z}^\beta=\frac{\dot{Z}^\beta-(Z_*)^\beta_b\peqsub{\vec{N}}{Z}^b}{\peqsub{\mathbf{N}}{Z}\smallcirc Z}\]
	  Now it is easy to check that $\peqsub{\mathbf{N}}{Z}=\varepsilon \peqsub{n}{Z}(\dot{Z}):=\varepsilon g(\dot{Z},\peqsub{\vec{n}}{Z})$ where $\peqsub{n}{Z}$ is the $g$-metrically equivalent $1$-form field $(\peqsub{n}{Z})_\alpha=g_{\alpha\beta}\peqsub{\vec{n}}{Z}^\beta$.

	  \subsubsection*{From diffeomorphisms to embeddings}\trassub
	  
	  Given a globally hyperbolic space-time $(M=\R\times\Sigma,g)$ and a diffeomorphism $Z\in\mathrm{Diff}(M)$, we have that $Z(\{t\}\times\Sigma)$ is a leaf for every $t$. We can then define for every $t$ the embedding $Z_t:\Sigma\to M$ given by $Z_t(\sigma)=Z(t,\sigma)$ and pullback some objects to $\Sigma$ in order to have a dynamical version of our theory at hand. We formalize this ideas in the following section.

  \subsection{The space of embeddings}\label{Mathematical background - Section - Space of embeddings}
    
    \subsubsection*{Smooth manifold}\trassub
    
      We consider $\mathrm{Emb}(\Sigma,M)\subset\Cinf{\Sigma,M}$ the \textbf{space of embeddings}\index{Embedding}\index{Space of embeddings} ---endowed with the Whitney topology--- from the $(n-1)$-manifold $\Sigma$ into the $n$-manifold $M$ (see \cite{michor1980manifolds} for a detailed discussion). We will consider that $M\cong [t_0,t_1]\times\Sigma$ is globally hyperbolic. This implies that the (possibly empty) boundary has three components $\partial M=\partial_0M\cup\peqsub{\partial}{\Sigma}M\cup\partial_1M$ where $\partial_iM\cong\{t_i\}\times\Sigma$ is the \textbf{temporal boundary} and $\peqsub{\partial}{\Sigma}M\cong [t_0,t_1]\times\partial\Sigma$ the \textbf{spatial boundary}. Notice that a Lorentzian metric $g$ of $M$ induces a Lorentzian metric $\peqsubfino{\jmath}{\partial}{-0.2ex}^*g$ through the inclusion $\peqsubfino{\jmath}{\partial}{-0.2ex}:\peqsub{\partial}{\Sigma}M\to M$ as well as a unitary, outer, normal vector field $\vec{\mathcal{V}}\in\Gamma(\peqsubfino{\jmath}{\partial}{-0.2ex}^*TM)$ which is, of course, space-like.\separ
      
      We recall that, intuitively, an embedding $X$ places a copy of $\Sigma$ inside $M$ as a nice hypersurface without self-intersections or corners. However, it is important to mention that, from a physical point of view, only the shape of $X(\Sigma)$ matters. This suggests to consider, instead of the space of embeddings, the \textbf{space of shapes}\index{Space of shapes}\label{Mathematical background - definition - shapce space} \cite{bauer2012almost} $\mathrm{Emb}(\Sigma,M)/\mathrm{Diff}(\Sigma)$ defined from the action
      \[\begin{array}{cccc}
        \mathrm{Diff}(\Sigma)\times\mathrm{Emb}(\Sigma,M)&\longrightarrow&\mathrm{Emb}(\Sigma,M)\\
        (\phi,X)\qquad & \longmapsto & X\smallcirc \phi
      \end{array}\]
      Nonetheless, such infinite dimensional manifold is much more difficult to handle so we will work with $\mathrm{Emb}(\Sigma,M)$ remembering that we have some sort of ``gauge freedom''. Actually we will deal with the smaller submanifold $\mathrm{Emb}_{g\textrm{-sl}}(\Sigma,M)$ of embeddings such that $X(\Sigma)$ is space-like and, when necessary, with \[\mathrm{Emb}^\partial_{g\textrm{-sl}}(\Sigma,M)=\Big\{X\in\mathrm{Emb}_{g\textrm{-sl}}(\Sigma,M)\ \ /\ \ f(\partial\Sigma)\subset \partial M\Big\}\]
      
      The idea now is to consider some objects defined over the space-time $M$ and see how they evolve when we ``advance in time'' through a given foliation. Imagine for instance that we know the whole history of the temperature $T$ in the universe.
      
      \mbox{}
     \vspace*{-1.5ex}
      
      \centerline{\includegraphics[width=0.88\linewidth]{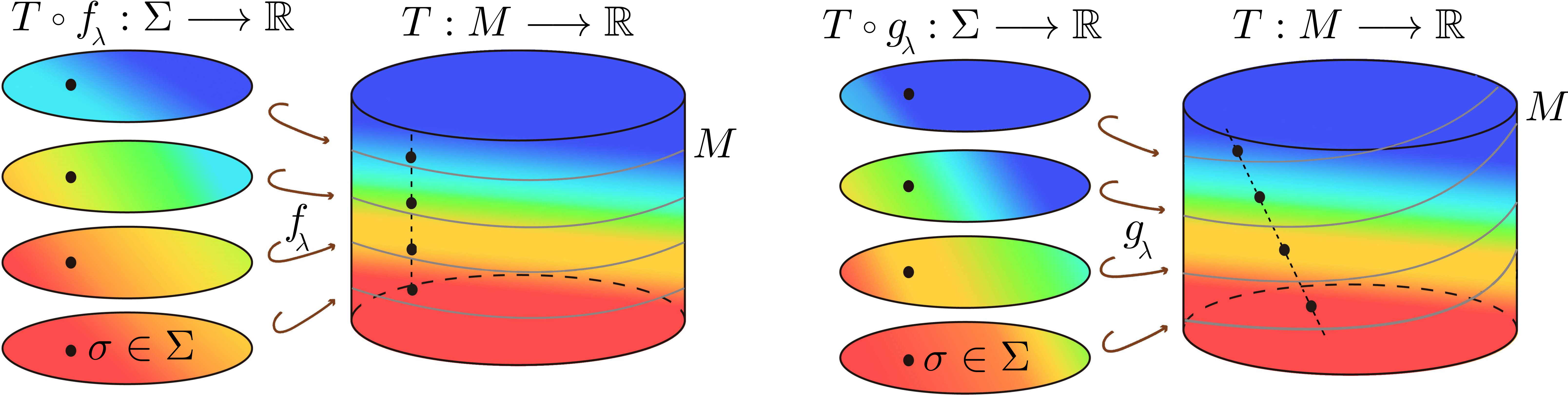}}
      
      \vspace*{2ex}
      
      The previous image gives an idea about how the evolution of the temperature depends on the chosen foliation $\{f_\lambda\}$ or $\{g_\lambda\}$ i.e.\ on the observers. The best way to deal with that is to realize that a foliation can be understood as a curve in the space of embeddings (an idea that will be later formalized) and that the evolution depends on the direction ---tangent vectors--- of such curve. Thus, we need to study the tangent bundle $T\mathrm{Emb}(\Sigma,M)$. Finally notice that not all curves of embeddings define a foliation as the embeddings might ``turn back'' or ``get stuck'' at some point. 
      
   \subsubsection*{Tangent space}\trassub
   
     We can define the tangent space of an infinite dimensional manifold as the equivalence class of curves going through a point with the ``same tangent vector''. In the case of a manifold of maps, a useful geometric interpretation is available: a single vector $\peqsub{\mathbb{V}}{\!X}\in \peqsub{T}{X}\mathrm{Emb}(\Sigma,M)$ is a vector field along the embedding $X:\Sigma\hookrightarrow M$.
   
   \begin{lemma}\mbox{}\label{Mathematical background - lemma - campos de vectores sobre Emb}
   	\begin{itemize}
   	  \item $\peqsub{T}{X}\mathrm{Emb}(\Sigma,M)=\Gamma(X^*TM)=\Big\{V:\Sigma\to TM \ \ /\ \ \pi\smallcirc V=X\Big\}$\vspace*{-1.5ex}
   	  \item $\peqsub{T}{X}\mathrm{Emb}^\partial_{g\textrm{-sl}}(\Sigma,M) = \Gamma^\partial\!(X^*TM):=\Big\{\peqsub{\mathbb{V}}{\!X}\in \Gamma(X^*TM)\ \ /\ \ \left.\peqsub{\mathbb{V}}{\!X}\right|_{\partial \Sigma}\in \Gamma(X^*T\peqsub{\partial}{\Sigma}M) \Big\}$
    \end{itemize}
	\end{lemma}
    \begin{proof}\mbox{}\\
   	Let $\peqsub{\mathbb{V}}{\!X}\in \peqsub{T}{X}\mathrm{Emb}(\Sigma,M)$ be given by the curve of embeddings $\{X_\lambda\}$ i.e.\ $X_0=X$ and $\dot{X}_0=\peqsub{\mathbb{V}}{\!X}$. For every $\sigma\in \Sigma$ we consider the curve $\{X_\lambda(\sigma)\}$ of $M$ that defines, in turn, a tangent vector of $M$ \[\peqsub{\mathbb{V}}{\!X}(\sigma):=\left.\frac{\mathrm{d}}{\mathrm{d}\lambda}\right|_{\lambda=0}X_\lambda(\sigma)\in \peqsub{T}{X(\sigma)}M\]
   	In particular, we have that $\peqsub{\mathbb{V}}{\!X}:\Sigma\rightarrow TM$ is a vector field along the embedding $X$ i.e.\ such that $\pi\smallcirc \peqsub{\mathbb{V}}{\!X}=X$. Those vector fields are in fact sections of the pullback bundle $X^*TM$.\separ
   	
   	Conversely, given $s\in\Gamma(X^*TM)$ we have that $s(\sigma)\in \peqsub{T}{X(\sigma)}M$ is given by a curve of $M$, thus the vector field can be realized as a curve of embeddings. Therefore $s$ defines a vector of $\peqsub{T}{X}\mathrm{Emb}(\Sigma,M)$.  
   	
   	\vspace*{2ex}
   	 
   	   \centerline{\includegraphics[width=0.7\linewidth]{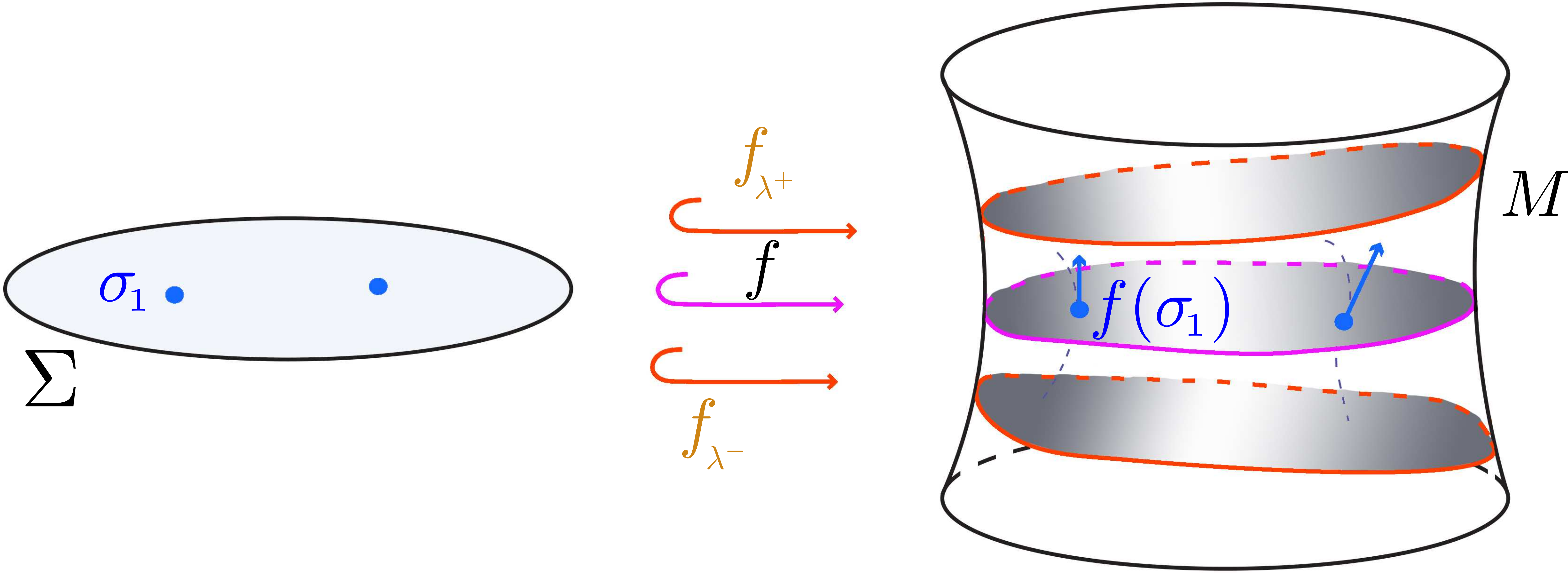}}
   	
   	\vspace*{2ex}
   	
   	For the second statement notice that if $X\in\mathrm{Emb}^\partial_{g\textrm{-sl}}(\Sigma,M)$ then $X(\partial \Sigma)\subset\peqsub{\partial}{\Sigma}M$. In particular the curve $X_\lambda(\sigma)$ for $\sigma\in\partial\Sigma$ lies entirely at the boundary so $v=\dot{X}_0\in \peqsub{T}{X(\sigma)}\peqsub{\partial}{\Sigma}M$.
   \end{proof}

\newpage
   
The previous lemma can be restated in terms of the commutativity of the following diagrams
   
   \begin{center}
   	\begin{tikzcd}
   		& &  TM\arrow{d}{\scalebox{1.2}{\ensuremath{\pi}}}\\
   		\Sigma \arrow{rru}{\scalebox{1.2}{\ensuremath{\peqsub{\mathbb{V}}{\!X}}}} \arrow[swap,hook]{rr}{\scalebox{1.2}{\ensuremath{X}}} &  & M
   	\end{tikzcd}  \qquad\qquad
   	\begin{tikzcd}
   		&  & T\partial_\Sigma M\arrow{d}{\scalebox{1.2}{\ensuremath{\pi}}}\\
   		\partial\Sigma \arrow{rru}{\scalebox{1.2}{\ensuremath{\left.\peqsub{\mathbb{V}}{\!X}\right|_{\partial\Sigma}}}} \arrow[swap,hook]{rr}{\scalebox{1.2}{\ensuremath{\left.X\right|_{\partial\Sigma}}}}  &  & \partial_\Sigma M
   	\end{tikzcd}
   \end{center}
The condition defined by the second diagram implies
   \begin{equation}
   g\left(\peqsub{\mathbb{V}}{\!X},\peqsubfino{\vec{\nu}}{X}{-0.1ex}\right)=0\qquad\text{ over }\qquad\partial\Sigma\label{nuperp}
   \end{equation}
   where $\peqsubfino{\vec{\nu}}{X}{-0.1ex}=\vec{\mathcal{V}}\smallcirc X\smallcirc\imath_\partial$. For generic elements of $\peqsub{T}{X}\mathrm{Emb}_{g\textrm{-sl}}(\Sigma,M)$ only the left diagram applies.

    \subsubsection*{Embedding-dependent objects}\trassub
    
    We devote this section to gather some relevant embedding-dependent geometrical objects. We will use subscripts to explicitly show the dependence on the embeddings.\separ
      
      $\blacktriangleright$ By restriction\separprevia
      
      Given $F\in\Cinf{M}$ a smooth function (like the aforementioned example of the temperature), we define for every $X\in\mathrm{Emb}(\Sigma,M)$ the map $\peqsubfino{f}{X}{-0.1ex}\in\Cinf{\Sigma}$ as $\peqsubfino{f}{X}{-0.1ex}:=F\smallcirc X$. The same procedure applies for more general tensor fields (like the normal vector field $\mathcal{V}$). However, notice that now we do not obtain an object over $\Sigma$, as it was the case for $\peqsubfino{f}{X}{-0.1ex}$, but we end up with a tensor field along the embedding $X:\Sigma\hookrightarrow M$. Indeed, given a vector field $V\in\mathfrak{X}(M)$ we can define $\peqsub{v}{X}\in\Gamma(X^*TM)$ by $\peqsub{v}{X}:=V\smallcirc X$.\separ
      
      Although $f:\mathrm{Emb}(\Sigma,M)\to\Cinf{\Sigma}$ is not a smooth map over the space of the embeddings, we can in fact consider $v$ as a vector field $v:\mathrm{Emb}(\Sigma,M)\to T\mathrm{Emb}(\Sigma,M)$ of the space $\mathrm{Emb}(\Sigma,M)$ because we have $\peqsub{v}{X}\in\Gamma(X^*TM)=\peqsub{T}{X}\mathrm{Emb}(\Sigma,M)$ thanks to lemma \ref{Mathematical background - lemma - campos de vectores sobre Emb}. In fact, relying on this lemma, we can introduce more general vector fields $\mathbb{V}\in\mathfrak{X}(\mathrm{Emb}(\Sigma,M))$ by defining, for every $X\in\mathrm{Emb}(\Sigma,M)$, $\peqsub{\mathbb{V}}{\!X}$ as a vector field along $X$.\separpost

      $\blacktriangleright$ Vector fields\separprevia
      
      Given a globally hyperbolic manifold let us construct $\mathbbm{n}\in\mathfrak{X}(\mathrm{Emb}(\Sigma,M))$ following the idea of the previous paragraph i.e.\ by defining each single vector $\peqsub{\mathbbm{n}}{X}\in\peqsub{T}{X}\mathrm{Emb}(\Sigma,M)$ through lemma \ref{Mathematical background - lemma - campos de vectores sobre Emb}. Thus we take $\peqsub{\mathbbm{n}}{X}:=\peqsub{\vec{n}}{X}$ where $\peqsub{\vec{n}}{X}\in\Gamma(X^*TM)$ is the future directed, unitary, normal vector field to $X(\Sigma)$. Notice that $\peqsub{\vec{n}}{X}$ is time-like as the leaf $X(\Sigma)$ is space-like. We denote by $\peqsub{n}{X}\in\Gamma(X^*T^*\!M)$ the metrically equivalent $1$-form field along the embedding.
      
      \vspace*{0.5ex}
      
      \centerline{\includegraphics[width=0.7\linewidth]{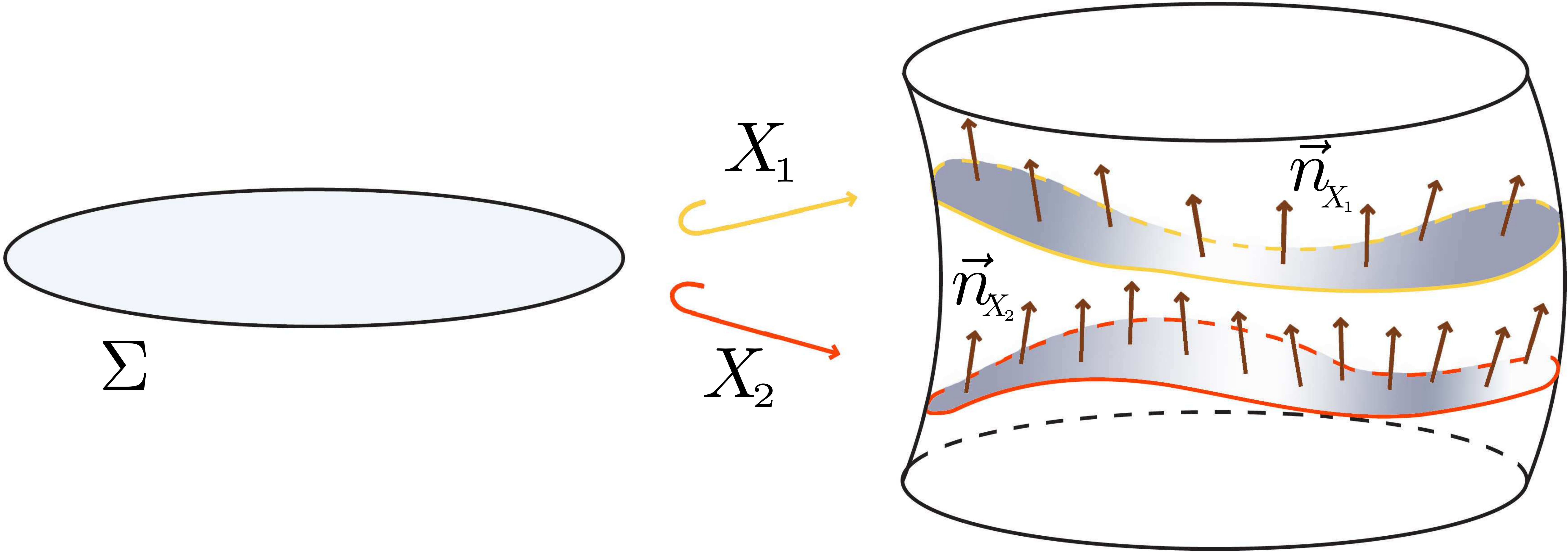}}
      
      \vspace*{.5ex}
      
      Another possibility is to define a vector field $\mathbbm{v}\in\mathfrak{X}(\mathrm{Emb}(\Sigma,M))$ in terms of a fixed vector field $v\in\mathfrak{X}(\Sigma)$ via $\peqsub{\mathbbm{v}}{X}:=X_*v$. Notice that, in fact, it is tangent to the leaf so $X_*v\in\peqsub{\mathfrak{X}}{X}^{\scriptscriptstyle\top}\!(M)$. Actually, the pushforward of $X$ plays such a relevant role that it deserves its own consideration.\separpost

      $\blacktriangleright$ Pushforward\separprevia
      
      For every $X\in\mathrm{Emb}(\Sigma,M)$ we define
      \begin{equation}\peqsub{\tau}{X}:=X_*\qquad\equiv\qquad\tensor*{(\peqsub{\tau}{X})}{*^{\alpha_1}_{b_1}^{\cdots}_{\cdots}_{b_r}^{\alpha_r}}=(X_*)^{\alpha_1}_{b_1}\cdots(X_*)^{\alpha_r}_{b_r}
      \end{equation}
      
      \centerline{\includegraphics[width=0.8\linewidth,clip,trim=0ex 0ex 0ex 0ex]{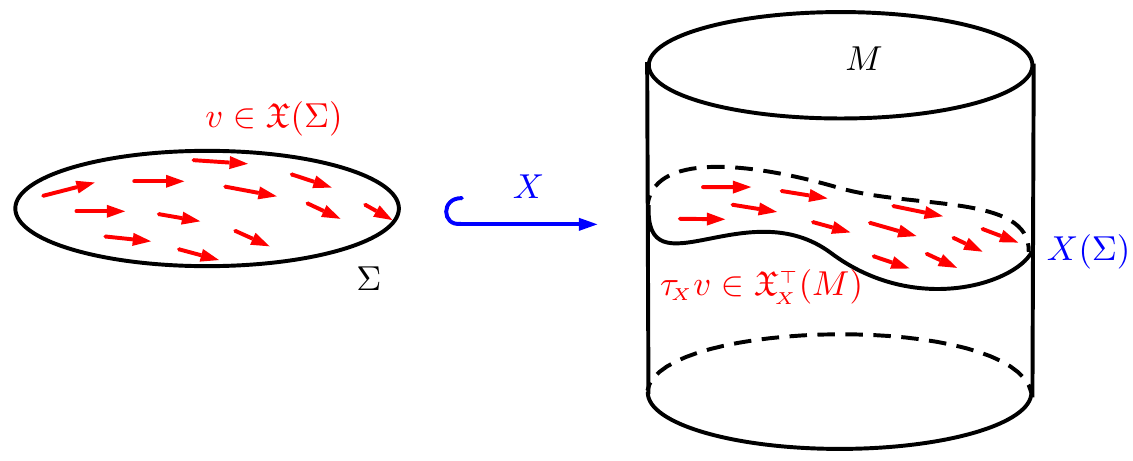}}
  
      Notice that it mixes indices of $\Sigma$ ($a,b\ldots$) with the ones of $M$ ($\alpha,\beta\ldots$). In particular it can act over $(r,0)$-tensor fields of $\Sigma$ as well as $(0,r)$-tensor fields of $M$. In the latter case we have that $\peqsub{\tau}{X}\beta=X^*\!\beta$ for $\beta\in\Omega^k(M)$ because $(X^*\!\beta)(v_1,\cdots,v_r)=\beta(X_*v_1,\cdots,X_*v_r)$.\separpost

      $\blacktriangleright$ Metric\separprevia
      
      We can define tensor fields over $\Sigma$ by pulling back $(0,r)$-tensor fields through $\tau$. The most important one being $\tau.g$ which we will denote, in order to be consistent with section \ref{Mathematical background - subection - pysics over space-time}, as
      \begin{equation}\begin{array}{cccc}
      \gamma:&\mathrm{Emb}_{g\textrm{-sl}}(\Sigma,M)&\longrightarrow&\mathrm{Met}(\Sigma)\\[1ex]
      &          X				  &  \longmapsto  & \peqsubfino{\gamma}{X}{-0.2ex}:=X^*\!g
      \end{array}
      \end{equation}
      $\peqsubfino{\gamma}{X}{-0.2ex}$ is indeed a Riemannian metric because $X(\Sigma)$ is space-like. Let us see what is the relation between $\peqsubfino{\gamma}{X}{-0.2ex}$ and the previously defined $\peqsubfino{\widetilde{\gamma}}{X}{-0.2ex}$. From the definition of $\peqsubfino{\gamma}{X}{-0.2ex}$ and $\peqsubfino{\widetilde{\gamma}}{X}{-0.2ex}$, and the fact that $(\peqsub{\Pi}{X})_*X_*=X_*$, we have
      \begin{equation}\label{Mathematical background - equation - gamma=tilde gamma X_*X_*}
      \peqsubfino{\gamma}{X}{-0.2ex}(v,w)=\peqsubfino{\widetilde{\gamma}}{X}{-0.2ex}(\peqsub{\tau}{X}v,\peqsub{\tau}{X}w)\qquad\equiv\qquad(\peqsubfino{\gamma}{X}{-0.2ex})_{bc}=(\peqsubfino{\widetilde{\gamma}}{X}{-0.2ex})_{\alpha\beta}(X_*)^\alpha_b(X_*)^\beta_c
      \end{equation}
      analogously, using $X^*\Pi^*=X^*$, we have that the inverse of $\peqsubfino{\gamma}{X}{-0.2ex}$ is related to $\peqsubfino{\zeta}{X}{-0.2ex}$ via
      \begin{equation}\label{Mathematical background - equation - eta=gamma^{-1}X_*X_*}
      \peqsubfino{\gamma}{X}{-0.2ex}^{-1}(\peqsub{\tau}{X}\omega_1,\peqsub{\tau}{X}\omega_2)=\peqsubfino{\zeta}{X}{-0.2ex}(\omega_1,\omega_2)\qquad\equiv\qquad\peqsubfino{\zeta}{X}{-0.2ex}^{\alpha\beta}=\peqsubfino{\gamma}{X}{-0.2ex}^{ab}(X_*)^\alpha_a(X_*)^\beta_b
      \end{equation}
      Once we have a metric $\peqsubfino{\gamma}{X}{-0.2ex}$ over $\Sigma$, we have also the associated embedding-dependent LC connection $\nabla^{(X)}$ (we will usually omit the superscript) as well as the metric volume form $\peqsubfino{{\mathrm{vol}_\gamma}}{\!X}{-0.2ex}$.\separpost
      
$\blacktriangleright$ ``Inverse'' of $\tau$\separprevia
    
    We can now lower all the $M$-indices with $g$ and raise the $\Sigma$-indices with $\peqsubfino{\gamma}{X}{-0.2ex}$ to obtain the pseudoinverse of $\tau$ that we denote $e$
    \begin{equation}
    \tensor*{(\peqsub{e}{X})}{*_{\alpha_1}^{b_1}^{\cdots}_{\cdots}^{b_r}_{\alpha_r}}=g_{\alpha_1\gamma_1}\cdots g_{\alpha_r\gamma_r}\tensor*{(\peqsub{\tau}{X})}{*^{\gamma_1}_{c_1}^{\cdots}_{\cdots}_{c_r}^{\gamma_r}}\peqsubfino{\gamma}{X}{-0.2ex}^{c_1b_1}\cdots\peqsubfino{\gamma}{X}{-0.2ex}^{c_rb_r}=\prod_{i=1}^rg_{\alpha_i\gamma_i}(X_*)^{\gamma_i}_{c_i}\peqsubfino{\gamma}{X}{-0.2ex}^{c_ib_i}
    \end{equation}
    It can act over $(0,r)$-tensor fields of $\Sigma$ as well as $(r,0)$-tensor fields of $M$. Using equations \eqref{Mathematical background - equation - gamma=tilde gamma X_*X_*} and \eqref{Mathematical background - equation - eta=gamma^{-1}X_*X_*} and the definitions of $\tau$ and $e$, it is easy to check that they are indeed pseudoinverses
    \begin{equation}\label{Mathematical background - equation - tau.e=Id e.tau=Id - cosas}
      \tensor*{(\peqsub{\tau}{X})}{*^{\alpha_1}_{b_1}^{\cdots}_{\cdots}^{\alpha_1}_{b_1}}\tensor*{(\peqsub{e}{X})}{*_{\alpha_1}^{c_1}^{\cdots}_{\cdots}_{\alpha_1}^{c_1}}=\prod_{i=1}^r\delta_{b_i}^{c_i}\qquad\qquad
      \tensor*{(\peqsub{e}{X})}{*_{\alpha_1}^{c_1}^{\cdots}_{\cdots}_{\alpha_1}^{c_1}}\tensor*{(\peqsub{\tau}{X})}{*^{\beta_1}_{c_1}^{\cdots}_{\cdots}^{\beta_1}_{c_1}}=\prod_{i=1}^r\Big(\delta_{\alpha_i}^{\beta_i}-\varepsilon (\peqsub{n}{X})_{\alpha_i}\peqsub{n}{X}^{\beta_i}\Big)
    \end{equation}
    
    $\blacktriangleright$ Decomposition\separprevia
    
    We have seen that, given a metric $g$ and a foliation we can decompose a vector field $V\in\mathfrak{X}(M)$ as
    \begin{align}
    &V=V^{\scriptscriptstyle\perp}\vec{n}+\vec{V}^{\scriptscriptstyle\top}&&\hspace*{-4ex}\text{where}\qquad\qquad\begin{array}{|l}
    V^{\scriptscriptstyle\perp}=\varepsilon g(V,\vec{n})\in\Cinf{M}\\[1ex] \vec{V}^{\scriptscriptstyle\top}=V-V^{\scriptscriptstyle\perp}\vec{n}\in\mathfrak{X}(M)\end{array}
    \end{align}
    We can, of course, consider the decomposition with respect to a single embedding in the normal direction $\peqsub{\vec{n}}{X}\in\Gamma(X^*TM)$ and the tangent direction. Indeed, take $V\in\Gamma(X^*TM)$ a vector field along $X\in\mathrm{Emb}(\Sigma,M)$, then
    \begin{align}\label{Mathematical background - equation - decomposition embedding}
    &V=\peqsub{V}{X}^{\scriptscriptstyle\perp}\peqsub{\vec{n}}{X}+\peqsub{\tau}{X}.\peqsub{\vec{v}}{X}^{\scriptscriptstyle\top}&&\hspace*{-4ex}\text{where}\qquad\qquad\begin{array}{|l}
    \peqsub{V}{X}^{\scriptscriptstyle\perp}=\varepsilon (g\smallcirc X)(V,\peqsub{\vec{n}}{X})\in\Cinf{\Sigma}\\[1ex] \peqsub{\tau}{X}.\peqsub{\vec{v}}{X}^{\scriptscriptstyle\top}:=V-\peqsub{V}{X}^{\scriptscriptstyle\perp} \peqsub{\vec{n}}{X}\end{array}
    \end{align}
    Notice that $\peqsub{\vec{v}}{X}^{\scriptscriptstyle\top}\in \mathfrak{X}(\Sigma)$ is well defined because $V-\peqsub{V}{X}^{\scriptscriptstyle\perp} \peqsub{\vec{n}}{X}\in\peqsub{\mathfrak{X}^{\scriptscriptstyle\top}}{X}(M)$ and $\peqsub{\tau}{X}$ is an isomorphism over its image.\separpost

    $\blacktriangleright$ Curvatures\separprevia
    
    The intrinsic curvatures (Riemann, Ricci and scalar) depend on the embedding through the metric $\peqsubfino{\gamma}{X}{-0.2ex}$. On the other hand, the dependence of $\peqsub{K}{X}$ on the embedding involves several geometric objects. It is given by
    \begin{equation}\label{Mathematical background - equation - K=tau nabla n}
    (\peqsub{K}{X})_{ab}=-(\peqsub{\tau}{X})^\beta_b\nabla^{(X)}_a (\peqsub{n}{X})_\beta
    \end{equation}
    which comes from the Weingarten equation \eqref{Mathematical background - equation - Weingarten} and the identities relating $\tau$ and $e$ \eqref{Mathematical background - equation - tau.e=Id e.tau=Id - cosas}.
    
    \subsubsection*{Variations}\index{Embeddings!Variations}\trassub
    
    It is now natural to wonder how an embedding-dependent object $T$ changes when we vary the embedding. The obvious approach of computing derivatives like in
    \[\frac{f(x+h)-f(x)}{h}\qquad f\in\Cinf{M}\]
    is not available because $\mathrm{Emb}(\Sigma,M)$ is not linear in general so we can not add a ``small'' element analogous to $h$ to an embedding $X$. We have, nonetheless, the linear infinitesimal version of the manifold i.e.\ the tangent bundle $T\mathrm{Emb}(\Sigma,M)$ which solves that problem. Indeed, it seems reasonable to consider a vector defined by a curve of embeddings $\{X_\lambda:\Sigma\hookrightarrow M\}_\lambda$ and compute the directional derivative $\left.\partial_\lambda\right|_0T(X_\lambda)$. Notice however that $T$ is in general a section of some bundle over $M$ so the direct derivation is not available either because we can not substract $T(X_{\lambda+h})-T(X_\lambda)$. Nonetheless, let us see how to make sense of the previous expression by using the covariant derivative.\separ
    
    The trick here is to rely on the powerful framework provided by \emph{convenient calculus}, which considers some special infinite dimensional manifolds, called \textbf{convenient manifolds}\index{Convenient manifold}, for which many of the powerful results available for finite dimensional manifolds are still valid. We will not get into the technical details (they can be looked up at \cite{kriegl1997convenient}) and only mention the following result that will be essential for our purposes:
    \begin{equation}\label{Mathematical background - equation - C(R,C(M,N)) cong C(R x M,N)}
    \mathcal{C}^\infty\hspace{-0.25ex}\Big(\R,\Cinf{M,N}\Big)\,\longleftrightarrow\ \Cinf{\R\times M,N} 
    \end{equation}
    for convenient manifolds $M$ and $N$. This expression implies, in particular, that a curve of embeddings $X:\R\to\mathrm{Emb}(\Sigma,M)$ can be understood as a map $X:\R\times \Sigma\to M$ such that $X(t,\cdot{}\,)\in\mathrm{Emb}(\Sigma,M)$ for every $t\in\R$. Following \cite{bauer2012almost}, if we pullback all the vector bundles over $M$ to $\R\times M$, where have defined a connection $\nabla$, and consider the vector field $\partial_t:=(\partial_t,0)\in\mathfrak{X}(\R\times M)$, we can define
    \begin{equation}\label{Mathematical background - equation - definition variation}
    \peqsub{D}{\left(X,\mathbb{Y}_{\!X}\!\right)}\,T:=\left.\nabla_{\!\raisemath{-2pt}{\!\scriptscriptstyle\partial_t}}\Big(T(X_t)\Big)\right|_0
    \end{equation}
    as the directional derivative of $T$ at $X$ in the $\peqsub{\mathbb{Y}}{\!X}\in\peqsub{T}{X}\mathrm{Emb}(\Sigma,M)$ direction, where $X_0=X$ and $\dot{X}_0=\peqsub{\mathbb{Y}}{\!X}$. Let us see the variation of the objects previously defined (see page \pageref{Appendix - section - embeddings} of the appendix for the proofs), starting with the ones defined by restriction $\peqsubfino{f}{X}{-0.1ex}:=F\smallcirc X$ and $\peqsub{v}{X}:=V\smallcirc X$
    \begin{equation}\peqsub{D}{\left(X,\mathbb{Y}_{\!X}\!\right)}\,\peqsubfino{f}{X}{-0.1ex}=\nabla_{\!\!\raisemath{-2pt}{\scriptscriptstyle\mathbb{Y}_{\!X}}} F\qquad\qquad\peqsub{D}{\left(X,\mathbb{Y}_{\!X}\!\right)}\,\peqsub{v}{X}=\nabla_{\!\raisemath{-2pt}{\scriptscriptstyle\mathbb{Y}_{\!X}}} V
    \end{equation}
    Let us make a very important remark that is a consequence of lemma \ref{Mathematical background - lemma - campos de vectores sobre Emb}: while at the LHS on the previous equations $\peqsub{\mathbb{Y}}{\!X}$ has to be understood as a single vector $\peqsub{\mathbb{Y}}{\!X}\in\peqsub{T}{X}\mathrm{Emb}(\Sigma,M)$ that indicates a direction on the space of embeddings, on the RHS it is a vector field along the embedding $\peqsub{\mathbb{Y}}{\!X}\in\Gamma(X^*TM)$ that indicates the direction in $M$ in which the whole embedding moves.
    
    \vspace*{1ex}
    
    \centerline{\includegraphics[width=.65\linewidth]{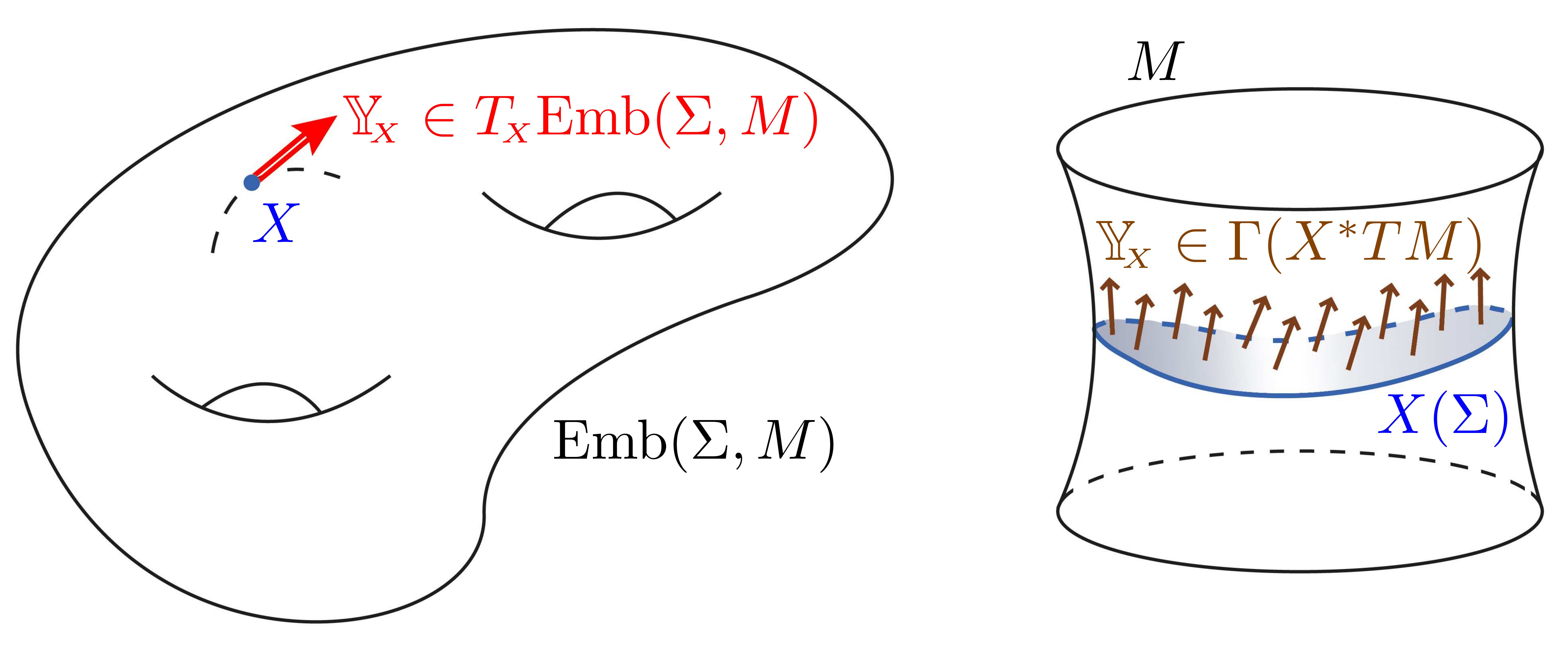}}
    
    \vspace*{1ex}
    
    It is convenient in the following to use the decomposition $\peqsub{\mathbb{Y}}{\!X}=\peqsub{Y}{X}^{\scriptscriptstyle\perp} \peqsub{\vec{n}}{X} + \peqsub{\tau}{X}\peqsub{\vec{Y}}{X}^{\scriptscriptstyle\top}$ because $(\peqsub{Y}{X}^{\scriptscriptstyle\perp},\peqsub{\vec{Y}}{X}^{\scriptscriptstyle\top})$ keeps track of the tangential and perpendicular parts of the variation. It will be useful to define
    \begin{equation}
    	\big(\peqsub{M}{\,\mathbb{Y}_{\!X}}\big)_{\!\raisemath{.7pt}{a}}^{\!\phantom{a}\raisemath{-1.5pt}{b}}:=\nabla_a \peqsub{Y}{X}^b-\peqsub{Y}{X}^{\scriptscriptstyle\perp} (\peqsub{K}{X})^b_a\qquad\qquad\qquad\big(\peqsub{m}{\,\mathbb{Y}_{\!X}}\big)_{\!\raisemath{.7pt}{a}}:=(\peqsub{K}{X})_{ab} \peqsub{Y}{X}^b+\varepsilon (\mathrm{d}\peqsub{Y}{X}^{\scriptscriptstyle\perp})_a
    \end{equation}
    where we denote $\peqsub{Y}{X}^a=(\peqsub{\vec{Y}}{X}^{\scriptscriptstyle\top})^a$ to simplify the notation as no confusion is possible. In fact, for simplicity, we will often omit the $X$ subscript.\separ
    
    Now we write down the variations of the objects that depend directly on the embedding
      \begin{equation}\label{Mathematical background - equations - variaciones}
      \begin{array}{lcl}
      \bullet\ \left(\peqsub{D}{\left(X,\mathbb{Y}_X\!\right)}\,n\right){}_{\!\alpha}=-e^b_\alpha \big(\peqsub{m}{\,\mathbb{Y}_{\!X}}\big)_{\!\raisemath{.7pt}{b}}&\quad
      &\bullet\ \left(\peqsub{D}{\left(X,\mathbb{Y}_X\!\right)}\,n\right){}^{\!\alpha}=-\tau^\alpha_b(\peqsub{m}{\,\mathbb{Y}_{\!X}})^b\\[1.5ex]
      \bullet\ \left(\peqsub{D}{\left(X,\mathbb{Y}_X\!\right)}e\right){}^{\!b}_{\!\alpha}=\varepsilon n_\alpha (\peqsub{m}{\,\mathbb{Y}_{\!X}})^b-e^c_\alpha \big(\peqsub{M}{\,\mathbb{Y}_{\!X}}\big)_{\!\raisemath{.7pt}{a}}^{\!\phantom{c}\raisemath{-1.5pt}{b}}&
      &\bullet\ \left(\peqsub{D}{\left(X,\mathbb{Y}_X\!\right)}\,\tau\right){}_{\!b}^{\!\alpha}=\varepsilon n^\alpha \big(\peqsub{m}{\,\mathbb{Y}_{\!X}}\big)_{\!\raisemath{.7pt}{b}}+\tau^\alpha_c\big(\peqsub{M}{\,\mathbb{Y}_{\!X}}\big)_{\!\raisemath{.7pt}{a}}^{\!\phantom{b}\raisemath{-1.5pt}{c}}\\[2ex]
      \bullet\ \left(\peqsub{D}{\left(X,\mathbb{Y}_X\!\right)}\,\gamma\right){}_{\!bc}=\big(\peqsub{M}{\,\mathbb{Y}_{\!X}}\big)_{\!\raisemath{.7pt}{bc}}+\big(\peqsub{M}{\,\mathbb{Y}_{\!X}}\big)_{\!\raisemath{.7pt}{cb}}
      \end{array}
      \end{equation}
      where the Latin abstract indices are raised and lowered with the metric $\peqsubfino{\gamma}{X}{-0.2ex}$ Notice that if we take $\peqsub{Y}{X}^{\scriptscriptstyle\perp}=0$ the variation is taken in a direction such that the shape $X(\Sigma)$ remains unchanged. In fact all the information is coded into the diffeomorphism of $\Sigma$ given by the flow of $\peqsub{\vec{Y}}{X}^{\scriptscriptstyle\top}\in\mathfrak{X}(\Sigma)$, which suggests that the variations are related to the Lie derivatives \cite[4.2]{bauer2012almost}. For instance we see that $\peqsub{D}{\left(X,\mathbb{Y}_X\!\right)}\,\gamma=\mathcal{L}_{\peqsub{\vec{Y}}{X}^{\scriptscriptstyle\top}}\gamma$ while the variation of $\peqsub{\vec{n}}{X}$ leads to the definition of the Weingarten map.\separ
      
      Finally, we list the variations of the objects that depend directly on the metric in terms of the variation of the metric itself. Why would we do so? There are several reasons, first notice that if the metric $\gamma$ is the pullback metric $X^*\!g$, then all the following objects depend on the embedding. Applying the previous equation leads to formulas with explicit dependence on the embeddings. However, in GR the metric is a variable itself instead of the embeddings (as we will see in chapter \ref{Chapter - General relativity}). Thus we will also be interested in the variations with respect to the metric.      
      \begin{equation}\label{Mathematical background - equations - variaciones2}
      \begin{array}{lcl}
      \displaystyle\bullet\ (D\gamma^{-1})^{dc}=-(D\gamma)^{dc}&\qquad
      &\displaystyle\bullet\ \peqsub{D}{Z}(\star_\gamma)_k=\left([D\gamma]_{n-k}-\dfrac{\mathrm{Tr}(D\gamma)}{2}\mathrm{Id}\right)(\star_\gamma)_k\\[2.2ex]
      \displaystyle\bullet\ D\peqsub{\mathrm{vol}}{\gamma}=\dfrac{\mathrm{Tr}(D\gamma)}{2}\peqsub{\mathrm{vol}}{\gamma}&
      &\displaystyle\bullet\ \tensor{(D\nabla)}{^c_a_b}=\frac{\gamma^{cd}}{2}\Big(\nabla_a\tensor{(D\gamma)}{_d_b}+\nabla_b\tensor{(D\gamma)}{_a_d}-\nabla_d(D\gamma)_{ab}\Big)
      \end{array}
      \end{equation}
      where $[D\gamma]_r$ is defined on equation \eqref{Appendix - equation - (Dg)_k} and $D$ might stand for the previously defined $\peqsub{D}{\left(X,\mathbb{Y}_{\!X}\!\right)}$ or $\peqsub{D}{\left(\gamma,Y_{\boldsymbol{\gamma}}\right)}$, that is the usual derivative of the space of metrics (which has a linear structure).
      
      \subsubsection*{The extrinsic curvature revisited}\index{Curvature!Extrinsic}\trassub
      
      Consider $(M=\R\times\Sigma,g)$ a globally hyperbolic space-time with some fixed foliation defined by $Z\in\mathrm{Diff}(M)$. Then we have that, thanks to \eqref{Mathematical background - equation - C(R,C(M,N)) cong C(R x M,N)}, $Z$ can be considered as a curve of embeddings  $\{Z_t:\{t\}\times\Sigma\to M\}$. Notice that, in particular, we can define for every $t$ all the previous objects associated with the inclusion $\jmath_t:\Sigma_t:=Z_t(\Sigma)\subset M\to M$ that we denote  $\tau_t$, $e_t$, $\gamma_t$ and so on. We can now state a result that will be very useful when dealing with General Relativity.
      \begin{equation}\label{Mathematical background - equation - 2K=L_ng}
      2K_t=-\jmath_t^*\big(\mathcal{L}_{\vec{n}_t}g\big)
      \end{equation}
      The proof is straightforward but we have to be careful about where the objects live (which indices correspond to them)
      \begin{align*}
      (\jmath_t^*\mathcal{L}_{\vec{n}_t}g)_{\bar{\alpha}\bar{\beta}}&=(\mathcal{L}_{\vec{n}_t}g)_{\alpha\beta}\big(\tau_t\big)^\alpha_{\bar{\alpha}}\big(\tau_t\big)^\beta_{\bar{\beta}}=\\
      &=\Big(\nabla_\alpha n_\beta+\nabla_\beta n_\alpha\Big)\big(\tau_t\big)^\alpha_{\bar{\alpha}}\big(\tau_t\big)^\beta_{\bar{\beta}}\overset{\eqref{Mathematical background - equation - nabla_a=tau nabla_alpha}}{=}\\
      &=\big(\tau_t\big)^\beta_{\bar{\beta}}\nabla_{\bar{\alpha}}n_\beta+\big(\tau_t\big)^\alpha_{\bar{\alpha}}\nabla_{\bar{\beta}}n_\alpha\overset{\eqref{Mathematical background - equation - K=tau nabla n}}{=}\\
      &=-(K_t)_{\bar{\beta}\bar{\alpha}}-(K_t)_{\bar{\alpha}\bar{\beta}}=-(K_t)_{\bar{\alpha}\bar{\beta}}
      \end{align*}

 \subsubsection*{Hypersurface deformation algebra}\trassub
 
   With the machinery developed so far, we can now obtain the \textbf{hypersurface deformation algebra}\index{Hypersurface deformation algebra} \cite{dirac1964lectures,kucha1976geometry}. First remember that given a vector field $\mathbb{V}\in\mathfrak{X}(\textrm{Emb}(\Sigma,M))$, the single vector $\peqsub{\mathbb{V}}{\!X}\in \peqsub{T}{X}\mathrm{Emb}_{g\textrm{-sl}}(\Sigma,M)$ can be considered as a vector field along the embedding $X$. According to equation \eqref{Mathematical background - equation - decomposition embedding} we have $\peqsub{\mathbb{V}}{\!X}=\peqsub{V}{X}^{\scriptscriptstyle\perp} \peqsub{\vec{n}}{X} + \peqsub{\tau}{X}\peqsub{\vec{v}}{X}^{\scriptscriptstyle\top}$. Such decomposition can be made over the space of embeddings by simply writing
   \[\mathbb{V}=V^{\scriptscriptstyle\perp} \mathbbm{n} + \tau.\vec{v}^{\scriptscriptstyle\top}\in\mathfrak{X}\big(\textrm{Emb}(\Sigma,M)\hspace*{-0.2ex}\big)\]
   This result is very important because it translates a decomposition over $M$ into a decomposition over the space of embeddings $\mathrm{Emb}(\Sigma,M)$. Note however that, although both addends are vector fields on the space of embeddings, $V^{\scriptscriptstyle\perp}:\textrm{Emb}(\Sigma,M)\to C^\infty(\Sigma)$ is not an element of $\Cinf{\mathrm{Emb}(\Sigma,M)}$ i.e.\ it is not a scalar in the space of embeddings. Analogously, $\vec{v}^{\scriptscriptstyle\top}:\textrm{Emb}(\Sigma,M)\to\mathfrak{X}(\Sigma)$ cannot be considered simply as a vector field over $\mathrm{Emb}(\Sigma,M)$.\separ

   We now define, for $\mathbb{V},\mathbb{W}\in\mathfrak{X}(\mathrm{Emb}(\Sigma,M))$, the Lie bracket
   \[[\mathbb{V},\mathbb{W}]:=\peqsub{D}{\mathbb{V}}\mathbb{W}-\peqsub{D}{\mathbb{W}}\mathbb{V}\in\mathfrak{X}\big(\mathrm{Emb}(\Sigma,M)\hspace*{-0.2ex}\big)\qquad\equiv\qquad\peqsub{[\mathbb{V},\mathbb{W}]}{X}:=\peqsub{D}{(X,\mathbb{V}_X)}\peqsub{\mathbb{W}}{X}-\peqsub{D}{(X,\mathbb{W}_X)}\peqsub{\mathbb{V}}{\!X}\]
   which is equivalent to consider $D$ as a torsion-free covariant derivative over $\mathrm{Emb}(\Sigma,M)$. In lemma  \ref{appendix - lemma - hypersurface deformation algebra}, using equations \eqref{Mathematical background - equations - variaciones} and the decomposition for $\mathbb{V}$ and $\mathbb{W}$, we obtain
   \begin{align}\label{Mathematical background - equation - hypersurface deformation algebra}
   \begin{split}
     [\mathbb{V},\mathbb{W}]&=\Big(\peqsub{D}{\mathbb{V}}\ \!\!W^{\scriptscriptstyle\perp}-\peqsub{D}{\mathbb{W}}\ \!\!V^{\scriptscriptstyle\perp}+\mathrm{d}V^{\scriptscriptstyle\perp}(\vec{w}^{\scriptscriptstyle\top})-\mathrm{d}W^{\scriptscriptstyle\perp}(\vec{v}^{\scriptscriptstyle\top})\Big)\mathbbm{n}+\\
     &\phantom{=}+\tau.\Big(\peqsub{D}{\mathbb{V}}\ \!\!\vec{w}^{\scriptscriptstyle\top}-\peqsub{D}{\mathbb{W}}\ \!\!\vec{v}^{\scriptscriptstyle\top}+\varepsilon\big(V^{\scriptscriptstyle\perp}\nabla^{\gamma}W^{\scriptscriptstyle\perp}-W^{\scriptscriptstyle\perp}\nabla^{\gamma}V^{\scriptscriptstyle\perp}\big)-[\vec{v}^{\scriptscriptstyle\top},\vec{w}^{\scriptscriptstyle\top}]\Big)\\[-5ex]\mbox{}
     \end{split}
   \end{align}
   which is the hypersurface deformation algebra. Notice that in the previous expression, somehow surprisingly, the pullback metric $\peqsubfino{\gamma}{X}{-0.2ex}$ is involved even though it is not a tensor field on the space of embeddings.

  \subsection{Symplectic geometry}\label{Mathematical background - subsection - symplectic}
    \subsubsection*{Introduction}\trassub
    
      We have seen that a metric is a non-degenerate symmetric $(0,2)$-tensor field on $M$. Now let us consider its antisymmetric counterpart which gives raise to symplectic geometry, an essential ingredient in many branches of science \cite{gotay1992symplectization} and of particular importance in Hamiltonian mechanics. A distinctive feature of symplectic geometry is the central role played by areas instead of lengths and angles as in Rimannian geometry.
      
      \begin{definitions}
        \item A \textbf{presymplectic manifold}\index{Presymplectic manifold} $(M,\Omega)$ is a manifold $M$ equipped with a $2$-form $\Omega$ which is closed $\mathrm{d}\Omega=0$ and of constant rank called a \textbf{presymplectic form}\index{Presymplectic form}.
        \item If $\Omega$ is closed $\mathrm{d}\Omega=0$ and non-degenerate, we say that $(M,\Omega)$ is a \textbf{symplectic manifold}\index{Symplectic manifold} and $\Omega$ is a \textbf{symplectic form}\index{Symplectic form}.
        \item A diffeomorphism $F:(M,\peqsub{\Omega}{M})\to(N,\peqsub{\Omega}{N})$ is a \textbf{symplectomorphism}\index{Symplectofmorphism} if $F^*\peqsub{\Omega}{N}=\peqsub{\Omega}{M}$.
      \end{definitions}
      
      The closedness of $\Omega$ can be naively considered to be similar to the requirement $\peqsub{\mathrm{d}}{\nabla}g=0$ because in abstract index notation they can be stated as
      \[\nabla_{[a}\Omega_{bc]}=0_{abc}\qquad\qquad\qquad\nabla^g_ag_{bc}=0_{abc}\]
      However their origin is very different: in the first one the metric is required to be compatible with the additional structure given by the connection $\nabla$, while $\mathrm{d}$ is canonically defined over forms by the differential structure of the manifold ($\mathrm{d}$ is prior to the definition of $\nabla$). Actually, $\mathrm{d}\Omega$ is an obstruction for the manifold to be symplectic in the same manner as $Riem$ is an obstruction for $(M,g)$ to be flat\index{Curvature!Symplectic}. We will see on the next section why it is convenient to restrict ourselves to the ``flat symplectic'' manifolds.\separ
      
      The non-degeneracy of $\Omega$, on the other hand, is analogous to the non-egdegeneracy of $g$ for finite dimensional manifolds. In particular we have the musical isomorphisms\index{Musical isomorphism} (formally analogous to the ones we had before) that allows us to raise and lower indices
      \[\mathfrak{X}(M) \updown{\peqsub{\flat}{\Omega}}{\peqsub{\sharp}{\Omega}}{\myrightleftarrows{\rule{1cm}{0cm}}}\ \Omega^1(M)\]
      For infinite dimensional manifold the non-degeneracy condition is more subtle as it only implies that $\peqsub{\flat}{\Omega}$ is injective. In that case we say that $\Omega$ is \textbf{weakly symplectic}\index{Weakly symplectic!Symplectic form}. If, moreover, $\peqsub{\flat}{\Omega}$ is an isomorphism, we can define its inverse $\peqsub{\sharp}{\Omega}$ and in that case we say that $\Omega$ is \textbf{strongly symplectic}\index{Strongly symplectic!Symplectic form}.\separ

      \subsubsection*{Algebra of observables}\trassub
      
      In classical mechanics it is relatively straightforward to get information from a system. We can ask questions such as where is the center of mass of a particle system or what is its average speed. Given a state space $M$, the appropriate way to ask for the center of mass is given by the function that for a specific state $p\in M$ of the system gives the weighted mean of the positions of all the particles, while the total momentum of the system is given by the sum of the momenta of the individual particles. Such functions are called \textbf{observables of the theory}\index{Observable!Classical}, therefore an observable is any smooth function $f:M\to\R$ taking a state and returning a number. Given two classical observables $f,g\in\Cinf{M}$ we have that, as $(\Cinf{M},+,\cdot{})$ is a ring. Thus we can construct more observables by addition $f+g$ or by multiplication $f\cdot{}g$. There is, however, yet another way to combine functions which plays an important role within the Hamiltonian form of the dynamics of the system called the Poisson bracket.\separ
      
      If $\peqsub{\sharp}{\Omega}$ exists, we can define the \textbf{Poisson bracket}\index{Poisson bracket} $\{\,,\}:\Cinf{M}\times\Cinf{M}\to\Cinf{M}$ as
      	\[\{f,g\}=\Omega\Big(\peqsub{\sharp}{\Omega}(\mathrm{d}f),\peqsub{\sharp}{\Omega}(\mathrm{d}g)\Big)=\Omega^{ab}(\mathrm{d}f)_a(\mathrm{d}g)_b\]
      The Poisson bracket is somehow more ``fundamental'' than the symplectic structure in the sense that from the latter we can define the former but it does not work the other way around\footnote{In this direction we have, nonetheless, an important results: a \textbf{Poisson manifold}\index{Poisson manifold} $M$, which is a manifold such that its algebra $\Cinf{M}$ is endowed with a Lie bracket satisfying the Leibniz rule, can be naturally partitioned into regularly immersed symplectic manifolds (not necessarily of the same dimension), called \textbf{symplectic leaves}.}. It is important to mention that in the presymplectic case we can only define an analog to the Poisson bracket, known as the \textbf{bracket of admissible functions}, over the functions $f$ such that $\mathrm{d}f\in \mathrm{Im}\peqsub{\flat}{\Omega}$. This is strongly related to the Dirac structures \cite{courant1990dirac,courant1988beyond} and the Dirac bracket \cite{dirac1950generalized}.\separ
      
      It is well known that $(\Cinf{M},\cdot{},\{\,,\})$ forms a \textbf{Poisson algebra}\index{Poisson algebra} i.e.\ it is an associative algebra with respect to the multiplication, a Lie algebra\index{Lie algebra} with respect to $\{\,,\}$ and the Poisson bracket obeys the Leibniz rule $\{f,g\cdot{}h\}=\{f,g\}\cdot{}h+g\cdot{}\{g,h\}$.\label{Lie alegbra with Poisson bracket}
  	
  	\subsubsection*{Dynamics}\trassub
  	
  	Symplectic geometry turns out to be a nice setting to define classical dynamics in a way that suggests how to arrive at its quantum counterpart. Let us begin by introducing a couple of definitions.
  	
  		\begin{definition}\mbox{}\\
  			Given some $H\in\Cinf{M}$ we define the \textbf{Hamilton's equation}\index{Hamilton's equation} 
      	\begin{equation}\label{Mathematical background - definition - X_H}
      	 \peqsubfino{{\imath_X}}{\!H}{-0.2ex}\hspace*{0.1ex}\Omega=\mathrm{d}H\qquad\qquad\equiv\qquad\qquad{\textstyle\peqsub{\flat}{\Omega}}\,\peqsub{X}{\!H}=\mathrm{d}H
      	\end{equation}
      	The solution $\peqsub{X}{\!H}$ is known as the \textbf{Hamiltonian vector field}\index{Hamiltonian vector field} of $H$.
      \end{definition}
      If $\Omega$ is symplectic then $\peqsub{X}{\!H}=\peqsub{\sharp}{\Omega}(\mathrm{d}H)$ or, equivalently $(\peqsub{X}{\!H})^a=\Omega^{ab}(\mathrm{d}H)_b$, is a solution to the Hamilton's equation. If, on the other hand, $\Omega$ is just presymplectic the solution is not unique or does not exist. The former case is easy to handle because the degeneracy of the solution comes from the non-injectivity of $\peqsub{\flat}{\Omega}$. Indeed, notice that we can add to a solution of \eqref{Mathematical background - definition - X_H} an element of
      \[\mathrm{ker}\,\Omega=\Big\{Y\in\mathfrak{X}(M)\ \  /\ \ \peqsub{\imath}{Y}\Omega={\textstyle\peqsub{\flat}{\Omega}}Y=0\Big\}\]
      The latter case, where a solution might not exist, is much more interesting and implies that $\mathrm{d}H$ is not in the range of $\peqsub{\flat}{\Omega}$ (in particular it is not surjective). In order to solve the Hamilton's equation we have to modify the equation \eqref{Mathematical background - definition - X_H}, the space $M$ where we want to solve it or both. In section \ref{Mathematical background - section - GNH} we will introduce a method to deal with this problem.
      
      \subsubsection*{Physical motivation}\trassub
      
        The mathematical motivation to study symplectic geometry is clear. Nonetheless, it is worth to take a brief detour to state also the physical motivation which, roughly speaking, is to introduce the Hamiltonian formalism. The interest in doing so is because it seems to be a prerequisite to perform the quantization of a theory. Besides, there are some problems whose solutions are easier to obtain in the Hamiltonian formulation, specially the celestial body problems \cite{arnol2013mathematical}.\separ
        
        Given some energy function $H\in\Cinf{M}$, called \textbf{Hamiltonian}\index{Hamiltonian}, we want to obtain the flow that the systems follows when some initial conditions are given or, equivalently, we want the \textbf{Hamiltonian vector field} $\peqsub{X}{\!H}\in\mathfrak{X}(M)$ associated to the flow. It is natural to require that such vector field depends linearly ---because Newton's laws are linear--- on $\mathrm{d}H\in\Omega^1(M)$, which encodes local changes of the energy. In particular we could look for the directions for which the energy does not change.\separ
        
        From the previous paragraph it is clear that we need a map $\Omega^1(M)\to\mathfrak{X}(M)$ that for every $\mathrm{d}H$ gives a solution $\peqsub{X}{H}$. In principle any $\Omega\in\mathfrak{T}^{2,0}(M)$ will do but we want to take into account some physical considerations.
        \begin{itemize}
        	\item \emph{Existence and uniqueness of solutions}. This means that $\Omega$ has to be non-degenerate. In that case we can equivalently consider, as it is more customary, $\Omega\in\mathfrak{T}^{0,2}(M)$. We have then $\Omega:\mathfrak{X}(M)\to\Omega^1(M)$ such that $\Omega(\peqsub{X}{\!H},\cdot{}\,)=\mathrm{d}H(\,\cdot{}\,)$.
        	\item \emph{Conservation of energy}. $H\in\Cinf{M}$ has to be constant along the flow of $\peqsub{X}{\!H}$.
        	\[0=\peqsub{X}{\!H}(H)=\mathrm{d}H(\peqsub{X}{\!H})=\Omega(\peqsub{X}{\!H},\peqsub{X}{\!H})\]
        	Although not every vector field is Hamiltonian, every vector at $T_pM$ belongs to a Hamiltonian vector field (because $\Omega$ is non-degenerate). Therefore, the previous equation forces $\Omega$ to be alternate i.e.\ a $2$-form field. Notice that this is one of the advantages to work with an element of $\mathfrak{T}^{0,2}(M)$ instead of with an element of $\mathfrak{T}^{2,0}(M)$. 
        	\item \emph{The physics is ``time independent''}. In an isolated system, given $H\in\Cinf{M}$ we require that the $2$-form $\Omega$ is preserved by the flow of $\peqsub{X}{\!H}$
        	\[0=\peqsub{\mathcal{L}}{\peqsubfino{X}{\!H}{-0.3ex}}\hspace*{0.2ex}\Omega\overset{\eqref{appendix - formula - L=di+id}}{=}\Big(\mathrm{d}\peqsub{\imath}{\peqsubfino{X}{\!H}{-0.3ex}}+\peqsub{\imath}{\peqsubfino{X}{\!H}{-0.3ex}}\mathrm{d}\Big)\Omega\overset{\eqref{Mathematical background - definition - X_H}}{=}\mathrm{d}(\mathrm{d}H)+\peqsub{\imath}{\peqsubfino{X}{\!H}{-0.3ex}}\mathrm{d}\Omega\overset{\eqref{appendix - equation - d^2=0}}{=}\peqsub{\imath}{\peqsubfino{X}{\!H}{-0.3ex}}\hspace*{0.1ex}\mathrm{d}\Omega\]
        	Thus it seems convenient to take $\mathrm{d}\Omega=0$ so that any Hamiltonian vector field $\peqsub{X}{\!H}$ preserves the symplectic form (analogously to a Killing vector field).
        \end{itemize}
    
    \centerline{\includegraphics[width=0.6\linewidth]{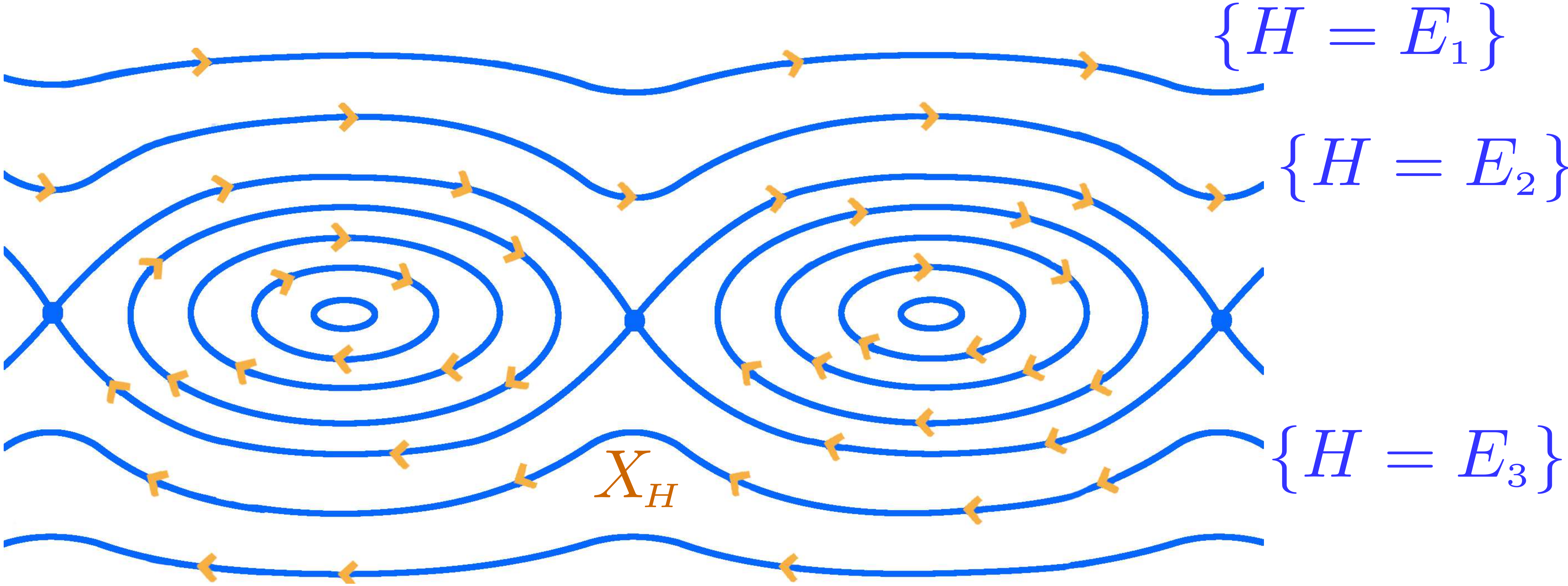}}

      \subsubsection*{Canonical symplectic form on \texorpdfstring{$\bm T^*\!\mathcal{Q}$}{T*Q}}\trassub
      
      It is well known that there are several obstructions for a manifold $M$ to have a symplectic structure. However, there are some important examples where we can define a \textbf{canonical} symplectic structure. The best known is the symplectic form $\Omega:\mathfrak{X}(T^*\!\mathcal{Q})\times\mathfrak{X}(T^*\!\mathcal{Q})\to\mathcal{C}^\infty(T^*\!\mathcal{Q})$ over the cotangent bundle $M=T^*\!\mathcal{Q}$ of a manifold $\mathcal{Q}$.\separ
      	
     We start by taking a patch $U\subset \mathcal{Q}$ small enough such that the tangent and cotangent bundle trivialize $TU=U\times F$ and $T^*\!U=U\times F^*$, where $F$ is the typical fiber of $T\hspace*{-0.2ex}\mathcal{Q}$ (thus $F^*$ is the typical fiber of $T^*\!\mathcal{Q})$. As the fiber is a vector space, we have of course that $T^*\!F=F\times F^*$. Thus, locally, $TT^*\!\mathcal{Q}$ is given by      	
      	\[T(T^*\hspace*{0.2ex}U)=T(\textcolor{red}{U}\times \textcolor{blue}{F^*})=T\textcolor{red}{U}\times T\textcolor{blue}{F^*}=(\textcolor{red}{U}\times\textcolor{blue}{F^*})\times(\textcolor{red}{F}\times\textcolor{blue}{F^*})\]
      	where we have gathered in the first parenthesis the base point and in the second the fiber. But now we have that the fiber is formed by both $F$ and $F^*$\!, so a typical vector of $T_{(q,\bm{p})}(T^*\!\mathcal{Q})$ will be of the form $(X_q,\bm{X_{\!p}})\in F\times F^*$. Thus we can define
      	\begin{equation}\label{eq:background canonical symplectic form}
      	  \Omega_{(\textcolor{red}{q},\textcolor{blue}{\bm{p}})}\Big(\!(\textcolor{red}{X_q},\textcolor{blue}{\bm{X_{\!p}}}),(\textcolor{red}{Y_q},\textcolor{blue}{\bm{Y_{\!\!p}}})\!\Big)=\textcolor{blue}{\bm{Y_{\!\!p}}}(\textcolor{red}{X_q})-\textcolor{blue}{\bm{X_{\!p}}}(\textcolor{red}{Y_q})
      	\end{equation}
      	where we have used the bold font to denote the elements of the dual $F^*$. We can see that $\Omega$ is clearly non-degenerate: if this expression is zero for every $Y=(Y_q,\bm{Y_{\!\!p}})\in\mathfrak{X}(T^*\!\mathcal{Q})$, it is also zero for every $Y=(Y_q,0)$ which leads immediately to $\bm{X_{\!p}}=0$. Now having that $\bm{Y_{\!\!p}}(X_q)=0$ for every $\bm{Y_{\!\!p}}$ implies that $X_q=0$ by lemma \ref{Appendix - equation - Hahn-Banach theorem}.\separ
      	
      	The closedness follows from the fact that $\Omega=\mathrm{d}\theta$ where $\theta\in\Omega^1(T^*\!\mathcal{Q})$ is the tautological $1$-form of $T^*\!\mathcal{Q}$ \cite{gotay1979presymplectic}. To define it consider $\pi:T^*\!\mathcal{Q}\to\mathcal{Q}$ the canonical projection, $\alpha\in T^*\!\mathcal{Q}$ and $W\in\mathfrak{X}(T^*\!\mathcal{Q})$. In particular notice that $W_\alpha\in T_\alpha(T^*\!\mathcal{Q})$. Then we define
      	\[\theta_{\alpha}(W_\alpha):=-\alpha(\pi_*W_\alpha)\]
      	
      \subsubsection*{Canonical symplectic form over the space of solutions}\trassub
      
      	There is yet another important example where a canonical symplectic form is available, namely the space of solutions of a covariant theory \cite{crnkovic1988symplectic,crnkovic1987covariant}. Let $(M,g)$ be an orientable and time oriented globally hyperbolic space-time and consider the the space of solutions to the Klein-Gordon equation $S=\{\varphi:M\to\R\ /\ \square\varphi=0\}$. Given $\varphi_1,\varphi_2\in S$, we define the symplectic form
        \begin{equation}\label{Mathematical background - equation - Omega(sol,sol)}
        \Omega(\varphi_1,\varphi_2)=\int_{\Sigma}\peqsubfino{{\mathrm{vol}_\gamma}}{\!X}{-0.3ex}\left[\varphi_2\peqsubfino{{\mathcal{L}_{\vec{n}}}}{\!X}{-0.3ex}\varphi_1-\varphi_1\peqsubfino{{\mathcal{L}_{\vec{n}}}}{\!X}{-0.3ex}\varphi_2\right]\smallcirc X
        \end{equation}
        The previous expression does not depend on the embedding $X\in\mathrm{Emb}^\partial_{g\textrm{-sl}}(\Sigma,M)$ thanks to lemma \ref{Appendix - lemma - Omega(sol,sol) embedding} and is trivially closed ($S$ is a vector space and $\Omega$ does not depend on the base point, only on the vectors). The non-degeneracy is a bit complicated to obtain from the previous expression, but it is trivial if we rewrite $\Omega$ in terms of the Cauchy data (equivalent to the space of solutions)
        \[\Omega\Big((q_1,p_1),(q_2,p_2)\Big)=\int_\Sigma\Big(q_2p_1-q_1p_2\Big)\peqsubfino{{\mathrm{vol}_\gamma}}{\!X}{-0.3ex}\]

  \section{Classical mechanics}\label{Mathematical background - section - Classical mechanics}
    
    Once we have established the basic mathematical grounds, putting emphasis on the geometric language, let us now use them to underline the importance of geometry in physics. We will briefly introduce the Newtonian mechanics as a necessary step towards Lagrangian mechanics, which is naturally defined over the tangent space. In order to take advantage of the canonical symplectic structure of the cotangent bundle, we need a map that links both spaces, namely the fiber derivative. This map allows us, in particular, to introduce the Hamiltonian framework, of central importance in the following chapters.
    
    \subsection{Newtonian framework}\trassub
    
      The Newton equation for a particle $m$ subject to a force $\vec{F}$ says
      \[\vec{F}(\vec{x})=m\deriv[2]{\vec{x}}{t}\qquad\qquad\equiv\qquad\qquad\begin{array}{|l}\displaystyle\deriv{\vec{x}}{t}=\vec{v}\\[2ex]\displaystyle \deriv{\vec{v}}{t}=\frac{\vec{F}(\vec{x})}{m}\end{array}\]
      The solutions to this system of ODE can be geometrically interpreted as the integral curves $\gamma:I\to T\R^3$ of the vector field
      \[X(\vec{x},\vec{v})=\sum_{i=1}^3\left(v^i\parcial{}{{x^i}}+\frac{F^i}{m}\parcial{}{{v^i}}\right)\]
       
    \subsection{Lagrangian framework}\label{Mathematical background - subsection - Lagrangian framework}\trassub
    
      The Lagrangian formulation allows us to deal with a lot of mechanical systems without knowing all the forces involved, in particular constraint forces that produce no work. Let us consider the manifold $\mathcal{Q}$, representing the positions of our system and known as the \textbf{configuration space}\index{Configuration space}, and define the \textbf{phase space of velocities}\index{Phase space of velocities} $T\hspace*{-0.2ex}\mathcal{Q}$ which is formed by all available positions and velocities. A \textbf{Lagrangian}\index{Lagrangian} is a smooth map $L:T\hspace*{-0.2ex}\mathcal{Q}\to\R$. It allows us also to define the \textbf{action}\index{Action} of our theory
      \[S:\mathcal{P}\to\R \quad\quad/\quad\quad S(\gamma)=\int_{t_0}^{t_1}\!L\Big(\gamma(t),\dot{\gamma}(t)\Big)\mathrm{d}t\]
	  where $\mathcal{P}$ is the space of smooth curves $\gamma:[t_0,t_1]\to\mathcal{Q}$ with fixed end points $\gamma(t_i)=Q_i$.
	  
	  Let us look for the stationary points of $S$. To that end consider a curve $\{\gamma_\tau\}_\tau$ over the space of curves $\mathcal{P}$ such that $\gamma_0=\gamma$ and $\delta\gamma=\left.\partial_\tau\gamma_\tau\right|_{\tau=0}\in T_\gamma\mathcal{P}$ (notice in particular that $\delta\gamma(t_j)=0$), then

	  \setlength{\figwidth}{0.65\linewidth}
	  \null\hfill\smash{\raisebox{-\dimexpr 5.5\baselineskip+\parskip\relax}%
	  	{\includegraphics[width=.28\linewidth]{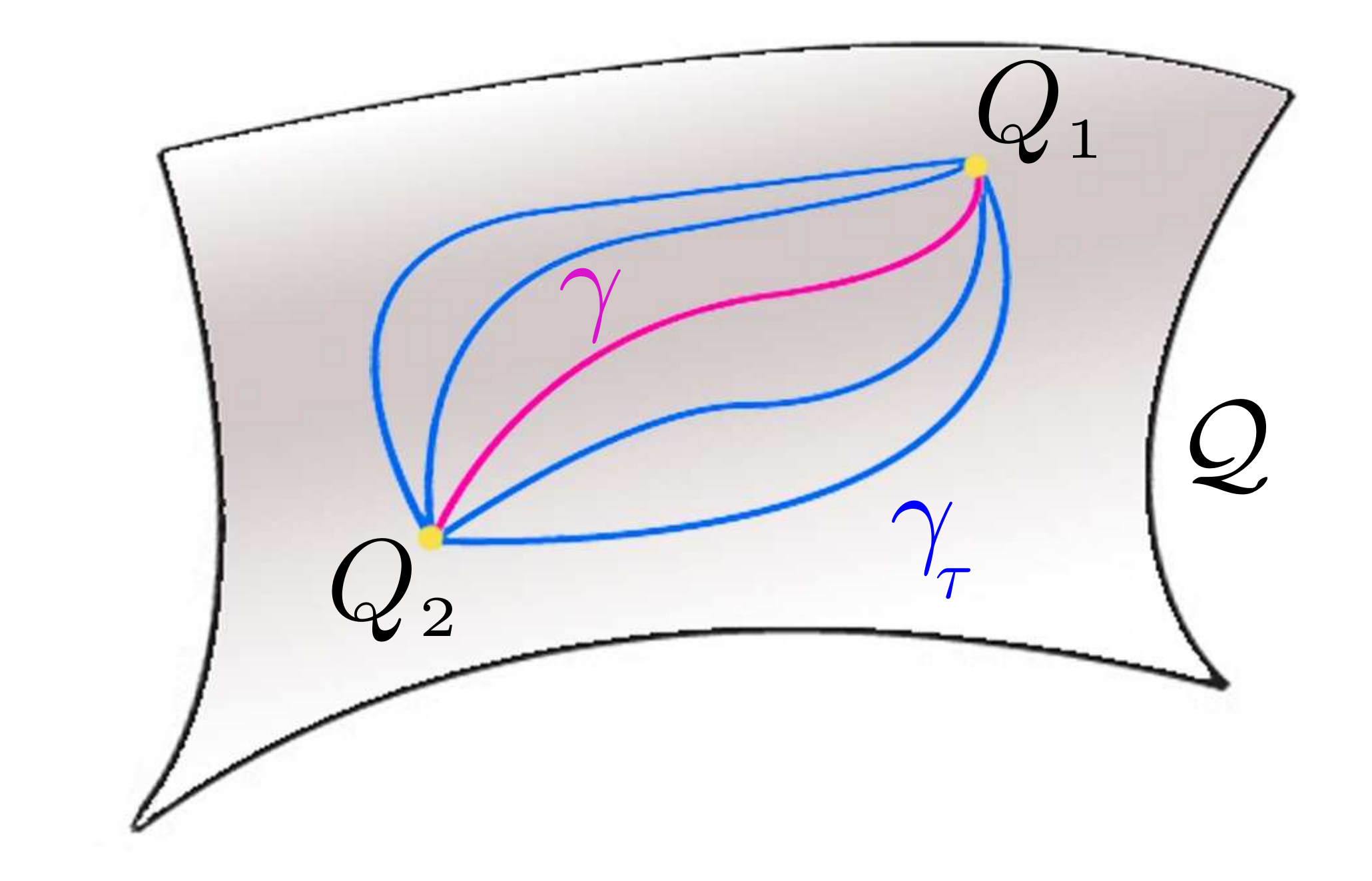}}}\strut \\[-\dimexpr 3.1\baselineskip+2\parskip\relax]
	  \begin{align*}
	    0&=\mathrm{d}_\gamma S(\delta\gamma)=\left.\deriv{}{\tau}\right|_{\tau=0}S(\gamma_\tau)=\int_{t_0}^{t_1}\left.\deriv{}{\tau}\right|_{\tau=0}L\Big(\gamma_\tau(t),\dot{\gamma}_\tau(t)\Big)\mathrm{d}t=\\
	    &=\int_{t_0}^{t_1}\left[D_1L\Big(\gamma(t),\dot{\gamma}(t)\Big)\ccdot\delta\gamma+D_2L\Big(\gamma(t),\dot{\gamma}(t)\Big)\ccdot\dot{\delta\gamma}\right]\mathrm{d}t=&&\hspace*{23ex}\mbox{}\\	    
	    &=\int_{t_0}^{t_1}\mathrm{d}t	\left[D_1L\Big(\gamma(t),\dot{\gamma}(t)\Big)-\deriv{}{t}D_2L\Big(\gamma(t),\dot{\gamma}(t)\Big)\right]\ccdot\delta\gamma     
	  \end{align*}
	  
	  We obtain then that $\gamma$ is a stationary point of $S$, $\mathrm{d}_\gamma S=0$, if and only if the curve $\gamma\in\mathcal{P}$ satisfies the \textbf{Euler-Lagrange equations}\index{Euler Lagrange equations}
	  \begin{equation}\label{Mathematical background - equation - EL equations}
	  	\deriv{}{t}\Big(D_2L(\gamma,\dot{\gamma})\ccdot\Big)=D_1L(\gamma,\dot{\gamma})\ccdot\ \ \equiv\ \  D_1\Big(D_2L(\gamma,\dot{\gamma})\ccdot\Big)\ccdot\dot{\gamma}+D_2\Big(D_2L(\gamma,\dot{\gamma})\ccdot\Big)\ccdot\ddot{\gamma}=D_1L(\gamma,\dot{\gamma})\ccdot
	  \end{equation}
	  It is very important to keep in mind that the central object is the set of equations of motion and not the Lagrangian itself (which is just a means to get some equations that might or not be of physical relevance). If fact, we can define different Lagrangians that give rise to the same equations.\separ
	  
	  Finally, notice that we can recover the Newtonian framework for conservative forces by taking $L(q,v)=\frac{1}{2}mv^2-V(q)$. Then the EL equation reads
	  \[m\ddot{q}=-V'(q)=:F(q)\]
	  	
    \subsubsection*{Fiber derivative}\trassub
    
	  The \textbf{fiber derivative}\index{Fiber derivative} links the tangent bundle to the cotangent bundle (also known as the \textbf{phase space}\index{Phase space} of the theory) in a way adapted to the physics of the system at hand. It is the strong bundle map $F\!L:T\hspace*{-0.2ex}\mathcal{Q}\to T^*\!\mathcal{Q}$, meaning that $F\!L(q,v):T_qQ\to\R$, which is given by
	  \begin{equation}\label{Mathematical background - equation - FL=D_2L}
		  F\!L(q,v)\big(q,w\big)=\left.\deriv{}{\tau}\right|_{\tau=0}L(q,v+\tau w)=D_2L(q,v)\ccdot w
	  \end{equation}
	  where the last equality is valid on a chart $T^*U=U\times F^*$ for some small $U\subset\mathcal{Q}$. Notice that in general this map is not surjective nor injective. The image $F\!L(T\hspace*{-0.2ex}\mathcal{Q})\subset T^*\!\mathcal{Q}$ is called the \textbf{primary constraint submanifold}\index{Primary constraint submanifold}. The image of the fiber derivative is usually called the \textbf{momenta}\index{Momentum} of the system
	  \[(q,\boldsymbol{p}):=F\!L(q,v)=\big(q,D_2L(q,v)\big)\in T^*_q\!\mathcal{Q}\qquad\quad\equiv\quad\qquad\boldsymbol{p}=D_2L(q,v)\ccdot\in T^*_q\!\mathcal{Q}\]
	   We use the bold font to emphasize that the momenta are elements of the dual. In the finite dimensional case the cotangent and tangent spaces can be identified and so, as expected, $p=\parcial{L}{v}$.
	  
	  \subsubsection*{Non-canonical symplectic form on \texorpdfstring{$\bm T\hspace*{-0.2ex}\mathcal{Q}$}{TQ}}\trassub
	  
	    Given a Lagrangian $L:T\hspace*{-0.2ex}\mathcal{Q}\to\R$, we can pullback the canonical symplectic form $\Omega$ of $T^*\!\mathcal{Q}$ to $T\hspace*{-0.2ex}\mathcal{Q}$ to obtain a (pre)symplectic form $\peqsub{\Omega}{L}=F\!L^*\Omega$. Its degeneracy will depend on the regularity of $F\!L$.\separ
	    
	    To obtain the explicit expression we use that the pushforward of a vector $(Y_q,Y_v)$ over $T\hspace*{-0.2ex}\mathcal{Q}$ is
	    \begin{align}\label{Mathematical background - equation - FL_*}
	    \begin{split}
	      F\!L_*\left(Y_q,Y_v\right)&=\left.\deriv{}{\tau}\right|_{\tau=0}F\!L(q_\tau,v_\tau)=\left.\deriv{}{\tau}\right|_{\tau=0}\big(q_\tau,D_2(q_\tau,v_\tau)\ccdot\big)=\\
	      &=\Big(Y_q,D_1\big(D_2L(q,v)\ccdot\big)\ccdot Y_q+D_2\big(D_2L(q,v)\ccdot\big)\ccdot Y_v\Big)
	    \end{split}
	    \end{align}
	    Therefore
	    \begin{align}\label{Mathematical background - equation - Omega^L(X,Y)}
	    \begin{split}
	      \peqsub{\Omega}{L}\Big((X_q,X_v),(Y_q,Y_v)\Big)&=\Omega\Big(F\!L_*(X_q,X_v),F\!L_*(Y_q,Y_v)\Big)\overset{\eqref{eq:background canonical symplectic form}}{=}\\
	      &=D_1\Big(D_2L(q,v)\ccdot X_q\Big)\ccdot Y_q+D_2\Big(D_2L(q,v)\ccdot X_q\Big)\ccdot Y_v-\\
	      &\phantom{=}-D_1\Big(D_2L(q,v)\ccdot Y_q\Big)\ccdot X_q-D_2\Big(D_2L(q,v)\ccdot Y_q\Big)\ccdot X_v
	    \end{split}
	    \end{align}
	    
	We say that $L$ is \textbf{hyperregular}\index{Hyperregular Lagrangian!Lagrangian} if $F\!L$ is a diffeomorphism. In that case, $\peqsub{\Omega}{L}$ is strongly symplectic. If $\peqsub{\Omega}{L}$ is just weakly symplectic (equivalently $D_2D_2L$ is weakly symplectic), we say that $L$ is \textbf{regular}\index{Regular Lagrangian!Lagrangian}. This implies that $F\!L$ is a local diffeomorphism if and only if $L$ is regular. Finally, $L$ is said to be \textbf{almost-regular}\index{Almost-regular!Lagrangian}\index{Lagrangian!Almost-regular} if $F\!L$ is a submersion onto its image and the fibers $F\!L^{-1}(F\!L(v))\subset T\mathcal{Q}$ are connected submanifolds. We refer the reader to \cite{gotay1979presymplectic,marsden2013introduction} for further details.
	  
	\subsubsection*{Dynamics over \texorpdfstring{$\bm T\hspace*{-0.2ex}\mathcal{Q}$}{TQ} - Symplectic-Lagrangian formalism}\trassub
	
	  We now define the \textbf{energy}\index{Energy} $E:T\hspace*{-0.2ex}\mathcal{Q}\to\R$ given by $E(q,v)=F\!L(q,v)\big(q,v\big)-L(q,v)$. Notice that for finite dimensional systems $E=pv-L$.\separ
	   
	  The dynamics of the system is given by the flow of the vector field $\peqsub{X}{\!E}\in\mathfrak{X}(T\hspace*{-0.2ex}\mathcal{Q})$ solving
	  \begin{equation}\label{Mathematical background - equation - symplectic lagrangian equation}
	    \peqsub{\imath}{\peqsubfino{X}{\!E}{-0.3ex}}\hspace*{0.2ex}\peqsub{\Omega}{L}=\mathrm{d}E\qquad\qquad\equiv\qquad\qquad(\peqsub{X}{\!E})^a(\peqsub{\Omega}{L})_{ab}=(\mathrm{d}E)_b
	   \end{equation}
	   If $\peqsub{\Omega}{L}$ is strongly symplectic, then $\peqsub{X}{\!E}=\peqsubfino{{\sharp_\Omega}}{\!L}{-0.2ex}(\mathrm{d}E)$, or equivalently $(\peqsub{X}{\!E})^a=\peqsub{\Omega}{L}^{ab}(\mathrm{d}E)_b$, is a solution. Let us see the relation between the previous equation and the EL equations. First we have
	   \begin{align*}
	     \mathrm{d}E\left(Y_q,Y_v\right)&=\left.\deriv{}{\tau}\right|_{\tau=0}E(q_\tau,v_\tau)\overset{\eqref{Mathematical background - equation - FL=D_2L}}{=}\left.\deriv{}{\tau}\right|_{\tau=0}\Big[D_2L(q_\tau,v_\tau))\ccdot v_\tau-L(q_\tau,v_\tau)\Big]=\\[1ex]
	     &=D_1\Big(D_2L(q,v)\ccdot v\Big)\ccdot Y_q+D_2\Big(D_2L(q,v)\ccdot v\Big)\ccdot Y_v-D_1L(q,v)\ccdot Y_q
	   \end{align*}
	   Now imposing it to be equal to $\peqsub{\Omega}{L}(X,Y)$, given by equation \eqref{Mathematical background - equation - Omega^L(X,Y)}, for every $Y$ we have
	   \begin{align*}
	     &D_2\Big(D_2L(q,v)\ccdot\big(X_q-v\big)\Big)\ccdot Y_v=0\\
	     &D_1\Big(D_2L(q,v)\ccdot Y_q\Big)\ccdot X_q+D_2\Big(D_2L(q,v)\ccdot Y_q\Big)\ccdot X_v-D_1L(q,v)\ccdot Y_q=D_1\Big(D_2L(q,v)\ccdot \big(X_q-v\big)\Big)\ccdot Y_q
	   \end{align*}
	   We see then that any solution to the EL equations \eqref{Mathematical background - equation - EL equations} is also a solution to the symplectic-Lagrangian equation \eqref{Mathematical background - equation - symplectic lagrangian equation} if we impose the second order condition $X_q=v$. The converse is not necessarily true if $\peqsub{\Omega}{L}$ is just presymplectic.\separ
	   
	   Notice that we can rewrite the symplectic-Lagrangian equation over $T^*\mathcal{Q}$ in terms of $\Omega$	   \[\mathrm{d}E=\peqsub{\imath}{\peqsubfino{X}{\!E}{-0.3ex}}\hspace*{0.2ex}\peqsub{\Omega}{L}=\peqsub{\imath}{\peqsubfino{X}{\!E}{-0.3ex}}\hspace*{0.2ex}F\!L^*\Omega\overset{\eqref{appendix - formula - i_x(f^*)=f^*i_{f_x}}}{=}F\!L^*\!\left(\peqsub{\imath}{\peqsubfino{F\!L_*\!X}{\!E}{-0.3ex}}\Omega\right)\]
	   so if there exists some $H$ such that $E=F\!L^*H=H\smallcirc F\!L$ we would recover, denoting $\peqsub{X}{\!H}=F\!L_*\peqsub{X}{\!E}$, the Hamilton's equation
	   \[F\!L^*\Big(\peqsubfino{{\imath_X}}{\!H}{-0.2ex}\hspace*{0.2ex}\Omega-\mathrm{d}H\Big)=0\quad\quad\equiv\quad\quad \Omega\Big(\peqsub{X}{\!H},FL_*Y\Big)-\mathrm{d}H(F\!L_*Y)=0\ \text{ for every }\ Y\in\mathfrak{X}(T\hspace*{-0.2ex}\mathcal{Q})\]
	   which leads us to the Hamiltonian framework \textbf{only} over $F\!L(T\hspace*{-0.2ex}\mathcal{Q})$.

	\subsection{Hamiltonian framework}
	  We define the \textbf{Hamiltonian}\index{Hamiltonian} as a map $H:F\!L(T\hspace*{-0.2ex}\mathcal{Q})\subset T^*\!\mathcal{Q}\to\R$ such that $H\smallcirc F\!L=E$. Now we might wonder if this is a good definition in the sense that it could be non-unique or not well defined. For instance we might have $v_1\neq v_2$ such that $F\!L(q,v_1)=F\!L(q,v_2)$ but $E(q,v_1)\neq E(q,v_2)$. If $L$ is almost-regular\index{Almost-regular!Lagrangian}\index{Lagrangian!Almost-regular}, then it can be proved \cite{gotay1979presymplectic} that the energy is constant along the fibers and, therefore, we can project it to obtain a Hamiltonian $H$. We will assume from now on that $L$ is almost-regular, nonetheless keep in mind that in the following chapters we will be working with concrete examples where the Hamiltonian $H$ will be built explicitly.\separ
	  
	  Let us consider now the (pre)symplectic manifold $(F\!L(T\hspace*{-0.2ex}\mathcal{Q}),\omega=i^*\Omega)$ where $i:F\!L(T\hspace*{-0.2ex}\mathcal{Q})\hookrightarrow T^*\!\mathcal{Q}$ is the natural inclusion. The Hamilton's equation \eqref{Mathematical background - definition - X_H}, that as we mentioned before prescribes the dynamics, then reads
	  \[\peqsubfino{{\imath_X}}{\!H}{-0.2ex}\hspace*{0.2ex}\omega=\mathrm{d}H\qquad\qquad\equiv\qquad\qquad{\textstyle\peqsub{\flat}{\omega}}\,\peqsub{X}{\!H}=\mathrm{d}H\]
	  First notice that if $F\!L$ is an immersion and $\peqsub{X}{\!E}$ satisfies the symplectic-Lagrangian equation \eqref{Mathematical background - equation - symplectic lagrangian equation}, then $\peqsub{X}{\!H}:=FL_*\peqsub{X}{\!E}$ satisfies the Hamilton's equation. To prove this we are going to show that $\omega\big(\peqsub{X}{\!H},Y\big)=\mathrm{d}H(Y)$ for every $Y\in\mathfrak{X}(F\!L(T\hspace*{-0.2ex}\mathcal{Q}))$. First notice that $i_*Y=F\!L_*\widetilde{Y}$ for some vector field $\widetilde{Y}\in\mathfrak{X}(T\hspace*{-0.2ex}\mathcal{Q})$ because $F\!L$ is a submersion.
	  \begin{align*}
	    \omega\big(\peqsub{X}{\!H},Y\big)&=\Omega\big(i_*\peqsub{X}{\!H},i_*Y\big)\overset{\star}{=}\Omega\big(FL_*\peqsub{X}{\!E},FL_*\widetilde{Y}\big)=\\
	    &=\peqsub{\Omega}{L}\big(\peqsub{X}{\!E},\widetilde{Y}\big)\overset{\eqref{Mathematical background - equation - symplectic lagrangian equation}}{=}\mathrm{d}E(\widetilde{Y})=FL^*\mathrm{d}H(\widetilde{Y})=\\
	    &=\mathrm{d}H(FL_*\widetilde{Y})=\mathrm{d}H(Y)
	  \end{align*}
	  where in the $\star$ we have used that, essentially, $\peqsub{X}{\!H}$ and $i_*\peqsub{X}{\!H}$ are the same (it just changes the ambient space).\separ
	  
	  Conversely, consider $\peqsub{X}{\!H}$ a solution to the Hamilton's equations, we already know that there exists some $\peqsub{X}{\!E}$ such that $F\!L_*\peqsub{X}{\!E}=\peqsub{X}{\!H}$. Let us see that $\peqsub{X}{\!E}$ is then a solution to the symplectic-Lagrangian equation.
	  \begin{align*}
	  \peqsub{\Omega}{L}\big(\peqsub{X}{\!E},\widetilde{Y}\big)-\mathrm{d}E(\widetilde{Y})&=\Omega\big(F\!L_*\peqsub{X}{\!E},F\!L_*\widetilde{Y}\big)-\mathrm{d}F\!L^*H(\widetilde{Y})=\\
	  &=\Omega\big(i_*\peqsub{X}{\!H},i_*Y\big)-F\!L^*\mathrm{d}H(\widetilde{Y})=\\
	  &=\omega\big(\peqsub{X}{\!H},Y\big)-\mathrm{d}H(Y)=0
	  \end{align*}
	  Notice that, of course, $\peqsub{X}{\!E}$ might not be unique. In fact, any vector field $\peqsub{X}{\!E}+Y$ where $Y\in\mathrm{ker}F\!L_*$ gives raise to the same $\peqsub{X}{\!H}$. This is related with the fact that the gauge freedom of in the Lagrangian setting might be larger than the one of the Hamiltonian framework.\separ
	  
	  Here we present a diagram summarizing the link among the objects presented in this section \ref{Mathematical background - section - Classical mechanics}. The dashed lines indicate that some hypothesis are required.
	  
	      \begin{center}
	      	\begin{tikzcd}%
	  	  S \arrow[d] & L  \arrow[l]\arrow[rr] & & F\!L \arrow[r] & E \arrow[r,dashed]\arrow[d]  & H\arrow[d]\arrow[l,bend right]\\
	  	  \text{EL equations}\arrow[rrrr] &  & & & \begin{minipage}{14.297ex}\centering{}Sympl-Lagran\\equation\end{minipage}\arrow[r,dashed]\arrow[llll,bend left,dashed,"L\text{ is regular}",swap] &\begin{minipage}{14.297ex}\centering{}Hamilton\\equation\end{minipage}\arrow[l,bend left]
	  \end{tikzcd}%
  		\end{center}
  	

  \section{GNH algorithm}\label{Mathematical background - section - GNH}
    We have seen that we can cast our problem in the Lagrangian or Hamiltonian setting, now we want to solve it in order to obtain the vector field and the submanifold where it is defined. The first approach to consistently solve this kind of systems, specially in the presence of constraints (when $F\!L$ is not surjective), is due to Dirac \cite{dirac1950generalized} circa 1950.  
  
  \subsection{Dirac's algorithm in a nutshell}\label{Mathematical background - subsection - Dirac algorithm}
  Dirac's algorithm allows us to deal with many constrained Hamiltonian systems. Very briefly, it consists of the following steps\vspace*{-1.1ex}
  \begin{itemize}
  	\item Given a Lagrangian of the theory $L:T\mathcal{Q}\to\mathbb{R}$, define the canonical momenta $p_k=\dfrac{\partial L}{\partial v_k}(q,v)$.\vspace*{-1ex}
  	\item Invert, if possible, the previous relations to obtain $v_k=v_k(q,p)$.
  	\item If this can be done, the Hamiltonian and Lagrangian formulations are completely equivalent. Otherwise some (primary) constraints relating positions and momenta appear $\phi_m(q,p)=0$.
  	\item Define the total Hamiltonian on the cotangent space $\peqsub{H}{T}=\sum v_kp_k-L+\sum\lambda_m\phi_m$, with $\lambda_m$ to be fixed. It determines (maybe non-uniquely) the dynamics $\dot{g}=\{g,\peqsub{H}{T}\}$. In particular 
  	\begin{equation}\label{equation hamiltoniana}
  	\dot{q_k}=\dfrac{\partial \peqsub{H}{T}}{\partial p_k}\qquad\qquad\dot{p_k}=-\dfrac{\partial \peqsub{H}{T}}{\partial q_k}
  	\end{equation}
  	\item The constraints have to be preserved in time $0=\dot{\phi}_m=\{\phi_m,\peqsub{H}{T}\}$. New (secondary) constraints may appear.
  	\item By iterating the previous step with all new constraints, some $\lambda_m$ might be fixed. This iteration could lead to infinitely many steps, or end up with an inconsistency or a tautology.
  \end{itemize}
  The final output of the algorithm is the total Hamiltonian and the constraints. Despite the success of this algorithm in many typical examples, it is difficult to apply it in some circumstances, for instance when dealing with field theories over manifolds with boundaries. In such cases, it is helpful to use other approaches that rely on a geometric description of the Hamiltonian dynamics. One such method uses the so-called GNH algorithm.
  
  \subsection{The GNH algorithm in a nutshell}
  The GNH algorithm, developed by Gotay, Nester and Hinds \cite{gotay1979presymplectic,gotay1980generalized,gotay1978presymplectic}, is a generalization of Dirac's algorithm that relies on global geometrical methods. It can be used both for mechanical systems with a finite number of degrees of freedom and for field systems with infinitely many. In the latter case, one should pay special attention to functional analysis issues, specially in the presence of boundaries. The GNH method can be sketched as follows
  \begin{itemize}
  	\item Let $L:T\mathcal{Q}\to\mathbb{R}$ be a Lagrangian and define the canonical momenta $\boldsymbol{p}=F\!L(q,v)$ where $F\!L:T\mathcal{Q}\to T^*\!\mathcal{Q}$ is the fiber derivative.
  	\item The energy $E:T\mathcal{Q}\to\mathbb{R}$ is defined by $E(q,v)=F\!L(q,v)(q,v)-L(q,v)$.
  	\item The Hamiltonian $H:F\!L(T\mathcal{Q})\subset T^*\!\mathcal{Q}\to\mathbb{R}$, defined \textbf{only} over $F\!L(T\mathcal{Q})\overset{i}{\hookrightarrow} T^*\!\mathcal{Q}$, is given by the equation $H\smallcirc F\!L=E$. If $F\!L$ is invertible, then $H$ carries the same information as $L$.
  	\item $(T^*\!\mathcal{Q},\Omega)$ is canonically a symplectic manifold while $(F\!L(T\mathcal{Q}),\omega=i^*\Omega)$ is, in general, a presymplectic manifold.
  	\item Wherever it makes sense, solve $\peqsub{\imath}{X}\omega=\mathrm{d}H$ for $X$.
  	\item $X$ is defined over $M_1\subset F\!L(T\mathcal{Q})$ but it is not necessary tangent to it. Take $M_2$ such that $X|_{M_2}$ is tangent to $M_1$. $X|_{M_2}$ is defined over $M_2$ but, in general, will not be tangent to it so we must iterate this process.
  	\item If finitely many steps are required to get tangency, then the last manifold is known as the \textbf{final constraint manifold}\index{Final constraint manifold}, and $X$ restricted to it defines the Hamiltonian dynamics of the system. Otherwise, there might be infinitely many conditions or an inconsistency.
  \end{itemize}

\vspace*{-.6ex}
\centerline{\includegraphics[width=.78\linewidth,clip,trim=0 0 0 0ex]{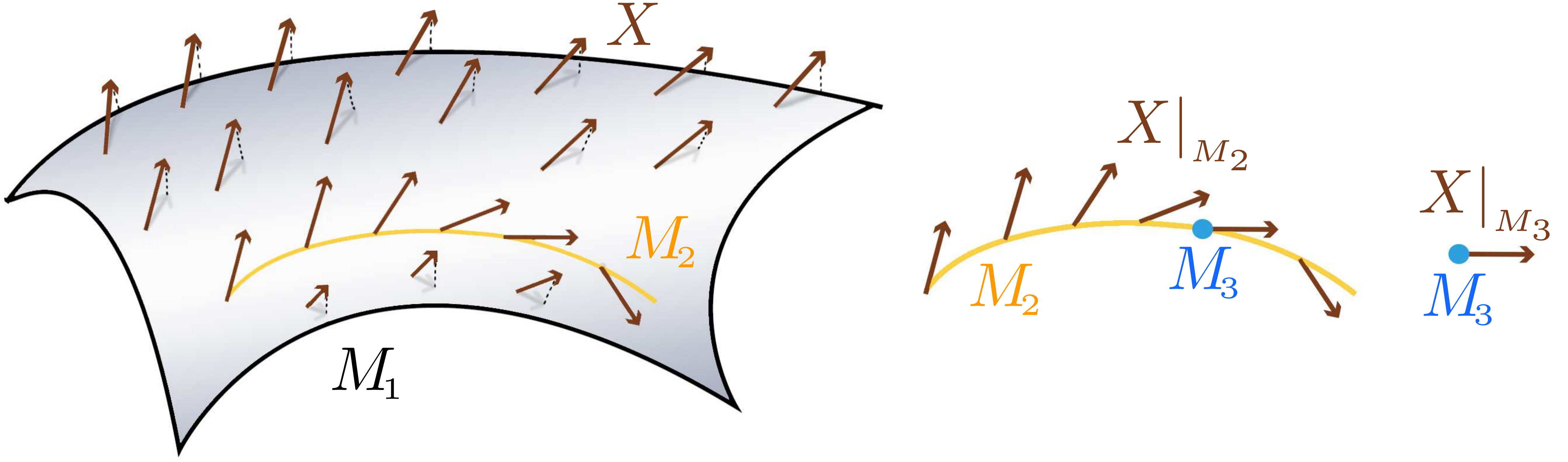}}

  The advantages of this procedure become clear even for finite dimensional problems because of its neat geometric interpretation. Besides, it can be applied to a wider range of problems. Indeed, notice that Dirac's algorithm requires the primary constraint space to be a submanifold of some symplectic manifold in order to have a Poisson bracket. Meanwhile, the GNH algorithm could be initiated at the fifth step with some (pre)symplectic structure $\omega$ that is not necessarily the pullback of $\Omega$. Furthermore, the last step can be relaxed to require only tangency to the closure $\overline{M_i}$ of $M_i$, which can help the algorithm to stop as we will see in section \ref{String masses - section - Alternative Hamiltonian formulation} of chapter \ref{Chapter - Scalar fields coupled to point masses}.\separ
  
  We will use extensively the GNH algorithm to solve the Hamiltonian equation in the following chapters. We refer the reader also to \cite{barbero2014hamiltonian} to see some more detailed applications of the algorithm.

    \section{Quantum mechanics}
    
    The history of quantum mechanics is fascinating and involves some of the best physicists along the years like Boltzmann, Planck, Einstein, Bohr, de Broglie, Born, Dirac, Heisenberg, Pauli, or Schrödinger. They struggled for decades to understand some experiments and observations that  could not be explained by classical physics. The most important ones involved the behavior of black bodies, the photoelectric effect, and the stability of atoms. The final explanation of them, an many other important phenomena, was provided by quantum mechanics.\separ
    
    It is now widely believed that the world behaves in a quantum manner and that the classical description refers to some type of ``effective'' or ``average'' behavior. The transit between classical and quantum descriptions of physical systems is not an easy one. On one hand it is difficult to derive all the features of the classical world that we see from ``microscopic'' quantum mechanical models. On the other, there is not such thing as a universal quantization procedure allowing us to associate a quantum model with any arbitrary classical system. Nonetheless, there are methods that, in some instances, lead to reasonable quantum dynamics.\separ
    
    If we assume that we already know how to describe the quantum behavior of a single particle, then the next desirable step would be to obtain a general theory that describes also the interaction or decay of several particles. This can be achieved through a mathematically well defined procedure called \textbf{second quantization}\index{Quantization!Second quantization}, which is also known as \textbf{Fock quantization}\index{Quantization!Fock quantization} (or Fock-Cook quantization). There is in fact a famous saying attributed to Edward Nelson
    \begin{quote}\centering
    	\emph{First quantization is a mystery, but second quantization is a functor!}
    \end{quote}
We use here the term first quantization to talk about of the single particle (one particle) quantization from a classical theory. As we mentioned before there is no standard procedure and this is why it is often said to be a mystery. On the other hand, the fact that the second quantization is a functor will be explained along this section.

    \subsection{Basic notions about quantization}\label{Mathematical background - subsection - First quantization}

    Given a classical system with configuration space $\mathcal{Q}$, we know that the Hamiltonian $H\in\Cinf{T^*\!\mathcal{Q}}$ encodes the dynamics of the system over the phase space $T^*\!\mathcal{Q}$, the space of all possible positions and momenta. In order to retrieve information from the system, we saw in section \ref{Mathematical background - subsection - symplectic} that we need to probe the system through the measurement of observables. The possible outcomes of the observations through an observable\index{Observable!Classical} $f\in\Cinf{T^*\!\mathcal{Q}}$ are the real numbers $\mathrm{Im}(f)\subset\R$.\separ
    
    It is important to mention that some observables can be considered as fundamental. A familiar example in the finite dimensional case is provided by the generalized position and momenta denoted as $q^i$ and $p_i$ (canonical coordinates over $T^*\!\mathcal{Q}$ which, seen as maps, are just the ``projections'') that satisfy the canonical commutation relations
    \begin{equation}\label{Mathematical background - equation - CCR}\index{Canonical commutation relations}
    \{q^i,p_i\}=\delta^i_j
    \end{equation}
    
    Meanwhile, in the quantum realm, answering and even asking questions about the system is much trickier than in the classical framework. It can be properly justified \cite{bratteli2012operator} that the previous classical ingredients have to be significantly changed  (see section \ref{Appendix - section - functional analysis} of the appendix for the definitions of the objects involved).
    
    \begin{itemize}
    	\item The phase space is now described by a complex (projective) Hilbert space $\mathfrak{h}$. The vectors of $\mathfrak{h}$, up to a phase, are called \textbf{states}.
    	\item The observables\index{Observable!Quantum} are self-adjoint operators $A:\mathfrak{h}\to\mathfrak{h}$. Notice in particular that the mathematical outputs of $A$ do not play the same role as those of a classical observable.
    	\item  In an actual physical observation with an experimental device, we do obtain real numbers. If $A$ models such experiment, the possible physical outputs of the device turn out to be the eigenvalues of $A$.
    \end{itemize}

\subsubsection*{Probabilistic interpretation}\trassub
	
	Consider an observable $A$ which, for the sake of simplicity, we assume that it has a discrete spectrum $\{\lambda_n\}\subset\R$ with non-degenerate eigenvectors $\{|\lambda_n\rangle\}_n$ (which form a complete basis of $\mathfrak{h}$ that we take orthonormal). If we pick an arbitrary state $|v\rangle\in\mathfrak{h}$, we know that we can expand it as
    \[|v\rangle=\sum_{n=1}^\infty c_n|\lambda_n\rangle\qquad\quad\text{ where }\quad\qquad c_n=\langle \lambda_n|v\rangle\]
    When we measure $|v\rangle$ with the device described by $A$ we obtain a real number which has to be an element $\lambda_n$ of the spectrum of $A$. The probability to obtain such output is precisely $|c_n|^2$. Notice that unless $|v\rangle$ is one of the eigenvectors, the measurement of $A$ will give different values when we repeat the experiment several times. It is thus natural to consider the expectation value and variance given by
    \[\langle A\rangle_{|v\rangle}=\frac{\langle v|A|v\rangle}{\langle v|v\rangle}\qquad\qquad \left(\Delta_{|v\rangle}A\right)^2=\Big\langle\big(A-\langle A\rangle_{|v\rangle}\big)^2\Big\rangle_{|v\rangle}\]

    \subsubsection*{From classical to quantum}\trassub
    
    We see that we can give a probabilistic interpretation to a quantum theory. However, it remains to know how to choose the Hilbert space $\mathfrak{h}$ and how to introduce quantum observables analogous to the classical ones. One of the first approaches to construct a quantum theory associated with a classical one is to reproduce at the quantum level some features of its classical formulation by defining a Poisson algebra\index{Poisson algebra} $(\mathrm{SAO}(\mathfrak{h}),\smallcirc,[\,,])$ as a \textbf{representation}\index{Representation} of the Poisson algebra $(\Cinf{T^*\!\mathcal{Q}},\cdot{},\{\,,\})$. A few more details can be found on section \ref{Appendix - section - functional analysis} of the appendix.

    To this end it would be ideal to have a functor\ $\ \raisemath{-.6ex}{\widehat{}}\ $\ from the category of symplectic manifolds to the category of Hilbert spaces such that for any observables $f,g\in\Cinf{T^*\!\mathcal{Q}}$ we have
	\begin{equation}\label{Mathematical background - equation - gorrificacion}
	\widehat{f},\widehat{g}\in\mathrm{SAO}(\mathfrak{h})\qquad\text{satisfying}\qquad[\widehat{f},\widehat{g}]=i\hbar\widehat{\{f,g\}}
	\end{equation}
    It is well known, in fact, that in general no such functor exists. We can, however, find a representation of the subalgebra generated by $\langle\mathrm{Id},q^i,p_i\rangle$ given by $\mathfrak{h}:=L^2(\R^n)$ and
    \begin{equation}\label{Mathematical background - equation - representation q p}
    \begin{array}{cccc}
    \widehat{q}^{\,i}:&\mathfrak{h}& \longrightarrow & \mathfrak{h}\\[1ex]
    & f&\longmapsto & q^i\cdot{}f
    \end{array}\qquad\qquad\qquad
    \begin{array}{cccc}
    \widehat{p}_i:&\mathfrak{h}& \longrightarrow & \mathfrak{h}\\
    & f&\longmapsto & \displaystyle-i\hbar\parc[f]{q^i}
    \end{array}
    \end{equation}
    It is easy to check that \eqref{Mathematical background - equation - gorrificacion} is satisfied. This representation is of great importance thanks to an important theorem due to Stone and Von Neumann that states that, under some reasonable hypothesis, all admissible representations are unitarily equivalent. However, this procedure gives no canonical procedure to represent more general observables (the obvious approach suffers from factor ordering ambiguities).

\subsubsection*{Dynamics}\trassub
    
    We recall that the dynamics over the Poisson algebra $(\Cinf{T^*\!\mathcal{Q}},\cdot{},\{\,,\})$ is given by the Hamiltonian $H$ through the Hamiltonian vector field, which can be written in terms of the Poisson bracket as
    \[\deriv{}{t}=\big\{\cdot{}\,,H\big\}\qquad\quad\equiv\quad\qquad\deriv{F}{t}=\big\{F,H\big\}\]
    This suggest to define the dynamics over the Hilbert space as
    \[i\hbar\deriv{}{t}=\big[\cdot{}\,,\widehat{H}\big]\qquad\quad\equiv\quad\qquad i\hbar\deriv{\widehat{A}}{t}=\big[\widehat{A},\widehat{H}\big]\]
    where $\widehat{H}$ is the quantum Hamiltonian, given as representation of the classical Hamiltonian $H$. This setting is known as the Heisenberg picture, where the observables carry all the evolution. In order to pass to the Schrödinger picture, where the states are the only objects that evolve, we can proceed as follows. Assume that the Hamiltonian is time independent, then there exists a formal solution to the last equation given by
    \[\widehat{A}(t)=\mathrm{Exp}\left(\frac{it}{\hbar}\widehat{H}\right)\widehat{A}(0)\mathrm{Exp}\left(-\frac{it}{\hbar}\widehat{H}\right)\]
    If we now compute the expectation value of this operator $\widehat{A}(t)$ we obtain
    \begin{align*}
    \langle \widehat{A}(t)\rangle_{|\psi\rangle}&=\langle \psi|\widehat{A}(t)|\psi\rangle=\left\langle \psi\left|\mathrm{Exp}\left(\frac{it}{\hbar}\widehat{H}\right)\right.\widehat{A}(0)\left.\mathrm{Exp}\left(-\frac{it}{\hbar}\widehat{H}\right)\right|\psi\right\rangle=\\
    &=\left\langle\left. \mathrm{Exp}\left(-\frac{it}{\hbar}\widehat{H}\right)\psi\right|\widehat{A}(0)\left|\mathrm{Exp}\left(-\frac{it}{\hbar}\widehat{H}\right)\right.\psi\right\rangle=\\
    &=\langle \psi(t)|\widehat{A}(t)|\psi(t)\rangle
    \end{align*}
    where we have defined
    \[|\psi(t)\rangle:=\mathrm{Exp}\left(-\frac{it}{\hbar}\widehat{H}\right)\left|\psi\right\rangle\qquad\qquad\longrightarrow\qquad\qquad\deriv{|\psi(t)\rangle}{t}=\widehat{H}|\psi(t)\rangle\]
    As we can see the time-dependent state $|\psi(t)\rangle$ satisfies the Schrödinger equation.
    
\subsubsection*{Algebraic interpretation}\trassub

The procedure to find a representation of the algebra of classical observables, which works reasonably well for quantum mechanics, is not suited for field theories. Let us quickly review the algebraic approach which suggests a better generalization.\separ

Let us study the classical harmonic oscillator. Using the representation \eqref{Mathematical background - equation - representation q p} we can construct the quantum Hamiltonian, without ambiguity, from the Hamiltonian $H$
\[H(q,p)=\frac{1}{2m}p^2+\frac{m}{2}\omega^2q^2\qquad\qquad\longrightarrow\qquad\qquad\widehat{H}=-\frac{\hbar^2}{2m}\parc[{}^2]{q^2}+\frac{m}{2}\omega^2q^2\]
With this Hamiltonian we could now solve the Schrödinger equation which, by separation of variables, turns into an eigenvalue problem
    \[\widehat{H}|\psi\rangle=E|\psi\rangle\]
It is well known that the eigenvalues are given by $E_n=\hbar\omega\left(n+\frac{1}{2}\right)$ while the eigenvectors $|\psi_n\rangle$ are given in terms of the Hermite polynomials. Instead of developing such solutions, we turn our attention into the algebraic approach. For that matter, we introduce the \textbf{creation}\index{Creation operator} and \textbf{annihilation}\index{Annihilation operator} operators given by
\[\widehat{a}^\dagger:=\sqrt{\frac{\omega}{2}}\widehat{q}-\frac{i}{\sqrt{2\omega}}\widehat{p}\qquad\qquad\qquad\widehat{a}:=\sqrt{\frac{\omega}{2}}\widehat{q}+\frac{i}{\sqrt{2\omega}}\widehat{p}\]
where we have taken units, to simplify the notation, such that $m=1$ and $\hbar=1$. It is important to notice that now $\widehat{q}$ and $\widehat{p}$ are not necessarily given by \eqref{Mathematical background - equation - representation q p}. They are to be thought as fundamental observables of the theory i.e.\ self-adjoint operators over the Hilbert space $\mathfrak{h}$ satisfying $[\widehat{q},\widehat{p}]=i\mathrm{Id}$.\separ

We can, of course, write $\widehat{q}$ and $\widehat{p}$ in terms of $\widehat{a}^\dagger$ and $\widehat{a}$, which allows us to write the Hamiltonian and the commutation relations
\[\widehat{H}=\omega\left(\widehat{a}^\dagger\widehat{a}+\frac{1}{2}\mathrm{Id}\right)\qquad\qquad\qquad[\widehat{a},\widehat{a}^\dagger]=\mathrm{Id}\qquad\qquad\qquad[\widehat{a},\widehat{H}]=\omega\widehat{a}\]
Besides, the eigenvectors $|\psi_n\rangle$ of the Hamiltonian $\widehat{H}$ are related through the creation and annihilation operators  by the equations
\[\widehat{a}^\dagger|\psi_n\rangle=\sqrt{n+1}|\psi_{n+1}\rangle\qquad\qquad\qquad\qquad\widehat{a}|\psi_n\rangle=\sqrt{n}|\psi_{n-1}\rangle\]
This behavior justifies their names: $\widehat{a}^\dagger$ creates a ``quantum of energy'' that is appended to the $n$-th oscillator while $\widehat{a}$ destroys it. Notice that if we know the ground state $|\psi_0\rangle$, i.e.\ an element of $\mathrm{ker}\,\widehat{a}$, then we can recover in this algebraic approach every single eigenvector and eigenvalue
\[|\psi_n\rangle=\frac{1}{\sqrt{n!}}(\widehat{a}^\dagger)^n|\psi_{0}\rangle\qquad\qquad \widehat{H}|\psi_n\rangle=\omega\left(n+\frac{1}{2}\right)|\psi_n\rangle\]
and the problem is completely solved. It is then reasonable to try to generalize such operators in the field theory case. To do so we need to relate them to the solutions of the theory.\separ

First we present the evolution of the operators in the Heisenberg representation
\[\widehat{a}(t):=\mathrm{Exp}\left(it\widehat{H}\right)\widehat{a}(0)\mathrm{Exp}\left(-it\widehat{H}\right)\qquad\quad\longrightarrow\qquad\quad i\deriv{\widehat{a}}{t}=[\widehat{a}(t),\widehat{H}]=\omega\widehat{a}(t)\]
whose solution is immediate
\[\widehat{a}(t)=e^{-i\omega t}\widehat{a}(0)\qquad\quad\longrightarrow\qquad\quad\widehat{a}^\dagger(t)=e^{i\omega t}\widehat{a}^\dagger(0)\]
Now we can write the evolution of $\widehat{q}(t)$ as
\[\widehat{q}(t)=\frac{1}{\sqrt{2\omega}}\Big(\widehat{a}(t)+\widehat{a}^\dagger(t)\Big)=\frac{1}{\sqrt{2\omega}}\Big(e^{-i\omega t}\widehat{a}(0)+e^{i\omega t}\widehat{a}^\dagger(0)\Big)\]
A crucial remark is in order now: the annihilation operator $\widehat{a}$\index{Annihilation operator}, in the Schrödinger picture, is precisely the ``positive frequency part'' of the Heisenberg position operator. This suggests that we have to split the solutions of our theory into positive and negative parts and pick one of them to have a candidate for the annihilation operator. This procedure is developed in the following section.

    \subsection{Fock quantization}\label{Mathematical background - subsection - second quantization}
 
 \subsubsection*{Algorithm \emph{à la Wald}}\trassub
 
 We now consider that we have a field theory and sketch the procedure to build a Hilbert space that represents the quantum counterpart to the initial classical theory. For more details see \cite{wald1994quantum}.
 
 \begin{enumerate}
 	\item $\mathcal{S}=\Big\{$ Real vector space of classical solutions to the Field Equations $\Big\}$
 	\item $\mathcal{S}^\C$ is a (non-canonical) complexification of $\mathcal{S}$ endowed with $\peqsubfino{\langle\hspace*{0.4ex},\hspace*{0.1ex}\rangle}{\Omega}{-0.2ex}$ a complex bilinear form.
 	\item $\peqsub{\mathcal{S}^\C}{+}$ is a subspace of ``positive frequency'' solutions where \[\peqsubfino{\langle\hspace*{0.4ex},\hspace*{0.1ex}\rangle}{+}{-0.3ex}:=\peqsubfino{\left.\peqsubfino{\langle\hspace*{0.4ex},\hspace*{0.1ex}\rangle}{\Omega}{-0.3ex}\right|}{\mathcal{S}^\C_+\times \mathcal{S}^\C_+}{-0.3ex}\]
 	is a complex scalar product.
 	\item $\left(\peqsub{\mathcal{S}^\C}{+},\peqsubfino{\langle\hspace*{0.4ex},\hspace*{0.1ex}\rangle}{+}{-0.3ex}\right)$ pre-Hilbert space $\updown{\text{Cauchy}}{\text{Completion}}{\longlongrightarrow{5.5}}\  \left(\mathfrak{h},\peqsubfino{\langle\hspace*{0.4ex},\hspace*{0.1ex}\rangle}{+}{-0.3ex}\right)\ 1$-particle Hilbert space.
 \end{enumerate}
This procedure allows us to built the $1$-particle Hilbert space that describes the quantum behavior of a particle. Once we have that, as we mentioned in the introduction of this section, the next desirable step is to obtain a general description for many particles. This is done in the following.
 
      \subsubsection*{Fock space: construction and properties}\trassub
      
       In this section we follow the nice lecture notes of Stéphane Attal \cite{attal}. Let $\mathfrak{h}$ be a complex separable Hilbert space. If, physically, it encodes all possibles states of $1$ particle, then
       \[\mathfrak{h}^{\otimes n}=\mathfrak{h}\otimes\overset{(n)}{\cdots}\otimes\mathfrak{h}\]
       represents the states of $n$ distinguishable particles. This new spaces are obtained after completion of the pre-Hilbert space of finite linear combinations of elements of the form $v_1\otimes\cdots\otimes v_n$ with the scalar product
      \begin{equation}\label{Mathematical background - equation - producto escalar Fock}
        \peqsub{\langle v_1\otimes\cdots\otimes v_n,w_1\otimes\cdots\otimes w_n\rangle}{\otimes}=\langle v_1,w_1\rangle\cdots\langle v_n,w_n\rangle
      \end{equation}
      For $n=0$ we set $\mathfrak{h}^{\otimes 0}=\C$. The element $1\in\mathfrak{h}^{\otimes 0}=\C$, usually denoted by $|0\rangle$, plays an important role and is called the \textbf{vacuum vector}\index{Vacuum vector}.\separ
      
      In physics one usually deals with systems of indistinguishable particles (bosons or fermions) that, as an ensemble, posses some exchange symmetries. The states of such system are thus described by subspaces of $\mathfrak{h}^{\otimes n}$. We define the symmetrized and antisymmetrized tensor products\allowdisplaybreaks[0]
      \begin{align*}
        & u_1\smallcirc\cdots\smallcirc u_n=\frac{1}{n!}\sum_{\sigma\in S_n}u_{\sigma(1)}\otimes\cdots\otimes u_{\sigma(n)}\\
        & u_1\wedge\cdots\wedge u_n=\frac{1}{n!}\sum_{\sigma\in S_n}(-1)^\sigma u_{\sigma(1)}\otimes\cdots\otimes u_{\sigma(n)}
      \end{align*}\allowdisplaybreaks
      where $S_n$ is the permutation group of $\{1,\ldots,n\}$ and $(-1)^\sigma$ the signature of $\sigma$. Notice that we divide by $n!$, the number of elements in $S_n$, so that (anti)symmetrizying an already (anti)symmetric vector is just the identity.\separ
      
      The closed subspace of $\mathfrak{h}^{\otimes n}$ generated by the elements of the form $u_1\smallcirc\cdots\smallcirc u_n$ is denoted by $\mathfrak{h}^{\smallcirc n}$, while the one generated by $u_1\wedge\cdots\wedge u_n$ is denoted $\mathfrak{h}^{\wedge n}$. It is straightforward to prove
      \begin{align*}
        &\peqsub{\langle v_1\smallcirc\cdots\smallcirc v_n, w_1\smallcirc\cdots\smallcirc w_n\rangle}{\otimes}=\frac{1}{n!}\textrm{per}\Big(\!\langle v_i,w_j\rangle\!\Big)\qquad
        \peqsub{\langle v_1\wedge\cdots\wedge v_n, w_1\wedge\cdots\wedge w_n\rangle}{\otimes}=\frac{1}{n!}\det\Big(\!\langle v_i,w_j\rangle\!\Big)
      \end{align*}
      where $\textrm{per}(A)$ is the permanent of a matrix (computed as the determinant but only with plus signs). It is much more covenient to modify the scalar product over $\mathfrak{h}^{\smallcirc n}$ and $\mathfrak{h}^{\wedge n}$ so that no factorial arises. Thus we consider
      \begin{align*}
      &\peqsub{\langle v_1\smallcirc\cdots\smallcirc v_n, w_1\smallcirc\cdots\smallcirc w_n\rangle}{\smallcirc}:=\textrm{per}\Big(\langle v_i,w_j\rangle\Big)\qquad
      \peqsub{\langle v_1\wedge\cdots\wedge v_n, w_1\wedge\cdots\wedge w_n\rangle}{\wedge}:=\det\Big(\langle v_i,w_j\rangle\Big)
      \end{align*}
      
      Let us define the full, symmetric and antisymmetric Fock spaces by
      \begin{align*}
        \peqsub{\mathcal{F}}{\!\!\otimes}(\mathfrak{h})=\left(\bigoplus_{n=0}^\infty\mathfrak{h}^{\otimes n},\peqsub{\langle\,,\ \!\!\rangle}{\otimes}\right)\qquad\qquad
        \peqsub{\mathcal{F}}{\!\!\smallcirc}(\mathfrak{h})=\left(\bigoplus_{n=0}^\infty\mathfrak{h}^{\smallcirc n},\peqsub{\langle\,,\ \!\!\rangle}{\smallcirc}\right)\qquad\qquad
        \peqsub{\mathcal{F}}{\!\!\wedge}(\mathfrak{h})=\left(\bigoplus_{n=0}^\infty\mathfrak{h}^{\wedge n},\peqsub{\langle\,,\ \!\!\rangle}{\wedge}\right)
      \end{align*}
      where the scalar products are extended by linearity and $\peqsub{\langle v_1\otimes\cdots\otimes v_n,w_1\otimes\cdots\otimes w_k\rangle}{\otimes}=0$ if $k\neq n$. Sometimes, when there is no need to specify which case we are considering or it is clear from the context, we will simply write $\mathfrak{h}_n$, $\langle\,,\ \!\!\rangle$ and $\mathcal{F}(\mathfrak{h})$.
      
      \begin{remarks}
        \item $\mathcal{F}(\mathfrak{h})$ is a separable Hilbert space.
        \item $\mathcal{F}(\C)=\ell^2(\N):=\{a:\N\to\C\}$ because $\C^{\otimes n}=\C$.
        \item The elements of $\mathcal{F}(\mathfrak{h})$ are formal series $F=\sum f_n$ such that $f_n\in\mathfrak{h}_n$ and
        \begin{align*}
          |F|^2=\sum_{n=0}^\infty\langle f_n,f_n\rangle<\infty
        \end{align*}      
      \end{remarks}
      
      \begin{definition}\label{def:background second quantization map}\mbox{}\\
      Given a map $T:\mathfrak{h}_I\to\mathfrak{h}_F$ between two Hilbert spaces we define the \textbf{second quantization}\index{Quantization!Second quantization} of $T$ as the map $\mathcal{F}(T):\mathcal{F}(\mathfrak{h}_I)\to\mathcal{F}(\mathfrak{h}_F)$ such that
      \begin{align*}
        \mathcal{F}(T)(v_1\otimes\cdots\otimes v_n)=T(v_1)\otimes\cdots\otimes T(v_n)
      \end{align*}
  \end{definition}

      \begin{properties}
      	\item If $|T|>1$ then $\mathcal{F}(T)$ is not bounded.
      	\item $\mathcal{F}(T_1\smallcirc T_2)=\mathcal{F}(T_1)\smallcirc\mathcal{F}(T_2)$
      	\item $\mathcal{F}(T)^*=\mathcal{F}(T^*)$. In particular, if $T$ is unitary so is $\mathcal{F}(T)$\label{property:background unitariedad del lift}.
      \end{properties}
      These last two properties tell us that the promotion of operators from $\mathfrak{h}$ to $\mathcal{F}(\mathfrak{h})$ via the second quantization behaves well.

    \subsubsection*{Coherent states}\trassub
    
      Notice that in the previous section we have defined a functor $\mathcal{F}:\mathrm{Hilbert}\to\mathrm{Hilbert}$, called the \textbf{second quantization functor}\index{Quantization!Second quantization functor}, taking a separable $1$-particle Hilbert space $\mathfrak{h}$ to another one $\mathcal{F}(\mathfrak{h})$ suitable to describe the physics of a system comprising an arbitrary number of particles. We also know how to lift a map between two Hilbert spaces so it seems natural to try to mimic these procedures to lift also elements from $\mathfrak{h}$ to $\mathcal{F}(\mathfrak{h})$. This is easy if one realizes that formally, up to the prefactors, $\mathcal{F}(\mathfrak{h})$ is defined as $e^{\otimes\mathfrak{h}}$, which motivates the following definition.
      
      \begin{definition}\mbox{}\\
        We define $\varepsilon:\mathfrak{h}\to\mathcal{F}(\mathfrak{h})$ by $\displaystyle\varepsilon(v)=\sum_{n=0}^\infty\frac{v^{\otimes n}}{n!}$, called \textbf{the coherent state associated with $\bm v\in\mathfrak{h}$}\index{Quantization!Coherent state}\index{Coherent state}.
      \end{definition}
      As usual we consider $v^{\otimes 0}=1=|0\rangle\in\C$. It is customary to talk about coherent states only in the symmetric case so we stick to this case from now on. Notice that $\varepsilon(v)$ can be considered as an element of $\mathcal{F}(\mathfrak{h})$ labeled by an element of $\mathfrak{h}$. Although not all elements of the Fock can be obtained in that way, we will see that they form an overcomplete set. Thus, to define some objects and prove some statements about $\mathcal{F}(\mathfrak{h})$, it will be enough to do it over the coherent states.
      
      \begin{properties}
      	\item $\peqsub{\langle\varepsilon(v),\varepsilon(w)\rangle}{\smallcirc}=e^{\langle v,w\rangle}$\label{Mathematical background - property - <e(v),e(w)>=e<u,v>}
      	\item If $T:\mathfrak{h}_I\to\mathfrak{h}_F$ then $\mathcal{F}(T)\big(\peqsub{\varepsilon}{I}(v)\big)=\peqsub{\varepsilon}{F}(Tv)$ that can be expressed as a commutative diagram\label{Mathematical background - property - F(T)(e(v))=e(TV)}
      	
      	\centerline{\begin{tikzpicture}[ampersand replacement=\&]
      		\matrix (m) [matrix of math nodes,row sep=3em,column sep=4em,minimum width=2em]
      		{\mathcal{F}(\mathfrak{h}_I) \& \mathcal{F}(\mathfrak{h}_F) \\
      			\mathfrak{h}_I  \& \mathfrak{h}_F \\};
      		\path[-stealth]
      		(m-2-1) edge node [left] {$\peqsub{\varepsilon}{I}$} (m-1-1)
      		(m-2-1.east|-m-2-2) edge node [below] {$T$} (m-2-2)
      		(m-1-1.east|-m-1-2) edge node [above] {$\mathcal{F}(T)$}	(m-1-2)
      		(m-2-2) edge node [right] {$\peqsub{\varepsilon}{F}$} (m-1-2);
      		\end{tikzpicture}}
      \end{properties}
      \begin{proof}\mbox{}\vspace*{-2.5ex}
      	\begin{align*}
      	  \refconchapp{Mathematical background - property - <e(v),e(w)>=e<u,v>}&&\left\langle\varepsilon(v),\varepsilon(w)\right\rangle_{\smallcirc}&=\left\langle\sum_{n=0}^\infty\frac{v^{\smallcirc n}}{n!},\sum_{k=0}^\infty\frac{w^{\smallcirc k}}{k!}\right\rangle_{\!\!\smallcirc}=\sum_{n,k=0}^\infty\frac{1}{n!k!}\left\langle v^{\smallcirc n},w^{\smallcirc k}\right\rangle_{\!\smallcirc}=\\
      	  &&&=\sum_{n=0}^\infty\frac{1}{(n!)^2}\left\langle v^{\smallcirc n},w^{\smallcirc n}\right\rangle_{\!\smallcirc}=\sum_{n=0}^\infty\frac{n!}{(n!)^2}\langle v,w\rangle^n=\sum_{n=0}^\infty\frac{\langle v,w\rangle^n}{n!}=e^{\langle v,w\rangle}\\
      	  \refconchapp{Mathematical background - property - F(T)(e(v))=e(TV)}&&\mathcal{F}(T)\Big(\peqsub{\varepsilon}{I}(v)\Big)&=\sum_{n=0}^\infty\frac{\mathcal{F}(T)\big(v^{\smallcirc n}\big)}{n!}=\sum_{n=0}^\infty\frac{(Tv)^{\smallcirc n}}{n!}=\peqsub{\varepsilon}{F}(Tv)
      	\end{align*}
      	\mbox{}\vspace*{-6.5ex}
      	
      \end{proof}      
      
      We have mentioned that the Fock space $\mathcal{F}(\mathfrak{h})$ is, in a sense, the exponential $e^{\otimes\mathfrak{h}}$. Let us see that it behaves indeed like an exponential taking addition to multiplication.
      \begin{proposition}\label{Mathematical background - proposition - e(h1+h2)=e(h1)xe(h2)}\mbox{}\\
      	There exists a unique unitary isomorphism $U:\peqsub{\mathcal{F}}{\!\!\smallcirc}(\mathfrak{h}_I\oplus\mathfrak{h}_F)\to\peqsub{\mathcal{F}}{\!\!\smallcirc}(\mathfrak{h}_I)\otimes\peqsub{\mathcal{F}}{\!\!\smallcirc}(\mathfrak{h}_F)$ such that
        \begin{align*}
          U\Big(\varepsilon(u_I\oplus u_F)\Big)=\peqsub{\varepsilon}{I}(u_I)\otimes \peqsub{\varepsilon}{F}(u_F)\qquad\equiv\qquad e^{\mathfrak{h}_I\oplus\mathfrak{h}_F}\cong e^{\mathfrak{h}_I}\otimes e^{\mathfrak{h}_F}
        \end{align*}
      \end{proposition}
      Using the fact that the coherent states are dense and that they behave well under the tensor product, it can be proved that $U$ can be extended to the whole Fock space \cite{attal}. The unitarity follows from property \refconchap{Mathematical background - property - <e(v),e(w)>=e<u,v>}.

    \subsubsection*{Creation and annihilation}\trassub
    
      We define $a^*:\mathfrak{h}\to\mathcal{L}(\peqsub{\mathcal{F}}{\!\!\smallcirc}(\mathfrak{h}))$ given by\allowdisplaybreaks[0]
      \begin{align*}
        &a^*(u)|0\rangle=u\\
        &a^*(u)\Big(v_1\smallcirc\cdots\smallcirc v_n\Big)=u\smallcirc v_1\smallcirc\cdots\smallcirc v_n
      \end{align*}\allowdisplaybreaks
      We see that $a^*(u)$, called the \textbf{creation operator}\index{Creation operator} associated to $u$, takes elements from $\mathfrak{h}^{\smallcirc n}$ to $\mathfrak{h}^{\smallcirc (n+1)}$, which physically means that it creates a particle at the state $u\in\mathfrak{h}$. We also define $a^*:\mathfrak{h}\to\mathcal{L}(\peqsub{\mathcal{F}}{\!\!\smallcirc}(\mathfrak{h}))$ given by
      \begin{align*}
        &a(u)|0\rangle=0\\
        &a(u)\Big(v_1\smallcirc\cdots\smallcirc v_n\Big)=\langle u,v_1\rangle v_2\smallcirc\cdots \smallcirc v_n+\cdots +\langle u,v_n\rangle v_1\smallcirc\cdots\smallcirc v_{n-1}
      \end{align*}
      $a(u)$, known as the \textbf{annihilation operator}\index{Annihilation operator} associated to $u$, reduces the number of particles by projecting over $u\in\mathfrak{h}$. In particular if the states $v_1,\ldots,v_n$ of the $n$ particles $v_1\smallcirc\cdots\smallcirc v_n$ are orthogonal to $u$, then $a^*(u)$ kills all the particles.
      
      \begin{properties}
      	\item $a^*$ is linear whereas $a$ is antilinear.
      	\item $a(u)^*=a^*(u)$ i.e.\ the adjoint operator of $a(u)$ is $a^*(u)$.
      	\item $a(u)\Big(\varepsilon(v)\Big)=\langle u,v\rangle\varepsilon(v)$.\label{property:background autovector estado coherente}
      	\item $\displaystyle a^*(u)\Big(\varepsilon(v)\Big)=\left.\deriv{}{\lambda}\right|_{\lambda=0}\varepsilon(v+\lambda u)$
      \end{properties}
        \begin{proof}\mbox{}\\
       	  The first property is immediate while the second can be found in \cite{attal}. The two last ones are direct computations\vspace*{-2.5ex}       	  
       	  \begin{align*}
       	    a(u)\Big(\varepsilon(v)\Big)&=a(u)\left(\sum_{n=0}^\infty\frac{v^{\smallcirc n}}{n!}\right)=\sum_{n=0}^\infty\frac{a(u)\Big(v^{\smallcirc n}\Big)}{n!}=\\
       	    &=\sum_{n=1}^\infty\frac{n\langle u,v\rangle  v^{\smallcirc (n-1)}}{n!}=\langle u,v\rangle\sum_{n=1}^\infty\frac{v^{\smallcirc (n-1)}}{(n-1)!}=\langle u,v\rangle\varepsilon(v)\\
        	\left.\deriv{}{\lambda}\right|_{\lambda=0}\varepsilon(v+\lambda u)&=\sum_{n=0}^\infty\left.\deriv{}{\lambda}\right|_{\lambda=0}\frac{(v+\lambda u)^{\smallcirc n}}{n!}=\sum_{n=1}^\infty\frac{n u\smallcirc v^{\smallcirc(n-1)}}{n!}=\\
       	    &=u\smallcirc \sum_{m=0}^\infty\frac{v^{\smallcirc m}}{m!}=u\smallcirc \varepsilon(v)=a^*(u)\Big(\varepsilon(v)\Big)\\[-9ex]
       	  \end{align*} 
        \end{proof}
    
      Property \refconchap{property:background autovector estado coherente} tells us that the coherent states are eigenvectors of the annihilation operator. In fact, sometimes this property is taken as its definition. 
   
      \subsubsection*{Evolution}\trassub
      
        With all the previous ingredients we can already introduce dynamics. So let us consider a Hamiltonian $h:\mathfrak{h}\to\mathfrak{h}$ over the $1$-particle Hilbert space i.e.\ a self-adjoint operator that encodes the dynamics of the system. Usually it will be given as the square root of minus a Laplace operator $h=\sqrt{-\Delta}$. The evolution is then given by the Schrödinger equation whose solutions are given by the (strongly continuous) one-parameter group  $\{e^{-ith}\}_t$ of unitary operators associated to the self-adjoint Hamiltonian $h$.
      \begin{align*}
        &i\deriv{}{t}v_t=h(v_t)\qquad\qquad\longrightarrow\qquad\qquad v_t=e^{-ith}v_0
      \end{align*}
      The converse is also true: given a strongly continuous one-parameter group  $\{U_t\}_t$ of unitary operators, we have a self-adjoint operator $h$ such that
      \begin{align*}
        i\deriv{}{t}U_tv=h(U_tv)\qquad\longrightarrow\qquad h=i\left.\deriv{}{t}\right|_{t=0}U_t
      \end{align*}
      Now notice that property \refconchap{property:background unitariedad del lift} ensures that the lift $\mathcal{F}(e^{-ith})$, which by definition is given by
      \begin{align*}
        &\mathcal{F}(e^{-ith})|0\rangle=|0\rangle\in\C\\
        &\mathcal{F}(e^{-ith})\Big(v_1\smallcirc \cdots\smallcirc v_n\Big)=\left(e^{-ith}v_1\right)\smallcirc \cdots\smallcirc \left(e^{-ith}v_n\right)
      \end{align*}
      forms a one-parameter group  $\{\mathcal{F}(e^{-ith})\}_t$ of unitary operators. It can also be proved that if $\{U_t\}_t$ is strongly continuous, then so is $\{\mathcal{F}(e^{-ith})\}_t$. Thus we have a self-adjoint operator 
      \begin{align}\label{eq:backgroud Hamiltonian Fock}
        H_h=i\left.\deriv{}{t}\right|_{t=0}\mathcal{F}(e^{-ith})
      \end{align}
      on $\peqsub{\mathcal{F}}{\!\!\smallcirc}(\mathfrak{h})$, known as the \textbf{second quantization} of the Hamiltonian $h$. We see that
      \begin{align*}
      &H_h|0\rangle=0\\
      &H_h\Big(v_1\smallcirc \cdots\smallcirc v_n\Big)=(hv_1)\smallcirc v_2\cdots\smallcirc v_n+v_1\smallcirc(hv_2)\smallcirc v_3\smallcirc\cdots\smallcirc v_n+\cdots +v_1\smallcirc\cdots\smallcirc v_{n-1}\smallcirc (hv_n)
      \end{align*}
      Notice that if $h$ is the identity then $\left.(H_{\mathrm{Id}})\right|_{\mathfrak{h}_n}=n\hspace*{0.2ex}\mathrm{Id}_{\mathfrak{h}_n}$ which is known as the \textbf{number operator}\index{Number operator}.\separ
      
      One of the most important features about the coherent sates is that their evolution is determined by the evolution of the $1$-particle label as the following proposition shows.
      \begin{proposition}\label{proposition:background evolucion coherent state}\mbox{}\\
      	$\mathcal{F}(e^{-ith})\varepsilon(v_0)=\varepsilon(e^{-ith}v_0)$
      \end{proposition}
      \begin{proof}\mbox{}\vspace*{-3.5ex}
      	\begin{align*}
      	  \mathcal{F}(e^{-ith})\varepsilon(v_0)&=\mathcal{F}(e^{-ith})\left(\sum_{n=0}^\infty\frac{v_0^{\smallcirc n}}{n!}\right)=\sum_{n=0}^\infty\frac{\mathcal{F}(e^{-ith})(v_0^{\smallcirc n})}{n!}=\\
      	  &=\sum_{n=0}^\infty\frac{(e^{-ith}v_0)^{\smallcirc n}}{n!}=\varepsilon\left(e^{-ith}v_0\right)\\[-8ex]
      	\end{align*}
      \end{proof}

    \subsubsection*{Passing dynamics}\label{Mathematical background - subsection - Parsing dynamics}\trassub
    
      Let $T:D(T)\subset\mathfrak{h}_I\to\mathfrak{h}_F$ be some map between Hilbert spaces defined over a dense set $D(T)$. Consider now some Hamiltonian $\peqsub{h}{I}$ over $\mathfrak{h}_I$ that encodes the dynamics in $\mathfrak{h}_I$ and, hence, in $\mathcal{F}(\mathfrak{h}_I)$. Let us see how we can translate some of the dynamics into $\mathcal{F}(\mathfrak{h}_F)$.\separ
      
      $\blacktriangleright$ Through the adjoint $T^*:D(T^*)\subset\mathfrak{h}_F\to\mathfrak{h}_I$ of $T$\separprevia
      
      Consider the adjoint of $T$, which exists because $D(T)$ is dense, and define $\peqsub{h}{F}:=T\peqsub{h}{I}T^*$ over $\mathfrak{h}_F$. It is easy to prove that it is positive and self-adjoint, but its domain turns out to be quite complicated (and maybe even zero!)
      \begin{align*}
        D(\peqsub{h}{F})=\Big\{y\in D(T^*)\ \ \ /\ \ \ T^*y\in D(\peqsub{h}{I})\ \text{ and }\ \peqsub{h}{I}T^*y\in D(T)\Big\}
      \end{align*}
      As $\peqsub{h}{F}$ is self-adjoint we have the induced unitary evolution in $\mathfrak{h}_F$ given by $\{e^{-it\peqsub{h}{F}}\}$. We can now lift this group to $\mathcal{F}(\mathfrak{h}_F)$ and get
      \begin{align*}
        H_{\peqsub{h}{F}}=i\left.\deriv{}{t}\right|_0\mathcal{F}\left(e^{-it\peqsub{h}{F}}\right)
      \end{align*}
      Notice that if $T^*T=\mathrm{Id}$ then $e^{-it\peqsub{h}{F}}=Te^{-it\peqsub{h}{I}}T^*$ and thus $\mathcal{F}\left(e^{-it\peqsub{h}{F}}\right)=\mathcal{F}\left(T\right)\mathcal{F}\left(e^{-it\peqsub{h}{F}}\right)\mathcal{F}\left(T\right)^*$. This result is natural since in this case $T$ is unitary.\separpost
      
      $\blacktriangleright$ Coherent sate of the image through $T$\separprevia
      
      Consider $v_0\in\mathfrak{h}_I\setminus\{0\}$ to be some initial state of a particle and denote the time dependent states $v_t=e^{-it\peqsub{h}{I}}v_0\in\mathfrak{h}_I$. Then we define the coherent state $\psi_t=\peqsub{\varepsilon}{F}(Tv_t)\in\mathcal{F}(\mathfrak{h}_F)$. Deriving we obtain
      \begin{align*}
        i\deriv{\psi_t}{t}&=i\deriv{}{t}\peqsub{\varepsilon}{F}(Tv_t)=i\sum_{n=0}^\infty\frac{\deriv{}{t}(Tv_t)^{\smallcirc n}}{n!}=i\sum_{n=1}^\infty\frac{n\deriv{(Tv_t)}{t}\smallcirc (Tv_t)^{\smallcirc(n-1)}}{n!}\updown{T\text{ is}}{\text{linear}}{=}\\
        &=T\left(i\deriv{v_t}{t}\right)\smallcirc \sum_{m=0}^\infty\frac{(Tv_t)^{\smallcirc m}}{m!}\updown{\text{Schr}}{\text{eq.}}{=}T(\peqsub{h}{I}v_t)\smallcirc\peqsub{\varepsilon}{F}(Tv_t)\updown{\text{def.}a^*}{\refconchap{property:background autovector estado coherente}}{=}a^*(T\peqsub{h}{I}v_t)\frac{a(Tv_t)}{|Tv_t|^2}\psi_t
      \end{align*}
      Thus we have some sort of Schrödinger equation 
      \begin{align}\label{eq:background Schro map T}
        i\deriv{}{t}\peqsub{\varepsilon}{F}(Te^{-it\peqsub{h}{I}}v_0)=\frac{a^*(T\peqsub{h}{I}e^{-it\peqsub{h}{I}}v_0)a(Te^{-it\peqsub{h}{I}}v_0)}{|Te^{-it\peqsub{h}{I}}v_0|^2}\peqsub{\varepsilon}{F}(Te^{-it\peqsub{h}{I}}v_0)
      \end{align}
      Before studying the consequences of this equation, let us focus on the following particular case: consider $T=\mathrm{Id}:\mathfrak{h}\to\mathfrak{h}$ and some Hamiltonian $h$. We obtain
      \begin{align}\label{eq:background Schro a a^*}
      \begin{split}
        H_h\varepsilon(v_0)&\overset{\eqref{eq:backgroud Hamiltonian Fock}}{=}i\left.\deriv{}{t}\right|_{t=0}\mathcal{F}(e^{-ith})(\varepsilon(v_0))\overset{\ref{proposition:background evolucion coherent state}}{=}\\[1.5ex]
        &\overset{\phantom{\eqref{eq:backgroud Hamiltonian Fock}}}{=}\hspace*{-0,5ex}i\left.\deriv{}{t}\right|_{t=0}\varepsilon(e^{-ith}v_0)\updown{\eqref{eq:background Schro map T}}{t=0}{=}\frac{a^*(hv_0)a(v_0)}{|v_0|^2}\varepsilon(v_0)
        \end{split}
      \end{align}
      Furthermore, let us now assume $\mathfrak{h}=\C$. First recall that $\mathcal{F}(\C)=\ell^2(\N)$. Second, notice that any Hamiltonian $h:\C\to\C$ has to be $h=\omega \mathrm{Id}$ for some $\omega\in\C$ because $h$ is a $1$-dimensional vector space. As $h$ is self-adjoint by definition, $\omega$ has to be real. Therefore $a^*(hv)=a^*(\omega v)=\omega v a^*(1)$ and $a(v)=\overline{v}a(1)$ for every $v\in\C$. Thus the previous equations reads simply 
      \begin{align*}
        H_h\varepsilon(v)=\omega a^*(1)a(1) \varepsilon(v) \qquad\longrightarrow\qquad H_h=\omega N
      \end{align*}
      where $N=a^*(1)a(1)$ is called the \textbf{number operator}\index{Number operator}.\separ
      
      Notice that this procedure cannot be carried over equations \eqref{eq:background Schro map T} and \eqref{eq:background Schro a a^*} because $h$ is in general more complicated. We have, however, that the evolution of the latter is unitary because $H_h$ is self-adjoint by construction, while for the former the unitarity is not assured because we have no Hamiltonian associated with this dynamics. If we compute
      \begin{align}\label{Mathematical background - equation - evolution unitaria producto escalar}\begin{split}
        i\deriv{}{t}\peqsub{\langle\peqsub{\varepsilon}{F}(Tv_t),\peqsub{\varepsilon}{F}(Tw_t)\rangle}{\!\smallcirc}&=i\deriv{}{t}e^{\langle Tv_t,Tw_t\rangle}=e^{\langle Tv_t,Tw_t\rangle}i\deriv{}{t}\langle Tv_t,Tw_t\rangle\updown{T\text{ is}}{\text{linear}}{=}\\
        &\hspace*{-6ex}=\peqsub{\langle\peqsub{\varepsilon}{F}(Tv_t),\peqsub{\varepsilon}{F}(Tw_t)\rangle}{\!\smallcirc}\left(\left\langle -Ti\deriv{}{t}v_t,Tw_t\right\rangle+\left\langle Tv_t,Ti\deriv{}{t}w_t\right\rangle\right)\overset{\eqref{eq:backgroud Hamiltonian Fock}}{=}\\
        &\hspace*{-6ex}=\peqsub{\langle\peqsub{\varepsilon}{F}(Tv_t),\peqsub{\varepsilon}{F}(Tw_t)\rangle}{\!\smallcirc}\Big(\langle Tv_t,T\peqsub{h}{I}w_t\rangle-\langle T\peqsub{h}{I}v_t,Tw_t\rangle\Big)\end{split}
      \end{align}
      where in the second equality we have used that the scalar product is sesquilinear. We know that the evolution is unitary if and only of the scalar product is preserved for all coherent states so in general we cannot assure that the evolution is unitary. Notice, for instance, that if $T=\mathrm{Id}$ then the evolution is unitary because $\peqsub{h}{I}$ is self-adjoint.\newpage
      
      $\blacktriangleright$ Summary\separprevia
      
      Given a map $T:\mathfrak{\peqsub{h}{I}}\to\mathfrak{h}_F$ we have endowed $\mathfrak{h}_F$ with two different dynamics that can be promoted to $\mathcal{F}(\mathfrak{h}_F)$. The first one is considering $\peqsub{h}{F}=T\peqsub{h}{I}T^*$, thus
      \begin{align*}
        &i\deriv{}{t}\peqsub{\varepsilon}{F}(e^{-it\peqsub{h}{F}}w_0)=\frac{a^*(\peqsub{h}{F}e^{-it\peqsub{h}{F}}w_0)a(e^{-it\peqsub{h}{F}}w_0)}{|w_0|^2}\peqsub{\varepsilon}{F}(e^{-it\peqsub{h}{F}}w_0)
      \end{align*}
      Notice that the functional expression is the same as equation \eqref{eq:background Schro map T} with $T=\mathrm{Id}$ and $h=\peqsub{h}{F}$. If we take some $v_0$ and define $w_0:=Tv_0\neq0$ we have
      \begin{align*}
      &i\deriv{}{t}\peqsub{\varepsilon}{F}(e^{-it\peqsub{h}{F}}Tv_0)=\frac{a^*(\peqsub{h}{F}e^{-it\peqsub{h}{F}}Tv_0)a(e^{-it\peqsub{h}{F}}Tv_0)}{|Tv_0|^2}\peqsub{\varepsilon}{F}(e^{-it\peqsub{h}{F}}Tv_0)
      \end{align*}\separ
      
      The second way is given by \eqref{eq:background Schro map T} itself
      \begin{align*} &i\deriv{}{t}\peqsub{\varepsilon}{F}(Te^{-it\peqsub{h}{I}}v_0)=\frac{a^*(T\peqsub{h}{I}e^{-it\peqsub{h}{I}}v_0)a(Te^{-it\peqsub{h}{I}}v_0)}{|Te^{-it\peqsub{h}{I}}v_0|^2}\peqsub{\varepsilon}{F}(Te^{-it\peqsub{h}{I}}v_0)
      \end{align*}
      We see that the former makes the image $Tv_0$ evolve with $\peqsub{h}{F}=T\peqsub{h}{I}T^*$ while the latter takes the image of the evolution of $v_0$. If we evaluate both expressions at $t=0$ we get
      \begin{align*}
      &i\left.\deriv{}{t}\right|_{t=0}\peqsub{\varepsilon}{F}(e^{-it\peqsub{h}{F}}Tv_0)=\frac{a^*(\peqsub{h}{F}Tv_0)a(Tv_0)}{|Tv_0|^2}\peqsub{\varepsilon}{F}(Tv_0)\overset{\eqref{eq:background Schro a a^*}}{=}H_{\peqsub{h}{F}}\varepsilon(v_0)\\
      &i\left.\deriv{}{t}\right|_{t=0}\peqsub{\varepsilon}{F}(Te^{-it\peqsub{h}{I}}v_0)=\frac{a^*(T\peqsub{h}{I}v_0)a(Tv_0)}{|Tv_0|^2}\peqsub{\varepsilon}{F}(Tv_0)
      \end{align*}
      Once again we see that if $T^*T=\mathrm{Id}$ both expressions are equivalent. However if $T^*T$ is not the identity but commutes with $\peqsub{h}{I}$, then both evolutions are unitary and, in general, different.

   \subsubsection*{Bogoliubov coefficients}\index{Bogoliubov coefficients}\trassub
   
   Let us finish this introductory chapter by studying some important kind of examples related to the space of solutions of a field theory (in this case the Klein-Gordon equation), closely related to the algorithm \emph{à la Wald} previously defined.\separ
   
   Consider a globally hyperbolic space-time $(M\cong\R\times\Sigma,g)$ and the vector space of solutions to the KG equation $\mathcal{S}=\{\varphi:M\to\C\ /\ \square\varphi=0\}$. We have also the vector space of Cauchy data, which is given by $\mathcal{C}=\Cinf{\Sigma}\times\Cinf{\Sigma}'$ over $\Sigma$. The elements of the dual $\Cinf{\Sigma}'$ that we will use are of the form
   \[\bm{p}(f)=\int_\Sigma f \cdot{}p\, \peqsub{\mathrm{vol}}{\gamma}=\peqsubfino{(f,\sqrt{\gamma}\,p)}{\Sigma}{-0.4ex}\]
   so they can be identified with densities $\sqrt{\gamma}\,p$ for some $p\in\Cinf{\Sigma}$ and some $\gamma\in\mathrm{Met}(\Sigma)$ once we fixed some auxiliary volume metric $\peqsub{\mathrm{vol}}{\Sigma}$.\separ
   
   For every embedding $X:\Sigma\to M$ with $X(\Sigma)\subset M$ a Cauchy hypersurface\index{Cauchy hypersurface} we define $\peqsub{\Phi}{X}:\mathcal{C}\to\mathcal{S}$\label{Mathematical background - definition isomorfismo Cauchy->Sol} such that $\peqsub{\Phi}{X}(q,p)$ is the solution to the KG equation with Cauchy data $(q\smallcirc\!X,p\smallcirc\!X)$ over $X(\Sigma)$. This is in fact an isomorphism because the KG equation is a well posed initial value problem. Notice that given $(q,p)\in\mathcal{C}$, we can build different solutions $\peqsub{\Phi}{X_I}(q,p)$ and $\peqsub{\Phi}{X_F}(q,p)$ over different Cauchy hypersurfaces. Let us see now how they can be related.\separ
   
   Being $\peqsub{\Phi}{X}$ an isomorphism, we can define the linear isomorphism $\peqsub{\mathcal{T}}{F\leftarrow I}=\peqsub{\Phi}{X_F}\smallcirc\peqsub{\Phi}{X_I}^{-1}:\mathcal{S}\to\mathcal{S}$ as follows:\newpage
   \begin{itemize}
   	\item Consider a solution $\peqsub{\phi}{I}\in\mathcal{S}$ on $M\cong\R\times\Sigma$.
   	\item Retrieve the induced Cauchy data over $\peqsub{X}{I}(\Sigma)$, namely the position and momentum at $\Sigma$, which are given by $q:=\peqsub{X}{I}^*\peqsub{\phi}{I}$ and $p:=\sqrt{\peqsub{\gamma}{I}}\peqsub{X}{I}^*(\mathcal{L}_{\vec{n}_I}\peqsub{\phi}{I})$.
   	\item Construct the solution $\peqsub{\phi}{F}:=\peqsub{\mathcal{T}}{F\leftarrow I}(\peqsub{\phi}{I})$ with the Cauchy data $(q,p)$ over $\peqsub{X}{F}(\Sigma)$. In particular we have $q=\peqsub{X}{F}^*\peqsub{\phi}{F}$ and $p:=\sqrt{\peqsub{\gamma}{F}}\peqsub{X}{F}^*(\mathcal{L}_{\vec{n}_F}\peqsub{\phi}{F})$.
   \end{itemize}
   We recall that $\peqsub{\vec{n}}{X}\in\Gamma(X^*TM)$, defined on section \ref{Mathematical background - Section - Space of embeddings}, is future directed and orthonormal to the hypersurface $X(\Sigma)$.\separ
   
   For stationary space-times,  $g=-\mathrm{d}t^2+\gamma$, the KG equation can be solved by separation of variables
   \[u(t,x)=\sum_{n=0}^\infty\Big(a_n \varphi^+_n(t,x)+a^*_n \varphi^-_n(t,x)\Big)\qquad\qquad\varphi^\eta_k(t,x)=\left\{\begin{array}{lcc}e^{-i\eta t\omega_k}Q_k(x)&& k\neq0\\(1-i\eta t)Q_0&& k=0\end{array}\right.\]
   where $Q_k$ is the $k$-th normal mode of the eigenvalue problem $\Delta_\gamma Q=-\omega^2Q$ with some boundary conditions ($\{Q_k\}$ is a complete set of eigenvector by the Sturm-Liouville theorem). Notice now that, on one hand, $\peqsub{\mathcal{T}}{F\leftarrow I}$ is linear and $\peqsub{\phi}{F}=\peqsub{\mathcal{T}}{F\leftarrow I}(\peqsub{\phi}{I})$ while, on the other hand, $\peqsub{\phi}{F}$ is an element of $\mathcal{S}$ that can be decomposed in normal modes. Thus we have these two equivalent expressions
   \begin{align}
   &\peqsub{\phi}{F}=\peqsub{\mathcal{T}}{F\leftarrow I}(\peqsub{\phi}{I})=\sum_{n=0}^\infty\Big(a^I_n \peqsub{\mathcal{T}}{F\leftarrow I}(\varphi^+_n)+(a^I_n)^* \peqsub{\mathcal{T}}{F\leftarrow I}(\varphi^-_n)\Big)\label{Mathematical background - equation - phi_F= T(phi_I)}\\
   &\peqsub{\phi}{F}=\sum_{n=0}^\infty\Big(a^F_n \varphi^+_n+(a^F_n)^* \varphi^-_n\Big)\label{Mathematical background - equation - phi_F= base}
   \end{align}
   Notice that $\peqsub{\mathcal{T}}{F\leftarrow I}(\varphi_{n}^\eta)\in\mathcal{S}$ so that we can expand them in terms of $\varphi_{\bar{k}}^\eta$
   \[\peqsub{\mathcal{T}}{F\leftarrow I}(\varphi_{n}^+)=\sum_{m=0}^\infty\Big(\gamma^{FI}_{nm}\varphi_m^++(\beta^{FI}_{nm})^*\varphi_m^-\Big)\qquad\qquad\peqsub{\mathcal{T}}{F\leftarrow I}(\varphi_{n}^-)=\sum_{m=0}^\infty\Big((\gamma^{FI}_{nm})^*\varphi_m^-+\beta^{FI}_{nm}\varphi_m^+\Big)\]
   where $\gamma,\beta,\gamma^*,\beta^*$ can be considered as operators ---infinite matrices--- over $\ell^2(\C)$. Plugging the previous equation in \eqref{Mathematical background - equation - phi_F= T(phi_I)} and equating it to \eqref{Mathematical background - equation - phi_F= base} leads to 
   \begin{equation}\label{Mathematical background - equation - comparaciones coeficientes}
   \sum_{n=0}^\infty\Big(\varphi^+_na^F_n + \varphi^-_n(a^F_n)^*\Big)=\sum_{m,n=0}^\infty\left(\big( \gamma^{FI}_{nm}a^I_n+ \beta^{FI}_{nm}(a^I_n)^*\big)\varphi^+_m+\big( \beta^{FI}_{nm}(a^I_n)^*  + \gamma^{FI}_{nm}a^I_n\big)^*\varphi^-_m\right)
   \end{equation}
   In order to compare the coefficients of both sides of the equation, let us introduce a bilinear product with the help of the canonical symplectic structure over the space of solutions \eqref{Mathematical background - equation - Omega(sol,sol)}
   \begin{equation}\label{Mathematical background - equation - forma bilinear bogoluilov}
   \llangle \peqsub{\phi}{I},\peqsub{\phi}{F}\rrangle:=-i\Omega(\overline{\peqsub{\phi}{I}},\peqsub{\phi}{F})=i\int_\Sigma X^*(\overline{\peqsub{\phi}{I}})\peqsub{p}{F,X}\peqsub{\mathrm{vol}}{\Sigma}-i\int_\Sigma X^*(\peqsub{\phi}{F})\overline{\peqsub{p}{I,X}}\peqsub{\mathrm{vol}}{\Sigma}
   \end{equation}
   where $X:\Sigma\to M$ is some fixed embedding and $p_{i,X}:=\sqrt{\peqsub{\gamma}{X}}X^*(\mathcal{L}_{\peqsub{n}{X}}\phi_i)$. Notice that the bilinear product does not depend on the embedding $X$ because $\Omega$ does not depend either. It is long but straightforward to prove the following lemma.
   
   \begin{lemma}\ $\llangle\varphi^\eta_k,\varphi^\xi_k\rrangle=\dfrac{\eta+\xi}{2}\delta_{kl}$
   \end{lemma}
   %
   
   %
   
   Contracting \eqref{Mathematical background - equation - comparaciones coeficientes} with the orthogonal basis on the left (to avoid the conjugation of the coefficients), we conclude that\allowdisplaybreaks[0]
   \begin{align*}
   &a^F_n =\sum_{m=0}^\infty\Big(\gamma^{FI}_{nm}a^I_n + \beta^{FI}_{nm} (a^I_n)^*\Big)&&(a^F_n)^*=\sum_{m=0}^\infty\Big((\beta^{FI}_{nm})^*a^I_n+(\gamma^{FI}_{nm})^*(a^I_n)^*\Big)\\[1ex]
   &a^F=\gamma^{FI}.a^I+\beta^{FI}.(a^I)^*&&(a^F)^*=(\beta^{FI})^*.a^I+(\gamma^{FI})^*.(a^I)^*
   \end{align*}\allowdisplaybreaks
   Recall that $\gamma^{FI},\beta^{FI}$, as operators (infinite dimensional matrices), relate the coefficients $a^I$ to the coefficients $a^F$. Once we have that, we can formally construct the creation and annihilation operators $\widehat{a}^I,(\widehat{a}^I)^\dagger$ and $\widehat{a}^I,(\widehat{a}^I)^\dagger$ over the Fock space. They will then be related by $\gamma^{FI},\beta^{FI}$ seen as maps between quantum operators. Likewise we have an operator $\peqsub{\mathcal{T}}{F\leftarrow I}$ acting over states of the Fock space $\widehat{\peqsub{\phi}{I}}$. The unitarity of such operator is characterized by the properties of the $\beta^{FI}$ coefficients, known as the \textbf{Bogoliubov coefficients}\index{Bogoliubov coefficients}, thanks to a result of Shale \cite{shale1962linear}.
   
   \begin{theorem}\mbox{}\\	
   	$\peqsub{\mathcal{T}}{F\leftarrow I}$ is unitary over the Fock space if and only if $\beta^{FI}$ is a trace class operator i.e.\ $\displaystyle\sum_{m,n}^\infty|\beta^{FI}_{mn}|^2<\infty$.
   \end{theorem}
	This result will be crucial to understand to unitarity implementation of some systems in the next chapter.
   

%% file: 3_string_masses.tex


\chapter{Scalar fields coupled to point-masses}\label{Chapter - Scalar fields coupled to point masses}

	\thispagestyle{empty}
\starwars[Master Yoda]{Judge me by my size, do you?}{The Empire Strikes Back}

  \section{Introduction}
  
    The measurement problem is one of the most difficult questions in quantum physics. Although the postulates of quantum mechanics provide a clear mathematical approach to this matter (as we mentioned in the previous chapter) there are important unresolved issues from the physical point of view. Consider for instance a quantum system modeled by a Hilbert space $\mathcal{H}_{\mathrm{syst}}$ and that we are interested in a measurement device modeled by a Hilbert space $\mathcal{H}_{\mathrm{mes}}$. It is then postulated, in what some authors called the \emph{zeroth principle} \cite{zurek2009quantum}, that the whole system can be modeled by the tensor product $\mathcal{H}_{\mathrm{syst}}\otimes\mathcal{H}_{\mathrm{mes}}$. The dynamics is then given by a Hamiltonian of the form $H=H_{\mathrm{syst}}+H_{\mathrm{mes}}+H_{\mathrm{int}}$, where $H_{\mathrm{syst}}$ describes the dynamics of the system, $H_{\mathrm{mes}}$ the dynamics of the detector, and $H_{\mathrm{int}}$ encodes the interaction between them. It is important to notice that the zeroth principle is a strong assumption that might or might not be in correspondence with the physical world.\separ
    
    Sidney Coleman, an American theoretical physicist, said during a QFT lecture at Harvard that
    \begin{quote}\centering%
    	\begin{minipage}{.72\linewidth}
    		\emph{The career of a young theoretical physicist consists of treating the harmonic oscillator in ever-increasing levels of abstraction.}
    		\end{minipage}
    \end{quote}
    and we have somehow taken this statement to the next level. Indeed, in this chapter we will deal with a realistic harmonic oscillator described by a string (that models more realistically a spring) with two masses attached to its ends, a system that we have throughly studied in some of our papers \cite{margalefboundary,margalef2017functional,margalef2015quantization}. Our motivation is not to get a better understanding of the harmonic oscillator, but to understand if ---and how--- we could use the attached masses as measurement devices i.e.\ what information of the string we can retrieve by looking at the masses. Although our approach may look simple, it is actually closer to an actual experiment. In particular, they can be thought of as generalizations of the Unruh-DeWitt particle detectors and similar devices used in the discussion of quantum field theories in curved space-times and accelerated frames.\separ
    
    The addition of point particles at the boundary will make this problem rather interesting and, in fact, will require the use of some technical machinery (GNH, measure theory, Fock quantization\ldots) that will be introduced in this chapter. We will see that the precise construction of the Fock space is important in order to discuss the possible factorization of
    the Hilbert space $\mathcal{H}$ of the whole system as $\mathcal{H}_{\mathrm{masses}}\otimes\mathcal{H}_{\mathrm{field}}$ that would account for a clean separation between quantum point particle and field degrees of freedom.

  \section{Lagrangian formulation}
    Consider first a point mass $\peqsub{m}{0}$ moving in a straight line and attached to a spring which has zero rest length and spring constant $\peqsub{k}{0}$. Thus, its Lagrangian $L_0:T\R\to\R$ is given by 
    \begin{equation}
      \peqsub{L}{0}(q,v)=\peqsub{K}{0}(q,v)-\peqsub{V}{0}(q,v)=\frac{\peqsub{m}{0}v^2}{2}-\frac{\peqsub{k}{0}q^2}{2}
    \end{equation}
    
    where $q$ is the deviation from its equilibrium position and $v$ its velocity. Considering now infinitely many of these masses joined one after another, using some mechanical considerations and taking the continuous limit, a candidate for the Lagrangian of a string can be obtained. Namely, we have that $L:T\mathcal{C}^1[0,\ell]\to\R$ given by
    \begin{equation}
      L(Q,V)=K(Q,V)-V(Q,V)=\frac{\rho}{2}\left\langle V,V\right\rangle-\frac{\gamma}{2}\left\langle Q',Q'\right\rangle
    \end{equation}
    corresponds to the Lagrangian of a string in $1+1$ dimensions of unstretched length $\ell$, linear mass density $\rho$ and Young's modulus $\gamma$ (which measures the resistance of the string to being elastically deformed). $Q(x)$ denotes the deviation of the string point $x$ from its equilibrium position while $\langle\,,\rangle$ denotes the usual scalar product of the Hilbert space $L^2[0,\ell]$ defined with the help of the usual Lebesgue measure $\peqsub{\mu}{L}$.\separ
    
    \setlength{\figwidth}{0.75\linewidth}
    \null\hfill\smash{\raisebox{-\dimexpr 3.7\baselineskip+\parskip\relax}%
    	{\includegraphics[width=.22\linewidth]{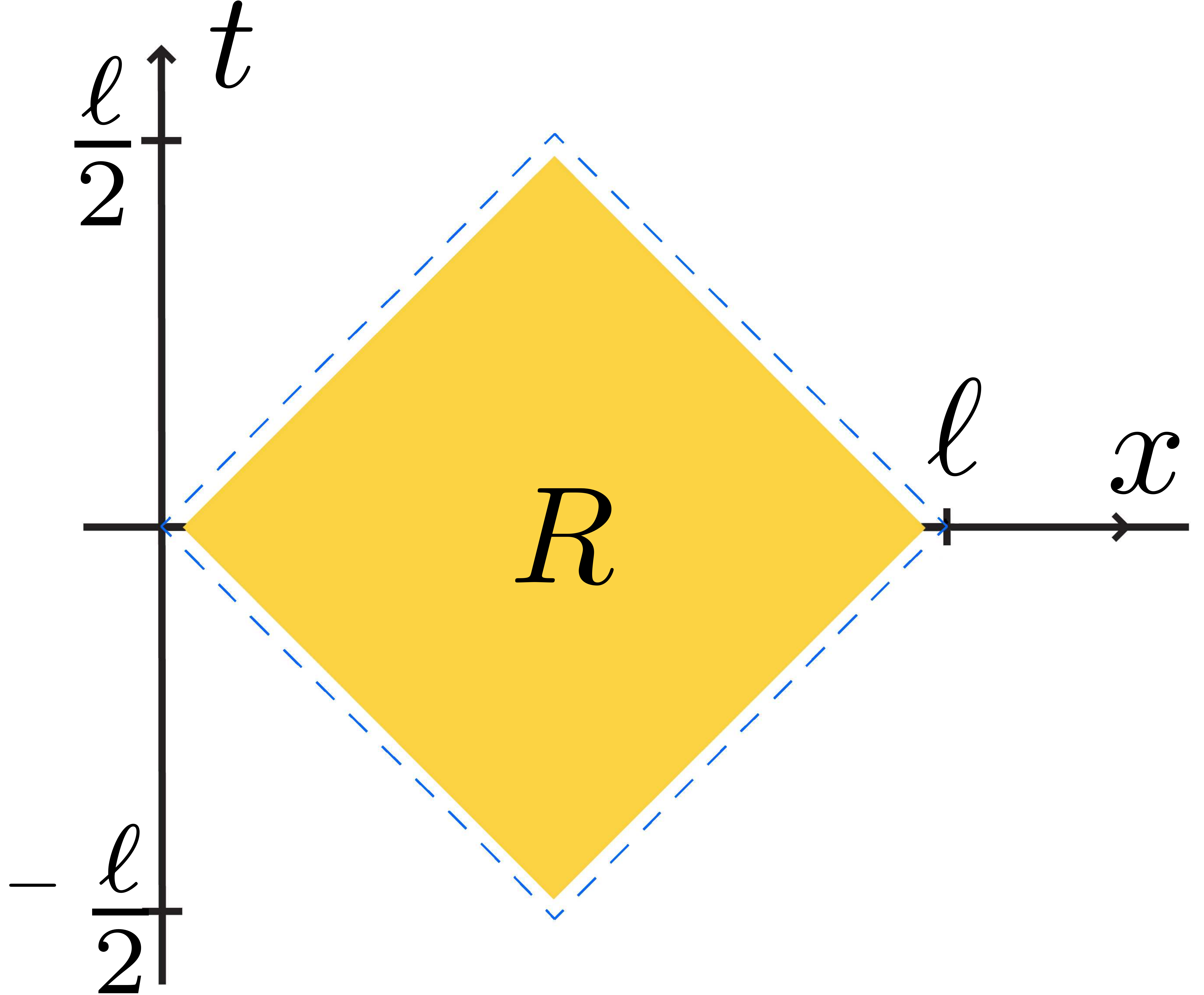}}}\strut \\[-\dimexpr 3\baselineskip+2\parskip\relax]
    
    \parshape=3 
    0pt \figwidth 0pt \figwidth 0pt \figwidth
    Once we consider initial conditions for the string, the problem can only be solved in the region $R=\{(t,x)\in\R^2\ /\ \ell\geq x-t\geq0\ \text{ and }\ \ell\geq t+x\geq0\}$.\separ
      
     \parshape=4 
     0pt \figwidth 0pt \figwidth 0pt \figwidth
     0pt \linewidth
    In order to get a unique solution in $[0,\ell]\times\R$ we have to specify some boundary conditions. The ones that arise naturally are the Dirichlet and Robin boundary conditions with some constant $\peqsub{k}{0}\geq0$ (Neumann is a particular case of the latter with $\peqsub{k}{0}=0$) that, for homogeneous problems, read at $x=0$ as follows
    \begin{equation*}
	  \text{Dirichlet:}\ \ \ Q(t,0)=0 \qquad\qquad \qquad 
	  \text{Robin:}\ \ \  Q'(t,0)-\peqsub{k}{0}Q(t,0)=0
	\end{equation*}
	Clearly the Dirichlet condition states that the string is fixed at its end while, as we will see, the Robin boundary condition is equivalent to having a spring of zero rest length and constant $\peqsub{k}{0}$ attached to its end. We will, however, consider a more general problem which contains as subcases the Dirichlet and Robin boundary conditions. Namely, a string with two masses $\peqsub{m}{0}$ and $\peqsub{m}{\ell}$ attached at its ends, and both of them attached to springs of zero rest length.
	
	\vspace{2ex}
	
	\centerline{\includegraphics[width=0.35\linewidth]{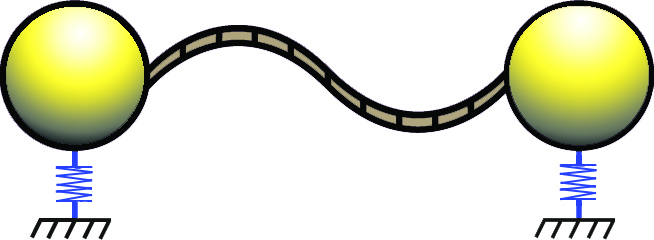}}
	
	\vspace{1.5ex}
	
	The Lagrangian of such system is the sum of the Lagrangian of the string, the Lagrangians of each mass and some coupling terms. Namely, defining $\mathcal{Q}=\R\times\mathcal{C}^\infty[0,\ell]\times\R\times\R^2$,  we have $L:T\hspace*{-0.2ex}\mathcal{Q}\to\R$ given by
	\begin{equation}\label{eq cuerda_masas lagrangiano 3 terminos}
	  L\big(\mathrm{q},\peqsub{\mathrm{v}}{\!\mathrm{q}}\hspace*{0.3ex}\big)=\frac{\rho}{2}\left\langle V,V\right\rangle-\frac{\gamma}{2}\left\langle Q',Q'\right\rangle+\!\sum_{j\in\{0,\ell\}}\left[\frac{m_jv_j^2}{2}-\frac{k_jq_j^2}{2}\right]+\!\sum_{j\in\{0,\ell\}} \lambda_j\Big(Q(j)-\varepsilon_j q_j\Big)
	\end{equation}
	for a given position $\mathrm{q}=(q_0,Q,q_\ell,\lambda_0,\lambda_\ell)\in\mathcal{Q}$ and velocity $\peqsub{\mathrm{v}}{\!\mathrm{q}}=(v_0,V,v_\ell,\nu_0,\nu_\ell)\in \peqsub{T}{\mathrm{q}}\mathcal{Q}$. We have included new variables $\lambda_j$, as Lagrange multipliers, and new binary constants $\varepsilon_j$, to turn the coupling on and off.

  \section{Action of the system}\label{section cuerda_masas action of the theroy}
    We consider now the action\index{Action} $S:\mathcal{C}^1([t_0,t_1],\mathcal{Q})_{\mathrm{q}_0}^{\mathrm{q}_1}\to\R$ given by
    \begin{equation}
      S(c)=\int_{t_0}^{t_1}\mathrm{d}\tau\, L\big(c(\tau),\dot{c}(\tau)\big)
    \end{equation}
    which is defined for a curve $c:[t_0,t_1]\to\mathcal{Q}$ of the space of positions  with prescribed endpoints $\mathrm{q}_i$. In particular $(c(\tau),\dot{c}(\tau))\in T\hspace*{-0.2ex}\mathcal{Q}$ for every $\tau\in[t_0,t_1]$ and $c(t_i)=\mathrm{q}_i$. The stationary points of the action are determined (where $\sigma(x)$ denotes the sign of $x\in\R$) by the following conditions.
    
    $\blacktriangleright$ $\peqsub{D}{\!(Q,V)}S=0$ for every direction $V\in \peqsub{T}{Q}\mathcal{C}^1[0,\ell]$ leads to\vspace*{-1.5ex}
    \begin{align*}
      &\rho\ddot{Q}(t,x)-\gamma Q''(t,x)=0 && (t,x)\in\R\times[0,\ell]\\
      &\lambda_j(t)=(-1)^{\sigma(j)+1}\gamma\,Q'(t,j) && t\in\R,\ \ j\in\{0,\ell\}\\
    \intertext{$\blacktriangleright$  $\peqsub{D}{\!(q_j,v_j)}S=0$ leads to the conditions}
      &m_j\ddot{q}_j(t)+k_jq_j(t)+\varepsilon_j\lambda_j(t)=0 &&t\in\R,\ \ j\in\{0,\ell\}\\
    \intertext{$\blacktriangleright$ $\peqsub{D}{\!(\lambda_j,\nu_j)}S=0$ imposes the conditions.}
    &Q(t,j)=\varepsilon_jq_j(t) &&t\in\R,\ \ j\in\{0,\ell\}
    \end{align*} 
    
    Notice that our system includes some important subcases
    \begin{itemize}
    	\item If $\varepsilon_j=0$ we have the wave equation for $Q$ over $[0,\ell]$ with Dirichlet boundary conditions $Q(t,j)=0$. The masses $m_j$ are harmonic oscillators decoupled from the string, and the Lagrange multipliers are fixed by the string.
    	\item If $\varepsilon_j=1$ but we remove the masses $m_j=0$, then $\lambda_j(t)=-k_jq_j(t)=-k_jQ(t,j)$. Therefore we have now the wave equation for $Q$ over $[0,\ell]$ with Robin boundary conditions
    	\[Q'(t,j)+(-1)^{\sigma(j)+1}k_jQ(t,j)=0\]
    	Both $\lambda_j$ and $q_j$ are determined by the well posed problem for $Q$.
    	\item If in this last case we also remove the springs $k_j=0$, then the Robin case reduces to the Neumann one.
    	\item If $\rho=0=\gamma$ we see that $Q$ is arbitrary except at the boundaries but, in particular, its derivatives at the boundary remain unfixed. This implies that $\lambda_j$ can be arbitrarily fixed and, thus, each mass is an harmonic oscillators subject to the time dependent force $\varepsilon_j\lambda_j(t)$.
    	\item We will see on section \ref{subsection cuerda_masas classical solutions} that when we take $k_j=0$ and $\rho=0$, the system behaves like two masses joined by a spring (harmonic oscillator around its center of mass). 
    \end{itemize}
    
    Notice that if $\varepsilon_jm_j\neq0$, although the point masses are subject to some dynamics, they are not independent classical degrees of freedom as they are fixed by continuity of the string. In fact, up to some constants, $\ddot{q}_j$ can be replaced by $Q(t,j)''$ and, thus, removing the explicit time derivatives.\separ
    
    It is important to mention here that we have defined the Lagrangian \eqref{eq cuerda_masas lagrangiano 3 terminos} with the help of some Lagrange multipliers. Nonetheless, there are other several ways to include boundary conditions in our theory by cleverly redefining our Lagrangian. We could, for instance, restrict the domain of the Lagrangian or choose a different functional form of the Lagrangian. Indeed, we have seen that the Dirichlet case can be considered through Lagrange multipliers. Alternatively we could have restricted the action to the functions $\mathcal{C}^\infty_0[0,\ell]$ vanishing at the boundary. We deduce also from the previous discussion that the Robin boundary conditions, on the other hand, can be obtained by including the boundary term $k_jQ(j)^2$ in the Lagrangian without the need to enlarge the configuration space with Lagrange multipliers.

  \section{Fiber derivative}\label{section cuerda_masas fiber derivative}
    For a given $\mathrm{q}=(q_0,Q,q_\ell,\lambda_0,\lambda_\ell)\in\mathcal{Q}=\R\times\mathcal{C}^\infty[0,\ell]\times\R\times\R^2$ a typical point of $\peqsub{T}{\mathrm{q}}^*\mathcal{Q}$ is of the form     $\bm{\peqsubfino{\mathrm{p}}{\!\mathrm{q}}{-0.1ex}}=(\bm{p}_0,\bm{P},\bm{p}_\ell,\bm{\pi}_0,\bm{\pi}_\ell)$ where $\boldsymbol{P}\in C^\infty[0,\ell]'$ is a distribution, i.e.\ a continuous linear functional $\boldsymbol{P}: C^\infty[0,\ell]\to\R$, and $\boldsymbol{p}_j,\bm{\pi}_j\in\R'=\R$, that we will denote $p_j$ and $\pi_j$ because we reserved the bold font for elements of the dual. The phase space $T^*\!\mathcal{Q}$ is equipped with the symplectic form \eqref{eq:background canonical symplectic form}, which in this case is given by
   \begin{align*}
   \Omega_{\boldsymbol{\mathrm{p}_{\mathrm{q}}}}(Y,Z)
   &=\Omega_{(\mathrm{q},\bm{\mathrm{p}})}\!\left(\!\rule{0ex}{3ex}(\vecc{Y}{Q},\vecc[j]{Y}{q},\peqsub{Y}{\bm{\lambda}j},\vecc{\bm{Y}}{\!\!\!P},\vecc[j]{\bm{Y}}{\!\!\!p},\peqsub{\bm{Y}}{\!\!\!\bm{\pi}j}),(\vecc{Z}{Q},\vecc[j]{Z}{q},\peqsub{Z}{\bm{\lambda}j},\vecc{\bm{Z}}{P},\vecc[j]{\bm{Z}}{p},\peqsub{\bm{Z}}{\bm{\pi}j})\!\right)=\\
   &=\vecc{\boldsymbol{Z}}{P}\!\left(\vecc{Y}{Q}\right)-
   \vecc{\boldsymbol{Y}}{\!\!\!P}\!\left(\vecc{Z}{Q}\right)+\sum_{j\in\{0,\ell\}}\Big[\vecc[j]{\bm{Z}}{p}(\vecc[j]{Y}{q})-\vecc[j]{\bm{Y}}{\!\!\!p}(\vecc[j]{Z}{q})+\peqsub{\bm{Z}}{\bm{\pi}j}(\peqsub{Y}{\bm{\lambda}j})-\peqsub{\bm{Y}}{\!\!\!\bm{\pi}j}(\peqsub{Z}{\bm{\lambda}j})\Big]
   \end{align*}   
   As we will see, over the image of the fiber derivative $\pi_j=0$ and the distribution $\boldsymbol{P}$ can be defined in terms of a map $P\in\mathcal{C}^\infty[0,\ell]$ by $\boldsymbol{P}(V)=\left\langle P,V\right\rangle$. Thus, in what follows, it is convenient to work with $\widetilde{\mathcal{P}}=\{(q_0,Q,q_\ell,\lambda_0,\lambda_\ell,p_0,P,p_\ell)\}$ which can be included in $T^*\!\mathcal{Q}$ via the previous representation $\jmath:\widetilde{\mathcal{P}}\hookrightarrow T^*\!\mathcal{Q}$. To compute the induced (pre)symplectic form $\omega=\jmath^*\Omega$ we need the pushforward of a generic vector field $Y=(\vecc{Y}{Q},\vecc[j]{Y}{q},\peqsub{Y}{\bm{\lambda}j},\vecc{Y}{P},\vecc[j]{Y}{p})\in\mathfrak{X}(\widetilde{\mathcal{P}})$ which is given by
   \begin{equation*}
     \jmath_*Y=\Big(\vecc{Y}{Q},\vecc[j]{Y}{q},\peqsub{Y}{\bm{\lambda}j},\left\langle\vecc{Y}{P},\,\cdot{}\,\right\rangle,\vecc[j]{Y}{p},0\Big)\in\mathfrak{X}(T^*Q)_{\ }
   \end{equation*}
   
   Thus the induced form $\omega(Y,Z)=(\Omega\smallcirc\jmath)(\jmath_*Y,\jmath_*Z)$ for $Y,Z\in\mathfrak{X}(\widetilde{\mathcal{P}})$ reduces to
   \begin{align*}
   \peqsub{\omega}{(Q,q_j,\lambda_j,P,p_j)}&(Y,Z)
   =\left\langle\vecc{Z}{P},\vecc{Y}{Q}\right\rangle-
   \left\langle\vecc{Y}{P},\vecc{Z}{Q}\right\rangle+\sum_{j\in\{0,\ell\}}\left[\vecc[j]{Z}{p}\vecc[j]{Y}{q}-\vecc[j]{Y}{p}\vecc[j]{Z}{q}\right]
   \end{align*}
   
    We compute now the fiber derivative\index{Fiber derivative} $F\!L:T\hspace*{-0.2ex}\mathcal{Q}\to T^*\!\mathcal{Q}$. The map $F\!L(\mathrm{q},\peqsub{\mathrm{v}}{\!\mathrm{q}}):T_\mathrm{q}\mathcal{Q}\to\R$ acting on $(\mathrm{q},\peqsub{\mathrm{w}}{\!\mathrm{q}}\hspace*{0.3ex})=((Q,q_j,\lambda_j),(W,w_j,\omega_j))$ is given by
    \begin{align}
      F\!L(\mathrm{q},\peqsub{\mathrm{v}}{\!\mathrm{q}}\hspace*{0.3ex})&\Big(\mathrm{q},\peqsub{\mathrm{w}}{\!\mathrm{q}}\Big)=\left.\deriv{}{\tau}\right|_{\tau=0}L(\mathrm{q},\peqsub{\mathrm{v}}{\!\mathrm{q}}+\tau\peqsub{\mathrm{w}}{\!\mathrm{q}})=\rho\langle V,W\rangle+\sum_{j\in\{0,\ell\}}m_jv_jw_j
    \end{align}
    Thus we have $F\!L(\mathrm{q},(V,v_j,\nu_j))=(\mathrm{q},(\rho\langle V,\cdot{}\,\rangle,m_j v_j,0))$ which allows us to define the canonical conjugate momenta $\bm{P}=\rho\langle V,\cdot{}\,\rangle\in \mathcal{C}^\infty[0,\ell]'$, $p_j=m_jv_j\in\R'$ and $\pi_j=0$. By using the representation $\jmath:\widetilde{\mathcal{P}}\hookrightarrow T^*\!\mathcal{Q}$, we may consider directly $P=\rho V$ and obtain $F\!L(T\hspace*{-0.2ex}\mathcal{Q})\cong\mathcal{P}$ where
    \begin{align*}
      &\mathcal{P}=\left\{(\mathrm{q},\peqsubfino{\mathrm{p}}{\mathrm{q}}{-0.2ex}\hspace*{0.3ex})=\Big((Q,q_j,\lambda_j),(P,p_j) \Big)\in \widetilde{\mathcal{P}}\ \ /\ \ P\in\rho\,\mathcal{C}^\infty(\R),\ \ p_j\in m_j\R\right\}
    \end{align*}
    Notice that the image of the fiber derivative depends strongly on whether the constants are zero or not.

  \section{Hamiltonian formulation}\label{section cuerda_masas Hamiltonian formulation}
    \subsection*{Obtaining the Hamiltonian}\trassub
    
      The energy\index{Energy} $E: T\hspace*{-0.2ex}\mathcal{Q}\to\R$, defined as $E(\mathrm{q},\peqsub{\mathrm{v}}{\!\mathrm{q}})=F\!L(\mathrm{q},\peqsub{\mathrm{v}}{\!\mathrm{q}}\hspace*{0.3ex})\big(\mathrm{q},\peqsub{\mathrm{v}}{\!\mathrm{q}}\big)-L(\mathrm{q},\peqsub{\mathrm{v}}{\!\mathrm{q}}\hspace*{0.3ex})$, has the form
      \begin{align}
        E(\mathrm{q},\peqsub{\mathrm{v}}{\!\mathrm{q}}\hspace*{0.3ex})&=\frac{\rho}{2}\langle V,V\rangle+\frac{\gamma}{2}\left\langle Q',Q'\right\rangle+\sum_{j\in\{0,\ell\}}\left[\frac{m_jv_j^2}{2}+\frac{k_jq_j^2}{2}-\lambda_j\Big(Q(j)-\varepsilon_jq_j\Big)\right]
      \end{align}
  
      The Hamiltonian\index{Hamiltonian} $H:F\!L(T\hspace*{-0.2ex}\mathcal{Q})\subset T^*\!\mathcal{Q}\to\R$ is given by the implicit equation $H\smallcirc F\!L=E$. Using the identification $F\!L(T\hspace*{-0.2ex}\mathcal{Q})\cong\mathcal{P}$ we have $H:\mathcal{P}\to\R$ given by
      \begin{align}
        H(\mathrm{q},\peqsubfino{\mathrm{p}}{\mathrm{q}}{-0.2ex}\hspace*{0.3ex})=\frac{\langle P,P\rangle}{2\rho}+\frac{\gamma}{2}\left\langle Q',Q'\right\rangle+\sum_{j\in\{0,\ell\}}\left[\frac{p^2_j}{2m_j}+\frac{k_jq_j^2}{2}-\lambda_j\Big(Q(j)-\varepsilon_jq_j\Big)\right]
      \end{align}
      where we assume that if $\rho=0$ then $P=0$ and analogously for $m_j$ and $p_j$.

    \subsection*{GNH algorithm}\trassub
    
    Once we have the Hamiltonian, we want to solve the equation $\peqsub{\imath}{Y}\omega=\mathrm{d}H$ to obtain the Hamiltonian vector field\index{Hamiltonian vector field} $Y\in\mathfrak{X}(\mathcal{P})$. We do so using the GNH algorithm explained in section \ref{Mathematical background - section - GNH} of chapter \ref{Chapter - Mathematical background}.\separ
    
      $\bullet$ Compute the differential $\mathrm{d}H:T\mathcal{P}\to\R$ of $H$.\separprevia
      
      Given $Z=(\vecc{Z}{Q},\vecc[j]{Z}{q},\peqsub{Z}{\bm{\lambda}j},\vecc{Z}{P},\vecc[j]{Z}{p})\in\mathfrak{X}(\mathcal{P})$
      \begin{align*}
        \mathrm{d}_{(\mathrm{q},\peqsubfino{\mathrm{p}}{\mathrm{q}}{-0.2ex})}H(\mathrm{Z})&=\frac{1}{\rho}\langle P, \vecc{Z}{P}\rangle+\gamma\left\langle Q',\vecc{Z}{Q}'\right\rangle+\sum_{j\in\{0,\ell\}}\left[\frac{p_j\vecc[j]{Z}{p}}{m_j}+k_jq_j\vecc[j]{Z}{q}\right]-\\
        &\phantom{=}-\sum_{j\in\{0,\ell\}}\left[\peqsub{Z}{\bm{\lambda}j}\Big(Q(j)-\varepsilon_jq_j\Big)+\lambda_j\Big(\vecc{Z}{Q}(j)-\varepsilon_j\vecc[j]{Z}{q}\Big)\right]=\\
        &=\frac{1}{\rho}\langle P, \vecc{Z}{P}\rangle-\gamma\left\langle Q'',\vecc{Z}{Q}\right\rangle+\sum_{j\in\{0,\ell\}}\left[\frac{p_j\vecc[j]{Z}{p}}{m_j}+(k_jq_j+\varepsilon_j\lambda_j)\vecc[j]{Z}{q}\right]-\\
        &\phantom{=}-\sum_{j\in\{0,\ell\}}\left[\Big(Q(j)-\varepsilon_jq_j\Big)\peqsub{Z}{\bm{\lambda}j}+\left(\lambda_j-\gamma(-1)^{\sigma(j)+1}Q'(j)\right)\vecc{Z}{Q}(j)\right]
      \end{align*}
      
      $\bullet$ Compute the Hamiltonian vector field $\peqsub{Y}{\!H}\in\mathfrak{X}(\mathcal{P})$.\separprevia
      
      To obtain $\peqsub{Y}{\!H}$ we have to solve,  over some space $\mathcal{P}_2\subset\mathcal{P}$, the Hamiltonian equation $\peqsub{\imath}{Y}\omega(Z)=\mathrm{d}H(Z)$ for every $Z=(\vecc{Z}{Q},\vecc[j]{Z}{q},\peqsub{Z}{\bm{\lambda}j},\vecc{Z}{P},\vecc[j]{Z}{p})\in\mathfrak{X}(\mathcal{P})$. Thus
      \begin{align*}
        &\left\langle\vecc{Z}{P},\vecc{Y}{Q}\right\rangle-
        \left\langle\vecc{Y}{P},\vecc{Z}{Q}\right\rangle+\sum_{j\in\{0,\ell\}}\left[\vecc[j]{Z}{p}\vecc[j]{Y}{q}-\vecc[j]{Y}{p}\vecc[j]{Z}{q}\right]=\frac{1}{\rho}\langle P, \vecc{Z}{P}\rangle-\gamma\left\langle Q'',\vecc{Z}{Q}\right\rangle+\\
        &\ +\sum_{j\in\{0,\ell\}}\left[\frac{p_j\vecc[j]{Z}{p}}{m_j}+(k_jq_j+\varepsilon_j\lambda_j)\vecc[j]{Z}{q}-\Big(Q(j)-\varepsilon_jq_j\Big)\peqsub{Z}{\bm{\lambda}j}-\Big(\lambda_j-\gamma(-1)^{\sigma(j)+1}Q'(j)\Big)\vecc{Z}{Q}(j)\right]
      \end{align*}
	  We should be specially careful when some of the coupling constants vanish. For instance, the condition $\rho=0$ forces both $P$ and $\vecc{Z}{P}$ to be zero and, hence, no condition over $\vecc{Y}{Q}$ appears. In other words, the statement ``for every $Z\in\mathfrak{X}(\mathcal{P})$'' is misleading unless we specify which constants are nonzero. Assuming $\rho$ and $m_j$ non-zero we get
	  \begin{align*}
	    &\vecc{Y}{Q}=\frac{P}{\rho} && \vecc{Y}{P}=\gamma Q''\\
	    &\vecc[j]{Y}{q}=\frac{p_j}{m_j} && \vecc[j]{Y}{p}=-k_jq_j-\varepsilon_j\lambda_j\\
	    &\peqsub{Y}{\bm{\lambda}j} \text{ arbitrary}
	  \end{align*}
	  over $\mathcal{P}_2=\{\mathcal{C}^j_1=0\}\cap\{\mathcal{C}^j_2=0\}$ with $\mathcal{C}^j_1=\lambda_j-\gamma(-1)^{\sigma(j)+1}Q'(j)$ and $\mathcal{C}^j_2=Q(j)-\varepsilon_jq_j$.\separpost

	  $\bullet$ Require $\peqsub{Y}{\!H}$ to be tangent to $\mathcal{P}_2$.\separprevia
	  
	  Let us determine the subspace $\mathcal{P}_3\subset\mathcal{P}_2$ where $\peqsub{Y}{\!H}$ is tangent to $\mathcal{P}_2$ i.e.\ where the constraints are preserved. For the first pair we have $0=Y(\mathcal{C}^j_1)=\mathrm{d}\mathcal{C}^j_1(Y)$ which fixes the value of $\peqsub{Y}{\bm{\lambda}j}$ to be
	  \begin{equation*}
	    \peqsub{Y}{\bm{\lambda}j}=\gamma(-1)^{\sigma(j)+1}\vecc{Y}{Q}'(j)=\frac{\gamma}{\rho}(-1)^{\sigma(j)+1}P'(j)\qquad\quad j\in\{0,\ell\}
	  \end{equation*}
	  The second pair of constraints $\mathcal{C}_2^j$, which enforces the masses to be glued to the string, imposes the additional pair of conditions
	  \begin{equation*}
	    \frac{P(j)}{\rho}=\vecc{Y}{Q}(j)=\varepsilon_j\vecc[j]{Y}{q}=\varepsilon_j\frac{p_j}{m_j}\qquad\quad j\in\{0,\ell\}
	  \end{equation*}
	  which ties the momentum of the masses to the momentum of the string at the boundary. This gives us a new pair of constraints $\{\mathcal{C}^j_3=0\}$ where $\mathcal{C}^j_3=P(j)/\rho-\varepsilon_jp_j/m_j$. Thus $\mathcal{P}_3=\mathcal{P}_2\cap\{\mathcal{C}^j_3=0\}$.\separpost
	  
	  $\bullet$ Require $\peqsub{Y}{\!H}$ to be tangent to $\mathcal{P}_3$.\separprevia
	  
	  Let us determine the subspace $\mathcal{P}_4\subset\mathcal{P}_3$ where $\peqsub{Y}{\!H}$ is tangent to $\mathcal{P}_3$. Notice that we only have to impose the tangency condition of $\peqsub{Y}{\!H}$ to $\{\mathcal{C}^j_3=0\}$, which leads to a new pair of constraints $\{\mathcal{C}^j_4=0\}$ where
	  \begin{align*}
	    \mathcal{C}^j_4=\frac{\gamma Q''(j)}{\rho}+(-1)^{\sigma(j)+1}\frac{\varepsilon_j^2\gamma}{m_j}Q'(j)+\frac{\varepsilon_jk_j}{m_j}Q(j)
	  \end{align*}
	  Thus we obtain that $\mathcal{P}_4=\mathcal{P}_3\cap\{\mathcal{C}^j_4=0\}$.\separpost
	  
	  $\bullet$ Iterate this process.\separprevia
	  
	  If we denote the derivative of order $2k$ by $\Delta^{(k)}$, then iterating we get the set of conditions
	  \begin{align*}
	    &\Delta^{(k)}\left(\frac{\gamma}{\rho}Q''(j)+(-1)^{\sigma(j)+1}\frac{\varepsilon_j^2\gamma}{m_j}Q'(j)+\frac{\varepsilon_jk_j}{m_j}Q(j)\right)=0\\
		&\Delta^{(k)}\left(\frac{\gamma}{\rho}P''(j)+(-1)^{\sigma(j)+1}\frac{\varepsilon_j^2\gamma}{m_j}P'(j)+\frac{\varepsilon_jk_j}{m_j}P(j)\right)=0
	  \end{align*}
	  for every $k\in\{0,1,2,\ldots\}$ and $\mathcal{P}_{k+1}=\mathcal{P}_k\cap\{\mathcal{C}^j_{k+1}=0\}$ It is straightforward to adapt this procedure when some of the constants are zero.

	\section{Classical solutions to the problem}\label{subsection cuerda_masas classical solutions}
	  In order to perform the Fock quantization, we need to consider the space of classical solutions to the equations of our system, which were obtained in section \ref{section cuerda_masas action of the theroy}
	  \begin{align}\begin{aligned}\label{eq:string system of equations}
	    &\rho\ddot{Q}(t,x)-\gamma Q''(t,x)=0 && (t,x)\in\R\times[0,\ell]\\
	    &m_j\ddot{q}_j(t)+k_jq_j(t)+\varepsilon_j\lambda_j(t)=0 &&t\in\R,\ \ j\in\{0,\ell\}\\
	    &\lambda_j(t)=(-1)^{\sigma(j)+1}\gamma\,Q'(t,j) && t\in\R,\ \ j\in\{0,\ell\}\\
	    &Q(t,j)=\varepsilon_jq_j(t) && t\in\R,\ \ j\in\{0,\ell\}
	  \end{aligned}\end{align}      
	  
	  Assuming that $\varepsilon_j=1$, which is the interesting case, we have that the dynamics of $\lambda_j$ and $q_j$ are given by the ones of $Q$ at the boundary. So the previous set of equations is equivalent to the following one
	  \begin{align}\begin{aligned}\label{eq cuerda_masas ecuacion de ondas}
	    &\rho\ddot{Q}(t,x)-\gamma\,Q''(t,x)=0 && (t,x)\in\R\times[0,\ell]\\
	    &m_j\ddot{Q}(t,j)+k_jQ(t,j)+(-1)^{\sigma(j)+1}\gamma\,Q'(t,j)=0\qquad &&t\in\R,\ \ j\in\{0,\ell\}
	  \end{aligned}\end{align}
	  Notice that the equations for the boundary are not \emph{standard} boundary conditions as they involve second order time derivatives. It is interesting to note also that these equations can be derived directly from the Lagrangian $L:T\mathcal{C}^\infty[0,\ell]\to\R$ given by
	  \begin{equation}\label{eq cuerda_masas lagrangiano 1 termino}
	  L\big(Q,V\big)=\frac{\rho}{2}\left\langle V,V\right\rangle-\frac{\gamma}{2}\left\langle Q',Q'\right\rangle+\sum_{j\in\{0,\ell\}}\left[\frac{m_jV(j)^2}{2}-\frac{k_jQ(j)^2}{2}\right]
	  \end{equation}
	  
	  Before solving \eqref{eq cuerda_masas ecuacion de ondas}, let us consider the case where we remove the external springs $k_j=0$ and make the string massless $\rho=0$. The equations then read\allowdisplaybreaks[0]
	  \begin{align}
	    &\gamma\,Q''(t,x)=0 && (t,x)\in\R\times[0,\ell]\\
	    &m_j\ddot{Q}(t,j)+(-1)^{\sigma(j)+1}\gamma\,Q'(t,j)=0 &&t\in\R,\ \ j\in\{0,\ell\}
	  \end{align}\allowdisplaybreaks
	  Thus, for each $t\in\R$, $Q(t,\cdot{}\,)$ is a straight line of slope $Q'(t,\cdot{}\,)=Q(t,\ell)-Q(t,0)$. The equations for the center of mass $C(t)$ and the distance between the masses $d(t)=Q(t,\ell)-Q(t,0)$ are
	  \begin{align*}
	    &\ddot{C}(t)=\deriv[2]{}{t}\left[\frac{\peqsub{m}{\ell}\,Q(t,\ell)+\peqsub{m}{0}\, Q(t,0)}{\peqsub{m}{\ell}+\peqsub{m}{0}}\right]=\frac{\gamma\,Q'(t,\ell)-\gamma\,Q'(t,0)}{\peqsub{m}{\ell}+\peqsub{m}{0}}=0\\
	    &\ddot{d}(t)=-\frac{\gamma\,Q'(t,\ell)}{\peqsub{m}{\ell}}-\frac{\gamma\,Q'(t,0)}{\peqsub{m}{0}}=-\gamma\frac{\peqsub{m}{\ell}+\peqsub{m}{0}}{\peqsub{m}{\ell}\,\peqsub{m}{0}}\Big(Q(t,\ell)-Q(t,0)\Big)=-\frac{\gamma}{\mu}d
	  \end{align*}
	  Which are exactly the equation for two masses connected by a spring, being $\mu$ the reduced mass of the system. Notice that taking $k_j=0=\rho$ in the Lagrangians \eqref{eq cuerda_masas lagrangiano 3 terminos} or \eqref{eq cuerda_masas lagrangiano 1 termino} does not lead to the typical Lagrangian for two masses joined by a spring. Indeed, notice that we do still have a variable $Q$ describing the shape of the string whose equation, as we mentioned before, turns out to be that of a straight line joining the two masses $Q(t,x)=Q(t,\ell)x/\ell+Q(t,0)(1-x/\ell)$. Besides, each mass moves according to
	  \begin{align}
	    \begin{split}
	      Q(t,j)&=\frac{\peqsub{m}{\ell} Q(0,\ell)+\peqsub{m}{0}Q(0,0)}{\peqsub{m}{0}+\peqsub{m}{\ell}}+t\frac{\peqsub{m}{\ell} \dot{Q}(0,\ell)+\peqsub{m}{0}\dot{Q}(0,0)}{\peqsub{m}{0}+\peqsub{m}{\ell}}+\\
	      &\phantom{=}+(-1)^{\sigma(j)+1}\frac{\mu}{m_j}\left([Q(0,\ell)-Q(0,0)]\cos(\omega t)+\frac{\dot{Q}(0,\ell)-\dot{Q}(0,0)}{\omega}\cos(\omega t)\right)
	    \end{split}
	  \end{align}
	  where the first line corresponds to the motion of the center of mass and the second one to the oscillations with frequency $\omega=(\gamma/\mu)^{1/2}$ around such center.\separ
	  
	  Now we assume $\rho\cdot{}\gamma\neq0$ to solve \eqref{eq cuerda_masas ecuacion de ondas}. By separation of variables $Q(t,x)=T(t)X(x)$ we get
	  \begin{align}\label{eq cuerda_masas ecuacion interior T}
	    &\frac{\rho}{\gamma} \ddot{T}(t)=-\lambda T(t)    && t\in\R\\
	    &X''(x)=-\lambda X(x)     && x\in[0,\ell]\label{eq cuerda_masas ecuacion interior X}\\
	    &X'(j)=(-1)^{\sigma(j)+1}(\mu_j\lambda-r_j)X(j) && j\in\{0,\ell\}\label{eq cuerda_masas ecuacion borde X}
	  \end{align}
	  for some $\lambda\in\R$ to be determined. We have introduced the constants $\mu_j=m_j/\rho$ and $r_j=k_j/\gamma$. Notice that, unless $m_j=0$ (Robin boundary conditions), the eigenvalue $\lambda$ appears in \eqref{eq cuerda_masas ecuacion borde X}, which means that the eigenvalue problem for $X$ \textbf{is not} of the Sturm-Liouville type. Thus we cannot apply theorem \ref{Appendix - theorem - sturm liouville}, which states roughly speaking that \emph{there are infinitely many eigenvalues increasing to infinite with orthogonal eigenvectors}. Nonetheless, in order to completely solve \eqref{eq cuerda_masas ecuacion de ondas} by separation of variables, we need to be sure that the eigenvectors form a complete set. We will deal with this problem once we obtain the eigenvectors.\separ
	  
	  In order to obtain the eigenvectors, we have to solve \eqref{eq cuerda_masas ecuacion interior X} for $\lambda$ and for $X$. Let us suppose first that $\lambda=-\omega^2<0$, then $X_\lambda(x)=A\,\mathrm{Exp}(\omega x)+B\,\mathrm{Exp}(-\omega x)$. If we plug it into \eqref{eq cuerda_masas ecuacion borde X} we obtain the trivial solution $X_\lambda=0$. Actually, this conclusion is obvious if we realize that the solution to \eqref{eq cuerda_masas ecuacion interior T} cannot be exponential due to the conservation of energy. The $\lambda=0$ case is similar although now we find a nontrivial solution if and only if both $k_j=0$. With no springs the center of mass is free to move with constant velocity. For the sake of concreteness we assume that at least one $k_j\neq0$, so no zero mode arises, but it is clear how to include such case.\separ
	  
	  Finally we consider $\lambda=\omega^2>0$, in which case $X_\lambda(x)=A\sin(\omega x)+B\cos(\omega x)$. Plugging this into \eqref{eq cuerda_masas ecuacion borde X} we obtain a system of two equations for $A,B$ which has nonzero solutions $A=r_0-\mu_0\omega^2$ and $B=\omega$ if and only if
	  \begin{equation}\label{eq cuerda_masas ecuacion para omega}
	    \Big(\omega^2(\mu_0+\mu_\ell)-(r_0+r_\ell)\Big)\omega\cos(\omega\ell)-\Big((\mu_0\omega^2-r_0)(\mu_\ell\omega^2-r_\ell)-\omega^2\Big)\sin(\omega\ell)=0
	  \end{equation}
	  It is clear that $\omega_{\peqsub{m}{0}}=\peqsub{m}{0}\pi/\ell$ is a solution if and only if
	  \begin{equation}
	  \peqsub{m}{0}=\frac{\ell}{\pi}\sqrt{\frac{r_0+r_\ell}{\mu_\ell+\mu_\ell}}\in\N
	  \end{equation}
	  Except for this possible solution, the solutions to \eqref{eq cuerda_masas ecuacion para omega} are those $\omega$ such that
	  \begin{equation}\label{eq cuerda_masas ecuacion para omega alternativa}
	  \omega\cotan(\omega\ell)=\frac{(\mu_0\omega^2-r_0)(\mu_\ell\omega^2-r_\ell)-\omega^2}{\omega^2(\mu_0+\mu_\ell)-(r_0+r_\ell)}
	  \end{equation}
	  which includes the possibility of having $\mu_j=0$ and $r_j=0$. In such case the solutions are simply $\omega_m=m\pi/\ell$.
	  
	  \begin{lemma}\label{lemma cuerda_masas infinitas soluciones crecientes}\mbox{}\\
	     	If $\mu_j\geq0$ and $r_j\geq0$, then there is exactly one solution $\omega_m\in I_m=[m\pi/\ell,(m+1)\pi/\ell)$ to \eqref{eq cuerda_masas ecuacion para omega alternativa} for every $m\in\N\setminus\{\peqsub{m}{0}\}$. If $\peqsub{m}{0}\in\N$ then there are two solutions in $I_{\peqsub{m}{0}}$. 
	  \end{lemma}
	  \begin{proof}\mbox{}\\
	     	It is easy to check that the LHS is decreasing over each open interval $\overset{\ \smallcirc}{I}_m=(m\pi/\ell,(m+1)\pi/\ell)$ therefore, if we prove that the RHS is increasing over $(0,\infty)\setminus\{\omega_{\peqsub{m}{0}}\}$, the proof will be complete.\separ
	  
	    Let $F(\omega)$ denote the RHS of \eqref{eq cuerda_masas ecuacion para omega alternativa}. Its derivative is given by
	    \begin{align*}
	      F'(\omega)&=2\omega\frac{\mu_0\mu_\ell(\mu_0+\mu_\ell)\omega^4-2\mu_0\mu_\ell(r_0+r_\ell)\omega^2+\mu_0r_\ell^2+\mu_\ell r_0^2+r_\ell+r_0}{[(\mu_0+\mu_\ell)\omega^2-(r_0+r_\ell)]^2}
	    \end{align*}
	    If $\mu_0+\mu_\ell=0$ then $F'(\omega)=2\omega/(r_0+r_\ell)>0$. Otherwise we can complete squares in the numerator
	    \begin{align*}
	      &\left(\sqrt{\mu_0\mu_\ell(\mu_0+\mu_\ell)}\omega^2-(r_0+r_\ell)\sqrt{\frac{\mu_0\mu_\ell}{\mu_0+\mu_\ell}}\right)^2-(r_0+r_\ell)^2\frac{\mu_0\mu_\ell}{\mu_0+\mu_\ell}+\mu_0r_\ell^2+\mu_\ell r_0^2+r_\ell+r_0=\\
	      &\phantom{=}=\left(\sqrt{\mu_0\mu_\ell(\mu_0+\mu_\ell)}\omega^2-(r_0+r_\ell)\sqrt{\frac{\mu_0\mu_\ell}{\mu_0+\mu_\ell}}\right)^2+\frac{(\mu_0r_\ell-\mu_\ell r_0)^2}{\mu_0+\mu_\ell}+r_0+r_\ell
	    \end{align*}
	    which is always positive. Therefore the RHS is increasing as we wanted to prove.
	  \end{proof}
	  
	  The asymptotic behavior of $\omega_m$ depends strongly on the vanishing of the constants.
	  \begin{align}\label{eq cuerda_masas omega_m assimpt sin masas}
	    &\blacktriangleright\ \text{ If both } \mu_j=0&&\omega_m=\frac{m\pi}{\ell}+\frac{r_0+r_\ell}{\pi \ m}+\mathcal{O}\left(\frac{1}{m^3}\right)\\ \label{eq cuerda_masas omega_m assimpt una masa}
	    &\blacktriangleright\ \text{ If only } \mu_\ell>0&&\omega_m=\frac{(2m+1)\pi}{2\ell}+\frac{1+\mu_\ell r_0}{\mu_\ell\pi\ m}-\frac{1+\mu_\ell r_0}{2\mu_\ell\pi \ m^2}+\mathcal{O}\left(\frac{1}{m^3}\right)\\
	    &\blacktriangleright\ \text{ If both } \mu_j>0&&\omega_m=\frac{m\pi}{\ell}+\frac{\mu_0+\mu_\ell}{\mu_0\mu_\ell\pi \ m}+\mathcal{O}\left(\frac{1}{m^3}\right)\label{eq cuerda_masas omega_m assimpt con masas}
	  \end{align}
	  
	  Summarizing, we have infinitely many eigenvalues $\{-\omega_m^2\}$ with their corresponding eigenvectors $\{X_m\}$. The standard procedure will be now to prove that they form a complete set in an appropriate functional space so that the solution can be expanded into its Fourier coefficients.
	  \begin{align*}
	    (\omega_m^2-\omega_n^2)\langle X_m,&X_n\rangle\overset{\eqref{eq cuerda_masas ecuacion interior X}}{=}-\langle X_m'',X_n\rangle+\langle X_m,X_n''\rangle=\\
	    &=-\left[X_m'X_n\right]_0^\ell+\langle X_m',X_n'\rangle+\left[X_mX_n'\right]_0^\ell-\langle X_m',X_n'\rangle\overset{\eqref{eq cuerda_masas ecuacion borde X}}{=}\\
	    &=-\left[(-1)^{\sigma(j)+1}(\mu_j\omega_m^2-r_j)X_mX_n\right]_{j=0}^\ell+\left[(-1)^{\sigma(j)+1}(\mu_j\omega_n^2-r_j)X_mX_n\right]_{j=0}^\ell=\\
	    &=\left[(-1)^{\sigma(j)}\mu_j\left(\omega_m^2-\omega_n^2\right)X_mX_n\right]_0^\ell=\\
	    &=-(\omega_m^2-\omega_n^2)\sum_{j\in\{0,\ell\}}\mu_jX_m(j)X_n(j)
	  \end{align*}
	  which is in general different from zero and, thus, they are not $L^2$-orthogonal, let alone an orthogonal basis. In particular, notice that the Laplacian (second derivative) is \textbf{not} a symmetric operator. It is, however, easy to realize that if we consider the modified scalar product
	  \begin{equation}\label{eq cuerda_masas producto escalar modificado}
	    \llangle f,g\rrangle=\mu_0f(0)g(0)+\langle f,g\rangle+\mu_\ell f(\ell)g(\ell)
	  \end{equation}
	  then $\{X_m\}$ are orthogonal with respect to it. If we denote
	  \begin{equation}
	    \widetilde{X}_m=\frac{1}{\llangle X_m,X_m\rrangle^{\raisemath{0.5ex}{\scriptscriptstyle 1\hspace*{-0.2ex}/\hspace*{-0.2ex}2}}}X_m
	  \end{equation}
	  the normalized eigenvectors, then $\displaystyle Q(t,x)=\sum T_m(t)\widetilde{X}_m(x)$ solves \eqref{eq cuerda_masas ecuacion de ondas} with initial conditions $Q(0,x)=q(x)$ and $\dot{Q}(0,x)=p(x)$ if $T_m$ satisfies
	  \begin{align*}
	    \ddot{T}_m(t)=-\omega_m^2 T(t)\\
	    T_m(0)=\llangle \widetilde{X}_m,q\rrangle\\
	    \dot{T}_m(0)=\llangle \widetilde{X}_m,p\rrangle
	  \end{align*}
	  We have constructed several solutions to \eqref{eq cuerda_masas ecuacion de ondas} but it remains to be known if we have found them all i.e.\ if every solution is of this form or, equivalently, if $\{\widetilde{X}_m\}$ is complete (see definition \refconchap{Appendix - definition - complete}).\separ
	  
	  One might be tempted to adapt the proof of the Sturm-Liouville theorem over $(L^2[0,\ell],\llangle\,,\rrangle)$ but notice that $\llangle\,,\rrangle$ is not well defined over $L^2[0,\ell]$ because $f(0)$ is meaningless (maps are identified up to zero measure sets). Of course we could restrict ourselves to some functional spaces where such values were meaningful like $\mathcal{C}^0[0,\ell]$ or $H^1[0,\ell]$, but we have found a nicer alternative which does not force us to constraint the regularity of the solutions from the beginning. Namely, changing the measure space, which allows us to rewrite our problem as a true Sturm-Liouville problem.

	\subsection*{Changing the measure space}\trassub
	
	  Looking at \eqref{eq cuerda_masas producto escalar modificado}, it seems natural to introduce the measure given by $\mu=\alpha_0\delta_0+\peqsub{\mu}{L}+\alpha_\ell\delta_\ell$ where $\peqsub{\mu}{L}$ is the Lebesgue measure of $[0,\ell]$ and $\delta_x(\Omega)$ is $1$ if $x\in\Omega$ and $0$ otherwise (see section \ref{Appendix - Section - Measure theory} of appendix \ref{appendix}). For technical reasons that will be clear later we have introduced the constants $\alpha_j>0$, to be appropriately chosen, instead of $\mu_j$. We consider now the space of maps, defined up to a set of $\mu$-measure zero, given by
	  \begin{equation*}
	    L^2_\mu[0,\ell]=\Big\{f:[0,\ell]\to\overline{\R}\ \quad \mu\text{-measurable} \, \ \ /\, \ \ \peqsubfino{\langle f,f\rangle}{\!\mu}{-0.2ex}<\infty\Big\}
	  \end{equation*}
	  where we define the scalar product
	  \begin{equation}
	    \peqsubfino{\langle f,g\rangle}{\!\mu}{-0.3ex}=\int_{[0,\ell]} f\cdot{}g\,\mathrm{d}\mu
	  \end{equation}
	  It is clear that indeed $(L^2_\mu[0,\ell],\peqsubfino{\langle\,,\rangle}{\!\mu}{-0.2ex})$ is a real Hilbert space. Notice that $\mu(\{j\})=\alpha_j$ if $j\in\{0,\ell\}$ so now, given some $f\in L^2_\mu[0,\ell]$ the value $f(0)$ is well defined. In fact we can split the integral of $\peqsubfino{\langle\,,\rangle}{\!\mu}{-0.2ex}$ in three parts corresponding to $\delta_0,\peqsub{\mu}{L},\delta_\ell$ as
	  \begin{equation}
	    \peqsubfino{\langle f,g\rangle}{\!\mu}{-0.3ex}=\alpha_0f(0)g(0)+\langle f,g\rangle+\alpha_\ell f(\ell)g(\ell)
	  \end{equation}
	  Actually, the space itself can also be split. If we define $\mathbb{L}=\R\times L^2(0,\ell)\times\R$ endowed with the scalar product $\llangle(a,f,c),(b,g,d)\rrangle_{\mathbb{L}}=\alpha_0ab+\langle f,g\rangle+\alpha_\ell cd$ we have the following result.
	  \begin{lemma}\label{lemma cuerda_masas ismorfismo espacio de medida}\mbox{}\\
	     	$(L^2_\mu[0,\ell],\peqsubfino{\langle\,,\rangle}{\!\mu}{-0.3ex}\hspace*{0.1ex})$ and $(\mathbb{L},\llangle\,,\rrangle_{\mathbb{L}})$ are isomorphic as Hilbert spaces.
	  \end{lemma}
	  \begin{proof}\mbox{}\\
	    	 $\Phi:L^2[0,\ell]\to\mathbb{L}$ defined by $\Phi(f)=(f(0),f|_{(0,\ell)},f(\ell))$ is clearly an isomorphism of vector spaces and $\llangle\Phi(f),\Phi(g)\rrangle_{\mathbb{L}}=\peqsubfino{\langle f,g\rangle}{\!\mu}{-0.3ex}$.
	  \end{proof}
	  
	  When convenient this isomorphism will be used without making it explicit.\separ
	  
	  It is possible to define differential calculus in this new measure space by introducing the so-called \textbf{Radon-Nikodym (RN) derivative} of a given $F\in L^2_\mu[0,\ell]$ (see section \ref{Appendix - Section - Measure theory} of appendix \ref{appendix}). It is defined by
	  \[\deriv{F}{\mu}:=\deriv{\peqsub{\nu}{\!F}}{\mu}\]
	  where $\peqsub{\nu}{\!F}$ is its Lebesgue-Stieltjes measure $\peqsub{\nu}{\!F}[x,y]=F(y)-F(x)$. In order to be able to derive the measure $\peqsub{\nu}{\!F}$ with respect to $\mu$ the former has to be $\mu$-a.c. i.e.\ $\peqsub{\nu}{\!F}\ll\mu$ (see definition \ref{Appendix - definition - a.c.}). In that case we say also that $F$ is $\mu$-a.c.\index{Absolutely continuous!Measure}\index{Absolutely continuous}\separ
	  
	  If $F$ is $\mu$-a.c., splitting it as before shows that no condition arises over the boundary values $F(0)$ and $F(1)$. However, we see that the measure $F|_{(0,\ell)}\peqsub{\mu}{L}$ has to be $\peqsub{\mu}{L}$-a.c., which is equivalent to $F|_{(0,\ell)}\in H^1(0,\ell)$, i.e.\ functions of $L^2(0,\ell)$ with distributional derivative in $L^2(0,\ell)$. An element $G\in H^1(0,\ell)$ is absolutely continuous in the usual calculus sense and, besides, has well defined boundary limits
	  \[G(0+)=\lim_{x\to 0^+}G(x)\qquad \qquad G(\ell-)=\lim_{x\to\ell^-}G(x)\]
	  which of course need not to be equal to the values at the boundary $G(0)$ and $G(1)$. Sometimes we will denote $\peqsubfino{\gamma}{0}{-0.2ex}(G)=G(0+)$ and $\peqsubfino{\gamma}{\ell}{-0.2ex}(G)=G(1-)$ where $\gamma_j$ is known as the \textbf{trace operator}\index{Trace operator}.\separ
	  
	  The previous paragraph can be summarized analogously to lemma \ref{lemma cuerda_masas ismorfismo espacio de medida} if we define
	  \begin{equation*}
	    H^1_\mu[0,\ell]:=\left\{F\in L^2_\mu[0,\ell]\ \ \mu\text{-a.c.} \ \ /\ \ \deriv{F}{\mu}\in L^2_\mu[0,\ell]\right\}\qquad\qquad\qquad \mathbb{H}^1=\R\times H^1(0,\ell)\times\R
	  \end{equation*}
	
	  \begin{lemma}\label{lemma cuerda_masas ismorfismo H^1}\mbox{}\\
	     	$(H^1_\mu[0,\ell],\peqsubfino{\langle\,,\rangle}{\!\mu}{-0.2ex}\hspace*{0.1ex})$ and $(\mathbb{H}^1,\llangle\,,\rrangle_{\mathbb{L}})$ are isomorphic as Hilbert spaces.
	  \end{lemma}\mbox{}
	  
	  \begin{lemma}\label{lemma cuerda_masas RN derivative}\mbox{}\\
	     	\mbox{}\vspace*{-4.5ex}
	     	
	    Given $F\in H^1_\mu[0,\ell]$ we have\ \  $\displaystyle\deriv{F}{\mu}(x)=\left\{\begin{array}{ll}\dfrac{F(0+)-F(0)}{\alpha_0}& x=0\\[2ex] \displaystyle\deriv{F}{x}(x)&x\in(0,\ell)\\[2.7ex]
	    \dfrac{F(\ell)-F(\ell-)}{\alpha_\ell}&x=\ell\end{array}\right.$
	  \end{lemma}
	  \begin{proof}
	    \begin{align*}
	      F(y)-F(x)&\overset{\eqref{Appendix - equation - LS-measure}}{=}\peqsub{\nu}{\!F}[x,y]\overset{\eqref{Appendix - equation - m(A)=int A}}{=}\int_{[x,y]}\mathrm{d}\peqsub{\nu}{\!F}\overset{\ref{Appendix - theorem - RN derivative}}{=}\\
	      &\overset{\phantom{\eqref{Appendix - equation - LS-measure}}}{=}\int_{[x,y]}\deriv{F}{\mu}\mathrm{d}\mu=\\[1ex]
	      &\overset{\phantom{\eqref{Appendix - equation - LS-measure}}}{=}\alpha_0\int_{[x,y]}\deriv{F}{\mu}\mathrm{d}\delta_0+\int_{[x,y]}\deriv{F}{\mu}\mathrm{d}\peqsub{\mu}{L}+\alpha_\ell\int_{[x,y]}\deriv{F}{\mu}\mathrm{d}\delta_\ell=\\[1ex]
	      &\overset{\phantom{\eqref{Appendix - equation - LS-measure}}}{=}\alpha_0\deriv{F}{\mu}(0)\delta_0[x,y]+\int_{[x,y]}\deriv{F}{\mu}\mathrm{d}\peqsub{\mu}{L}+\alpha_\ell\deriv{F}{\mu}(\ell)\mathrm\delta_\ell[x,y]
	    \end{align*}
	    If $x,y\in(0,\ell)$ then the first and last term vanish. Thus $\deriv{F}{\mu}$ is the RN derivative with respect to $\peqsub{\mu}{L}$ so it is the standard weak derivate that we usually denote as $\deriv{F}{x}$. If $y=\ell$ and we make $x\to \ell^-$ we see that $\delta_0[x,y]=0$ and $\delta_\ell[x,y]=1$. The integral is computed over an interval whose length tends to zero and, hence, vanishes at the limit. Therefore $F(\ell)-F(\ell^-)=\alpha_\ell\deriv{F}{\mu}(0)$. An analogous reasoning for $x=0$ and $y\to 0^+$ leads to the required result.
	  \end{proof}
	  
	  Notice that the continuity at the boundary is equivalent to the vanishing of the RN derivative at the boundary. Besides, with the help of the trace operators $\gamma_j$, we have the more compact formula for the boundary
	  \begin{equation}\label{eq cuerda_masas RN derivative traza}
	    \alpha_j\deriv{F}{\mu}(j)=(-1)^{\sigma(j)}\Big(\gamma_j(F)-F(j)\Big)
	  \end{equation}
	  In order to perform integration by parts we need to know the derivative of the product which does not follow the usual Leibniz rule. For that purpose we introduce $K:[0,\ell]\to\R$ with $K|_{(0,\ell)}=0$ and $K(j)=(-1)^{\sigma(j)}\alpha_j$.
	  \begin{lemma}\label{lemma cuerda_masas mu-derivada del producto}\mbox{}\\
	     	Given $F,G\in H^1_\mu[0,\ell]$ then $\displaystyle\deriv{(FG)}{\mu}=\deriv{F}{\mu}G+F\deriv{G}{\mu}+K\deriv{F}{\mu}\deriv{G}{\mu}$ $\mu$-a.e. in $[0,\ell]$.
	  \end{lemma}
	  \begin{proof}\mbox{}\\
	     	For $x\in(0,\ell)$ it is clear that the standard Leibniz rule holds. If $x=0$ we have
	     	\begin{align*}
	   	  \deriv{(FG)}{\mu}(0)&=\frac{F(0+)G(0+)-F(0)G(0)}{\alpha_0}=\\
	   	  &=\frac{F(0+)-F(0)}{\alpha_0}G(0+)+F(0)\frac{G(0+)-G(0)}{\alpha_0}=\\
	   	  &=\deriv{F}{\mu}(0)\left(\alpha_0\deriv{G}{\mu}(0)+G(0)\right)+F(0)\deriv{G}{\mu}(0)=\\
	   	  &=\deriv{F}{\mu}(0)G(0)+F(0)\deriv{G}{\mu}(0)+K(0)\deriv{F}{\mu}(0)\deriv{G}{\mu}(0)
	     	\end{align*}
	     	And analogously for $x=\ell$.
	  \end{proof}
	  
	\subsection*{The Laplacian operator}\trassub
	
	  We recall that our goal is to define a self-adjoint Laplace operator $\Delta_\mu$ over $L^2_\mu[0,\ell]$ such that the eigenvalue problem for $\Delta_\mu$ with some boundary conditions (which would be of the Sturm-Liouville type) is in correspondence with our initial eigenvalue problem \eqref{eq cuerda_masas ecuacion interior X}-\eqref{eq cuerda_masas ecuacion borde X}.\separ
	  
	  A naive way to do this would be to define $\Delta_\mu$ simply as the second RN derivative with respect to $\mu$. However, this would not work entirely because more freedom is required as we will explain in the following. This additional freedom can be achieved noticing that, as our space is $\R\times L^2(0,\ell)\times\R$, the Laplacian could act differently over each factor. Thus let us consider the following more general Laplacian defined over the largest possible domain, which turns out to be
	  \begin{align*}
	    &\domDelta=\left\{u\in L^2_\mu[0,\ell]\ \ / \ \ \exists \deriv{u}{\mu}(x)\ \forall x\in[0,\ell],\ \ \deriv{u}{\mu}\in H^1_\mu[0,\ell]\right\}\\[2.5ex]
	    &\!\!\begin{array}{cccc}
	      \Delta_\mu:&\domDelta\subset L^2_\mu[0,\ell]&\longrightarrow & L^2_\mu[0,\ell]\\[1.1ex]
	                 &        u                         &   \longmapsto  & (1+C)\displaystyle\deriv[2]{u}{\mu}
	    \end{array}
	  \end{align*}
	  where $C:[0,\ell]\to\R$ is $0$ over $(0,\ell)$ and $C(j)=c_j\in\R\setminus\{-1\}$ are to be determined. Integrating by parts according to lemma \ref{lemma cuerda_masas mu-derivada del producto} and taking into account that $K\deriv{C}{\mu}=-C$, it follows
	  \begin{align}\label{eq cuerda_masas integracion por partes mu}
	  \begin{split}
	    &\peqsubfino{\left\langle\Delta_\mu u,v\right\rangle}{\!\mu}{-0.1ex}=\peqsubfino{\left\langle\deriv[2]{F}{\mu},(1+C)G\right\rangle}{\!\!\!\mu}{0.2ex}=\\
	    &\hspace*{2ex}=-\peqsubfino{\left\langle\deriv{u}{\mu},\deriv{v}{\mu}\right\rangle}{\!\!\!\mu}{0.2ex}-\sum_{j\in\{0,\ell\}}\left[(-1)^{\sigma(j)}\deriv{u}{\mu}(j)v(j)+\alpha_j\deriv[2]{u}{\mu}(j)K(j)\left(\deriv{v}{\mu}(j)-\frac{C(j)}{K(j)}v(j)\right)\right]
	  \end{split}
	  \end{align}
	  Thus $\Delta_\mu$ is not symmetric over $\domDelta$. However if we consider, for instance, elements of
	  \begin{align}\label{eq cuerda_masas dominio robin}
	  \domRobin=\left\{u\in H^1_\mu[0,\ell]\ \ /\ \ \deriv{u}{\mu}(j)-(-1)^{\sigma(j)}\frac{c_j}{\alpha_j}u(j)=0\right\}
	  \end{align}
	  then the last term of \eqref{eq cuerda_masas integracion por partes mu} vanishes and the middle one becomes symmetric in $u,v$. Therefore $\left.\Delta_\mu\right|_{\mathcal{D}}$ is symmetric over $\mathcal{D}:=\domDelta\cap\domRobin$ and, in fact, it can be showed following the method of \cite{evans1970non} that it is self-adjoint and that $\mathcal{D}$ is dense in $L^2_\mu[0,\ell]$. Notice that the boundary conditions defined by the maps in $\domRobin$ are of the Robin type. It seems therefore natural to consider the eigenvalue problem with Robin boundary conditions
	  \begin{align}\label{eq cuerda_masas robin laplaciano}
	    &\Delta_\mu u=-\lambda u && \mu-\text{a.e.\ in }[0,\ell]\\
	    &\deriv{u}{\mu}(j)=(-1)^{\sigma(j)}\frac{c_j}{\alpha_j}u(j)&&j\in\{0,\ell\}\label{eq cuerda_masas robin mu-conditions}
	  \end{align}
	  which defines a complete set of eigenvectors $\{\widehat{X}^\mu_m\}$ with eigenvalues $\{\widehat{\omega}_m^2\}$. Using equation \eqref{eq cuerda_masas RN derivative traza} we see that \eqref{eq cuerda_masas robin mu-conditions} is equivalent to
	  \begin{align}\label{eq cuerda_masas borde-limite}
	  &\gamma_j(u)=\left(1+c_j\right)u(j)&&j\in\{0,\ell\}
	  \end{align}
	  which ties the values at the boundary with their limits. On the other hand the first equation can be split in three. Over $(0,\ell)$ we clearly have $u''=-\lambda u$ because the RN derivative over such interval is just the usual weak derivative. Thus we recover equation \eqref{eq cuerda_masas ecuacion interior X}. For $x=j\in\{0,\ell\}$ we can remove all references to the values at the boundary using equation \eqref{eq cuerda_masas RN derivative traza} and \eqref{eq cuerda_masas borde-limite}
	  \begin{align*}
	    -\lambda \frac{\gamma_j(u)}{1+c_j}&=-\lambda u(j)=\Delta_\mu u(j)=\\
	    &=\big(1+c_j\big)\frac{(-1)^{\sigma(j)}}{\alpha_j}\left(\gamma_j\left(\deriv{u}{\mu}\right)-(-1)^{\sigma(j)}\frac{c_j}{\alpha_j(1+c_j)}\gamma_j(u)\right)
	  \end{align*}
	  or equivalently
	  \begin{equation}
	    \gamma_j(u')=(-1)^{\sigma(j)+1}\left(\lambda\frac{\alpha_j}{(1+c_j)^2}-\frac{c_j}{\alpha_j(1+c_j)}\right)\gamma_j(u)
	  \end{equation}
	  
	  Clearly if we take $c_j=\frac{\alpha_jr_j}{1-\alpha_jr_j}$ and $\alpha_j>0$ satisfying\footnote{If $\mu_j=0$, then we get $\alpha_j=0$ and, in fact, we should remove its corresponding $\R$ factor. If $\mu_j>0$ then clearly $\alpha_j>0$. Finally notice that $\alpha_j=\mu_j$ if and only if $r_j=0$ i.e.\ there is no spring at $x=j\in\{0,\ell\}$.} $\alpha_j(1-\alpha_jr_j)^2=\mu_j$ then we obtain a formula identical to \eqref{eq cuerda_masas ecuacion borde X} replacing $X(j)$ by $\gamma_j(X)$.\separ
	  
	  Let us recap what we have done so far. We started with a Sturm-Liouville problem \eqref{eq cuerda_masas robin laplaciano}-\eqref{eq cuerda_masas robin mu-conditions} over $\mathcal{D}\subset L^2_\mu[0,\ell]$, which has infinitely many eigenvalues $\{\widehat{\omega}_m^2\}$ with normalized eigenvectors $\{\widehat{X}_m^\mu\}\subset\mathcal{D}$. Its solutions can be considered as triples over $\R\times L^2(0,\ell)\times\R$ and we adjust the constants in such a way that the middle factor of such solutions satisfies \eqref{eq cuerda_masas ecuacion interior X}-\eqref{eq cuerda_masas ecuacion borde X} where the boundary conditions are applied to the values of the trace. Doing so forces the eigenvalues $\widehat{\omega}_m$ to be equal to the previously computed $\omega_m$ on lemma \ref{lemma cuerda_masas infinitas soluciones crecientes}, and the eigenvectors are related by
	  \begin{equation}\label{eq cuerda_masas mu-autovectores=autovectores-viejos}
	    \left.\widehat{X}^\mu_m\right|_{(0,\ell)}=\frac{\left.X_m\right|_{(0,\ell)}}{g_m}\qquad\  \gamma_j(\widehat{X}^\mu_m)=\frac{\gamma_j(X_m)}{g_m}=\frac{X_m(j)}{g_m}\qquad\  \widehat{X}_m^\mu(j)=(1-\alpha_jr_j)\frac{X_m(j)}{g_m}\ \ \
	  \end{equation}
	  where $g_m$ guarantees that $\widehat{X}^\mu_m$ is normalized and is given by
	  \begin{equation}
	    g_m^2=\frac{1}{2}\left(r_0+(\ell+\mu_0)\omega_m^2+(\mu_0\omega_m^2-r_0)^2\ell+(\mu_\ell\omega_m^2+r_\ell)\frac{\omega_m^2+(\mu_0\omega_m^2-r_0)^2}{\omega_m^2+(\mu_\ell\omega_m^2-r_\ell)^2}\right)
	  \end{equation}
	  Actually, when we restrict ourselves to $\domRobin$ (so equation \eqref{eq cuerda_masas borde-limite} holds) and take into account that $\alpha_j(1-r_j\alpha_j)^2=\mu_j$, we obtain an interesting relation between the two formulations of the same problem
	  \begin{equation}
	    \llangle u,v\rrangle=\!\sum_{j\in\{0,\ell\}}\mu_j\gamma_j(u)\gamma_j(v)+\langle u,v\rangle=\!\sum_{j\in\{0,\ell\}}\alpha_ju(j)v(j)+\langle u,v\rangle=\peqsubfino{\left\langle u,v\right\rangle}{\!\mu}{-0.1ex}
	  \end{equation}
	  
	  We cannot stress enough that the physics of the initial problem is contained in the middle factor of $L^2_\mu[0,\ell]\cong\R\times L^2(0,\ell)\times\R$ and that the positions of the masses are given by the traces $\gamma_j(u)$ and not by the values at the boundary $u(j)$ corresponding to the $\R$ factors. To help us understand the role $u(j)$ plays in this setting, let us think about  Lagrange multipliers: they do help us to get the desired dynamics (so does $u(j)$ in our case), the dynamics of such multipliers are determined by the dynamics of the system (see equation \eqref{eq cuerda_masas borde-limite}), but usually we do not care about them (here only $\gamma_j(u)$ is physically relevant).
	        
	  \begin{theorem}\mbox{}\\
	    The eigenvalues $\{X_m\}$ form a complete set over $L^2[0,\ell]$. 
	  \end{theorem}
	  \begin{proof}\mbox{}\\
	     	The operator $\Delta_\mu$ is self-adjoint over the dense subset $\mathcal{D}=\domDelta\cap\domRobin\subset L^2_\mu[0,\ell]$, thus their eigenvectors $\{\widehat{X}_m\}$ form a complete set of $L^2_\mu[0,\ell]$. Take a map $f\in H^1[0,\ell]$ and consider \textbf{the} map $F\in \domRobin\subset L^2_\mu[0,\ell]$ such that $F|_{(0,\ell)}=f|_{(0,\ell)}$. $F(0)$ and $F(1)$ are then given by \eqref{eq cuerda_masas borde-limite} where the traces are well defined as $f\in H^1[0,\ell]$. Then we have
	     	\begin{equation}\label{eq cuerda_masas expansion de f}
	   	  F=\sum_m\peqsubfino{\langle F,\widehat{X}_m^\mu\rangle}{\!\mu}{-0.1ex}\,\widehat{X}_m^\mu\qquad\Longrightarrow\qquad f=\sum_m \frac{\peqsubfino{\langle F,\widehat{X}_m^\mu\rangle}{\!\mu}{-0.1ex}}{g_m}X_m
	     	\end{equation}
	     	We have on one hand that $\{\peqsubfino{\langle F,\widehat{X}_m^\mu\rangle}{\!\mu}{-0.1ex}\}\in \ell^2(\N)$ and, on the other hand, using the asymptotic behavior of $\omega_m$ obtained in \eqref{eq cuerda_masas omega_m assimpt sin masas}-\eqref{eq cuerda_masas omega_m assimpt con masas}, we can prove
	     	\begin{align}\label{eq cuerda_masas desarollo 1/g}
	     	\frac{1}{g_m}=\frac{\sqrt{2}\ell^{3/2}}{\mu_0\pi^2m^2}+\mathcal{O}\left(\frac{1}{m^3}\right)
	     	\end{align}
	     	Thus $\{\peqsubfino{\langle F,\widehat{X}_m^\mu\rangle}{\!\mu}{-0.1ex}/g_m\}\in \ell^2(\N)$ which ends the proof because $H^1[0,\ell]$ is dense in $L^2[0,\ell]$.
	  \end{proof}
	  
	  One might wonder what would happen if we extended $f$ arbitrarily as $(a,f,b)$. We could proceed as in the previous proof and it might seem that the coefficients of $f$ in the $\{X_m\}$ basis would depend on $a,b$. Of course this is not possible as the representations must be unique but it is interesting to understand why. Let us consider the map $F:[0,\ell]\to\R$ given by
	  \begin{align}\label{eq cuerda_masas F=(1,0,0)}
	    F(x)=\left\{ \begin{array}{ll}1&\ \ \text{if }x=0\\ 0 &\ \ \text{if }x\in(0,\ell]
	    \end{array}\right.
	  \end{align}
	  which can also be thought as $F=(1,0,0)\in\R\times L^2(0,\ell)\times\R$. If we expand $F$
	  \begin{align}\begin{split}\label{eq cuerda_masas expansion F=(1,0,0)}
	    (1,0,0)&=F=\sum_m\peqsubfino{\langle F,\widehat{X}_m^\mu\rangle}{\!\mu}{-0.1ex} \widehat{X}_m^\mu=\sum_m\peqsubfino{\langle F,\widehat{X}_m^\mu\rangle}{\!\mu}{-0.1ex} \Big(\widehat{X}_m^\mu(0),\left.\widehat{X}_m^\mu\right|_{(0,\ell)},\widehat{X}_m^\mu(\ell)\Big)=\\
	    &=\sum_m \alpha_0\widehat{X}^\mu_m(0)\Big(\widehat{X}_m^\mu(0),\left.\widehat{X}_m^\mu\right|_{(0,\ell)},\widehat{X}_m^\mu(\ell)\Big)
	  \end{split}\end{align}
	  and analogously for $G=(0,1,0)$ and $H=(0,0,1)$, we get the following identities
	  \begin{align*}
	    &\alpha_j\sum_m\widehat{X}_m^\mu(j)^2=1              && \mu_j\sum_m\frac{X_m(j)^2}{g_m}=1                    &&j\in\{0,\ell\}\\
	    &\sum_m\widehat{X}_m^\mu(0)\widehat{X}_m^\mu(\ell)=0 && \sum_m\frac{X_m(0)}{g_m}\frac{X_m(\ell)}{g_m}=0 &&\\
	    &\sum_m\widehat{X}_m^\mu(j)\widehat{X}_m^\mu(x)=0    && \sum_m\frac{X_m(j)}{g_m}\frac{X_m(x)}{g_m}=0    && x\in(0,\ell)\\
	    &\sum_m\left\langle\,\cdot{}\,,\left.\widehat{X}_m^\mu\right|_{(0,\ell)}\right\rangle\widehat{X}_m^\mu(j)=0    && \sum_m\left\langle\,\cdot{}\,,\frac{X_m}{g_m}\right\rangle\frac{X_m(j)}{g_m}=0\\
	    &f=\sum_m \left\langle f,\left.\widehat{X}^\mu_m\right|_{(0,\ell)}\right\rangle\left.\widehat{X}^\mu_m\right|_{(0,\ell)} && f=\sum_m \left\langle f,\frac{X_m}{g_m}\right\rangle\frac{X_m}{g_m}&&
	  \end{align*}
	  Notice that without taking advantage of the fact that $\{\widehat{X}^\mu_m\}$ forms a basis, it would be very difficult to obtain these identities as the $\omega_m$ are only defined implicitly and they even appear in the expression of $g_m$. Some remarks are in order now.
		  \begin{remarks}
	  	  	\item These identities show some sort of orthogonality between the factors.
	  		\item The last expression is equivalent to \eqref{eq cuerda_masas expansion de f} once we applied the rest of the identities.
	  		\item The first and second columns are related by \eqref{eq cuerda_masas mu-autovectores=autovectores-viejos}.
		  \end{remarks}

	\section{Alternative Lagrangian formulation}
	  We have just defined a new Laplace operator $\Delta_\mu$ in such a way that we recover the eigenvalue problem obtained when we performed the separation of variables at the beginning of section \ref{subsection cuerda_masas classical solutions}. It is straightforward to check that \eqref{eq cuerda_masas ecuacion de ondas} is equivalent to the restriction of
	  \begin{align}\label{eq cuerda_masas u=Delta_mu}
	  \begin{aligned}
	    &\rho\ddot{Q}(t,x)-\gamma\Delta_\mu Q(t,x)=0&&\quad t\in\R,\ x\in[0,\ell]\ \mu-\text{a.e.}\\[1ex]
	    &\deriv{Q}{\mu}(t,j)+(-1)^{\sigma(j)+1}\frac{c_j}{\alpha_j}Q(t,j)=0 &&\quad t\in\R,\ j\in\{0,\ell\}
	    \end{aligned}
	  \end{align}
	  to the middle factor of $L^2_\mu[0,\ell]\cong\R\times L^2[0,1]\times\R$. Notice that \eqref{eq cuerda_masas u=Delta_mu} describes the Klein-Gordon equation on the interval $[0,\ell]$ subject to Robin boundary conditions written in terms of the RN derivative.\separ
	  
	  Once we have the equations, we would like to obtain a Lagrangian leading to them. Let us first consider the action coming from the kinetic energy minus the potential energy in terms of the scalar product $\peqsubfino{\langle\cdot{}\,,\cdot{}\rangle}{\!\mu}{-0.1ex}$. Computing the EL equations, integrating by parts and applying \eqref{eq cuerda_masas integracion por partes mu}, gives rise to too many boundary terms. Some of them are removed if we restrict the domain to $\domRobin\times L^2_\mu[0,\ell]$ while the others can be removed by adding some counterterms. After these considerations it is natural to consider the Lagrangian $L^{\!\,\mu}:\mathcal{M}_1\to\R$
	  \begin{equation}
	    L^{\!\,\mu}(Q,V)=\frac{\rho}{2}\peqsubfino{\langle V,V\rangle}{\!\mu}{-0.1ex}-\frac{\gamma}{2}\peqsubfino{\left\langle\deriv{Q}{\mu},\deriv{Q}{\mu}\right\rangle}{\!\!\!\mu}{0.2ex}-\frac{\gamma}{2}\!\sum_{j\in\{0,\ell\}}\frac{c_j}{\alpha_j}Q^2(j)
	  \end{equation}
	  with domain $\mathcal{M}_1:=\peqsub{T}{H^1_{\!\mu\hspace*{-0.1ex},\hspace*{-0.1ex}R}}L^2_\mu[0,\ell]=\domRobin\times L^2_\mu[0,\ell]$. It is straightforward to see that the variations of its associated action, once we applied \eqref{eq cuerda_masas integracion por partes mu} over $\domRobin$, lead to the required equations.\separ
	  
	  As the equations of our theory are given by a proper wave equation, the Hamiltonian formulation is straightforward (see the appendix of \cite{barbero2014hamiltonian}). However the concrete spaces where such dynamics is consistent with the boundary conditions are tricky to be determined. Thus we need to proceed as in sections \ref{section cuerda_masas fiber derivative} and \ref{section cuerda_masas Hamiltonian formulation} for this new Lagrangian $L^{\!\,\mu}$.
	  
	\section{Alternative fiber derivative}
	  Our space $\mathcal{M}_1=\domRobin\times L^2_\mu[0,\ell]\subset TL^2_\mu[0,\ell]$ has typical elements $(Q,V)$. On the other hand a typical element of $\peqsub{T}{H^1_{\mu,R}}^*L^2_\mu[0,\ell]=\domRobin\times L^2_\mu[0,\ell]'\subset T^*\!L^2_\mu[0,\ell]$ will be of the form $(Q,\bm{P})$.\separ
	  
	  As usual, the phase space $T^*\!L^2_\mu[0,\ell]$ is equipped with the symplectic form \eqref{eq:background canonical symplectic form} which we denote $\Omega^{\!\,\mu}:\mathfrak{X}(T^*\!L^2_\mu[0,\ell])\times \mathfrak{X}(T^*\!L^2_\mu[0,\ell])\to\mathcal{C}^\infty(T^*\!L^2_\mu[0,\ell])$ and is given by
	  \begin{align*}
	    \Omega^{\!\,\mu}_{(Q,\bm{P})}(Y,Z)
	    &=\Omega^{\!\,\mu}_{(Q,\bm{P})}\left(\rule{0ex}{3ex}(\vecc{Y}{Q},\vecc{\boldsymbol{Y}}{\!\!P}),(\vecc{Z}{Q},\vecc{\boldsymbol{Z}}{P})\right)=\vecc{\boldsymbol{Z}}{P}\left(\vecc{Y}{Q}\right)-\vecc{\boldsymbol{Y}}{\!\!P}\left(\vecc{Z}{Q}\right)
	  \end{align*}   
	  Applying the Riesz representation theorem $\boldsymbol{P}$ (and $\vecc{\boldsymbol{Y}}{\!\!P}$) can be expressed in terms of a map $P\in L^2_\mu[0,\ell]$ by $\boldsymbol{P}(V)=\peqsubfino{\left\langle P,V\right\rangle}{\!\mu}{-0.1ex}$. This allows us to identify the cotangent space $T^*\!L^2_\mu[0,\ell]$ with the tangent space $TL^2_\mu[0,\ell]$. Over the latter we can consider the induced symplectic form
	  \begin{align}\label{eq cuerda_masas omega^mu}
	    \omega^{\!\,\mu}_{(Q,P)}(Y,Z)
	    &:=\Omega^{\!\,\mu}_{(Q,\langle P,\,\cdot{}\,\rangle)}\left(\rule{0ex}{3ex}(\vecc{Y}{Q},\langle\vecc{Y}{P},\,\cdot{}\,\rangle),(\vecc{Z}{Q},\langle\vecc{Z}{P},\,\cdot{}\,\rangle)\right)=\peqsubfino{\langle\vecc{Z}{P},\vecc{Y}{Q}\rangle}{\!\mu}{-0.1ex}-\peqsubfino{\langle\vecc{Y}{P},\vecc{Z}{Q}\rangle}{\!\mu}{-0.1ex}
	  \end{align}   
	  Notice that this symplectic form has some boundary contributions hidden in the scalar product even though there exist no independent degrees of freedom at the boundary. The fiber derivative\index{Fiber derivative} $F\!L^{\!\,\mu}:\mathcal{M}_1\subset TL^2_\mu[0,\ell]\to T^*\!L^2_\mu[0,\ell]$ is now given by
	  \begin{align}
	    F\!L^{\!\,\mu}(Q,V)&\Big(Q,W\Big)=\left.\deriv{}{\tau}\right|_{\tau=0}L^{\!\,\mu}(Q,V+\tau W)=\rho\peqsubfino{\langle V,W\rangle}{\!\mu}{-0.1ex}
	  \end{align}
	  Thus we have $F\!L^{\!\,\mu}(Q,V)=(Q,\rho\peqsubfino{\langle V,\,\cdot{}\,\rangle}{\!\mu}{-0.1ex})$ which allows us to define the canonically conjugate momenta $\bm{P}=\rho\peqsubfino{\langle V,\,\cdot{}\,\rangle}{\!\mu}{-0.1ex}$. Identifying as before the tangent and cotangent spaces, we have  $P=\rho V$ and, then, $F\!L^{\!\,\mu}$ is simply the inclusion of $F\!L^{\!\,\mu}(\mathcal{M}_1)\cong \mathcal{M}_1$ in $TL^2_\mu[0,\ell]$.

  \section{Alternative Hamiltonian formulation}\label{String masses - section - Alternative Hamiltonian formulation}
    \subsection*{Obtaining the Hamiltonian}\trassub
    
      The energy\index{Energy} $E^{\!\,\mu}:\mathcal{M}_1\to\R$, which is defined as $E^{\!\,\mu}(Q,V)=F\!L^{\!\,\mu}(Q,V)\left(\rule{0ex}{2.1ex}Q,V\right)-L^{\!\,\mu}(Q,V)$, has the same functional expression as the Lagrangian but changing the two minus signs for pluses. The Hamiltonian\index{Hamiltonian} $H^{\!\,\mu}:F\!L^{\!\,\mu}(\mathcal{M}_1)\subset T^*\!L^2_\mu[0,\ell]\to\R$, defined implicitly by $H^{\!\,\mu}\smallcirc F\!L^{\!\,\mu}=E^{\!\,\mu}$, can be obtained through the identification $F\!L^{\!\,\mu}(\mathcal{M}_1)\cong\mathcal{M}_1$. We then have $H^{\!\,\mu}:\mathcal{M}_1\to\R$ given by
      \begin{align}
        H^{\!\,\mu}(Q,P)=\frac{\peqsubfino{\langle P,P\rangle}{\!\mu}{-0.1ex}}{2\rho}+\frac{\gamma}{2}\left\langle\deriv{Q}{\mu},\deriv{Q}{\mu}\right\rangle_{\!\!\mu}+\frac{\gamma}{2}\!\sum_{j\in\{0,\ell\}}\frac{c_j}{\alpha_j}Q(j)^2
      \end{align}
    
    \subsection*{GNH algorithm}\trassub
    
    $\bullet$ Compute the differential $\mathrm{d}H^{\!\,\mu}:T\mathcal{M}_1\to\R$ of $H^{\!\,\mu}$.\separprevia
    
    Using equations \eqref{eq cuerda_masas integracion por partes mu} and \eqref{eq cuerda_masas RN derivative traza} we get
      \begin{align}\label{eq cuerda_masas dH^mu}
        \mathrm{d}_{(Q,P)}H^{\!\,\mu}(\vecc{Z}{Q},\vecc{Z}{P})=\frac{1}{\rho}\peqsubfino{\langle P,\vecc{Z}{P}\rangle}{\!\mu}{-0.1ex}+\gamma\left\langle\deriv{Q}{\mu},\deriv{\vecc{Z}{Q}}{\mu}\right\rangle_{\!\!\mu}+\gamma\!\sum_{j\in\{0,\ell\}}\frac{c_j}{2\alpha_j}Q(j)\vecc{Z}{Q}(j)
      \end{align}
      
      $\bullet$ Compute the Hamiltonian vector field\index{Hamiltonian vector field} $\peqsub{Y}{\!H}\in\mathfrak{X}(\mathcal{M}_1)$.\separprevia
      
      Let us solve, over some subspace $\mathcal{M}_2\subset\mathcal{M}_1=\domRobin\times L^2_\mu[0,\ell]$, the Hamiltonian equation $\imath_Y\omega^{\!\,\mu}(Z)-\mathrm{d}H^{\!\,\mu}(Z)=0$ for every $Z=(\vecc{Z}{Q},\vecc{Z}{P})\in\mathfrak{X}(\mathcal{M}_1)$. Notice that as $\mathcal{M}_1$ is a vector space we have $\vecc{Y}{Q}(Q,P)\in\domRobin$ and $\vecc{Y}{P}(Q,P)\in L^2_\mu[0,\ell]$ for every $(Q,P)\in\mathcal{M}_1=\domRobin\times L^2_\mu[0,\ell]$.\separ
      
      Comparing equations \eqref{eq cuerda_masas omega^mu} and \eqref{eq cuerda_masas dH^mu} for $\vecc{Z}{Q}=0$ and for every $\vecc{Z}{P}\in L^2_\mu[0,\ell]$ we get
      \begin{align*}
        &\vecc{Y}{Q}(Q,P)=\frac{P}{\rho}
      \end{align*}      
      In particular the domain has to be restricted as $P=\rho\vecc{Y}{Q}\in\domRobin$. Now the last equation that must be solved is
      \begin{align*}
        -\peqsubfino{\langle\vecc{Y}{P},\vecc{Z}{Q}\rangle}{\!\mu}{-0.1ex}=\gamma\left\langle\deriv{Q}{\mu},\deriv{\vecc{Z}{Q}}{\mu}\right\rangle_{\!\!\mu}+\gamma\!\sum_{j\in\{0,\ell\}}\frac{c_j}{2\alpha_j}Q(j)\vecc{Z}{Q}(j)
      \end{align*}
      for all $\vecc{Z}{Q}\in\domRobin$. Thus we should write the right hand side as a scalar product $\peqsubfino{\langle\cdot{}\,,\vecc{Z}{Q}\rangle}{\!\mu}{-0.1ex}$. Following \cite{barbero2014hamiltonian} it is easy to prove that in order to do that we must require $Q\in\domDelta$ so that we can integrate the previous expression by parts. Furthermore, as we also have that $Q\in\domRobin$, we can apply equation \eqref{eq cuerda_masas integracion por partes mu} and obtain
      \begin{align*}
        \peqsubfino{\langle\vecc{Y}{P},\vecc{Z}{Q}\rangle}{\!\mu}{-0.1ex}=\gamma\left\langle\Delta_\mu Q,\vecc{Z}{Q}\right\rangle_{\!\!\mu}
      \end{align*}
      for all $\vecc{Z}{Q}\in\domRobin$. As this space is dense in $L^2_\mu[0,\ell]$ the Hamiltonian vector field is given by
      \begin{align*}
      &\peqsub{Y}{\!H}(Q,P)=\left(\frac{P}{\rho},\gamma\Delta_\mu Q\right)\in \domRobin\times L^2_\mu[0,\ell]\\ &\text{For every }(Q,P)\in\Big(\domDelta\cap\domRobin\Big)\times\domRobin=:\mathcal{M}_2
      \end{align*}
      So we have $\peqsub{Y}{\!H}:\mathcal{M}_2\to\mathcal{M}_1$ and we need to find $\mathcal{M}_3=\{(Q,P)\in\mathcal{M}_2\ /\ \peqsub{Y}{\!H}(Q,P)\in \peqsub{T}{(Q,P)}\overline{\mathcal{M}_2}\}$. If we demand exact tangency we should proceed as in section \ref{section cuerda_masas Hamiltonian formulation} but for our purposes it is enough to require tangency to the closure of the constrained space. Using the techniques employed in \cite{barbero2014hamiltonian}, it can be proved that $\mathcal{M}_2=(\domDelta\cap\domRobin)\times\domRobin$ is dense in $\mathcal{M}_1=\domRobin\times L^2_\mu[0,\ell]$ hence $\mathcal{M}_3=\mathcal{M}_2$ and the algorithm stops.

  \section{Fock quantization}
  
  We have seen that the positions of the point particles attached at the ends of the string are  not independent physical degrees of freedom. This indicates that the Fock space for this  system will not have the form of a tensor product of different Hilbert spaces associated with the masses and the string. Notice that this shows some sort of ``entanglement'' stronger than the usual one. Indeed, a state $v\in\mathcal{H}_1\otimes\mathcal{H}_2$ is entangled if it cannot be written as $v_1\otimes v_2$. For instance $v\otimes v+w\otimes w$ is entangled in general. Here we have that the space itself cannot be written as a tensor product!\separ
  
  This ``entanglement'' gives raise to some physical questions that we would like to address. For instance, if we think about this model as two masses connected by a physical spring (with ``internal degrees of freedom'' as we mentioned in the introduction), how does one recover the situation where the string just models an ideal spring? What is the origin of the $L^2(\R)\otimes L^2(\R)$ Hilbert space that one would use to describe this system? As a first step towards addressing these questions it is important to understand in detail why the Fock space does not factorize.
    
    \subsection*{Construction of the Fock space}\trassub
    
      We have already introduced in section \ref{Mathematical background - subsection - second quantization} of chapter \ref{Chapter - Mathematical background} the procedure to construct the Fock space. It can be summarized as follows
      \begin{enumerate}
      	\item Let $\mathcal{S}=\{\text{\emph{Real vector space of Hamilt.\ solutions to the linear field equations}}\}$ endowed with the bilinear form $\langle\,,\rangle_\Omega$ induced by the symplectic structure.
      	\item Let $\mathcal{S}^\C$ be a (non-canonical) complexification of $\mathcal{S}$ with the induced complexified sesquilinear form $\langle\,,\rangle_\Omega^\C$.
      	\item Let $\mathcal{S}^\C_+$ be a subspace (called of positive frequency solutions) such that $\langle\,,\rangle_\Omega^\C$ restricted to it is a proper scalar product denoted as $\peqsubfino{\langle\,,\rangle}{+}{-0.2ex}$.
      	\item $(\mathcal{S}^\C_+\,,\peqsubfino{\langle\,,\rangle}{\!+}{-0.2ex})$ is a pre-Hilbert space. Upon Cauchy completion we obtain the $1$-particle Hilbert space denoted $(\mathfrak{h},\peqsubfino{\langle\,,\rangle}{\!+}{-0.2ex})$.
      	\item We define the Fock space as $\displaystyle \mathcal{F}(\mathfrak{h})=\bigoplus_{n=0}^\infty\mathfrak{h}^{\smallcirc n}$ with the scalar product given by \eqref{Mathematical background - equation - producto escalar Fock}.
      \end{enumerate}
  
  	  In the case we are considering, the Hamiltonian description of the previous section has produced the linear manifold of $L^2_\mu[0,\ell]\times L^2_\mu[0,\ell]$ given by $\mathcal{M}_2=(\domDelta\cap\domRobin)\times\domRobin$, where the classical Hamiltonian dynamics takes place, together with a Hamiltonian vector field $\peqsub{Y}{\!H}$ tangent to the closure of $\mathcal{M}_2$. The space of solutions to the field equations $\mathcal{S}$ consists of the integral curves of the Hamiltonian vector field $\peqsub{Y}{\!H}$ that lie over $\mathcal{M}_2$. As usual, those curves in $\mathcal{S}$ are in correspondence with the space of initial data $\mathcal{M}_2$ under an isomorphism $\Phi_0$ (see page \pageref{Mathematical background - definition isomorfismo Cauchy->Sol}). In particular we can define the analog of \eqref{Mathematical background - equation - forma bilinear bogoluilov} but over the space of Cauchy data i.e.\ using $\omega^{\!\,\mu}$ given by equation \eqref{eq cuerda_masas omega^mu}, to obtain \[\peqsub{\big\langle(Q_1,P_1),(Q_2,P_2)\big\rangle}{\!\Omega}=-i\omega^\mu\big((Q_1,P_1),(Q_2,P_2)\big)\]
  	  
  	  For the second step we consider the complexified vector space $\mathcal{S}^\C=\mathcal{M}_2^\C$ where vector addition is defined componentwise as the standard sum of real functions, while multiplication by scalars is defined relying on the complex structure
  	  \begin{align*}
  	    \begin{array}{cccc}
  	    	J:&\mathcal{M}_2^\C=\mathcal{M}_2\times\mathcal{M}_2 & \longrightarrow & \mathcal{M}_2^\C=\mathcal{M}_2\times\mathcal{M}_2\\
  	    	  &  \Big((Q_1,P_1),(Q_2,P_2)\Big)                   &  \longmapsto    & \Big(-(Q_2,P_2),(Q_1,P_1)\Big)
  	    \end{array}
  	  \end{align*}
  	  as follows: $(a+ib)\cdot{}(V):=a V+bJ(V)$, for every $a,b\in\R$ and every $V\in\mathcal{M}_2\times\mathcal{M}_2$. Of course we can simply think that the elements of the complexified space $\mathcal{M}_2^\C$ are complex functions in $\mathcal{M}_2$ with the standard sum and multiplication by complex scalars. The complexified sesquilinear form is just the natural extension of $\langle\,,\rangle_\Omega$.\separ
  	  
  	  The third step requires the selection of a subspace of the space of solutions $\mathcal{S}^\C=\mathcal{M}_2^\C$ so that $\langle\,,\rangle_\Omega^\C$ becomes a scalar product. To do that we consider the general solution to \eqref{eq cuerda_masas u=Delta_mu} for some Cauchy data $(Q,P)=(u(\,\cdot{}\ \!,0),\dot{u}(\,\cdot{}\ \!,0))\in\mathcal{M}_2^\C$ for $t=0$, which is given by
  	  \begin{align*}
  	    u(t,x)=\frac{1}{2}\sum_n\left[\Big(\omega_n Q_n-iP_n\Big)\frac{e^{i\omega_n t}}{\omega_n}+\Big(\omega_n Q_n+iP_n\Big)\frac{e^{-i\omega_n t}}{\omega_n}\right]\widehat{X}^\mu_n(x)
  	  \end{align*}
      where $Q_n=\peqsubfino{\langle \widehat{X}^\mu_n,Q\rangle}{\!\mu}{-0.1ex}\in\C$ and $P_n=\peqsubfino{\langle \widehat{X}^\mu_n,P\rangle}{\!\mu}{-0.1ex}\in\C$ are the Fourier coefficients of the initial data in the basis $\{\widehat{X}^\mu_n\}$. The exponentials suggest that in order to keep the positive frequency part we have to kill the prefactors of the ``negative'' exponentials. Thus we require that for all $n\in\N$
      \begin{align*}
        0&=\omega_nQ_n+iP_n=\peqsubfino{\langle \omega_n\widehat{X}^\mu_n,Q\rangle}{\!\mu}{-0.1ex}+i\peqsubfino{\langle \widehat{X}^\mu_n,P\rangle}{\!\mu}{-0.1ex}=\\
        &=\peqsubfino{\langle \sqrt{-\Delta_\mu}\,\widehat{X}^\mu_n,Q\rangle}{\!\mu}{-0.1ex}+i\peqsubfino{\langle \widehat{X}^\mu_n,P\rangle}{\!\mu}{-0.1ex}=\peqsubfino{\left\langle \widehat{X}^\mu_n,\sqrt{-\Delta_\mu}\,Q+iP\right\rangle}{\!\!\!\mu}{-0.1ex}
      \end{align*}
      where $h=\sqrt{-\Delta_\mu}$ is defined as the diagonal operator with $\omega_n$ as its eigenvalues, which is of course self-adjoint (recall that $\Delta_\mu$ is the diagonal operator with $\{-\omega^2_n\}$ as its eigenvalues). The fact that the previous expression is zero for every $n\in\N$ is tantamount to
      \begin{align}
        \sqrt{-\Delta_\mu}\,Q+iP=0 
      \end{align}
      and therefore the space of positive frequencies is given by
      \begin{align*}
        \mathcal{S}^\C_+&=\Big\{(Q,P)\in\mathcal{M}_2^\C\ /\ \sqrt{-\Delta_\mu}\,Q+iP=0\Big\}=\\
        &=\Big\{(Q,i\sqrt{-\Delta_\mu}\,Q)\in\mathcal{M}_2^\C=\left(\domDelta\cap\domRobin\right)^\C\times\domRobin^\C\Big\}\cong\\
        &\cong\Big\{Q\in\left(\domDelta\cap\domRobin\right)^\C\Big\}=\left(\domDelta\cap\domRobin\right)^\C
      \end{align*}
      with the scalar product
      \begin{align*}
        \peqsubfino{\left\langle Q_1,Q_2\right\rangle}{\!+}{-0.2ex}&:=-i\omega^\mu_\C\Big((\overline{Q_1},-i\sqrt{-\Delta_\mu}\,\overline{Q_1}),(Q_2,i\sqrt{-\Delta_\mu}\,Q_2)\Big)=\\
        &=-i\peqsubfino{\left\langle i\sqrt{-\Delta_\mu}\,Q_2, \overline{Q_1}\right\rangle}{\!\!\!\mu}{-0.1ex}+i\peqsubfino{\left\langle -i\sqrt{-\Delta_\mu}\,\overline{Q_1},Q_2 \right\rangle}{\!\!\!\mu}{-0.1ex}=\\
        &=\peqsubfino{\left\langle \sqrt{-\Delta_\mu}\,Q_2, \overline{Q_1}\right\rangle}{\!\!\!\mu}{-0.1ex}+\peqsubfino{\left\langle \sqrt{-\Delta_\mu}\,\overline{Q_1},Q_2 \right\rangle}{\!\!\!\mu}{-0.1ex}=2\peqsubfino{\left\langle \overline{Q_1}, \sqrt{-\Delta_\mu}\,Q_2\right\rangle}{\!\!\!\mu}{-0.1ex}=2\sum_n (\overline{Q_1})_n\omega_n(Q_2)_n     
      \end{align*}
      We have then obtained an expression for $\peqsubfino{\langle\,,\rangle}{\!+}{-0.2ex}$ in terms of the Fourier coefficients of $Q_j$ which shows, in particular, that it is indeed a scalar product. Notice that although $\{\widehat{X}_n^\mu\}$ are still orthogonal with respect to this new scalar product, they are not longer of norm $1$. We thus normalize them
      \begin{align*}
        \widetilde{X}^+_n:=\frac{1}{\peqsubfino{\langle \widehat{X}^\mu_n,\widehat{X}^\mu_n\rangle}{\!+}{-0.2ex}^{\raisemath{0.5ex}{\scriptscriptstyle 1\hspace*{-0.2ex}/\hspace*{-0.2ex}2}}}\widehat{X}^\mu_n=\frac{1}{\sqrt{2\omega_n}}\widehat{X}^\mu_n
      \end{align*}
      So step 3 is finished and we end up with a pre-Hilbert space $\mathcal{S}^\C_+=(\domDelta\cap\domRobin)^\C$, with an orthonormal basis $\{\widetilde{X}^+_n\}$, and the scalar product $\peqsubfino{\langle\,,\rangle}{\!+}{-0.2ex}$.\separ
      
      The $1$-particle Hilbert space of the fourth step is simply
      \begin{align}
        \left(\mathfrak{h}=\left\{\psi=\sum_n\psi_n\widetilde{X}^+_n\ /\ \{\psi_m\}\in\ell^2(\C)\right\},\peqsubfino{\langle\,,\rangle}{\!+}{-0.2ex}\right)
      \end{align}

      Finally the Hilbert space of our quantum field theory is given by the symmetric Fock space $\mathcal{F}(\mathfrak{h})$.\separ
      
      Before considering dynamics over the Fock space, we need to understand how the presence of a boundary affects the Fock construction.
      
    \subsection*{Factorization}\trassub
    
      Had we begun with the $1$-particle Hilbert space $\mathfrak{h}_0=L_\mu[0,\ell]$, we would have had the Fock space\allowdisplaybreaks[0]
      \begin{align*}
        \mathcal{F}(\mathfrak{h}_0)&=\mathcal{F}(L_\mu[0,\ell])\cong \\
        &\cong \mathcal{F}(\R\oplus L^2(0,\ell)\oplus\R)\cong \mathcal{F}(\R)\otimes \mathcal{F}(L^2(0,\ell))\otimes \mathcal{F}(\R)
      \end{align*}\allowdisplaybreaks
      
      where we have used the exponential behavior of the Fock space given by proposition \ref{Mathematical background - proposition - e(h1+h2)=e(h1)xe(h2)}. In that case we clearly have two factors $\mathcal{F}(\R)$ associated with the boundary. We have, however, that $\mathfrak{h}$ is the Cauchy completion of $\mathcal{S}^\C_+=(\domDelta\cap\domRobin)^\C$ which entangles the behavior of the boundary and the bulk so we should not expect to have $\mathfrak{h}\cong\C\oplus\mathfrak{h}_{\textrm{string}}\oplus\C$.\separ
      
      Let us consider the map $F\in L^2_\mu[0,\ell]$ given by
      \begin{align}
      	F(x)=\left\{ \begin{array}{ll}1&\ \ \text{if }x=0\\ 0 &\ \ \text{if }x\in(0,\ell]
      	\end{array}\right.
      \end{align}
      which was already considered in \eqref{eq cuerda_masas F=(1,0,0)}. We now proceed as in \eqref{eq cuerda_masas expansion F=(1,0,0)} but using instead the scalar product $\peqsubfino{\langle\,,\rangle}{\!+}{-0.2ex}$ and the basis $\{\widetilde{X}^+_n\}$.
      \begin{align*}
        \peqsubfino{\left\langle F,\widetilde{X}^+_n\right\rangle}{\!\!+}{0.1ex}&=2\left\langle F,\sqrt{-\Delta_\mu}\frac{\widehat{X}^\mu_n}{\sqrt{2\omega_n}}\right\rangle_{\!\!\mu}=\sqrt{2\omega_n}\left\langle F,\widehat{X}^\mu_n\right\rangle_{\!\!\mu}=\\
        &=\sqrt{2\omega_n}\alpha_0\widehat{X}_n^\mu(0)=\sqrt{2\omega_n}\alpha_0(1-\alpha_0r_0)\frac{X_n(0)}{g_n}=\\
        &=\sqrt{2\alpha_0\mu_0}\,\frac{\sqrt{\omega_n}X_n(0)}{g_n}=\sqrt{2\alpha_0\mu_0}\,\frac{\omega_n^{3/2}}{g_n}
      \end{align*}
      where we have used the definitions of $\mu_0$, $\peqsubfino{\langle\,,\rangle}{\!+}{-0.2ex}$ and $\peqsubfino{\langle \,,\,\!\rangle}{\!\mu}{-0.1ex}$, also equation \eqref{eq cuerda_masas mu-autovectores=autovectores-viejos} and the explicit solution $X_n(x)=(r_0-\mu_0\omega_n^2)\sin(\omega_nx)+\omega_n\cos(\omega_nx)$ given in section \ref{subsection cuerda_masas classical solutions}. Now using the asymptotic behaviors \eqref{eq cuerda_masas desarollo 1/g} and \eqref{eq cuerda_masas omega_m assimpt sin masas}-\eqref{eq cuerda_masas omega_m assimpt con masas} we obtain
      \begin{align*}
        \peqsubfino{\left\langle F,\widetilde{X}^+_n\right\rangle}{\!\!+}{0.1ex}^2=\frac{4\alpha_0}{\pi \mu_0\,n}+\mathcal{O}\left(\frac{1}{n^2}\right)
      \end{align*}
      As the sequence of coefficients $\{\langle F,\widetilde{X}^+_n\rangle_{\!+}\}$ of $F$ does not belong to $\ell^2(\C)$ we have that $F\notin\mathfrak{h}$ therefore
      
      \centerline{$\mathfrak{h}$ is \textbf{not} of the form $\C\oplus\mathfrak{h}_{\textrm{string}}\oplus\C$}\separ
      
      This implies in particular that $\mathcal{F}(\mathfrak{h})$ has no natural decomposition with factors corresponding to the boundary. Notice that it might be possible to find factorizations in other ways although it seems unlikely that they would have a true physical interest.

    \subsection*{Dynamics at the boundary}\trassub
        
        We have obtained that the Hilbert space does not factorize in a way that allows us to isolate the masses. Nonetheless, we know that our system models, in a more realistic way, the simple mechanical system consisting of two masses connected by a spring. In the usual treatment of such system one makes, more or less implicitly, the assumption that the internal degrees of freedom of the spring are irrelevant. So the question that arises naturally is: how can we then get the simplified models where the string configurations seem to play no role and only the masses are relevant?\separ
        
      We will answer this question, following our work in \cite{margalefboundary}, by obtaining a way to concentrate only on the dynamics of the point particles attached to the string. To do so we proceed as in section \ref{Mathematical background - subsection - second quantization} of chapter \ref{Chapter - Mathematical background}, where we introduced the coherent states as well as the creation and annihilation operators\index{Annihilation operator}, and make use of the trace operator. Meanwhile, the dynamics over the Fock space is determined by the quantum Hamiltonian $H_h$, given by equation \eqref{eq:backgroud Hamiltonian Fock} with $h=\sqrt{-\Delta_\mu}$, which is guaranteed to be unitary thanks to the self-adjointness of $\Delta_\mu$ in $\domRobin$.\separ

   Notice that, on one hand, $h$ defines some dynamics over $\mathfrak{h}$ while, on the other hand, we have the trace operator
   \[\begin{array}{cccc}
   \gamma:&\mathfrak{h}&\to     & \C^2\\
   &    f       &\mapsto & \big(f(0+),f(1-)\big)
   \end{array}\]
   from the Hilbert space to the boundary. On page \pageref{Mathematical background - subsection - Parsing dynamics} we explain how to take dynamics from one Hilbert space to another relying on property \refconchap{Mathematical background - property - F(T)(e(v))=e(TV)}, which in this case reads
    	
    	      	\centerline{\begin{tikzpicture}[ampersand replacement=\&]
    		\matrix (m) [matrix of math nodes,row sep=3em,column sep=4em,minimum width=2em]
    		{\mathcal{F}(\mathfrak{h}) \& \mathcal{F}(\C^2) \\
    			\mathfrak{h}  \& \C^2 \\};
    		\path[-stealth]
    		(m-2-1) edge node [left] {$\peqsub{\varepsilon}{I}$} (m-1-1)
    		(m-2-1.east|-m-2-2) edge node [below] {$\gamma$} (m-2-2)
    		(m-1-1.east|-m-1-2) edge node [above] {$\mathcal{F}(\gamma)$}	(m-1-2)
    		(m-2-2) edge node [right] {$\peqsub{\varepsilon}{F}$} (m-1-2);
    		\end{tikzpicture}}
    
    This diagram tells us in particular that the dynamics can be taken (see proposition \ref{proposition:background evolucion coherent state}) from the boundary to its Fock counterpart. This might seem at first sight incompatible with what we mentioned before about the non-factorization of the Fock space $\mathcal{F}(\mathfrak{h})$ because if we are able to define some coherent states at the boundary with a well behaved evolution, we end up in practice with a factorization. The solution to this apparent contradiction is, as we will see in the next paragraph, that the dynamics over $\mathcal{F}(\C^2)$ is not unitary (so the evolution is not ``well behaved'') although it is unitary over the whole system. Notice that this is somehow equivalent to the fact that, classically, the total energy is conserved but the energy of each of the masses is not.\separ
    
    For the sake of concreteness we considered the map $\peqsubfino{\gamma}{0}{-0.2ex}:\mathfrak{h}\to\C$, the trace operator over the left point of the boundary, given by $\peqsubfino{\gamma}{0}{-0.2ex}(f)=f(0+)$. The evolution of the norm of a coherent state is given by equation \eqref{Mathematical background - equation - evolution unitaria producto escalar}, that for $T=\peqsubfino{\gamma}{0}{-0.2ex}$ and Hamiltonian $h$ reads
    \begin{align*}
      i\deriv{}{t}\peqsub{\|\varepsilon(\peqsubfino{\gamma}{0}{-0.2ex}v_t)\|^2}{\smallcirc}&=\peqsub{\|\varepsilon(\peqsubfino{\gamma}{0}{-0.2ex}v_t)\|^2}{\smallcirc}\Big(\langle \peqsubfino{\gamma}{0}{-0.2ex}v_t,\peqsubfino{\gamma}{0}{-0.2ex}hv_t\rangle-\langle \peqsubfino{\gamma}{0}{-0.2ex}hv_t,\peqsubfino{\gamma}{0}{-0.2ex}v_t\rangle\Big)=\\
      &=\peqsub{\|\varepsilon(\peqsubfino{\gamma}{0}{-0.2ex}v_t)\|^2}{\smallcirc}\Big(\overline{\peqsubfino{\gamma}{0}{-0.2ex}(v_t)}\peqsubfino{\gamma}{0}{-0.2ex}(hv_t)-\overline{\peqsubfino{\gamma}{0}{-0.2ex}(hv_t)}\peqsubfino{\gamma}{0}{-0.2ex}(v_t)\Big)=\\
      &=2i\peqsub{\|\varepsilon(\peqsubfino{\gamma}{0}{-0.2ex}v_t)\|^2}{\smallcirc}\,\mathrm{Im}\Big(\overline{\peqsubfino{\gamma}{0}{-0.2ex}(v_t)}\peqsubfino{\gamma}{0}{-0.2ex}(hv_t)\Big)=2i\peqsub{\|\varepsilon(\peqsubfino{\gamma}{0}{-0.2ex}v_t)\|^2}{\smallcirc}\,|\peqsubfino{\gamma}{0}{-0.2ex}(v_t)|^2\mathrm{Im}\big(\peqsubfino{\omega}{0}{-0.2ex}(v_t)\big)
      \end{align*}
      where we have introduced
      \[\peqsubfino{\omega}{0}{-0.2ex}(v_t):=\frac{\peqsubfino{\gamma}{0}{-0.2ex}(hv_t)}{\peqsubfino{\gamma}{0}{-0.2ex}(v_t)}\in\C\]
      This complex number is, in general, non-real (as can be easily seen by taking a sum of normal modes \cite{margalefboundary}) so the previous expression is different from zero, showing that the dynamics is not unitary. This might seem a disaster as unitarity is one of the fundamental properties of quantum mechanics. However, as we mentioned before, it should not be surprising at all the fact that in a subsystem unitarity is not preserved (in the same way that we do not expect the energy to be conserved). In fact, the use of trace operators, natural for coupled systems like this one, offers the interesting possibility of introducing a sort of non-unitary dynamics that may illuminate issues related to the collapse of the wave function or the surprising quantum behavior of gravitational systems in the presence of boundaries.
      
    \section{Unitary implementation}
    
    We have seen in the previous section that the inclusion of the masses prevents the boundary from having unitary dynamics. If we remove such masses the problem disappears and, in principle, we can define some unitary dynamics as in the case of inertial foliations. Nonetheless, we have seen in chapter \ref{Chapter - Mathematical background} that it is more natural to consider arbitrary foliations to control the evolution, hence we have to study, as we did in \cite{margalef2017functional}, which foliations give raise to unitary evolution.\separ
    
      In what follows we consider the scalar field equation with Dirichlet ($k_j\to\infty$) or Robin boundary conditions. This is equivalent to taking equations \eqref{eq cuerda_masas ecuacion de ondas} and remove the masses $m_j=0$. Thus we can rewrite \eqref{eq cuerda_masas ecuacion interior T}-\eqref{eq cuerda_masas ecuacion borde X} to obtain\allowdisplaybreaks[0]
      \begin{align}
      &\frac{\rho}{\gamma} \ddot{T}_k(t)=-\omega_k^2 T_k(t)    && t\in\R\\
      &X_k''(x)=-\omega_k^2 X_k(x)     && x\in[0,\ell]\\
      &X_k'(j)=(-1)^{\sigma(j)}r_jX_k(j) && j\in\{0,\ell\}
      \end{align}\allowdisplaybreaks
      Now we have a proper Sturm-Liouville problem and the standard procedure applies. In particular we get that the general solution is given by
      \[Q(t,x)=\Big(a_0(1-it)+a_0^*(1+it)\Big)X_0+\sum_{n=1}^\infty\left(a_n e^{-it\omega_n}+a^*_n e^{it\omega_n}\right)X_n(x)\]
      where the constants $a_k$ are determined by the initial conditions. The normal modes were already obtained on page \pageref{eq cuerda_masas ecuacion para omega} but, for what follows, it is much more convenient to rewrite the solutions in terms of exponentials. To do so, we take $\alpha_k^{j\pm}=\omega_k\mp i r_j$ if we are in the Robin case and $\alpha_k^{j\pm}=\mp i/2$ for the Dirichlet one. The normal modes are then given by
      \begin{align*}
      	&X_k(x)=\frac{1}{c_k}\Big(\alpha_k^{0+}e^{i\omega_kx}+\alpha_k^{0-}e^{i\omega_kx}\Big) &&k\in\N\\
      	&X_0(x)=\frac{1}{\sqrt{2}}&&\text{Only in the Neumann case}\\
      	&\mathrlap{c_k^2=4\omega_k\ell|\alpha_k^{0+}|^2+2\sin(\omega_k\ell)\Big[(\alpha^{0+}_k)^2e^{i\omega_k\ell}+(\alpha^{0-}_k)^2e^{-i\omega_k\ell}\Big]}
      \end{align*}
      where $c_k$ has been defined so that we have the normalizations $\langle X_k,X_k\rangle=\frac{1}{2\omega_k}$ and $\langle X_0,X_0\rangle=\frac{1}{2}$. The non-zero allowed frequencies $\omega_k$ are solutions to the (real) equation
      \[e^{i\omega_k\ell}\alpha_k^{0+}\alpha_k^{\ell+}-e^{-i\omega_k\ell}\alpha_k^{0-}\alpha_k^{\ell-}=0\]
      whose asymptotic behavior for $k\to\infty$, already mentioned on equation \eqref{eq cuerda_masas omega_m assimpt sin masas}, is given by
      \[\omega_k=\frac{\pi}{\ell}k+\frac{r_0+r_\ell}{k\pi}+O\left(\frac{1}{k^3}\right)\]
       
       Now, following the ideas developed in section \ref{Mathematical background - subsection - second quantization} of chapter \ref{Chapter - Mathematical background}, we can obtain the Bogoliubov coefficients. For that, notice first that we are dealing with the simple case where $\Sigma=I$ and $M=\R\times I$, then an embedding is given by $X=(\peqsub{t}{X},\peqsub{x}{X})$ and the normal vector field to $X(\Sigma)$ is given by \[\peqsub{\vec{n}}{X}=\frac{1}{\sqrt{\det\peqsubfino{\gamma}{X}{-0.2ex}}}\left(\frac{\mathrm{d}\peqsub{x}{X}}{\mathrm{d}\sigma},\frac{\mathrm{d}\peqsub{t}{X}}{\mathrm{d}\sigma}\right)\]
       This is so because
       \[\peqsubfino{\gamma}{X}{-0.2ex}:=X^*\peqsubfino{\eta}{\mathrm{Mkw}}{-0.2ex}=X^*(-\mathrm{d}t^2+\mathrm{d}x^2)=\left[-\left(\frac{\mathrm{d}\peqsub{t}{X}}{\mathrm{d}\sigma}\right)^2+\left(\frac{\mathrm{d}\peqsub{x}{X}}{\mathrm{d}\sigma}\right)^2\right]\mathrm{d}\sigma^2=\sqrt{\det\peqsub{\gamma}{X}}\mathrm{d}\sigma^2\]
       Using the definition of the Bogoliubov coefficients to this $1+1$ dimensional case, it is a long but straightforward computation to obtain that the Bogoliubov coefficients\index{Bogoliubov coefficients} are given by four integrals (coming from the product of two solutions, each one involving two exponentials, see \cite{margalef2017functional})
       \begin{align}\label{String masses - equation - beta}
       \begin{split}
       \beta^{IF}_{lm}&=\frac{\alpha^{0+}_l\alpha^{0+}_m}{c_lc_m}I_{lm}(\peqsub{t}{I},\peqsub{x}{I},\peqsub{t}{F},\peqsub{x}{F})+\frac{\alpha^{0+}_l\alpha^{0-}_m}{c_lc_m}J_{lm}(\peqsub{t}{I},\peqsub{x}{I},\peqsub{t}{F},\peqsub{x}{F})-\\
       &\phantom{=}-\frac{\alpha^{0-}_l\alpha^{0+}_m}{c_lc_m}J_{ml}(\peqsub{t}{F},\peqsub{x}{F},\peqsub{t}{I},\peqsub{x}{I})-\frac{\alpha^{0-}_l\alpha^{0-}_m}{c_lc_m}	I_{lm}(\peqsub{t}{I},-\peqsub{x}{I},\peqsub{t}{F},-\peqsub{x}{F})
       \end{split}
       \end{align}
       
       where
       \begin{align*}
       	I_{lm}(\peqsub{t}{I},\peqsub{x}{I},\peqsub{t}{F},\peqsub{x}{F})&=\int_\Sigma \Big[\omega_l(\peqsub{t'}{I}+ \peqsub{x'}{I})-\omega_m(\peqsub{t'}{F}+ \peqsub{x'}{F})\Big]e^{i\omega_l(\peqsub{t}{I}+\peqsub{x}{I})+i\omega_m(\peqsub{t}{F}+\peqsub{x}{F})}\mathrm{d}\sigma\\
       	J_{lm}(\peqsub{t}{I},\peqsub{x}{I},\peqsub{t}{F},\peqsub{x}{F})&=\int_\Sigma \Big[\omega_l(\peqsub{t'}{I}+ \peqsub{x'}{I})+\omega_m(\peqsub{t'}{F}- \peqsub{x'}{F})\Big]e^{i\omega_l(\peqsub{t}{I}+\peqsub{x}{I})+i\omega_m(\peqsub{t}{F}-\peqsub{x}{F})}\mathrm{d}\sigma
       \end{align*}
       The $J$ integrals can be computed directly as the integrand is of the form $e^{f(x)}f'(x)\mathrm{d}x$. The $I$ integrals cannot be computed directly but it would be enough to have an estimation. For that, following the ideas in \cite{torre1999functional}, we perform the change of variable $u(\sigma)=(1-\tau)(\peqsub{t}{I}+\peqsub{x}{I})+\tau(\peqsub{t}{F}+\peqsub{x}{F})$ where $\tau=\omega_m/(\omega_l+\omega_m)$, and integrate by parts. Thus we get
      \begin{align*}
      I_{lm}(\peqsub{t}{I},\peqsub{x}{I},\peqsub{t}{F},\peqsub{x}{F})&=-i\left[\frac{(\peqsub{t'}{I}+\peqsub{x'}{I})(1-\tau)-(\peqsub{t'}{F}+\peqsub{x'}{F})\tau}{(\peqsub{t'}{I}+\peqsub{x'}{I})(1-\tau)+(\peqsub{t'}{F}+\peqsub{x'}{F})\tau}e^{i\omega_l\peqsub{t}{I}+i\omega_m\peqsub{t}{F}}e^{i\omega_l\peqsub{x}{I}+i\omega_m\peqsub{x}{F}}\right]_0^{\ell}+O\left(\frac{1}{\omega_l+\omega_m}\right)
      \end{align*}
      Plugging these computations into equation \eqref{String masses - equation - beta}, using the decomposition $X_k=X_k^++X_k^-$ of the solution into positive and negative frequencies and taking into account that
      \[\mathrm{Im}\big[X^+_l(j)X^-_m(j)\big]=\frac{\omega_m-\omega_l}{\omega_m+\omega_l}\mathrm{Im}\big[X^+_l(j)X^+_m(j)\big]\qquad\qquad j\in\{0,\ell\}\]
      we obtain
      \begin{align*}
      \beta^{IF}_{lm}&=\frac{4\omega_l\omega_m}{(\omega_l+\omega_m)^2}\Bigg[\Big(\rule{0ex}{3ex}\mathrm{Im}\left[X^+_l(\peqsub{x}{I})X^+_m(\peqsub{x}{F})\right]\frac{N_I\omega_l-N_F\omega_m+A(\omega_m-\omega_l)}{\omega_m+\omega_l}-\\
      &\hspace*{17ex}-i \mathrm{Re}\left[X^+_l(\peqsub{x}{I})X^+_m(\peqsub{x}{F})\right] B\Big)f_\tau e^{i\omega_l\peqsub{t}{I}+i\omega_m\peqsub{t}{F}}\Bigg]_0^\ell+O\left(\frac{1}{\omega_m\omega_l}\right)
      \end{align*}
      where
      \begin{align*}
      	&\mathrlap{\frac{1}{f_\tau}=\Big(\peqsub{x'}{I}(1-\tau)+\peqsub{x'}{F}\tau\Big)^2-\Big(\peqsub{t'}{I}(1-\tau)+\peqsub{t'}{F}\tau\Big)^2>0}\\
      	&N_I=(\peqsub{x'}{I})^2-(\peqsub{t'}{I})^2>0&&
      	N_F=(\peqsub{x'}{F})^2-(\peqsub{t'}{F})^2>0\\
      	&B=\peqsub{t'}{I}\peqsub{x'}{F}-\peqsub{t'}{F}\peqsub{x'}{I}&&
      	A=\peqsub{x'}{I}\peqsub{x'}{F}-\peqsub{t'}{I}\peqsub{t'}{F}>0
      \end{align*}
      Notice that the square modulus $|\beta^{IF}_{lm}|^2$ has three parts, the square of the real part, the square of the imaginary part, and the cross term. The series of the last two terms are always convergent, while the first one diverges unless $B=0$ (details can be found in \cite{margalef2017functional}). Therefore, we have that
      \[\sum_{lm}|\beta^{IF}_{lm}|^2\]
      converges if and only if
      \[\frac{\peqsub{t'}{I}}{\peqsub{x'}{I}}=\frac{\peqsub{t'}{F}}{\peqsub{x'}{F}}\qquad\text{over the boundary}\]
      along the dynamics. This implies that the slope of the embeddings remains constant. In particular, we recover the result of \cite{torre1999functional} that with no boundary $\Sigma=\mathbb{S}^1$, the evolution is always unitary.\separ
      
      Summarizing, we have manage to obtain a characterization of the equivalence classes of space-like embeddings that admit unitary evolution. We see that they are labeled by the pair of values \[\left(\frac{t'(0)}{x'(0)},\frac{t'(\ell)}{x'(\ell)}\right)\]
      in the sense that for any two embeddings $X_1$, $X_2$ with this pair of slopes at the boundary, the field dynamics between $X_1(\Sigma)$ and $X_2(\Sigma)$ can be unitarily implemented. In particular,  notice that any inertial (free) observer is always labeled by the pair $(0,0)$. 

%% file: 4_parametrized_theories.tex
\chapter{Parametrized theories}\label{Chapter - Parametrized theories}\thispagestyle{empty}

\starwars[Darth Vader]{I find your lack of faith disturbing.}{A New Hope}

  \section{Introduction}
    In 1964 Paul A.M.\ Dirac delivered a four day long course at the Yeshiva University. Among other things he discussed the problem of having a preferential time in a relativistic theory and developed a method to solve it by considering all possible time parameters \cite{dirac1964lectures}.\separ

    Specifically, given an action\index{Action} $S:\mathcal{C}^1(I,Q)\rightarrow \R$  defined by a Lagrangian $L:TQ\to\R$ we have
    \begin{align*}
      S[q]&=\int_{t=t_i}^{t_f}  L\left(q(t),\frac{\mathrm{d}q}{\mathrm{d}t}(t)\right)\mathrm{d}t\overset{t=t(\tau)}{=}\int_{\tau=t^{-1}(t_i)}^{t^{-1}(t_f)}  L\left((q\smallcirc t)(\tau),\left[\frac{\mathrm{d}t}{\mathrm{d}\tau}\right]^{-1}\frac{\mathrm{d}(q\smallcirc t)}{\mathrm{d}\tau}(\tau)\right)\frac{\mathrm{d}t}{\mathrm{d}\tau}(\tau)\mathrm{d}\tau
    \end{align*}
    This motivates the definition of a new Lagrangian $L_{rep}:T(Q\times\R)\to\R$ that includes time as a canonical variable, and a new (reparametrized) action  $S_{rep}:\mathcal{C}^1(I,Q\times \R)\rightarrow \R$ given by
	\begin{align}\label{Parametrized CM - equation - parametrized particle Lagrangian}
	  L_{rep}\Big(q,t;v,w\Big)&=w\, L\left(q,\frac{v}{w}\right)\hspace*{12ex} S_{rep}[q,t]=\int_{\tau_0}^{\tau_f}  L_{rep}\left(q,t;\frac{\mathrm{d}q}{\mathrm{d}\tau},\frac{\mathrm{d}t}{\mathrm{d}\tau}\right)\mathrm{d}\tau
	\end{align}

	We have that this new action is invariant under time reparametrization
	\[S_{rep}[q\smallcirc T,t\smallcirc T]=S_{rep}[q,t]\qquad\qquad T\in\mathrm{Diff}(\R)\]
	and, as a consequence, there is a freedom to choose the time parameter $\tau$.\separ
	
	This idea works nicely for relativistic mechanics. However, if we want to use it in the framework of general relativity or deal with field theories, we should be much more careful. In order to have the possibility of dealing with arbitrary foliations or, better, select them dynamically, new ideas are necessary in order to be able to use the Hamiltonian framework. We have already revisited in section \ref{Mathematical background - section - GNH} of chapter \ref{Chapter - Mathematical background} the first working approach of this kind, developed by Dirac \cite{dirac1950generalized,dirac1964lectures}, as well as the extended approach developed by Gotay, Nester, and Hinds \cite{gotay1979presymplectic,gotay1980generalized,gotay1978presymplectic}.
  
  \section{What is a parametrized field theory?}
    The parametrization of a theory is a procedure to introduce diffeomorphism invariance when background geometric objects are present. When this is done, these objects vary in a very specific way, namely, by pullbacks through diffeomorphisms. For the sake of concreteness, consider $(M,g)$ a space-time, and $S:\Omega^k(M)\to\R$ the action given by
	\begin{equation}
	  S(\beta)=\int_M\beta\wedge\peqsub{\star}{g}\beta
	\end{equation}
	where $\peqsub{\star}{g}$ is the Hodge star operator associated with $g$, the sole background object. The parametrized theory is defined by the action\index{Action!Parametrized} $S_p:\Omega^k(M)\times\mathrm{Diff}(M)\to\R$ given by
	\begin{equation}\label{param action EM}
	  S_p(\beta,Z)=\int_M\beta\wedge\peqsub{\star}{Z^*\!g}\beta
	\end{equation}
  The action $S_p$ of a parametrized field theory is analogous to the aforementioned $S_{rep}$. Indeed, $S_{p}$ is invariant under diffeomorphisms
  \[S_{p}(Y^*\beta,Z\smallcirc Y)=S_{p}(\beta,Z)\qquad\qquad Y\in\mathrm{Diff}(M)\]
  allowing us to \emph{dynamically} choose the foliation.
	
  \section{Why parametrize field theories?}
	There is, certainly, some aesthetic pleasure in obtaining the dynamics of a given theory for any possible foliation. It is worth noticing that a parametrized theory is a generalization of the original one because $Z=\mathrm{Id}$ deparametrizes it. Besides, this type of theories have been used in the context of loop quantum gravity, where diffeomorphisms play a central role \cite{laddha2008polymer}. Nonetheless, there are at least two more fundamental reasons for studying parametrized field theories.\separ
	
	First of all, it allows any theory to be invariant under diffeomorphisms (gauge symmetry). So we have at our disposal a large amount of toy models for general relativity. Besides, a natural question comes up: is general relativity an already parametrized theory \cite{torre1992general}? If it were, it would be conceivable to deparametrize it and find a simpler version.\separ
	
	Secondly, and most important, by following the ideas partly developed by Isham and Kucha\v{r} \cite{isham1985representations,isham1985representations2}, one could try to apply algebraic methods to quantize theories similar to general relativity and, hopefully, learn something about the quantization of gravity. 


\section{Parametrized classical mechanics}

Before dealing with the general theory given by the action \eqref{param action EM} it is really worth it to study thoroughly the motivational theory: the so-called parametrized classical mechanics \eqref{Parametrized CM - equation - parametrized particle Lagrangian}.\separ

It is not hard to prove that \eqref{Parametrized CM - equation - parametrized particle Lagrangian} is a particular case of \eqref{param action EM} taking $k=0$, $\Sigma=\{\star\}$ to be just one point, and $g$ to be ``time-independent'' (meaning that $X^*\!g$ does not depend on the embedding), then we recover the parametrized version of the Lagrangian $\frac{mv^2}{2}$. Nonetheless, in this chapter we will work with the slightly more general Lagrangian 
\[\frac{mv^2}{2}-W(q)\]
for some potential $W$, because it adds almost no difficulty but is closer to the systems usually studied in classical mechanics books. Its parametrized version is given by
\begin{align*}
L(q,T\ ;\,v,\uptau)=\frac{mv^2}{2\uptau}-\uptau W(q)\qquad\qquad \mathcal{D}:=\Big\{(q,T\ ;\,v,\uptau)\in T(\mathcal{Q}\times\R)\ \ /\ \tau>0\Big\}
\end{align*}
where $\mathcal{Q}$ is the space of positions of the system. Notice that the time and its ``velocity'' can be reinterpreted in the embedding language once we realize that $\peqsub{V}{X}^{\scriptscriptstyle\perp}=\uptau\in\R$ and $X(\star)=t\in\R$.

\subsection{Variations of the action}\trassub

Let us compute the differential of the action associated with the previous Lagrangian. For that we take a curve $(q(s),T(s))$ with initial tangent vector $(\delta q,\delta T)$, then
\begin{align*}
\peqsub{\mathrm{d}}{(q,T)}S(\delta q,\delta T)&=\int\left[-\dot{T}W'(q)\delta q+0+\frac{m\dot{q}}{\dot{T}}\dot{\delta q}+\left(-\frac{m\dot{q}^2}{2\dot{T}^2}-W(q)\right)\!\dot{\delta T}\right]\mathrm{d}s=\\
&=\int\left[-\left(\frac{\mathrm{d}}{\mathrm{d}s}\frac{m\dot{q}}{\dot{T}}+\dot{T}W'(q)\right)\delta q+\frac{\mathrm{d}}{\mathrm{d}s}\left(\frac{m\dot{q}^2}{2\dot{T}^2}+W(q)\right)\!\delta T\right]\mathrm{d}s=\\
&=\int\left[-\left(\frac{\mathrm{d}}{\mathrm{d}s}\frac{m\dot{q}}{\dot{T}}+\dot{T}W'(q)\right)\delta q+\left(\frac{\dot{q}}{\dot{T}}\frac{\mathrm{d}}{\mathrm{d}s}\frac{m\dot{q}}{\dot{T}}+W'(q)\dot{q}\right)\!\delta T\right]\mathrm{d}s=\\
&=\int\left(\frac{\mathrm{d}}{\mathrm{d}s}\frac{m\dot{q}}{\dot{T}}+W'(q)\dot{T}\right)\left(\frac{\dot{q}}{\dot{T}}\delta T-\delta q\right)\mathrm{d}s
\end{align*}
which leads to the single equation
\begin{equation}\label{Parametrized CM - equation - Lagrangian equation}
m\dfrac{\mathrm{d}}{\mathrm{d}s}\left(\dfrac{\dot{q}}{\dot{T}}\right)+\dot{T}W'(q)=0
\end{equation}
Notice that, as expected for a parametrized theory, if $(q,T)$ is a solution to these equations then so is $(q\smallcirc\widetilde{T},T\smallcirc\widetilde{T})$ for every $\widetilde{T}\in\mathrm{Diff}(\R)$.\separ

Finally, notice that if we (un)do the change of variable $t=T(s)$ we have that $\dot{q}/\dot{T}=\frac{\mathrm{d}q}{\mathrm{d}t}$. Thus, the first equation becomes the classical equation of motion for a particle \[m\frac{\mathrm{d}^2q}{\mathrm{d}t^2}(t)+W'\big(q(t)\big)=0\]

\subsection{Fiber derivative}

We compute now the fiber derivative $F\!L:\mathcal{D}\subset T(\mathcal{Q}\times\R)\to T^*(\mathcal{Q}\times\R)$ which is given by:
\begin{align*}
\bullet\ \ F\!L&\left(\rule{0ex}{2.5ex}q,t\ ;\,v,\uptau\right)(q,t\ ;\,\bar{v},0)=\left.\frac{\mathrm{d}}{\mathrm{d}\lambda}\right|_{\lambda=0}L(q,t\ ;\,v+\lambda\bar{v},\uptau)=\\
&=\left.\frac{\mathrm{d}}{\mathrm{d}\lambda}\right|_{\lambda=0}\left(\frac{m(v+\lambda\bar{v})^2}{2\uptau}-\uptau W(q)\right)=\frac{mv}{\uptau}\bar{v}\\
\bullet\ \ F\!L&\left(\rule{0ex}{2.5ex}q,t\ ;\,v,\uptau\right)(q,t\ ;\,0,\bar{\uptau})=\left.\frac{\mathrm{d}}{\mathrm{d}\lambda}\right|_{\lambda=0}L(q,t\ ;\,v,\uptau+\lambda\bar{\uptau})=\\
&=\left.\frac{\mathrm{d}}{\mathrm{d}\lambda}\right|_{\lambda=0}\left(\frac{mv^2}{2(\uptau+\lambda\bar{\uptau})}-(\uptau+\lambda\bar{\uptau}) W(q)\right)=-\left(\frac{mv^2}{2\uptau^2}+W(q)\right)\bar{\uptau}
\end{align*}
that allows us to define the canonical momenta
\begin{align*}
&p=\frac{mv}{\uptau}\\
&\pi=-\mathcal{H}(q,p)\qquad\text{ where }\qquad\mathcal{H}(q,p):=\frac{p^2}{2m}+W(q)
\end{align*}
Notice that the equation for $\pi$ does not involve velocities, hence it defines a constraint on the cotangent bundle. The primary constraint submanifold is then $\peqsub{\mathcal{P}}{\mathrm{CM}}=F\!L(\mathcal{D})$ where
\[\peqsub{\mathcal{P}}{\mathrm{CM}}=\Big\{(q,t\ ;\,p,\pi)\in T^*\mathcal{D}\ \ /\ \ \pi=-\mathcal{H}(q,p)\Big\}\cong\left\{\left(q,t\ ;\,p\right)\right\}=:\mathcal{P}\]
We define the inclusion $\peqsubfino{\jmath}{\mathrm{CM}}{-0.2ex}:\mathcal{P}\hookrightarrow T^*\mathcal{D}$ that allows us to pullback the canonical symplectic form $\Omega$ to $\mathcal{P}$ in order to define $\omega:=\peqsub{\jmath}{\mathrm{CM}\,}^*\Omega$.
\begin{align}
&\bullet\ \ \Omega_{(q,t\, ;\,p,\pi)}=\mathrm{d}q\wedge \mathrm{d}p+\mathrm{d}t\wedge \mathrm{d}\pi\\[1.5ex]
\begin{split}&\bullet\ \ \omega_{(q,t,p)}=\mathrm{d}q\wedge \mathrm{d}p-\mathrm{d}t\wedge\mathrm{d}\mathcal{H}=\\
&\hspace*{10.47ex}=\mathrm{d}q\wedge \mathrm{d}p-\mathrm{d}t\wedge\left(\frac{p}{m}\mathrm{d}p+W'(q)\mathrm{d}q\right)
\end{split}\label{Parametrized CM - equation - omega}
\end{align}

\subsection{Hamiltonian Formulation}\label{Parametrized CM - section - Hamiltonian}
\subsubsection*{Obtaining the Hamiltonian}\trassub

The energy $E:T(\mathcal{Q}\times\R)\rightarrow \R$ is given by $E(q,t,v,\tau)=F\!L(q,t,v,\tau)(q,t,v,\tau)-L(q,t,v,\tau)$ and it has to be zero because the Lagrangian is homogeneous of degree $1$. This means that if we derive the equation $L(q,t;\lambda v,\lambda\uptau)=\lambda L(q,t;v,\uptau)$ with respect to $\lambda$ and evaluate at $\lambda=0$ we obtain $F\!L(q,t;v\uptau)(q,t;v\uptau)=L(q,t;v\uptau)$. Nonetheless, in this case it is as easy to obtain the same result by a direct computation.
\begin{align*}
E(q,t\ ;\,v,\uptau)&=F\!L\left(\rule{0ex}{2.5ex}q,t\ ;\,v,\uptau\right)(q,t\ ;\,v,\uptau)-L(q,t\ ;\,v,\uptau)=\\
&=\frac{mv}{\uptau}v-\left(\frac{mv^2}{2\uptau^2}+W(q)\right)\uptau-\frac{mv^2}{2\uptau}+\uptau W(q)=0
\end{align*}
Now the Hamiltonian is defined as the function $H:\mathcal{P}\rightarrow \R$ such that $H\smallcirc F\!L=E$, which here is clearly $H=0$. Notice that this does not imply that the dynamics is trivial, it does imply however that the dynamics is purely gauge in the sense that it goes along the degenerate direction of the presymplectic form $\omega$.

\subsubsection*{GNH algorithm}\trassub

Let us find the Hamiltonian vector field $Y\in\mathfrak{X}(\mathcal{P})$, which is of the form $Y=\vecc{Y}{q}\partial_q+\vecc{Y}{p}\partial_p+Y_t\partial_t$, that solves the equation $\peqsub{\imath}{Y}\omega=\mathrm{d}H=0$. Using the explicit expression given by \eqref{Parametrized CM - equation - omega} we have
\begin{align*}
0&=\peqsub{i}{Y}\omega\overset{\eqref{Appendix - equation - regla de Leibniz producto interior}}{=}\vecc{Y}{q}\Big(\mathrm{d}p+W'(q)\mathrm{d}t\Big)+Y_t\left(-\frac{p}{m}\mathrm{d}p-W'(q)\mathrm{d}q\right)+\vecc{Y}{p}\left(-\mathrm{d}q+\frac{p}{m}\mathrm{d}t\right)=\\
&=\left(\vecc{Y}{q}-\frac{p}{m}Y_t\right)\Big(\mathrm{d}p+W'(q)\mathrm{d}t\Big)-\Big(\vecc{Y}{p}+W'(q)Y_t\Big)\left(\mathrm{d}q-\frac{p}{m}\mathrm{d}t\right)
\end{align*}

Thus we obtain
\[\vecc{Y}{q}=\frac{p}{m}Y_t\qquad \qquad \vecc{Y}{p}=-W'(q)Y_t\qquad\qquad Y_t\ \text{ arbitrary}\]

This vector field is defined over $\mathcal{P}$ but it can be lifted to a vector field tangent to $\peqsub{\mathcal{P}}{\mathrm{CM}}\subset T^*(\mathcal{Q}\times\R)$ (denoted again by $Y$), hence now it has one more component $Y_\pi$. A simple computation shows that a vector field $Y=\vecc{Y}{q}\partial_q+Y_t\partial_t+\vecc{Y}{p}\partial_p+Y_\pi\partial_\pi$ is tangent to $\peqsub{\mathcal{P}}{\mathrm{CM}}$ if and only if 
\[Y_\pi=-\frac{p}{m}\vecc{Y}{p}-W'(q)\vecc{Y}{q}\]
This component is zero if we plug in the corresponding components $\vecc{Y}{q},\vecc{Y}{p}$ of the Hamiltonian vector field $Y$.

\subsubsection*{GNH algorithm revisited}\trassub

The key idea of the previous computation was to take advantage of the fact that we can write the symplectic form in some coordinates. We will see in the following chapter that this is not so for the infinite dimensional case. That is why we consider that it will be very useful to take a small detour and obtain the previous Hamiltonian vector field in an alternative way that can be easily implemented in the following chapters.\separ

We start by studying the phase space where the dynamics takes place. A typical point of the phase space $T^*\mathcal{D}$ is of the form \[\boldsymbol{\mathrm{p}}_{(q,t)}=(q,t;,\boldsymbol{p},\boldsymbol{\pi}) \in T^*\mathcal{D} \]
where  $\boldsymbol{p}\in \mathcal{Q}'$ and $\boldsymbol{\pi}\in \R'$ are elements of the dual of $\mathcal{Q}$ and $\R$ respectively. The phase space $T^*\mathcal{D}$ is equipped with the symplectic form
\begin{align*}
\Omega_{\boldsymbol{\mathrm{p}}_{(q,t)}}(Y,Z)
&=\Omega_{(q,t;,\boldsymbol{p},\boldsymbol{\pi})}\left(\rule{0ex}{3ex}(\vecc{Y}{q},\vecc{Y}{t},,\vecc{\boldsymbol{Y}}{\!\!\!p},\peqsub{\boldsymbol{Y}}{\!\!\pi}),(\vecc{Z}{q},\vecc{Z}{t},\vecc{\boldsymbol{Z}}{p},\peqsub{\boldsymbol{Z}}{\pi})\right)\overset{\eqref{eq:background canonical symplectic form}}{=}\\
&=\vecc{\boldsymbol{Z}}{p}\!\left(\vecc{Y}{q}\right)-\vecc{\boldsymbol{Y}}{\!\!\!p}\!\left(\vecc{Z}{q}\right)+\peqsub{\boldsymbol{Z}}{\pi}\!\left(\vecc{Y}{t}\right)-\peqsub{\boldsymbol{Y}}{\!\!\pi}\!\left(\vecc{Z}{t}\right)
\end{align*}
for $Y,Z\in\mathfrak{X}(T^*\mathcal{D})$ where we have omitted the base point for simplicity. In particular we have $\vecc{Y}{q}\in T\mathcal{Q}$, $\vecc{Y}{t}\in \R$, $\vecc{\boldsymbol{Y}}{\!\!\!p}\in T^*\!\mathcal{Q}$ and $\peqsub{\boldsymbol{Y}}{\!\!\pi}\in\R'$, and analogously for the $Z$ components.\separ

In our case at hand, we have that $\boldsymbol{p}$ and $\boldsymbol{\pi}$ can be written, via the Riesz representation theorem, in terms of some $p\in T\mathcal{Q}$ and $\pi\in\R$ as
\begin{align}\label{Parametrized CM - equation - p=p. and pi=pi.}
&\boldsymbol{p}(v)=p\cdot{}v&&\boldsymbol{\pi}(\tau)=\pi\cdot{}\tau
\end{align}
with the usual dot product of $\R^n$ and $\R$ respectively. In the infinite dimensional case we would have that only some elements can be represented like that but, lucky enough, those are the ones that we will be interested in.\separ

Let $\widetilde{\mathcal{P}}=\{\peqsub{\mathrm{p}}{\mathrm{q}}:=(q,t,p,\pi)\}\overset{\jmath}{\hookrightarrow }T^*\mathcal{D}$ considered as a subset of $T^*\mathcal{D}$ via the previous representations (in this case we have an equivalence thanks to the Riesz representation theorem). We define the induced form $\widetilde{\Omega}:=\jmath^*\Omega$ given by  $\widetilde{\Omega}(Y,Z)=(\Omega\smallcirc\jmath)(\jmath_*Y,\jmath_*Z)$. In order to obtain its explicit expression we compute first $\jmath_*Y\in\mathfrak{X}^{\scriptscriptstyle \top}_\jmath\!(\mathcal{D})$ for any $Y=(\vecc{Y}{q},\vecc{Y}{t},\vecc{Y}{p},\peqsub{Y}{\pi})\in\mathfrak{X}(\widetilde{\mathcal{P}})$. From \eqref{Parametrized CM - equation - p=p. and pi=pi.} we obtain
\[\jmath_*Y=	\big(\vecc{Y}{q}\coma\vecc{Y}{t}\coma\vecc{Y}{p}\cdot{}\coma\peqsub{Y}{\pi}\cdot{}\big)\in\mathfrak{X}(T^*\mathcal{D})\]
so we have
\begin{align}\label{Parametrized CM - equation - tilde Omega}
\widetilde{\Omega}_{\peqsub{\mathrm{p}}{\mathrm{q}}}(Y,Z)&=\vecc{Z}{p}\cdot{}\vecc{Y}{q}-\vecc{Y}{p}\cdot{}\vecc{Z}{q}+\peqsub{Y}{\!\pi}\cdot{}\vecc{Y}{t}-\peqsub{Y}{\! \pi}\cdot{}\vecc{Z}{t}
\end{align}

To obtain $\omega:=\peqsub{\jmath}{\mathrm{CM}\,}^*\widetilde{\Omega}$ (see equation \eqref{Parametrized CM - equation - omega}) we compute first the push-forward $(\peqsubfino{\jmath}{\mathrm{CM}}{-0.2ex})_*Y$ for every $Y=(\vecc{Y}{q},\vecc{Y}{t},\vecc{Y}{p})\in\mathfrak{X}(\mathcal{P})$ because, by definition, $\omega(Y,Z):=(\widetilde{\Omega}\smallcirc \peqsubfino{\jmath}{\mathrm{CM}}{-0.2ex})((\peqsubfino{\jmath}{\mathrm{CM}}{-0.2ex})_*Y,(\peqsubfino{\jmath}{\mathrm{CM}}{-0.2ex})_*Z)$. It is easy to obtain
\[(\peqsubfino{\jmath}{\mathrm{CM}}{-0.2ex})_*Y=\left(\vecc{Y}{q}\coma\vecc{Y}{t}\coma\vecc{Y}{p}\coma-\parc[\mathcal{H}]{q}\vecc{Y}{q}-\parc[\mathcal{H}]{t}\vecc{Y}{t}-\parc[\mathcal{H}]{p}\vecc{Y}{p}\right)\in\mathfrak{X}(\mathcal{P})\]	
Thus
\begin{align}\label{Parametrized CM - equation - omega final}
\begin{split}
&\omega_{(q,t,p)}\Big((\vecc{Y}{q},\vecc{Y}{t},\vecc{Y}{p}),(\vecc{Z}{q},\vecc{Z}{y},\vecc{Z}{p})\Big)
=\\
&\qquad=\vecc{Y}{q}\cdot{}\vecc{Z}{p}-\vecc{Z}{q}\cdot{}\vecc{Y}{p}-\left(W'(q)\vecc{Z}{q}+\frac{p}{m}\vecc{Z}{p}\right)\vecc{Y}{t}+\left[W'(q)\vecc{Y}{q}+\frac{p}{m}\vecc{Y}{p}\right]\vecc{Z}{t}=\\
&\qquad=\left(\vecc{Y}{q}-\frac{p}{m}\vecc{Y}{t}\right)\cdot{}\vecc{Z}{p}-\left(\vecc{Y}{p}+W'(q)\vecc{Y}{t}\right)\cdot{}\vecc{Z}{q}+\left[W'(q)\left(\vecc{Y}{q}-\frac{p}{m}\vecc{Y}{t}\right)+\frac{p}{m}\left(\vecc{Y}{p}+W'(q)\vecc{Y}{t}\right)\right]\vecc{Z}{t}
\end{split}
\end{align}

The Hamiltonian vector field $Y\in\mathfrak{X}(\mathcal{P})$ is given by the Hamilton equation
\begin{equation}\label{Parametrized CM - equation - Hamiltonina equation =0}
\omega_{(q,t,p)}\Big((\vecc{Y}{q},\vecc{Y}{t},\vecc{Y}{p})\coma(\vecc{Z}{q},\vecc{Z}{t},\vecc{Z}{p})\Big)=\mathrm{d}H(\vecc{Z}{q},\vecc{Z}{t},\vecc{Z}{p})=0\end{equation}
for every $Z\in\mathfrak{X}(\mathcal{P})$. From \eqref{Parametrized CM - equation - omega final} it is clear that we obtain again
\begin{align}\label{Parametrized CM - equation - Hamiltonian vector field}
\vecc{Y}{q}=\frac{p}{m}\vecc{Y}{t}\qquad\qquad\qquad \vecc{Y}{p}=-W'(q)\vecc{Y}{t}\qquad\qquad\qquad \vecc{Y}{t}\ \text{ arbitrary}
\end{align}

\subsubsection*{Gauge orbits}\trassub

We mentioned at the beginning of section \ref{Parametrized CM - section - Hamiltonian} that the fact that $H=0$ does not imply that the dynamics is trivial, because we can have pure gauge dynamics i.e.\ evolution in the degenerate directions of the presymplectic form. Nonetheless, some of the solutions obtained in that way might be physically equivalent.\separ

A \textbf{gauge orbit}\index{Gauge orbit} of a given solution is the space of all their physically equivalent solutions. Usually they are not easy to handle or they are not convenient because some nice properties are lost, that is why we have not paid too much attention to this issue. However, in this simple example, it is doable and it will also be useful to gain some intuition for more general settings.

To begin with, notice that the most general solution to the Hamiltonian equation is given by \eqref{Parametrized CM - equation - Hamiltonian vector field} which can be rewritten as
\begin{align*}
Y&=\frac{p}{m}Y_t\partial_q-W'(q)Y_t\partial_p+Y_t\partial_t=\\
&=Y_t\left(\frac{p}{m}\partial_q-W'(q)\partial_p+\partial_t\right)
\end{align*}
so we see then that the vector field
\[Y^1=\frac{p}{m}\partial_q-W'(q)\partial_p+\partial_t\]
generates all the Hamiltonian vector fields because any other will be of the form $Y^N=NY^1$. Here we take $N:=Y_t$ to simplify the notation but also to stress that it is analogous to the lapse. Notice that $Y^1$ and $Y^N$ are collinear with $N$ as the proportional factor. Thus the integral curves of $Y^N$ are ``contained'' in the ones of $Y^1$ although they might not be the same if, for instance, $N$ vanishes because in that case an orbit of $Y^1$ would be cut into several orbits for $Y^N$. We can say that $Y^1$ is maximal because its orbits are maximal among all possible Hamiltonian vector fields.\separ

The idea then is to identify the points that lay in the same orbit for every possible choice of $N=\vecc{Y}{t}$. As $Y^1$ is maximal, it is enough to identify the points connected by the flow $\Phi^1:D\subset I\times \mathcal{P}\rightarrow \mathcal{P}$ of $Y^1$. Thus we consider the equivalence relation
\[(q_0,p_0,\tau_0)\sim (q_1,p_1,\tau_1) \qquad \Longleftrightarrow \qquad \exists s\in\R\ \ \/\ \ \Phi^1\Big(s,(q_0,p_0,\tau_0)\Big)=(q_1,p_1,\tau_1)\]With this characterization it is clear that the \textbf{reduced phase space}\index{Reduced phase space}, defined as the quotient space $\mathfrak{p}:=\mathcal{P}/\!\!\sim\,$, is the set of orbits of the vector field $Y^1$.

\subsubsection*{Reduced phase space}\trassub

Once we have the reduced phase space, we would like to recover the non-gauge dynamics. For that we need to define a symplectic form and a Hamiltonian. Notice first that $H$ is always zero over $\mathcal{P}$, thus it is well defined over the quotient space and, of course, also vanishes. This implies that if we define a true symplectic form, there will be no dynamics over the reduced phase space. Unfortunately, there is no natural way to define a symplectic form over the quotient space. Despite this fact, having a vanishing Hamiltonian, all that matters is the existence of any non-degenerate symplectic form because then the Hamiltonian vector field has to be zero.\separ

In order to prove the existence of one symplectic form we assume the physical hypothesis that $Y^1$ is complete (otherwise the system reaches the infinity in a finite amount of time). This is necessary to ensure that all the orbits exist forever and thus, they all cross every hypersurface $\mathcal{P}_\tau=\mathcal{P}\cap\{t=\tau\}$. This is what is known as the \textbf{gauge fixing} i.e.\ we select a surface containing one and only one representative of each class. In particular we have
\[\mathfrak{p}\ \ \cong\ \ \mathcal{P}_\tau \ \ \cong\ \ \R^2\]

In fact, we can define on $\mathfrak{p}$ two global coordinates $(q_\tau,p_\tau)$ such that for any orbit $c\in\mathfrak{p}$ we have $\{(q_\tau(c),p_\tau(c),\tau)\}:=c\,\cap\mathcal{P}_\tau$. It will be useful to introduce the following maps
\[\begin{array}{cccc}
Z_\tau: & \mathfrak{p} & \longrightarrow & \mathcal{P}\\
&     c        &    \longmapsto  &  \big(q_\tau(c),p_\tau(c),\tau\big)
\end{array}\qquad\qquad
\begin{array}{cccc}
z_\tau: & \mathfrak{p} & \longrightarrow & \R^2\\
&     c            &    \longmapsto  &  \big(q_\tau(c),p_\tau(c)\big)
\end{array}\]

Finally the pullback of $\omega$ through the map $\mathfrak{p}\cong\mathcal{P}_\tau\hookrightarrow \mathcal{P}$ reads simply $\omega_\tau=\mathrm{d}q_\tau\wedge \mathrm{d}p_\tau$, showing by the way that these coordinates are canonical, and we can thus endow the quotient (in a non-canonical way) with this symplectic form. The initial data for the theory would be then $(\mathfrak{p},\omega_\tau,H=0)$. Alternatively we can consider $\mathcal{P}_\tau$ for any $\tau\in\R$, then the theory is given by $(\mathcal{P}_\tau,\omega_\tau,H=0)$.

\subsubsection*{Observables}\trassub

An observable of the reduced space phase $\mathfrak{p}$ is a map $F\in\mathcal{C}^\infty(\mathfrak{p})$. In order to follow the usual quantization schemes, we need to polarize the space into positions and momenta and, if possible, with canonical coordinates. As we have seen, we have at our disposal the canonical coordinates given by $z_\tau:\mathfrak{p}\to\R^2$. So if some observer considers such coordinates for a fixed $\tau_0$, its reduced phase space would be $\R^2$ and he could measure the observable $f_{\tau_0}:=F\smallcirc z_{\tau_0}^{-1}\in\mathcal{C}^\infty(R^2)$. Letting $\tau_0$ vary we obtain a $1$-parameter family $\{f_\tau\}_\tau$ of observables each associated to a different observer and in general $f_{\tau_1}\neq f_{\tau_2}$ for $\tau_1\neq\tau_2$. It may happen, however, that for some observable $F$, we have $f_\tau$ independent of $\tau$.\separ

In particular it is clear that the constants of motion of the un-parametrized theory (maps that are constants along the integral curves) allow us to construct well defined observables in the parametrized setting. For instance
\[h_\tau(q_\tau,p_\tau):=h(q_\tau,p_\tau)\qquad\text{ where }\qquad h(q,p)=\frac{p^2}{2m}+W(q)\]
allows us to obtain a well defined observable of $\mathfrak{p}$ given by 
\[H(c)=h\smallcirc z_\tau(c)=\frac{p^2_\tau}{2m}+W(q_\tau)\]
As the function $h$ is nothing but the energy of $y^1=\frac{p}{m}\partial_q-W'(q)\partial_p$, then it is indeed constant along the orbits of $y^1$ so $H$ is well defined. On the other hand if we consider $f=\pi_q$, the projection over the first component, then $f_\tau(q_\tau,p_\tau)=q_\tau$ is an observable for a $\tau$-observer that clearly depends on the chosen $\tau$.

\subsection{Symplectic-Lagrangian Formulation}

We mentioned on section \ref{Mathematical background - subsection - Lagrangian framework} of chapter \ref{Chapter - Mathematical background} that we can pull-back the symplectic form of the cotangent bundle to the tangent bundle and solve the symplectic-Lagrangian equation \eqref{Mathematical background - equation - symplectic lagrangian equation}. It is quite long to develop the symplectic-Lagrangian theory in a general setting like the one we introduced in the previous chapter, however, for this simpler case it is worth to do the effort to show some of the similarities and differences with the Hamiltonian framework.

\subsubsection*{GNH Algorithm}\trassub

We define the $2$-form defined as the pullback of the canonical symplectic form through the fiber derivative $\varpi:=F\!L^*\Omega$. From the definition of the momenta $p$ and $\pi$ we obtain
\begin{align*}
\varpi:&=\mathrm{d}q\wedge \mathrm{d}\!\left(\frac{mv}{\uptau}\right)-\mathrm{d}t\wedge \mathrm{d}\!\left(\frac{m v^2}{2\uptau^2}+W(q)\right)=\\
&=\mathrm{d}q\wedge \left(\frac{m}{\uptau}\mathrm{d}v-\frac{mv}{\uptau^2}\mathrm{d}\uptau\right)-\mathrm{d}t\wedge \left(\frac{m v}{\uptau^2}\mathrm{d}v-\frac{m v^2}{\uptau^3}\mathrm{d}\uptau+W'(q)\mathrm{d}q\right)
\end{align*}

We have to solve \eqref{Mathematical background - equation - symplectic lagrangian equation} that reads $\imath_Y\varpi=\mathrm{d}E=0$:
\begin{align*}
\imath_Y\varpi&=\vecc{Y}{q}\frac{m}{\uptau}\left(\mathrm{d}v-\frac{v}{\uptau}\mathrm{d}\uptau\right)+\frac{m}{\uptau}\left(-\vecc{Y}{v}+\frac{v}{\uptau}\peqsub{Y}{\uptau}\right)\mathrm{d}q-\vecc{Y}{t} \left(\frac{m v}{\uptau^2}\mathrm{d}v-\frac{m v^2}{\uptau^3}\mathrm{d}\uptau+W'(q)\mathrm{d}q\right)+\\
&\ \phantom{=}+\mathrm{d}t \left(\frac{m v}{\uptau^2}\vecc{Y}{v}-\frac{m v^2}{\uptau^3}\peqsub{Y}{\uptau}+W'(q)\vecc{Y}{q}\right)=\\
&=-\frac{m}{\uptau}\left(\vecc{Y}{v}-\frac{v}{\uptau}\peqsub{Y}{\uptau}+\frac{\uptau}{m}W'(q)\vecc{Y}{t}\right)\!\left(\mathrm{d}q-\frac{ v}{\uptau}\mathrm{d}t\right)+\frac{m}{\uptau}\left(\vecc{Y}{q}-\frac{v}{\uptau}\vecc{Y}{t}\right)\!\left(\mathrm{d}v-\frac{v}{\uptau}\mathrm{d}\uptau+W'(q)\mathrm{d}t\right)
\end{align*}
Hence we obtain the following solutions
\begin{align*}
&\vecc{Y}{q}=\frac{v}{\uptau}\vecc{Y}{t}\\
&\vecc{Y}{v}=\frac{v}{\uptau}\peqsub{Y}{\uptau}-\frac{\uptau}{m}W'(q)\vecc{Y}{t}\\
&\vecc{Y}{t}\quad\text{arbitrary}\\
&\peqsub{Y}{\uptau}\quad\text{arbitrary}
\end{align*}
where the only constraint comes from the domain of the Lagrangian that imposes that $\uptau>0$. Of course this result is consistent with the previous ones, as $F\!L_*\peqsub{X}{L}=\peqsub{X}{H}$.\separ

Notice that the previous solution, two linear equations for four unknowns, are not of second order as in the Lagrangian formulation. In general such conditions, that here reduces to have $\vecc{Y}{q}=v$ and $\vecc{Y}{t}=\uptau$, are necessary to recover the Euler-Lagrange equations. We could impose them but, in the parametrized case, there is no need as we recover them anyway taking into account that only the quotient $\vecc{Y}{q}/\vecc{Y}{t}$ appears in the Euler-Lagrange equation \eqref{Parametrized CM - equation - Lagrangian equation}, and such quotient is of ``second order''.
\[\frac{\vecc{Y}{q}}{\vecc{Y}{t}}=\frac{v}{\uptau}\quad\longlongrightarrow{3}\quad m\frac{\mathrm{d}}{\mathrm{d}s}\left(\frac{\dot{q}}{\dot{t}}\right)+W'(q)\dot{t}=m\frac{\mathrm{d}}{\mathrm{d}s}\left(\frac{v}{\uptau}\right)+W'(q)\vecc{Y}{t}=\frac{m}{\uptau}\left(\vecc{Y}{v}-\frac{v}{\uptau}\peqsub{Y}{\uptau}+\frac{\uptau}{m}W'(q)\vecc{Y}{t}\right)=0\]

%% file: 5_parametrized_EM.tex
\chapter{Parametrized electromagnetism}\label{Chapter - Parametrized EM}

\starwars[Obi-Wan Kenobi]{Your eyes can deceive you. Don’t trust them.}{A New Hope}

  \section{Introduction}\label{Parametrized theories - section - introduction}
  
In the previous chapter we have introduced the parametrized field theories and mentioned that they provide us with interesting examples of relatively simple diff-invariant models. In fact they have been used as a test bed to understand the quantization of general relativity and related theories \cite{isham1985representations,isham1985representations2,kuchavr1971canonical,villasenor2010quantization,hajivcek1996symplectic}.
One of our initial motivations was to study the interplay between ordinary gauge symmetries and diffeomorphism invariance. In the canonical treatment of general relativity the standard use of projections onto Cauchy surfaces gives rise to the so-called hypersurface deformation algebra\index{Hypersurface deformation algebra}, given by \eqref{Mathematical background - equation - hypersurface deformation algebra}, that replaces the algebra of four-dimensional diffeomorphisms. Many of the difficulties encountered in the quest for a quantum theory of gravity have their origin in the fact that the hypersurface deformation algebra is very hard to quantize.\separ

To tackle this problem Isham and Kucha\v{r} \cite{isham1985representations,isham1985representations2} proposed an approach that, in the case of the scalar field, led to the recovery of the full Lie algebra of four dimensional diffeomorphisms in terms of the Poisson brackets of some functions defined in the full phase space (i.e. not only on the primary constraint submanifold in phase space). However, it is not straightforward how to extend their procedure to parametrized gauge theories due to the non-trivial role played by gauge symmetries in this framework \cite{kuchar1987canonical,stone1992representation,rosenbaum2008space}. It was in fact suggested by Torre \cite{torre1992covariant} that such task should require a different approach. We obtained in \cite{margalef2016hamiltonianEM} that in fact the same methods we developed in \cite{margalef2016hamiltonian} sufficed to understand parametrized electromagnetism (EM) and found that the Gauss law plays a special role to recover the Dirac hypersurface deformation algebra. Precisely, this chapter is devoted to explain this result by developing the parametrized theory of EM.

\section{Action of the theory}\label{Parametrized theories - section - action}
	
	Let $(M\cong\R\times\Sigma,g)$ be a globally hyperbolic space-time of dimension $n$. Let us consider, for some $k\in\{0,\ldots,n\}$, the actions\index{Action!Electromagnetism}\footnote{It is customary to call EM theory just to the case $k=1$ but we will allow a generic $k\in\N$ as it hardly changes the computations.} $\peqsub{S}{\mathrm{d}},\peqsub{S}{\delta},\peqsub{S}{\square},\peqsub{S}{\partial}:\Omega^k(M)\times\mathrm{Diff}(M)\to\R$ given by
	\begin{align*}
	  &\peqsub{S}{\mathrm{d}}(A,Z)=\frac{\varepsilon}{2}\int_M\mathrm{d}A\wedge\peqsubfino{{\star_g}}{\!Z}{-0.2ex}\mathrm{d}A&&\peqsub{S}{\partial}(A,Z)=\frac{1}{2}\int_{\partial M}\peqsub{b}{Z}^2\,\peqsub{A}{\partial}\wedge\peqsubfino{{\star_{g^{\scriptscriptstyle\partial}}}}{\!\!\!\!\!Z}{-0.2ex}\,\peqsub{A}{\partial}\\[1.5ex]
	  &\peqsub{S}{\delta}(A,Z)=\frac{\varepsilon}{2}\int_M\delta A\wedge\peqsubfino{{\star_g}}{\!Z}{-0.2ex}\delta A
	  && \peqsub{S}{\square}(A,Z)=\peqsub{S}{\mathrm{d}}(A,Z)+\peqsub{S}{\delta}(A,Z)
	\end{align*}
	where $\peqsub{A}{\partial}=\peqsub{\jmath}{\partial}^*A$, $\peqsubfino{g}{Z}{-0.2ex}=Z^*\!g$, $\peqsubfino{g}{Z}{-0.2ex}^\partial=Z^*\peqsub{\jmath}{\partial}^*g$ and $\peqsub{b}{Z}=B\smallcirc Z$ for some fixed $B\in\Cinf{\partial M}$.  Notice that for $0$-forms $\peqsub{S}{\delta}=0$ and for $n$-forms $\peqsub{S}{\mathrm{d}}=0=\peqsub{S}{\partial}$. Let us mention again that all the previous actions are invariant under diffeomorphisms
	\[S(Y^*\!A,Z\smallcirc Y)=S(A,Z)\]
	Besides, as $\mathrm{d}^2=0$ and $\delta^2=0$, we have that
	\[\peqsub{S}{\mathrm{d}}(A+\mathrm{d}f,Z)=\peqsub{S}{\mathrm{d}}(A,Z)\qquad\qquad \peqsub{S}{\delta}(A+\delta\beta,Z)=\peqsub{S}{\delta}(A,Z)\]
	However, no such gauge symmetry exists for $\peqsub{S}{\partial}$. This boundary action, studied in the context of condensed matter \cite{balachandran1994maxwell,balachandran1996edge,balachandran1995edge}, allows us to consider some Robin-like boundary conditions although it comes with some difficulties that will be explained throughout this chapter (to be compared with section \ref{Applications - Section - MCS boundaries} of chapter \ref{Chapter - parametrized MCS}).\separ
	
	Let us now consider
	\begin{align*}
	  &\Omega^k_\partial(M)=\Big\{A\in\Omega^k(M)\ /\ \peqsub{\jmath}{\partial}^*\!\left(A\right)=0\Big\}&&\text{Dirichlet boundary conditions}\\
	  &\Omega^k_{\partial\star}(M)=\Big\{A\in\Omega^k(M)\ /\ \peqsub{\jmath}{\partial}^*\!\left(\peqsubfino{{\star_g}}{\!Z}{-0.2ex}A\right)=0\Big\}&&\text{Neumann boundary conditions}
	\end{align*}
	In the following we will denote with a superscript $\partial$ and $\partial\star$ the restrictions to these subspaces. Notice, in particular, that $S^\partial_\partial=0$.\separ
	
	Taking into account that, on one hand, $\peqsub{{\star_g}}{Z}$ establishes the isomorphism $\Omega^k_\partial(M)\cong\Omega^{n-k}_{\partial\star}(M)$ and, on the other, the fact that
	\[\peqsub{S}{\delta}(A,Z)\updown{\eqref{Appendix - definition - delta=star d star}}{\eqref{appendix property star delta=d star}}{=}\varepsilon\peqsub{S}{\mathrm{d}}(\peqsubfino{{\star_g}}{\!Z}{-0.2ex}A,Z)\]
	from now we will only focus on the actions $\peqsub{S}{\mathrm{d}}$, $\peqsub{S}{\partial}$ and their corresponding restrictions to $\Omega^k_\partial(M)$.
	
    \subsection*{Variations of the action}\trassub
    
      Given $(A,\peqsub{V}{\!A})\in \peqsub{T}{A}\Omega^k(M)=\Omega^k(M)\times \Omega^k(M)$, using equation \eqref{appendix equation integracion por partes} it is easy to obtain
        \begin{align*}
          \peqsub{D}{\!(A,\peqsub{V}{\!A})}(\peqsub{S}{\mathrm{d}}+\peqsub{S}{\partial})(A,Z)&=\varepsilon \int_M\peqsubfino{{\big\langle\hspace*{0.1ex}\peqsub{V}{\!A}\hspace*{0.3ex},\hspace*{0.05ex}\peqsub{\delta}{\!Z}  \mathrm{d}A\big\rangle_{\!\!g}}}{\!Z}{-0.2ex}\peqsubfino{{\mathrm{vol}_g}}{\!Z}{-0.2ex}+\int_{\partial M}\big\langle\hspace*{0.1ex} \peqsub{\jmath}{\partial}^*\peqsub{V}{\!A}\hspace*{0.3ex},\hspace*{0.05ex}\peqsub{\jmath}{\partial}^*(\varepsilon\peqsubfino{{\imath_{\vec{\nu}}}}{\!Z}{-0.3ex}\mathrm{d}A+\peqsub{b}{Z}^2A)\hspace*{0.1ex}\big\rangle_{\!g^\partial_Z}\mathrm{vol}_{g^\partial_Z}
        \end{align*}        
        where $\peqsub{\vec{\nu}}{Z}$ is the unitary $\peqsubfino{g}{\!Z}{-0.2ex}$-normal vector field to the boundary $\partial M$. It is easy to check that $\peqsub{\vec{\nu}}{Z}=(Z^{-1})_*\vec{\nu}$ where $\vec{\nu}$ is the unitary $g$-normal vector field.\separ
        
        Now we compute the variation with respect to the diffeomorphisms. Applying lemma \ref{appendix lemma variacion embedding energia cinetica} to the bulk and to the boundary, we have that the variation in the direction of $\peqsub{\mathbb{V}}{\!Z}\in \peqsub{T}{Z}\mathrm{Diff}(M)=\mathfrak{X}(M)$ (where we denote $\peqsub{\mathbb{V}}{\!Z}^\partial$ the vector field over the boundary such that $(\peqsub{\jmath}{\partial})_*\peqsub{\mathbb{V}}{\!Z}^\partial=\peqsub{\mathbb{V}}{\!Z}$) is given by 
\begin{align*}
&\peqsub{D}{(Z,\mathbb{V}_{\!Z})}(\peqsub{S}{\mathrm{d}}+\peqsub{S}{\partial})(A,Z)=-\varepsilon\!\int_M \peqsubfino{{\big\langle\hspace*{0.1ex}\peqsubfino{{\mathcal{L}_{\vec{\mathbb{V}}}}}{\!Z}{-0.2ex}\mathrm{d}A\hspace*{0.3ex},\hspace*{0.05ex}\mathrm{d}A\hspace*{0.1ex}\big\rangle_{\!g}}}{\!Z}{-0.2ex}\peqsub{{\mathrm{vol}_g}}{\!Z}-\!\int_{\partial M}\peqsubfino{{\big\langle \hspace*{0.1ex}\peqsubfino{{\mathcal{L}_{\vec{\mathbb{V}}^\partial}}}{\!\!\!\!\!Z}{-0.4ex}\,\peqsub{A}{\partial}\hspace*{0.3ex},\hspace*{0.05ex}\peqsub{b}{Z}^2\peqsub{A}{\partial}\hspace*{0.1ex}\big\rangle_{\!\!g^\partial}}}{\!\!\!\!\!Z}{-0.3ex}\,\mathrm{vol}_{g^\partial_{\!Z}}\updown{\eqref{appendix - equation - d^2=0}}{\eqref{appendix - formula - L=di+id}\eqref{appendix equation integracion por partes}}{=}\\
&=\!-\varepsilon\!\int_M\!\peqsubfino{{\big\langle\hspace*{0.1ex}\peqsub{\imath}{\mathbb{V}_{\!Z}} \mathrm{d}A\hspace*{0.3ex},\hspace*{0.05ex}\peqsub{\delta}{\!Z} \mathrm{d}A\big\rangle_{\!\!g}}}{\!Z}{-0.2ex}\peqsubfino{{\mathrm{vol}_g}}{\!Z}{-0.2ex}-\!\int_{\partial M}\!\Big[\peqsubfino{{\big\langle\hspace*{0.1ex}\peqsub{\imath}{\mathbb{V}_{\!Z}} \mathrm{d}A\hspace*{0.3ex},\hspace*{0.05ex}\varepsilon\peqsubfino{{\imath_{\vec{\nu}}}}{\!Z}{-0.3ex} \mathrm{d}A\big\rangle_{\!\!g}}}{\!Z}{-0.2ex}+\peqsubfino{{\big\langle \hspace*{0.1ex}\peqsubfino{{\mathcal{L}_{\vec{\mathbb{V}}^\partial}}}{\!\!\!\!\!Z}{-0.4ex}\,\peqsub{A}{\partial}\hspace*{0.3ex},\hspace*{0.05ex}\peqsub{b}{Z}^2\peqsub{A}{\partial}\hspace*{0.1ex}\big\rangle_{\!\!g^\partial}}}{\!\!\!\!\!Z}{-0.3ex}\,\Big]\mathrm{vol}_{g^\partial_{\!Z}}\updown{\eqref{Appendix - lemma - g=gamma+nu nu}}{\eqref{appendix - formula - f^*d=df^*}\eqref{appendix - formula - i_x(f^*)=f^*i_{f_x}}}{=}\\
&=\!-\varepsilon\!\int_M\!\peqsubfino{{\big\langle\hspace*{0.1ex}\peqsub{\imath}{\mathbb{V}_{\!Z}} \mathrm{d}A\hspace*{0.3ex},\hspace*{0.05ex}\peqsub{\delta}{\!Z} \mathrm{d}A\big\rangle_{\!\!g}}}{\!Z}{-0.2ex}\peqsubfino{{\mathrm{vol}_g}}{\!Z}{-0.2ex}-\!\int_{\partial M}\!\Big[\peqsubfino{{\big\langle\hspace*{0.1ex}\peqsub{\imath}{\mathbb{V}^\partial_{\!Z}}\mathrm{d} \peqsub{A}{\partial}\hspace*{0.3ex},\hspace*{0.05ex}\varepsilon\peqsub{\jmath}{\partial}^*(\peqsubfino{{\imath_{\vec{\nu}}}}{\!Z}{-0.3ex} \mathrm{d}A)\big\rangle_{\!\!g^\partial}}}{\!\!\!\!\!Z}{-0.3ex}\,+\peqsubfino{{\big\langle \hspace*{0.1ex}\peqsubfino{{\mathcal{L}_{\vec{\mathbb{V}}^\partial}}}{\!\!\!\!\!Z}{-0.4ex}\,\peqsub{A}{\partial}\hspace*{0.3ex},\hspace*{0.05ex}\peqsub{b}{Z}^2\peqsub{A}{\partial}\hspace*{0.1ex}\big\rangle_{\!\!g^\partial}}}{\!\!\!\!\!Z}{-0.3ex}\,\Big]\mathrm{vol}_{g^\partial_{\!Z}}=\\
&=\!-\varepsilon\!\int_M\!\peqsubfino{{\big\langle\hspace*{0.1ex}\peqsub{\imath}{\mathbb{V}_{\!Z}} \mathrm{d}A\hspace*{0.3ex},\hspace*{0.05ex}\peqsub{\delta}{\!Z} \mathrm{d}A\big\rangle_{\!\!g}}}{\!Z}{-0.2ex}\peqsubfino{{\mathrm{vol}_g}}{\!Z}{-0.2ex}-\!\int_{\partial M}\!\Big[\peqsubfino{{\big\langle \hspace*{0.1ex}\peqsubfino{{\mathcal{L}_{\vec{\mathbb{V}}^\partial}}}{\!\!\!\!\!Z}{-0.4ex}\,\peqsub{A}{\partial}\hspace*{0.3ex},\hspace*{0.05ex}\varepsilon\peqsub{\jmath}{\partial}^*\big(\peqsubfino{{\imath_{\vec{\nu}}}}{\!Z}{-0.3ex} \mathrm{d}A)+\peqsub{b}{Z}^2\peqsub{A}{\partial}\big)\hspace*{0.1ex}\big\rangle_{\!\!g^\partial}}}{\!\!\!\!\!Z}{-0.3ex}\,-\peqsubfino{{\big\langle\hspace*{0.1ex}\mathrm{d}\peqsub{\imath}{\mathbb{V}^\partial_{\!Z}} \peqsub{A}{\partial}\hspace*{0.3ex},\hspace*{0.05ex}\varepsilon\peqsub{\jmath}{\partial}^*(\peqsubfino{{\imath_{\vec{\nu}}}}{\!Z}{-0.3ex} \mathrm{d}A)\big\rangle_{\!\!g^\partial}}}{\!\!\!\!\!Z}{-0.3ex}\,\Big]\mathrm{vol}_{g^\partial_{\!Z}}
\end{align*}
Thus, we have
\begin{align}\label{Parametrized theories - equation - Variacion accion}
\begin{split}
 D(\peqsub{S}{\mathrm{d}}+&\peqsub{S}{\partial})(A,Z)=\varepsilon \int_M\peqsubfino{{\big\langle\hspace*{0.1ex}\peqsub{V}{\!A}-\peqsub{\imath}{\mathbb{V}_{\!Z}} \mathrm{d}A\hspace*{0.3ex},\hspace*{0.05ex}\peqsub{\delta}{\!Z}  \mathrm{d}A\hspace*{0.1ex}\big\rangle_{\!\!g}}}{\!Z}{-0.2ex}\peqsubfino{{\mathrm{vol}_g}}{\!Z}{-0.2ex}+{}\\
 &\phantom{=}+\int_{\partial M}\Big[\peqsubfino{{\big\langle \hspace*{0.1ex}\peqsub{V}{\!A}^{\scriptscriptstyle\partial}-\peqsubfino{{\mathcal{L}_{\vec{\mathbb{V}}^\partial}}}{\!\!\!\!\!Z}{-0.4ex}\,\peqsub{A}{\partial}\hspace*{0.3ex},\hspace*{0.05ex}\varepsilon\peqsub{\jmath}{\partial}^*\big(\peqsubfino{{\imath_{\vec{\nu}}}}{\!Z}{-0.3ex} \mathrm{d}A\big)+\peqsub{b}{Z}^2\peqsub{A}{\partial}\hspace*{0.1ex}\big\rangle_{\!\!g^\partial}}}{\!\!\!\!\!Z}{-0.3ex}\,+\peqsubfino{{\big\langle\hspace*{0.1ex}\mathrm{d}\peqsub{\imath}{\mathbb{V}^\partial_{\!Z}} \peqsub{A}{\partial}\hspace*{0.3ex},\hspace*{0.05ex}\varepsilon\peqsub{\jmath}{\partial}^*(\peqsubfino{{\imath_{\vec{\nu}}}}{\!Z}{-0.3ex} \mathrm{d}A)\big\rangle_{\!\!g^\partial}}}{\!\!\!\!\!Z}{-0.3ex}\,\Big]\mathrm{vol}_{g^\partial_{\!Z}}
\end{split}
\end{align}
The first two terms lead to the equations
\begin{equation*}
\peqsub{S}{\mathrm{d}}+\peqsub{S}{\partial}:\ 
\begin{array}{|l}
\peqsub{\delta}{\!Z}\mathrm{d}A=0\\[1ex]
\varepsilon\peqsub{\jmath}{\partial}^*\big(\peqsubfino{{\imath_{\vec{\nu}}}}{\!Z}{-0.3ex} \mathrm{d}A)+\peqsub{b}{Z}^2\peqsubfino{A}{\partial}{-0.2ex}=0
\end{array} \qquad\text{ or }\qquad
\peqsub{S}{\mathrm{d}}^{\scriptscriptstyle\partial}+ \peqsub{S}{\partial}^{\scriptscriptstyle\partial}:\ \begin{array}{|l} \peqsub{\delta}{\!Z}\mathrm{d}A=0\\[1ex]
\peqsubfino{A}{\partial}{-0.2ex}=0\end{array}
\end{equation*}
Using \eqref{appendix - formula - i_x(f^*)=f^*i_{f_x}}, \eqref{appendix - formula - f^*d=df^*}, \ref{Appendix - lemma - volz^*g=z^*vol_g and star_z^*g=z^*star_g} and the definitions of $\peqsub{b}{Z}$ and $\peqsubfino{\vec{\nu}}{\!Z}{-0.3ex}$, it is easy to check that if $(A,Z)$ is a solution then so is $(Y^*\!A,Z\smallcirc Y)$ for every $Y\in\mathrm{Diff}(M)$.\separ

Let us now study what happens with the last term. First notice that in the Dirichlet and Neumann cases, it trivially vanishes, however in the Robin case we obtain
\[\int_{\partial M}\peqsubfino{{\big\langle\hspace*{0.1ex}\mathrm{d}\peqsub{\imath}{\mathbb{V}^\partial_{\!Z}} \peqsub{A}{\partial}\hspace*{0.3ex},\hspace*{0.05ex}\varepsilon\peqsub{\jmath}{\partial}^*(\peqsubfino{{\imath_{\vec{\nu}}}}{\!Z}{-0.3ex} \mathrm{d}A)\big\rangle_{\!\!g^\partial}}}{\!\!\!\!\!Z}{-0.3ex}\,\mathrm{vol}_{g^\partial_{\!Z}}=\int_{\partial M}\peqsubfino{{\big\langle\hspace*{0.1ex}\mathrm{d}\peqsub{\imath}{\mathbb{V}^\partial_{\!Z}} \peqsub{A}{\partial}\hspace*{0.3ex},\hspace*{0.05ex}\peqsub{b}{Z}^2\peqsub{A}{\partial}\big\rangle_{\!\!g^\partial}}}{\!\!\!\!\!Z}{-0.3ex}\,\mathrm{vol}_{g^\partial_{\!Z}}=\int_{\partial M}\peqsubfino{{\big\langle\hspace*{0.1ex}\peqsub{\imath}{\mathbb{V}^\partial_{\!Z}} \peqsub{A}{\partial}\hspace*{0.3ex},\hspace*{0.05ex}\delta(\peqsub{b}{Z}^2\peqsub{A}{\partial})\big\rangle_{\!\!g^\partial}}}{\!\!\!\!\!Z}{-0.3ex}\,\mathrm{vol}_{g^\partial_{\!Z}}\]
The \textbf{additional} condition that appears due to this last term over the fields $(A,Z)$ at the boundary can be written, using the notation $\langle\imath_{-}\alpha,\beta\rangle(\vec{v})=\langle\imath_{\vec{v}}\alpha,\beta\rangle$, as the vanishing of
\begin{equation}\label{Parametrized EM - equation - additional condition}
\peqsubfino{{\big\langle\hspace*{0.1ex}\peqsub{\imath}{-} \peqsub{A}{\partial}\hspace*{0.3ex},\hspace*{0.05ex}\delta(\peqsub{b}{Z}^2\peqsub{A}{\partial})\big\rangle_{\!\!g^\partial}}}{\!\!\!\!\!Z}{-0.3ex}\in\Omega^1(\partial M)
\end{equation}
It is interesting to mention that $\peqsub{\imath}{-} \peqsub{A}{\partial}$ is always zero if $k=0$ i.e.\ for the parametrized scalar field. We will study this particular case with due care in chapter \ref{Chapter - parametrized scalar}.

	\section{Lagrangian formulation}\label{Parametrized theories - section - Lagrangian}
	  For the Lagrangian formulation we have to perform the $n+1$ decomposition of the action. For that we need to decompose the differential $\mathrm{d}A\in\Omega^{k+1}(M)$ with respect to the unitary normal vector field
	  \[\vec{n}=\frac{\partial_t-\vec{N}}{\mathbf{N}}\]
	  According to equations \eqref{Mathematical background - equation - descomposicion campo vectorial}-\eqref{Mathematical background - equation - d^top} and lemma \ref{appendix - lemma - descomposicion dA} we have
	  \[\mathrm{d}A=n\wedge\peqsubfino{(\mathrm{d}A)}{\!\perp}{-0.3ex}+(\mathrm{d}A)^{\!\scriptscriptstyle\top}\qquad\text{ where }\qquad\begin{array}{l}\displaystyle\peqsubfino{(\mathrm{d}A)}{\!\perp}{-0.3ex}=\varepsilon\frac{\mathcal{L}_{\partial_t}A^{\scriptscriptstyle\top}-\mathcal{L}_{\vec{N}}A^{\scriptscriptstyle\top}}{\mathbf{N}}-\frac{\mathrm{d}^{\!\scriptscriptstyle\top}\!(\mathbf{N}\peqsub{A}{\perp})}{\mathbf{N}}\\[2ex]
	  \displaystyle(\mathrm{d}A)^{\!\scriptscriptstyle\top}=\mathrm{d}^{\!\scriptscriptstyle\top}\!\!A^{\scriptscriptstyle\top}
	  \end{array}\]

	\begin{align*}
	\varepsilon&\mathrm{d}A\wedge\star\mathrm{d}A=\varepsilon\peqsubfino{{\big\langle n\wedge\peqsubfino{(\mathrm{d}A)}{\!\perp}{-0.3ex}+(\mathrm{d}A)^{\!\scriptscriptstyle\top},n\wedge\peqsubfino{(\mathrm{d}A)}{\!\perp}{-0.3ex}+(\mathrm{d}A)^{\!\scriptscriptstyle\top} \big\rangle_{\!\!g}}}{\!Z}{-0.2ex}\peqsubfino{{\mathrm{vol}_g}}{\!Z}{-0.2ex}\overset{\eqref{Appendix - equation - (v wedge beta,omega)=(beta,i_v omega)}}{=}\\
	&=\varepsilon\Big[\peqsubfino{{\varepsilon\big\langle \peqsubfino{(\mathrm{d}A)}{\!\perp}{-0.3ex},\peqsubfino{(\mathrm{d}A)}{\!\perp}{-0.3ex}\big\rangle_{\!\!g}}}{\!Z}{-0.2ex} + 0+ \peqsubfino{{\big\langle \mathrm{d}^{\!\scriptscriptstyle\top}\!\!A^{\scriptscriptstyle\top},\mathrm{d}^{\!\scriptscriptstyle\top}\!\!A^{\scriptscriptstyle\top}\big\rangle_{\!\!g}}}{\!Z}{-0.2ex}\,\Big]\peqsubfino{{\mathrm{vol}_g}}{\!Z}{-0.2ex}\updown{\hspace*{0ex}\eqref{Appendix - lemma - vol=nu wedge vol}}{\hspace*{5ex}\mathclap{g=\varepsilon n\otimes n+\gamma\raisebox{.5ex}{\!\!\!\textasciitilde}}}{\hspace*{-3ex}=}\\
	&=\Big[\big\langle \peqsubfino{(\mathrm{d}A)}{\!\perp}{-0.3ex},\peqsubfino{(\mathrm{d}A)}{\!\perp}{-0.3ex}\big\rangle_{\gamma\raisebox{.5ex}{\!\!\!\textasciitilde}} +\varepsilon \big\langle \mathrm{d}^{\!\scriptscriptstyle\top}\!\!A^{\scriptscriptstyle\top},\mathrm{d}^{\!\scriptscriptstyle\top}\!\!A^{\scriptscriptstyle\top}\big\rangle_{\gamma\raisebox{.5ex}{\!\!\!\textasciitilde}}\Big]\varepsilon n\wedge\mathrm{vol}_{\gamma\raisebox{.5ex}{\!\!\!\textasciitilde}}\overset{\eqref{Mathematical background - equation - n=varepsilon N dt}}{=}\\
	&=\mathbf{N}\mathrm{d}t\wedge\left[\left\langle\varepsilon\frac{\dot{A}^{\scriptscriptstyle\top}\!-\mathcal{L}_{\vec{N}}A^{\scriptscriptstyle\top}}{\mathbf{N}}-\frac{\mathrm{d}^{\!\scriptscriptstyle\top}\!(\mathbf{N}\peqsub{A}{\perp})}{\mathbf{N}},\varepsilon\frac{\dot{A}^{\scriptscriptstyle\top}\!-\mathcal{L}_{\vec{N}}A^{\scriptscriptstyle\top}}{\mathbf{N}}-\frac{\mathrm{d}^{\!\scriptscriptstyle\top}\!(\mathbf{N}\peqsub{A}{\perp})}{\mathbf{N}}\right\rangle_{\!\!\gamma\raisebox{.5ex}{\!\!\!\textasciitilde}}+\varepsilon \big\langle\mathrm{d}^{\!\scriptscriptstyle\top}\!\!A^{\scriptscriptstyle\top},\mathrm{d}^{\!\scriptscriptstyle\top}\!\!A^{\scriptscriptstyle\top}\rangle_{\gamma\raisebox{.5ex}{\!\!\!\textasciitilde}}  \right]\!\mathrm{vol}_{\gamma\raisebox{.5ex}{\!\!\!\textasciitilde}}=\\
	&=\mathrm{d}t\wedge\left[\frac{1}{\mathbf{N}}\big\langle\dot{A}^{\scriptscriptstyle\top}\!-\mathcal{L}_{\vec{N}}A^{\scriptscriptstyle\top}\!-\mathrm{d}^{\!\scriptscriptstyle\top}\!(\varepsilon\mathbf{N}\peqsub{A}{\perp}),\dot{A}^{\scriptscriptstyle\top}\!-\mathcal{L}_{\vec{N}}A^{\scriptscriptstyle\top}\!-\mathrm{d}^{\!\scriptscriptstyle\top}\!(\varepsilon\mathbf{N}\peqsub{A}{\perp})\big\rangle_{\gamma\raisebox{.5ex}{\!\!\!\textasciitilde}}   +\varepsilon\mathbf{N} \big\langle\mathrm{d}^{\!\scriptscriptstyle\top}\!\!A^{\scriptscriptstyle\top},\mathrm{d}^{\!\scriptscriptstyle\top}\!\!A^{\scriptscriptstyle\top}\big\rangle_{\gamma\raisebox{.5ex}{\!\!\!\textasciitilde}} \right]\!\mathrm{vol}_{\gamma\raisebox{.5ex}{\!\!\!\textasciitilde}}
	\end{align*}
	
	To break the boundary term we use the $1$-form field $\theta$ which is metrically equivalent to the unitary vector field $g^\partial$-normal to the foliation at the boundary. It is easy to check that
	\[\vec{\theta}=\frac{\vec{n}-\varepsilon\peqsub{\nu}{\perp}\vec{\nu}}{|\nu^{\scriptscriptstyle\top}|}\qquad\text{ where }\qquad\begin{array}{l}
	\vec{\nu}=\varepsilon \peqsub{\nu}{\perp}\vec{n}+\tau.\vec{\nu}^{\scriptscriptstyle\top}\\[1ex]
	1=g(\vec{\nu},\vec{\nu})=\varepsilon\peqsub{\nu}{\perp}^2+\gamma(\vec{\nu}^{\scriptscriptstyle\top\!},\vec{\nu}^{\scriptscriptstyle\top\!})=\varepsilon\peqsub{\nu}{\perp}^2+|\nu^{\scriptscriptstyle\top}|^2\end{array}\]
	is the unitary vector field tangent to the boundary and normal to the foliation. Thus we have (omitting the subscript $Z$ for simplicity)\allowdisplaybreaks[0]
	\begin{align*} 
	&\langle \peqsub{\jmath}{\partial}^*A,\peqsub{\jmath}{\partial}^*A\rangle_{g^\partial}\mathrm{vol}_{g^\partial}\updown{\ref{Appendix - lemma - vol=nu wedge vol}}{\ref{Appendix - lemma - g=gamma+nu nu}}{=}\Big[\peqsubfino{\langle A,A\rangle}{\!g}{-0.2ex}-\peqsubfino{\langle \imath_{\vec{\nu}}A,\imath_{\vec{\nu}}A\rangle}{\!g}{-0.2ex}\Big]\varepsilon\theta\wedge\mathrm{vol}_{\gamma^\partial\raisebox{.5ex}{\hspace*{-2.2ex}\textasciitilde}}=\\
	&\ \ =\Big[\varepsilon\langle \peqsub{A}{\perp},\peqsub{A}{\perp}\rangle_{\gamma\raisebox{.5ex}{\!\!\!\textasciitilde}}+\langle A^{\scriptscriptstyle\top\!},A^{\scriptscriptstyle\top\!}\rangle_{\gamma\raisebox{.5ex}{\!\!\!\textasciitilde}}-\varepsilon\peqsubfino{\big\langle (\imath_{\vec{\nu}}A)}{\!\perp}{-0.3ex},\peqsubfino{(\imath_{\vec{\nu}}A)}{\!\perp}{-0.3ex}\big\rangle_{\gamma\raisebox{.5ex}{\!\!\!\textasciitilde}}-\big\langle (\imath_{\vec{\nu}}A)^{\scriptscriptstyle\!\top\!},(\imath_{\vec{\nu}}A)^{\scriptscriptstyle\!\top}\hspace*{-0.1ex}\rangle_{\gamma\raisebox{.5ex}{\!\!\!\textasciitilde}}\Big]\mathbf{N}_\theta\mathrm{d}t\wedge\mathrm{vol}_{\gamma^\partial\raisebox{.5ex}{\hspace*{-2.2ex}\textasciitilde}}\overset{\star}{=}\\
	&\ \ =\mathbf{N}_\theta\mathrm{d}t\wedge\Big[\varepsilon\langle \peqsub{A}{\perp},\peqsub{A}{\perp}\rangle_{\gamma\raisebox{.5ex}{\!\!\!\textasciitilde}}+\langle A^{\scriptscriptstyle\top\!},A^{\scriptscriptstyle\top\!}\rangle_{\gamma\raisebox{.5ex}{\!\!\!\textasciitilde}}-\varepsilon\big\langle \imath_{\vec{\nu}}(\peqsubfino{A}{\!\perp}{-0.3ex}),\imath_{\vec{\nu}}(\peqsubfino{A}{\!\perp}{-0.3ex})\big\rangle_{\gamma\raisebox{.5ex}{\!\!\!\textasciitilde}}-\big\langle \imath_{\vec{\nu}}(A^{\scriptscriptstyle\!\top\!}),\imath_{\vec{\nu}}(A^{\scriptscriptstyle\!\top\!})\hspace*{-0.1ex}\rangle_{\gamma\raisebox{.5ex}{\!\!\!\textasciitilde}}\Big]\mathrm{vol}_{\gamma^\partial\raisebox{.5ex}{\hspace*{-2.2ex}\textasciitilde}}\overset{\ref{Appendix - lemma - g=gamma+nu nu}}{=}\\
	&\ \ =\mathbf{N}_\theta \mathrm{d}t\wedge \Big[\varepsilon\big\langle \peqsub{\jmath}{\partial}^*(\peqsub{A}{\perp}),\peqsub{\jmath}{\partial}^*(\peqsub{A}{\perp})\big\rangle_{\gamma^\partial\raisebox{.5ex}{\hspace*{-2.2ex}\textasciitilde}}\ +\big\langle \peqsub{\jmath}{\partial}^*(A^{\scriptscriptstyle\top\!}),\peqsub{\jmath}{\partial}^*(A^{\scriptscriptstyle\top\!})\big\rangle_{\gamma^\partial\raisebox{.5ex}{\hspace*{-2.2ex}\textasciitilde}}\ \Big]\mathrm{vol}_{\gamma^\partial\raisebox{.5ex}{\hspace*{-2.2ex}\textasciitilde}}
	\end{align*}\allowdisplaybreaks
	where in the $\star$ equality we have used $\peqsubfino{(\imath_{\vec{\nu}}A)}{\!\perp}{-0.3ex}=\varepsilon\imath_{\vec{n}}\imath_{\vec{\nu}}A=-\varepsilon\imath_{\vec{\nu}}\imath_{\vec{n}}A=-\imath_{\vec{\nu}}\peqsubfino{A}{\!\perp}{-0.3ex}$ and analogously for $A^{\scriptscriptstyle\top\!}=\varepsilon\imath_{\vec{n}}(n\wedge A)$.\separ

	Through the equivalence \eqref{Mathematical background - equation - C(R,C(M,N)) cong C(R x M,N)} we can consider $A^{\scriptscriptstyle\top}\in\Omega^k(I\times\Sigma)$ and $\peqsub{A}{\perp}\in\Omega^{k-1}(I\times\Sigma)$ as curves over $\Omega^k(\Sigma)$ and $\Omega^{k-1}(\Sigma)$, meanwhile $\widetilde{\gamma}$ can be pulled-back to a metric $\peqsubfino{\gamma}{X}{-0.2ex}\in\mathrm{Met}(\Sigma)$ (analogously for the boundary) which allows us to read the Lagrangian
	\[\begin{array}{cccc}
	L: & \mathcal{D}\subset T\Big(\Omega^{k-1}(\Sigma)\times \Omega^k(\Sigma)\times\mathrm{Emb}(\Sigma,M)\Big) & \longrightarrow & \R\\
	&           \mathbf{v}_{(\peqsubfino{q}{\perp}{-0.2ex},q,X)}=(\peqsubfino{q}{\perp}{-0.2ex},q,X;v_{\!\scriptscriptstyle\perp},v,\peqsub{\mathbb{V}}{\!X})                   &  \longmapsto    & L(\mathbf{v}_{(\peqsubfino{q}{\perp}{-0.2ex},q,X)})
	\end{array}\]
	The explicit expression can be obtained taking into account the decomposition \eqref{Mathematical background - equation - decomposition embedding} which says that $\peqsub{\mathbb{V}}{\!X}=\peqsub{V}{X}^{{\raisemath{0.2ex}{\!\scriptscriptstyle\perp}}}\peqsub{\vec{n}}{X}+\peqsub{\tau}{X}.\peqsub{\vec{v}}{X}^{\,\scriptscriptstyle\top}$ where $\peqsub{V}{X}^{{\raisemath{0.2ex}{\!\scriptscriptstyle\perp}}}=\varepsilon \peqsub{\vec{n}}{X}(\peqsub{\mathbb{V}}{\!X})$, that corresponds to the lapse $\mathbf{N}$, and $\peqsub{\vec{v}}{X}^{\,a}=(\peqsub{e}{X})^a_\alpha V^\alpha$, that corresponds to the shift $\vec{N}$. Finally, at the boundary we have, using the definition of $\vec{\theta}$, $\mathbf{N}_\theta=\varepsilon\peqsub{\vec{\theta}}{X}(\peqsub{\mathbb{V}}{\!X})=\peqsub{V}{X}^{{\raisemath{0.2ex}{\!\scriptscriptstyle\perp}}}/|\vec{\nu}^{\hspace*{0.1ex}\scriptscriptstyle\top}\!|$ because $\vec{\nu}\perp\peqsub{\mathbb{V}}{\!X}$. Therefore we find
	
	\begin{align}\label{Parametrized theories - equation - Lagrangian}
	\begin{split}
	L(\mathbf{v}_{(\peqsubfino{q}{\perp}{-0.2ex},q,X)})&=\frac{1}{2}\left\llangle\frac{v-\mathcal{L}_{\peqsub{\vec{v}}{X}^{\scriptscriptstyle\top}}q-\varepsilon\mathrm{d}(\peqsub{V}{X}^{{\raisemath{0.2ex}{\!\scriptscriptstyle\perp}}}\peqsubfino{q}{\perp}{-0.2ex})}{\peqsub{V}{X}^{{\raisemath{0.2ex}{\!\scriptscriptstyle\perp}}}},\frac{v-\mathcal{L}_{\peqsub{\vec{v}}{X}^{\scriptscriptstyle\top}}q-\varepsilon\mathrm{d}(\peqsub{V}{X}^{{\raisemath{0.2ex}{\!\scriptscriptstyle\perp}}}\peqsubfino{q}{\perp}{-0.2ex})}{\peqsub{V}{X}^{{\raisemath{0.2ex}{\!\scriptscriptstyle\perp}}}}\right\rrangle_{\!\!\peqsub{V}{X}^{{\raisemath{0.2ex}{\!\scriptscriptstyle\perp}}}}+\frac{\varepsilon}{2}{\big\llangle \mathrm{d}q,\mathrm{d}q\big\rrangle}_{\peqsub{V}{X}^{{\raisemath{0.2ex}{\!\scriptscriptstyle\perp}}}}+{}\\
	&\phantom{=}-\frac{\varepsilon}{2}\big\llangle \peqsub{b}{X}\,\peqsubfino{q^{\scriptscriptstyle\partial}}{\perp}{-0.2ex},\peqsub{b}{X}\,\peqsubfino{q^{\scriptscriptstyle\partial}}{\perp}{-0.2ex}\big\rrangle_{\peqsub{V}{X}^{{\raisemath{0.2ex}{\!\scriptscriptstyle\perp}}}/|\vec{\nu}^{\hspace*{0.1ex}\scriptscriptstyle\top}\!|}-\frac{1}{2}\big\llangle \peqsub{b}{X}\, \peqsubfino{q}{\partial}{-0.2ex},\peqsub{b}{X}\, \peqsubfino{q}{\partial}{-0.2ex}\big\rrangle_{\peqsub{V}{X}^{{\raisemath{0.2ex}{\!\scriptscriptstyle\perp}}}/|\vec{\nu}^{\hspace*{0.1ex}\scriptscriptstyle\top}\!|}
	\end{split}
	\end{align}
	where 
	\[\big\llangle \alpha,\beta\big\rrangle_{\!f}=\int_\Sigma f\peqsubfino{{\big\langle \alpha,\beta\big\rangle_{\!\gamma}}}{\!X}{-0.1ex}\peqsubfino{{\mathrm{vol}_\gamma}}{\!\!X}{-0.4ex}=\int_\Sigma f\sqrt{\peqsubfino{\gamma}{\!X}{-.3ex}}\,\peqsubfino{{\big\langle \alpha,\beta\big\rangle_{\!\gamma}}}{\!X}{-0.1ex}\peqsubfino{\mathrm{vol}}{\Sigma}{-0.2ex}\]
	in the usual scalar product \eqrefconchap{Mathematical background - property - (a,b)=int <a,b>vol} with \textbf{weight} $f\in\Cinf{\Sigma}$ and analogously for the boundary (we remind the reader that the $\partial$ symbol attached to a variable means that we are taking the pullback through $\peqsubfino{\jmath}{\partial}{-.2ex}$). Notice that those scalar products depend on the embedding through the metric $\peqsubfino{\gamma}{\!X}{-0.2ex}=X^*g$ and, in our case, also through the map $f$.\separ
	
	Finally notice that the Lagrangian is defined on the open subset \[\mathcal{D}=T\Big(\Omega^{k-1}(\Sigma)\times\Omega^k(\Sigma)\Big)\times\Big\{(X,\peqsub{\mathbb{V}}{\!X})\in T\mathrm{Emb}(\Sigma,M)\ /\ \varepsilon \peqsub{\vec{n}}{X}(\peqsub{\mathbb{V}}{\!X})>0\Big\}\]

	\section{Fiber derivative}\label{Parametrized theories - section - Fiber derivative}
	Before computing the fiber derivative associated with the previous Lagrangian let us study its target space i.e.\ the phase space of the theory.
	
	\subsection*{Geometric arena}\trassub
	
	A typical point of the phase space $T^*\mathcal{D}$ is of the form \[\boldsymbol{\mathrm{p}}_{(\peqsubfino{q}{\perp}{-0.2ex},q,X)}=(\peqsubfino{q}{\perp}{-0.2ex},q,X;\peqsubfino{\boldsymbol{p}}{\!\perp}{-0.2ex},\boldsymbol{p},\peqsubfino{\boldsymbol{P}}{\!\!X}{-0.2ex}) \in T^*\Big(\mathcal{C}^\infty(\Sigma)\times \Omega^k(\Sigma)\times\mathrm{Emb}(\Sigma,M)\Big) \]
	where  $\peqsubfino{\boldsymbol{p}}{\!\perp}{-0.2ex}\in C^\infty(\Sigma)'$ and $\boldsymbol{p}\in \Omega^k(\Sigma)'$ are elements of the dual of $\Omega^{k-1}(\Sigma)$ and $\Omega^k(\Sigma)$ respectively. Meanwhile $\peqsubfino{\boldsymbol{P}}{\!X}{-0.2ex}:\Gamma^\partial\!(X^*TM)\rightarrow\R$ is a continuous linear functional (see lemma \ref{Mathematical background - lemma - campos de vectores sobre Emb}). The phase-space $T^*\mathcal{D}$ is equipped with the symplectic form
	\begin{align*}
	\Omega_{\boldsymbol{\mathrm{p}}_{(\peqsubfino{q}{\perp}{-0.2ex},q,X)}}(Y,Z)
	&=\Omega_{(\peqsubfino{q}{\perp}{-0.2ex},q,X;\peqsubfino{\boldsymbol{p}}{\!\perp}{-0.2ex},\boldsymbol{p},\peqsubfino{\boldsymbol{P}}{\!X}{-0.2ex})}\left(\rule{0ex}{3ex}(\vecc[\scalebox{0.4}{$\perp$}]{Y}{q},\vecc{Y}{q},\vecc{Y}{X},\vecc[\scalebox{0.4}{$\perp$}]{\boldsymbol{Y}}{\!\!\!p},\vecc{\boldsymbol{Y}}{\!\!\!p},\vecc{\boldsymbol{Y}}{\!\!P}),(\vecc[\scalebox{0.4}{$\perp$}]{Z}{q},\vecc{Z}{q},\vecc{Z}{X},\vecc[\scalebox{0.4}{$\perp$}]{\boldsymbol{Z}}{p},\vecc{\boldsymbol{Z}}{p},\vecc{\boldsymbol{Z}}{P})\right)\overset{\eqref{eq:background canonical symplectic form}}{=}\\
	&=\vecc[\scalebox{0.4}{$\perp$}]{\boldsymbol{Z}}{p}\!\left(\vecc[\scalebox{0.4}{$\perp$}]{Y}{q}\right)-
	\vecc[\scalebox{0.4}{$\perp$}]{\boldsymbol{Y}}{\!p}\!\left(\vecc[\scalebox{0.4}{$\perp$}]{Z}{q}\right)+\vecc{\boldsymbol{Z}}{p}\!\left(\vecc{Y}{q}\right)-\vecc{\boldsymbol{Y}}{\!\!\!p}\!\left(\vecc{Z}{q}\right)+\vecc{\boldsymbol{Z}}{P}\!\left(\vecc{Y}{X}\right)-\vecc{\boldsymbol{Y}}{\!\!P}\!\left(\vecc{Z}{X}\right)
	\end{align*}
	for $Y,Z\in\mathfrak{X}(T^*\mathcal{D})$ where we have omitted the base point for simplicity. In particular we have $\vecc[\scalebox{0.4}{$\perp$}]{Y}{q}\in C^\infty(\Sigma)$, $\vecc{Y}{q}\in \Omega^k(\Sigma)$, $\vecc{Y}{X}\in \Gamma^\partial(X^*TM)$, $\vecc[\scalebox{0.4}{$\perp$}]{\boldsymbol{Y}}{\!\!\!p}\in C^\infty(\Sigma)'$, $\vecc{\boldsymbol{Y}}{\!\!\!p}\in \Omega^k(\Sigma)'$, and $\vecc{\boldsymbol{Y}}{\!\!P}:\Gamma^\partial(X^*TM)\rightarrow\R$, and analogously for the $Z$ components.\separ
	
	In our case at hand, we will prove that the primary constraint submanifold $F\!L(\mathcal{D})$ is not all $T^*\mathcal{D}$. In particular we will get that $\peqsubfino{\boldsymbol{p}}{\!\perp}{-0.1ex}$ is zero, $\boldsymbol{p}$ is defined in terms of an antisymmetric covector field density and the usual pairing with the dual, while $\peqsubfino{\boldsymbol{P}}{\!\!X}{-0.2ex}$ is given by two antisymmetric covector fields densities $\peqsub{P}{\!X}:\Gamma^\partial\!(X^*TM)\rightarrow C^\infty(\Sigma)$ and $\peqsub{P}{\partial X}:\Gamma^\partial\!(X^*T\partial_\Sigma M)\rightarrow C^\infty(\partial\Sigma)$ along $X$. More precisely we will have for some $v\in\Omega^k(\Sigma)$ and $\peqsub{\mathbb{V}}{\!X}\in \peqsub{T}{\!X}\mathrm{Emb}(\Sigma,M)$ the expressions
	\begin{align*}
	\boldsymbol{p}(v)&=\int_\Sigma (v,p)\peqsub{\mathrm{vol}}{\Sigma}\\
	\peqsubfino{\boldsymbol{P}}{\!X}{-0.2ex}(\peqsub{\mathbb{V}}{\!X})&=  \int_\Sigma   \big(\peqsub{\mathbb{V}}{\!X},\peqsub{P}{\!X}\big)\,\peqsub{\mathrm{vol}}{\Sigma}+\int_{\partial\Sigma} \big(\peqsub{\mathbb{V}}{\!X}^{\scriptscriptstyle\partial},\peqsub{P}{\partial X}\big)\,\peqsub{\mathrm{vol}}{\partial\Sigma}
	\end{align*}
	where $(\alpha,V)=\frac{1}{k!}\alpha_{a_1\cdots a_k}V^{a_1\cdots a_k}$. Notice that $p$, $\peqsub{P}{\!X}$, and $\peqsub{P}{\partial X}$ are densities (see page \pageref{Mathematical background - subsection - densities}) because they depend on the chosen volume forms $\peqsub{\mathrm{vol}}{\Sigma}\in\mathrm{Vol}(\Sigma)$ and $\peqsub{\mathrm{vol}}{\partial\Sigma}\in\mathrm{Vol}(\partial\Sigma)$. These distributions can also be written in terms of the metric volume forms of $\peqsubfino{\gamma}{X}{-0.2ex}=X^*\!g$ and $\peqsubfino{\gamma}{\partial X}{-0.2ex}:=(X|_{\partial M})^*\peqsub{\jmath}{\partial}^*\gamma$.
	\begin{align}
	\boldsymbol{p}(v)&=\int_\Sigma (v,p)\,\peqsub{\mathrm{vol}}{\Sigma}=\int_\Sigma \frac{(v,p)}{\sqrt{\peqsubfino{\gamma}{X}{-0.2ex}}}\,\peqsubfino{{\mathrm{vol}_\gamma}}{\!\!X}{-0.4ex}\label{Parametrized theories - equation - p=int p}
	\\
	\peqsubfino{\boldsymbol{P}}{\!X}{-0.2ex}(\peqsub{\mathbb{V}}{\!X})&=\int_\Sigma   \big(\peqsub{\mathbb{V}}{\!X},\peqsub{P}{\!X}\big)\,\peqsub{\mathrm{vol}}{\Sigma}+\!\int_{\partial\Sigma} \big(\peqsub{\mathbb{V}}{\!X}^{\scriptscriptstyle\partial},\peqsub{P}{\partial X}\big)\,\peqsub{\mathrm{vol}}{\partial\Sigma}
	=\int_\Sigma   \frac{\big(\peqsub{\mathbb{V}}{\!X},\peqsub{P}{\!X}\big)}{\sqrt{\peqsubfino{\gamma}{X}{-0.2ex}}}\,\peqsubfino{{\mathrm{vol}_\gamma}}{\!\!X}{-0.4ex}+\!\int_{\partial\Sigma} \frac{\big(\peqsub{\mathbb{V}}{\!X}^{\scriptscriptstyle\partial},\peqsub{P}{\partial X}\big)}{\sqrt{\peqsub{\gamma}{\partial X}}} \,\peqsubfino{{\mathrm{vol}_\gamma}}{\!\partial\!X}{-0.4ex}\label{Parametrized theories - equation - P=int P}
	\end{align}
	Let $\peqsub{\widetilde{\mathcal{P}}}{\mathrm{EM}}=\{\peqsub{\mathrm{p}}{\mathrm{q}}:=(\peqsubfino{q}{\perp}{-0.2ex},q,X,\peqsubfino{p}{\perp}{-0.2ex},p,\peqsub{P}{\!X},\peqsub{P}{\partial X})\}\overset{\jmath}{\hookrightarrow }T^*\mathcal{D}$ considered as a subset of $T^*\mathcal{D}$ via the previous representations. We define the induced form $\widetilde{\Omega}^{\scriptscriptstyle\mathrm{EM}}:=\jmath^*\Omega$ given by  $\widetilde{\Omega}^{\scriptscriptstyle\mathrm{EM}}(Y,Z)=(\Omega\smallcirc\jmath)(\jmath_*Y,\jmath_*Z)$. In order to compute its explicit expression we see that we have to compute first $\jmath_*Y\in\mathfrak{X}^{\scriptscriptstyle \top}_\jmath\!(\mathcal{D})$ for any $Y=(\vecc[\scalebox{0.4}{$\perp$}]{Y}{q},\vecc{Y}{q},\vecc{Y}{X},\vecc[\scalebox{0.4}{$\perp$}]{Y}{p},\vecc{Y}{p},\vecc{Y}{P},\vecc{Y}{P}^{\scriptscriptstyle\partial})\in\mathfrak{X}(\peqsub{\widetilde{\mathcal{P}}}{\mathrm{EM}})$. From \eqref{Parametrized theories - equation - p=int p}, \eqref{Parametrized theories - equation - P=int P} and the analog for $\peqsubfino{p}{\perp}{-0.2ex}$ (which will turn out to be zero) we obtain
	\[\jmath_*Y=\left(\vecc[\scalebox{0.4}{$\perp$}]{Y}{q}\coma\vecc{Y}{q}\coma\vecc{Y}{X}\coma\int_\Sigma \big(\,\!\cdot{}\!\coma\vecc[\scalebox{0.4}{$\perp$}]{Y}{p}\big)\peqsub{\mathrm{vol}}{\Sigma}\coma\int_\Sigma \big(\,\!\cdot{}\!\coma\vecc{Y}{p}\big)\peqsub{\mathrm{vol}}{\Sigma}\coma\int_\Sigma\big(\,\!\cdot{}\!\coma\vecc{Y}{P}\big)\peqsub{\mathrm{vol}}{\Sigma}+\int_{\partial\Sigma}\big(\,\!\cdot{}\!\coma\vecc{Y}{P}^{\scriptscriptstyle\partial}\big)\peqsub{\mathrm{vol}}{\partial\Sigma}\right)\in\mathfrak{X}(T^*\mathcal{D})\]
	so we have
	\begin{align}\label{Parametrized theories - equation - tilde Omega}
	\begin{split}
	\widetilde{\Omega}^{\scriptscriptstyle\mathrm{EM}}_{\peqsub{\mathrm{p}}{\mathrm{q}}}(Y,Z)&=\int_\Sigma\Big[\Big(\vecc[\scalebox{0.4}{$\perp$}]{Y}{q},\vecc[\scalebox{0.4}{$\perp$}]{Z}{p}\Big)-\Big(\vecc[\scalebox{0.4}{$\perp$}]{Z}{q},\vecc[\scalebox{0.4}{$\perp$}]{Y}{p}\Big)+\Big(\vecc{Y}{q},\vecc{Z}{p}\Big)-\Big(\vecc{Z}{q},\vecc{Y}{p}\Big)+\vecc{Z}{P}(\vecc{Y}{X})-\vecc{Y}{P}(\vecc{Z}{X})\Big]\peqsub{\mathrm{vol}}{\Sigma}+{}\\
	&\phantom{=}+\int_{\partial\Sigma}\left[\vecc{Z}{P}^{\scriptscriptstyle\partial}(\vecc{Y}{X})-\vecc{Y}{P}^{\scriptscriptstyle\partial}(\vecc{Z}{X})\right]\peqsub{\mathrm{vol}}{\partial\Sigma}     
	\end{split}
	\end{align}

	\subsection*{Volume forms}\trassub
	
	It is important to notice that we have fixed two auxiliary volume forms in $\Sigma$ and $\partial\Sigma$: $\peqsub{\mathrm{vol}}{\Sigma}$ and $\peqsub{\mathrm{vol}}{\partial\Sigma}$ respectively. Each of them is related to the metric volume form by the corresponding determinant, while the unitary $\peqsubfino{\gamma}{X}{-0.2ex}$-normal vector field to the boundary \[\frac{\peqsubfino{\vec{\nu}^{\,\scriptscriptstyle\top}}{\!X}{-0.2ex}}{|\peqsubfino{\vec{\nu}^{\,\scriptscriptstyle\top}}{\!X}{-0.2ex}|}\in\mathfrak{X}(\Sigma)\quad\text{ where }\quad\peqsubfino{\vec{\nu}}{\!X}{-0.2ex}=\varepsilon \peqsubfino{\nu^{\scriptscriptstyle\perp}}{\!X}{-0.2ex}\vec{n}+\peqsubfino{\tau}{\!X}{-0.2ex}.\peqsubfino{\vec{\nu}^{\,\scriptscriptstyle\top}}{\!X}{-0.2ex}\]
	relates both metric volumes via lemma \ref{Appendix - lemma - vol=nu wedge vol}. Notice that although $\peqsubfino{\vec{\nu}}{\!X}{-0.2ex}\in\mathfrak{X}^{\scriptscriptstyle\perp}_\jmath\hspace*{-0.2ex}(M)$ has norm $1$, $\peqsubfino{\vec{\nu}^{\,\scriptscriptstyle\top}}{\!X}{-0.2ex}\in\mathfrak{X}(\Sigma)$ in general has not. Thus we have
	\[\peqsub{\mathrm{vol}}{\Sigma}=\frac{1}{\sqrt{\peqsubfino{\gamma}{\!X}{-0.2ex}}}\peqsubfino{{\mathrm{vol}_\gamma}}{\!X}{-0.2ex}=\frac{1}{\sqrt{\peqsubfino{\gamma}{\!X}{-0.2ex}}}\frac{\peqsubfino{\nu}{\!X}{-0.2ex}}{|\peqsubfino{\vec{\nu}^{\,\scriptscriptstyle\top}}{\!X}{-0.2ex}|}\wedge\mathrm{vol}_{\gamma^\partial_{\!X}}=\frac{\sqrt{\peqsubfino{\gamma^{\,\scriptscriptstyle\partial}}{\!X}{-0.2ex}}}{\sqrt{\peqsubfino{\gamma}{\!X}{-0.2ex}}}\frac{\peqsubfino{\nu}{\!X}{-0.2ex}}{|\peqsubfino{\vec{\nu}^{\,\scriptscriptstyle\top}}{\!X}{-0.2ex}|}\wedge\peqsub{\mathrm{vol}}{\partial\Sigma}\]
	In particular the following expression does not depend on the embedding (because the LHS of the previous equation does not either)
	\begin{equation}\label{Parametrized theories - equation - gamma/gamma nu/nu}
	\frac{\sqrt{\peqsubfino{\gamma^{\,\scriptscriptstyle\partial}}{\!X}{-0.2ex}}}{\sqrt{\peqsubfino{\gamma}{\!X}{-0.2ex}}}\frac{\peqsubfino{\nu}{\!X}{-0.2ex}}{|\peqsubfino{\vec{\nu}^{\,\scriptscriptstyle\top}}{\!X}{-0.2ex}|}
	\end{equation}
	
	\subsection*{Computation of the fiber derivative}\trassub
	
	The fiber derivative, given by equation \eqref{Mathematical background - equation - FL=D_2L}, is computed by taking an initial point of the tangent bundle $\mathbf{v}_{(\peqsubfino{q}{\perp}{-0.2ex},q,X)}=(\peqsubfino{q}{\perp}{-0.2ex},q,X;\peqsubfino{v}{\!\perp}{-0.2ex},v,\peqsubfino{\mathbb{V}}{\!X}{-0.1ex})$ and some initial velocities $\mathbf{w}^1_{(\peqsubfino{q}{\perp}{-0.2ex},q,X)}=(\peqsubfino{q}{\perp}{-0.2ex},q,X;\peqsubfino{w}{\!\perp}{-0.2ex},0,0)$, $\mathbf{w}^2_{(\peqsubfino{q}{\perp}{-0.2ex},q,X)}=(\peqsubfino{q}{\perp}{-0.2ex},q,X;0,w,0)$, and $\mathbf{w}^3_{(\peqsubfino{q}{\perp}{-0.2ex},q,X)}=(\peqsubfino{q}{\perp}{-0.2ex},q,X;0,0,\peqsubfino{\mathbb{W}}{\!X}{-0.1ex})$. The first two are immediate while the last one, much more complicated, is computed in lemma \ref{Appendix - equation - FL(w3)=Pi}.
	\begin{align*}
	FL&(\mathbf{v}_{(\peqsubfino{q}{\perp}{-0.2ex},q,X)})\!\left(\hspace*{-0.1ex}\mathbf{w}^1_{(\peqsubfino{q}{\perp}{-0.2ex},q,X)}\hspace*{-0.1ex}\right)=0\\
	FL&(\mathbf{v}_{(\peqsubfino{q}{\perp}{-0.2ex},q,X)})\!\left(\hspace*{-0.1ex}\mathbf{w}^2_{(\peqsubfino{q}{\perp}{-0.2ex},q,X)}\hspace*{-0.1ex}\right)=\left\llangle\frac{w}{\peqsub{V}{X}^{{\raisemath{0.2ex}{\!\scriptscriptstyle\perp}}}},\frac{v-\mathcal{L}_{\peqsub{\vec{v}}{X}^{\scriptscriptstyle\top}}q-\varepsilon\mathrm{d}(\peqsub{V}{X}^{{\raisemath{0.2ex}{\!\scriptscriptstyle\perp}}}\peqsubfino{q}{\perp}{-0.2ex})}{\peqsub{V}{X}^{{\raisemath{0.2ex}{\!\scriptscriptstyle\perp}}}}\right\rrangle_{\!\!\peqsub{V}{X}^{{\raisemath{0.2ex}{\!\scriptscriptstyle\perp}}}}\hspace*{-0.3ex}=\left(\hspace*{-0.3ex} w\hspace*{.2ex},\hspace*{.1ex}\sqrt{\peqsubfino{\gamma}{X}{-0.2ex}}\peqsub{\sharp}{\gamma}\frac{v-\mathcal{L}_{\peqsub{\vec{v}}{X}^{\scriptscriptstyle\top}}q-\varepsilon\mathrm{d}(\peqsub{V}{X}^{{\raisemath{0.2ex}{\!\scriptscriptstyle\perp}}}\peqsubfino{q}{\perp}{-0.2ex})}{\peqsub{V}{X}^{{\raisemath{0.2ex}{\!\scriptscriptstyle\perp}}}}\right)\\
	FL&(\mathbf{v}_{(\peqsubfino{q}{\perp}{-0.2ex},q,X)})\!\left(\hspace*{-0.1ex}\mathbf{w}^3_{(\peqsubfino{q}{\perp}{-0.2ex},q,X)}\hspace*{-0.1ex}\right)=-\int_\Sigma \mathbb{W}^\alpha\Big(\varepsilon(\peqsub{n}{X})_\alpha\peqsubfino{\mathcal{H}}{\!\perp}{-0.1ex}+(\peqsubfino{e}{X}{-0.2ex})^b_\alpha\mathcal{H}_b\Big)\peqsub{\mathrm{vol}}{\Sigma}\,-{}\\
	&\hspace*{30ex}-\int_{\partial\Sigma}\mathbb{W}^\alpha\Big(\varepsilon(\peqsub{\theta}{X})_\alpha\peqsubfino{\mathcal{H}}{\!\perp}{-0.1ex}^{\scriptscriptstyle\partial\,B}+\varepsilon (\peqsub{n}{X})_\alpha\peqsubfino{\mathcal{H}}{\!\perp}{-0.1ex}^{\scriptscriptstyle\partial}+ (\peqsubfino{e}{X}{-0.2ex})^b_\alpha\mathcal{H}^{\scriptscriptstyle\partial}_b\Big)\peqsub{{\mathrm{vol}}}{\partial\Sigma}
	\end{align*}
	where $(\alpha,V)=\frac{1}{k!}\alpha_{a_1\cdots a_k}V^{a_1\cdots a_k}$ is the natural pairing. We can now define the canonical momenta
	\begin{align*}
	&\hspace*{2.2ex}p=\sqrt{\peqsubfino{\gamma}{X}{-0.2ex}}\,\peqsubfino{{\sharp_{\gamma}}}{\!X}{-0.2ex}\frac{v-\mathcal{L}_{\peqsub{\vec{v}}{X}^{\scriptscriptstyle\top}}q-\varepsilon\mathrm{d}(\peqsub{V}{X}^{{\raisemath{0.2ex}{\!\scriptscriptstyle\perp}}}\peqsubfino{q}{\perp}{-0.2ex})}{\peqsub{V}{X}^{{\raisemath{0.2ex}{\!\scriptscriptstyle\perp}}}}\\
	&\left.\begin{array}{l}
	\peqsubfino{p}{\perp}{-0.2ex}=0\\[1.4ex]
	(\peqsub{P}{\!X})_\alpha=-\varepsilon(\peqsub{n}{X})_\alpha\peqsubfino{\mathcal{H}}{\!\perp}{-0.1ex}(\peqsubfino{q}{\perp}{-0.2ex},q,X,p)-(\peqsubfino{e}{X}{-0.2ex})^b_\alpha\mathcal{H}_b(q,p)\\[1.9ex]
	(\peqsub{P^{\scriptscriptstyle\partial}}{X})_\alpha=-\varepsilon(\peqsub{\theta}{X})_\alpha\peqsubfino{\mathcal{H}}{\!\perp}{-0.1ex}^{\scriptscriptstyle\partial\,B}(\peqsubfino{q}{\perp}{-0.2ex},q,X)-\varepsilon(\peqsub{n}{X})_\alpha\peqsubfino{\mathcal{H}}{\!\perp}{-0.1ex}^{\scriptscriptstyle\partial}(q,p)-(\peqsubfino{e}{X}{-0.2ex})^b_\alpha\mathcal{H}^{\scriptscriptstyle\partial}_b(\peqsubfino{q}{\perp}{-0.2ex},q,X,p)
	\end{array}\right\}\text{ Constraints}
	\end{align*}
	where $\peqsubfino{{\sharp_{\gamma}}}{\!X}{-0.2ex}$ is the musical isomorphism \eqref{Mathematical background - equation - musical ishomorphisms} of $\peqsubfino{\gamma}{X}{-0.2ex}$ and
	\begin{align}\label{Parametrized theories - definitions - H's}
	\begin{split}
	&\mathcal{H}(q,p)=\Big(\imath_{-}\mathrm{d}q,p\Big)+\Big(\imath_{-}q,\delta p\Big)\in\Omega^1(\Sigma)\\
	&\peqsubfino{\mathcal{H}}{\!\perp}{-0.1ex}(\peqsubfino{q}{\perp}{-0.2ex},q,X,p)=\frac{1}{2\sqrt{\peqsubfino{\gamma}{X}{-0.2ex}}}\Big(p\hspace*{0.1ex},\hspace*{0.1ex}\underline{p}\Big)-\frac{\varepsilon\sqrt{\peqsubfino{\gamma}{X}{-0.2ex}}}{2}\big\langle\mathrm{d}q,\mathrm{d}q\big\rangle_{\!\peqsubfino{\gamma}{\!X}{-0.4ex}}+\varepsilon\Big(\peqsubfino{q}{\perp}{-0.2ex}\hspace*{0.1ex},\hspace*{0.1ex}\delta p\Big)\in\Cinf{\Sigma}\\
	&\mathcal{H}^{\scriptscriptstyle\partial}(q,p)=\left(\frac{\sqrt{\peqsubfino{\gamma^{\scriptscriptstyle\partial}}{\!X}{-0.3ex}}}{\sqrt{\peqsubfino{\gamma}{\!X}{-0.3ex}}}\frac{\peqsub{\nu}{X}}{|\peqsub{\vec{\nu}}{X}^{\,\scriptscriptstyle\top}|}\wedge\imath_{-}q,p\right)\in\Omega^1(\partial\Sigma)\\
	&\peqsubfino{\mathcal{H}}{\!\perp}{-0.1ex}^{\scriptscriptstyle\partial}(\peqsubfino{q}{\perp}{-0.2ex},p)=\varepsilon\,\left(\frac{\sqrt{\peqsubfino{\gamma^{\scriptscriptstyle\partial}}{\!X}{-0.3ex}}}{\sqrt{\peqsubfino{\gamma}{\!X}{-0.3ex}}}\frac{\peqsub{\nu}{X}}{|\peqsub{\vec{\nu}}{X}^{\,\scriptscriptstyle\top}|}\wedge\peqsubfino{q}{\perp}{-0.2ex},p\right)\in\Cinf{\partial\Sigma}\\
	&\peqsubfino{\mathcal{H}}{\!\perp}{-0.1ex}^{\scriptscriptstyle\partial\,B}(\peqsubfino{q}{\perp}{-0.2ex},q,X)=\frac{\sqrt{\peqsubfino{\gamma^{\scriptscriptstyle\partial}}{\!X}{-0.3ex}}}{2}\peqsub{b}{X}^2\Big[ \varepsilon\big\langle\peqsubfino{q^{\scriptscriptstyle\partial}}{\perp}{-0.2ex},\peqsubfino{q^{\scriptscriptstyle\partial}}{\perp}{-0.2ex}\big\rangle_{\!\peqsubfino{\gamma^{\scriptscriptstyle\partial}}{\!X}{-0.4ex}}\!+\big\langle \peqsubfino{q}{\partial}{-0.2ex},\peqsubfino{q}{\partial}{-0.2ex}\big\rangle_{\!\peqsubfino{\gamma^{\scriptscriptstyle\partial}}{\!X}{-0.4ex}}\Big]\in\Cinf{\partial\Sigma}
	\end{split}
	\end{align}
	We recall the useful notation
	\[\big(\imath_{-}\alpha,V\big)\in\Omega^1(\Sigma)\qquad\quad\text{given by}\quad\qquad\big(\imath_{-}\alpha,V\big)(\vec{v}):=\big(\imath_{\vec{v}}\alpha,V\big)\]
	and $\underline{p}=\flat_\gamma p$. In particular $(\delta p)^{a_2\cdots a_{k}}=-\nabla_cp^{ca_2\cdots a_k}$. It is important to realize that $\mathcal{H}$, $\mathcal{H}^{\scriptscriptstyle\partial}$ and $\peqsubfino{\mathcal{H}}{\!\perp}{-0.1ex}^{\scriptscriptstyle\partial}$ are independent of the embeddings because \eqref{Parametrized theories - equation - gamma/gamma nu/nu} does not depend on the embedding and neither does $\delta p$ according to lemma \ref{Appendix - lemma - divergencia densidad} and the fact that $p$ is a density. Furthermore, notice that in the definition of $(\peqsub{P^{\scriptscriptstyle\partial}}{X})_\alpha$, the $1$-form $\theta$ could be written in terms of $n$ and $e$, but it is much more convenient for the computations to leave the three terms as they have appeared.\separ
	
	 The primary constraint submanifold $\peqsub{\mathcal{P}}{\mathrm{EM}}:=FL(TQ)$ of the parametrized electromagnetism is then given by
	\begin{align*}
	\peqsub{\mathcal{P}}{\mathrm{EM}}&=\left\{(\peqsubfino{q}{\perp}{-0.2ex},q,X;\peqsubfino{p}{\perp}{-0.2ex},p,\,\peqsub{P}{\!X},\,\peqsub{P}{\!X}^{\scriptscriptstyle\partial})\in\peqsub{\widetilde{\mathcal{P}}}{\mathrm{EM}}\ \ /\ \ \peqsubfino{p}{\perp}{-0.2ex}=0\ \begin{array}{l} \peqsub{P}{\!X}=-\varepsilon\peqsubfino{\mathcal{H}}{\!\perp}{-0.1ex} n-e.\mathcal{H}\\ \peqsub{P}{\!X}^{\scriptscriptstyle\partial}=-\varepsilon\peqsubfino{\mathcal{H}}{\!\perp}{-0.1ex}^{\scriptscriptstyle\partial\,B}\theta-\varepsilon\peqsubfino{\mathcal{H}}{\!\perp}{-0.1ex}^{\scriptscriptstyle\partial}n-e.\mathcal{H}^{\scriptscriptstyle\partial}\end{array}\right\}=\\
	&=\Big\{(\peqsubfino{q}{\perp}{-0.2ex},q,X;0,p\coma-\varepsilon\peqsubfino{\mathcal{H}}{\!\perp}{-0.1ex} n-e.\mathcal{H}\coma -\varepsilon\peqsubfino{\mathcal{H}}{\!\perp}{-0.1ex}^{\scriptscriptstyle\partial\,B}\theta-\varepsilon\peqsubfino{\mathcal{H}}{\!\perp}{-0.1ex}^{\scriptscriptstyle\partial} n-e.\mathcal{H}^{\scriptscriptstyle\partial}) \Big\}\cong\\
	&\cong\Big\{(\peqsubfino{q}{\perp}{-0.2ex},q,X,p)\Big\}=:\mathcal{P}
	\end{align*}
	Notice that $\peqsub{\mathcal{P}}{\mathrm{EM}}$ is a ``graph'' analogous to $\{(x,y,f(y,x))\in\R^3\}$ where the role of $f$ is played by the maps $\peqsub{P}{X}$ and $\peqsub{P^{\scriptscriptstyle\partial}}{X}$. We define the inclusion $\peqsubfino{\jmath}{\mathrm{EM}}{-0.2ex}:\mathcal{P}\hookrightarrow \peqsub{\widetilde{\mathcal{P}}}{\mathrm{EM}}$ that allows us to pullback the induced form $\widetilde{\Omega}^{\scriptscriptstyle\mathrm{EM}}=\jmath^*\Omega$ of $\peqsub{\widetilde{\mathcal{P}}}{\mathrm{EM}}$, given by equation \eqref{Parametrized theories - equation - tilde Omega}, to $\mathcal{P}$ in order to define $\omega^{\scriptscriptstyle\mathrm{EM}}:=\peqsub{\jmath}{\mathrm{EM}\,}^*\widetilde{\Omega}^{\scriptscriptstyle\mathrm{EM}}$.\separ
	
	To compute $\omega^{\scriptscriptstyle\mathrm{EM}}$ we need the push-forward $(\peqsubfino{\jmath}{\mathrm{EM}}{-0.2ex})_*Y$ for every $Y=(\vecc[\scalebox{0.4}{$\perp$}]{Y}{q},\vecc{Y}{q},\vecc{Y}{X},\vecc{Y}{p})\in\mathfrak{X}(\mathcal{P})$ because $\omega^{\scriptscriptstyle\mathrm{EM}}(Y,Z):=(\widetilde{\Omega}^{\scriptscriptstyle\mathrm{EM}}\smallcirc \peqsubfino{\jmath}{\mathrm{EM}}{-0.2ex})((\peqsubfino{\jmath}{\mathrm{EM}}{-0.2ex})_*Y,(\peqsubfino{\jmath}{\mathrm{EM}}{-0.2ex})_*Z)$. In is interesting to take the analogy with the graph manifold a bit further and notice that if we consider a vector $v=(v_x,v_y)\in\R^2$, it can be lifted to a vector
	\[\left(v_x\coma v_y\coma \parc[f]{x}v_x+\parc[f]{y}v_y\right)\in\R^3\]
	
	\centerline{\quad\qquad\qquad\qquad\includegraphics[clip,trim=5ex 0ex 0ex 20ex,width=.58\linewidth]{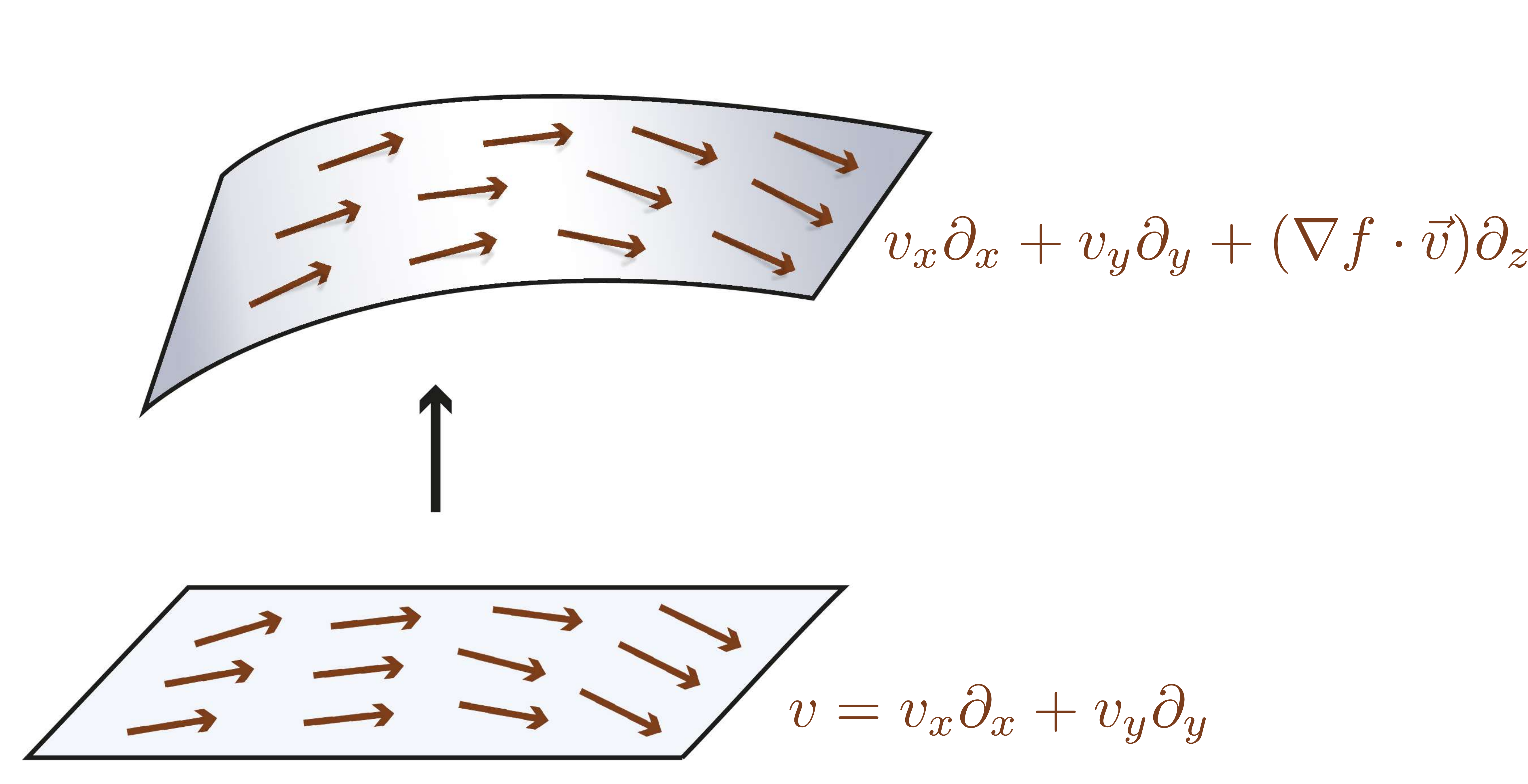}}
	
	\vspace*{3ex}
	
	In our case we have
	\[(\peqsubfino{\jmath}{\mathrm{EM}}{-0.2ex})_*Y=\Big(\vecc[\scalebox{0.4}{$\perp$}]{Y}{q},\vecc{Y}{q},\vecc{Y}{X},\vecc{Y}{p},-\peqsub{D}{Y}\big(\varepsilon n \peqsubfino{\mathcal{H}}{\!\perp}{-0.1ex}+e.\mathcal{H}\big),-\peqsub{D}{Y}\big(\varepsilon \theta \peqsubfino{\mathcal{H}}{\!\perp}{-0.1ex}^{\scriptscriptstyle\partial\,B}+\varepsilon n\peqsubfino{\mathcal{H}}{\!\perp}{-0.1ex}^{\scriptscriptstyle\partial}+e.\mathcal{H}^{\scriptscriptstyle\partial}\big)\Big)\in\mathfrak{X}(\peqsub{\widetilde{\mathcal{P}}}{\mathrm{EM}})\]	
	where the $\peqsub{D}{Y}$ means that we take the variations with respect to all the variables $(\peqsubfino{q}{\perp}{-0.2ex},q,X,p)$, as in the chain rule, in the direction $Y=(\vecc[\scalebox{0.4}{$\perp$}]{Y}{q},\vecc{Y}{q},\vecc{Y}{X},\vecc{Y}{p})$. Notice that the variables $(\peqsubfino{q}{\perp}{-0.2ex},q,p)$ belong to vector spaces, so their variations pose no problem. However, the variation with respect to the last variable, the embedding, is much trickier because $\mathrm{Emb}(\Sigma,M)$ is non-linear in general (see section \ref{Mathematical background - Section - Space of embeddings} of chapter \ref{Chapter - Mathematical background}). Remembering that $\omega^{\scriptscriptstyle\mathrm{EM}}(Y,Z):=(\widetilde{\Omega}^{\scriptscriptstyle\mathrm{EM}}\smallcirc \peqsubfino{\jmath}{\mathrm{EM}}{-0.2ex})((\peqsubfino{\jmath}{\mathrm{EM}}{-0.2ex})_*Y,(\peqsubfino{\jmath}{\mathrm{EM}}{-0.2ex})_*Z)$ we have
	\begin{align}\label{Parametrized theories - equation - omega final}
	\begin{split}
	&\omega^{\scriptscriptstyle\mathrm{EM}}_{(\peqsubfino{q}{\perp}{-0.2ex},q,X,p)}\Big((\vecc[\scalebox{0.4}{$\perp$}]{Y}{q},\vecc{Y}{q},\vecc{Y}{X},\vecc{Y}{p}),(\vecc[\scalebox{0.4}{$\perp$}]{Z}{q},\vecc{Z}{q},\vecc{Z}{X},\vecc{Z}{p})\Big)
	=\\ &\qquad=\int_\Sigma\left[\Big(\vecc{Y}{q}\coma \vecc{Z}{p}\Big)-\Big(\vecc{Z}{q}\coma \vecc{Y}{p}\Big)-\peqsub{D}{Z}(\varepsilon\peqsubfino{\mathcal{H}}{\!\perp}{-0.1ex}n+e.\mathcal{H})_\alpha\vecc{Y}{X}^\alpha+\peqsub{D}{Y}(\varepsilon\peqsubfino{\mathcal{H}}{\!\perp}{-0.1ex}n+e.\mathcal{H})_\alpha\vecc{Z}{X}^\alpha\right]\peqsub{\mathrm{vol}}{\Sigma}-{}\\
	&\qquad\phantom{=}-\int_{\partial\Sigma}\left[\peqsub{D}{Z}(\varepsilon \theta \peqsubfino{\mathcal{H}}{\!\perp}{-0.1ex}^{\scriptscriptstyle\partial\,B}+\varepsilon\peqsubfino{\mathcal{H}}{\!\perp}{-0.1ex}^{\scriptscriptstyle\partial}n+e.\mathcal{H}^{\scriptscriptstyle\partial})_\alpha\vecc{Y}{X}^\alpha-\peqsub{D}{Y}(\varepsilon \theta \peqsubfino{\mathcal{H}}{\!\perp}{-0.1ex}^{\scriptscriptstyle\partial\,B}+\varepsilon\peqsubfino{\mathcal{H}}{\!\perp}{-0.1ex}^{\scriptscriptstyle\partial}n+e.\mathcal{H}^{\scriptscriptstyle\partial})_\alpha\vecc{Z}{X}^\alpha\right]\peqsub{\mathrm{vol}}{\partial\Sigma}     
	\end{split}
	\end{align}
	In page \pageref{Appendix - computation - omega} of the appendix we give the key steps that lead to
	\begin{align}\label{Parametrized theories - equation - omega final explicita}
	\begin{split}
	&\omega^{\scriptscriptstyle\mathrm{EM}}_{(\peqsubfino{q}{\perp}{-0.2ex},q,X,p)}\Big((\vecc[\scalebox{0.4}{$\perp$}]{Y}{q},\vecc{Y}{q},\vecc{Y}{X},\vecc{Y}{p}),(\vecc[\scalebox{0.4}{$\perp$}]{Z}{q},\vecc{Z}{q},\vecc{Z}{X},\vecc{Z}{p})\Big)=\\
	&\ =\int_\Sigma\varepsilon Z^{{\raisemath{0.2ex}{\scriptscriptstyle\perp\!}}} \Big(\vecc[\perp]{Y}{q}\hspace*{0.1ex}-\mathcal{L}_{\vec{y}^{\scriptscriptstyle\top}}\peqsubfino{q}{\perp}{-0.2ex}+\peqsub{\imath}{\mathrm{d}Y^{\!\scriptscriptstyle\perp}}q\coma\delta p\Big)\peqsub{\mathrm{vol}}{\Sigma}+{}\\
	&\ \phantom{=}+\int_\Sigma\left(\vecc{Y}{q}-\mathcal{L}_{\vec{y}^{\scriptscriptstyle\top}}q-Y^{{\raisemath{0.2ex}{\scriptscriptstyle\perp\!}}}\frac{\underline{p}}{\sqrt{\gamma}}-\varepsilon\mathrm{d}( Y^{{\raisemath{0.2ex}{\scriptscriptstyle\perp\!}}}\peqsubfino{q}{\perp}{-0.2ex})\coma\vecc{Z}{p}-\mathcal{L}_{\vec{z}^{\scriptscriptstyle\top}}p-\varepsilon\sqrt{\gamma}\,\sharp_\gamma\delta(Z^{{\raisemath{0.2ex}{\scriptscriptstyle\perp\!}}}\mathrm{d}q)\right)\peqsub{\mathrm{vol}}{\Sigma}-{}\\
	&\ \phantom{=}-\int_{\partial\Sigma}\varepsilon Z^{{\raisemath{0.2ex}{\scriptscriptstyle\perp\!}}}\peqsubfino{{\left\langle\peqsub{\jmath}{\partial}^*\!\!\left(\vecc{Y}{q}-\mathcal{L}_{\vec{y}^{\scriptscriptstyle\top}}q-Y^{{\raisemath{0.2ex}{\scriptscriptstyle\perp\!}}}\frac{\underline{p}}{\sqrt{\gamma}}-\varepsilon\mathrm{d}( Y^{{\raisemath{0.2ex}{\scriptscriptstyle\perp\!}}}\peqsubfino{q}{\perp}{-0.2ex})\right)\!\!\coma\jmath^*(\imath_{\vec{\nu}}\mathrm{d}q)+\peqsub{\nu}{\perp}\frac{\underline{p}^{\scriptscriptstyle\partial}}{\sqrt{\gamma}}-\varepsilon b^2\peqsubfino{q}{\partial}{-0.2ex}\right\rangle_{\!\!\!\gamma}}}{\!\partial}{-.3ex}\frac{\peqsubfino{{\mathrm{vol}_\gamma}}{\!\partial\!X}{-0.4ex}}{|\vec{\nu}^{\,\scriptscriptstyle\top}|}+{}\\
	&\ \phantom{=}+\int_{\partial\Sigma}\varepsilon Z^{{\raisemath{0.2ex}{\scriptscriptstyle\perp\!}}} \peqsubfino{{\left\langle\peqsub{\jmath}{\partial}^*\big(\vecc[\perp]{Y}{q}\hspace*{0.1ex}-\mathcal{L}_{\vec{y}^{\scriptscriptstyle\top}}\peqsubfino{q}{\perp}{-0.2ex}+\peqsub{\imath}{\mathrm{d}Y^{\!\scriptscriptstyle\perp}}q\big)\coma\peqsub{\jmath}{\partial}^*\!\left(\frac{\imath_{\vec{\nu}}\underline{p}}{\sqrt{\gamma}}\right)+b^2\peqsubfino{q^{\scriptscriptstyle\partial}}{\perp}{-0.2ex}\right\rangle_{\!\!\!\gamma}}}{\!\partial}{-.3ex}\frac{\peqsubfino{{\mathrm{vol}_\gamma}}{\!\partial\!X}{-0.4ex}}{|\vec{\nu}^{\,\scriptscriptstyle\top}|}-(Y\leftrightarrow Z)
	\end{split}
	\end{align}
	It is interesting to mention that only the lapse, and not the shift, appears in the boundary integrals.
	
	\section{Hamiltonian formulation}\label{Parametrized theories - section - Hamiltonian}
	
	\subsection*{Obtaining the Hamiltonian}\trassub
	
	As the Lagrangian \eqref{Parametrized theories - equation - Lagrangian} is homogeneous of degree $1$ in the velocities, it is clear that the energy $E=D_2L-L=0$, thus the Hamiltonian is also zero. Notice that this does not imply that the dynamics is trivial, it does imply however that the dynamics is purely gauge in the sense that it evolves in the degenerate directions of the presymplectic form $\omega^{\scriptscriptstyle\mathrm{EM}}$.
	
	\subsection*{GNH algorithm}\trassub
	
	The Hamiltonian vector field $Y\in\mathfrak{X}(\mathcal{P})$ is given by the Hamilton equation
	\begin{equation}\label{Parametrized theories - equation - Hamiltonina equation =0}
	\omega^{\scriptscriptstyle\mathrm{EM}}_{(\peqsubfino{q}{\perp}{-0.2ex},q,X,p)}\Big((\vecc[\scalebox{0.4}{$\perp$}]{Y}{q},\vecc{Y}{q},\vecc{Y}{X},\vecc{Y}{p})\coma(\vecc[\scalebox{0.4}{$\perp$}]{Z}{q},\vecc{Z}{q},\vecc{Z}{X},\vecc{Z}{p})\Big)=\mathrm{d}H(\vecc[\scalebox{0.4}{$\perp$}]{Z}{q},\vecc{Z}{q},\vecc{Z}{X},\vecc{Z}{p})=0\end{equation}
	for every $Z\in\mathfrak{X}(\mathcal{P})$. Taking in particular $(0,0,0,\vecc{Z}{p})$, $(0,\vecc{Z}{q},0,0)$, $(\vecc[\perp]{Z}{q},0,0,0)$ and $(0,0,Z^{{\raisemath{0.2ex}{\scriptscriptstyle\perp\!}}}\mathbbm{n},0)$ vanishing at the boundary leads to the conditions on the bulk
	\begin{align}\label{Parametrized theories - equation - H vector field delta p=0}
	\begin{split}
	&\vecc{Y}{q}=\mathcal{L}_{\vec{y}^{\scriptscriptstyle\top}}q+Y^{{\raisemath{0.2ex}{\scriptscriptstyle\perp\!}}}\frac{\underline{p}}{\sqrt{\gamma}}+\varepsilon\mathrm{d}( Y^{{\raisemath{0.2ex}{\scriptscriptstyle\perp\!}}}\peqsubfino{q}{\perp}{-0.2ex})\\[1ex]
	&\vecc{Y}{p}=\mathcal{L}_{\vec{y}^{\scriptscriptstyle\top}}p+\varepsilon\sqrt{\gamma}\,\sharp_\gamma\delta(Y^{{\raisemath{0.2ex}{\scriptscriptstyle\perp\!}}}\mathrm{d}q)\\[1.8ex]
	&Y^{{\raisemath{0.2ex}{\scriptscriptstyle\perp\!}}}\delta p=0\\[1ex]
	&\Big(\vecc[\perp]{Y}{q}\hspace*{0.1ex}-\mathcal{L}_{\vec{y}^{\scriptscriptstyle\top}}\peqsubfino{q}{\perp}{-0.2ex}+\peqsub{\imath}{\mathrm{d}Y^{\!\scriptscriptstyle\perp}}q\hspace*{0.1ex},\hspace*{0.1ex}\delta p\Big)=0
	\end{split}
	\end{align}
	Let us focus on a remarkable feature of the previous expressions: there is a bifurcation in the outcome of the GNH algorithm. First notice that in the non-parametrized version we have by construction $Y^{{\raisemath{0.2ex}{\scriptscriptstyle\perp\!}}}>0$ in order to have a foliation, thus $\delta p=0$ ---known as the \textbf{Gauss constraint}\index{Gauss constraint}--- and the component $\vecc[\perp]{Y}{q}$ is arbitrary, which is the typical gauge freedom of electromagnetism.\separ
	
	In the parametrized version, on the other hand, we have to solve the equation $Y^{{\raisemath{0.2ex}{\scriptscriptstyle\perp\!}}}\delta p=0$ for $Y^{{\raisemath{0.2ex}{\scriptscriptstyle\perp\!}}}$. The solution depends, obviously, on the support of $\delta p$: wherever it vanishes, $Y^{{\raisemath{0.2ex}{\scriptscriptstyle\perp\!}}}$ is arbitrary, wherever it does not vanish, $Y^{{\raisemath{0.2ex}{\scriptscriptstyle\perp\!}}}=0$. Therefore $\delta p=0$ is no longer a constraint! What is happening is that if $\delta p\neq0$ then the foliation cannot ``advance in time'' but there are still solutions to the field equations.\newpage
	
	$\blacktriangleright$ Wherever $\delta p\neq0$ (in particular $k\geq1$)
	\begin{align}
	\begin{split}
	&\vecc[\perp]{Y}{q}\hspace*{0.1ex}=\mathcal{L}_{\vec{y}^{\scriptscriptstyle\top}}\peqsubfino{q}{\perp}{-0.2ex}\\[1ex]
	&\vecc{Y}{q}=\mathcal{L}_{\vec{y}^{\scriptscriptstyle\top}}q\\[1ex]
	&\vecc{Y}{p}=\mathcal{L}_{\vec{y}^{\scriptscriptstyle\top}}p\\[1ex]
	&Y^{{\raisemath{0.2ex}{\scriptscriptstyle\perp\!}}}=0\\[.5ex]
	&y^{{\raisemath{-0.1ex}{\scriptscriptstyle\top\!}}}\text{ arbitrary}
	\end{split}\label{EM - equation - solucion delta p not 0}
	\end{align}
	Here we have the action of the group of diffeomorphism $\mathrm{Diff}(\Sigma)$ acting through its infinitesimal version, the Lie derivative. It is interesting to notice that the fact that $y^{{\raisemath{-0.1ex}{\scriptscriptstyle\top\!}}}$ is arbitrary reflects the ``gauge freedom'' we mentioned when we introduced the shape space in page \pageref{Mathematical background - definition - shapce space}.\separ
	
	In this case the GNH algorithm stops because we have found the most general solution with no constraint.\separ

	$\blacktriangleright$ Wherever $\delta p =0$
	\begin{align}\label{Parametrized theories - equations - Hamiltonian vector field delta p=0}
	\begin{split}
	&\vecc[\perp]{Y}{q}\hspace*{0.1ex}\text{ arbitrary}\\[.9ex]
	&\vecc{Y}{q}=\mathcal{L}_{\vec{y}^{\scriptscriptstyle\top}}q+Y^{{\raisemath{0.2ex}{\scriptscriptstyle\perp\!}}}\frac{\underline{p}}{\sqrt{\gamma}}+\varepsilon\mathrm{d}( Y^{{\raisemath{0.2ex}{\scriptscriptstyle\perp\!}}}\peqsubfino{q}{\perp}{-0.2ex})\\[1ex]
	&\vecc{Y}{p}=\mathcal{L}_{\vec{y}^{\scriptscriptstyle\top}}p+\varepsilon\sqrt{\gamma}\,\sharp_\gamma\delta(Y^{{\raisemath{0.2ex}{\scriptscriptstyle\perp\!}}}\mathrm{d}q)\\[1.7ex]
	&Y^{{\raisemath{0.2ex}{\scriptscriptstyle\perp\!}}}\text{ arbitrary}\\[.8ex]
	&y^{{\raisemath{-0.1ex}{\scriptscriptstyle\top\!}}}\text{ arbitrary}
	\end{split}
	\end{align}
	Now we have, not only the ``gauge freedom'' given by the arbitrariness of $y^{{\raisemath{-0.1ex}{\scriptscriptstyle\top\!}}}$ that leaves unchanged the shape of the embedding, but we also have that $Y^{{\raisemath{0.2ex}{\scriptscriptstyle\perp\!}}}$ is also arbitrary, which allows to recover the invariance under space-time diffeomorphism of the ``physical solutions''. Furthermore, we have the usual gauge freedom of the electromagnetism corresponding to the $\peqsub{q}{\perp}$ factor.\separ
		 
	In this case, the GNH algorithm goes on by requiring that the Hamiltonian vector field $Y\in\mathfrak{X}(\mathcal{P})$ that we have found in the first step is tangent to
	\[\mathcal{P}_2=\Big\{(\peqsubfino{q}{\perp}{-0.2ex},q,X,p)\in\mathcal{P}\ \ /\ \ \delta p=0\Big\}\]
	Notice that this is always the case because
	\[\peqsub{D}{Y}\big(\delta p\big)\overset{\eqref{Appendix - lemma - divergencia densidad}}{=}\delta\big(\peqsub{D}{Y}p\big)=\delta\vecc{Y}{p}=\delta\mathcal{L}_{\vec{y}^{\scriptscriptstyle\top}}p+\varepsilon\sqrt{\gamma}\,\sharp_\gamma\delta^2(Y^{{\raisemath{0.2ex}{\scriptscriptstyle\perp\!}}}\mathrm{d}q)\updown{\eqref{Appendix - definition - delta^2=0}}{\eqref{Appendix - definition - delta^2=0}}{=}\mathcal{L}_{\vec{y}^{\scriptscriptstyle\top}}(\delta p)\]
	where in the last equality we have used the fact that the Lie derivative is a variation in the direction $\mathbb{Y}=\tau.\vec{y}$, so it commutes with the codifferential as in the first equality. Of course we have that if $\delta p=0$ at some point it will remain so for ever (so the algorithm also stops at this first step for the bulk) and, therefore, the two cases are complementary.\separ
	
	In order to shed some light on the previous bifurcation let us consider a simple finite dimensional analog. Consider the homogeneous system
	\[\begin{pmatrix}
	x & 0\\ 0&1\end{pmatrix}\begin{pmatrix}a\\b\end{pmatrix}=\begin{pmatrix}0\\0\end{pmatrix}\]
	whose solution is $ax=0$ and $b=0$. Notice that $a=0$ if $x\neq0$ and $a$ is arbitrary otherwise. Of course here we see that the rank of the matrix depends on $x$ and, therefore, the dimensionality of the kernel depends also on $x$ due to the rank theorem. In the infinite dimensional case, however, such theorem does not hold and the rank (which is infinite) cannot measure the dimensionality of the kernel. We see nonetheless that we obtain in fact a different number of solutions depending on the point of $\mathcal{P}$.\separ
	
	$\blacktriangleright$ Boundary\separprevia
	
	Let us now keep going with the GNH algorithm to see what happens at the boundary. We have already solved \eqref{Parametrized theories - equation - Hamiltonina equation =0} at the bulk, so only the boundary integrals remain. Note that if $\delta p\neq0$ all of them are zero due to \eqref{EM - equation - solucion delta p not 0} and the fact that $Y^{{\raisemath{0.2ex}{\scriptscriptstyle\perp\!}}}=0$, thus we are done. Let us then focus on the case $\delta p=0$, in which we still have that some of the integrals are zero by continuity of \eqref{Parametrized theories - equations - Hamiltonian vector field delta p=0}, so we have to solve 
	\begin{align*}
	0&=\int_{\partial\Sigma} \peqsubfino{{\left\langle\peqsub{\jmath}{\partial}^*\!\left(\vecc{Z}{q}-\mathcal{L}_{\vec{z}^{\scriptscriptstyle\top}}q-Z^{{\raisemath{0.2ex}{\scriptscriptstyle\perp\!}}}\frac{\underline{p}}{\sqrt{\gamma}}-\varepsilon\mathrm{d}( Z^{{\raisemath{0.2ex}{\scriptscriptstyle\perp\!}}}\peqsubfino{q}{\perp}{-0.2ex})\right)\!\coma Y^{{\raisemath{0.2ex}{\scriptscriptstyle\perp\!}}}\left(\peqsub{\jmath}{\partial}^*(\imath_{\vec{\nu}}\mathrm{d}q)+\peqsub{\nu}{\perp}\frac{\underline{p}^{\scriptscriptstyle\partial}}{\sqrt{\gamma}}-\varepsilon b^2\peqsubfino{q}{\partial}{-0.2ex}\right)\right\rangle_{\!\!\!\gamma}}}{\!\partial}{-.3ex}\frac{\peqsubfino{{\mathrm{vol}_\gamma}}{\!\partial\!X}{-0.4ex}}{|\vec{\nu}^{\,\scriptscriptstyle\top}|}+{}\\
	&\phantom{=}+\int_{\partial\Sigma}  \peqsubfino{{\left\langle Z^{{\raisemath{0.2ex}{\scriptscriptstyle\perp\!}}}\peqsub{\jmath}{\partial}^*\big(\vecc[\perp]{Y}{q}\hspace*{0.1ex}-\mathcal{L}_{\vec{y}^{\scriptscriptstyle\top}}\peqsubfino{q}{\perp}{-0.2ex}+\peqsub{\imath}{\mathrm{d}Y^{\!\scriptscriptstyle\perp}}q\big)-Y^{{\raisemath{0.2ex}{\scriptscriptstyle\perp\!}}}\peqsub{\jmath}{\partial}^*\big(\vecc[\perp]{Z}{q}\hspace*{0.1ex}-\mathcal{L}_{\vec{z}^{\scriptscriptstyle\top}}\peqsubfino{q}{\perp}{-0.2ex}+\peqsub{\imath}{\mathrm{d}Z^{\!\scriptscriptstyle\perp}}q\big)\!\coma\peqsub{\jmath}{\partial}^*\!\left(\frac{\imath_{\vec{\nu}}\underline{p}}{\sqrt{\gamma}}\right)+b^2\peqsubfino{q^{\scriptscriptstyle\partial}}{\perp}{-0.2ex}\right\rangle_{\!\!\!\gamma}}}{\!\partial}{-.3ex}\frac{\peqsubfino{{\mathrm{vol}_\gamma}}{\!\partial\!X}{-0.4ex}}{|\vec{\nu}^{\,\scriptscriptstyle\top}|}
	\end{align*}
	for every $Z$. Here we have again bifurcations although now there are many more possibilities. Notice that the case $Y^{{\raisemath{0.2ex}{\scriptscriptstyle\perp\!}}}=0$ is easy to solve and less interesting from a physical point of view, so we will assume from now on that $Y^{{\raisemath{0.2ex}{\scriptscriptstyle\perp\!}}}$ is arbitrary at the boundary. Let us sketch a solution to the previous equation.\separ
	
	$\blacktriangleright$ Dirichlet $\peqsub{\mathcal{P}}{D1}:=\Big\{(\peqsubfino{q}{\perp}{-0.2ex},q,X,p)\in\mathcal{P}\ \ /\ \ \delta p=0\ \text{ and }\ \peqsubfino{q^{\scriptscriptstyle\partial}}{\perp}{-0.2ex}=0\Big\}$\separprevia
	
	In this case the component $\peqsubfino{q}{\perp}{-0.2ex}$ of the tangent vectors to $\peqsub{\mathcal{P}}{D1}$ is also zero at the boundary and the previous equation reduces to
	\begin{align*}
	0&=\int_{\partial\Sigma} \peqsubfino{{\left\langle\peqsub{\jmath}{\partial}^*\!\left(\vecc{Z}{q}-\mathcal{L}_{\vec{z}^{\scriptscriptstyle\top}}q-Z^{{\raisemath{0.2ex}{\scriptscriptstyle\perp\!}}}\frac{\underline{p}}{\sqrt{\gamma}}-\varepsilon Z^{{\raisemath{0.2ex}{\scriptscriptstyle\perp\!}}}\mathrm{d} \peqsubfino{q}{\perp}{-0.2ex}\right)\!\coma Y^{{\raisemath{0.2ex}{\scriptscriptstyle\perp\!}}}\left(\peqsub{\jmath}{\partial}^*(\imath_{\vec{\nu}}\mathrm{d}q)+\peqsub{\nu}{\perp}\frac{\underline{p}^{\scriptscriptstyle\partial}}{\sqrt{\gamma}}-\varepsilon b^2\peqsubfino{q}{\partial}{-0.2ex}\right)\right\rangle_{\!\!\!\gamma}}}{\!\partial}{-.3ex}\frac{\peqsubfino{{\mathrm{vol}_\gamma}}{\!\partial\!X}{-0.4ex}}{|\vec{\nu}^{\,\scriptscriptstyle\top}|}+{}\\
	&\phantom{=}+\int_{\partial\Sigma}  \peqsubfino{{\left\langle Z^{{\raisemath{0.2ex}{\scriptscriptstyle\perp\!}}}\peqsub{\jmath}{\partial}^*\big(-\mathcal{L}_{\vec{y}^{\scriptscriptstyle\top}}\peqsubfino{q}{\perp}{-0.2ex}+\peqsub{\imath}{\mathrm{d}Y^{\!\scriptscriptstyle\perp}}q\big)-Y^{{\raisemath{0.2ex}{\scriptscriptstyle\perp\!}}}\peqsub{\jmath}{\partial}^*\big(-\mathcal{L}_{\vec{z}^{\scriptscriptstyle\top}}\peqsubfino{q}{\perp}{-0.2ex}+\peqsub{\imath}{\mathrm{d}Z^{\!\scriptscriptstyle\perp}}q\big)\!\coma\peqsub{\jmath}{\partial}^*\!\left(\frac{\imath_{\vec{\nu}}\underline{p}}{\sqrt{\gamma}}\right)\right\rangle_{\!\!\!\gamma}}}{\!\partial}{-.3ex}\frac{\peqsubfino{{\mathrm{vol}_\gamma}}{\!\partial\!X}{-0.4ex}}{|\vec{\nu}^{\,\scriptscriptstyle\top}|}
	\end{align*}
	If we want $Y^{{\raisemath{0.2ex}{\scriptscriptstyle\perp\!}}}$ to be arbitrary we obtain the conditions
	\begin{align*}
	&\peqsub{\jmath}{\partial}^*(\imath_{\vec{\nu}}\mathrm{d}q)+\peqsub{\nu}{\perp}\frac{\underline{p}^{\scriptscriptstyle\partial}}{\sqrt{\gamma}}-\varepsilon b^2\peqsubfino{q}{\partial}{-0.2ex}=0\\
	&\peqsub{\jmath}{\partial}^*\!\left(\imath_{\vec{\nu}}\underline{p}\right)=0
	\end{align*}
	
	$\blacktriangleright$ Dirichlet $\peqsub{\mathcal{P}}{D2}:=\Big\{(\peqsubfino{q}{\perp}{-0.2ex},q,X,p)\in\mathcal{P}\ \ /\ \ \delta p=0\text{ and }\peqsubfino{q}{\partial}{-0.2ex}=0\Big\}$\separprevia
	
	Now the component $q$ of the tangent vectors to $\peqsub{\mathcal{P}}{D2}$ is zero at the boundary. In particular we have the condition that
	\begin{equation}\vecc{Y}{q}=\mathcal{L}_{\vec{y}^{\scriptscriptstyle\top}}q+Y^{{\raisemath{0.2ex}{\scriptscriptstyle\perp\!}}}\frac{\underline{p}}{\sqrt{\gamma}}+\varepsilon\mathrm{d}( Y^{{\raisemath{0.2ex}{\scriptscriptstyle\perp\!}}}\peqsubfino{q}{\perp}{-0.2ex})
	\end{equation}
	is zero at the boundary. Meanwhile, the equation that we have to solve reduces to
	\begin{align*}
	0&=\int_{\partial\Sigma} \peqsubfino{{\left\langle\peqsub{\jmath}{\partial}^*\!\left(-\mathcal{L}_{\vec{z}^{\scriptscriptstyle\top}}q-Z^{{\raisemath{0.2ex}{\scriptscriptstyle\perp\!}}}\frac{\underline{p}}{\sqrt{\gamma}}-\varepsilon\mathrm{d}( Z^{{\raisemath{0.2ex}{\scriptscriptstyle\perp\!}}}\peqsubfino{q}{\perp}{-0.2ex})\right)\!\coma Y^{{\raisemath{0.2ex}{\scriptscriptstyle\perp\!}}}\left(\peqsub{\jmath}{\partial}^*(\imath_{\vec{\nu}}\mathrm{d}q)+\peqsub{\nu}{\perp}\frac{\underline{p}^{\scriptscriptstyle\partial}}{\sqrt{\gamma}}\right)\right\rangle_{\!\!\!\gamma}}}{\!\partial}{-.3ex}\frac{\peqsubfino{{\mathrm{vol}_\gamma}}{\!\partial\!X}{-0.4ex}}{|\vec{\nu}^{\,\scriptscriptstyle\top}|}+{}\\
	&\phantom{=}+\int_{\partial\Sigma}  \peqsubfino{{\left\langle Z^{{\raisemath{0.2ex}{\scriptscriptstyle\perp\!}}}\peqsub{\jmath}{\partial}^*\big(\vecc[\perp]{Y}{q}\hspace*{0.1ex}-\mathcal{L}_{\vec{y}^{\scriptscriptstyle\top}}\peqsubfino{q}{\perp}{-0.2ex}\big)-Y^{{\raisemath{0.2ex}{\scriptscriptstyle\perp\!}}}\peqsub{\jmath}{\partial}^*\big(\vecc[\perp]{Z}{q}\hspace*{0.1ex}-\mathcal{L}_{\vec{z}^{\scriptscriptstyle\top}}\peqsubfino{q}{\perp}{-0.2ex}\big)\!\coma\peqsub{\jmath}{\partial}^*\!\left(\frac{\imath_{\vec{\nu}}\underline{p}}{\sqrt{\gamma}}\right)+b^2\peqsubfino{q^{\scriptscriptstyle\partial}}{\perp}{-0.2ex}\right\rangle_{\!\!\!\gamma}}}{\!\partial}{-.3ex}\frac{\peqsubfino{{\mathrm{vol}_\gamma}}{\!\partial\!X}{-0.4ex}}{|\vec{\nu}^{\,\scriptscriptstyle\top}|}
	\end{align*}
	We can now impose the (sufficient) conditions\allowdisplaybreaks[0]
	\begin{align*}
	&\peqsub{\jmath}{\partial}^*\!\left(\frac{\imath_{\vec{\nu}}\underline{p}}{\sqrt{\gamma}}\right)+b^2\peqsubfino{q^{\scriptscriptstyle\partial}}{\perp}{-0.2ex}=0\\
	&\peqsub{\jmath}{\partial}^*(\imath_{\vec{\nu}}\mathrm{d}q)+\peqsub{\nu}{\perp}\frac{\underline{p}^{\scriptscriptstyle\partial}}{\sqrt{\gamma}}=0
	\end{align*}\allowdisplaybreaks
	
	$\blacktriangleright$ Dirichlet $\peqsub{\mathcal{P}}{D3}:=\Big\{(\peqsubfino{q}{\perp}{-0.2ex},q,X,p)\in\mathcal{P}\ \ /\ \ \delta p=0\ \ \peqsubfino{q^{\scriptscriptstyle\partial}}{\perp}{-0.2ex}=0\ \text{ and } \ \peqsubfino{q}{\partial}{-0.2ex}=0\Big\}$
	
	\begin{align*}
	&\peqsub{\jmath}{\partial}^*\!\left(\imath_{\vec{\nu}}\underline{p}\right)=0\\
	&\peqsub{\jmath}{\partial}^*(\imath_{\vec{\nu}}\mathrm{d}q)+\peqsub{\nu}{\perp}\frac{\underline{p}^{\scriptscriptstyle\partial}}{\sqrt{\gamma}}=0\\
	&\peqsub{\jmath}{\partial}^*\!\Big(\mathcal{L}_{\vec{y}^{\scriptscriptstyle\top}}q\Big)+Y^{{\raisemath{0.2ex}{\scriptscriptstyle\perp\!}}}\frac{\underline{p}^{\scriptscriptstyle\partial}}{\sqrt{\gamma}}+\varepsilon Y^{{\raisemath{0.2ex}{\scriptscriptstyle\perp\!}}} \mathrm{d}\peqsubfino{q^{\scriptscriptstyle\partial}}{\perp}{-0.2ex}=0
	\end{align*}
	
	$\blacktriangleright$ Robin-like $\mathcal{P}$
	
	\begin{align*}
	&\peqsub{\jmath}{\partial}^*(\imath_{\vec{\nu}}\mathrm{d}q)+\peqsub{\nu}{\perp}\frac{\underline{p}^{\scriptscriptstyle\partial}}{\sqrt{\gamma}}-\varepsilon b^2\peqsubfino{q}{\partial}{-0.2ex}=0\\
	&\peqsub{\jmath}{\partial}^*\!\left(\frac{\imath_{\vec{\nu}}\underline{p}}{\sqrt{\gamma}}\right)+b^2\peqsubfino{q^{\scriptscriptstyle\partial}}{\perp}{-0.2ex}=0
	\end{align*}
	
	In all the previous cases we should check that the dynamics preserved the corresponding boundary conditions. In general an infinite chain of conditions appear. We see how the inclusion of the boundary makes the theory much richer but also much more difficult to handle.

%% file: 6_parametrized_scalar.tex
\chapter{Parametrized scalar field revisited}\thispagestyle{empty}\label{Chapter - parametrized scalar}

\starwars[Princess Leia]{Aren't you a little short for a storm trooper?}{A New Hope}

\section{Introduction}\label{scalar- section - scalar field}

In the previous chapter we study the parametrized EM field theory for a general $k$-form $A$. It is of great interest to consider the simplest case of a parametrized field theory, namely the scalar field $k=0$. We already saw in the previous chapter that in that case the variations with respect to the diffeomorphism add no dynamics. Furthermore, the study of the boundary conditions can be carried out in a more systematic way (see \cite{margalef2016hamiltonian} for some more details).

\section{Action of the theory}

We consider the parametrized scalar field with boundary $S:\Omega^0(M)\times\mathrm{Diff}(M)\to\R$ given by
\begin{align*}
&S(\varphi,Z)=\frac{\varepsilon}{2}\int_M\mathrm{d}\varphi\wedge\peqsubfino{{\star_g}}{\!Z}{-0.2ex}\mathrm{d}\varphi+\frac{1}{2}\int_{\partial M}\peqsub{b}{Z}^2\,\peqsubfino{\varphi}{\!\partial}{-0.4ex}\wedge\peqsubfino{{\star_{g^{\scriptscriptstyle\partial}}}}{\!\!\!\!\!Z}{-0.2ex}\,\peqsubfino{\varphi}{\!\partial}{-0.4ex}
\end{align*}
where $\peqsubfino{\varphi}{\!\partial}{-0.4ex}=\peqsub{\jmath}{\partial}^*\varphi$, $\peqsubfino{g}{Z}{-0.2ex}=Z^*\!g$, $\peqsubfino{g}{Z}{-0.2ex}^\partial=Z^*\peqsub{\jmath}{\partial}^*g$ and $\peqsub{b}{Z}=B\smallcirc Z$ for some fixed $B\in\Cinf{\partial M}$. We saw in chapter \ref{Chapter - Parametrized EM} that the variations of the action\index{Action!Scalar field} lead to the equations
\begin{align*}
D(\peqsub{S}{\mathrm{d}}+\peqsub{S}{\partial})(\varphi,Z)&=\varepsilon \int_M\peqsubfino{{\big\langle\hspace*{0.1ex}V_\varphi-\peqsub{\imath}{\mathbb{V}_{\!Z}} \mathrm{d}\varphi\hspace*{0.3ex},\hspace*{0.05ex}\peqsub{\delta}{\!Z}  \mathrm{d}\varphi\hspace*{0.1ex}\big\rangle_{\!\!g}}}{\!Z}{-0.2ex}\peqsubfino{{\mathrm{vol}_g}}{\!Z}{-0.2ex}+\int_{\partial M}\peqsubfino{{\big\langle \hspace*{0.1ex}V_\varphi^\partial+\peqsubfino{{\mathcal{L}_{\vec{\mathbb{V}}^\partial}}}{\!\!\!\!\!Z}{-0.4ex}\,\peqsubfino{\varphi}{\!\partial}{-0.4ex}\hspace*{0.3ex},\hspace*{0.05ex}\varepsilon\peqsub{\jmath}{\partial}^*\big(\peqsubfino{{\imath_{\vec{\nu}}}}{\!Z}{-0.3ex} \mathrm{d}\varphi)+\peqsub{b}{Z}^2\peqsubfino{\varphi}{\!\partial}{-0.4ex}\hspace*{0.1ex}\big\rangle_{\!\!g^\partial}}}{\!\!\!\!\!Z}{-0.3ex}\mathrm{vol}_{g^\partial_{\!Z}}
\end{align*}
where the last term of \eqref{Parametrized theories - equation - Variacion accion} vanishes because the interior product of a $0$-form is zero. Then we obtain the equations
\begin{equation*}
\peqsub{S}{\mathrm{d}}+\peqsub{S}{\partial}:\ 
\begin{array}{|l}
\peqsub{\delta}{\!Z}\mathrm{d}\varphi=0\\[1ex]
\varepsilon\peqsub{\jmath}{\partial}^*\big(\peqsubfino{{\imath_{\vec{\nu}}}}{\!Z}{-0.3ex} \mathrm{d}\varphi)+\peqsub{b}{Z}^2\peqsubfino{\varphi}{\!\partial}{-0.4ex}=0
\end{array} \qquad\text{ or }\qquad
\peqsub{S}{\mathrm{d}}^{\scriptscriptstyle\partial}+ \peqsub{S}{\partial}^{\scriptscriptstyle\partial}:\ \begin{array}{|l} \peqsub{\delta}{\!Z}\mathrm{d}\varphi=0\\[1ex]
\peqsubfino{\varphi}{\!\partial}{-0.4ex}=0\end{array}
\end{equation*}
Using \eqref{appendix - formula - i_x(f^*)=f^*i_{f_x}}, \eqref{appendix - formula - f^*d=df^*}, \ref{Appendix - lemma - volz^*g=z^*vol_g and star_z^*g=z^*star_g} and the definitions of $\peqsub{b}{Z}$ and $\peqsubfino{\vec{\nu}}{\!Z}{-0.3ex}$, it is easy to check that if $(\varphi,Z)$ is a solution then so is $(Y^*\varphi,Z\smallcirc Y)$ for every $Y\in\mathrm{Diff}(M)$.\separ

\section{Hamiltonian formulation}

In section \ref{Parametrized theories - section - Hamiltonian} of chapter \ref{Chapter - Parametrized EM} we developed the Hamiltonian formulation of some parametrized theories. We obtain that a bifurcation showed up according to the support of $\delta p$. It is important to realize that now, for $k=0$, both $\peqsubfino{q}{\perp}{-0.2ex}$ and $\delta p$ vanish because $\delta$ and $\imath$ are zero over $0$-forms. In particular the bifurcation does not appear for the parametrized scalar field and we have (see equation \eqref{Parametrized theories - equations - Hamiltonian vector field delta p=0}) that the Hamiltonian vector field $Y=(\vecc{Y}{q},\vecc{Y}{p},\mathbb{Y})$ is given by
\begin{align}
\begin{split}
&\vecc{Y}{q}=\mathcal{L}_{\vec{y}^{\scriptscriptstyle\top}}q+Y^{{\raisemath{0.2ex}{\scriptscriptstyle\perp\!}}}\frac{\underline{p}}{\sqrt{\gamma}}\\[1ex]
&\vecc{Y}{p}=\mathcal{L}_{\vec{y}^{\scriptscriptstyle\top}}p+\varepsilon\sqrt{\gamma}\,\sharp_\gamma\delta(Y^{{\raisemath{0.2ex}{\scriptscriptstyle\perp\!}}}\mathrm{d}q)\\[1.7ex]
&Y^{{\raisemath{0.2ex}{\scriptscriptstyle\perp\!}}}\text{ arbitrary}\\[.8ex]
&y^{{\raisemath{-0.1ex}{\scriptscriptstyle\top\!}}}\text{ arbitrary}
\end{split}
\end{align}

For manifolds with no boundary we are done. However, for manifolds with boundaries we have to deal with the cases as at the end of section \ref{Parametrized theories - section - Hamiltonian} of chapter \ref{Chapter - Parametrized EM}. Notice that for the scalar field $\peqsubfino{q}{\perp}{-0.2ex}=0$ (besides, $\imath_{\vec{\nu}}\underline{p}=0$ because $p\in\Omega^0(\Sigma)$), so the remaining term in the Hamiltonian equation reads
	\begin{align}\label{scalar - equation - frontera forma simplectica}
0=\int_{\partial\Sigma} \peqsubfino{{\left\langle\peqsub{\jmath}{\partial}^*\!\left(\vecc{Z}{q}-\mathcal{L}_{\vec{z}^{\scriptscriptstyle\top}}q-Z^{{\raisemath{0.2ex}{\scriptscriptstyle\perp\!}}}\frac{\underline{p}}{\sqrt{\gamma}}\right)\!\coma Y^{{\raisemath{0.2ex}{\scriptscriptstyle\perp\!}}}\left(\peqsub{\jmath}{\partial}^*(\imath_{\vec{\nu}}\mathrm{d}q)+\peqsub{\nu}{\perp}\frac{\underline{p}^{\scriptscriptstyle\partial}}{\sqrt{\gamma}}-\varepsilon b^2\peqsubfino{q}{\partial}{-0.2ex}\right)\right\rangle_{\!\!\!\gamma}}}{\!\partial}{-.3ex}\frac{\peqsubfino{{\mathrm{vol}_\gamma}}{\!\partial\!X}{-0.4ex}}{|\vec{\nu}^{\,\scriptscriptstyle\top}|}
\end{align}
Now we have the following cases.\separ

$\blacktriangleright$ Dirichlet $\peqsub{\mathcal{P}}{D}:=\Big\{(q,X,p)\in\mathcal{P}\ \ /\ \ \peqsubfino{q}{\partial}{-0.2ex}=0\Big\}$\separprevia

Back in the previous chapter we obtain some sufficient conditions to ensure that the previous expression was zero. Nonetheless, we are going to prove that in the scalar field case they are not necessary and, actually, \eqref{scalar - equation - frontera forma simplectica} always vanishes. Indeed, first notice that the component $\vecc{Y}{q}$, which is zero over the boundary in order to be tangent to $\peqsub{\mathcal{P}}{D}$ can be rewritten as
\begin{align*}
\vecc{Y}{q}&=\mathcal{L}_{\vec{y}^{\scriptscriptstyle\top}}q+Y^{{\raisemath{0.2ex}{\scriptscriptstyle\perp\!}}}\frac{\underline{p}}{\sqrt{\gamma}}=(\mathrm{d}q)_ay^a+\varepsilon n_\alpha\mathbb{Y}^\alpha\frac{\underline{p}}{\sqrt{\gamma}}=\\
&=\left((\mathrm{d}q)_ae^a_\alpha+\varepsilon n_\alpha\frac{\underline{p}}{\sqrt{\gamma}}\right)\mathbb{Y}^\alpha
\end{align*}
As over the boundary $\mathbb{Y}=\varepsilon Y^{{\raisemath{0.2ex}{\scriptscriptstyle\perp\!}}} n^\alpha+\tau.\vec{y}^{\scriptscriptstyle\top}$ is tangent to the boundary, this means that the term inside the parentheses is perpendicular to the boundary (unless $\mathbb{Y}=0$). Equivalently it is proportional to the normal vector field
\[\varepsilon n_\alpha\frac{\underline{p}}{\sqrt{\gamma}}+(\mathrm{d}q)_ae^a_\alpha \propto \nu_\alpha\qquad\text{ at the boundary}\]
In fact it is easy to obtain the proportionality factor by multiplying by $\nu^\alpha$ to obtain
\begin{align}\label{scalar - equation - proportional to nu}
\varepsilon n_\alpha\frac{\underline{p}}{\sqrt{\gamma}}+(\mathrm{d}q)_ae^a_\alpha=\left(\peqsub{\jmath}{\partial}^*(\imath_{\vec{\nu}}\mathrm{d}q)+\peqsub{\nu}{\perp}\frac{\underline{p}^{\scriptscriptstyle\partial}}{\sqrt{\gamma}}\right)\nu_\alpha\qquad\text{ at the boundary}
\end{align}
Thus equation \eqref{scalar - equation - frontera forma simplectica} reads now
	\begin{align*}
&\int_{\partial\Sigma} \peqsubfino{{\left\langle\peqsub{\jmath}{\partial}^*\!\left(\vecc{Z}{q}-\left(\varepsilon n_\alpha\frac{\underline{p}}{\sqrt{\gamma}}+(\mathrm{d}q)_ae^a_\alpha\right)\mathbb{Z}^\alpha\right)\!\coma Y^{{\raisemath{0.2ex}{\scriptscriptstyle\perp\!}}}\left(\peqsub{\jmath}{\partial}^*(\imath_{\vec{\nu}}\mathrm{d}q)+\peqsub{\nu}{\perp}\frac{\underline{p}^{\scriptscriptstyle\partial}}{\sqrt{\gamma}}\right)\right\rangle_{\!\!\!\gamma}}}{\!\partial}{-.3ex}\frac{\peqsubfino{{\mathrm{vol}_\gamma}}{\!\partial\!X}{-0.4ex}}{|\vec{\nu}^{\,\scriptscriptstyle\top}|}
\end{align*}
which is zero using the fact that $\vecc{Z}{q}^{\scriptscriptstyle\partial}=0$ and equation \eqref{scalar - equation - proportional to nu}. We then obtain no further condition at the first step of the GNH algorithm. We see here that we have obtained again a bifurcation because if condition \eqref{scalar - equation - proportional to nu} holds then $\mathbb{Y}$ is arbitrary at the boundary, otherwise it is forced to be zero. Considering the former case which is the physically relevant, we have now to impose the tangency to the space
\[\peqsub{\mathcal{P}}{D}^2:=\left\{(q,X,p)\in\mathcal{P}\ \ /\ \ \peqsubfino{q}{\partial}{-0.2ex}=0\qquad \peqsub{\jmath}{\partial}^*\!\left(\varepsilon n_\alpha\frac{\underline{p}}{\sqrt{\gamma}}+(\mathrm{d}q)_ae^a_\alpha\right)=\left(\peqsub{\jmath}{\partial}^*(\imath_{\vec{\nu}}\mathrm{d}q)+\peqsub{\nu}{\perp}\frac{\underline{p}^{\scriptscriptstyle\partial}}{\sqrt{\gamma}}\right)\nu_\alpha\right\}\]
which leads to an infinite series of conditions that we are not going to write down explicitly because they are somewhat complicated and their explicit expression is not particularly illuminating.\separ

$\blacktriangleright$ Robin-like $\mathcal{P}$\separprevia

If we impose no restriction at the domain $\mathcal{P}$, then equation \eqref{scalar - equation - frontera forma simplectica} leads to the condition
\begin{align*}
& Y^{{\raisemath{0.2ex}{\scriptscriptstyle\perp\!}}}\left(\peqsub{\jmath}{\partial}^*(\imath_{\vec{\nu}}\mathrm{d}q)+\peqsub{\nu}{\perp}\frac{\underline{p}^{\scriptscriptstyle\partial}}{\sqrt{\gamma}}-\varepsilon b^2\peqsubfino{q}{\partial}{-0.2ex}\right)=0
\end{align*}
Such condition, wherever $Y^{{\raisemath{0.2ex}{\scriptscriptstyle\perp\!}}}\neq0$, can be rewritten as
\begin{align*}
&\left(\varepsilon n_\alpha\frac{\underline{p}^{\scriptscriptstyle\partial}}{\sqrt{\gamma}}+\peqsub{\jmath}{\partial}^*(e^a\alpha(\mathrm{d}q)_a)-\varepsilon b^2\peqsubfino{q}{\partial}{-0.2ex}\nu_\alpha\right)\nu^\alpha=0
\end{align*}
In this case we have that the term inside the parentheses has to be tangent to the boundary
\[\varepsilon \peqsub{\vec{n}}{X}\frac{\underline{p}^{\scriptscriptstyle\partial}}{\sqrt{\gamma}}+\peqsub{\tau}{X}.\nabla^{\gamma_X}q-\varepsilon \peqsub{b}{X}^2\peqsubfino{q}{\partial}{-0.2ex}\,\peqsub{\vec{\nu}}{X}\in \peqsub{T}{X\smallcirc\jmath_\partial}\peqsub{\partial}{\Sigma}M\]

It is interesting to note that for the Dirichlet case we have that
\begin{equation}\label{scalar - equation - n+tau}
\varepsilon \peqsub{\vec{n}}{X}\frac{\underline{p}^{\scriptscriptstyle\partial}}{\sqrt{\gamma}}+\peqsub{\tau}{X}.\nabla^{\gamma_X}q
\end{equation}
is normal to the boundary, while for the Neumann case ($b=0$) we have that it is tangent to the boundary. Meanwhile, for the Robin case with $b\neq0$ we have that \eqref{scalar - equation - n+tau} is not normal nor tangent to the boundary but moves towards the normal direction for larger $b$ without ever reaching it. This comes from the different origin of the Dirichlet and Robin boundary conditions: the former is imposed by restricting the domain of the theory (and hence the Hamilton equation is trivially satisfied at the boundary), while the latter appears when we solve the Hamiltonian equation. Finally note that only in the limit $b\to\infty$ do we recover the Dirichlet case, which shows once again the different nature of the both boundary conditions.

%% file: 7_parametrized_MCS.tex
\chapter[Parametrized Maxwell Chern-Simons]{Parametrized Maxwell-Chern-Simons}\thispagestyle{empty}\label{Chapter - parametrized MCS}
\starwars[Darth Vader]{When I left you, I was but the learner; now I am the master.}{A New Hope}

\section{Introduction}\label{Applications - Section - MCS boundaries}

We have seen in section \ref{Parametrized theories - section - action} of chapter \ref{Chapter - Parametrized EM} a way to introduce in the parametrized electromagnetism some Robin-like boundary conditions. However, this method includes some additional conditions over the solution and it is not easy to understand what they mean physically. We mentioned at that moment that there exists an alternative in the odd dimensional case, namely, the Maxwell-Chern-Simons theory. It is interesting to mention that these and similar theories are used in condensed matter \cite{balachandran1994maxwell,balachandran1996edge,balachandran1995edge} and massive gauge theories \cite{deser1982}.\separ

We consider the action\index{Action!Maxwell-Chern-Simons} $S:\Omega^k(M)\to\R$, for some odd $k$, over a space-time $M$ of dimension $n=2k+1$
\[S(A,Z)=\frac{\varepsilon\eta }{2}\int_M \mathrm{d}A\wedge\peqsub{{\star_g}}{\!Z}\mathrm{d}A+\frac{\mu }{2}\int_MA\wedge \mathrm{d}A\]
where $\mu,\eta\in\R$ are constants of the theory (in particular, we can set $\mu=0$ to recover the Maxwell theory or $\eta=0$ to obtain the Chern-Simons theory as we will do in the next section). We could have added the boundary term involving the $B$ but we have preferred to keep things clean and simple in order to understand properly the interaction of the Chern-Simons term with the boundary.

\section{Variations of the action}\trassub

The variations of the first integral of the action $\peqsub{S}{\mathrm{d}}$ are given by \eqref{Parametrized theories - equation - Variacion accion}. Meanwhile, notice that the second integral, $\peqsub{S}{\mathrm{CS}}$, does not depend on the diffeomorphisms so we have only to compute its variation with respect to the field $A$.
\begin{align*}
\peqsub{D}{\!(A,\peqsub{V}{\!A})}&\peqsub{S}{\mathrm{CS}}(A,Z)=\frac{\mu}{2}\int_M\Big(\peqsub{V}{\!A}\wedge\mathrm{d}A+A\wedge\mathrm{d}\peqsub{V}{\!A}\Big)\updown{\eqref{Appendix - property - wedge supercommutativity}}{\eqref{Appendix - equaion - star^2=Id}}{=}\\
&=-\frac{\mu}{2}\int_M\Big(\peqsub{V}{\!A}\wedge\peqsub{{\star_g}}{\!Z}\peqsub{{\star_g}}{\!Z}\mathrm{d}A+\mathrm{d}\peqsub{V}{\!A}\wedge\peqsub{{\star_g}}{\!Z}\peqsub{{\star_g}}{\!Z}A\Big)\overset{\eqref{Appendix - equation - definition Hodge (a,b)vol=a wedge*b}}{=}\\
&=-\frac{\mu}{2}\int_M\Big(\peqsubfino{{\big\langle\peqsub{V}{\!A}\coma\peqsub{{\star_g}}{\!Z}\mathrm{d}A\big\rangle_{\!\!g}}}{\!Z}{-0.2ex}+\peqsubfino{{\big\langle\mathrm{d}\peqsub{V}{\!A}\coma\peqsub{{\star_g}}{\!Z}A\big\rangle_{\!\!g}}}{\!Z}{-0.2ex}\Big)\peqsubfino{{\mathrm{vol}_g}}{\!Z}{-0.2ex}\updown{\eqref{appendix equation integracion por partes}}{\eqref{appendix property delta star=star d}}{=}\\
&=-\mu\frac{(-1)^k-1}{2}\int_M\peqsubfino{{\big\langle\peqsub{V}{\!A}\coma\delta\peqsub{{\star_g}}{\!Z}A\big\rangle_{\!\!g}}}{\!Z}{-0.2ex}\peqsubfino{{\mathrm{vol}_g}}{\!Z}{-0.2ex}-\frac{\mu}{2}\int_{\partial M}\peqsubfino{{\langle\peqsub{\jmath}{\partial}^*\peqsub{V}{\!A},\peqsub{\jmath}{\partial}^*\peqsubfino{{\imath_{\vec{\nu}}}}{\!Z}{-0.3ex}(\peqsub{{\star_g}}{\!Z}A)\big\rangle_{\!\!g^\partial}}}{\!\!\!\!\!Z}{-0.3ex}\,\mathrm{vol}_{g^\partial_{\!Z}}
\end{align*}
 which leads to
\begin{align*}
\begin{split}
\peqsub{D}{\!(A,\peqsub{V}{\!A})}(\peqsub{S}{\mathrm{d}}+\peqsub{S}{\mathrm{CS}})(A,Z)&= \int_M\peqsubfino{{\left\langle\hspace*{0.1ex}\peqsub{V}{\!A}\hspace*{0.3ex},\hspace*{0.05ex}\peqsub{\delta}{\!Z}\!\left(\varepsilon\eta\mathrm{d}A-\mu\frac{(-1)^k-1}{2}\peqsub{{\star_g}}{\!Z}A\right)\hspace*{0.1ex}\right\rangle_{\!\!g}}}{\!Z}{-0.2ex}\peqsubfino{{\mathrm{vol}_g}}{\!Z}{-0.2ex}+{}\\
&\phantom{=}+\int_{\partial M}\peqsubfino{{\left\langle \hspace*{0.1ex}\peqsub{V}{\!A}^{\scriptscriptstyle\partial}\hspace*{0.3ex},\hspace*{0.05ex}\peqsub{\jmath}{\partial}^*\peqsubfino{{\imath_{\vec{\nu}}}}{\!Z}{-0.3ex}\left(\varepsilon\eta \mathrm{d}A-\frac{\mu}{2}\peqsub{{\star_g}}{\!Z}A\right)\right\rangle_{\!\!\!g^\partial}}}{\!\!\!\!\!Z}{-0.3ex}\,\mathrm{vol}_{g^\partial_{\!Z}}
\end{split}
\end{align*}
Which lead to the equations
\begin{equation*}
\begin{array}{|l}
\peqsub{\delta}{\!Z}\!\left(\varepsilon\eta\mathrm{d}A-\mu\dfrac{(-1)^k-1}{2}\peqsub{{\star_g}}{\!Z}A\right)=0\\[1.5ex]
\displaystyle\varepsilon\eta\,\peqsub{\jmath}{\partial}^*\big(\peqsubfino{{\imath_{\vec{\nu}}}}{\!Z}{-0.3ex} \mathrm{d}A)=\frac{\mu}{2}\peqsub{\jmath}{\partial}^*\left(\peqsubfino{{\imath_{\vec{\nu}}}}{\!Z}{-0.3ex}\peqsub{{\star_g}}{\!Z}A\right)
\end{array} \qquad\text{ or }\qquad
\begin{array}{|l}
\peqsub{\delta}{\!Z}\!\left(\varepsilon\eta\mathrm{d}A-\mu\dfrac{(-1)^k-1}{2}\peqsub{{\star_g}}{\!Z}A\right)=0\\[1ex]
\peqsubfino{A}{\partial}{-0.2ex}=0\end{array}
\end{equation*}
Notice that there is a term which appears only if $k$ is odd. Besides, in the first set of equations we have some kind of Robin-like boundary conditions, while the second set is just the usual Dirichlet ones.\separ

We keep on by studying the variation with respect to the diffeomorphisms which is given again by equation \eqref{Parametrized theories - equation - Variacion accion}
\begin{align*}
\begin{split}
\peqsub{D}{(Z,\mathbb{V}_{\!Z})}(\peqsub{S}{\mathrm{d}}+&\peqsub{S}{\mathrm{CS}})(A,Z)=-\varepsilon\eta \int_M\peqsubfino{{\big\langle\hspace*{0.1ex}\peqsub{\imath}{\mathbb{V}_{\!Z}} \mathrm{d}A\hspace*{0.3ex},\hspace*{0.05ex}\peqsub{\delta}{\!Z}  \mathrm{d}A\hspace*{0.1ex}\big\rangle_{\!\!g}}}{\!Z}{-0.2ex}\peqsubfino{{\mathrm{vol}_g}}{\!Z}{-0.2ex}-\varepsilon\eta\int_{\partial M}\peqsubfino{{\big\langle \hspace*{0.1ex}\peqsub{\imath}{\mathbb{V}^\partial_{\!Z}}\mathrm{d}\peqsub{A}{\partial}\hspace*{0.3ex},\hspace*{0.05ex}\peqsub{\jmath}{\partial}^*\big(\peqsubfino{{\imath_{\vec{\nu}}}}{\!Z}{-0.3ex} \mathrm{d}A)\big\rangle_{\!\!g^\partial}}}{\!\!\!\!\!Z}{-0.3ex}\,\mathrm{vol}_{g^\partial_{\!Z}}
\end{split}
\end{align*}
Let us see that this variation is zero at the bulk if we impose the field equations.
\begin{align*}
&\varepsilon\eta\peqsubfino{{\big\langle\hspace*{0.1ex}\peqsub{\imath}{\mathbb{V}_{\!Z}} \mathrm{d}A\hspace*{0.3ex},\hspace*{0.05ex}\peqsub{\delta}{\!Z}  \mathrm{d}A\hspace*{0.1ex}\big\rangle_{\!\!g}}}{\!Z}{-0.2ex}=\mu\frac{(-1)^k-1}{2}\peqsubfino{{\big\langle\hspace*{0.1ex}\peqsub{\imath}{\mathbb{V}_{\!Z}} \mathrm{d}A\hspace*{0.3ex},\hspace*{0.05ex}\peqsub{\delta}{\!Z}  \peqsub{{\star_g}}{\!Z}A\hspace*{0.1ex}\big\rangle_{\!\!g}}}{\!Z}{-0.2ex}\overset{\eqref{appendix property delta star=star d}}{=}\mu\frac{-1+(-1)^{k}}{2}\peqsubfino{{\big\langle\hspace*{0.1ex}\peqsub{\imath}{\mathbb{V}_{\!Z}} \mathrm{d}A\hspace*{0.3ex},\hspace*{0.05ex} \peqsub{{\star_g}}{\!Z}\mathrm{d}A\hspace*{0.1ex}\big\rangle_{\!\!g}}}{\!Z}{-0.2ex}\overset{\eqref{Appendix - lemma - <i_F,star F>=0}}{=}0
\end{align*}
Considering the Dirichlet boundary condition, we have that no additional condition arises
\begin{align*}
\varepsilon\eta&\int_{\partial M}\peqsubfino{{\big\langle \hspace*{0.1ex}\peqsub{\imath}{\mathbb{V}^\partial_{\!Z}}\mathrm{d}\peqsub{A}{\partial}\hspace*{0.3ex},\hspace*{0.05ex}\peqsub{\jmath}{\partial}^*\big(\peqsubfino{{\imath_{\vec{\nu}}}}{\!Z}{-0.3ex} \mathrm{d}A)\big\rangle_{\!\!g^\partial}}}{\!\!\!\!\!Z}{-0.3ex}\,\mathrm{vol}_{g^\partial_{\!Z}}\overset{\eqref{appendix - formula - L=di+id}}{=}\\
&=\varepsilon\eta\int_{\partial M}\peqsubfino{{\big\langle \hspace*{0.1ex}\mathrm{d}\peqsub{\imath}{\mathbb{V}^\partial_{\!Z}}\peqsub{A}{\partial}\hspace*{0.3ex},\hspace*{0.05ex}\peqsub{\jmath}{\partial}^*\big(\peqsubfino{{\imath_{\vec{\nu}}}}{\!Z}{-0.3ex} \mathrm{d}A)\big\rangle_{\!\!g^\partial}}}{\!\!\!\!\!Z}{-0.3ex}\,\mathrm{vol}_{g^\partial_{\!Z}}\overset{\eqref{appendix equation integracion por partes}}{=}\\
&=\varepsilon\eta\int_{\partial M}\peqsubfino{{\big\langle \hspace*{0.1ex}\peqsub{\imath}{\mathbb{V}^\partial_{\!Z}}\peqsub{A}{\partial}\hspace*{0.3ex},\hspace*{0.05ex}\peqsub{\delta}{\!Z}\peqsub{\jmath}{\partial}^*\big(\peqsubfino{{\imath_{\vec{\nu}}}}{\!Z}{-0.3ex} \mathrm{d}A)\big\rangle_{\!\!g^\partial}}}{\!\!\!\!\!Z}{-0.3ex}\,\mathrm{vol}_{g^\partial_{\!Z}}=0
\end{align*}
Meanwhile, if we consider the MCS boundary conditions we obtain
\begin{align*}
\varepsilon\eta&\int_{\partial M}\peqsubfino{{\big\langle \hspace*{0.1ex}\peqsub{\imath}{\mathbb{V}^\partial_{\!Z}}\mathrm{d}\peqsub{A}{\partial}\hspace*{0.3ex},\hspace*{0.05ex}\peqsub{\jmath}{\partial}^*\big(\peqsubfino{{\imath_{\vec{\nu}}}}{\!Z}{-0.3ex} \mathrm{d}A)\big\rangle_{\!\!g^\partial}}}{\!\!\!\!\!Z}{-0.3ex}\,\mathrm{vol}_{g^\partial_{\!Z}}=\frac{\mu}{2}\int_{\partial M}\peqsubfino{{\big\langle \hspace*{0.1ex}\peqsub{\imath}{\mathbb{V}^\partial_{\!Z}}\mathrm{d}\peqsub{A}{\partial}\hspace*{0.3ex},\hspace*{0.05ex}\peqsub{\jmath}{\partial}^*\left(\peqsubfino{{\imath_{\vec{\nu}}}}{\!Z}{-0.3ex}\peqsub{{\star_g}}{\!Z}A\right)\big\rangle_{\!\!g^\partial}}}{\!\!\!\!\!Z}{-0.3ex}\,\mathrm{vol}_{g^\partial_{\!Z}}\updown{\eqref{appendix - formula - f^*d=df^*}\eqref{appendix - formula - i_x(f^*)=f^*i_{f_x}}}{\ref{Appendix - lemma - g=gamma+nu nu}\eqref{appendix - formula - i_x^2=0}}{=}\\
&=\frac{\mu}{2}\int_{\partial M}\peqsubfino{{\big\langle \hspace*{0.1ex}\peqsub{\imath}{\mathbb{V}^\partial_{\!Z}}\mathrm{d}A\hspace*{0.3ex},\hspace*{0.05ex}\peqsubfino{{\imath_{\vec{\nu}}}}{\!Z}{-0.3ex}\peqsub{{\star_g}}{\!Z}A\big\rangle_{\!\!g}}}{\!Z}{-0.3ex}\,\mathrm{vol}_{g^\partial_{\!Z}}\updown{\ref{Appendix - lemma - j*(a wedge star b)=(a,i_nu b)vol}}{\eqref{appendix property star delta=d star}}{=}-\frac{\mu}{2}\int_{\partial M}\peqsub{\jmath}{\partial}^*\Big(\peqsub{\imath}{\mathbb{V}^\partial_{\!Z}}\mathrm{d}A\wedge A\Big)=-\frac{\mu}{2}\int_{\partial M}\peqsub{\imath}{\mathbb{V}^\partial_{\!Z}}\mathrm{d}\peqsub{A}{\partial}\hspace*{0.3ex}\wedge \peqsub{A}{\partial}
\end{align*}
This means that if we impose the variations to be zero we get that
\begin{equation}
  \mu\,\big(\peqsub{\imath}{-}\mathrm{d}\peqsub{A}{\partial}\hspace*{0.1ex}\big)\wedge \peqsub{A}{\partial}\in\Omega^1(\partial M)
\end{equation}
has to be zero. This \textbf{additional condition} is somewhat analogous to \eqref{Parametrized EM - equation - additional condition}.

\section{Lagrangian formulation}\trassub

The same procedure that we follow to obtain the Lagrangian \eqref{Parametrized theories - equation - Lagrangian} allows us to obtain the MCS Lagrangian	\[\begin{array}{cccc}
\peqsub{L}{\mathrm{MCS}}: & \peqsub{\mathcal{D}}{\mathrm{MCS}}\subset T\Big(\Omega^{k-1}(\Sigma)\times \Omega^k(\Sigma)\times\mathrm{Emb}(\Sigma,M)\Big) & \longrightarrow & \R\\
&           \mathbf{v}_{(\peqsubfino{q}{\perp}{-0.2ex},q,X)}=(\peqsubfino{q}{\perp}{-0.2ex},q,X;v_{\!\scriptscriptstyle\perp},v,\peqsub{\mathbb{V}}{\!X})                   &  \longmapsto    & L(\mathbf{v}_{(\peqsubfino{q}{\perp}{-0.2ex},q,X)})
\end{array}\]
which is given by
\begin{align*}
   \peqsub{L}{\mathrm{MCS}}(\mathbf{v}_{(\peqsub{q}{\perp},q,X)})&=\frac{\eta}{2}\left\llangle\frac{v-\mathcal{L}_{\peqsub{\vec{v}}{X}^{\scriptscriptstyle\top}}q-\varepsilon\mathrm{d}(\peqsub{V}{X}^{{\raisemath{0.2ex}{\!\scriptscriptstyle\perp}}}\peqsubfino{q}{\perp}{-0.2ex})}{\peqsub{V}{X}^{{\raisemath{0.2ex}{\!\scriptscriptstyle\perp}}}},\frac{v-\mathcal{L}_{\peqsub{\vec{v}}{X}^{\scriptscriptstyle\top}}q-\varepsilon\mathrm{d}(\peqsub{V}{X}^{{\raisemath{0.2ex}{\!\scriptscriptstyle\perp}}}\peqsubfino{q}{\perp}{-0.2ex})}{\peqsub{V}{X}^{{\raisemath{0.2ex}{\!\scriptscriptstyle\perp}}}}\right\rrangle_{\!\!\peqsub{V}{X}^{{\raisemath{0.2ex}{\!\scriptscriptstyle\perp}}}}+\frac{\varepsilon\eta}{2}{\big\llangle \mathrm{d}q,\mathrm{d}q\big\rrangle}_{\peqsub{V}{X}^{{\raisemath{0.2ex}{\!\scriptscriptstyle\perp}}}}+\\
&\phantom{=}+\frac{\varepsilon\mu}{2}{\big\llangle \peqsubfino{q}{\perp}{-0.2ex},\peqsubfino{{\star_\gamma}}{\!X}{-0.2ex}(\mathrm{d}q)\big\rrangle}_{\peqsub{V}{X}^{{\raisemath{0.2ex}{\!\scriptscriptstyle\perp}}}}-\frac{\mu}{2}\left\llangle\frac{v-\mathcal{L}_{\peqsub{\vec{v}}{X}^{\scriptscriptstyle\top}}q-\varepsilon\mathrm{d}(\peqsub{V}{X}^{{\raisemath{0.2ex}{\!\scriptscriptstyle\perp}}}\peqsubfino{q}{\perp}{-0.2ex})}{\peqsub{V}{X}^{{\raisemath{0.2ex}{\!\scriptscriptstyle\perp}}}},\peqsubfino{{\star_\gamma}}{\!X}{-0.2ex}q\right\rrangle_{\!\!\peqsub{V}{X}^{{\raisemath{0.2ex}{\!\scriptscriptstyle\perp}}}}
\end{align*}
where we recall that
\[\big\llangle \alpha,\beta\big\rrangle_{\!f}=\int_\Sigma f\peqsubfino{{\big\langle \alpha,\beta\big\rangle_{\!\gamma}}}{\!X}{-0.1ex}\peqsubfino{{\mathrm{vol}_\gamma}}{\!\!X}{-0.4ex}=\int_\Sigma f\sqrt{\peqsubfino{\gamma}{\!X}{-.3ex}}\,\peqsubfino{{\big\langle \alpha,\beta\big\rangle_{\!\gamma}}}{\!X}{-0.1ex}\peqsubfino{\mathrm{vol}}{\Sigma}{-0.2ex}\]
and that the Lagrangian is defined on the open subset \[\peqsub{\mathcal{D}}{\mathrm{MCS}}=T\Big(\Omega^{k-1}(\Sigma)\times\Omega^k(\Sigma)\Big)\times\Big\{(X,\peqsub{\mathbb{V}}{\!X})\in T\mathrm{Emb}(\Sigma,M)\ /\ \varepsilon \peqsub{\vec{n}}{X}(\peqsub{\mathbb{V}}{\!X})>0\Big\}\]

\section{Geometric arena}\trassub
	
A typical point of the phase space $T^*\mathcal{D}$ is of the form \[\boldsymbol{\mathrm{p}}_{(\peqsubfino{q}{\perp}{-0.2ex},q,X)}=(\peqsubfino{q}{\perp}{-0.2ex},q,X;\peqsubfino{\boldsymbol{p}}{\!\perp}{-0.2ex},\boldsymbol{p},\peqsubfino{\boldsymbol{P}}{\!\!X}{-0.2ex}) \in T^*\Big(\Omega^{k-1}(\Sigma)\times \Omega^k(\Sigma)\times\mathrm{Emb}(\Sigma,M)\Big) \]
where  $\peqsubfino{\boldsymbol{p}}{\!\perp}{-0.2ex}\in C^\infty(\Sigma)'$, $\boldsymbol{p}\in \Omega^k(\Sigma)'$, and $\peqsubfino{\boldsymbol{P}}{\!X}{-0.2ex}:\Gamma^\partial\!(X^*TM)\rightarrow\R$ can be represented (see section \ref{Parametrized theories - section - Fiber derivative} of chapter \ref{Chapter - Parametrized EM}) as
\begin{align}
\boldsymbol{p}(v)&=\int_\Sigma (v,p)\,\peqsub{\mathrm{vol}}{\Sigma}=\int_\Sigma \frac{(v,p)}{\sqrt{\peqsubfino{\gamma}{X}{-0.2ex}}}\,\peqsubfino{{\mathrm{vol}_\gamma}}{\!\!X}{-0.4ex}\label{Applications - equation - p=int p}
\\
\peqsubfino{\boldsymbol{P}}{\!X}{-0.2ex}(\peqsub{\mathbb{V}}{\!X})&=\int_\Sigma   \big(\peqsub{\mathbb{V}}{\!X},\peqsub{P}{\!X}\big)\,\peqsub{\mathrm{vol}}{\Sigma}+\!\int_{\partial\Sigma} \big(\peqsub{\mathbb{V}}{\!X}^{\scriptscriptstyle\partial},\peqsub{P}{\partial X}\big)\,\peqsub{\mathrm{vol}}{\partial\Sigma}
=\int_\Sigma   \frac{\big(\peqsub{\mathbb{V}}{\!X},\peqsub{P}{\!X}\big)}{\sqrt{\peqsubfino{\gamma}{X}{-0.2ex}}}\,\peqsubfino{{\mathrm{vol}_\gamma}}{\!\!X}{-0.4ex}+\!\int_{\partial\Sigma} \frac{\big(\peqsub{\mathbb{V}}{\!X}^{\scriptscriptstyle\partial},\peqsub{P}{\partial X}\big)}{\sqrt{\peqsub{\gamma}{\partial X}}} \,\peqsubfino{{\mathrm{vol}_\gamma}}{\!\partial\!X}{-0.4ex}\label{Applications - equation - P=int P}
\end{align}
Let $\peqsub{\widetilde{\mathcal{P}}}{\mathrm{MCS}}=\{\peqsub{\mathrm{p}}{\mathrm{q}}:=(\peqsubfino{q}{\perp}{-0.2ex},q,X,\peqsubfino{p}{\perp}{-0.2ex},p,\peqsub{P}{\!X},\peqsub{P}{\partial X})\}\overset{\jmath}{\hookrightarrow }T^*\mathcal{D}$ considered as a subset of $T^*\mathcal{D}$ via the previous representations. We define the induced form $\widetilde{\Omega}^{\scriptscriptstyle\mathrm{MCS}}:=\jmath^*\Omega$ which is given by
\begin{align}\label{Applications - equation - tilde Omega}
\begin{split}
\widetilde{\Omega}^{\scriptscriptstyle\mathrm{MCS}}_{\peqsub{\mathrm{p}}{\mathrm{q}}}(Y,Z)&=\int_\Sigma\Big[\Big(\vecc[\scalebox{0.4}{$\perp$}]{Y}{q},\vecc[\scalebox{0.4}{$\perp$}]{Z}{p}\Big)-\Big(\vecc[\scalebox{0.4}{$\perp$}]{Z}{q},\vecc[\scalebox{0.4}{$\perp$}]{Y}{p}\Big)+\Big(\vecc{Y}{q},\vecc{Z}{p}\Big)-\Big(\vecc{Z}{q},\vecc{Y}{p}\Big)+\vecc{Z}{P}(\vecc{Y}{X})-\vecc{Y}{P}(\vecc{Z}{X})\Big]\peqsub{\mathrm{vol}}{\Sigma}+\\
&\phantom{=}+\int_{\partial\Sigma}\left[\vecc{Z}{P}^{\scriptscriptstyle\partial}(\vecc{Y}{X})-\vecc{Y}{P}^{\scriptscriptstyle\partial}(\vecc{Z}{X})\right]\peqsub{\mathrm{vol}}{\partial\Sigma}     
\end{split}
\end{align}

\section{Fiber derivative}\trassub

The fiber derivative, given by equation \eqref{Mathematical background - equation - FL=D_2L}, is computed taking an initial point of the tangent bundle $\mathbf{v}_{(\peqsubfino{q}{\perp}{-0.2ex},q,X)}=(\peqsubfino{q}{\perp}{-0.2ex},q,X;\peqsubfino{v}{\!\perp}{-0.2ex},v,\peqsubfino{\mathbb{V}}{\!X}{-0.1ex})$ and some initial velocities $\mathbf{w}^1_{(\peqsubfino{q}{\perp}{-0.2ex},q,X)}=(\peqsubfino{q}{\perp}{-0.2ex},q,X;\peqsubfino{w}{\!\perp}{-0.2ex},0,0)$, $\mathbf{w}^2_{(\peqsubfino{q}{\perp}{-0.2ex},q,X)}=(\peqsubfino{q}{\perp}{-0.2ex},q,X;0,w,0)$, and $\mathbf{w}^3_{(\peqsubfino{q}{\perp}{-0.2ex},q,X)}=(\peqsubfino{q}{\perp}{-0.2ex},q,X;0,0,\peqsubfino{\mathbb{W}}{\!X}{-0.1ex})$. The first two are immediate while the last one can be obtained following the ideas of lemma \ref{Appendix - equation - FL(w3)=Pi}.
\begin{align*}
FL&(\mathbf{v}_{(\peqsubfino{q}{\perp}{-0.2ex},q,X)})\!\left(\hspace*{-0.1ex}\mathbf{w}^1_{(\peqsubfino{q}{\perp}{-0.2ex},q,X)}\hspace*{-0.1ex}\right)=0\\
FL&(\mathbf{v}_{(\peqsubfino{q}{\perp}{-0.2ex},q,X)})\!\left(\hspace*{-0.1ex}\mathbf{w}^2_{(\peqsubfino{q}{\perp}{-0.2ex},q,X)}\hspace*{-0.1ex}\right)=\left\llangle\frac{w}{\peqsub{V}{X}^{{\raisemath{0.2ex}{\!\scriptscriptstyle\perp}}}},\eta\frac{v-\mathcal{L}_{\peqsub{\vec{v}}{X}^{\scriptscriptstyle\top}}q-\varepsilon\mathrm{d}(\peqsub{V}{X}^{{\raisemath{0.2ex}{\!\scriptscriptstyle\perp}}}\peqsubfino{q}{\perp}{-0.2ex})}{\peqsub{V}{X}^{{\raisemath{0.2ex}{\!\scriptscriptstyle\perp}}}}-\frac{\mu}{2}\peqsubfino{{\star_\gamma}}{\!X}{-0.2ex}q\right\rrangle_{\!\!\peqsub{V}{X}^{{\raisemath{0.2ex}{\!\scriptscriptstyle\perp}}}}\hspace*{-0.3ex}=\\
&\phantom{(\mathbf{v}_{(\peqsubfino{q}{\perp}{-0.2ex},q,X)})\!\left(\hspace*{-0.1ex}\mathbf{w}^2_{(\peqsubfino{q}{\perp}{-0.2ex},q,X)}\hspace*{-0.1ex}\right)}=\left(\hspace*{-0.3ex} w\hspace*{.2ex},\hspace*{.1ex}\sqrt{\peqsubfino{\gamma}{X}{-0.2ex}}\peqsub{\sharp}{\gamma}\left[\eta\frac{v-\mathcal{L}_{\peqsub{\vec{v}}{X}^{\scriptscriptstyle\top}}q-\varepsilon\mathrm{d}(\peqsub{V}{X}^{{\raisemath{0.2ex}{\!\scriptscriptstyle\perp}}}\peqsubfino{q}{\perp}{-0.2ex})}{\peqsub{V}{X}^{{\raisemath{0.2ex}{\!\scriptscriptstyle\perp}}}}-\frac{\mu}{2}\peqsubfino{{\star_\gamma}}{\!X}{-0.2ex}q\right]\right)\\
FL&(\mathbf{v}_{(\peqsubfino{q}{\perp}{-0.2ex},q,X)})\!\left(\hspace*{-0.1ex}\mathbf{w}^3_{(\peqsubfino{q}{\perp}{-0.2ex},q,X)}\hspace*{-0.1ex}\right)=-\int_\Sigma \mathbb{W}^\alpha\Big(\varepsilon(\peqsub{n}{X})_\alpha\peqsubfino{\mathcal{H}}{\!\perp}{-0.1ex}^{\scriptscriptstyle\mathrm{MCS}}+(\peqsubfino{e}{X}{-0.2ex})^b_\alpha\mathcal{H}_b\Big)\peqsub{\mathrm{vol}}{\Sigma}\,-\\
&\hspace*{30ex}-\int_{\partial\Sigma}\mathbb{W}^\alpha\Big(\varepsilon (\peqsub{n}{X})_\alpha\peqsubfino{\mathcal{H}}{\!\perp}{-0.1ex}^{\scriptscriptstyle\partial}+ (\peqsubfino{e}{X}{-0.2ex})^b_\alpha\mathcal{H}^{\scriptscriptstyle\partial}_b\Big)\peqsub{{\mathrm{vol}}}{\partial\Sigma}
\end{align*}
where $(\alpha,V)=\frac{1}{k!}\alpha_{a_1\cdots a_k}V^{a_1\cdots a_k}$ is the natural pairing and $\peqsubfino{{\sharp_{\gamma}}}{\!X}{-0.2ex}$ is the musical isomorphism \eqref{Mathematical background - equation - musical ishomorphisms} of $\peqsubfino{\gamma}{X}{-0.2ex}$. We can now define the canonical momenta
\begin{align*}
&\hspace*{2.2ex}p=\sqrt{\peqsubfino{\gamma}{X}{-0.2ex}}\peqsub{\sharp}{\gamma}\left[\eta\frac{v-\mathcal{L}_{\peqsub{\vec{v}}{X}^{\scriptscriptstyle\top}}q-\varepsilon\mathrm{d}(\peqsub{V}{X}^{{\raisemath{0.2ex}{\!\scriptscriptstyle\perp}}}\peqsubfino{q}{\perp}{-0.2ex})}{\peqsub{V}{X}^{{\raisemath{0.2ex}{\!\scriptscriptstyle\perp}}}}-\frac{\mu}{2}\peqsubfino{{\star_\gamma}}{\!X}{-0.2ex}q\right]\\
&\left.\begin{array}{l}
\peqsubfino{p}{\perp}{-0.2ex}=0\\[1.4ex]
(\peqsub{P}{\!X})_\alpha=-\varepsilon(\peqsub{n}{X})_\alpha\peqsubfino{\mathcal{H}}{\!\perp}{-0.1ex}^{\scriptscriptstyle\mathrm{MCS}}(\peqsubfino{q}{\perp}{-0.2ex},q,X,p)-(\peqsubfino{e}{X}{-0.2ex})^b_\alpha\mathcal{H}_b(q,p)\\[1.9ex]
(\peqsub{P^{\scriptscriptstyle\partial}}{X})_\alpha=-\varepsilon(\peqsub{n}{X})_\alpha\peqsubfino{\mathcal{H}}{\!\perp}{-0.1ex}^{\scriptscriptstyle\partial}(q,p)-(\peqsubfino{e}{X}{-0.2ex})^b_\alpha\mathcal{H}^{\scriptscriptstyle\partial}_b(\peqsubfino{q}{\perp}{-0.2ex},q,X,p)
\end{array}\right\}\text{ Constraints}
\end{align*}
with 
\begin{align}\label{Applications - definitions - H's}
\begin{split}
&\mathcal{H}(q,p)=\Big(\imath_{-}\mathrm{d}q,p\Big)+\Big(\imath_{-}q,\delta p\Big)\in\Omega^1(\Sigma)\\
&\peqsubfino{\mathcal{H}}{\!\perp}{-0.1ex}^{\scriptscriptstyle\mathrm{MCS}}(\peqsubfino{q}{\perp}{-0.2ex},q,X,p)=\frac{1}{2\eta\sqrt{\peqsubfino{\gamma}{X}{-0.2ex}}}\Big(\peqsubfino{\Lambda}{+}{0.4ex}\hspace*{0.1ex},\hspace*{0.1ex}\underline{\peqsubfino{\Lambda}{+}{0.4ex}}\Big)-\frac{\varepsilon\eta\sqrt{\peqsubfino{\gamma}{X}{-0.2ex}}}{2}\big\langle\mathrm{d}q,\mathrm{d}q\big\rangle_{\!\peqsubfino{\gamma}{\!X}{-0.4ex}}+\varepsilon\Big(\peqsubfino{q}{\perp}{-0.2ex}\hspace*{0.1ex},\hspace*{0.1ex}\delta \Lambda_-\Big)\in\Cinf{\Sigma}\\
&\mathcal{H}^{\scriptscriptstyle\partial}(q,p)=\left(\frac{\sqrt{\peqsubfino{\gamma^{\scriptscriptstyle\partial}}{\!X}{-0.3ex}}}{\sqrt{\peqsubfino{\gamma}{\!X}{-0.3ex}}}\frac{\peqsub{\nu}{X}}{|\peqsub{\vec{\nu}}{X}^{\,\scriptscriptstyle\top}|}\wedge\imath_{-}q,p\right)\in\Omega^1(\partial\Sigma)\\
&\peqsubfino{\mathcal{H}}{\!\perp}{-0.1ex}^{\scriptscriptstyle\partial}(\peqsubfino{q}{\perp}{-0.2ex},q,p)=\varepsilon\,\left(\frac{\sqrt{\peqsubfino{\gamma^{\scriptscriptstyle\partial}}{\!X}{-0.3ex}}}{\sqrt{\peqsubfino{\gamma}{\!X}{-0.3ex}}}\frac{\peqsub{\nu}{X}}{|\peqsub{\vec{\nu}}{X}^{\,\scriptscriptstyle\top}|}\wedge\peqsubfino{q}{\perp}{-0.2ex},p\right)\in\Cinf{\partial\Sigma}
\end{split}
\end{align}
where we have defined the densities
\begin{equation}
\Lambda_\pm=p\pm\frac{\mu}{2}\sqrt{\gamma}\,\peqsub{\sharp}{\gamma}\peqsubfino{{\star_\gamma}}{\!X}{-0.2ex}q
\end{equation}
The manifold of first constraints of the MCS theory is $\peqsub{\mathcal{P}}{\mathrm{MCS}}:=FL(TQ)$ which is then given by
\begin{align*}
\peqsub{\mathcal{P}}{\mathrm{MCS}}&=\left\{(\peqsubfino{q}{\perp}{-0.2ex},q,X;\peqsubfino{p}{\perp}{-0.2ex},p,\,\peqsub{P}{\!X},\,\peqsub{P}{\!X}^{\scriptscriptstyle\partial})\in\peqsub{\widetilde{\mathcal{P}}}{\mathrm{MCS}}\ \ /\ \ \peqsubfino{p}{\perp}{-0.2ex}=0\ \begin{array}{l} \peqsub{P}{\!X}=-\varepsilon\peqsubfino{\mathcal{H}}{\!\perp}{-0.1ex} n-e.\mathcal{H}\\ \peqsub{P}{\!X}^{\scriptscriptstyle\partial}=-\varepsilon\peqsubfino{\mathcal{H}}{\!\perp}{-0.1ex}^{\scriptscriptstyle\partial}n-e.\mathcal{H}^{\scriptscriptstyle\partial}\end{array}\right\}=\\
&=\Big\{(\peqsubfino{q}{\perp}{-0.2ex},q,X;0,p\coma-\varepsilon\peqsubfino{\mathcal{H}}{\!\perp}{-0.1ex} n-e.\mathcal{H}\coma -\varepsilon\peqsubfino{\mathcal{H}}{\!\perp}{-0.1ex}^{\scriptscriptstyle\partial} n-e.\mathcal{H}^{\scriptscriptstyle\partial}) \Big\}\cong\\
&\cong\Big\{(\peqsubfino{q}{\perp}{-0.2ex},q,X,p)\Big\}=:\mathcal{P}
\end{align*}
We define the inclusion $\peqsubfino{\jmath}{\mathrm{MCS}}{-0.2ex}:\mathcal{P}\hookrightarrow \peqsub{\widetilde{\mathcal{P}}}{\mathrm{MCS}}$ that allows us to pullback the induced form $\widetilde{\Omega}^{\scriptscriptstyle\mathrm{MCS}}=\jmath^*\Omega$ of $\peqsub{\widetilde{\mathcal{P}}}{\mathrm{MCS}}$, given by equation \eqref{Parametrized theories - equation - tilde Omega}, to $\mathcal{P}$ in order to define $\omega^{\scriptscriptstyle\mathrm{MCS}}:=\peqsub{\jmath}{\mathrm{MCS}\,}^*\widetilde{\Omega}^{\scriptscriptstyle\mathrm{MCS}}$ that, in this case, is given by
\begin{align}\label{Application - equation - omega final}
\begin{split}
&\omega^{\scriptscriptstyle\mathrm{MCS}}_{(\peqsubfino{q}{\perp}{-0.2ex},q,X,p)}\Big((\vecc[\scalebox{0.4}{$\perp$}]{Y}{q},\vecc{Y}{q},\vecc{Y}{X},\vecc{Y}{p}),(\vecc[\scalebox{0.4}{$\perp$}]{Z}{q},\vecc{Z}{q},\vecc{Z}{X},\vecc{Z}{p})\Big)
=\\ &\qquad=\int_\Sigma\left[\Big(\vecc{Y}{q}\coma \vecc{Z}{p}\Big)-\Big(\vecc{Z}{q}\coma \vecc{Y}{p}\Big)-\peqsub{D}{Z}(\varepsilon\peqsubfino{\mathcal{H}}{\!\perp}{-0.1ex}n+e.\mathcal{H})_\alpha\vecc{Y}{X}^\alpha+\peqsub{D}{Y}(\varepsilon\peqsubfino{\mathcal{H}}{\!\perp}{-0.1ex}n+e.\mathcal{H})_\alpha\vecc{Z}{X}^\alpha\right]\peqsub{\mathrm{vol}}{\Sigma}-{}\\
&\qquad\phantom{=}-\int_{\partial\Sigma}\left[\peqsub{D}{Z}(\varepsilon\peqsubfino{\mathcal{H}}{\!\perp}{-0.1ex}^{\scriptscriptstyle\partial}n+e.\mathcal{H}^{\scriptscriptstyle\partial})_\alpha\vecc{Y}{X}^\alpha-\peqsub{D}{Y}(\varepsilon\peqsubfino{\mathcal{H}}{\!\perp}{-0.1ex}^{\scriptscriptstyle\partial}n+e.\mathcal{H}^{\scriptscriptstyle\partial})_\alpha\vecc{Z}{X}^\alpha\right]\peqsub{\mathrm{vol}}{\partial\Sigma}     
\end{split}
\end{align}
To obtain the explicit expression we have to proceed as in page \pageref{Appendix - computation - omega} of the appendix but taking into account that, on one hand, the terms including $q\wedge\mathrm{d}q\in\Omega^3(\Sigma)$ are zero because $\Sigma$ has dimension $2$. On the other hand we use the fact that
\begin{equation}\label{Applications - equation - DLambda}
D\Lambda_\pm=\vecc{Y}{p}\pm\frac{\mu}{2}\sqrt{\gamma}\,\peqsub{\sharp}{\gamma}\peqsubfino{{\star_\gamma}}{\!X}{-0.2ex}\vecc{Y}{q}
\end{equation}
because it can be easily proved that $\sqrt{\gamma}\,\peqsub{\sharp}{\gamma}\peqsubfino{{\star_\gamma}}{\!X}{-0.2ex}$ does not depend on the embedding. With all these ideas in mind we obtain
\begin{align*}
\begin{split}
&\omega^{\scriptscriptstyle\mathrm{MCS}}_{(\peqsubfino{q}{\perp}{-0.2ex},q,X,p)}\Big((\vecc[\scalebox{0.4}{$\perp$}]{Y}{q},\vecc{Y}{q},\vecc{Y}{X},\vecc{Y}{p}),(\vecc[\scalebox{0.4}{$\perp$}]{Z}{q},\vecc{Z}{q},\vecc{Z}{X},\vecc{Z}{p})\Big)=\\
&=\int_\Sigma\varepsilon Z^{{\raisemath{0.2ex}{\scriptscriptstyle\perp\!}}} \Big(\vecc[\perp]{Y}{q}\hspace*{0.1ex}-\mathcal{L}_{\vec{y}^{\scriptscriptstyle\top}}\peqsubfino{q}{\perp}{-0.2ex}+\peqsub{\imath}{\mathrm{d}Y^{\!\scriptscriptstyle\perp}}q\coma\delta \Lambda_-\Big)\peqsub{\mathrm{vol}}{\Sigma}+{}\\
&+\!\int_\Sigma\!\left(\vecc{Y}{q}\!-\!\mathcal{L}_{\vec{y}^{\scriptscriptstyle\top}}q\!-\!Y^{{\raisemath{0.2ex}{\scriptscriptstyle\perp\!}}}\frac{\underline{\peqsubfino{\Lambda}{+}{0.4ex}}}{\eta\sqrt{\gamma}}\!-\!\varepsilon\mathrm{d}( Y^{{\raisemath{0.2ex}{\scriptscriptstyle\perp\!}}}\peqsubfino{q}{\perp}{-0.2ex})\coma\vecc{Z}{p}\!-\!\mathcal{L}_{\vec{z}^{\scriptscriptstyle\top}}p\!-\! Z^{{\raisemath{0.2ex}{\scriptscriptstyle\perp\!}}}\frac{\mu}{2\eta}{\star_\gamma}\Lambda_+\!-\!\varepsilon\sqrt{\gamma}\,\sharp_\gamma\delta\!\left(\eta Z^{{\raisemath{0.2ex}{\scriptscriptstyle\perp\!}}}\mathrm{d}q+\frac{\mu}{2} Z^{{\raisemath{0.2ex}{\scriptscriptstyle\perp\!}}}{\star_\gamma}\peqsubfino{q}{\perp}{-0.2ex}\right)\!\right)\!\peqsub{\mathrm{vol}}{\Sigma}-{}\\
&-\!\int_{\partial\Sigma}\varepsilon Z^{{\raisemath{0.2ex}{\scriptscriptstyle\perp\!}}}\peqsubfino{{\left\langle\peqsub{\jmath}{\partial}^*\!\!\left(\vecc{Y}{q}-\mathcal{L}_{\vec{y}^{\scriptscriptstyle\top}}q-Y^{{\raisemath{0.2ex}{\scriptscriptstyle\perp\!}}}\frac{\underline{\peqsubfino{\Lambda}{+}{0.4ex}}}{\eta\sqrt{\gamma}}-\varepsilon\mathrm{d}( Y^{{\raisemath{0.2ex}{\scriptscriptstyle\perp\!}}}\peqsubfino{q}{\perp}{-0.2ex})\right)\!\!\coma\peqsub{\jmath}{\partial}^*\imath_{\vec{\nu}}\left(\eta\mathrm{d}q+\frac{\mu}{2}{\star_\gamma}\peqsubfino{q}{\perp}{-0.2ex}\right)+\peqsub{\nu}{\perp}\frac{\underline{p}^{\scriptscriptstyle\partial}}{\sqrt{\gamma}}\right\rangle_{\!\!\!\gamma}}}{\!\partial}{-.3ex}\frac{\peqsubfino{{\mathrm{vol}_\gamma}}{\!\partial\!X}{-0.4ex}}{|\vec{\nu}^{\,\scriptscriptstyle\top}|}+{}\\
&+\!\int_{\partial\Sigma}\varepsilon Z^{{\raisemath{0.2ex}{\scriptscriptstyle\perp\!}}} \peqsubfino{{\left\langle\peqsub{\jmath}{\partial}^*\big(\vecc[\perp]{Y}{q}\hspace*{0.1ex}-\mathcal{L}_{\vec{y}^{\scriptscriptstyle\top}}\peqsubfino{q}{\perp}{-0.2ex}+\peqsub{\imath}{\mathrm{d}Y^{\!\scriptscriptstyle\perp}}q\big)\coma\peqsub{\jmath}{\partial}^*\!\left(\frac{\imath_{\vec{\nu}}\underline{p}}{\sqrt{\gamma}}\right)\right\rangle_{\!\!\!\gamma}}}{\!\partial}{-.3ex}\frac{\peqsubfino{{\mathrm{vol}_\gamma}}{\!\partial\!X}{-0.4ex}}{|\vec{\nu}^{\,\scriptscriptstyle\top}|}-(Y\leftrightarrow Z)
\end{split}
\end{align*}

\section{GNH algorithm}\trassub

As the Lagrangian $\peqsub{L}{\mathrm{MCS}}$ is homogeneous of degree $1$ in the velocities, we have again that the energy, and thus the Hamiltonian, is zero. We have once again that the dynamics is purely gauge in the sense that it goes along the degenerate direction of the presymplectic form $\omega^{\scriptscriptstyle\mathrm{MCS}}$.\separ

The Hamiltonian vector field $Y\in\mathfrak{X}(\mathcal{P})$ is given by the Hamilton equation
\begin{equation}\label{Applications - equation - Hamiltonina equation =0}
\omega^{\scriptscriptstyle\mathrm{MCS}}_{(\peqsubfino{q}{\perp}{-0.2ex},q,X,p)}\Big((\vecc[\scalebox{0.4}{$\perp$}]{Y}{q},\vecc{Y}{q},\vecc{Y}{X},\vecc{Y}{p})\coma(\vecc[\scalebox{0.4}{$\perp$}]{Z}{q},\vecc{Z}{q},\vecc{Z}{X},\vecc{Z}{p})\Big)=\mathrm{d}H(\vecc[\scalebox{0.4}{$\perp$}]{Z}{q},\vecc{Z}{q},\vecc{Z}{X},\vecc{Z}{p})=0\end{equation}
for every $Z\in\mathfrak{X}(\mathcal{P})$. Taking in particular $(0,0,0,\vecc{Z}{p})$, $(0,\vecc{Z}{q},0,0)$, $(\vecc[\perp]{Z}{q},0,0,0)$ and $(0,0,Z^{{\raisemath{0.2ex}{\scriptscriptstyle\perp\!}}}\mathbbm{n},0)$ vanishing at the boundary leads to the conditions on the bulk
\begin{align}\label{Applications - equation - H vector field delta p=0}
\begin{split}
&\vecc{Y}{q}=\mathcal{L}_{\vec{y}^{\scriptscriptstyle\top}}q+Y^{{\raisemath{0.2ex}{\scriptscriptstyle\perp\!}}}\frac{\underline{\peqsubfino{\Lambda}{+}{0.4ex}}}{\eta\sqrt{\gamma}}+\varepsilon\mathrm{d}( Y^{{\raisemath{0.2ex}{\scriptscriptstyle\perp\!}}}\peqsubfino{q}{\perp}{-0.2ex})\\[1ex]
&\vecc{Y}{p}=\mathcal{L}_{\vec{y}^{\scriptscriptstyle\top}}p+ Y^{{\raisemath{0.2ex}{\scriptscriptstyle\perp\!}}}\frac{\mu}{2\eta}\peqsubfino{{\star_\gamma}}{\!X}{-0.2ex}\Lambda_++\varepsilon\sqrt{\gamma}\,\sharp_\gamma\delta\!\left(\eta Y^{{\raisemath{0.2ex}{\scriptscriptstyle\perp\!}}}\mathrm{d}q+\frac{\mu}{2} Y^{{\raisemath{0.2ex}{\scriptscriptstyle\perp\!}}}\peqsubfino{{\star_\gamma}}{\!X}{-0.2ex}\peqsubfino{q}{\perp}{-0.2ex}\right)\\[1.8ex]
&Y^{{\raisemath{0.2ex}{\scriptscriptstyle\perp\!}}}\delta \Lambda_-=0\\[1ex]
&\Big(\vecc[\perp]{Y}{q}\hspace*{0.1ex}-\mathcal{L}_{\vec{y}^{\scriptscriptstyle\top}}\peqsubfino{q}{\perp}{-0.2ex}+\peqsub{\imath}{\mathrm{d}Y^{\!\scriptscriptstyle\perp}}q\hspace*{0.1ex},\hspace*{0.1ex}\delta \Lambda_-\Big)=0
\end{split}
\end{align}
Once again we have a bifurcation depending on the support of $\delta \Lambda_-$: wherever it vanishes, $Y^{{\raisemath{0.2ex}{\scriptscriptstyle\perp\!}}}$ is arbitrary, wherever it does not vanish, $Y^{{\raisemath{0.2ex}{\scriptscriptstyle\perp\!}}}=0$.\separ

$\blacktriangleright$ Wherever $\delta\Lambda_-\neq0$
\begin{align}
\begin{split}
&\vecc[\perp]{Y}{q}\hspace*{0.1ex}=\mathcal{L}_{\vec{y}^{\scriptscriptstyle\top}}\peqsubfino{q}{\perp}{-0.2ex}\\[1ex]
&\vecc{Y}{q}=\mathcal{L}_{\vec{y}^{\scriptscriptstyle\top}}q\\[1ex]
&\vecc{Y}{p}=\mathcal{L}_{\vec{y}^{\scriptscriptstyle\top}}p\\[1ex]
&Y^{{\raisemath{0.2ex}{\scriptscriptstyle\perp\!}}}=0\\[.5ex]
&y^{{\raisemath{-0.1ex}{\scriptscriptstyle\top\!}}}\text{ arbitrary}
\end{split}\label{MCS - equation - solucion delta Lambda not 0}
\end{align}
Here we have the action of the group of $\mathrm{Diff}(\Sigma)$ acting through its infinitesimal version, the Lie derivative. The GNH algorithm stops because we have found the most general solution with no constraint.\separ

$\blacktriangleright$ Wherever $\delta \Lambda_-=0$
\begin{align}\label{Applications - equations - Hamiltonian vector field delta lambda=0}
\begin{split}
&\vecc[\perp]{Y}{q}\hspace*{0.1ex}\text{ arbitrary}\\[.9ex]
&\vecc{Y}{q}=\mathcal{L}_{\vec{y}^{\scriptscriptstyle\top}}q+Y^{{\raisemath{0.2ex}{\scriptscriptstyle\perp\!}}}\frac{\underline{\peqsubfino{\Lambda}{+}{0.4ex}}}{\eta\sqrt{\gamma}}+\varepsilon\mathrm{d}( Y^{{\raisemath{0.2ex}{\scriptscriptstyle\perp\!}}}\peqsubfino{q}{\perp}{-0.2ex})\\[1ex]
&\vecc{Y}{p}=\mathcal{L}_{\vec{y}^{\scriptscriptstyle\top}}p+ Y^{{\raisemath{0.2ex}{\scriptscriptstyle\perp\!}}}\frac{\mu}{2\eta}\peqsubfino{{\star_\gamma}}{\!X}{-0.2ex}\Lambda_++\varepsilon\sqrt{\gamma}\,\sharp_\gamma\delta\!\left(\eta Y^{{\raisemath{0.2ex}{\scriptscriptstyle\perp\!}}}\mathrm{d}q+\frac{\mu}{2} Y^{{\raisemath{0.2ex}{\scriptscriptstyle\perp\!}}}\peqsubfino{{\star_\gamma}}{\!X}{-0.2ex}\peqsubfino{q}{\perp}{-0.2ex}\right)\\[1.8ex]
&Y^{{\raisemath{0.2ex}{\scriptscriptstyle\perp\!}}}\text{ arbitrary}\\[.8ex]
&y^{{\raisemath{-0.1ex}{\scriptscriptstyle\top\!}}}\text{ arbitrary}
\end{split}
\end{align}
In this case, the GNH algorithm goes on by requiring that the Hamiltonian vector field $Y\in\mathfrak{X}(\mathcal{P})$ that we have found in the first step is tangent to
\[\mathcal{P}_2=\Big\{(\peqsubfino{q}{\perp}{-0.2ex},q,X,p)\in\mathcal{P}\ \ /\ \ \delta \Lambda_-=0\Big\}\]
Notice that this is always the case because
\begin{align*}
\peqsub{D}{Y}(\delta\Lambda_-)&\overset{\eqref{Appendix - lemma - divergencia densidad}}{=}\delta(\peqsub{D}{Y}\Lambda_-)\overset{\eqref{Applications - equation - DLambda}}{=}\\
&\overset{\phantom{\eqref{Appendix - lemma - divergencia densidad}}}{=}\delta\left(\vecc{Y}{p}-\frac{\mu}{2}\sqrt{\gamma}\,\peqsubfino{{\star_\gamma}}{\!X}{-0.2ex}\vecc{Y}{q}\right)\overset{\eqref{Applications - equations - Hamiltonian vector field delta lambda=0}}{=}\\
&\overset{\phantom{\eqref{Appendix - lemma - divergencia densidad}}}{=}\delta\left(\mathcal{L}_{\vec{y}^{\scriptscriptstyle\top}}p+ \textcolor{red}{Y^{{\raisemath{0.2ex}{\scriptscriptstyle\perp\!}}}\frac{\mu}{2\eta}\peqsubfino{{\star_\gamma}}{\!X}{-0.2ex}\Lambda_+}-\frac{\mu}{2}\sqrt{\gamma}\,\peqsubfino{{\star_\gamma}}{\!X}{-0.2ex}\left[\mathcal{L}_{\vec{y}^{\scriptscriptstyle\top}}q+\textcolor{red}{Y^{{\raisemath{0.2ex}{\scriptscriptstyle\perp\!}}}\frac{\underline{\peqsubfino{\Lambda}{+}{0.4ex}}}{\eta\sqrt{\gamma}}}+\varepsilon\mathrm{d}( Y^{{\raisemath{0.2ex}{\scriptscriptstyle\perp\!}}}\peqsubfino{q}{\perp}{-0.2ex})\right]\right)\updown{\eqref{appendix property delta star=star d}}{\eqref{Appendix - definition - delta^2=0}}{=}\\
&\overset{\phantom{\eqref{Appendix - lemma - divergencia densidad}}}{=}\delta\left(\mathcal{L}_{\vec{y}^{\scriptscriptstyle\top}}p-\frac{\mu}{2}\sqrt{\gamma}\,\peqsubfino{{\star_\gamma}}{\!X}{-0.2ex}\mathcal{L}_{\vec{y}^{\scriptscriptstyle\top}}q\right)+\textcolor{red}{0}+0\overset{\eqref{Appendix - lemma - divergencia densidad}}{=}\mathcal{L}_{\vec{y}^{\scriptscriptstyle\top}}(\delta\Lambda_-)
\end{align*}
As it is clear from the context, we have omitted the underline in several places to simplify the notation. Notice that in the last equality we have used the fact that the Lie derivative is a variation in the direction $\mathbb{Y}=\tau.\vec{y}$ so it commutes with the codifferential as in the first equality. Of course we have that if $\delta \Lambda_-=0$ at some point it will remain so for ever (so the algorithm also stops at this first step) and, therefore, the two cases are complementary.\separ

$\blacktriangleright$ Boundary\separprevia

Let us now keep going with the GNH algorithm to see what happens at the boundary. We have already solved \eqref{Applications - equation - Hamiltonina equation =0} at the bulk, so only the boundary integrals remain. Note that if $\delta \Lambda_-\neq0$ all of them are zero due to \eqref{MCS - equation - solucion delta Lambda not 0} and the fact that $Y^{{\raisemath{0.2ex}{\scriptscriptstyle\perp\!}}}=0$, thus we are done. Let us then focus on the case $\delta \Lambda_-=0$, in which we still have that some of the integrals are zero by continuity of \eqref{Applications - equations - Hamiltonian vector field delta lambda=0}, so we have to solve
\begin{align*}
\begin{split}
0&=\int_{\partial\Sigma} \peqsubfino{{\left\langle\peqsub{\jmath}{\partial}^*\!\!\left(\vecc{Z}{q}-\mathcal{L}_{\vec{z}^{\scriptscriptstyle\top}}q-Z^{{\raisemath{0.2ex}{\scriptscriptstyle\perp\!}}}\frac{\underline{\peqsubfino{\Lambda}{+}{0.4ex}}}{\eta\sqrt{\gamma}}-\varepsilon\mathrm{d}( Z^{{\raisemath{0.2ex}{\scriptscriptstyle\perp\!}}}\peqsubfino{q}{\perp}{-0.2ex})\right)\!\!\coma Y^{{\raisemath{0.2ex}{\scriptscriptstyle\perp\!}}}\left[\peqsub{\jmath}{\partial}^*\imath_{\vec{\nu}}\left(\eta\mathrm{d}q+\frac{\mu}{2}\peqsubfino{{\star_\gamma}}{\!X}{-0.2ex}\peqsubfino{q}{\perp}{-0.2ex}\right)+\peqsub{\nu}{\perp}\frac{\underline{p}^{\scriptscriptstyle\partial}}{\sqrt{\gamma}}\right]\right\rangle_{\!\!\!\gamma}}}{\!\partial}{-.3ex}\frac{\peqsubfino{{\mathrm{vol}_\gamma}}{\!\partial\!X}{-0.4ex}}{|\vec{\nu}^{\,\scriptscriptstyle\top}|}+{}\\
&\phantom{=}+\int_{\partial\Sigma} \peqsubfino{{ \left\langle Z^{{\raisemath{0.2ex}{\scriptscriptstyle\perp\!}}}\peqsub{\jmath}{\partial}^*\big(\vecc[\perp]{Y}{q}\hspace*{0.1ex}-\mathcal{L}_{\vec{y}^{\scriptscriptstyle\top}}\peqsubfino{q}{\perp}{-0.2ex}+\peqsub{\imath}{\mathrm{d}Y^{\!\scriptscriptstyle\perp}}q\big)-Y^{{\raisemath{0.2ex}{\scriptscriptstyle\perp\!}}}\peqsub{\jmath}{\partial}^*\big(\vecc[\perp]{Z}{q}\hspace*{0.1ex}-\mathcal{L}_{\vec{z}^{\scriptscriptstyle\top}}\peqsubfino{q}{\perp}{-0.2ex}+\peqsub{\imath}{\mathrm{d}Z^{\!\scriptscriptstyle\perp}}q\big)\coma\peqsub{\jmath}{\partial}^*\!\left(\frac{\imath_{\vec{\nu}}\underline{p}}{\sqrt{\gamma}}\right)\right\rangle_{\!\!\!\gamma}}}{\!\partial}{-.3ex}\frac{\peqsubfino{{\mathrm{vol}_\gamma}}{\!\partial\!X}{-0.4ex}}{|\vec{\nu}^{\,\scriptscriptstyle\top}|}
\end{split}
\end{align*}
for every $Z$. Here we have again bifurcations with several possibilities. Notice that the case $Y^{{\raisemath{0.2ex}{\scriptscriptstyle\perp\!}}}=0$ is easy to solve and less interesting from a physical point of view, so we will assume from now on that $Y^{{\raisemath{0.2ex}{\scriptscriptstyle\perp\!}}}$ is arbitrary at the boundary.\separ

Notice that the CS parameter $\mu$ only appears once at the boundary. We now proceed to sketch how to obtain solutions to the previous equation.\separ

$\blacktriangleright$ Dirichlet $\peqsub{\mathcal{P}}{D1}:=\Big\{(\peqsubfino{q}{\perp}{-0.2ex},q,X,p)\in\mathcal{P}\ \ /\ \ \delta\Lambda_-=0\ \text{ and }\ \peqsubfino{q^{\scriptscriptstyle\partial}}{\perp}{-0.2ex}=0\Big\}$\separprevia

In this case the component $\peqsubfino{q}{\perp}{-0.2ex}$ of the tangent vectors to $\peqsub{\mathcal{P}}{D1}$ is also zero at the boundary. Besides, $\peqsubfino{{\star_\gamma}}{\!X}{-0.2ex}\peqsubfino{q}{\perp}{-0.2ex}=0$ at the boundary thanks to equation \eqref{Appendix - equation - definicion hodge}, so the previous equation reduces to
\begin{align*}
\begin{split}
0&=\int_{\partial\Sigma} \peqsubfino{{\left\langle\peqsub{\jmath}{\partial}^*\!\!\left(\vecc{Z}{q}-\mathcal{L}_{\vec{z}^{\scriptscriptstyle\top}}q-Z^{{\raisemath{0.2ex}{\scriptscriptstyle\perp\!}}}\frac{\underline{\peqsubfino{\Lambda}{+}{0.4ex}}}{\eta\sqrt{\gamma}}-\varepsilon\mathrm{d}( Z^{{\raisemath{0.2ex}{\scriptscriptstyle\perp\!}}}\peqsubfino{q}{\perp}{-0.2ex})\right)\!\!\coma Y^{{\raisemath{0.2ex}{\scriptscriptstyle\perp\!}}}\left[\eta\,\peqsub{\jmath}{\partial}^*\left(\imath_{\vec{\nu}}\mathrm{d}q\right)+\peqsub{\nu}{\perp}\frac{\underline{p}^{\scriptscriptstyle\partial}}{\sqrt{\gamma}}\right]\right\rangle_{\!\!\!\gamma}}}{\!\partial}{-.3ex}\frac{\peqsubfino{{\mathrm{vol}_\gamma}}{\!\partial\!X}{-0.4ex}}{|\vec{\nu}^{\,\scriptscriptstyle\top}|}+{}\\
&\phantom{=}+\int_{\partial\Sigma} \peqsubfino{{ \left\langle Z^{{\raisemath{0.2ex}{\scriptscriptstyle\perp\!}}}\peqsub{\jmath}{\partial}^*\big(-\mathcal{L}_{\vec{y}^{\scriptscriptstyle\top}}\peqsubfino{q}{\perp}{-0.2ex}+\peqsub{\imath}{\mathrm{d}Y^{\!\scriptscriptstyle\perp}}q\big)-Y^{{\raisemath{0.2ex}{\scriptscriptstyle\perp\!}}}\peqsub{\jmath}{\partial}^*\big(-\mathcal{L}_{\vec{z}^{\scriptscriptstyle\top}}\peqsubfino{q}{\perp}{-0.2ex}+\peqsub{\imath}{\mathrm{d}Z^{\!\scriptscriptstyle\perp}}q\big)\coma\peqsub{\jmath}{\partial}^*\!\left(\frac{\imath_{\vec{\nu}}\underline{p}}{\sqrt{\gamma}}\right)\right\rangle_{\!\!\!\gamma}}}{\!\partial}{-.3ex}\frac{\peqsubfino{{\mathrm{vol}_\gamma}}{\!\partial\!X}{-0.4ex}}{|\vec{\nu}^{\,\scriptscriptstyle\top}|}
\end{split}
\end{align*}
If we want $Y^{{\raisemath{0.2ex}{\scriptscriptstyle\perp\!}}}$ to be arbitrary we can impose the conditions
\begin{align*}
&\eta\,\peqsub{\jmath}{\partial}^*(\imath_{\vec{\nu}}\mathrm{d}q)+\peqsub{\nu}{\perp}\frac{\underline{p}^{\scriptscriptstyle\partial}}{\sqrt{\gamma}}=0\\
&\peqsub{\jmath}{\partial}^*\!\left(\imath_{\vec{\nu}}\underline{p}\right)=0
\end{align*}

$\blacktriangleright$ Dirichlet $\peqsub{\mathcal{P}}{D2}:=\Big\{(\peqsubfino{q}{\perp}{-0.2ex},q,X,p)\in\mathcal{P}\ \ /\ \ \delta\Lambda_-=0\text{ and }\peqsubfino{q}{\partial}{-0.2ex}=0\Big\}$\separprevia

Now the component $q$ of the tangent vectors to $\peqsub{\mathcal{P}}{D2}$ is zero at the boundary and the previous equation reduces to
\begin{align*}
\begin{split}
0&=\int_{\partial\Sigma} \peqsubfino{{\left\langle\peqsub{\jmath}{\partial}^*\!\!\left(-\mathcal{L}_{\vec{z}^{\scriptscriptstyle\top}}q-Z^{{\raisemath{0.2ex}{\scriptscriptstyle\perp\!}}}\frac{\underline{\peqsubfino{\Lambda}{+}{0.4ex}}}{\eta\sqrt{\gamma}}-\varepsilon\mathrm{d}( Z^{{\raisemath{0.2ex}{\scriptscriptstyle\perp\!}}}\peqsubfino{q}{\perp}{-0.2ex})\right)\!\!\coma Y^{{\raisemath{0.2ex}{\scriptscriptstyle\perp\!}}}\left[\peqsub{\jmath}{\partial}^*\imath_{\vec{\nu}}\left(\eta\mathrm{d}q+\frac{\mu}{2}\peqsubfino{{\star_\gamma}}{\!X}{-0.2ex}\peqsubfino{q}{\perp}{-0.2ex}\right)+\peqsub{\nu}{\perp}\frac{\underline{p}^{\scriptscriptstyle\partial}}{\sqrt{\gamma}}\right]\right\rangle_{\!\!\!\gamma}}}{\!\partial}{-.3ex}\frac{\peqsubfino{{\mathrm{vol}_\gamma}}{\!\partial\!X}{-0.4ex}}{|\vec{\nu}^{\,\scriptscriptstyle\top}|}+{}\\
&\phantom{=}+\int_{\partial\Sigma} \peqsubfino{{ \left\langle Z^{{\raisemath{0.2ex}{\scriptscriptstyle\perp\!}}}\peqsub{\jmath}{\partial}^*\big(\vecc[\perp]{Y}{q}\hspace*{0.1ex}-\mathcal{L}_{\vec{y}^{\scriptscriptstyle\top}}\peqsubfino{q}{\perp}{-0.2ex}\big)-Y^{{\raisemath{0.2ex}{\scriptscriptstyle\perp\!}}}\peqsub{\jmath}{\partial}^*\big(\vecc[\perp]{Z}{q}\hspace*{0.1ex}-\mathcal{L}_{\vec{z}^{\scriptscriptstyle\top}}\peqsubfino{q}{\perp}{-0.2ex}\big)\coma\peqsub{\jmath}{\partial}^*\!\left(\frac{\imath_{\vec{\nu}}\underline{p}}{\sqrt{\gamma}}\right)\right\rangle_{\!\!\!\gamma}}}{\!\partial}{-.3ex}\frac{\peqsubfino{{\mathrm{vol}_\gamma}}{\!\partial\!X}{-0.4ex}}{|\vec{\nu}^{\,\scriptscriptstyle\top}|}
\end{split}
\end{align*}
Again, as we want $Y^{{\raisemath{0.2ex}{\scriptscriptstyle\perp\!}}}$ to be arbitrary, we can impose the (sufficient) conditions
\begin{align*}
&\peqsub{\jmath}{\partial}^*\!\left(\imath_{\vec{\nu}}\underline{p}\right)=0\\
&\peqsub{\jmath}{\partial}^*\imath_{\vec{\nu}}\left(\eta\mathrm{d}q+\frac{\mu}{2}\peqsubfino{{\star_\gamma}}{\!X}{-0.2ex}\peqsubfino{q}{\perp}{-0.2ex}\right)+\peqsub{\nu}{\perp}\frac{\underline{p}^{\scriptscriptstyle\partial}}{\sqrt{\gamma}}=0\\
&\peqsub{\jmath}{\partial}^*\!\Big(\mathcal{L}_{\vec{y}^{\scriptscriptstyle\top}}q\Big)+Y^{{\raisemath{0.2ex}{\scriptscriptstyle\perp\!}}}\frac{\underline{p}^{\scriptscriptstyle\partial}}{\eta\sqrt{\gamma}}+\varepsilon\mathrm{d}( Y^{{\raisemath{0.2ex}{\scriptscriptstyle\perp\!}}}\peqsubfino{q}{\perp}{-0.2ex}^{\scriptscriptstyle\partial})=0
\end{align*}

$\blacktriangleright$ Dirichlet $\peqsub{\mathcal{P}}{D3}:=\Big\{(\peqsubfino{q}{\perp}{-0.2ex},q,X,p)\in\mathcal{P}\ \ /\ \ \delta\Lambda_-=0\ \ \peqsubfino{q^{\scriptscriptstyle\partial}}{\perp}{-0.2ex}=0\ \text{ and } \ \peqsubfino{q}{\partial}{-0.2ex}=0\Big\}$
\begin{align*}
&\peqsub{\jmath}{\partial}^*\!\left(\imath_{\vec{\nu}}\underline{p}\right)=0\\
&\eta\,\peqsub{\jmath}{\partial}^*\left(\imath_{\vec{\nu}}\mathrm{d}q\right)+\peqsub{\nu}{\perp}\frac{\underline{p}^{\scriptscriptstyle\partial}}{\sqrt{\gamma}}=0\\
&\peqsub{\jmath}{\partial}^*\!\Big(\mathcal{L}_{\vec{y}^{\scriptscriptstyle\top}}q\Big)+Y^{{\raisemath{0.2ex}{\scriptscriptstyle\perp\!}}}\frac{\underline{p}^{\scriptscriptstyle\partial}}{\eta\sqrt{\gamma}}+\varepsilon Y^{{\raisemath{0.2ex}{\scriptscriptstyle\perp\!}}}\mathrm{d} \peqsubfino{q}{\perp}{-0.2ex}^{\scriptscriptstyle\partial}=0
\end{align*}

$\blacktriangleright$ Robin-like $\mathcal{P}$
\begin{align*}
&\peqsub{\jmath}{\partial}^*\!\left(\imath_{\vec{\nu}}\underline{p}\right)=0\\
&\peqsub{\jmath}{\partial}^*\imath_{\vec{\nu}}\left(\eta\mathrm{d}q+\frac{\mu}{2}\peqsubfino{{\star_\gamma}}{\!X}{-0.2ex}\peqsubfino{q}{\perp}{-0.2ex}\right)+\peqsub{\nu}{\perp}\frac{\underline{p}^{\scriptscriptstyle\partial}}{\sqrt{\gamma}}=0\\
\end{align*}

In all the previous cases we should check that the dynamics preserved the corresponding boundary conditions. In general an infinite chain of conditions appear. We see how the inclusion of the boundary makes the theory much richer but also much more difficult to handle.

\section{Parametrized Chern-Simons with boundaries}\label{Applications - Section - CS boundaries}

\subsection*{Introduction}\trassub

It is quite interesting to study the Chern-Simons theory alone which is given by the action\index{Action!Chern-Simons}
 \[S(A,Z)=\frac{\mu }{2}\int_MA\wedge \mathrm{d}A\qquad\quad A\in\Omega^k(M)\ \ k \text{ odd}\]
 which corresponds to the Maxwell-Chern-Simons theory with $\eta=0$. However, we cannot just take $\eta$ to be zero in the Hamiltonian vector field because it appears in the denominator. Actually, as we will see, this comes from the fact that the phase space has only positions and no momenta.\separ
 
 An interesting feature of the previous action is that it has no background geometric object and, in particular, it does not depend on $Z$. It might seem therefore unnecessary to parametrize the theory, nonetheless if we did not when we consider the Lagrangian formulation, we would  need to fix a foliation and then the Lagrangian would be ``embedding-dependent''. This complicates quite a lot the GNH algorithm so it is wiser and conceptually better to parametrize the theory.

 \subsection*{Lagrangian formulation}\trassub
 
 The same procedure that we follow to obtain the Lagrangian \eqref{Parametrized theories - equation - Lagrangian} allows us to obtain the CS Lagrangian	\[\begin{array}{cccc}
 \peqsub{L}{\mathrm{CS}}: & \peqsub{\mathcal{D}}{\mathrm{CS}}\subset T\Big(\Omega^{k-1}(\Sigma)\times \Omega^k(\Sigma)\times\mathrm{Emb}(\Sigma,M)\Big) & \longrightarrow & \R\\
 &           \mathbf{v}_{(\peqsubfino{q}{\perp}{-0.2ex},q,X)}=(\peqsubfino{q}{\perp}{-0.2ex},q,X;v_{\!\scriptscriptstyle\perp},v,\peqsub{\mathbb{V}}{\!X})                   &  \longmapsto    & L(\mathbf{v}_{(\peqsubfino{q}{\perp}{-0.2ex},q,X)})
 \end{array}\]
 which is given by
 \begin{align*}
 \peqsub{L}{\mathrm{CS}}(\mathbf{v}_{(\peqsub{q}{\perp},q,X)})&=\frac{\varepsilon\mu}{2}{\big\llangle\peqsubfino{q}{\perp}{-0.2ex},\peqsubfino{{\star_\gamma}}{\!X}{-0.2ex}(\mathrm{d}q)\big\rrangle}_{\peqsub{V}{X}^{{\raisemath{0.2ex}{\!\scriptscriptstyle\perp}}}}-\frac{\mu}{2}\left\llangle\frac{v-\mathcal{L}_{\peqsub{\vec{v}}{X}^{\scriptscriptstyle\top}}q-\varepsilon\mathrm{d}(\peqsub{V}{X}^{{\raisemath{0.2ex}{\!\scriptscriptstyle\perp}}}\peqsubfino{q}{\perp}{-0.2ex})}{\peqsub{V}{X}^{{\raisemath{0.2ex}{\!\scriptscriptstyle\perp}}}},\peqsubfino{{\star_\gamma}}{\!X}{-0.2ex}q\right\rrangle_{\!\!\peqsub{V}{X}^{{\raisemath{0.2ex}{\!\scriptscriptstyle\perp}}}}
 \end{align*}
 We see that the Lagrangian depends indeed on the velocity of the embeddings. Although the metric appears explicitly, the Lagrangian does not depend on it. This is clear as we can rewrite the Lagrangian as
\[L_{\mathrm{CS}}\Big(\mathbf{v}_{(q_\perp,q,X)}\Big)=\frac{\mu}{2}\int_\Sigma\Big(\varepsilon V^\perp_X q_\perp\wedge \mathrm{d}q+\left(v-\mathcal{L}_{\vec{v}_X^\top}q-\varepsilon\mathrm{d}(V^\perp_X q_\perp)\right)\wedge q\Big)\]
 We could use this latter way of writing the Lagrangian but using the former allows us to reuse the computations of the Maxwell-Chern-Simons case.
 
 \subsection*{Geometric arena}\trassub
 
 We define, as in the MCS case, $\widetilde{\mathcal{P}}=\{\peqsub{\mathrm{p}}{\mathrm{q}}:=(\peqsubfino{q}{\perp}{-0.2ex},q,X,\peqsubfino{p}{\perp}{-0.2ex},p,\peqsub{P}{\!X},\peqsub{P}{\partial X})\}\overset{\jmath}{\hookrightarrow }T^*\mathcal{D}$ considered as a subset of $T^*\mathcal{D}$. We define the induced form $\widetilde{\Omega}^{\scriptscriptstyle\mathrm{CS}}:=\jmath^*\Omega$ which is given by
 \begin{align}\label{Applications - equation - tilde Omega CS}
 \begin{split}
 \widetilde{\Omega}^{\scriptscriptstyle\mathrm{CS}}_{\peqsub{\mathrm{p}}{\mathrm{q}}}(Y,Z)&=\int_\Sigma\Big[\Big(\vecc[\scalebox{0.4}{$\perp$}]{Y}{q},\vecc[\scalebox{0.4}{$\perp$}]{Z}{p}\Big)-\Big(\vecc[\scalebox{0.4}{$\perp$}]{Z}{q},\vecc[\scalebox{0.4}{$\perp$}]{Y}{p}\Big)+\Big(\vecc{Y}{q},\vecc{Z}{p}\Big)-\Big(\vecc{Z}{q},\vecc{Y}{p}\Big)+\vecc{Z}{P}(\vecc{Y}{X})-\vecc{Y}{P}(\vecc{Z}{X})\Big]\peqsub{\mathrm{vol}}{\Sigma}+{}\\
 &\phantom{=}+\int_{\partial\Sigma}\left[\vecc{Z}{P}^{\scriptscriptstyle\partial}(\vecc{Y}{X})-\vecc{Y}{P}^{\scriptscriptstyle\partial}(\vecc{Z}{X})\right]\peqsub{\mathrm{vol}}{\partial\Sigma}     
 \end{split}
 \end{align}
 
 \subsection*{Computation of the fiber derivative}\trassub
 
 The fiber derivative, given by equation \eqref{Mathematical background - equation - FL=D_2L}, is computed taking an initial point of the tangent bundle $\mathbf{v}_{(\peqsubfino{q}{\perp}{-0.2ex},q,X)}=(\peqsubfino{q}{\perp}{-0.2ex},q,X;\peqsubfino{v}{\!\perp}{-0.2ex},v,\peqsubfino{\mathbb{V}}{\!X}{-0.1ex})$ and some initial velocities $\mathbf{w}^1_{(\peqsubfino{q}{\perp}{-0.2ex},q,X)}=(\peqsubfino{q}{\perp}{-0.2ex},q,X;\peqsubfino{w}{\!\perp}{-0.2ex},0,0)$, $\mathbf{w}^2_{(\peqsubfino{q}{\perp}{-0.2ex},q,X)}=(\peqsubfino{q}{\perp}{-0.2ex},q,X;0,w,0)$ and $\mathbf{w}^3_{(\peqsubfino{q}{\perp}{-0.2ex},q,X)}=(\peqsubfino{q}{\perp}{-0.2ex},q,X;0,0,\peqsubfino{\mathbb{W}}{\!X}{-0.1ex})$. The first two are immediate while the last one can be obtained following the ideas of lemma \ref{Appendix - equation - FL(w3)=Pi}.
 \begin{align*}
 FL&(\mathbf{v}_{(\peqsubfino{q}{\perp}{-0.2ex},q,X)})\!\left(\hspace*{-0.1ex}\mathbf{w}^1_{(\peqsubfino{q}{\perp}{-0.2ex},q,X)}\hspace*{-0.1ex}\right)=0\\
 FL&(\mathbf{v}_{(\peqsubfino{q}{\perp}{-0.2ex},q,X)})\!\left(\hspace*{-0.1ex}\mathbf{w}^2_{(\peqsubfino{q}{\perp}{-0.2ex},q,X)}\hspace*{-0.1ex}\right)=\left\llangle\frac{w}{\peqsub{V}{X}^{{\raisemath{0.2ex}{\!\scriptscriptstyle\perp}}}},-\frac{\mu}{2}\peqsubfino{{\star_\gamma}}{\!X}{-0.2ex}q\right\rrangle_{\!\!\peqsub{V}{X}^{{\raisemath{0.2ex}{\!\scriptscriptstyle\perp}}}}\hspace*{-0.3ex}=\left(\hspace*{-0.3ex} w\hspace*{.2ex},\hspace*{.1ex}-\frac{\mu}{2}\sqrt{\peqsubfino{\gamma}{X}{-0.2ex}}\,\peqsub{\sharp}{\gamma}\peqsubfino{{\star_\gamma}}{\!X}{-0.2ex}q\right)\\
 FL&(\mathbf{v}_{(\peqsubfino{q}{\perp}{-0.2ex},q,X)})\!\left(\hspace*{-0.1ex}\mathbf{w}^3_{(\peqsubfino{q}{\perp}{-0.2ex},q,X)}\hspace*{-0.1ex}\right)=-\int_\Sigma \mathbb{W}^\alpha\Big(\varepsilon(\peqsub{n}{X})_\alpha\peqsubfino{\mathcal{H}}{\!\perp}{-0.1ex}^{\scriptscriptstyle\mathrm{CS}}+(\peqsubfino{e}{X}{-0.2ex})^b_\alpha\mathcal{H}_b\Big)\peqsub{\mathrm{vol}}{\Sigma}\,-\\
 &\hspace*{30ex}-\int_{\partial\Sigma}\mathbb{W}^\alpha\Big(\varepsilon (\peqsub{n}{X})_\alpha\peqsubfino{\mathcal{H}}{\!\perp}{-0.1ex}^{\scriptscriptstyle\partial}+ (\peqsubfino{e}{X}{-0.2ex})^b_\alpha\Big)\mathcal{H}^{\scriptscriptstyle\partial}_b\peqsub{{\mathrm{vol}}}{\partial\Sigma}
 \end{align*}
 where $(\alpha,V)=\frac{1}{k!}\alpha_{a_1\cdots a_k}V^{a_1\cdots a_k}$ is the natural pairing. We can now define the canonical momenta
 \begin{align*}
 &\left.\begin{array}{l}
 \peqsubfino{p}{\perp}{-0.2ex}=0\\[.8ex]
 \displaystyle p=-\frac{\mu}{2}\sqrt{\peqsubfino{\gamma}{X}{-0.2ex}}\,\peqsub{\sharp}{\gamma}\peqsubfino{{\star_\gamma}}{\!X}{-0.2ex}q\\[2.2ex]
 (\peqsub{P}{\!X})_\alpha=-\varepsilon(\peqsub{n}{X})_\alpha\peqsubfino{\mathcal{H}}{\!\perp}{-0.1ex}^{\scriptscriptstyle\mathrm{CS}}(\peqsubfino{q}{\perp}{-0.2ex},q)-(\peqsubfino{e}{X}{-0.2ex})^b_\alpha\mathcal{H}_b(q)\\[1.9ex]
 (\peqsub{P^{\scriptscriptstyle\partial}}{X})_\alpha=-\varepsilon(\peqsub{n}{X})_\alpha\peqsubfino{\mathcal{H}}{\!\perp}{-0.1ex}^{\scriptscriptstyle\partial}(q)-(\peqsubfino{e}{X}{-0.2ex})^b_\alpha\mathcal{H}^{\scriptscriptstyle\partial}_b(\peqsubfino{q}{\perp}{-0.2ex},q)
 \end{array}\right\}\text{ Constraints}
 \end{align*}
 where $\peqsubfino{{\sharp_{\gamma}}}{\!X}{-0.2ex}$ is the musical isomorphism \eqref{Mathematical background - equation - musical ishomorphisms} of $\peqsubfino{\gamma}{X}{-0.2ex}$ and
 \begin{align}\label{Applications - definitions - H's CS}
 \begin{split}
 &\mathcal{H}(q)=-\frac{\mu}{2}\Big(\imath_{-}\mathrm{d}q,\sqrt{\peqsubfino{\gamma}{X}{-0.2ex}}\,\peqsub{\sharp}{\gamma}\peqsubfino{{\star_\gamma}}{\!X}{-0.2ex}q\Big)-\frac{\mu}{2}\Big(\imath_{-}q,\delta \big[\sqrt{\peqsubfino{\gamma}{X}{-0.2ex}}\,\peqsub{\sharp}{\gamma}\peqsubfino{{\star_\gamma}}{\!X}{-0.2ex}q\big]\Big)\in\Omega^1(\Sigma)\\
 &\peqsubfino{\mathcal{H}}{\!\perp}{-0.1ex}^{\scriptscriptstyle\mathrm{CS}}(\peqsubfino{q}{\perp}{-0.2ex},q)=-\varepsilon\mu\Big(\peqsubfino{q}{\perp}{-0.2ex}\hspace*{0.1ex},\hspace*{0.1ex}\delta \big[\sqrt{\peqsubfino{\gamma}{X}{-0.2ex}}\,\peqsub{\sharp}{\gamma}\peqsubfino{{\star_\gamma}}{\!X}{-0.2ex}q\big]\Big)\in\Cinf{\Sigma}\\
 &\mathcal{H}^{\scriptscriptstyle\partial}(q)=-\frac{\mu}{2}\left(\frac{\sqrt{\peqsubfino{\gamma^{\scriptscriptstyle\partial}}{\!X}{-0.3ex}}}{\sqrt{\peqsubfino{\gamma}{\!X}{-0.3ex}}}\frac{\peqsub{\nu}{X}}{|\peqsub{\vec{\nu}}{X}^{\,\scriptscriptstyle\top}|}\wedge\imath_{-}q\coma\sqrt{\peqsubfino{\gamma}{X}{-0.2ex}}\,\peqsub{\sharp}{\gamma}\peqsubfino{{\star_\gamma}}{\!X}{-0.2ex}q\right)\in\Omega^1(\partial\Sigma)\\
 &\peqsubfino{\mathcal{H}}{\!\perp}{-0.1ex}^{\scriptscriptstyle\partial}(\peqsubfino{q}{\perp}{-0.2ex},q)=-\frac{\varepsilon\mu}{2}\left(\frac{\sqrt{\peqsubfino{\gamma^{\scriptscriptstyle\partial}}{\!X}{-0.3ex}}}{\sqrt{\peqsubfino{\gamma}{\!X}{-0.3ex}}}\frac{\peqsub{\nu}{X}}{|\peqsub{\vec{\nu}}{X}^{\,\scriptscriptstyle\top}|}\wedge\peqsubfino{q}{\perp}{-0.2ex}\coma\sqrt{\peqsubfino{\gamma}{X}{-0.2ex}}\,\peqsub{\sharp}{\gamma}\peqsubfino{{\star_\gamma}}{\!X}{-0.2ex}q\right)\in\Cinf{\partial\Sigma}
 \end{split}
 \end{align}
 The last canonical momenta can be obtained by direct computation or imposing 
 \begin{equation}\label{MCS - equation - p=-sqrt star q}
 p=-\frac{\mu}{2}\sqrt{\peqsubfino{\gamma}{X}{-0.2ex}}\,\peqsub{\sharp}{\gamma}\peqsubfino{{\star_\gamma}}{\!X}{-0.2ex}q
 \end{equation}
 in \eqref{Applications - definitions - H's}. In particular $\Lambda_+=0$ and $\Lambda_-=-\mu\sqrt{\peqsubfino{\gamma}{X}{-0.2ex}}\,\peqsub{\sharp}{\gamma}\peqsubfino{{\star_\gamma}}{\!X}{-0.2ex}q$. Notice that none of the functions defined in \eqref{Applications - definitions - H's CS} depend on the embeddings because, as we saw on  equations \eqref{Parametrized theories - equation - gamma/gamma nu/nu} and \eqref{Applications - equation - DLambda}, the quantities
 \begin{equation}\label{MCS - equation - independece embedding}
 \frac{\sqrt{\peqsubfino{\gamma^{\scriptscriptstyle\partial}}{\!X}{-0.3ex}}}{\sqrt{\peqsubfino{\gamma}{\!X}{-0.3ex}}}\frac{\peqsub{\nu}{X}}{|\peqsub{\vec{\nu}}{X}^{\,\scriptscriptstyle\top}|}\qquad\qquad\text{ and }\qquad\qquad\sqrt{\peqsubfino{\gamma}{X}{-0.2ex}}\,\peqsub{\sharp}{\gamma}\peqsubfino{{\star_\gamma}}{\!X}{-0.2ex}
 \end{equation}
 are independent of the embeddings. In particular we have (see equation \eqref{Applications - equation - DLambda}) that
 \begin{equation}\label{MCS - equation - Yp=-sqrt star Y_q}
 \vecc{Y}{p}=-\frac{\mu}{2}\sqrt{\peqsubfino{\gamma}{X}{-0.2ex}}\,\peqsub{\sharp}{\gamma}\peqsubfino{{\star_\gamma}}{\!X}{-0.2ex}\vecc{Y}{q} 
 \end{equation}
 The manifold of first constraints of the CS theory is $\peqsub{\mathcal{P}}{\mathrm{CS}}:=FL(TQ)$ which is then given by
 \begin{align*}
 \peqsub{\mathcal{P}}{\mathrm{CS}}&=\left\{(\peqsubfino{q}{\perp}{-0.2ex},q,X;\peqsubfino{p}{\perp}{-0.2ex},p,\,\peqsub{P}{\!X},\,\peqsub{P}{\!X}^{\scriptscriptstyle\partial})\in\widetilde{\mathcal{P}}\ \ /\ \ \begin{array}{ll}\peqsubfino{p}{\perp}{-0.2ex}=0& \peqsub{P}{\!X}=-\varepsilon\peqsubfino{\mathcal{H}}{\!\perp}{-0.1ex} n-e.\mathcal{H}\\ p=-\frac{\mu}{2}\sqrt{\peqsubfino{\gamma}{X}{-0.2ex}}\,\peqsub{\sharp}{\gamma}\peqsubfino{{\star_\gamma}}{\!X}{-0.2ex}q&\peqsub{P}{\!X}^{\scriptscriptstyle\partial}=-\varepsilon\peqsubfino{\mathcal{H}}{\!\perp}{-0.1ex}^{\scriptscriptstyle\partial}n-e.\mathcal{H}^{\scriptscriptstyle\partial}\end{array}\right\}=\\
 &=\Big\{(\peqsubfino{q}{\perp}{-0.2ex},q,X;0,-\frac{\mu}{2}\sqrt{\peqsubfino{\gamma}{X}{-0.2ex}}\,\peqsub{\sharp}{\gamma}\peqsubfino{{\star_\gamma}}{\!X}{-0.2ex}q\coma-\varepsilon\peqsubfino{\mathcal{H}}{\!\perp}{-0.1ex} n-e.\mathcal{H}\coma -\varepsilon\peqsubfino{\mathcal{H}}{\!\perp}{-0.1ex}^{\scriptscriptstyle\partial} n-e.\mathcal{H}^{\scriptscriptstyle\partial}) \Big\}\cong\\
 &\cong\Big\{(\peqsubfino{q}{\perp}{-0.2ex},q,X)\Big\}=:\mathcal{P}
 \end{align*}
 Notice that, as we mentioned in the introduction, the manifold of first constraints has only positions and no momenta whatsoever. We define the inclusion $\peqsubfino{\jmath}{\mathrm{CS}}{-0.2ex}:\mathcal{P}\hookrightarrow \widetilde{\mathcal{P}}$ that allows us to pullback the induced form $\widetilde{\Omega}^{\scriptscriptstyle\mathrm{CS}}=\jmath^*\Omega$ of $\widetilde{\mathcal{P}}$, given by equation \eqref{Parametrized theories - equation - tilde Omega}, to $\mathcal{P}$ in order to define $\omega^{\scriptscriptstyle\mathrm{CS}}:=\peqsub{\jmath}{\mathrm{CS}\,}^*\widetilde{\Omega}^{\scriptscriptstyle\mathrm{CS}}$ that, in this case, is given by
 \begin{align}\label{Application - equation - omega final CS}
 \begin{split}
 &\omega^{\scriptscriptstyle\mathrm{CS}}_{(\peqsubfino{q}{\perp}{-0.2ex},q,X)}\Big((\vecc[\scalebox{0.4}{$\perp$}]{Y}{q},\vecc{Y}{q},\vecc{Y}{X}),(\vecc[\scalebox{0.4}{$\perp$}]{Z}{q},\vecc{Z}{q},\vecc{Z}{X})\Big)=\\
 &=\int_\Sigma\Big[\Big(\vecc{Y}{q}\coma -\frac{\mu}{2}\peqsub{D}{Z}(\sqrt{\peqsubfino{\gamma}{X}{-0.2ex}}\,\peqsub{\sharp}{\gamma}\peqsubfino{{\star_\gamma}}{\!X}{-0.2ex}q)\Big)-\Big(\vecc{Z}{q}\coma -\frac{\mu}{2}\peqsub{D}{Y}(\sqrt{\peqsubfino{\gamma}{X}{-0.2ex}}\,\peqsub{\sharp}{\gamma}\peqsubfino{{\star_\gamma}}{\!X}{-0.2ex}q)\Big)-\peqsub{D}{Z}(\varepsilon\peqsubfino{\mathcal{H}}{\!\perp}{-0.1ex}n+e.\mathcal{H})_\alpha\vecc{Y}{X}^\alpha+\\
 & +\peqsub{D}{Y}(\varepsilon\peqsubfino{\mathcal{H}}{\!\perp}{-0.1ex}n+e.\mathcal{H})_\alpha\vecc{Z}{X}^\alpha\Big]\peqsub{\mathrm{vol}}{\Sigma}-\!\int_{\partial\Sigma}\left[\peqsub{D}{Z}(\varepsilon\peqsubfino{\mathcal{H}}{\!\perp}{-0.1ex}^{\scriptscriptstyle\partial}n+e.\mathcal{H}^{\scriptscriptstyle\partial})_\alpha\vecc{Y}{X}^\alpha\!-\peqsub{D}{Y}(\varepsilon\peqsubfino{\mathcal{H}}{\!\perp}{-0.1ex}^{\scriptscriptstyle\partial}n+e.\mathcal{H}^{\scriptscriptstyle\partial})_\alpha\vecc{Z}{X}^\alpha\right]\!\peqsub{\mathrm{vol}}{\partial\Sigma}     
 \end{split}
 \end{align}
 The explicit expression of $\omega^{\scriptscriptstyle\mathrm{CS}}$ can be obtained using similar techniques to the ones used to compute $\omega^{\scriptscriptstyle\mathrm{MCS}}$. An alternative way is to consider the expressions
 \begin{align*}
 &\bullet\ \delta\Lambda_-\peqsub{\mathrm{vol}}{\Sigma}\overset{\eqref{MCS - equation - p=-sqrt star q}}{=}-\frac{\mu}{2}\sqrt{\gamma}(\delta\peqsubfino{{\star_\gamma}}{\!X}{-0.2ex}q)\peqsub{\mathrm{vol}}{\Sigma}\overset{\eqref{appendix property delta star=star d}}{=}-\frac{\mu}{2}(\peqsubfino{{\star_\gamma}}{\!X}{-0.2ex}\mathrm{d}q)\peqsub{\mathrm{vol}}{\gamma}\overset{\eqref{Appendix - equation - definition Hodge (a,b)vol=a wedge*b}}{=}\\
 &\qquad=-\frac{\mu}{2}\wedge\peqsubfino{{\star_\gamma}}{\!X}{-0.2ex}\peqsubfino{{\star_\gamma}}{\!X}{-0.2ex}(\mathrm{d}q)\overset{\eqref{Appendix - equaion - star^2=Id}}{=}-\frac{\mu}{2}\,\mathrm{d}q\\
 &\bullet\ \vecc{Y}{p}-\mathcal{L}_{\vec{y}^{\scriptscriptstyle\top}}p- Y^{{\raisemath{0.2ex}{\scriptscriptstyle\perp\!}}}\frac{\mu}{2\eta}\peqsubfino{{\star_\gamma}}{\!X}{-0.2ex}\Lambda_+-\varepsilon\sqrt{\gamma}\,\sharp_\gamma\delta\!\left(\eta Y^{{\raisemath{0.2ex}{\scriptscriptstyle\perp\!}}}\mathrm{d}q+\frac{\mu}{2} Y^{{\raisemath{0.2ex}{\scriptscriptstyle\perp\!}}}\peqsubfino{{\star_\gamma}}{\!X}{-0.2ex}\peqsubfino{q}{\perp}{-0.2ex}\right)\updown{\eqref{MCS - equation - p=-sqrt star q}}{\eqref{MCS - equation - Yp=-sqrt star Y_q}}{=}\\
 &\qquad =-\frac{\mu}{2}\sqrt{\peqsubfino{\gamma}{X}{-0.2ex}}\,\peqsub{\sharp}{\gamma}\peqsubfino{{\star_\gamma}}{\!X}{-0.2ex}\vecc{Y}{q} +\frac{\mu}{2}\mathcal{L}_{\vec{y}^{\scriptscriptstyle\top}}\Big(\sqrt{\peqsubfino{\gamma}{X}{-0.2ex}}\,\peqsub{\sharp}{\gamma}\peqsubfino{{\star_\gamma}}{\!X}{-0.2ex}q\Big)-0-\varepsilon\sqrt{\gamma}\,\sharp_\gamma\delta\!\left(0+\frac{\mu}{2}\peqsubfino{{\star_\gamma}}{\!X}{-0.2ex}( Y^{{\raisemath{0.2ex}{\scriptscriptstyle\perp\!}}}\peqsubfino{q}{\perp}{-0.2ex})\right)\updown{\eqref{Applications - equation - DLambda}}{\eqref{appendix property delta star=star d}}{=}\\
 &\qquad =-\frac{\mu}{2}\sqrt{\peqsubfino{\gamma}{X}{-0.2ex}}\,\peqsub{\sharp}{\gamma}\left[\peqsubfino{{\star_\gamma}}{\!X}{-0.2ex}\vecc{Y}{q} -\peqsubfino{{\star_\gamma}}{\!X}{-0.2ex}\mathcal{L}_{\vec{y}^{\scriptscriptstyle\top}}q-\varepsilon\peqsubfino{{\star_\gamma}}{\!X}{-0.2ex}\mathrm{d}(Y^{{\raisemath{0.2ex}{\scriptscriptstyle\perp\!}}}\peqsubfino{q}{\perp}{-0.2ex})\right]=\\
 &\qquad =-\frac{\mu}{2}\sqrt{\peqsubfino{\gamma}{X}{-0.2ex}}\,\peqsub{\sharp}{\gamma}\peqsubfino{{\star_\gamma}}{\!X}{-0.2ex}\left[\vecc{Y}{q} -\mathcal{L}_{\vec{y}^{\scriptscriptstyle\top}}q-\varepsilon\mathrm{d}(Y^{{\raisemath{0.2ex}{\scriptscriptstyle\perp\!}}}\peqsubfino{q}{\perp}{-0.2ex})\right]
 \end{align*}
 and plug them into $\omega^{\scriptscriptstyle\mathrm{MCS}}$. Both methods lead to
\begin{align*}
\begin{split}
	&\omega^{\scriptscriptstyle\mathrm{CS}}_{(\peqsubfino{q}{\perp}{-0.2ex},q,X)}\Big((\vecc[\scalebox{0.4}{$\perp$}]{Y}{q},\vecc{Y}{q},\vecc{Y}{X}),(\vecc[\scalebox{0.4}{$\perp$}]{Z}{q},\vecc{Z}{q},\vecc{Z}{X})\Big)=\\
	&=-\frac{\varepsilon\mu}{2}\int_\Sigma Z^{{\raisemath{0.2ex}{\scriptscriptstyle\perp\!}}} \Big[\vecc[\perp]{Y}{q}\hspace*{0.1ex}-\mathcal{L}_{\vec{y}^{\scriptscriptstyle\top}}\peqsubfino{q}{\perp}{-0.2ex}+\peqsub{\imath}{\mathrm{d}Y^{\!\scriptscriptstyle\perp}}q\Big]\mathrm{d}q-{}\\
	&\phantom{=}+\frac{\mu}{2}\int_\Sigma\!\Big[\vecc{Y}{q}-\mathcal{L}_{\vec{y}^{\scriptscriptstyle\top}}q-\varepsilon\mathrm{d}( Y^{{\raisemath{0.2ex}{\scriptscriptstyle\perp\!}}}\peqsubfino{q}{\perp}{-0.2ex})\Big]\wedge\Big[\vecc{Z}{q} -\mathcal{L}_{\vec{z}^{\scriptscriptstyle\top}}q-\varepsilon\mathrm{d}(Z^{{\raisemath{0.2ex}{\scriptscriptstyle\perp\!}}}\peqsubfino{q}{\perp}{-0.2ex})\Big]-{}\\
	&\phantom{=}-\frac{\varepsilon\mu}{2}\int_{\partial\Sigma} Z^{{\raisemath{0.2ex}{\scriptscriptstyle\perp\!}}}\peqsubfino{{\left\langle\peqsub{\jmath}{\partial}^*\!\!\left(\vecc{Y}{q}-\mathcal{L}_{\vec{y}^{\scriptscriptstyle\top}}q-\varepsilon\mathrm{d}( Y^{{\raisemath{0.2ex}{\scriptscriptstyle\perp\!}}}\peqsubfino{q}{\perp}{-0.2ex})\right)\!\!\coma\peqsub{\jmath}{\partial}^*\Big[\imath_{\vec{\nu}}\!\left(\peqsubfino{{\star_\gamma}}{\!X}{-0.2ex}\peqsubfino{q}{\perp}{-0.2ex}\right)-\peqsub{\nu}{\perp} \peqsubfino{{\star_\gamma}}{\!X}{-0.2ex}q\Big]\right\rangle_{\!\!\gamma}}}{\!\partial}{-.3ex}\frac{\peqsubfino{{\mathrm{vol}_\gamma}}{\!\partial\!X}{-0.4ex}}{|\vec{\nu}^{\,\scriptscriptstyle\top}|}-{}\\
	&\phantom{=}-\frac{\varepsilon\mu}{2}\int_{\partial\Sigma} Z^{{\raisemath{0.2ex}{\scriptscriptstyle\perp\!}}} \peqsubfino{{\left\langle\peqsub{\jmath}{\partial}^*\big(\vecc[\perp]{Y}{q}\hspace*{0.1ex}-\mathcal{L}_{\vec{y}^{\scriptscriptstyle\top}}\peqsubfino{q}{\perp}{-0.2ex}+\peqsub{\imath}{\mathrm{d}Y^{\!\scriptscriptstyle\perp}}q\big)\coma\peqsub{\jmath}{\partial}^*\imath_{\vec{\nu}}\!\left(\peqsubfino{{\star_\gamma}}{\!X}{-0.2ex}q\right)\right\rangle_{\!\!\gamma}}}{\!\partial}{-.3ex}\frac{\peqsubfino{{\mathrm{vol}_\gamma}}{\!\partial\!X}{-0.4ex}}{|\vec{\nu}^{\,\scriptscriptstyle\top}|}-(Y\leftrightarrow Z)
\end{split}
\end{align*}
 where we have used that for $k$-forms over a manifold of dimension $2k+1$ we have \[\Big(\alpha\coma\sqrt{\peqsubfino{\gamma}{X}{-0.2ex}}\,\peqsub{\sharp}{\gamma}\peqsubfino{{\star_\gamma}}{\!X}{-0.2ex}\beta\Big)\peqsub{\mathrm{vol}}{\Sigma}=\langle\alpha,\peqsubfino{{\star_\gamma}}{\!X}{-0.2ex}\beta\rangle\peqsubfino{{\mathrm{vol}_\gamma}}{X}{-0.2ex}=\alpha\wedge\peqsubfino{{\star_\gamma}}{\!X}{-0.2ex}\peqsubfino{{\star_\gamma}}{\!X}{-0.2ex}\beta=-\alpha\wedge\beta\]
 It is important to realize that the integral over the bulk does not depend, as expected, on the metric. The boundary terms might seem that depend on the metric, but notice that the ones involving $\imath_{\vec{\nu}}$ can be written in terms of the expressions given in \eqref{MCS - equation - independece embedding}, which do not depend on the embedding. The term involving $\peqsub{\nu}{\perp}$ is a bit trickier as we have to realize that it can be written, using the fact that $0=\nu_\alpha\mathbb{Z}^\alpha=\varepsilon Z^{{\raisemath{0.2ex}{\scriptscriptstyle\perp\!}}}\peqsub{\nu}{\perp}+Z^a\nu_a$, as
 \[\varepsilon z^a\peqsub{\jmath}{\partial}^*\Big[\sqrt{\gamma}\peqsubfino{{\star_\gamma}}{\!X}{-0.2ex}q\Big]\frac{\sqrt{\peqsubfino{\gamma}{\partial}{-0.2ex}}}{\sqrt{\gamma}}\frac{\nu_a}{|\vec{\nu}^{\,\scriptscriptstyle\top}|}\peqsubfino{\mathrm{vol}}{\partial\Sigma}{-0.4ex} \]
 which is explicitly independent using \eqref{MCS - equation - independece embedding}. 

 \subsubsection*{GNH algorithm}\trassub
 
 The Hamiltonian is, once again, zero. Thus the Hamiltonian vector field $Y\in\mathfrak{X}(\mathcal{P})$ is given by the Hamilton equation
 \begin{equation}
 \omega^{\scriptscriptstyle\mathrm{CS}}_{(\peqsubfino{q}{\perp}{-0.2ex},q,X)}\Big((\vecc[\scalebox{0.4}{$\perp$}]{Y}{q},\vecc{Y}{q},\vecc{Y}{X})\coma(\vecc[\scalebox{0.4}{$\perp$}]{Z}{q},\vecc{Z}{q},\vecc{Z}{X})\Big)=\mathrm{d}_{(\peqsubfino{q}{\perp}{-0.2ex},q,X)}H(\vecc[\scalebox{0.4}{$\perp$}]{Z}{q},\vecc{Z}{q},\vecc{Z}{X})=0\end{equation}
 for every $Z\in\mathfrak{X}(\mathcal{P})$. We have clearly the conditions at the bulk
 \begin{align}
 \begin{split}
 &\vecc{Y}{q}=\mathcal{L}_{\vec{y}^{\scriptscriptstyle\top}}q+\varepsilon\mathrm{d}( Y^{{\raisemath{0.2ex}{\scriptscriptstyle\perp\!}}}\peqsubfino{q}{\perp}{-0.2ex})\\[1ex]
 &Y^{{\raisemath{0.2ex}{\scriptscriptstyle\perp\!}}}\mathrm{d}q=0\\[1ex]
 &\Big[\vecc[\perp]{Y}{q}\hspace*{0.1ex}-\mathcal{L}_{\vec{y}^{\scriptscriptstyle\top}}\peqsubfino{q}{\perp}{-0.2ex}+\peqsub{\imath}{\mathrm{d}Y^{\!\scriptscriptstyle\perp}}q\Big]\mathrm{d}q=0
 \end{split}
 \end{align}
 Once again we have a bifurcation but now it depends on the support of $\mathrm{d}q$: wherever it vanishes, $Y^{{\raisemath{0.2ex}{\scriptscriptstyle\perp\!}}}$ is arbitrary, wherever it does not vanish, $Y^{{\raisemath{0.2ex}{\scriptscriptstyle\perp\!}}}=0$.\separ
 
 $\blacktriangleright$ Wherever $\mathrm{d}q\neq0$
 \begin{align}
 \begin{split}
 &\vecc[\perp]{Y}{q}\hspace*{0.1ex}=\mathcal{L}_{\vec{y}^{\scriptscriptstyle\top}}\peqsubfino{q}{\perp}{-0.2ex}\\[1ex]
 &\vecc{Y}{q}=\mathcal{L}_{\vec{y}^{\scriptscriptstyle\top}}q\\[1ex]
 &Y^{{\raisemath{0.2ex}{\scriptscriptstyle\perp\!}}}=0\\[.5ex]
 &y^{{\raisemath{-0.1ex}{\scriptscriptstyle\top\!}}}\text{ arbitrary}
 \end{split}
 \end{align}
The GNH algorithm stops because we have found the most general solution with no constraint.\separ
 
 $\blacktriangleright$ Wherever $\mathrm{d}q=0$
 \begin{align}
 \begin{split}
 &\vecc[\perp]{Y}{q}\hspace*{0.1ex}\text{ arbitrary}\\[.9ex]
 &\vecc{Y}{q}=\mathcal{L}_{\vec{y}^{\scriptscriptstyle\top}}q+\varepsilon\mathrm{d}( Y^{{\raisemath{0.2ex}{\scriptscriptstyle\perp\!}}}\peqsubfino{q}{\perp}{-0.2ex})\\[1ex]
 &Y^{{\raisemath{0.2ex}{\scriptscriptstyle\perp\!}}}\text{ arbitrary}\\[.8ex]
 &y^{{\raisemath{-0.1ex}{\scriptscriptstyle\top\!}}}\text{ arbitrary}
 \end{split}
 \end{align}
 Notice that the Lie derivative $\mathcal{L}_{\vec{y}^{\scriptscriptstyle\top}}q=\mathrm{d}\imath_{\vec{y}^{\scriptscriptstyle\top}}q+\imath_{\vec{y}^{\scriptscriptstyle\top}}\mathrm{d}q$ so actually
 \[\vecc{Y}{q}=\mathrm{d}\imath_{\vec{y}^{\scriptscriptstyle\top}}q+\varepsilon\mathrm{d}( Y^{{\raisemath{0.2ex}{\scriptscriptstyle\perp\!}}}\peqsubfino{q}{\perp}{-0.2ex})=\mathrm{d}(\mathbb{Y}^\alpha q_\alpha)\]
 where we define $q_\alpha=\varepsilon n_\alpha\peqsub{q}{\perp}+e_\alpha^aq_a$. In this case, the GNH algorithm goes on by requiring that the Hamiltonian vector field $Y\in\mathfrak{X}(\mathcal{P})$ that we have found in the first step is tangent to
 \[\mathcal{P}_2=\Big\{(\peqsubfino{q}{\perp}{-0.2ex},q,X)\in\mathcal{P}\ \ /\ \ \mathrm{d}q=0\Big\}\]
 which is trivial as 
 \begin{align*}
 \peqsub{D}{Y}(\mathrm{d}q)&=\mathrm{d}(\peqsub{D}{Y}q)=\mathrm{d}\vecc{Y}{q}=\mathrm{d}^2(\mathbb{Y}^\alpha q_\alpha)=0
 \end{align*}
 
 $\blacktriangleright$ Boundary\separprevia
 
 To obtain the solution for the boundary, first notice that once again only the case $\mathrm{d}q=0$ is non-trivial. So we have to solve
 \begin{align*}
 \begin{split}
 0&=-\int_{\partial\Sigma} Y^{{\raisemath{0.2ex}{\scriptscriptstyle\perp\!}}}\peqsubfino{{\left\langle\peqsub{\jmath}{\partial}^*\!\!\left(\vecc{Z}{q}-\mathcal{L}_{\vec{z}^{\scriptscriptstyle\top}}q-\varepsilon\mathrm{d}( Z^{{\raisemath{0.2ex}{\scriptscriptstyle\perp\!}}}\peqsubfino{q}{\perp}{-0.2ex})\right)\!\!\coma\peqsub{\jmath}{\partial}^*\Big[\imath_{\vec{\nu}}\!\left(\peqsubfino{{\star_\gamma}}{\!X}{-0.2ex}\peqsubfino{q}{\perp}{-0.2ex}\right)-\peqsub{\nu}{\perp} \peqsubfino{{\star_\gamma}}{\!X}{-0.2ex}q\Big]\right\rangle_{\!\!\gamma}}}{\!\partial}{-.3ex}\frac{\peqsubfino{{\mathrm{vol}_\gamma}}{\!\partial\!X}{-0.4ex}}{|\vec{\nu}^{\,\scriptscriptstyle\top}|}+{}\\
 &\phantom{=}+\int_{\partial\Sigma} \peqsubfino{{ \left\langle Z^{{\raisemath{0.2ex}{\scriptscriptstyle\perp\!}}}\peqsub{\jmath}{\partial}^*\big(\vecc[\perp]{Y}{q}\hspace*{0.1ex}-\mathcal{L}_{\vec{y}^{\scriptscriptstyle\top}}\peqsubfino{q}{\perp}{-0.2ex}+\peqsub{\imath}{\mathrm{d}Y^{\!\scriptscriptstyle\perp}}q\big)-Y^{{\raisemath{0.2ex}{\scriptscriptstyle\perp\!}}}\peqsub{\jmath}{\partial}^*\big(\vecc[\perp]{Z}{q}\hspace*{0.1ex}-\mathcal{L}_{\vec{z}^{\scriptscriptstyle\top}}\peqsubfino{q}{\perp}{-0.2ex}+\peqsub{\imath}{\mathrm{d}Z^{\!\scriptscriptstyle\perp}}q\big)\coma\peqsub{\jmath}{\partial}^*\imath_{\vec{\nu}}\!\left(\peqsubfino{{\star_\gamma}}{\!X}{-0.2ex}q\right)\right\rangle_{\!\!\gamma}}}{\!\partial}{-.3ex}\frac{\peqsubfino{{\mathrm{vol}_\gamma}}{\!\partial\!X}{-0.4ex}}{|\vec{\nu}^{\,\scriptscriptstyle\top}|}
 \end{split}
 \end{align*}
 and we can obtain sufficient conditions like
 \[\left\{\begin{array}{l}
 \!\peqsub{\jmath}{\partial}^*\Big[\imath_{\vec{\nu}}\!\left(\peqsubfino{{\star_\gamma}}{\!X}{-0.2ex}\peqsubfino{q}{\perp}{-0.2ex}\right)-\peqsub{\nu}{\perp} \peqsubfino{{\star_\gamma}}{\!X}{-0.2ex}q\Big]=0\\[2ex]
 \!\peqsub{\jmath}{\partial}^*\imath_{\vec{\nu}}\!\Big[\peqsubfino{{\star_\gamma}}{\!X}{-0.2ex}q\Big]
 \end{array}\right.\qquad\qquad
 \left\{\begin{array}{l}
 \!\peqsub{\jmath}{\partial}^*\Big[\imath_{\vec{\nu}}\!\left(\peqsubfino{{\star_\gamma}}{\!X}{-0.2ex}\peqsubfino{q}{\perp}{-0.2ex}\right)\Big]=0\\[2ex]
 \!\peqsub{\jmath}{\partial}^*\Big[\peqsubfino{{\star_\gamma}}{\!X}{-0.2ex}q\Big]=0
 \end{array}\right.\]
 
 

%% file: 8_GR.tex

\chapter{General relativity via GNH algorithm}\thispagestyle{empty}\label{Chapter - General relativity}
\starwars[Obi-Wan Kenobi]{If you strike me down, I shall become more powerful than you could possibly imagine.}{A New Hope}

  \section{Introduction}
  
  General relativity is of one of the most successful and accurate theories in present day science. It was developed by Einstein in 1915 and describes the gravitational effects as a consequence of the curvature of space-time. We saw already in section \ref{Mathematical background - subection - pysics over space-time} of chapter \ref{Chapter - Mathematical background} a brief introduction, so I will devote this initial section to motivate and explain this chapter.\separ
  
  Short after Dirac developed his procedure to deal with constrained Hamiltonian systems (see \ref{Mathematical background - subsection - Dirac algorithm} of chapter \ref{Chapter - Mathematical background} and \cite{dirac1964lectures}) he tried to apply it to the general theory of relativity. However, he found some difficulties that prevented him to give the complete Hamiltonian description. Some years later, in 1959, Arnowit, Deser, and Misner completed the task and presented in \cite{arnowitt2008republication} what now is called the ADM formalism of GR. Although it might seem that we are a bit late to say anything new about the Hamiltonian formulation of GR, we strongly believe that the use of the GNH algorithm sheds some light to the subject. For instance, the inclusion of boundaries is always problematic in the usual setting while in the GNH context, as we have learned in the previous chapters, poses no conceptual problem (it is more complicated at the functional analysis level). Furthermore, this study will allow us to get into some generalizations like unimodular gravity, which is very relevant to study the problem of time \cite{isham1993canonical,kuchavr1991does,smolin2009quantization} and the cosmological constant problem \cite{henneaux1989cosmological}\index{Cosmological constant}.\separ
  
  This chapter is then devoted to develop for the first time the GNH algorithm in GR and unimodular gravity to obtain the well known results but in a more straightforward way. We will apply the same methods we have used along the thesis with no need to adapt them to these cases. Finally, notice that parametrizing GR does nothing because there is no background geometric object.

  \section{The Einstein-Hilbert action}\label{GR - section - HE action}
  Let $M\cong I\times\Sigma$ be a globally hyperbolic space-time (see section \ref{Mathematical background - subection - pysics over space-time} of chapter \ref{Chapter - Mathematical background}) of dimension $n+1$ where $\Sigma$ is compact. Consider the action for General Relativity $S:\peqsub{\mathrm{Met}}{\partial}(M)\to\R$ given by
  \begin{equation}\label{GR - equation - accion}
  S(g)=\int_M \Big[R(g)-2\Lambda\Big]\peqsub{\mathrm{vol}}{g}+2\int_{\partial_\Sigma M} K(g)\peqsub{\mathrm{vol}}{\partial g}
  \end{equation}
  where $\peqsubfino{g}{\partial}{-0.3ex}=\peqsub{\imath}{\partial}^*g$ with $\peqsub{\imath}{\partial}:\peqsub{\partial}{\Sigma}M\hookrightarrow M$ and $\peqsub{\mathrm{Met}}{\partial}(M)$ is the space of metrics of $M$ such that $\peqsubfino{g}{\partial}{-0.3ex}$ is some fixed metric $\peqsubfino{g}{\partial,0}{-0.3ex}\in\mathrm{Met}(\peqsub{\partial}{\Sigma}M)$.
  
  \subsection*{Variations of the action}\trassub
  
  The variation $DS:T\peqsub{\mathrm{Met}}{\partial}(M)\to T\R$ is given, for some $h\in T_g\peqsub{\mathrm{Met}}{\partial}(M)=\mathfrak{T}^{2,0}_{\mathrm{sym}}(M)$, by
  \begin{equation}\label{GR - equation - Variacion accion}
    \peqsub{D}{(g,h)}S=\int_M\Big\{ \peqsub{D}{(g,h)}R(g)\peqsub{\mathrm{vol}}{g}+\big[R(g)-2\Lambda\big]\peqsub{D}{(g,h)}\peqsub{\mathrm{vol}}{g}\Big\}+2\int_{\partial_\Sigma M} \peqsub{D}{(g,h)}K(g)\peqsub{\mathrm{vol}}{\partial g}
   \end{equation}
  Notice that the volume of the boundary is fixed because $\peqsub{\mathrm{Met}}{\partial}(M)$ is fixed.\separ
  
  First notice that the variation of the volume is given by equation \eqref{Appendix - equation - variacion volumen} with $\peqsub{D}{(g,h)}g=h$. This means that $\peqsub{D}{(g,h)}\peqsub{\mathrm{vol}}{g}=\frac{1}{2}\peqsub{\mathrm{Tr}}{g}h\,\peqsub{\mathrm{vol}}{g}$. The variation of the scalar curvature in the direction $h\in T_g\mathrm{Met}(M)$, computed in lemma \ref{Appendix - lemma - DR}, is given by
  \[\peqsub{D}{(g,h)}R(g)=-h^{ab}\mathrm{Ric}(g)_{ab}+\nabla^d\Big(\nabla^bh_{bd}-\nabla_d\peqsub{\mathrm{Tr}}{g}h\Big)\]
  Finally, it is easy to construct a local foliation where the boundary is the first leaf. For instance, take a vector field defined on a neighborhood of the boundary that is transversal to the boundary. Then the flow of such vector field defines, for sufficiently small times, the required foliation. With that we obtain that $2K=\mathcal{L}_{\vec{\nu}}g$ over the boundary (see equation \eqref{Mathematical background - equation - 2K=L_ng}). Besides, we have that the metric $g$ is constant over the boundary so
 \begin{align*}
\peqsub{D}{(g,h)}K(g)&=\peqsub{D}{(g,h)}\Big[\peqsubfino{g}{\partial}{-0.3ex}^{ab}K_{ab}(g)\Big]=\frac{1}{2}\peqsubfino{g}{\partial}{-0.3ex}^{ab}\peqsub{D}{(g,h)}(\mathcal{L}_{\nu}g)_{ab}=\\
&=\frac{1}{2}\peqsubfino{g}{\partial}{-0.3ex}^{ab}\peqsub{D}{(g,h)}\Big(\nu^c\nabla_cg_{ab}+g_{cb}\nabla_a\nu^c+g_{ac}\nabla_b\nu^c\Big)=\\
&=\frac{1}{2}\peqsubfino{g}{\partial}{-0.3ex}^{ab}\nu^c\nabla_ch_{ab}+0+0=\frac{1}{2}\peqsubfino{g}{\partial}{-0.3ex}^{ab}\nu^d\nabla_dh_{ab}
\end{align*}
  Plugging the three variations we already have into equation \eqref{GR - equation - Variacion accion} leads to
 \begin{align*}
  \peqsub{D}{(g,h)}&S=\int_M\!\Big\{ \big(\peqsub{D}{(g,h)}R\big)\peqsub{\mathrm{vol}}{g}+\Big[R-2\Lambda\Big]\peqsub{D}{(g,h)}\peqsub{\mathrm{vol}}{g}\Big\}+2\int_{\partial_\Sigma M}\big(\peqsub{D}{(g,h)}K\big)\peqsub{\mathrm{vol}}{\partial g}=\\
  &=\int_M\!\left\{ -h^{ab}\mathrm{Ric}_{ab}+\left[\frac{R}{2}-\Lambda\right]g_{ab}h^{ab}\right\}\peqsub{\mathrm{vol}}{g}+\int_{\partial_\Sigma M}\!\nu^d\Big[\peqsubfino{g}{\partial}{-0.3ex}^{ab}\nabla_dh_{ab}+g^{ab}\nabla_{[a}h_{d]b}\Big]\peqsub{\mathrm{vol}}{g_\partial}=\\
  &=-\int_M \Big[G_{ab}+\Lambda g_{ab}\Big]h^{ab}\peqsub{\mathrm{vol}}{g}
  \end{align*}
  where in the last line we have used the definition of the Einstein tensor $G=\mathrm{Ric}-\frac{R}{2}g$, the decomposition of $g$ over the space-like boundary $\peqsub{\partial}{\Sigma}M$ given by $g^{ab}=\nu^a\nu^b+\peqsubfino{g}{\partial}{-0.3ex}^{ab}$, and the fact that $h$ is zero over the boundary (in particular its tangential derivatives vanish).\separ
  
  If $\peqsub{D}{(g,h)}S=0$ for every $h\in T_g\mathrm{Met}(M)$ we obtain the Einstein equation with cosmological constant
  \begin{equation}\label{GR - equation - Einstein eq}
  \mathrm{Ric}(g)-\frac{1}{2}R(g)\, g+\Lambda\, g=0
  \end{equation}
  
  Notice that the boundary term in the action has been included in order to cancel the boundary integral that arises after the integration by parts. By doing this, we assure that the boundary conditions are prescribed by the functional space we are working on (in this case $\peqsub{\mathrm{Met}}{\partial}(M)$, that gives Dirichlet-like boundary conditions).
    
  \section{Lagrangian formulation}\label{GR - section - Lagrangian formulation}
  
  We saw on section \ref{Mathematical background - subection - pysics over space-time} of chapter \ref{Chapter - Mathematical background} that, given a foliation $Z:\R\times\Sigma\to\R\times\Sigma$, a metric $g$ can be decomposed as
  \[g=\Big(\varepsilon\mathbf{N}^2+N(\vec{N})\Big)\mathrm{d}t\otimes\mathrm{d}t+N\otimes\mathrm{d}t+\mathrm{d}t\otimes N+\widetilde{\gamma}\]
  and that the triple $(\mathbf{N},\vec{N},\widetilde{\gamma})\in \Cinf{M}\times\mathfrak{X}(M)\times\mathfrak{T}^{2,0}_{sym}(M)$ contains the same information as $g$. We would like to use the identification \eqref{Mathematical background - equation - C(R,C(M,N)) cong C(R x M,N)} and a curve of embeddings to see these objects as curves over $\Cinf{\Sigma}\times\mathfrak{X}(\Sigma)\times\mathrm{Met}(M)$. To do that, first notice that a curve of embeddings $Z:\R\to\mathrm{Emb}(\Sigma,M)$ can be considered as $\{Z_t:\{t\}\times\Sigma\to M\}$ and that for every $t\in\R$ we have (see lemma \ref{appendix - lemma - escalar de curvatura})
  \begin{align}\label{GR - equation - R(n+1)=R(n)+...}
  R^{(n+1)}(g)=R^{(n)}(\gamma_t)+\varepsilon \mathrm{Tr}_{\gamma_t}(K_t)^2-2\varepsilon \langle K_t,K_t\rangle_{\gamma_t}-2\varepsilon\mathrm{div}(\vec{n_t}\,\mathrm{div}\,\vec{n_t}-\nabla_{\vec{n_t}}\vec{n_t})
  \end{align}
  where $\gamma_t=Z_t^*g(=Z_t^*\widetilde{\gamma})$, $K_t$ is the extrinsic curvature of $\Sigma_t:=Z_t(\Sigma)$, $R^{(n)}(\gamma_t)$ the intrinsic curvature, and $R^{(n+1)}(g)$ the intrinsic curvature of $M$ evaluated at $Z_t(\sigma)$.\separ
  
  For simplicity we restrict ourselves to space-times without boundary, though the next procedure can be applied, if used with due care, to space-times with boundary. Plugging \eqref{GR - equation - R(n+1)=R(n)+...} into \eqref{GR - equation - accion} and taking into account \eqref{Mathematical background - equation - n=varepsilon N dt} and lemma \ref{Appendix - lemma - vol=nu wedge vol} leads to
  \begin{align*}
  S(g)=\int_{\R}\mathrm{d}t\int_\Sigma Z_t^*\left\{\mathbf{N}\sqrt{\gamma}\Big[R^{(n)}_\gamma+\varepsilon \mathrm{Tr}(K)^2-2\varepsilon \langle K,K\rangle_\gamma-2\Lambda\Big]\right\}\peqsub{\mathrm{vol}}{\Sigma}
  \end{align*}
  where the pullback $Z_t^*$ brings the objects from $\Sigma_t$ to $\Sigma$. We see that we cannot read the Lagrangian directly from this expression as we need to express $K$ in terms of the variables $(\mathbf{N},\vec{N},\widetilde{\gamma})$ and their ``velocities'' (Lie derivatives in the $\partial_t$ direction). To do so let us compute the velocity of $\widetilde{\gamma}$ with respect to $\partial_t^\alpha=\mathbf{N}n^\alpha+N^\alpha$. First notice that
  \begin{align*}
  (\peqsub{\mathcal{L}}{\partial_t}g)_{\alpha\beta}&=(\peqsub{\mathcal{L}}{\mathbf{N}n}g)_{\alpha\beta}+(\mathcal{L}_{\vec{N}}g)_{\alpha\beta}=\\
  &=n_\alpha(\mathrm{d}\mathbf{N})_\beta+n_\beta(\mathrm{d}\mathbf{N})_\alpha+\mathbf{N}(\mathcal{L}_{\vec{n}}g)_{\alpha\beta}+g_{\sigma\beta}\nabla_\alpha N^\sigma+g_{\alpha\sigma}\nabla_\beta N^\sigma=\\
  &=n_\alpha(\mathrm{d}\mathbf{N})_\beta+n_\beta(\mathrm{d}\mathbf{N})_\alpha+2\mathbf{N}(e_{\jmath_t})^{\bar{\mu}}_\alpha(e_{\jmath_t})^{\bar{\nu}}_\beta (K_t)_{{\bar{\mu}}{\bar{\nu}}}+g_{\sigma\beta}\nabla_\alpha (\tau_{\jmath_t})^\sigma_{\bar{\sigma}}N_t^{\bar{\sigma}}+g_{\alpha\sigma}\nabla_\beta (\tau_{\jmath_t})_{\bar{\sigma}}^\sigma N_t^{\bar{\sigma}}
  \end{align*}
  where $\jmath_t:\Sigma_t\subset M\hookrightarrow M$ is just the inclusion and the bar indexes are the ones of $\Sigma_t=Z_t(\Sigma)$. Notice that in the last equality we have used that $\vec{N}\in\mathfrak{X}(M)$ is tangent to the foliation so there exists, for every $t\in\R$, a vector field $\vec{N}_t\in\mathfrak{X}(\Sigma_t)$ such that $(\jmath_t)_*\vec{N}_t=\vec{N}$. Besides we define the analog for the shift $\mathbf{N}_t:=\jmath_t^*\mathbf{N}\in\Cinf{\Sigma_t}$ and the metric $\widetilde{\gamma}_t:=\jmath^*_t\widetilde{\gamma}\in\mathrm{Met}(\Sigma_t)$. Finally we have
  \begin{align*}
  \mathcal{V}_{\bar{\alpha}\bar{\beta}}&:=\big(\peqsub{\mathcal{L}}{\partial_t}\widetilde{\gamma}\big)_{\bar{\alpha}\bar{\beta}}=\peqsub{\mathcal{L}}{\partial_t}\Big((\tau_{\jmath_t})^\alpha_{\bar{\alpha}}(\tau_{\jmath_t})^\beta_{\bar{\beta}}g_{\alpha\beta}\Big)\overset{\star}{=}(\tau_{\jmath_t})^\alpha_{\bar{\alpha}}(\tau_{\jmath_t})^\beta_{\bar{\beta}}(\peqsub{\mathcal{L}}{\partial_t}g)_{\alpha\beta}+0+0\overset{\dagger}{=}\\
  &\phantom{:}=0+0+2\delta^{\bar{\mu}}_{\bar{\alpha}}\delta^{\bar{\nu}}_{\bar{\beta}}\mathbf{N}_t(K_t)_{\bar{\mu}\bar{\nu}}+(\tau_{\jmath_t})^\beta_{\bar{\beta}}g_{\sigma\beta}\nabla_{\bar{\alpha}}(\tau_{\jmath_t})^\sigma_{\bar{\sigma}}N_t^{\bar{\sigma}}+(\tau_{\jmath_t})^\alpha_{\bar{\alpha}}g_{\alpha\sigma}\nabla_{\bar{\beta}}(\tau_{\jmath_t})_{\bar{\sigma}}^\sigma N_t^{\bar{\sigma}}\overset{\ref{Mathematical background - theorem - Gauss lemma}}{=}\\
  &=2\mathbf{N}_t(K_t)_{\bar{\alpha}\bar{\beta}}+\textcolor{red}{(\tau_{\jmath_t})^\beta_{\bar{\beta}}g_{\sigma\beta}}\Big[\textcolor{red}{(\tau_{\jmath_t})^\sigma_{\bar{\sigma}}}\overline{\nabla}_{\bar{\alpha}}N_t^{\bar{\sigma}}-K_{\bar{\alpha}\bar{\sigma}}n^\sigma\Big]+\textcolor{teal}{(\tau_{\jmath_t})^\alpha_{\bar{\alpha}}g_{\alpha\sigma}}\Big[\textcolor{teal}{(\tau_{\jmath_t})_{\bar{\sigma}}^\sigma}\overline{\nabla}_{\bar{\beta}}N_t^{\bar{\sigma}}-K_{\bar{\beta}\bar{\sigma}}n^\sigma\Big]\overset{\tau.n=0}{\hspace*{-3ex}=}\\
  &=2\mathbf{N}_t(K_t)_{\bar{\alpha}\bar{\beta}}+\textcolor{red}{(\widetilde{\gamma}_t)_{\bar{\beta}\bar{\sigma}}}\overline{\nabla}_{\bar{\alpha}}N_t^{\bar{\sigma}}-0+\textcolor{teal}{(\widetilde{\gamma}_t)_{\bar{\alpha}\bar{\sigma}}}\overline{\nabla}_{\bar{\beta}}N_t^{\bar{\sigma}}-0\overset{\eqref{Appendix - equation - derivada de lie abstract index}}{=}\\
  &=2\mathbf{N}_t(K_t)_{\bar{\alpha}\bar{\beta}}+(\mathcal{L}_{\vec{N}_t}\gamma_t)_{\bar{\alpha}\bar{\beta}}
  \end{align*}
  in the $\star$ equality we have used that $\mathcal{L}_{\partial_t}(\jmath_t)_*=0$ because $(\jmath_t)_*$ is precisely the projection in the $\partial_t$ direction. Meanwhile, in the $\dagger$ equality we have used the previously computed expression  for $\peqsub{\mathcal{L}}{\partial_t}g$, the fact that $\tau.n=0$, and equations \eqref{Mathematical background - equation - nabla_a=tau nabla_alpha} and \eqref{Mathematical background - equation - tau.e=Id e.tau=Id - cosas}. Thus, we have obtained that
  \begin{equation}
    (K_t)_{\bar{\alpha}\bar{\beta}}=\frac{1}{2\mathbf{N}_t}\Big(\mathcal{V}_{\bar{\alpha}\bar{\beta}}-\big(\mathcal{L}_{\vec{N}_t}\gamma_t\big)_{\bar{\alpha}\bar{\beta}}\Big)
  \end{equation}
  
  Pulling back everything to $\Sigma$ through $Z_t$ we can then define the GR Lagrangian in terms of a lapse\index{Lapse}, a shift\index{Shift}, and a spatial metric
  \[\begin{array}{cccc}
  L: & \mathcal{D}:=T\Big(\Cinf{\Sigma}\times\mathfrak{X}(\Sigma)\times\mathrm{Met}(\Sigma)\Big) & \longrightarrow & \R\\
  &           \mathbf{v}_{(\mathbf{N},\vec{N},\gamma)}=(\mathbf{N},\vec{N},\gamma;\peqsubfino{v}{\!\perp}{-0.2ex},v,\mathcal{V})                   &  \longmapsto    & L\left(\mathbf{v}_{(\mathbf{N},\vec{N},\gamma)}\right)
  \end{array}\]
  which is given by
\begin{align}\label{GR - equation - Lagrangian GR}
L\left(\mathbf{v}_{(\mathbf{N},\vec{N},\gamma)}\right)&=\int_\Sigma\peqsub{\mathrm{vol}}{\Sigma}\sqrt{\gamma}\,\mathbf{N}\Big[\peqsub{R^{(n)}}{\gamma}+\varepsilon \mathrm{Tr}(K)^2-2\varepsilon \langle K,K\rangle_\gamma-2\Lambda\Big]
\end{align}
where
\begin{align}\label{GR - equation - K=(v-Ln)/N} 
K\left(\mathbf{v}_{(\mathbf{N},\vec{N},\gamma)}\right)=\frac{\mathcal{V}-\mathcal{L}_{\vec{N}}\gamma}{2\mathbf{N}}
\end{align}
It is interesting to note that this Lagrangian was first derived in \cite{fischer1972einstein} using the DeWitt metric over the space of metrics.
  
  \section{Fiber derivative}\label{GR - section - fiber derivative}
  Before computing the fiber derivative associated with the previous Lagrangian let us study, as we did in chapter \ref{Chapter - Parametrized EM}, its target space i.e.\ the phase space of the theory.
  
  \subsection*{Geometric arena}\trassub
  
  A typical point of the phase space $T^*\mathcal{D}$ is of the form \[\boldsymbol{\mathrm{p}}_{(\mathbf{N},\vec{N},\gamma)}=(\mathbf{N},\vec{N},\gamma;\peqsubfino{\boldsymbol{p}}{\!\perp}{-0.2ex},\boldsymbol{p},\boldsymbol{\pi}) \in T^*\Big(\Cinf{\Sigma}\times\mathfrak{X}(\Sigma)\times\mathrm{Met}(\Sigma)\Big) \]
  where  $\peqsubfino{\boldsymbol{p}}{\!\perp}{-0.2ex}\in C^\infty(\Sigma)'$, $\boldsymbol{p}\in \mathfrak{X}(\Sigma)'$, and $\boldsymbol{\pi}\in\mathrm{Met}(\Sigma)'$ are elements of the dual of $\Cinf{\Sigma}$, $\mathfrak{X}(\Sigma)$ and $\mathrm{Met}(\Sigma)$ respectively. The phase space $T^*\mathcal{D}$ is equipped with the symplectic form
  \begin{align*}
  \Omega_{\boldsymbol{\mathrm{p}}_{(\mathbf{N},\vec{N},\gamma)}}(Y,Z)
  &=\Omega_{(\mathbf{N},\vec{N},\gamma;\peqsubfino{\boldsymbol{p}}{\!\perp}{-0.2ex},\boldsymbol{p},\boldsymbol{\pi})}\left(\rule{0ex}{3ex}(\peqsubfino{Y}{\!\boldsymbol{\mathbf{N}}}{-0.1ex},\peqsubfino{Y}{\!\boldsymbol{\vec{N}}}{-0.1ex},\peqsubfino{Y}{\!\boldsymbol{\gamma}}{-0.1ex},\vecc[\scalebox{0.4}{$\perp$}]{\boldsymbol{Y}}{\!\!\!p},\vecc{\boldsymbol{Y}}{\!\!\!p},\peqsubfino{\boldsymbol{Y}}{\!\!\!\pi}{-0.1ex}),(\peqsubfino{Z}{\boldsymbol{\mathbf{N}}}{-0.1ex},\peqsubfino{Z}{\!\boldsymbol{\vec{N}}}{-0.1ex},\peqsubfino{Z}{\gamma}{-0.1ex},\vecc[\scalebox{0.4}{$\perp$}]{\boldsymbol{Z}}{p},\vecc{\boldsymbol{Z}}{p},\peqsubfino{\boldsymbol{Z}}{\!\pi}{-0.1ex})\right)\overset{\eqref{eq:background canonical symplectic form}}{=}\\
  &=\vecc[\scalebox{0.4}{$\perp$}]{\boldsymbol{Z}}{p}\!\left(\peqsubfino{Y}{\!\boldsymbol{\mathbf{N}}}{-0.1ex}\right)-
  \vecc[\scalebox{0.4}{$\perp$}]{\boldsymbol{Y}}{\!p}\!\left(\peqsubfino{Z}{\boldsymbol{\mathbf{N}}}{-0.1ex}\right)+\vecc{\boldsymbol{Z}}{p}\!\left(\peqsubfino{Y}{\!\boldsymbol{\vec{N}}}{-0.1ex}\right)-\vecc{\boldsymbol{Y}}{\!\!\!p}\!\left(\peqsubfino{Z}{\!\boldsymbol{\vec{N}}}{-0.1ex}\right)+\peqsubfino{\boldsymbol{Z}}{\!\pi}{-0.1ex}\!\left(\peqsubfino{Y}{\!\boldsymbol{\gamma}}{-0.1ex}\right)-\peqsubfino{\boldsymbol{Y}}{\!\!\!\pi}{-0.1ex}\!\left(\peqsubfino{Z}{\gamma}{-0.1ex}\right)
  \end{align*}
  for $Y,Z\in\mathfrak{X}(T^*\mathcal{D})$ where we have omitted the base point for simplicity. In particular we have $\peqsubfino{Y}{\!\boldsymbol{\mathbf{N}}}{-0.1ex}\in C^\infty(\Sigma)$, $\peqsubfino{Y}{\!\boldsymbol{\vec{N}}}{-0.1ex}\in \mathfrak{X}(\Sigma)$, $\peqsubfino{Y}{\!\boldsymbol{\gamma}}{-0.1ex}\in\mathfrak{T}^{2,0}_{sym}(\Sigma)$, $\vecc[\scalebox{0.4}{$\perp$}]{\boldsymbol{Y}}{\!\!\!p}\in C^\infty(\Sigma)'$, $\vecc{\boldsymbol{Y}}{\!\!\!p}\in \mathfrak{X}(\Sigma)'$, and $\peqsubfino{\boldsymbol{Y}}{\!\!\!\pi}{-0.1ex}\in\mathrm{Met}(\Sigma)$.\separ
  
  As in the case of a parametrized theory, we will prove that the primary constraint submanifold $F\!L(T\mathcal{D})$ is not all $T^*\mathcal{D}$. In particular we will get that $\peqsubfino{\boldsymbol{p}}{\!\perp}{-0.1ex}$ and $\boldsymbol{p}$ are both zero, while $\boldsymbol{\pi}$ is given by a symmetric $(0,2)$ tensor field $\pi$. More precisely we will have for a given $T\in\mathfrak{T}^{2,0}_{sym}(\Sigma)$
  \begin{align*}
  \boldsymbol{\pi}(T)&=  \int_\Sigma   \big(\pi,T\big)\,\peqsub{\mathrm{vol}}{\Sigma}
  \end{align*}
  where $(\alpha,T)=\frac{1}{k!}\alpha_{a_1\cdots a_k}T^{a_1\cdots a_k}$. Notice that $\pi$ is a density (see page \pageref{Mathematical background - subsection - densities}) because it depends on the chosen volume form $\peqsub{\mathrm{vol}}{\Sigma}\in\mathrm{Vol}(\Sigma)$. We could, of course, rewrite it in terms of the metric volume form of $\gamma$.
  \begin{align}
  \boldsymbol{\pi}(T)&=\int_\Sigma  \big(\pi,T\big)\,\peqsub{\mathrm{vol}}{\Sigma}=\int_\Sigma   \frac{\big(\pi,T\big)}{\sqrt{\peqsubfino{\gamma}{X}{-0.2ex}}}\,\peqsubfino{{\mathrm{vol}_\gamma}}{\!\!X}{-0.4ex}\label{GE - equation - pi=int pi}
  \end{align}
  Let $\peqsub{\widetilde{\mathcal{P}}}{\mathrm{GR}}=\{\peqsub{\mathrm{p}}{\mathrm{q}}:=(\mathbf{N},\vec{N},\gamma,\peqsubfino{p}{\perp}{-0.2ex},p,\pi)\}\overset{\jmath}{\hookrightarrow }T^*\mathcal{D}$ considered as a subset of $T^*\mathcal{D}$ via the previous representation. We define the induced form $\widetilde{\Omega}^{\scriptscriptstyle\mathrm{GR}}:=\jmath^*\Omega$ given by  $\widetilde{\Omega}^{\scriptscriptstyle\mathrm{GR}}(Y,Z)=(\Omega\smallcirc\jmath)(\jmath_*Y,\jmath_*Z)$. In order to compute its explicit expression we see that we have to compute first $\jmath_*Y\in\mathfrak{X}^{\scriptscriptstyle \top}_\jmath\!(\mathcal{D})$ for any $Y=(\peqsubfino{Y}{\!\boldsymbol{\mathbf{N}}}{-0.1ex},\peqsubfino{Y}{\!\boldsymbol{\vec{N}}}{-0.1ex},\peqsubfino{Y}{\!\boldsymbol{\gamma}}{-0.1ex},\vecc[\scalebox{0.4}{$\perp$}]{Y}{p},\vecc{Y}{p},\peqsubfino{Y}{\!\boldsymbol{\pi}}{-0.1ex})\in\mathfrak{X}(\peqsub{\widetilde{\mathcal{P}}}{\mathrm{GR}})$. From \eqref{GE - equation - pi=int pi} and the analogs for $\peqsubfino{p}{\perp}{-0.2ex}$ and $p$ we obtain
  \[\jmath_*Y=\left(\peqsubfino{Y}{\!\boldsymbol{\mathbf{N}}}{-0.1ex}\coma\peqsubfino{Y}{\!\boldsymbol{\vec{N}}}{-0.1ex}\coma\peqsubfino{Y}{\!\boldsymbol{\gamma}}{-0.1ex}\coma\int_\Sigma (\vecc[\scalebox{0.4}{$\perp$}]{Y}{p}\coma\cdot{}\, )\peqsub{\mathrm{vol}}{\Sigma}\coma\int_\Sigma (\vecc{Y}{p}\coma\cdot{}\, )\peqsub{\mathrm{vol}}{\Sigma}\coma\int_\Sigma(\peqsubfino{Y}{\!\boldsymbol{\pi}}{-0.1ex}\coma\cdot{}\, )\peqsub{\mathrm{vol}}{\Sigma}\right)\]
  so we have
  \begin{align}\label{GR - equation - tilde Omega}
  \widetilde{\Omega}^{\scriptscriptstyle\mathrm{GR}}_{\peqsub{\mathrm{p}}{\mathrm{q}}}(Y,Z)&=\int_\Sigma\Big[\Big(\vecc[\scalebox{0.4}{$\perp$}]{Z}{p}\coma\peqsubfino{Y}{\!\boldsymbol{\mathbf{N}}}{-0.1ex}\Big)-\Big(\vecc[\scalebox{0.4}{$\perp$}]{Y}{p}\coma\peqsubfino{Z}{\boldsymbol{\mathbf{N}}}{-0.1ex}\Big)+\Big(\vecc{Z}{p}\coma\peqsubfino{Y}{\!\boldsymbol{\vec{N}}}{-0.1ex}\Big)-\Big(\vecc{Y}{p}\coma\peqsubfino{Z}{\!\boldsymbol{\vec{N}}}{-0.1ex}\Big)+\Big(\peqsubfino{Z}{\boldsymbol{\pi}}{-0.1ex}\coma\peqsubfino{Y}{\!\boldsymbol{\gamma}}{-0.1ex}\Big)-\Big(\peqsubfino{Y}{\!\boldsymbol{\pi}}{-0.1ex}\coma\peqsubfino{Z}{\boldsymbol{\gamma}}{-0.1ex}\Big)\Big]\peqsub{\mathrm{vol}}{\Sigma}
  \end{align}

  \subsection*{Computation of the fiber derivative}\trassub
  
  The fiber derivative, given by equation \eqref{Mathematical background - equation - FL=D_2L}, is computed taking an initial point of the tangent bundle $\mathbf{v}_{(\mathbf{N},\vec{N},\gamma)}=(\mathbf{N},\vec{N},\gamma;\peqsubfino{v}{\!\perp}{-0.2ex},v,\peqsubfino{\mathbb{V}}{\!X}{-0.1ex})$ and some initial velocities $\mathbf{w}^1_{(\mathbf{N},\vec{N},\gamma)}=(\mathbf{N},\vec{N},\gamma;\peqsubfino{w}{\perp}{-0.2ex},0,0)$, $\mathbf{w}^2_{(\mathbf{N},\vec{N},\gamma)}=(\mathbf{N},\vec{N},\gamma;0,w,0)$, and $\mathbf{w}^3_{(\mathbf{N},\vec{N},\gamma)}=(\mathbf{N},\vec{N},\gamma;0,0,\mathcal{W})$. Now the three are immediate\allowdisplaybreaks[0]
  \begin{align*}
  &FL\left(\peqsub{\mathbf{v}}{(\mathbf{N},\vec{N},\gamma)}\right)\!\left(\peqsub{\mathbf{w}}{(\mathbf{N},\vec{N},\gamma)}^1\right)=0&&\Longrightarrow \quad \peqsubfino{p}{\perp}{-0.2ex}=0\\[1.5ex]
  &FL\left(\peqsub{\mathbf{v}}{(\mathbf{N},\vec{N},\gamma)}\right)\!\left(\peqsub{\mathbf{w}}{(\mathbf{N},\vec{N},\gamma)}^2\right)=0&&\Longrightarrow \quad p=0\\[1ex]
  &FL\left(\peqsub{\mathbf{v}}{(\mathbf{N},\vec{N},\gamma)}\right)\!\left(\peqsub{\mathbf{w}}{(\mathbf{N},\vec{N},\gamma)}^3\right)=-\!\int_\Sigma\peqsub{\mathrm{vol}}{\Sigma}\mathcal{W}_{ab}\varepsilon\sqrt{\gamma}\Big[K^{ba}-\gamma^{ba}K^c_{\ c}\Big]&&\Longrightarrow\quad\pi^{ab}=-\varepsilon\sqrt{\gamma}[K^{ab}-\gamma^{ab}K^c_{\ c}]
  \end{align*}\allowdisplaybreaks
  We have the canonical momenta $\peqsub{p}{\perp}$, $p$ and $\pi$. Notice that $\pi$ and $K$ are related by
  \begin{equation}\label{GR - equation - pi=k... K=pi...}
  \pi^{ab}=-\varepsilon\sqrt{\gamma}\big[K^{ab}-K\gamma^{ab}\big]\qquad\quad\equiv\quad\qquad K^{ab}=\frac{-\varepsilon}{\sqrt{\gamma}}\left[\pi^{ab}-\frac{\pi}{n-1}\gamma^{ab}\right]
  \end{equation}
  The constrain manifold is
  \begin{align*}
  F\!L(\mathcal{D})&=\left\{(\mathbf{N},\vec{N},\gamma,\peqsubfino{p}{\perp}{-0.2ex},p,\pi)\in \peqsub{\widetilde{\mathcal{P}}}{\mathrm{GR}}\ /\quad \peqsubfino{p}{\perp}{-0.2ex}=0\quad p=0\right\}\cong\\
  &\cong\{(\mathbf{N},\vec{N},\gamma,\pi)\}=\peqsub{\mathcal{P}}{\mathrm{GR}}
  \end{align*}
  We define the inclusion $\peqsubfino{\jmath}{\mathrm{GR}}{-0.2ex}:\peqsub{\mathcal{P}}{\mathrm{GR}}\hookrightarrow \peqsub{\widetilde{\mathcal{P}}}{\mathrm{GR}}$ that allows to pullback the induced form $\widetilde{\Omega}^{\scriptscriptstyle\mathrm{GR}}=\jmath^*\Omega$ of $\peqsub{\widetilde{\mathcal{P}}}{\mathrm{GR}}$, given by equation \eqref{GR - equation - tilde Omega}, to $\peqsub{\mathcal{P}}{\mathrm{GR}}$ in order to define $\omega^{\scriptscriptstyle\mathrm{GR}}:=\peqsub{\jmath}{\mathrm{GR}\,}^*\widetilde{\Omega}^{\scriptscriptstyle\mathrm{GR}}$.\separ
  
  To compute $\omega^{\scriptscriptstyle\mathrm{GR}}$ we need the pushforward $(\peqsubfino{\jmath}{\mathrm{GR}}{-0.2ex})_*Y$ for every $Y=(\vecc[\scalebox{0.4}{$\perp$}]{Y}{q},\vecc{Y}{q},\vecc{Y}{X},\vecc{Y}{p})\in\mathfrak{X}(\peqsub{\mathcal{P}}{\mathrm{GR}})$ because $\omega^{\scriptscriptstyle\mathrm{GR}}(Y,Z):=(\widetilde{\Omega}^{\scriptscriptstyle\mathrm{GR}}\smallcirc \peqsubfino{\jmath}{\mathrm{GR}}{-0.2ex})((\peqsubfino{\jmath}{\mathrm{GR}}{-0.2ex})_*Y,(\peqsubfino{\jmath}{\mathrm{GR}}{-0.2ex})_*Z)$ but in this case is trivial leading to the (weakly) symplectic form 
  \begin{align}\label{GR - equation - omega final}
  \omega^{\scriptscriptstyle\mathrm{GR}}_{(\mathbf{N},\vec{N},\gamma,\pi)}&\Big((\peqsub{Y}{\boldsymbol{\mathbf{N}}},\peqsub{Y}{\boldsymbol{\vec{N}}},\peqsub{Y}{\boldsymbol{\gamma}},\peqsub{Y}{\boldsymbol{\pi}}),(\peqsub{Z}{\boldsymbol{\mathbf{N}}},\peqsub{Z}{\boldsymbol{\vec{N}}},\peqsub{Z}{\boldsymbol{\gamma}},\peqsub{Z}{\boldsymbol{\pi}})\Big)=\int_\Sigma\Big[\Big(\peqsubfino{Z}{\boldsymbol{\pi}}{-0.1ex}\coma\peqsubfino{Y}{\!\boldsymbol{\gamma}}{-0.1ex}\Big)-\Big(\peqsubfino{Y}{\!\boldsymbol{\pi}}{-0.1ex}\coma\peqsubfino{Z}{\boldsymbol{\gamma}}{-0.1ex}\Big)\Big]\peqsub{\mathrm{vol}}{\Sigma}
  \end{align}

  \section{Hamiltonian Formulation}\label{GR - section - Hamiltonian formulation}
    \subsection*{Obtaining the Hamiltonian}\trassub
    
  Let us now compute the energy\index{Energy} $E:T\mathcal{D}\to\R$ given by $E(q,v)=F\!L(q,v)\big(q,v\big)-L(q,v)$.
  \begin{align*}
  E&\left(\mathbf{v}_{(\mathbf{N},\vec{N},\gamma)}\right)=FL\left(\mathbf{v}_{(\mathbf{N},\vec{N},\gamma)}\right)\left(\mathbf{v}_{(\mathbf{N},\vec{N},\gamma)}\right)-L\left(\mathbf{v}_{(\mathbf{N},\vec{N},\gamma)}\right)=\\
  &=\int_\Sigma\peqsub{\mathrm{vol}}{\Sigma}\sqrt{\gamma}\left(-\varepsilon\mathcal{V}_{ab}\Big[K^{ba}-\gamma^{ab}K^c_{\ c}\Big]-\mathbf{N}\Big[\peqsub{R^{(n)}}{\gamma}+\varepsilon K^{a}_{\ a}K^b_{\ b}-\varepsilon K^d_{\ b}K^b_{\ d}-2\Lambda\Big]\right)\overset{\eqref{GR - equation - K=(v-Ln)/N}}{=}\\
  &=\int_\Sigma\peqsub{\mathrm{vol}}{\Sigma}\sqrt{\gamma}\left(-\varepsilon\Big[\mathbf{N}K_{ab}+(\mathcal{L}_{\vec{N}}\gamma)_{ab}\Big]\Big[K^{ba}-\gamma^{ab}K^c_{\ c}\Big]-\mathbf{N}\Big[\peqsub{R^{(n)}}{\gamma}-2\Lambda\Big]\right)=\\
  &=\int_\Sigma\peqsub{\mathrm{vol}}{\Sigma}\left(\Big[\mathbf{N}K_{ab}+(\mathcal{L}_{\vec{N}}\gamma)_{ab}\Big]\pi^{ab}-\sqrt{\gamma}\mathbf{N}\Big[\peqsub{R^{(n)}}{\gamma}-2\Lambda\Big]\right)
  \end{align*}
  Notice that $\pi$ can be written in terms of $K$ which, in turns, depend on $\mathcal{V}$ so we could leave everything in terms of the velocities as it should be for the energy. Nonetheless, we are interested in the Hamiltonian and it is better to express it like that. The Hamiltonian $H:F\!L(\mathcal{D})\subset T^*\!\mathcal{D}\to\R$ is given by the implicit equation $H\smallcirc F\!L=E$. Using the identification $F\!L(\mathcal{D})\cong\peqsub{\mathcal{P}}{\mathrm{GR}}$ we can obtain $H:\peqsub{\mathcal{P}}{\mathrm{GR}}\to\R$ simply by writing $K$ in terms of $\pi$
  \begin{align}\label{GR - equation - Hamiltonian}
  \begin{split}
  H&\left(\mathbf{N},\vec{N},\gamma;\pi\right)=\\
  &=\int_\Sigma\peqsub{\mathrm{vol}}{\Sigma}\left(\pi^{ab}(\mathcal{L}_{\vec{N}}\gamma)_{ab}-\frac{\varepsilon \mathbf{N}}{\sqrt{\gamma}}\pi^{ab}\pi^{cd}\left(\gamma_{ac}\gamma_{bd}-\frac{\gamma_{ab}\gamma_{cd}}{n-1}\right)-\sqrt{\gamma}\mathbf{N}\Big[\peqsub{R^{(n)}}{\gamma}-2\Lambda\Big]\right)\updown{\pi}{\mathrm{sym.}}{=}\\
  &=\int_\Sigma\peqsub{\mathrm{vol}}{\Sigma}\left(2\pi^{ab}\nabla_aN_b+\mathbf{N}\peqsubfino{\mathcal{H}}{\perp}{-0.2ex}\right)\overset{\eqref{appendix equation integracion por partes}}{=}\int_\Sigma\peqsub{\mathrm{vol}}{\Sigma}\left(-2N^c\gamma_{bc}\nabla_a\pi^{ab}+\mathbf{N}\peqsubfino{\mathcal{H}}{\perp}{-0.2ex}\right)=\\
  &=\int_\Sigma\peqsub{\mathrm{vol}}{\Sigma}\left(N^a\mathcal{H}_a+\mathbf{N}\peqsubfino{\mathcal{H}}{\perp}{-0.2ex}\right)
  \end{split}
  	\end{align}
  where we have defined
  \begin{align}
  \mathcal{H}_c(\gamma,\pi)&=-2\gamma_{bc}\nabla_a\pi^{ab}\label{GR - equation - H_tangente}\\
  \peqsubfino{\mathcal{H}}{\perp}{-0.2ex}(\gamma,\pi)&=-\frac{\varepsilon}{\sqrt{\gamma}}\left[\gamma_{ac}\gamma_{bd}-\frac{1}{n-1}\gamma_{ab}\gamma_{cd}\right]\pi^{ab}\pi^{cd}-\sqrt{\gamma}\Big(R^{(n)}_\gamma-2\Lambda\Big)\label{GR - equation - H_perpendicular}
  \end{align}
  
   \subsection*{GNH algorithm}\trassub
  
  Once we have the Hamiltonian, we want to solve the equation $\peqsub{\imath}{Y}\omega^{\scriptscriptstyle\mathrm{GR}}=\mathrm{d}H$ to obtain the Hamiltonian vector field $Y\in\mathfrak{X}(\peqsub{\mathcal{P}}{\mathrm{GR}})$. We do so using the GNH algorithm explained in section \ref{Mathematical background - section - GNH} of chapter \ref{Chapter - Mathematical background}.\separ
  
  $\bullet$ Compute the differential $\mathrm{d}H:T\peqsub{\mathcal{P}}{\mathrm{GR}}\to\R$ of $H$.\separprevia
  
  It is long, but not very hard, to obtain the differential. The details of the computation can be found on lemma \ref{Appendix - lemma - dH} of the appendix.
  \begin{align}\label{GR - equation - dH}
  \begin{split}
  &\mathrm{d}_{(\mathbf{N},\vec{N},\gamma;\pi)}H\Big( \peqsub{Z}{\boldsymbol{\mathbf{N}}},\peqsub{Z}{\boldsymbol{\vec{N}}},\peqsub{Z}{\boldsymbol{\gamma}},\peqsub{Z}{\boldsymbol{\pi}}\Big)=\int_\Sigma\peqsub{\mathrm{vol}}{\Sigma}\,\Bigg\{\peqsub{Z}{\boldsymbol{\vec{N}}}^a\mathcal{H}_a+\peqsub{Z}{\boldsymbol{\mathbf{N}}}\peqsubfino{\mathcal{H}}{\perp}{-0.2ex}+(\peqsub{Z}{\boldsymbol{\pi}})^{ab}\Big[(\mathcal{L}_{\vec{N}}\gamma)_{ab}+2\mathbf{N}K_{ab}\Big]-\\
  &\hspace*{30ex}-(\peqsub{Z}{\boldsymbol{\gamma}})_{ab}\Big[(\mathcal{L}_{\vec{N}}\pi)^{ab}-\mathbf{N}\beta^{ab}+\sqrt{\gamma}\Big(\nabla^a\nabla^b-\gamma^{ab}\nabla_d\nabla^d\Big)\mathbf{N}\Big]\Bigg\}
  \end{split}
  \end{align}
  where
  \begin{equation}\label{GR - equation - beta}
  \beta^{ab}=\sqrt{\gamma}R^{ac}-\frac{2\varepsilon}{\sqrt{\gamma}}\left(\pi^{ad}\pi^c_{\ d}-\frac{\pi}{n-1}\pi^{ac}\right)+\frac{\varepsilon}{\sqrt{\gamma}}\left(\pi_{bd}\pi^{bd}-\frac{\pi^2}{n-1}\right)\gamma^{ac}+\frac{\gamma^{ac}}{2}\peqsubfino{\mathcal{H}}{\perp}{-0.2ex}
  \end{equation}

$\bullet$ Compute the Hamiltonian vector field $Y\in\mathfrak{X}(\peqsub{\mathcal{P}}{\mathrm{GR}})$.\separprevia
  
  We now obtain the equations of $Y$ by solving, for every $Z$, the equation $\omega^{\scriptscriptstyle\mathrm{GR}}(Y,Z)=\mathrm{d}H(Z)$. Thus from \eqref{GR - equation - omega final} and \eqref{GR - equation - dH} we obtain
  \begin{align*}
  &\int_\Sigma\left[(\peqsub{Z}{\boldsymbol{\pi}})^{ab}(\peqsub{Y}{\boldsymbol{\gamma}})_{ab}-(\peqsub{Y}{\boldsymbol{\pi}})^{ab}(\peqsub{Z}{\boldsymbol{\gamma}})_{ab}\right]\peqsub{\mathrm{vol}}{\Sigma}=\int_\Sigma\peqsub{\mathrm{vol}}{\Sigma}\Bigg\{\peqsub{Z}{\boldsymbol{\vec{N}}}^a\mathcal{H}_a+\peqsub{Z}{\boldsymbol{\mathbf{N}}}\peqsubfino{\mathcal{H}}{\perp}{-0.2ex}+(\peqsub{Z}{\boldsymbol{\pi}})^{ab}\Big[(\mathcal{L}_{\vec{N}}\gamma)_{ab}+2\mathbf{N}K_{ab}\Big]-\\
  &\hspace*{38ex}-(\peqsub{Z}{\boldsymbol{\gamma}})_{ab}\Big[(\mathcal{L}_{\vec{N}}\pi)^{ab}-\mathbf{N}\beta^{ab}+\sqrt{\gamma}\Big(\nabla^a\nabla^b-\gamma^{ab}\nabla_d\nabla^d\Big)\mathbf{N}\Big]\Bigg\}
  \end{align*}
  We obtain directly, using \eqref{GR - equation - pi=k... K=pi...} to write everything in terms of $\pi$, that
  \begin{align}\label{GR - equation - Hamiltonian equations pi}
  \begin{split}
  &\peqsubfino{Y}{\!\boldsymbol{\mathbf{N}}}{-0.1ex},\peqsubfino{Y}{\!\boldsymbol{\vec{N}}}{-0.1ex}\quad\text{ arbitrary}\\
  &(\peqsub{Y}{\boldsymbol{\gamma}})_{ab}=(\mathcal{L}_{\vec{N}}\gamma)_{ab}-2\varepsilon\frac{\mathbf{N}}{\sqrt{\gamma}}\left[\pi_{ab}-\frac{\pi}{n-1}\gamma_{ab}\right]\\
  &(\peqsub{Y}{\boldsymbol{\pi}})^{ab}=(\mathcal{L}_{\vec{N}}\pi)^{ab}-\mathbf{N}\beta^{ab}+\sqrt{\gamma}\Big(\nabla^a\nabla^b-\gamma^{ab}\nabla_d\nabla^d\Big)\mathbf{N}\\%
  &\hspace*{-1.2ex}\left.\begin{array}{l}%
  \peqsubfino{\mathcal{H}}{\perp}{-0.2ex}=0\\[0.8ex]
  \mathcal{H}_a=0_a
  \end{array}\right\}\text{ Constraints}
  \end{split}
  \end{align}
 The solutions are only defined over $\peqsub{\mathcal{P}}{\mathrm{GR}}^2:=\{\peqsubfino{\mathcal{H}}{\perp}{-0.2ex}=0\,, \ \mathcal{H}_a=0_a\}\subset\peqsub{\mathcal{P}}{\mathrm{GR}}$.\separpost

 $\bullet$ Require $\peqsub{Y}{\!H}$ to be tangent to $\peqsub{\mathcal{P}}{\mathrm{GR}}^2$.\separprevia
 
We have now to determine the space $\peqsub{\mathcal{P}}{\mathrm{GR}}^3\subset\peqsub{\mathcal{P}}{\mathrm{GR}}^2$ where $Y$ is tangent to $\peqsub{\mathcal{P}}{\mathrm{GR}}^2$. This is equivalent to require that the dynamics preserves the constraints. In lemma \ref{Appendix - lemma - DH and DH_perp} we prove that the variations of $\peqsubfino{\mathcal{H}}{\perp}{-0.2ex}$ and $\mathcal{H}_a$ are given by\allowdisplaybreaks[0]
\begin{align*}
&\blacktriangleright\  (D\mathcal{H})_a=(\mathcal{L}_{\vec{N}}\mathcal{H})_a+\peqsubfino{\mathcal{H}}{\perp}{-0.2ex}\nabla_a\mathbf{N}\\[1ex]
&\blacktriangleright\  (D\peqsubfino{\mathcal{H}}{\perp}{-0.2ex})=\mathcal{L}_{\vec{N}}\peqsubfino{\mathcal{H}}{\perp}{-0.2ex}+\varepsilon\mathbf{N}\nabla^a\mathcal{H}_a+2\varepsilon\mathcal{H}_a\nabla^a\mathbf{N}
\end{align*}\allowdisplaybreaks
If we have the conditions $\peqsubfino{\mathcal{H}}{\perp}{-0.2ex}=0$ and $\mathcal{H}_a=0_a$, then all the derivatives are also zero and, hence, their variations vanish. Thus $\peqsub{\mathcal{P}}{\mathrm{GR}}^3=\peqsub{\mathcal{P}}{\mathrm{GR}}^2$ and the algorithm stops.

\subsection*{ADM variables}\trassub
  
  It is customary in the literature to use the so-called ADM variables i.e.\ $(\mathbf{N},\vec{N},\gamma,K)$ instead of $(\mathbf{N},\vec{N},\gamma,\pi)$. Then the equations of motion \eqref{GR - equation - Hamiltonian equations pi} can be rewritten, as we prove in lemma \ref{Appendix - lemma - Y_K}, as
  \begin{align*}
  &\peqsubfino{Y}{\!\boldsymbol{\mathbf{N}}}{-0.1ex},\peqsubfino{Y}{\!\boldsymbol{\vec{N}}}{-0.1ex}\quad\text{ arbitrary}\\
  &(\peqsub{Y}{\boldsymbol{\gamma}})_{ab}=(\mathcal{L}_{\vec{N}}\gamma)_{ab}+2\mathbf{N}K_{ab}\\
  &(\vecc{Y}{K})_{ab}=(\mathcal{L}_{\vec{N}}K)_{ab}-\varepsilon\nabla_a\nabla_b\mathbf{N}+\mathbf{N}\Bigg(\varepsilon R_{ab}-KK_{ab}+2K_{ac}\tensor{K}{^d_b}-\frac{\varepsilon\gamma_{ab}}{n-1}\left[2\Lambda-\frac{\peqsub{\mathcal{H}}{\perp}}{2\sqrt{\gamma}}\right]\Bigg)\\
  &\peqsubfino{\mathcal{H}}{\perp}{-0.2ex}=0\\
  &\mathcal{H}_a=0_a
  \end{align*}
  which can be rewritten in a more geometric flavor as
    \begin{align*}  \label{GR - equation - Hamiltonian equations K}
  &\peqsubfino{Y}{\!\boldsymbol{\mathbf{N}}}{-0.1ex},\peqsubfino{Y}{\!\boldsymbol{\vec{N}}}{-0.1ex}\quad\text{ arbitrary}\\
  &\peqsub{Y}{\boldsymbol{\gamma}}=\mathcal{L}_{\vec{N}}\gamma+2\mathbf{N}K\\
  &\vecc{Y}{K}=\mathcal{L}_{\vec{N}}K-\varepsilon\mathrm{Hess}(\mathbf{N})+\mathbf{N}\left(\varepsilon \mathrm{Ric}-\mathrm{Tr}_\gamma(K)K+2 (K\times K)-\frac{\varepsilon}{n-1}\left[2\Lambda-\frac{\peqsubfino{\mathcal{H}}{\perp}{-0.2ex}}{2\sqrt{\gamma}}\right]\gamma\right)\\
  &\peqsubfino{\mathcal{H}}{\perp}{-0.2ex}=0\\
  &\mathcal{H}_a=0_a
  \end{align*}
  where we define the cross product for symmetric tensors as
  \begin{equation}
    (T\times R)_{ab}=T_{ac}\tensor{R}{^c_b}
    \end{equation}

  \section{Unimodular gravity}\label{GR - section - unimodular}
  
  We have studied in the previous section the standard Hilbert-Einstein action with cosmological constant $\Lambda$\index{Cosmological constant}. Such action, and the resulting theory, is invariant under diffeomorphisms and can be rendered into its Hamiltonian formulation as we have shown. However, the need to explain the precise value of the cosmological constant $\Lambda$ suggested some approaches where it does not play the role of a constant anymore. One of the most interesting ones is unimodular gravity \cite{henneaux1989cosmological,smolin2009quantization,bufalo2015unimodular}, where we consider the action  $\peqsub{S}{\mathrm{uni}}:\mathrm{Met}(M)\times\R\to\R$
  \[\peqsub{S}{\mathrm{uni}}(g,\lambda)=\int_M R(g)\peqsub{\mathrm{vol}}{g}-2 \int_{M}\lambda\big(\peqsub{\mathrm{vol}}{g}-\epsilon\peqsub{\mathrm{vol}}{0}\big)\]
  for some fixed volume form $\mathrm{vol}_0$. We have included the parameter $\epsilon=1$ so that we can study the limit $\epsilon\to0$. The role of the cosmological constant will be played by $\lambda$, as we will see in a moment, which is a variable that will be forced to be constant by the equations of motion.\separ
  
  It is important to mention that $\peqsub{S}{\mathrm{uni}}$ is not invariant under the full group of diffeomorphisms, but only under the subgroup
  \[\peqsub{\mathrm{Diff}}{0}(M)=\Big\{Z\in\mathrm{Diff}(M)\ \ /\ \ Z^*\peqsub{\mathrm{vol}}{0}=\peqsub{\mathrm{vol}}{0}\Big\}\]
  
  \subsection*{Variations of the action}\trassub
  
  The variations of the action are easy to obtain using the computations of section \ref{GR - section - HE action} with $\Lambda=0$. Indeed, given $h\in T_g\mathrm{Met}(M)$ and $\peqsub{v}{\lambda}\in T_\lambda\Cinf{M}$  we have
  \begin{align*}
   \peqsub{D}{(g,h)}\peqsub{S}{\mathrm{uni}}&=\int_M G_{ab}h^{ab}\peqsub{\mathrm{vol}}{g}-2\int_M\lambda\peqsub{D}{(g,h)}\peqsub{\mathrm{vol}}{g}\overset{\eqref{Appendix - equation - variacion volumen}}{=}\\
   &=\int_M\Big[G_{ab}-\lambda g_{ab}\Big]h^{ab}\peqsub{\mathrm{vol}}{g}\\[1ex]
   \peqsub{D}{(\lambda,\peqsub{v}{\lambda})}\peqsub{S}{\mathrm{uni}}&=-2\int_M\ \peqsub{v}{\lambda}\big(\peqsub{\mathrm{vol}}{g}-\epsilon\peqsub{\mathrm{vol}}{0}\big)
   \end{align*}
   which leads to the equations
   \begin{align}\begin{split}
    &\mathrm{Ric}(g)-\frac{1}{2}R(g)\,g=\lambda\, g\\
    &\peqsub{\mathrm{vol}}{g}=\epsilon\peqsub{\mathrm{vol}}{0}\end{split}
   \end{align}
   Notice that the first equation is formally analog to \eqref{GR - equation - Einstein eq} although now $\lambda$ is not constant but a variable of the theory. Nonetheless, as the divergence of $G$ is zero due to equation \eqref{Appendix - equation - div Ric=grad R}, we have that the first equation implies
   \begin{equation}\label{GR - equation - lambda=cte}
   \mathrm{d}\lambda=0\qquad\quad\longrightarrow\quad\qquad\lambda=\mathrm{constant}
   \end{equation}
   Meanwhile, the second equation can be rewritten using $\peqsub{\mathrm{vol}}{g}=\det(\varepsilon g)\peqsub{\mathrm{vol}}{0}$ which implies that $\det(\varepsilon g)=:\epsilon$ is constant.
  
  \subsection*{Lagrangian formulation}\trassub
  
  In order to obtain the Lagrangian formulation notice that the fixed volume $\mathrm{vol}_0$ can be written as $\mathrm{vol}_0=\mathrm{d}t\wedge\peqsub{\mathrm{vol}}{\Sigma}$ for some fixed volume $\peqsub{\mathrm{vol}}{\Sigma}$ and function $t$. Therefore we have that
  \[\sqrt{\varepsilon g}\,\peqsub{\mathrm{vol}}{0}=\peqsub{\mathrm{vol}}{g}\overset{\ref{Appendix - lemma - vol=nu wedge vol}}{=}\varepsilon n\wedge\peqsub{\mathrm{vol}}{\gamma}\overset{\eqref{Mathematical background - equation - n=varepsilon N dt}}{=}\mathbf{N}\sqrt{\gamma}\,\mathrm{d}t\wedge\peqsub{\mathrm{vol}}{\Sigma}\qquad\quad\longrightarrow\quad\qquad\sqrt{\varepsilon g}=\mathbf{N}\sqrt{\gamma}\]
  Taking this equality into account and proceeding as in section \ref{GR - section - Lagrangian formulation} we obtain
    \begin{align*}
  S(g,\lambda)=\int_{\R}\mathrm{d}t\int_\Sigma Z_t^*\left\{\mathbf{N}\sqrt{\gamma}\left[R^{(n)}_\gamma+\varepsilon \mathrm{Tr}(K)^2-2\varepsilon \langle K,K\rangle_\gamma-2\lambda\left(1-\frac{\epsilon}{\mathbf{N}\sqrt{\gamma}}\right)\right]\right\}\peqsub{\mathrm{vol}}{\Sigma}
  \end{align*}
  The Lagrangian can be read off from the previous expression using equation \eqref{GR - equation - K=(v-Ln)/N} 
  \[\begin{array}{cccc}
  \peqsub{L}{\mathrm{uni}}: & \mathcal{D}:=T\Big(\Cinf{\Sigma}\times\Cinf{\Sigma}\times\mathfrak{X}(\Sigma)\times\mathrm{Met}(\Sigma)\Big) & \longrightarrow & \R\\
  &           \mathbf{v}_{(\lambda,\mathbf{N},\vec{N},\gamma)}=(\lambda,\mathbf{N},\vec{N},\gamma;\peqsub{v}{\lambda},\peqsubfino{v}{\!\perp}{-0.2ex},v,\mathcal{V})                   &  \longmapsto    & \peqsub{L}{\mathrm{uni}}\left(\mathbf{v}_{(\lambda,\mathbf{N},\vec{N},\gamma)}\right)
  \end{array}\]
  which is given by
  \begin{align}\label{GR - equation - Lagrangian GR unimod}
  \peqsub{L}{\mathrm{uni}}\left(\mathbf{v}_{(\lambda,\mathbf{N},\vec{N},\gamma)}\right)&=\int_\Sigma\peqsub{\mathrm{vol}}{\Sigma}\left\{\sqrt{\gamma}\,\mathbf{N}\left[\peqsub{R^{(n)}}{\gamma}+\varepsilon \mathrm{Tr}(K)^2-2\varepsilon \langle K,K\rangle_\gamma\right]-2\lambda\Big(\mathbf{N}\,\sqrt{\gamma}-\epsilon\Big)\right\}
  \end{align}
  where we recall that
  \begin{align}
  K\left(\mathbf{v}_{(\lambda,\mathbf{N},\vec{N},\gamma)}\right)=\frac{\mathcal{V}-\mathcal{L}_{\vec{N}}\gamma}{2\mathbf{N}}
  \end{align}

  \subsection*{Fiber derivative}\trassub
  
  Using the same arguments and notations as in section \ref{GR - section - fiber derivative} (but now with one more variable), we have that the geometric arena is $\peqsub{\widetilde{\mathcal{P}}}{\mathrm{uni}}=\{\peqsub{\mathrm{p}}{\mathrm{q}}:=(\lambda,\mathbf{N},\vec{N},\gamma,\peqsub{p}{\lambda},\peqsubfino{p}{\perp}{-0.2ex},p,\pi)\}\overset{\jmath}{\hookrightarrow }T^*\mathcal{D}$ considered as a subset of $T^*\mathcal{D}$. The induced symplectic form over $\peqsub{\widetilde{\mathcal{P}}}{\mathrm{uni}}$ reads
  \begin{align}\begin{split}
  \widetilde{\Omega}^{\scriptscriptstyle\mathrm{uni}}_{\peqsub{\mathrm{p}}{\mathrm{q}}}(Y,Z)&=\int_\Sigma\Big[\Big(\vecc[\scalebox{0.4}{$\lambda$}]{Z}{p}\coma\peqsubfino{Y}{\lambda}{-0.1ex}\Big)-\Big(\vecc[\scalebox{0.4}{$\lambda$}]{Y}{p}\coma\peqsubfino{Z}{\lambda}{-0.1ex}\Big)+\Big(\vecc[\scalebox{0.4}{$\perp$}]{Z}{p}\coma\peqsubfino{Y}{\!\boldsymbol{\mathbf{N}}}{-0.1ex}\Big)-\Big(\vecc[\scalebox{0.4}{$\perp$}]{Y}{p}\coma\peqsubfino{Z}{\boldsymbol{\mathbf{N}}}{-0.1ex}\Big)+\\
  &\qquad\quad+\Big(\vecc{Z}{p}\coma\peqsubfino{Y}{\!\boldsymbol{\vec{N}}}{-0.1ex}\Big)-\Big(\vecc{Y}{p}\coma\peqsubfino{Z}{\!\boldsymbol{\vec{N}}}{-0.1ex}\Big)+\Big(\peqsubfino{Z}{\boldsymbol{\pi}}{-0.1ex}\coma\peqsubfino{Y}{\!\boldsymbol{\gamma}}{-0.1ex}\Big)-\Big(\peqsubfino{Y}{\!\boldsymbol{\pi}}{-0.1ex}\coma\peqsubfino{Z}{\boldsymbol{\gamma}}{-0.1ex}\Big)\Big]\peqsub{\mathrm{vol}}{\Sigma}\end{split}\label{GR - equation - tilde Omega uni}
  \end{align}
  The fiber derivative, in this case, is given by
  \begin{align*}
  &F\peqsub{L}{\mathrm{uni}}\left(\peqsub{\mathbf{v}}{(\lambda,\mathbf{N},\vec{N},\gamma)}\right)\!\left(\peqsub{\mathbf{w}}{(\lambda,\mathbf{N},\vec{N},\gamma)}^0\right)\!=0\\
  &F\peqsub{L}{\mathrm{uni}}\left(\peqsub{\mathbf{v}}{(\lambda,\mathbf{N},\vec{N},\gamma)}\right)\!\left(\peqsub{\mathbf{w}}{(\lambda,\mathbf{N},\vec{N},\gamma)}^1\right)\!=0\\[1.5ex]
  &F\peqsub{L}{\mathrm{uni}}\left(\peqsub{\mathbf{v}}{(\lambda,\mathbf{N},\vec{N},\gamma)}\right)\!\left(\peqsub{\mathbf{w}}{(\lambda,\mathbf{N},\vec{N},\gamma)}^2\right)\!=0\\[1ex]
  &F\peqsub{L}{\mathrm{uni}}\left(\peqsub{\mathbf{v}}{(\lambda,\mathbf{N},\vec{N},\gamma)}\right)\!\left(\peqsub{\mathbf{w}}{(\lambda,\mathbf{N},\vec{N},\gamma)}^3\right)\!=-\!\int_\Sigma\peqsub{\mathrm{vol}}{\Sigma}\mathcal{W}_{ab}\varepsilon\sqrt{\gamma}\Big[K^{ba}-\gamma^{ba}K^c_{\ c}\Big]
  \end{align*}
  We have that the canonical momenta $\peqsubfino{p}{\!\perp}{-0.2ex}$, $p$ and $\peqsubfino{p}{\lambda}{-0.2ex}$ are zero, while $\pi$ is again related to $K$ by
  \begin{equation}\label{GR - equation - pi=k... K=pi... uni}
  \pi^{ab}=-\varepsilon\sqrt{\gamma}\big[K^{ab}-K\gamma^{ab}\big]\qquad\quad\equiv\quad\qquad K^{ab}=\frac{-\varepsilon}{\sqrt{\gamma}}\left[\pi^{ab}-\frac{\pi}{n-1}\gamma^{ab}\right]
  \end{equation}
  The first constraint manifold is
  \begin{align*}
  F\!\peqsub{L}{\mathrm{uni}}(\mathcal{D})&=\left\{(\lambda,\mathbf{N},\vec{N},\gamma,\peqsubfino{p}{\lambda}{-0.2ex},\peqsubfino{p}{\!\perp}{-0.2ex},p,\pi)\in \peqsub{\widetilde{\mathcal{P}}}{\mathrm{uni}}\ /\quad \peqsubfino{p}{\lambda}{-0.2ex}=0\quad \peqsubfino{p}{\perp}{-0.2ex}=0\quad p=0\right\}\cong\\
  &\cong\{(\lambda,\mathbf{N},\vec{N},\gamma,\pi)\}=\peqsub{\mathcal{P}}{\mathrm{uni}}
  \end{align*}
  We define the inclusion $\peqsubfino{\jmath}{\mathrm{uni}}{-0.2ex}:\peqsub{\mathcal{P}}{\mathrm{uni}}\hookrightarrow \peqsub{\widetilde{\mathcal{P}}}{\mathrm{uni}}$ that allows us to pullback the induced form $\widetilde{\Omega}^{\scriptscriptstyle\mathrm{uni}}$ of $\peqsub{\widetilde{\mathcal{P}}}{\mathrm{uni}}$, given by equation \eqref{GR - equation - tilde Omega uni}, to $\peqsub{\mathcal{P}}{\mathrm{uni}}$ in order to define $\omega^{\scriptscriptstyle\mathrm{uni}}:=\peqsub{\jmath}{\mathrm{uni}\,}^*\widetilde{\Omega}^{\scriptscriptstyle\mathrm{uni}}$ which is given by
  \begin{align}\label{GR - equation - omega final uni}
  \omega^{\scriptscriptstyle\mathrm{uni}}_{(\lambda,\mathbf{N},\vec{N},\gamma,\pi)}&\Big((\peqsub{Y}{\boldsymbol{\lambda}},\peqsub{Y}{\boldsymbol{\mathbf{N}}},\peqsub{Y}{\boldsymbol{\vec{N}}},\peqsub{Y}{\boldsymbol{\gamma}},\peqsub{Y}{\boldsymbol{\pi}}),(\peqsub{Z}{\boldsymbol{\lambda}},\peqsub{Z}{\boldsymbol{\mathbf{N}}},\peqsub{Z}{\boldsymbol{\vec{N}}},\peqsub{Z}{\boldsymbol{\gamma}},\peqsub{Z}{\boldsymbol{\pi}})\Big)=\int_\Sigma\Big[\Big(\peqsubfino{Z}{\boldsymbol{\pi}}{-0.1ex}\coma\peqsubfino{Y}{\!\boldsymbol{\gamma}}{-0.1ex}\Big)-\Big(\peqsubfino{Y}{\!\boldsymbol{\pi}}{-0.1ex}\coma\peqsubfino{Z}{\boldsymbol{\gamma}}{-0.1ex}\Big)\Big]\peqsub{\mathrm{vol}}{\Sigma}
  \end{align}

  \subsection*{Obtaining the Hamiltonian}\trassub
  
  The energy\index{Energy} and the Hamiltonian can be obtained from the GR case by taking
  \[\Lambda =\lambda\left(1-\frac{\varepsilon}{\mathbf{N}\,\sqrt{\gamma}}\right)\]
  because there is no differentiation involving this $\Lambda$ term. Thus we obtain
  \begin{align*}
  &\peqsub{E}{\mathrm{uni}}\left(\mathbf{v}_{(\lambda,\mathbf{N},\vec{N},\gamma)}\right)=F\!\peqsub{L}{\mathrm{uni}}\left(\mathbf{v}_{(\lambda,\mathbf{N},\vec{N},\gamma)}\right)\left(\mathbf{v}_{(\lambda,\mathbf{N},\vec{N},\gamma)}\right)-\peqsub{L}{\mathrm{uni}}\left(\mathbf{v}_{(\lambda,\mathbf{N},\vec{N},\gamma)}\right)=\\
  &\ =\int_\Sigma\peqsub{\mathrm{vol}}{\Sigma}\sqrt{\gamma}\left(-\varepsilon\mathcal{V}_{ab}\Big[K^{ba}-\gamma^{ab}K^c_{\ c}\Big]-\mathbf{N}\left[\peqsub{R^{(n)}}{\gamma}+\varepsilon K^{a}_{\ a}K^b_{\ b}-\varepsilon K^d_{\ b}K^b_{\ d}-2\lambda\left(1-\frac{\epsilon}{\mathbf{N}\,\sqrt{\gamma}}\right)\right]\right)=\\
  &\ =\int_\Sigma\peqsub{\mathrm{vol}}{\Sigma}\left(\Big[\mathbf{N}K_{ab}+(\mathcal{L}_{\vec{N}}\gamma)_{ab}\Big]\pi^{ab}-\sqrt{\gamma}\mathbf{N}\peqsub{R^{(n)}}{\gamma}+2\lambda\Big(\mathbf{N}\,\sqrt{\gamma}-\epsilon\Big)\right)
  \end{align*}
  Using the identification $F\!\peqsub{L}{\mathrm{uni}}(\mathcal{D})\cong\peqsub{\mathcal{P}}{\mathrm{uni}}$ we have that the Hamiltonian $\peqsub{H}{\mathrm{uni}}:\peqsub{\mathcal{P}}{\mathrm{uni}}\to\R$ is given by
  \begin{align*}
  \peqsub{H}{\mathrm{uni}}&\left(\lambda,\mathbf{N},\vec{N},\gamma,\pi\right)=\int_\Sigma\peqsub{\mathrm{vol}}{\Sigma}\left(\lambda\peqsub{\mathcal{H}}{\lambda}+N^a\mathcal{H}_a+\mathbf{N}\peqsubfino{\mathcal{H}}{\perp}{-0.2ex}\right)
  \end{align*}
  where we have defined
  \begin{align}
  &\mathcal{H}_c(\gamma,\pi)=-2\gamma_{bc}\nabla_a\pi^{ab}\label{GR - equation - H_tangente uni}\\
  &\peqsubfino{\mathcal{H}}{\perp}{-0.2ex}(\gamma,\pi)=-\frac{\varepsilon}{\sqrt{\gamma}}\left[\gamma_{ac}\gamma_{bd}-\frac{1}{n-1}\gamma_{ab}\gamma_{cd}\right]\pi^{ab}\pi^{cd}-\sqrt{\gamma}\peqsub{R^{(n)}}{\gamma}\label{GR - equation - H_perpendicular uni}\\
  &\peqsub{\mathcal{H}}{\lambda}(\mathbf{N},\gamma)=2\Big(\mathbf{N}\,\sqrt{\gamma}-\epsilon\Big)\label{GR - equation - H_lambda uni}
  \end{align}

     \subsection*{GNH algorithm}\trassub
  
  Let us now solve the equation $\peqsub{\imath}{Y}\omega^{\scriptscriptstyle\mathrm{uni}}=\mathrm{d}\peqsub{H}{\mathrm{uni}}$ to obtain the Hamiltonian vector field $Y\in\mathfrak{X}(\peqsub{\mathcal{P}}{\mathrm{uni}})$. We do so by using the GNH algorithm explained in section \ref{Mathematical background - section - GNH} of chapter \ref{Chapter - Mathematical background}.\separ
  
  $\bullet$ Compute the differential $\mathrm{d}\peqsub{H}{\mathrm{uni}}:T\peqsub{\mathcal{P}}{\mathrm{uni}}\to\R$ of $\peqsub{H}{\mathrm{uni}}$.\separprevia
  
  The differential of $\peqsub{H}{\mathrm{uni}}$ can be easily computed once we realize that
  \[\peqsub{H}{\mathrm{uni}}=\peqsub{H}{\mathrm{GR}}+\int_\Sigma\peqsub{\mathrm{vol}}{\Sigma}\,\lambda\peqsub{\mathcal{H}}{\lambda}\qquad\quad\text{and}\quad\qquad D\peqsub{\mathcal{H}}{\lambda}\overset{\eqref{Appendix - equation - variacion volumen}}{=}2\peqsub{Z}{\boldsymbol{\mathbf{N}}}\sqrt{\gamma}+\mathbf{N}\sqrt{\gamma}\,\mathrm{Tr}(\peqsub{Z}{\boldsymbol{\gamma}})\]
  with $\Lambda=0$. Now using \eqref{GR - equation - dH} and the previous equations we have
  \begin{align*}
  &\mathrm{d}_{(\lambda,\mathbf{N},\vec{N},\gamma,\pi)}\peqsub{H}{\mathrm{uni}}\Big(\peqsub{Z}{\boldsymbol{\lambda}},\peqsub{Z}{\boldsymbol{\mathbf{N}}},\peqsub{Z}{\boldsymbol{\vec{N}}},\peqsub{Z}{\boldsymbol{\gamma}},\peqsub{Z}{\boldsymbol{\pi}}\Big)=\int_\Sigma\peqsub{\mathrm{vol}}{\Sigma}\,\Bigg\{\peqsub{Z}{\boldsymbol{\lambda}}\peqsub{\mathcal{H}}{\lambda}+\peqsub{Z}{\boldsymbol{\vec{N}}}^a\mathcal{H}_a+\peqsub{Z}{\boldsymbol{\mathbf{N}}}\Big(\peqsubfino{\mathcal{H}}{\perp}{-0.2ex}+2\lambda\sqrt{\gamma}\Big)+\\
  &\quad+(\peqsub{Z}{\boldsymbol{\pi}})^{ab}\Big[(\mathcal{L}_{\vec{N}}\gamma)_{ab}+2\mathbf{N}K_{ab}\Big]-(\peqsub{Z}{\boldsymbol{\gamma}})_{ab}\Big[(\mathcal{L}_{\vec{N}}\pi)^{ab}-\mathbf{N}\peqsub{\beta}{\mathrm{uni}}^{ab}+\sqrt{\gamma}\Big(\nabla^a\nabla^b-\gamma^{ab}\nabla_d\nabla^d\Big)\mathbf{N}\Big]\Bigg\}
  \end{align*}
  where
  \[\peqsub{\beta}{\mathrm{uni}}^{ab}=\sqrt{\gamma}R^{ac}-\frac{2\varepsilon}{\sqrt{\gamma}}\left(\pi^{ad}\pi^c_{\ d}-\frac{\pi}{n-1}\pi^{ac}\right)+\frac{\varepsilon}{\sqrt{\gamma}}\left(\pi_{bd}\pi^{bd}-\frac{\pi^2}{n-1}\right)\gamma^{ac}+\frac{\gamma^{ac}}{2}\Big(\peqsubfino{\mathcal{H}}{\perp}{-0.2ex}+2\lambda\sqrt{\gamma}\Big)\]

  $\bullet$ Compute the Hamiltonian vector field $Y\in\mathfrak{X}(\peqsub{\mathcal{P}}{\mathrm{uni}})$.\separprevia
  
  We now obtain the equations of $Y$ by solving, for every $Z$, the equation $\omega^{\scriptscriptstyle\mathrm{uni}}(Y,Z)=\mathrm{d}\peqsub{H}{\mathrm{uni}}(Z)$.
  \begin{align*}
  &\int_\Sigma\left[(\peqsub{Z}{\boldsymbol{\pi}})^{ab}(\peqsub{Y}{\boldsymbol{\gamma}})_{ab}-(\peqsub{Y}{\boldsymbol{\pi}})^{ab}(\peqsub{Z}{\boldsymbol{\gamma}})_{ab}\right]\peqsub{\mathrm{vol}}{\Sigma}=\int_\Sigma\peqsub{\mathrm{vol}}{\Sigma}\,\Bigg\{\peqsub{Z}{\boldsymbol{\lambda}}\peqsub{\mathcal{H}}{\lambda}+\peqsub{Z}{\boldsymbol{\vec{N}}}^a\mathcal{H}_a+\peqsub{Z}{\boldsymbol{\mathbf{N}}}\Big(\peqsubfino{\mathcal{H}}{\perp}{-0.2ex}+2\lambda\sqrt{\gamma}\Big)+\\
  &\quad+(\peqsub{Z}{\boldsymbol{\pi}})^{ab}\Big[(\mathcal{L}_{\vec{N}}\gamma)_{ab}+2\mathbf{N}K_{ab}\Big]-(\peqsub{Z}{\boldsymbol{\gamma}})_{ab}\Big[(\mathcal{L}_{\vec{N}}\pi)^{ab}-\mathbf{N}\peqsub{\beta}{\mathrm{uni}}^{ab}+\sqrt{\gamma}\Big(\nabla^a\nabla^b-\gamma^{ab}\nabla_d\nabla^d\Big)\mathbf{N}\Big]\Bigg\}
  \end{align*}
  We obtain directly, using \eqref{GR - equation - pi=k... K=pi... uni} to write everything in terms of $\pi$, that
  \begin{align}\label{GR - equation - Hamiltonian equations pi uni}
  \begin{split}
  &\peqsubfino{Y}{\!\boldsymbol{\lambda}}{-0.1ex},\peqsubfino{Y}{\!\boldsymbol{\mathbf{N}}}{-0.1ex},\peqsubfino{Y}{\!\boldsymbol{\vec{N}}}{-0.1ex}\quad\text{ arbitrary}\\
  &(\peqsub{Y}{\boldsymbol{\gamma}})_{ab}=(\mathcal{L}_{\vec{N}}\gamma)_{ab}-2\varepsilon\frac{\mathbf{N}}{\sqrt{\gamma}}\left[\pi_{ab}-\frac{\pi}{n-1}\gamma_{ab}\right]=(\mathcal{L}_{\vec{N}}\gamma)_{ab}+2\mathbf{N}K_{ab}\\
  &(\peqsub{Y}{\boldsymbol{\pi}})^{ab}=(\mathcal{L}_{\vec{N}}\pi)^{ab}-\mathbf{N}\peqsub{\beta}{\mathrm{uni}}^{ab}+\sqrt{\gamma}\Big(\nabla^a\nabla^b-\gamma^{ab}\nabla_d\nabla^d\Big)\mathbf{N}\\%
  &\hspace*{-1.2ex}\left.\begin{array}{l}%
  \peqsub{\mathcal{H}}{\lambda}=0\\[0.8ex]
  \peqsubfino{\mathcal{H}}{\perp}{-0.2ex}+2\lambda\sqrt{\gamma}=0\\[0.8ex]
  \mathcal{H}_a=0_a
  \end{array}\right\}\text{ Constraints}
  \end{split}
  \end{align}
  We have to look for solutions
  over $\peqsub{\mathcal{P}}{\mathrm{uni}}^2:=\{\peqsub{\mathcal{H}}{\lambda}=0\,,\ \peqsubfino{\mathcal{H}}{\perp}{-0.2ex}+2\lambda\sqrt{\gamma}=0\,, \ \mathcal{H}_a=0_a\}\subset\peqsub{\mathcal{P}}{\mathrm{uni}}$.\separpost

  $\bullet$ Require $\peqsub{Y}{\!H}$ to be tangent to $\peqsub{\mathcal{P}}{\mathrm{uni}}^2$.\separprevia
  
  We have now to determine the space $\peqsub{\mathcal{P}}{\mathrm{uni}}^3\subset\peqsub{\mathcal{P}}{\mathrm{uni}}^2$ where $Y$ is tangent to $\peqsub{\mathcal{P}}{\mathrm{uni}}^2$. This is equivalent to require that the dynamics preserves the constraints. In lemma \ref{Appendix - lemma - DH DH_perp Dlambda uni} we prove that the variations of the constraints are given by
	\begin{align*}
	&\blacktriangleright\ (D\mathcal{H})_a=\big(\peqsub{\mathcal{H}}{\perp}+2\lambda\sqrt{\gamma}\big)\nabla_a\mathbf{N}+\peqsub{\mathcal{H}}{\lambda}\nabla_a\lambda+2\epsilon\nabla_a\lambda\\[1ex]
	&\blacktriangleright\ D\Big[\peqsub{\mathcal{H}}{\perp}+2\lambda\sqrt{\gamma}\Big]=\mathcal{L}_{\vec{N}}\Big[\peqsub{\mathcal{H}}{\perp}+2\lambda\sqrt{\gamma}\Big]+\varepsilon\mathbf{N}\nabla^a\mathcal{H}_a+2\varepsilon\mathcal{H}_a\nabla^a\mathbf{N}+2\sqrt{\gamma}\Big(\peqsub{Y}{\boldsymbol{\lambda}}-\mathcal{L}_{\vec{N}}\lambda\Big)\\
	&\blacktriangleright\ D\peqsub{\mathcal{H}}{\lambda}=\mathcal{L}_{\vec{N}}\peqsub{\mathcal{H}}{\lambda}+2\sqrt{\gamma}\left[\peqsub{Y}{\boldsymbol{\mathbf{N}}}-\mathcal{L}_{\vec{N}}\mathbf{N}+\frac{\varepsilon\pi}{(n-1)\sqrt{\gamma}}\mathbf{N}^2\right]
\end{align*}
  Thus, over $\peqsub{\mathcal{P}}{\mathrm{uni}}^2$, we obtain
  \begin{align}\label{GR - equation - Hamiltonian equations pi uni 2}
\begin{split}
&\peqsubfino{Y}{\!\boldsymbol{\vec{N}}}{-0.1ex}\quad\text{ arbitrary}\\
&\peqsubfino{Y}{\!\boldsymbol{\lambda}}{-0.1ex}=\mathcal{L}_{\vec{N}}\lambda\\
&\peqsubfino{Y}{\!\boldsymbol{\mathbf{N}}}{-0.1ex}=\mathcal{L}_{\vec{N}}\mathbf{N}-\frac{\varepsilon\pi}{(n-1)\sqrt{\gamma}}\mathbf{N}^2=\mathcal{L}_{\vec{N}}\mathbf{N}-K\mathbf{N}^2\\
&(\peqsub{Y}{\boldsymbol{\gamma}})_{ab}=(\mathcal{L}_{\vec{N}}\gamma)_{ab}-2\varepsilon\frac{\mathbf{N}}{\sqrt{\gamma}}\left[\pi_{ab}-\frac{\pi}{n-1}\gamma_{ab}\right]=(\mathcal{L}_{\vec{N}}\gamma)_{ab}+2\mathbf{N}K_{ab}\\
&(\peqsub{Y}{\boldsymbol{\pi}})^{ab}=(\mathcal{L}_{\vec{N}}\pi)^{ab}-\mathbf{N}\peqsub{\beta}{\mathrm{uni}}^{ab}+\sqrt{\gamma}\Big(\nabla^a\nabla^b-\gamma^{ab}\nabla_d\nabla^d\Big)\mathbf{N}\\%
&\hspace*{-1.2ex}\left.\begin{array}{l}%
\peqsub{\mathcal{H}}{\lambda}=0\\[0.8ex]
\peqsubfino{\mathcal{H}}{\perp}{-0.2ex}+2\lambda\sqrt{\gamma}=0\\[0.8ex]
\mathcal{H}_a=0_a\\[0.8ex]
(\mathrm{d}\lambda)_a=0_a
\end{array}\right\}\text{ Constraints}
\end{split}
\end{align}
So we have $\peqsub{\mathcal{P}}{\mathrm{uni}}^3:=\{\peqsub{\mathcal{H}}{\lambda}=0\,,\ \peqsubfino{\mathcal{H}}{\perp}{-0.2ex}+2\lambda\sqrt{\gamma}=0\,, \ \mathcal{H}_a=0_a\,, \ (\mathrm{d}\lambda)_a=0_a\}\subset\peqsub{\mathcal{P}}{\mathrm{uni}}^2$.\separpost

$\bullet$ Require $\peqsub{Y}{\!H}$ to be tangent to $\peqsub{\mathcal{P}}{\mathrm{uni}}^3$.\separprevia

  Let us compute the variation of $\mathrm{d}\lambda$.
  \begin{align*}
    D(\mathrm{d}\lambda)&=\mathrm{d}(D\lambda)=\mathrm{d}\peqsub{Y}{\boldsymbol{\lambda}}=\\
    &=\mathrm{d}(\mathcal{L}_{\vec{N}}\lambda)\overset{\eqref{appendix - formula - L=di+id}}{=}\mathrm{d}(\imath_{\vec{N}}\mathrm{d}+\mathrm{d}\imath_{\vec{N}})\lambda\updown{\eqref{appendix - formula - L=di+id}}{\eqref{appendix - equation - d^2=0}}{=}\\
    &=(\mathcal{L}_{\vec{N}}-\imath_{\vec{N}}\mathrm{d})(\mathrm{d}\lambda)+0\overset{\eqref{appendix - equation - d^2=0}}{=}\mathcal{L}_{\vec{N}}(\mathrm{d}\lambda)
  \end{align*}
  The vector field is tangent to $\peqsub{\mathcal{P}}{\mathrm{uni}}^3$ and, therefore, the algorithm stops.
  
  \subsection*{Interpretation}\trassub
  
  We end up this section with some useful remarks that will help us to understand the previous results. For further details see, for instance, \cite{bufalo2015unimodular}.
  
  \begin{itemize}
  	\item $\lambda$ is constant, as expected from equation \eqref{GR - equation - lambda=cte}, up to a gauge transformation.
  	\item The evolution of $\mathbf{N}$ is not arbitrary anymore as it is restricted by $\peqsub{Y}{\boldsymbol{\mathbf{N}}}$. This condition comes from the requirement that $\peqsub{\mathcal{H}}{\lambda}$ is preserved (see lemma \ref{Appendix - lemma - DH DH_perp Dlambda uni}) which is equivalent to require that $\varepsilon\det(g)=\mathbf{N}^2\det(\gamma)$ is constant.
   \item We can remove $\lambda$ from our theory using the constraint $\peqsubfino{\mathcal{H}}{\perp}{-0.2ex}+2\lambda\sqrt{\gamma}=0$. Once we do that, we recover the equations of GR for $\gamma$ and $\pi$ (notice that $\peqsub{\beta}{\mathrm{uni}}$ is then equal to $\beta$ over $\peqsub{\mathcal{P}}{\mathrm{GR}}^2$).
   \item Unimodular gravity is ``contained'' in standard GR but they are not directly equivalent. We can on one hand drop $\peqsub{Y}{\boldsymbol{\mathbf{N}}}$ from the unimodular version in order to recover the full diffeomorphism group or, on the other hand, we can make a partial gauge fixation in the standard GR version to pass to the unimodular version.
  \end{itemize}

  

%% file: 9_conclusions.tex

\chapter{Conclusions}\thispagestyle{empty}

  \section{Summary}
  \starwarssection[Han Solo]{Great, kid! Don't get cocky.}{A New Hope}
  
  \vspace*{1ex}
  
  Let us begin this last chapter recalling the \emph{motto} we mentioned at the beginning of this thesis
  \begin{quote}\centering%
  	\begin{minipage}{.72\linewidth}\centering
  		\emph{Boundaries, GNH, and parametrized theories.\\It takes three to tango.}
  	\end{minipage}
  \end{quote}
  Those three elements have been the motivation and the vertebral axis of this thesis. In chapter \ref{Chapter - Mathematical background} we presented a review of some of the most important geometrical notions used throughout the thesis. Of capital importance was the description of the space of embeddings given in section \ref{Mathematical background - Section - Space of embeddings}, as well as a discussion of the meaning of the GNH algorithm that we outlined in section \ref{Mathematical background - section - GNH}.\separ
  
  In chapter \ref{Chapter - Scalar fields coupled to point masses}, based on our papers \cite{margalef2015quantization,margalef2017functional,margalefboundary}, we began to think about boundaries and how they can be used to measure what happens in the bulk. To do so we studied what seemed to be a very simple case: a string with a mass attached to each end. This system is, however, surprisingly rich and hard to deal with. The inclusion of the masses prevents us from using the standard Sturm-Liouville theory which suggests that novel ideas are needed. In fact, we circumvented this problem by relying on non-trivial measures and their associated Radon-Nikodym derivatives. The main result of this chapter is that the Fock space of the whole system, that we obtained thanks to the GNH algorithm, is not of the form $\mathcal{H}_{\mathrm{syst}}\otimes\mathcal{H}_{\mathrm{mes}}$ with a factor associated with the bulk and a factor associated with the boundary (measure device at the boundary). This suggests some sort of strong entanglement of the boundary and the bulk, which should come as no surprise after a short reflection because at the classical level the positions of the masses are completely determined by the configuration of the string (continuity conditions). Besides, we show that we can define some dynamics over the boundary with the help of the trace operators. Such dynamics is not unitary, which is to be expected in the same way that the energy is conserved in the whole system but not on the subsystems. We end up this section studying the unitary implementation of the scalar field with standard boundary conditions obtaining a complete characterization of the possible unitary evolutions through foliations. This result generalizes the ones known for boundaryless systems.\separ
  
  The following chapter is devoted to motivate and describe in detail the parametrization procedure. The main idea is to include diffeomorphisms as variables in such a way that the resulting theory is diff-invariant. The importance of this procedure has been pinpointed out several times, but it is worth it to make it explicit once again. First, let us mention that it is a generalization of the unparametrized theory which allows us to describe any field theory in an arbitrary foliation. Second, it is a perfect generator of toy-models for GR. Finally, some algebraic quantization methods can be applied in this simpler context with the hope to learn something about the quantization of the full GR. After the description of the parametrized theories the chapter continues with the warm-up exercise of studying parametrized classical mechanics. We carefully develop the ideas and the implementation of the GNH algorithm, outlining the steps where due care is needed in the infinite dimensional case.\separ
  
  With our hand and mind loosened up with the preparatory exercise of chapter \ref{Chapter - Parametrized theories}, we are ready to jump into the parametrized electromagnetic field with boundaries. Chapter \ref{Chapter - Parametrized EM} is devoted to its study, which by the way generalizes our work \cite{margalef2016hamiltonianEM} as here we include boundaries. The most important result of this chapter is the identification of sectors, closely related to the Gauss law of electromagnetism, where a bifurcation appears in the dynamics. The same bifurcation is identified at the boundary although, in this case, its careful study is much trickier and not so interesting for our purposes. Nonetheless, a discussion of similar nature can be carried out in a simpler example: the parametrized scalar field with boundaries. We do so in chapter \ref{Chapter - parametrized scalar}, and also in \cite{margalef2016hamiltonian}, where we obtain a complete characterization of the sectors associated with the behavior of the boundaries. The consistency of the dynamics forces some additional conditions that can be explicitly derived for some concrete examples.\separ
  
  The next natural step is to study some theories with interesting and non-standard behaviors at the boundary. Chapter \ref{Chapter - parametrized MCS} is devoted to the study of the parametrized Maxwell-Chern-Simons theory, a generalization of parametrized EM in $2+1$ dimensions. We develop, using the same methods as in the previous chapters, the Hamiltonian formulation of the theory and identified the different boundary conditions that naturally appear. This is important in order to understand some features that are supposed to play a role in the quantum description of those theories. We end up the chapter with an analog study of Chern-Simons theory.\separ
  
  Chapter \ref{Chapter - General relativity}, the last one, is devoted to the general theory of gravitation. We derive, using the full machinery developed in this thesis, the Hamiltonian formulation of GR. It obviously coincides with the well known ADM formulation but our novel approach through the GNH algorithm is simpler to use and easier to interpret. After that, we proceed to study the Hamiltonian formulation of unimodular gravity, which is much less known but useful in several contexts because it provides an interesting perspective on the problem of time and the origin of the cosmological constant.\separ
  
  Finally, in the appendix we include two sections with a brief introduction to functional analysis and measure theory. After that, we list geometric formulas that are extensively used throughout this thesis. Finally, in the last section we provide some computational details.

  \section{Future work}
  \starwarssection[Darth Vader]{Join me, and together we can rule the galaxy as father and son.}{The Empire Strikes Back}
  
  \vspace*{-2ex}
  
  \subsection*{Boundaries}\trassub
  
  We mentioned in the introduction how boundaries were one of the unifying threads of this thesis. We have, indeed, discussed several physical models where they play a fundamental role. However, there is still much to be understood. For instance, despite the diff-invariance of the action $\peqsub{S}{\mathrm{d}}$ of a parametrized theory, its variation $\peqsub{D}{(Z,\mathbb{V}_{\!Z})}\peqsub{S}{\mathrm{d}}$ with respect to diffeomorphisms is non-zero in general. This is due to the fact that the diffeomorphisms themselves are dynamical variables of the action! It is thus natural to wonder if there exists a physical interpretation of the additional condition \eqref{Parametrized EM - equation - additional condition} or if it can be related to the fact that some degrees of freedom live at the boundary. An analogous condition appears in the Maxwell-Chern-Simons case, which can shed some light over some condensed matter problems.\separ
  
  The number of open questions at the quantum level is even larger than at the classical one. For instance, as we saw in chapter \ref{Chapter - Scalar fields coupled to point masses}, it is not always possible to associate Hilbert spaces  to boundaries. This problem can be of great importance in quantum gravity (black hole entropy, holographic models\ldots).
  
  \subsection*{More strings}\trassub
  
  A natural generalization of the models mixing fields and point particles that we presented in chapter \ref{Chapter - Scalar fields coupled to point masses} is to consider relativistic versions of them. A natural requirement would be to demand that the space-time trajectories of the points of the string are time-like curves and, hence, can be taken as the world lines of physical point particles. Notice that this is not what is done in standard string theory!\separ
  
  An alternative way to look at the restriction that we want to consider is to think about it as the causality condition employed in the Causal Dynamical Triangulations approach to quantum gravity. A natural way to attack the problem of writing an action principle for a causal string of length $\ell$ is to introduce appropriate geometric objects, in this case embeddings from the interval $[0,\ell]$ to the space-time where the string moves. The fact that the embeddings are geometric objects with distinctive features renders the problem of writing appropriate relativistic actions quite a non-trivial one. The goal is to build the most useful relativistic dynamics for these models, study the resulting field equations and consider possible extensions where the strings are coupled to point particle objects (while respecting relativistic invariance), all of this without losing sight of the final goal: quantize the resulting models.

  \subsection*{Unimodular gravity}\trassub
  
  We have learned throughout this work that the parametrization of a theory enriches it in several ways and helps to understand it better. Nonetheless, as we saw in chapter \ref{Chapter - General relativity}, the theory of general relativity has no background geometric objects and, therefore, its parametrization is trivial and uninteresting. One alternative idea might be to consider the DeWitt metric defined over the space of metrics. However, it is not a background object of the space-time and, thus, it is hard to give a geometrical interpretation to what it means to parametrize it. Another possible approach would be to consider unimodular gravity, which is defined over a space-time $M$ with a fixed (background) volume form $\mathrm{vol}_0$.\separ 
  
  In section \ref{GR - section - unimodular} of chapter \ref{Chapter - General relativity} we introduce and study unimodular gravity, which is as the general theory of relativity but with the addition of a constraint forcing the metric volume form to coincide with a fixed one $\mathrm{vol}_g=\mathrm{vol}_0$. It is well known that the field equations for such modification of general relativity are physically equivalent to the standard Einstein equations. The only significant difference is the fact that the cosmological constant becomes an integration constant of the theory, in the sense that its constancy can be seen as a consequence of the unimodular field equations. The Hamiltonian formulation of this model was developed and understood by Henneaux and Teitelboim \cite{henneaux1989cosmological} and since then has been revisited quite often \cite{smolin2009quantization}. I have studied this problem, with the help of the GNH algorithm, recovering the known result about its Hamiltonian description. The question that arises then is: could the parametrized version shed some light on some of the unsolved riddles of the general relativity? This question was considered by Kucha\v{r} \cite{kuchavr1991does} and my plan is to extend his analysis and consider the different action principles proposed for this model in \cite{henneaux1989cosmological}.

  

%% file: A_more_mathematics.tex
\begin{appendix}
	
	\renewcommand{\barrita}{- }
\  \chapter{Ancillary mathematical material}\label{appendix}\thispagestyle{empty}
   
    \section{Functional analysis}\label{Appendix - section - functional analysis}
    
    \starwarssection[Obi-Wan Kenobi]{An elegant weapon for a more civilized age.}{A New Hope}
    
    \subsection*{Historical introduction}\trassub
    
    At the beginning of the 20th century the world of mathematics was moving towards abstraction and axiomatization. Functional analysis, which has its origins in the study of the ODEs, PDEs, and integral equations, started to become a discipline of its own with prominent names associated to this early period: Fredholm, Lebesgue, Fréchet, Riesz, and Hilbert among others. Nonetheless, it was probably not until the 20's when it was finally elevated to a full new branch of mathematics thanks to the works of Banach, Hahn, Steinhaus, or Schauder.\separ
    
    Although historically it was developed more or less at the same time as linear algebra, nowadays functional analysis can be considered as an infinite dimensional version of linear algebra dealing, for instance, with functions instead of vectors from a finite-dimensional space. Although most of our intuitive concepts of linear algebra are still valid, there are important differences that make functional analysis both hard and interesting, for instance:

\begin{itemize}
	\item  There is a notion of length in these infinite dimensional spaces called \textbf{norm} but, unlike the case of $\R^n$ where all of them are equivalent (in the sense that they define the same topology), here we have to be careful because in general they are not equivalent.
	\item  A linear operator cannot always be represented by an infinite dimensional matrix.
	\item We still have the notion of an eigenvalue $\lambda$ and eigenvector $v$ of an operator $T$ given by the condition $(T-\lambda\mathrm{Id})v=0$. However, it turns that $T-\lambda\mathrm{Id}$ is much more interesting than in the finite dimensional case, notice for example that injectivity and surjectivity are not equivalent anymore.
	\item Many important operators, several of them defined in the context of quantum mechanics, are not continuous.
\end{itemize}

\subsubsection{Basic notions}\trassub
 
 We assume that the reader is familiar with the concept of norm, scalar product, orthogonality, Cauchy sequence, and convergent sequence.
 
 \begin{definitions}
 	\item We say that the normed $\C$-vector space $(\mathcal{B},\|\cdot{}\|)$ is a \textbf{Banach space}\index{Banach space} if it is complete with respect to its norm i.e.\ every Cauchy sequence of $\mathcal{B}$ is convergent in $\mathcal{B}$.
 	\item A \textbf{pre-Hilbert space}\index{Hilbert space!Pre-Hilbert space} is a $\C$-vector space $\mathcal{H}$ endowed with a sesquilinear inner product $(\,,)$.
 	\item Given a  pre-Hilbert space $(\mathcal{H},(\,,))$ we say that it is a \textbf{Hilbert space} if the norm $\|v\|:=(v,v)$ is complete.
 	\item Let $(\mathcal{H},(\,,))$ be a Hilbert space, we say that a set of elements $\{v_a\}_{a\in A}\subset\mathcal{H}$ is a \textbf{complete orthonormal basis}\index{Complete orthonormal basis} if they are orthonormal and every element $w\in\mathcal{H}$ can be written as\label{Appendix - definition - complete}
 	\[w=\sum_{a\in A}\alpha_av_a\]
 	A Hilbert space that admits a countable complete orthonormal basis is said to be \textbf{separable}\index{Hilbert space!separable}.
 \end{definitions}

      \begin{wrapfigure}{r}{0.28\textwidth}
	\vspace{-4ex}
	\centering\includegraphics[width=1.05\linewidth]{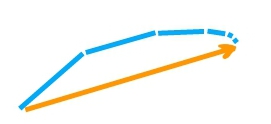}
	\vspace*{-10ex}
\end{wrapfigure}   
The completeness of a space can be understood with the image on the right. If a particle moves along an infinitely broken path but the total amount of distance traveled is finite, then it has a well-defined net displacement.\separ

The most typical examples of Hilbert spaces are the space $\ell^2(\N)$ of sequences and the space $L^2[a,b]$ of square integrable functions with their respective scalar product 
\[(x,y)=\sum_{i=1}^\infty \overline{x_n}y_n\qquad\qquad\qquad \langle f,g\rangle=\int_{[a,b]}\overline{f(x)}g(x)\mathrm{d}x\]
Given a Banach space $\mathcal{B}$ we define the \textbf{dual space}\index{Banach space!dual space}\index{Hilbert space!dual space} $\mathcal{B}'$ as the set of continuous functions from $\mathcal{B}$ to $\R$ (or $\C$). The Hahn-Banach theorems says, in one of its many versions \cite{brezis2010functional} that $\mathcal{B}'$ is not trivial.
\begin{theorem}\mbox{}\\
	Let $v\in\mathcal{B}$ be an element of a Banach space, then there exists an $f\in\mathcal{B}'$ such that $f(v)\neq0$.
\end{theorem}
This theorem has an immediate corollary that we will use several times when solving the GNH algorithm.

\begin{corollary}\label{Appendix - equation - Hahn-Banach theorem}\mbox{}\\
	Let $\mathfrak{B}$ a Banach space and $v\in\mathfrak{B}$. If $f(v)=0$ for every $f\in\mathfrak{B}'$ then $v=0$.
\end{corollary}
For a Hilbert space it is easy to define several elements of the dual, namely $T_u:\mathcal{H}\to\mathcal{H}$ given by $T_u(v)=(u,v)$. In fact, the Riesz representation theorem says that every element of the dual can be realized in that way.
\begin{theorem}\mbox{}\\
	Given a Hilbert space $(\mathcal{H},\langle\,,\rangle)$ and $f\in\mathcal{H}'$, there exists a unique $u\in\mathcal{H}$ such that $f(\cdot{})=\langle\,\cdot\,,u\rangle$.
\end{theorem}

\subsubsection{Adjoint operators}\trassub

Given a continuous operator $A:\mathcal{H}\to\mathcal{H}$ we define the \textbf{adjoint operator} as the unique operator $A^*:\mathcal{H}\to\mathcal{H}$ such that
\[\langle Av_1,v_2\rangle=\langle v_1,A^*v_2\rangle\qquad\qquad v_1,v_2\in\mathcal{H}\]
$A$ is said to be \textbf{self-adjoint} if $A=A^*$. We denote $\mathrm{SAO}(\mathcal{H})$ the set of all self-adjoint operators of $A$. It is well known that its spectrum (the set of eigenvalues of $A$) is real.\separ

We can define the adjoint of a densely defined unbounded operator $A:D(A)\subset\mathcal{H}\to\mathcal{H}$ (where $D(A)$ is a dense linear subspace). By definition, the domain of $A^*$ is
\[D(A^*):=\Big\{y\in\mathcal{H}\ \ /\ \ \exists z\in\mathcal{H} \text{ with }\langle Ax,y\rangle=\langle x,z\rangle \text{ for every }x\in\mathcal{H}\Big\}\]
     we define thus $A^*(y):=z$.

     \subsubsection{Sturm-Liouville theory}\trassub
        
        Let us now state the important Sturm-Liouville theorem \cite{al2008sturm}. For that consider the space
        \[\mathcal{H}:=\left\{f\in L^2[a,b]\ \ /\ \ \begin{array}{l}
        \alpha u(a)+\beta u'(a)=0\\
        \gamma u(b)+\delta u'(b)=0\end{array}\right\} \]
        for some real constants such that $(\alpha,\beta)\neq(0,0)\neq(\gamma,\delta)$. We define, for some smooth real  functions $p,q:[a,b]\to\R$ such that $p>0$, the operator $L:\mathcal{H}\to\mathcal{H}$ given by
        \[Lu=-\deriv{}{x}\!\left(p(x)\deriv{u}{x}\right)+q(x)u(x)\]
        
        \begin{theorem}\label{Appendix - theorem - sturm liouville}\mbox{}\\
        	 $L$ is a self-adjoint operator on $(\mathcal{H},\langle\,,\rangle)$. In particular its spectrum $\{\lambda_n\}$ is formed by non-degenerate and real eigenvalues such that $\lambda_n\to\infty$. Furthermore, the eigenvectors form a complete orthogonal basis.
        \end{theorem}

    \section{Measure theory}\label{Appendix - Section - Measure theory}
    
    \starwarssection[Han Solo]{She may not look like much, but she’s got it where it counts, kid.}{A New Hope}
		
		\subsection*{Historical introduction}\trassub
		
		Determining the size of an object ---like the length of a circle, the area of a parallelepiped, the volume of a cone, or the mass of a sphere--- has been one of the fundamental questions of classical geometry. Two were the main strategies to handle this problem. The first one was to break the object into pieces, moving around them to form a simpler object, which presumably has the same size. The second strategy was to obtain bounds by inscribing and circumscribing our object into another geometric objects whose size is known. This idea was used extensively by Archimedes to approximate the area of many bodies like the circle. Both methods rely of the prior existence of a measure of the size of the bodies as well as the possibility to rearrange such bodies preserving the size. However, with the development of analytic geometry such notions and intuitions become obsolete. For instance, the interval $[0,1]$ of length $1$ can be mapped bijectively into the interval $[0,2]$ which has length $2$, suggesting that the strategy of breaking into pieces and reassembling must be refined in some sense. This kind of problem were partially solved by the introduction of differential calculus by Leibniz and Newton. Nonetheless, there is a more dramatic example known as the Banach-Tarski paradox \cite{banach1924decomposition}, a theorem that states that a unit ball can be broken into five pieces in such a way that, after a proper rearrangement, we obtain two disjoint unit balls.
		
		\subsubsection{Measure}\trassub
		
		The first attempt to go beyond the initial ideas developed by Leibniz and Newton was due to Borel in 1889 by studying functions whose domain consists of subsets of $\R^n$. He introduced two important definitions. The first one was a \textbf{$\boldsymbol{\sigma}$-algebra} $\Sigma$ of a set $X$ which is a subset $\Sigma\subset \mathcal{P}(X)$ satisfying 
		\begin{itemize}
			\item $X\in\Sigma$
			\item If $A\in\Sigma$ then $X\setminus A\in\Sigma$.
			\item If $A_n\in\Sigma$ for every $n\in\N$ then $\displaystyle\bigcup_{n\in\N}A_n\in\Sigma$.
		\end{itemize}
		The pair $(X,\Sigma)$ is known as \textbf{measurable space}. The other important concept that he introduced was the notion of \textbf{$\boldsymbol{\sigma}$-additivity}: assign to a disjoint union of sets a value that coincides with the sum of the values that assigns to each individual set
		\[m\left(\bigsqcup_{i\in\N}B_i\right)=\sum_{i\in\N}m(B_i)\]
		His ideas were later simplified and extended by Lebesgue \cite{lebesgue1902integrale} in his doctoral thesis in 1902, where he also developed the theory now known as Lebesgue integration and differentiation. In order to define the \textbf{Lebesgue measure}\index{Measure!Lebesgue measure} $\peqsub{\mu}{L}$ assigning to a set $A$ a measure $\peqsub{\mu}{L}(A)\in[-\infty,+\infty]$, he started considering the interval $[0,1]$ and assumed that the open intervals $(a,b)\subset[0,1]$ have measure $\peqsub{\mu}{L}(a,b)=b-a$. As any open set $U$ is, by definition of a topology, the union of a disjoint sequence of intervals $I_n$ (that we know how to measure), we define
		\[\peqsub{\mu}{L}(U):=\sum_{n\in\N}\peqsub{\mu}{L}(I_n)\qquad\qquad \peqsub{\mu}{L}(\varnothing ):=0\]
		Now a closed set $C$ is the complement of an open set $U$ so we define $\peqsub{\mu}{L}(C):=1-\peqsub{\mu}{L}(U)$. Finally, he introduced the outer/inner measure of a set $C$ as the infimum/supremum of the measures of open sets containing/contained in $C$. If both values are equal, then $C$ is said to be \textbf{measurable}\index{Measurable set} with measure this common value. The question that might arises now is if all sets are measurable\ldots\, which is not the case as Vitali proved in 1905! He constructed a set (using the axiom of choice) which was not measurable \cite{vitali1905sul}. It is interesting to note that the set of measurable sets form a $\sigma$-algebra and that $\peqsub{\mu}{L}$ is $\sigma$-additive.\separ
		
		From the properties of the Lebesgue measure we can obtain the definition of a general measure.
		
		\begin{definition}\mbox{}\\
			Given a $\sigma$-algebra $\Sigma$ over a set $X$, a function $\mu:\Sigma\to(-\infty,\infty]$ is a \textbf{measure}\index{Measure} if it is is $\sigma$-additive and $\mu(\varnothing)=0$.
		\end{definition}
		
		Given some $x\in X$ we can define a very important measure $\delta_x$, known as the \textbf{Dirac measure}, which is given by
		\begin{equation}
		  \delta_x(A)=\left\{\begin{array}{ll}
		  1 &\text{if }x\in A\\[1ex]
		  0 &\text{if }x\notin A
		  \end{array}\right.
		\end{equation}
		Given a well behaved \cite{royden1968real} function $F:[0,1]\to\R$, we can define another important measure $\peqsub{\nu}{F}$ called \textbf{Lebesgue-Stieltjes measure}. It generalizes the Lebesgue measure by defining
		\begin{equation}\label{Appendix - equation - LS-measure}
			\peqsub{\nu}{F}[a,b)=F(b)-F(a)
		\end{equation}
		 instead of just $b-a$. Notice that if $F=\mathrm{Id}$ we recover the Lebesgue measure.

		\subsubsection{Integration}\trassub
		
		Once we have a measure $m$ over $[0,1]$ it is easy to define the integral of $f:[0,1]\to\R$ as
		\[\int_{[0,1]}f\,\mathrm{d}\mu:=\mu\left(A^+_f\right)-\mu\left(A^-_f\right)\]
		where $A^+_f$ is the region under the graph where $f$ is positive and $A^-_f$ the region over the graph where it is negative. We use the notation $\mathrm{d}\mu$ to stress that it is some sort of infinitesimal version of the measure $\mu$, in particular we have
		\begin{equation}\label{Appendix - equation - m(A)=int A}
		  \mu(A)=\int_A\mathrm{d}\mu
		\end{equation} Of course, as not all sets are measurable, not all maps can be integrated; the function needs to be \text{measurable}\index{Measurable function} i.e.\ for any measurable set $U$ the preimage $f^{-1}(U)$ is measurable. This definition of integration, known as \textbf{Lebesgue integral}, generalizes the Riemann integral and avoids some if its limitations (specially in the interchange of integrals and limits). Of course it satisfies all the natural properties like linearity and monotonicity.

      \subsection*{Radon-Nikodym derivative}
      
      Imagine now that we have two measures: $\alpha$ which measures the area of a land and $\rho$ which measures its price on the market. It seems natural to think that they are somehow related because if we take an area and we increase it, its price will also increase. Nonetheless, different regions of equal area might have different prices depending on several factors like the number of gold mines or the crops that grow on it. The following image
      
      \centerline{\includegraphics[width=.5\linewidth]{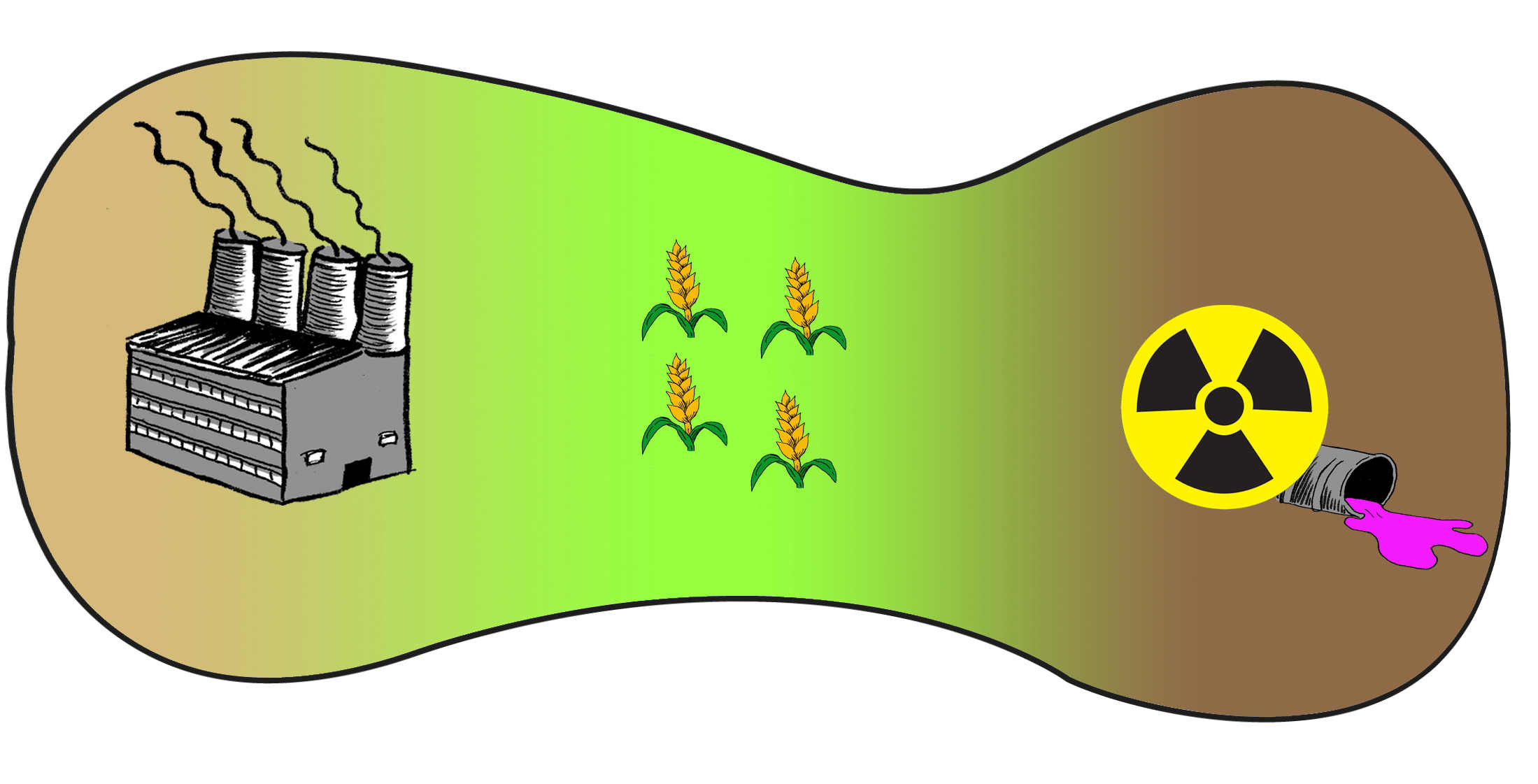}}
      
      shows some sort of infinitesimal price (a density of the price so to speak) that we denote $p$, the greener areas are more expensive while the dark ones are cheaper. Then, we have two ways to measure the price of a land $\Omega$: the first one is directly using $\rho$ to obtain $\rho(\Omega)$, while the second is measuring the area but taking into account the infinitesimal price
      \[\int_\Omega p\,\mathrm{d}\alpha=\rho(\Omega)\overset{\eqref{Appendix - equation - m(A)=int A}}{=}\int_\Omega\mathrm{d}\rho\qquad\quad\equiv\quad\qquad p\,\mathrm{d}\alpha=\mathrm{d}\rho\]
      Where the equation on the right is the infinitesimal notation which, by definition, is given by the equation on the left. The map $p$ is known as the \textbf{Radon-Nikodym derivative} and is usually denoted, motivated by the infinitesimal notation, by
      \[p=\frac{\mathrm{d}\rho}{\mathrm{d}\alpha}\]
      Notice that we are ``dividing'' by $\mathrm{d}\alpha$ thus we have to be cautious if it vanishes as we should also have that the numerator vanishes. In this case we see that if $\alpha(A)=0$ i.e.\ the land $A$ has zero area, then its price is obviously zero and the previous quotient ``makes sense''. Notice that the opposite does not hold necessarily: we could have a land of non-zero area but so devalued (for instance due to pollution) that it is worth noting. These ideas are formalized in the following definition and theorem.

      \begin{definition}\label{Appendix - definition - a.c.}\mbox{}\\
      	Given a measure $\mu$ we say that another measure $\nu$ is $\mu$ absolutely continuous\index{Absolutely continuous}, usually denoted $\mu$-a.c., if whenever $\mu(A)=0$ then $\nu(A)$ is also zero. This suggests the notation $\nu\ll \mu$.
      \end{definition}

      \begin{theorem}\label{Appendix - theorem - RN derivative}\mbox{}\\
      	Given a measurable space $(X,\Sigma)$ and a measure $\nu$ which is $\mu$-a.c., then there exists a measurable map $f:X\to(-\infty,\infty]$ such that $\mathrm{d}\nu=f\mathrm{d}\mu$.
      \end{theorem}
      The equality $\mathrm{d}\nu=f\mathrm{d}\mu$ uses the infinitesimal notation and it actually means that for every $A\subset X$ measurable we have
      \[\nu(A)=\int_A f\mathrm{d}\mu\]
      The RN derivative of $\nu$ with respect to $\mu$ can be understood as the $\mu$-anti-integral in the same sense that  the usual derivative is inverse to the Riemann integral (fundamental theorem of calculus).

    \section{Useful identities}\label{appendix section useful identities}
    
\starwarssection[Master Yoda]{That is why you fail.}{The Empire Strikes Back}

    \subsection*{k-forms}\index{k-form field}\vspace*{-2ex}
    
    \begin{flalign}\label{Appendix - equation - wedge indeces}
      \beta_{[a_1\cdots a_k]}=\frac{1}{k!}\sum_{\sigma\in S_k}(-1)^\sigma\beta_{\sigma(a_1)\cdots\sigma(a_k)}&&
     \end{flalign}
     

    \subsection*{Wedge product}\index{Wedge product}\vspace*{-2ex}
    
    \begin{flalign}
    &(\alpha\wedge\beta)_{a_1\cdots a_k a_{k+1}\cdots a_{k+m}}=\frac{(k+m)!}{k!\,m!}\alpha_{[a_1\cdots a_k}\beta_{a_{k+1}\cdots a_{k+m}]}&\label{Appendix - definition - wedge}\\
    &\alpha\wedge \beta=(-1)^{|\alpha||\beta|}\beta\wedge \alpha\label{Appendix - property - wedge supercommutativity}\\
    &f^*(\alpha\wedge \beta)=(f^*\alpha)\wedge(f^*\beta)\qquad f:M\to N 
    \end{flalign}
    

    \subsection*{Contraction}\index{Contraction}\vspace*{-2ex}
    
        \begin{flalign}\tensor{(C^\alpha_\beta T)}{^{a_1}^\cdots^{a_{r-1}}_{b_1}_\cdots_{b_{s-1}}}=\tensor{T}{^{a_1}^\cdots^{\overset{(\alpha)}{c}}^\cdots^{a_{r-1}}_{b_1}_\cdots_{\underset{(\beta)}{c}}_\cdots_{b_{s-1}}}&&
        \end{flalign}
    

    \subsection*{Lie derivative}\index{Lie derivative}\vspace*{-2ex}
      
    \begin{flalign}
    &\peqsub{\mathcal{L}}{V}f=V(f)\label{Appendix - equation - derivada de lie funcion}&\\
    &\peqsub{\mathcal{L}}{V}(T\otimes R)=(\peqsub{\mathcal{L}}{V}T)\otimes R+T\otimes(\peqsub{\mathcal{L}}{V}R)\label{Appendix - equation - Lebniz Lie}&\\
    &(\peqsub{\mathcal{L}}{V}W)^a=[V,W]^a=V^b\nabla_bW^a-W^b\nabla_bV^a\label{Appendix - equation - Lie bracket}\\
    &(\peqsub{\mathcal{L}}{V}\omega)(W)=\peqsub{\mathcal{L}}{V}(\omega(W))-\omega(\peqsub{\mathcal{L}}{V}W) \qquad \omega\in\Omega^1(M)\label{Appendix - equation - Lie 1-form}\\[2ex]
    \begin{split}&(\peqsub{\mathcal{L}}{V} T)(\alpha_1,\ldots,\alpha_r,W_1,\ldots,W_s)=V\left(\rule{0ex}{2.5ex}T(\alpha_1\ldots\alpha_r,W_1\ldots W_s)\right)-{}\\
    &\hspace*{7ex}-\sum_{k=1}^rT(\alpha_1\ldots \peqsub{\mathcal{L}}{V}\alpha_k\ldots\alpha_r,W_1\ldots W_s)-\sum_{k=1}^rT(\alpha_1\ldots\alpha_r,W_1\ldots \peqsub{\mathcal{L}}{V}W_k\ldots W_s)\end{split}\\[2ex]
    \begin{split}&\tensor*{(\peqsub{\mathcal{L}}{V} T)}{^{a_1}^{\cdots}^{a_r}_{b_1}_{\cdots}_{b_s}}= V^c\nabla_c\tensor*{T}{^{a_1}^{\cdots}^{a_r}_{\ b_1}_{\cdots}_{b_s}}-\sum_{k=1}^r \tensor*{T}{^{a_1}^{\cdots}^{\overset{\left.k\right)}{\rule{0ex}{1.5ex}c}}^{\cdots}^{a_r}_{\ b_1\ }_{\ \cdots\ }_{\ b_s}}\nabla_c V^{a_k}+{}\\
    &\hspace*{30ex}+\sum_{k=1}^s  \tensor*{T}{^{a_1\ }^{\ \cdots\ }^{\ a_r}_{\ b_1}_{\cdots}_{\underset{\rule{0ex}{2ex}\left.k\right)}{c}}_{\cdots}_{b_s}}\nabla_{b_k}V^{c}+w\tensor*{T}{*^{a_1}_{b_1}^{\cdots}_{\cdots}^{a_r}_{b_s}}\nabla_cV^{c}\end{split}\label{Appendix - equation - derivada de lie abstract index}
    \end{flalign}
    

    \subsection*{Interior derivative}\index{Interior derivative}\vspace*{-2ex}
    
    \begin{flalign}
    &\peqsub{\imath}{V}\alpha(W_1,\ldots,W_{k-1})=\alpha(V,W_1,\ldots,W_{k-1})&\\
    &\peqsub{\imath}{fV}\alpha=f\peqsub{\imath}{V}\alpha&\\
    &(\peqsub{\imath}{V}\beta)_{a_1\cdots a_{k-1}}=V^b\beta_{ba_1\cdots a_{k-1}}\\
    &\peqsub{\imath}{V}(\alpha\wedge\beta)=(\peqsub{\imath}{V}\alpha)\wedge\beta+(-1)^{|\alpha|}\alpha\wedge(\peqsub{\imath}{V}\beta)\label{Appendix - equation - regla de Leibniz producto interior}\\
    &\peqsub{\imath}{V}(f^*\alpha)=f^*(\peqsub{\imath}{f_*V}\alpha)\label{appendix - formula - i_x(f^*)=f^*i_{f_x}}\\
    &\peqsub{\imath}{V}\peqsub{\imath}{V}=0\label{appendix - formula - i_x^2=0}
    \end{flalign}
    

    \subsection*{Exterior derivative}\index{Exterior derivative}\vspace*{-2ex}
    
    \begin{flalign}
    &\mathrm{d}f(V)=V^a\nabla_af=\peqsub{\nabla}{V}f=V(f)&\\
    \begin{split}
    	&\mathrm{d}\alpha(V_1,\ldots,V_{k+1})=\sum_{j=1}^{k+1}(-1)^{j+1}\peqsub{\nabla}{V_j}\Big(\alpha(V_1,\ldots,\widehat{V_j},\ldots,V_{k+1}\Big)+{}\\
    	&\hspace*{20ex}+\sum_{i<j}(-1)^{i+j}\alpha\Big([V_i,V_j],V_1,\ldots,\widehat{V_i},\ldots,\widehat{V_j},\ldots,V_{k+1}\Big)
    \end{split}\\
    &(\mathrm{d}\beta)_{a_1\cdots a_{k+1}}=(k+1)\nabla_{[a_1}\beta_{a_2\cdots a_{k+1}]}\\
    &\mathrm{d}^2=0\label{appendix - equation - d^2=0}\\
    &\mathrm{d}(\alpha\wedge\beta)=(\mathrm{d}\alpha)\wedge\beta+(-1)^{|\alpha|}\alpha\wedge(\mathrm{d}\beta)\label{Appendix - equation - derivada de lie L_(fV)}\\
    &\peqsub{\mathcal{L}}{fV}\beta=f\peqsub{\mathcal{L}}{V}\beta+\mathrm{d}f\wedge\peqsub{\imath}{V}\beta\\
    &\peqsub{\mathcal{L}}{V}=\mathrm{d}\peqsub{\imath}{V}+\peqsub{\imath}{V}\mathrm{d}\label{appendix - formula - L=di+id}\\
    &\peqsub{\mathrm{d}}{M}f^*=f^*\peqsub{\mathrm{d}}{N}\label{appendix - formula - f^*d=df^*}\\
    &\peqsub{\mathcal{L}}{V}\mathrm{d}=\mathrm{d}\peqsub{\mathcal{L}}{V}
    \end{flalign}
    

    
    

    \subsection*{Exterior covariant derivative}\index{Exterior covariant derivative}\vspace*{-2ex}
    
    \begin{flalign}
    \begin{split}
    &(\peqsub{\mathrm{d}}{\nabla}T)(V_1,\ldots,V_{k+1})=\sum_{j=1}^{k+1}(-1)^{j+1}\peqsub{\nabla}{V_j}\Big(T(V_1,\ldots,\widehat{V_j},\ldots,V_{k+1}\Big)+{}\\
    &\hspace*{20ex}+\sum_{i<j}(-1)^{i+j}T\Big([V_i,V_j],V_1,\ldots,\widehat{V_i},\ldots,\widehat{V_j},\ldots,V_{k+1}\Big)
    \end{split}&\\
    \begin{split}
    &\tensor*{(\peqsub{\mathrm{d}}{\nabla}T)}{^{a_1}^{\cdots}^{a_r}_{b_1}_{\cdots}_{b_s}}= \tensor*{(\mathrm{d}T)}{*^{a_1}_{b_1}^{\cdots}_{\cdots}^{a_r}_{b_s}}+\sum_{k=1}^r \tensor*{T}{^{a_1}^{\cdots}^{\overset{\left.k\right)}{\rule{0ex}{1.5ex}d}}^{\cdots}^{a_r}_{\ b_1\ }_{\ \cdots\ }_{\ b_s}}\tensor{(A^\nabla)}{^{a_k}_d}-{}\\
    &\hspace*{30ex}-\sum_{k=1}^s  \tensor*{T}{^{a_1\ }^{\ \cdots\ }^{\ a_r}_{\ b_1}_{\cdots}_{\underset{\rule{0ex}{2ex}\left.k\right)}{d}}_{\cdots}_{b_s}}\tensor{(A^\nabla)}{^d_{b_k}}+w\tensor*{T}{^{a_1}^{\cdots}^{a_r}_{\ b_1}_{\cdots}_{b_s}}\tensor{(A^\nabla)}{^d_d}\end{split}
    \end{flalign}
    

    \subsection*{Covariant derivative}\index{Covariant derivative}\vspace*{-2ex}

    \begin{flalign}
    &(\peqsub{\nabla}{V}\omega)(W)=V(\omega(W))-\omega(\peqsub{\nabla}{V}W)&\\
    \begin{split}&(\peqsub{\nabla}{W}T)(V_1,\ldots,V_r,\omega_1,\ldots,\omega_s)=W\left(\rule{0ex}{2.5ex}T(V_1\ldots V_r,\omega_1\ldots\omega_s)\right) -{}\\
    &\hspace*{10ex}-\sum_{i=1}^sT(V_1\ldots\overset{\left.i\right)}{\peqsub{\nabla}{W}V_i}\ldots V_r,\omega_1\ldots\omega_s)-\sum_{i=1}^rT(V_1\ldots V_r,\omega_1\ldots\overset{\left.i\right)}{\peqsub{\nabla}{W}\omega_i}\ldots\omega_s)\end{split}&\\
    &\mathrm{div}_\omega V=\nabla_aV^a\\
    &\tensor*{(\nabla_cT)}{^{a_1}^{\cdots}^{a_r}_{b_1}_{\cdots}_{b_s}}= \partial_c\tensor*{T}{^{a_1}^{\cdots}^{a_r}_{\ b_1}_{\cdots}_{b_s}}+\sum_{k=1}^r \tensor*{T}{^{a_1}^{\cdots}^{\overset{\left.k\right)}{\rule{0ex}{1.5ex}d}}^{\cdots}^{a_r}_{\ b_1\ }_{\ \cdots\ }_{\ b_s}}\tensor{\Gamma}{^{a_k}_c_d}-\sum_{k=1}^s  \tensor*{T}{^{a_1\ }^{\ \cdots\ }^{\ a_r}_{\ b_1}_{\cdots}_{\underset{\rule{0ex}{2ex}\left.k\right)}{d}}_{\cdots}_{b_s}}\tensor{\Gamma}{^d_c_{b_k}}+w\tensor*{T}{^{a_1}^{\cdots}^{a_r}_{\ b_1}_{\cdots}_{b_s}}\tensor{\Gamma}{^d_c_d}&\label{Appendix - equation - covariant derivative}\\
    &\tensor{\Gamma}{^\lambda_\mu_\nu}=\frac{g^{\lambda\rho}}{2}\Big(\partial_\mu g_{\rho\nu}+\partial_\nu g_{\mu\rho}-\partial_\rho g_{\mu\nu}\Big)
    \end{flalign}
        

    \subsection*{Metric}\index{Metric}\vspace*{-2ex}

    \begin{flalign}
      &\peqsubfino{\langle T,R\rangle}{\!g}{-0.2ex}=g\big(T,R\big)=\frac{1}{(r+s)!}g_{a_1c_1}\ \overset{(s)}{\cdots}\ g_{a_sc_s}g^{b_1d_1}\ \overset{(r)}{\cdots}\ g^{b_rd_r}\tensor{T}{^{a_1}^\cdots^{a_s}_{b_1}_\cdots_{b_s}}\tensor{R}{^{c_1}^\cdots^{c_s}_{d_1}_\cdots_{d_s}}\\
      &\peqsubfino{\langle\alpha\wedge \beta,\omega\rangle}{\!g}{-0.2ex}=\peqsubfino{\langle\beta,\imath_{\vec{v}}\omega\rangle}{\!g}{-0.2ex}\qquad \alpha\in\Omega^1(M)\text{ and }\vec{v}=\#_g\alpha&\label{Appendix - equation - (v wedge beta,omega)=(beta,i_v omega)}\\
      &\tensor*{\delta}{*^{a_1}_{b_1}^{\dots}_\dots^{a_k}_{b_k}} 
      = \begin{vmatrix}
      \delta^{a_1}_{b_1} & \cdots & \delta^{a_1}_{b_k} \\
      \vdots & \ddots & \vdots  \\
      \delta^{a_k}_{b_1} & \cdots & \delta^{a_k}_{b_k}
      \end{vmatrix}\\
      &\tensor*{\delta}{*^{a_1}_{b_1}^{\dots}_\dots^{a_k}_{b_k}} =\sum_{j=1}^k(-1)^{j-1}\delta^{a_1}_{b_j}\tensor*{\delta}{*^{a_2}_{b_1}^{\dots}_\dots^{\ \, a_j}_{b_{j-1}}^{a_{j+1}}_{b_{j+1}}^{\dots}_\dots^{a_k}_{b_k}} \label{Appendix - label - descomposicion delta}\\
      &\beta_{[b_1 \dots b_k]}=\frac{1}{k!} \delta^{a_1 \dots a_k}_{b_1 \dots b_k} \beta_{a_1 \dots a_k}\label{Appendix - equation - delta antisim}\\
      &\delta^{a_1 \dots a_s}_{b_1 \dots b_s} = \frac{1}{(n-s)!}\mathrm{vol}^{a_1 \dots a_s \, c_{s+1} \dots c_n}\mathrm{vol}_{b_1 \dots b_s \, c_{s+1} \dots c_n}\label{Appendix - equation - delta=vol vol}\\
      &V_{a_1\cdots a_k}:=(\peqsub{{\flat}}{g}V)_{a_1\cdots a_k}=g_{a_1b_1}\cdots g_{a_kb_k}V^{b_1\cdots b_k}\\
      &\beta^{b_1\cdots b_k}:=(\peqsub{\#}{g}\beta)^{b_1\cdots b_k}=g^{a_1b_1}\cdots g^{a_kb_k}\beta_{a_1\cdots a_k}
    \end{flalign}


    \subsection*{Curvature}\index{Curvature}\vspace*{-2ex}

    \begin{flalign} &\peqsub{\nabla}{\!V}\peqsub{\nabla}{\!W}s-\peqsub{\nabla}{\!W}\peqsub{\nabla}{\!V}s-\peqsub{\nabla}{\![V,W]}s=Riem^\nabla(V,W)s&\\
    \begin{split}&(\nabla_d \nabla_c -\nabla_c\nabla_d)\tensor*{T}{^{a_1}_{\, b_1}^\cdots_\cdots^{a_r}_{b_s}}=\tensor{Riem}{^{a_1}_f_d_c}\tensor*{T}{^f_{\, b_1}^{a_2}^\cdots_{\ \cdots\ }^{a_r}_{b_s}}+\cdots +\tensor{Riem}{^{a_r}_f_d_c}\tensor*{T}{^{a_1}_{\, b_1}^\cdots_{\ \cdots\ }^{a_{r-1}}^f_{b_s}}-{}\\
    &\hspace*{25ex}-\tensor{Riem}{^f_{b_1}_d_c}\tensor*{T}{^{a_1}_{\, f}_{b_2}^{\ \cdots\ }_\cdots^{a_r}_{b_s}} - \cdots - \tensor{Riem}{^f_{b_s}_d_c}\tensor*{T}{^{a_1}_{\, b_1}^{\ \cdots\ }_\cdots^{a_r}_{b_{s-1}}_f}\end{split}\label{Appendix - definition - Riemann}&\\
    &\riemann(V,U)=-\riemann(U,V)\\
    &Riem_{abcd}=-Riem_{bacd}=-Riem_{abdc}=Riem_{cdab}\label{Appendix - property - simetrias Riemann}\\
    &\mathrm{Ric}_{ab}=\tensor{\riemann}{^c_a_c_b}\\
    &R=g^{ab}\mathrm{Ric}_{ab}\\
    &\peqsub{\nabla}{V} \riemann(W,U)+\peqsub{\nabla}{W}\riemann(U,V)+\peqsub{\nabla}{U}\riemann(V,W)=0\\
    &\nabla_c\tensor{Riem}{^a_b_d_e}+\nabla_e\tensor{Riem}{^a_b_c_d}+\nabla_d\tensor{Riem}{^a_b_e_c}=\tensor{0}{^a_b_c_d_e}\\
    &\nabla^a\tensor{\mathrm{Ric}}{_a_b}=\frac{1}{2}\nabla_aR\label{Appendix - equation - div Ric=grad R}\\
    &\riemann(V,W)U+\riemann(U,V)W+\riemann(W,U)V=0\\
    &\tensor{Riem}{^a_b_c_d}+\tensor{Riem}{^a_d_b_c}+\tensor{Riem}{^a_c_d_b}=\tensor{0}{^a_b_c_d}\\
    &\tensor{Riem}{^\rho_\sigma_\mu_\nu}=\partial_\mu\tensor{\Gamma}{^\rho_\sigma_\nu}-\partial_\nu\tensor{\Gamma}{^\rho_\sigma_\mu}+\tensor{\Gamma}{^\rho_\mu_\lambda}\tensor{\Gamma}{^\lambda_\sigma_\nu}-\tensor{\Gamma}{^\rho_\nu_\lambda}\tensor{\Gamma}{^\lambda_\sigma_\mu}
    \end{flalign}

    
    

    
    

    \subsection*{Killing vector field}\index{Killing vector field}\vspace*{-2ex}
    
    \begin{flalign}
      &\peqsub{(\mathcal{L}}{\vec{V}}g)_{ab}=\nabla_aV_b+\nabla_bV_a&
    \end{flalign}
    

    \subsection*{Hodge star operator}\index{Hodge star operator}\vspace*{-2ex}

    \begin{flalign}
    &\alpha\wedge\peqsub{\star}{g}\beta=\peqsubfino{\langle\alpha,\beta\rangle}{\!g}{-0.2ex}\peqsub{\mathrm{vol}}{g}\label{Appendix - equation - definition Hodge (a,b)vol=a wedge*b}\\
    &(\peqsub{\star}{g}\beta)_{a_{k+1}\cdots a_n}=\frac{1}{k!}\beta_{a_1\cdots a_k}\tensor{\mathrm{vol}}{^{a_1}^\cdots ^{a_k}_{a_{k+1}}_{\cdots}_{a_n}}\label{Appendix - equation - definicion hodge}\\
    &\peqsub{\star}{g} 1=\peqsub{\mathrm{vol}}{g}\label{Appendix - equation - star 1=vol}\\
    &\star_{n-k}\star_k=(-1)^{k(n-k)+s}\,\mathrm{Id}\label{Appendix - equaion - star^2=Id}\\
    &\peqsub{\star}{g}\peqsub{\mathrm{vol}}{g}=(-1)^s\\
    &\delta_k=(-1)^{n(k+1)+1+s}\star_{n-k+1} \mathrm{d}_{n-k}\star_k\label{Appendix - definition - delta=star d star}&\\
    &(\delta p)_{a_1\cdots a_{k-1}}=-\nabla^bp_{ba_1\dots a_{k-1}}\label{Appendix - definition - delta=-nabla}\\
    &\delta^2=0\label{Appendix - definition - delta^2=0}\\
    &\delta_{n-k}\star_{k}=(-1)^{k+1}\star_{k+1}\mathrm{d}_{k}\label{appendix property delta star=star d}\\
    &\star_{k-1}\delta_k=(-1)^{k}\mathrm{d}_{n-k}\star_k\label{appendix property star delta=d star}\\
    &\peqsub{\star}{g}\alpha=\imath_{\alpha}\peqsub{\mathrm{vol}}{g}\text{ for }\alpha\in\Omega^1(M)\label{appendix property star alpha=i_alpha vol}\\
    \begin{split}
    &\int_M \peqsubfino{\langle\mathrm{d}_{k-1}\alpha,\beta\rangle}{\!g}{-0.2ex}\peqsub{\mathrm{vol}}{g}=\int_M\peqsubfino{\langle\alpha,\delta_k\beta\rangle}{\!g}{-0.2ex}\peqsub{\mathrm{vol}}{g}+\int_{\partial M}\peqsub{\jmath}{\partial}^*\peqsubfino{\langle\alpha,\imath_\nu(\beta)\rangle}{\!g}{-0.2ex}\mathrm{vol}_{g_\partial}\updown{\eqref{Appendix - lemma - g=gamma+nu nu}}{\eqref{appendix - formula - i_x^2=0}}{=}\\
    &\phantom{\int_M\peqsubfino{\langle \mathrm{d}_{k-1}\alpha,\beta\rangle}{\!g}{-0.2ex}\peqsub{\mathrm{vol}}{g}}=\int_M\peqsubfino{\langle\alpha,\delta_k\beta\rangle}{\!g}{-0.2ex}\peqsub{\mathrm{vol}}{g}+\int_{\partial M}\peqsubfino{\langle\peqsub{\jmath}{\partial}^*\alpha,\peqsub{\jmath}{\partial}^*\imath_\nu(\beta)\rangle}{\!g}{-0.2ex}\mathrm{vol}_{g_\partial}
    \end{split}\label{appendix equation integracion por partes}
    \end{flalign}


\subsection*{Hypersurfaces}\index{Hypersurfaces}\vspace*{-2ex}

\begin{flalign}
&\jmath:(\overline{M},\jmath^*g)\hookrightarrow(M,g)\\
&\overline{\nabla}:\mathfrak{X}(\overline{M})\times\mathfrak{X}(\overline{M})\to\mathfrak{X}(\overline{M})\\
&\nabla^{(\jmath)}:\mathfrak{X}(\overline{M})\times\Gamma(\jmath^*TM)\to\Gamma(\jmath^*TM)\qquad\text{with}\qquad\peqsub{\nabla^{(\jmath)}}{V}(W\smallcirc\jmath)=\nabla_{\jmath_*\!V}W\\
&\nabla_b\equiv\nabla^{(\jmath)}_b=(\jmath_*)^\beta_b\nabla_\beta&\label{Appendix - equation - definition nabla_b}\\
&\jmath_*\big(\mathcal{W}(V)\big)=-\peqsub{\nabla}{V}\vec{n}\\ &(\jmath_*)_b^\beta\tensor{\mathcal{W}}{_a^b}=-\nabla_an^\beta\label{Appendix - equation - nabla n=K}\\
&K(V,U)=\overline{g}(\mathcal{W}V,U)\\
&K_{ab}=\overline{g}_{ac}\tensor{\mathcal{W}}{_b^c}\\
&\peqsub{\nabla}{V}(\jmath_*W)=\jmath_*\peqsub{\overline{\nabla}}{V}W+\varepsilon K(V,W)\vec{n}\\ &\nabla_a\big((\jmath_*)^\beta_bW^b\big)=(\jmath_*)^\beta_b\overline{\nabla}_aW^b+\varepsilon K_{ab}W^b n^\beta\label{Appendix - equation - Gauss lemma indexes}\\ &\overline{g}_{ab}\tensor{\overline{Riem}}{^b_c_d_e}-\varepsilon \tensor{K}{_c_e}K_{da}+\varepsilon \tensor{K}{_c_a}K_{de}=g_{\alpha\beta}\tensor{Riem}{^\beta_\gamma_\delta_\sigma}(\jmath_*)^\gamma_{c}(\jmath_*)^\delta_d(\jmath_*)^\sigma_e(\jmath_*)^\alpha_a
\end{flalign}


\subsection*{Embeddings}\index{Embedding}\vspace*{-2ex}

\begin{flalign}
&\tensor*{(\peqsub{\tau}{X})}{*^{\alpha_1}_{b_1}^{\cdots}_{\cdots}_{b_r}^{\alpha_r}}=(X_*)^{\alpha_1}_{b_1}\cdots(X_*)^{\alpha_r}_{b_r}&\\
&\tensor*{(\peqsub{e}{X})}{*_{\alpha_1}^{b_1}^{\cdots}_{\cdots}^{b_r}_{\alpha_r}}=\prod_{i=1}^rg_{\alpha_i\gamma_i}(X_*)^{\gamma_i}_{c_i}\peqsubfino{\gamma}{X}{-0.2ex}^{c_ib_i}\\
&    \tensor*{(\peqsub{\tau}{X})}{*^{\alpha_1}_{b_1}^{\cdots}_{\cdots}^{\alpha_1}_{b_1}}\tensor*{(\peqsub{e}{X})}{*_{\alpha_1}^{c_1}^{\cdots}_{\cdots}_{\alpha_1}^{c_1}}=\prod_{i=1}^r\delta_{b_i}^{c_i}\label{Appendix - equation - e.tau=delta}\\
&\tensor*{(\peqsub{e}{X})}{*_{\alpha_1}^{c_1}^{\cdots}_{\cdots}_{\alpha_1}^{c_1}}\tensor*{(\peqsub{\tau}{X})}{*^{\beta_1}_{c_1}^{\cdots}_{\cdots}^{\beta_1}_{c_1}}=\prod_{i=1}^r\Big(\delta_{\alpha_i}^{\beta_i}-\varepsilon (\peqsub{n}{X})_{\alpha_i}\peqsub{n}{X}^{\beta_i}\Big)\label{Appendix - equation - e.tau=delta-nn}\\
&g_{\alpha\beta}=\varepsilon n_\alpha n^\beta+\widetilde{\gamma}_{\alpha\beta}\label{Appendix - equation - g=nn+gamma}\\
&\peqsub{\mathbb{Y}}{\!X}=\peqsub{Y}{X}^{\scriptscriptstyle\perp} \peqsub{\vec{n}}{X} + \peqsub{\tau}{X}\peqsub{\vec{Y}}{X}^{\scriptscriptstyle\top}\label{Appendix - equation - decomposition Y=perp+top}\\
&\!(\peqsub{K}{X})_{ab}=-(\peqsub{\tau}{X})^\beta_b\nabla^{(X)}_a (\peqsub{n}{X})_\beta\\
&2K_t=-\jmath_t^*\big(\mathcal{L}_{\vec{n}_t}g\big)
\end{flalign}


\subsection*{Variations}\index{Embeddings!Variations}\vspace*{-2ex}

\begin{flalign}
&\big(\peqsub{M}{\,\mathbb{Y}}\big)_{\!\raisemath{.7pt}{a}}^{\!\phantom{a}\raisemath{-1.5pt}{b}}:=\nabla_a Y^b-Y^{\scriptscriptstyle\perp} K^b_a&\label{Appendix - equation - definicion M}\\
&\big(\peqsub{m}{\,\mathbb{Y}}\big)_{\!\raisemath{.7pt}{a}}:=K_{ab} Y^b+\varepsilon (\mathrm{d}Y^{\scriptscriptstyle\perp})_a\label{Appendix - equation - definicion m}\\
&Z^{\scriptscriptstyle\perp}\big(\peqsub{M}{\,\mathbb{Y}}\big)_{\!\raisemath{.7pt}{a}}^{\!\phantom{a}\raisemath{-1.5pt}{b}}-Y^{\scriptscriptstyle\perp}\big(\peqsub{M}{\,\mathbb{Z}}\big)_{\!\raisemath{.7pt}{a}}^{\!\phantom{a}\raisemath{-1.5pt}{b}}=Z^{\scriptscriptstyle\perp}\nabla_aY^b-Y^{\scriptscriptstyle\perp}\nabla_aZ^b\label{Appendix - equation - perp M-perp M}\\
&Z^a(\peqsub{m}{\,\mathbb{Y}})_a-Y^a(\peqsub{m}{\,\mathbb{Z}})_a=\varepsilon Z^a\nabla_aY^{\scriptscriptstyle\perp}-\varepsilon Y^a\nabla_aZ^{\scriptscriptstyle\perp}\label{Appendix - equation - top m-top m}\\
&Z^a\big(\peqsub{M}{\,\mathbb{Y}}\big)_{\!\raisemath{.7pt}{a}}^{\!\phantom{a}\raisemath{-1.5pt}{b}}+Y^{\scriptscriptstyle\perp}(\peqsub{m}{\,\mathbb{Z}})^b=Z^a\nabla_aY^b+\varepsilon Y^{\scriptscriptstyle\perp}\nabla^bZ^{\scriptscriptstyle\perp}\label{Appendix - equation - top M+perp m}\\
&\peqsub{D}{\left(X,\mathbb{Y}_X\!\right)}\,F=\peqsubfino{{\nabla_\mathbb{\vec{Y}}}}{\!X}{-0.2ex}F\label{Appendix - equation - variacion funcion}\\
&\!\left(\peqsub{D}{\left(X,\mathbb{Y}_X\!\right)}\,\tau\right){}_{\!b}^{\!\alpha}=\varepsilon n^\alpha \big(\peqsub{m}{\,\mathbb{Y}_{\!X}}\big)_{\!\raisemath{.7pt}{b}}+\tau^\alpha_c\big(\peqsub{M}{\,\mathbb{Y}_{\!X}}\big)_{\!\raisemath{.7pt}{a}}^{\!\phantom{b}\raisemath{-1.5pt}{c}}\label{Appendix - equation - Dtau}\\
&\!\left(\peqsub{D}{\left(X,\mathbb{Y}_X\!\right)}e\right){}^{\!b}_{\!\alpha}=\varepsilon n_\alpha (\peqsub{m}{\,\mathbb{Y}_{\!X}})^b-e^c_\alpha \big(\peqsub{M}{\,\mathbb{Y}_{\!X}}\big)_{\!\raisemath{.7pt}{c}}^{\!\phantom{c}\raisemath{-1.5pt}{b}}\label{Appendix - equation - De}\\
&\!\left(\peqsub{D}{\left(X,\mathbb{Y}_X\!\right)}\,n\right){}^{\!\alpha}=-\tau^\alpha_b(\peqsub{m}{\,\mathbb{Y}_{\!X}})^b\label{Appendix - equation - Dvec(n)}\\
&\!\left(\peqsub{D}{\left(X,\mathbb{Y}_X\!\right)}\,n\right){}_{\!\alpha}=-e^b_\alpha \big(\peqsub{m}{\,\mathbb{Y}_{\!X}}\big)_{\!\raisemath{.7pt}{b}}\label{Appendix - equation - Dn}\\
&\!\left(\peqsub{D}{\left(X,\mathbb{Y}_X\!\right)}\,\gamma\right){}_{\!bc}=\big(\peqsub{M}{\,\mathbb{Y}_{\!X}}\big)_{\!\raisemath{.7pt}{bc}}+\big(\peqsub{M}{\,\mathbb{Y}_{\!X}}\big)_{\!\raisemath{.7pt}{cb}}\label{Appendix - equation - Dgamma}\\
&\!\left(D\gamma^{-1}\right)^{dc}=-(D\gamma)^{dc}\label{Appendix - equation - variacion inversa metrica}\\
&\peqsub{D}{Z}(\star_\gamma)_k=\left([D\gamma]_{n-k}-\dfrac{\mathrm{Tr}(D\gamma)}{2}\mathrm{Id}\right)(\star_\gamma)_k\label{Appendix - equation - variacion Hodge}\\
&D\peqsub{\mathrm{vol}}{\gamma}=\dfrac{\mathrm{Tr}(D\gamma)}{2}\peqsub{\mathrm{vol}}{\gamma}\label{Appendix - equation - variacion volumen}\\
&D\sqrt{\gamma}=\dfrac{\mathrm{Tr}(D\gamma)}{2}\sqrt{\gamma}\label{Appendix - equation - variacion determinante}\\
&\tensor{(D\nabla)}{^c_a_b}=\frac{\gamma^{cd}}{2}\Big(\nabla_a\tensor{(D\gamma)}{_d_b}+\nabla_b\tensor{(D\gamma)}{_a_d}-\nabla_d(D\gamma)_{ab}\Big)\label{Appendix - equation - variacion nabla}
\end{flalign}


    \subsection*{Hypersurface deformation algebra}\index{Hypersurface deformation algebra}\vspace*{-2ex}
    
       \begin{flalign}
    \begin{split}
    &[\mathbb{V},\mathbb{W}]=\Big(\peqsub{D}{\mathbb{V}}\ \!\!W^{\scriptscriptstyle\perp}-\peqsub{D}{\mathbb{W}}\ \!\!V^{\scriptscriptstyle\perp}+\mathrm{d}V^{\scriptscriptstyle\perp}(\vec{w}^{\scriptscriptstyle\top})-\mathrm{d}W^{\scriptscriptstyle\perp}(\vec{v}^{\scriptscriptstyle\top})\Big)\mathbbm{n}+{}\\
    &\phantom{[\mathbb{V},\mathbb{W}]=}+\tau.\Big(\peqsub{D}{\mathbb{V}}\ \!\!\vec{w}^{\scriptscriptstyle\top}-\peqsub{D}{\mathbb{W}}\ \!\!\vec{v}^{\scriptscriptstyle\top}+\varepsilon\big(V^{\scriptscriptstyle\perp}\nabla^{\gamma}W^{\scriptscriptstyle\perp}-W^{\scriptscriptstyle\perp}\nabla^{\gamma}V^{\scriptscriptstyle\perp}\big)-[\vec{v}^{\scriptscriptstyle\top},\vec{w}^{\scriptscriptstyle\top}]\Big)\\[-5ex]\mbox{}
    \end{split}&
    \end{flalign}

    
    \subsection*{Symplectic geometry}\index{Symplectic form}\vspace*{-2ex}

\begin{flalign}
&\{f,g\}=\Omega^{ab}(\mathrm{d}f)_a(\mathrm{d}g)_b&\\
&\peqsubfino{{\imath_X}}{\!H}{-0.2ex}\hspace*{0.1ex}\Omega=\mathrm{d}H\\
&\Omega_{ab}\peqsub{X}{H}^a=(\mathrm{d}H)_b\\
&E(q,v)=F\!L(q,v)\big(q,v\big)-L(q,v)\\
&(\peqsub{X}{\!E})^a(\peqsub{\Omega}{L})_{ab}=(\mathrm{d}E)_b
\end{flalign}


    \section{Some boring computations}
    
  \starwarssection[Luke Skywalker]{I have a very bad feeling about this.}{A New Hope}
    
    In this section we gather several computations and proofs that are important enough to be given in detail but too long or not essential enough to be included in the main body of this thesis. This section is therefore not intended to be read linearly at all, but to be consulted whenever the reader needs more information. We have included titles to the subsections in order to indicate the context. Some colors have been included in the computations to help follow the reasoning.

    \subsection*{Metric}\trassub
    
        \begin{lemma}\label{Appendix - lemma - vol=nu wedge vol}\mbox{}\\
    	Given $(\jmath(\overline{M}),\jmath^*g)$ a hypersurface of $(M,g)$ we have that their metric volume forms are related by the unitary $g$-normal vector field $\nu\in\Gamma(\jmath^*TM)$ by
    	\[\jmath^*(\imath_{\vec{\nu}}\peqsub{\mathrm{vol}}{g})=\peqsub{{\mathrm{vol}_g}}{\!\partial}\]
    	In fact, this is equivalent to write $\peqsub{\mathrm{vol}}{g}=\varepsilon\nu\wedge\imath_{\vec{\nu}}\peqsub{\mathrm{vol}}{g}$ over the hypersurface where $\varepsilon=\nu(\vec{\nu})=\pm1$.
    \end{lemma}
    \begin{proof}\mbox{}\\
    	First notice that $\nu\wedge\peqsub{{\mathrm{vol}_g}}{\!\partial}$ is an $n$-form, so there exists some $f\in\Cinf{\partial M}$ such that $\peqsub{\mathrm{vol}}{g}=f\,\nu\wedge\peqsub{{\mathrm{vol}_g}}{\!\partial}$. Take now an orthonormal basis $\mathcal{B}=\{\vec{\nu},w_2,\ldots,w_n\}$ at some point of the boundary, in particular notice that $\{w_2,\ldots,w_n\}$ is an orthonormal basis of the boundary. We have then
    	\[1=\peqsub{\mathrm{vol}}{g}(\vec{\nu},w_2,\ldots,w_n)=f\,(\nu\wedge\peqsub{{\mathrm{vol}_g}}{\!\partial})(\vec{\nu},w_2,\ldots,w_n)=f\,\nu(\vec{\nu})\peqsub{{\mathrm{vol}_g}}{\!\partial}(w_2,\ldots,w_n)=\varepsilon f\]
    	where we have used that $\imath_{\vec{\nu}}\peqsub{{\mathrm{vol}_g}}{\partial}=0$ and that a metric volume evaluated over an orthonormal basis is always 1.
    \end{proof}

\begin{lemma}\label{Appendix - lemma - g=gamma+nu nu}\mbox{}\\
Given $(\jmath(\overline{M}),\jmath^*g)$ a hypersurface of $(M,g)$ we have	\begin{align*}
(\jmath^*g)^{-1}\Big(\jmath^*T,\jmath^*R\Big)=g^{-1}\big(T,R\big)-g^{-1}\big(\imath_{\vec{\nu}}T,\imath_{\vec{\nu}}R\big)
\end{align*}
\end{lemma}
\begin{proof}\mbox{}\\
We have $g=\tilde{g}+\varepsilon\nu\otimes\nu$ where $\nu$ is the $1$-form field metrically equivalent to the unitary vector field $\vec{\nu}$ normal to the foliation, and $\varepsilon=\nu(\vec{\nu})=\pm1$.
\begin{align*}
(\jmath^*g)^{-1}&\Big(\jmath^*T,\jmath^*R\Big)=\frac{1}{k!}(\jmath^*g)^{\bar{\alpha}_1\bar{\beta}_1}\cdots (\jmath^*g)^{\bar{\alpha}_k\bar{\beta}_k}(\jmath^*T)_{\bar{\alpha}_1\cdots\bar{\alpha}_k}(\jmath^*R)_{\bar{\beta}_1\cdots\bar{\beta}_r}=\\
&=\frac{1}{k!}(\jmath^*g)^{\bar{\alpha}_1\bar{\beta}_1}\tau_{\bar{\alpha}_1}^{\alpha_1}\tau_{\bar{\beta}_1}^{\beta_1}\cdots (\jmath^*g)^{\bar{\alpha}_k\bar{\beta}_k}\tau_{\bar{\alpha}_k}^{\alpha_k}\tau_{\bar{\beta}_r}^{\beta_r}T_{\alpha_1\cdots\alpha_k}R_{\beta_1\cdots\beta_r}=\\
&=\frac{1}{k!}\widetilde{g}^{\alpha_1\beta_1}\cdots \widetilde{g}^{\alpha_r\beta_r}T_{\alpha_1\cdots\alpha_k}R_{\beta_1\cdots\beta_r}=\\
&=\frac{1}{k!}(g^{\alpha_1\beta_1}-\nu^{\alpha_1}\nu^{\beta_1})\cdots (g^{\alpha_r\beta_r}-\nu^{\alpha_r}\nu^{\beta_r})T_{\alpha_1\cdots\alpha_k}R_{\beta_1\cdots\beta_r}\overset{\eqref{appendix - formula - i_x^2=0}}{=}\\
&=\frac{1}{k!}g^{\alpha_1\beta_1}\cdots g^{\alpha_r\beta_r}T_{\alpha_1\cdots\alpha_k}R_{\beta_1\cdots\beta_r}-\frac{1}{k!}\sum_{j=1}^r\left(\prod_{i\neq j}^rg^{\alpha_i\beta_i}\right)\nu^{\alpha_j}\nu^{\beta_j}T_{\alpha_1\cdots\alpha_k}R_{\beta_1\cdots\beta_r}=\\
&=g^{-1}\big(T,R\big)-\frac{1}{k!}\sum_{j=1}^r\left(\prod_{i\neq j}^rg^{\alpha_i\beta_i}\right)(-1)^{j-1}(-1)^{j-1}(\imath_{\vec{\nu}}T)_{\alpha_1\cdots\hat{\alpha}_j\cdots\alpha_k}(\imath_{\vec{\nu}}R)_{\beta_1\cdots\hat{\beta}_j\cdots\beta_r}=\\
&=g^{-1}\big(T,R\big)-\frac{1}{k}\sum_{j=1}^rg^{-1}\big(\imath_{\vec{\nu}}T,\imath_{\vec{\nu}}R\big)=\\
&=g^{-1}\big(T,R\big)-g^{-1}\big(\imath_{\vec{\nu}}T,\imath_{\vec{\nu}}R\big)
\end{align*}
\mbox{}\vspace*{-7ex}

\end{proof}

    \subsection*{Hodge dual operator}\trassub
    
    \begin{lemma}\label{Appendix - lemma - j*(a wedge star b)=(a,i_nu b)vol}\mbox{}\\
    	Given $(\jmath(\overline{M}),\jmath^*g)$ a hypersurface of $(M,g)$, $\alpha\in\Omega^k(M)$ and $\beta\in\Omega^{k+1}(M)$, then
    	\begin{align*}
    	\jmath^*(\alpha\wedge\star_{k+1}\beta)=\peqsubfino{\langle\alpha,\imath_{\vec{\nu}}\beta\rangle}{\!g}{-0.2ex}\peqsub{{\mathrm{vol}_g}}{\!\partial}
    	\end{align*}
    \end{lemma}
    \begin{proof}
    	\begin{align*}
    	&\tensor{(\alpha\wedge\star\beta)}{_{a_1}_\cdots_{a_{n-1}}}\overset{\eqref{Appendix - equation - wedge indeces}}{=}\frac{(n-1)!}{k!(n-k-1)!}\tensor{\alpha}{_{[a_1}_\cdots_{a_{k}}}\tensor{(\star\beta)}{_{a_{k+1}}_\cdots_{a_{n-1}]}}\updown{\eqref{Appendix - equation - definicion hodge}}{\eqref{Appendix - equation - delta antisim}}{=}\\
    	&=\frac{1}{k!(n-k-1)!}\tensor*{\delta}{*^{c_1}_{a_1}^\cdots_\cdots^{c_{n-1}}_{a_{n-1}}}\tensor{\alpha}{_{c_1}_\cdots_{c_{k}}}\frac{1}{(k+1)!}\tensor{\beta}{_{b_{1}}_\cdots_{b_{k+1}}}\tensor{\mathrm{vol}}{^{b_1}^\cdots^{b_{k+1}}_{c_{k+1}}_\cdots_{c_{n-1}}}\overset{\eqref{Appendix - equation - delta=vol vol}}{=}\\
    	&=\frac{1}{k!(k+1)!\textcolor{red}{(n-k-1)!}}\frac{1}{1!}\textcolor{red}{\mathrm{vol}^{c_1\cdots c_{n-1}d}}\mathrm{vol}_{a_1\cdots a_{n-1}d}\tensor{\alpha}{_{c_1}_\cdots_{c_{k}}}\tensor{\beta}{_{b_{1}}_\cdots_{b_{k+1}}}\textcolor{red}{\tensor{\mathrm{vol}}{^{b_1}^\cdots^{b_{k+1}}_{c_{k+1}}_\cdots_{c_{n-1}}}}\updown{\ref{Appendix - lemma - vol=nu wedge vol}}{\eqref{Appendix - equation - delta=vol vol}}{=}\\
    	&=\frac{1}{k!(k+1)!}(\nu\wedge\imath_{\vec{\nu}}\mathrm{vol})_{da_1\cdots a_{n-1}}\tensor{\alpha}{_{c_1}_\cdots_{c_{k}}}\tensor{\beta}{^{b_{1}}^\cdots^{b_{k+1}}}\textcolor{red}{\tensor*{\delta}{*^{d}_{b_1}^{c_1}_{b_2}^\cdots^{c_k}_{b_{k+1}}}}\updown{\eqref{Appendix - equation - wedge indeces}}{\eqref{Appendix - equation - delta antisim}}{=}\\
    	&=\frac{1}{k!}\frac{n!}{1!(n-1)!}\nu_{[d}(\imath_{\vec{\nu}}\mathrm{vol})_{a_1\cdots a_{n-1}]}\tensor{\alpha}{_{c_1}_\cdots_{c_{k}}}\tensor{\beta}{^{[d}^{c_{1}}^\cdots^{c_{k}]}}\overset{\eqref{Appendix - equation - delta antisim}}{=}\\
    	&=\frac{n}{k!}\frac{1}{n!}\tensor*{\delta}{*^{d'}_d^{d_1}_{a_1}^\cdots_\cdots^{d_{n-1}}_{a_{n-1}}}\nu_{d'}(\imath_{\vec{\nu}}\mathrm{vol})_{d_1\cdots d_{n-1}}\tensor{\alpha}{_{c_1}_\cdots_{c_{k}}}\tensor{\beta}{^{d}^{c_{1}}^\cdots^{c_{k}}}\overset{\eqref{Appendix - label - descomposicion delta}}{=}\\
    	&=\frac{\nu_{d'}(\imath_{\vec{\nu}}\mathrm{vol})_{d_1\cdots d_{n-1}}}{k!(n-1)!}\!\left(\delta^{d'}_d\tensor*{\delta}{*^{d_1}_{a_1}^\cdots_\cdots^{d_{n-1}}_{a_{n-1}}}+\sum_{j=1}^n(-1)^j\delta^{d'}_{a_j}\tensor*{\delta}{*^{d_1}_{d}^{d_2}_{a_1}^\cdots_\cdots^{\ d_j}_{a_{j-1}}^{d_{j+1}}_{a_{j+1}}^\cdots_\cdots^{d_{n-1}}_{a_{n-1}}}\right)\!\tensor{\alpha}{_{c_1}_\cdots_{c_{k}}}\tensor{\beta}{^{d}^{c_{1}}^\cdots^{c_{k}}}\overset{\eqref{Appendix - equation - delta antisim}}{=}\\
    	&=\frac{1}{k!}\!\left(\nu_{d}(\imath_{\vec{\nu}}\mathrm{vol})_{a_1\cdots a_{n-1}}+\sum_{j=1}^n(-1)^j\nu_{a_j}(\imath_{\vec{\nu}}\mathrm{vol})_{d a_1\cdots a_{n-1}}\right)\!\tensor{\alpha}{_{c_1}_\cdots_{c_{k}}}\tensor{\beta}{^{d}^{c_{1}}^\cdots^{c_{k}}}=\\
    	&=\frac{1}{k!}\tensor{\alpha}{_{c_1}_\cdots_{c_{k}}}\tensor{(\imath_{\vec{\nu}}\beta)}{^{c_{1}}^\cdots^{c_{k}}}(\imath_{\vec{\nu}}\mathrm{vol})_{a_1\cdots a_{n-1}}+\frac{1}{k!}\sum_{j=1}^n(-1)^j\nu_{a_j}(\imath_{\vec{\nu}}\mathrm{vol})_{d a_1\cdots a_{n-1}}\tensor{\alpha}{_{c_1}_\cdots_{c_{k}}}\tensor{\beta}{^{d}^{c_{1}}^\cdots^{c_{k}}}=\\
    	&=\peqsubfino{\langle \alpha,\imath_{\vec{\nu}}\beta\rangle}{\!g}{-0.2ex}(\imath_{\vec{\nu}}\mathrm{vol})_{a_1\cdots a_{n-1}}+\sum_{j=1}^n\nu_{a_j}T_{a_1\cdots \hat{a}_j\cdots a_{n-1}}
    	\end{align*}
    	Using lemma \ref{Appendix - lemma - vol=nu wedge vol} and the fact that $\jmath^*\nu=0$, we obtain the desired result.
    \end{proof}

\begin{lemma}\label{Appendix - lemma - volz^*g=z^*vol_g and star_z^*g=z^*star_g}\mbox{}\\
	Let $Z\in\mathrm{Diff}(M)$ a diffeomorphism of $(M,g)$ then $\mathrm{vol}_{Z^*g}=Z^*\peqsub{\mathrm{vol}}{g}$ and $\star_{Z^*g}Z^*=Z^*\peqsub{\star}{g}$.
\end{lemma}
\begin{proof}\mbox{}\\
	There exists some $f$ such that $\mathrm{vol}_{Z^*g}=fZ^*\peqsub{\mathrm{vol}}{g}$. Taking $\{v_1,\ldots,v_n\}$ an orthonormal basis of $Z^*g$ then notice that $\{Z_*v_1,\ldots,Z_*v_n\}$ is an orthonormal basis for $g$ because we have that $g(Z_*v_i,Z_*v_j)=(Z^*g)(v_i,v_j)=\delta_{ij}$. Thus
	\[1=\mathrm{vol}_{Z^*g}(v_1,\ldots,v_n)=f(Z^*\peqsub{\mathrm{vol}}{g})(v_1,\ldots,v_n)=f\peqsub{\mathrm{vol}}{g}(Z_*v_1,\ldots,Z_*v_n)=f\]
	For the second statement we use the definition of the Hodge star
	\begin{align*}
	\alpha\wedge\star_{Z^*g}Z^*\!\beta&=\langle Z^*(Z^{-1})^*\alpha,Z^*\!\beta\rangle_{Z^*g}\mathrm{vol}_{Z^*g}=Z^*\Big(\langle (Z^{-1})^*\alpha,\beta\rangle_{g}\Big)Z^*\mathrm{vol}_{g}=\\
	&=Z^*\Big((Z^{-1})^*\alpha\wedge\peqsub{\star}{g}\beta\Big)=\alpha\wedge Z^*\!\peqsub{\star}{g}\beta
	\end{align*}
	\mbox{}\vspace*{-7ex}
	
\end{proof}

\begin{lemma}\label{Appendix - lemma - <i_F,star F>=0}\mbox{}\\
Let $F\in\Omega^{k+1}(M)$ with $\dim(M)=2k+1$. If $k$ is odd then $\langle \imath_{-}F,\star F\rangle=0$.
\end{lemma}
\begin{proof}\mbox{}\\
	Given $\beta=\langle \imath_{-}F,\star F\rangle\in\Omega^1(M)$, let us prove that $\star\beta$ is zero which, in turns, proves that $\beta$ is zero.
\begin{align*}
(\star\beta)_{a_1\cdots a_{2k}}&\overset{\eqref{Appendix - equation - definicion hodge}}{=}\frac{1}{1!}\beta_d\tensor{\mathrm{vol}}{^d_{a_1}_\cdots_{a_{2k}}}=F^{dc_{k+2}\cdots c_{2k+1}}(\star F)_{c_{k+2} \cdots c_{2k+1}}\tensor{\mathrm{vol}}{_d_{a_1}_\cdots_{a_{2k}}}\overset{\eqref{Appendix - equation - definicion hodge}}{=}\\
&\overset{\phantom{\eqref{Appendix - equation - definicion hodge}}}{=}F^{dc_{k+2}\cdots c_{2k+1}}\frac{1}{(k+1)!}F_{c_1\cdots c_{k+1}}\tensor{\mathrm{vol}}{^{c_1}^\cdots^{c_{k+1}}_{c_{k+2}}_\cdots_{c_{2k+1}}}\tensor{\mathrm{vol}}{_d_{a_1}_\cdots_{a_{2k}}}\overset{\eqref{Appendix - equation - delta=vol vol}}{=}\\
&\overset{\phantom{\eqref{Appendix - equation - definicion hodge}}}{=}\frac{1}{(k+1)!}\tensor{F}{^d_{c_{k+2}}_\cdots _{c_{2k+1}}}F_{c_1\cdots c_{k+1}}\tensor*{\delta}{*^{c_1}_d^{c_2}_{a_1}^\cdots_\cdots^{c_{2k+1}}_{a_{2k}}}\overset{\eqref{Appendix - label - descomposicion delta}}{=}\\
&\overset{\phantom{\eqref{Appendix - equation - definicion hodge}}}{=}\frac{1}{(k+1)!}\sum_{j=1}^{2k+1}(-1)^{j-1}\delta^{c_j}_d\tensor*{\delta}{*^{c_1}_{a_1}^\cdots_\cdots^{c_{j-1}}_{a_{j-1}}^{c_{j+1}}_{\ a_j}^\cdots_\cdots^{c_{2k+1}}_{\ a_{2k}}}\tensor{F}{^d_{c_{k+2}}_\cdots _{c_{2k+1}}}F_{c_1\cdots c_{k+1}}\overset{\dagger}{=}\\
&\overset{\phantom{\eqref{Appendix - equation - definicion hodge}}}{=}\frac{1}{(k+1)!}\sum_{j=1}^{k+1}(-1)^{j-1}\tensor*{\delta}{*^{c_1}_{a_1}^\cdots_\cdots^{c_{j-1}}_{a_{j-1}}^{c_{j}}_{a_j}^\cdots_\cdots^{c_{2k}}_{a_{2k}}}\tensor{F}{^d_{c_{k+1}}_\cdots _{c_{2k}}}F_{c_1\cdots c_{j-1}dc_{j}\cdots c_{k}}=\\
&\overset{\phantom{\eqref{Appendix - equation - definicion hodge}}}{=}\frac{1}{(k+1)!}\sum_{j=1}^{k+1}(-1)^{j-1}\tensor*{\delta}{*^{c_1}_{a_1}^\cdots_\cdots^{c_{2k}}_{a_{2k}}}\tensor{F}{^d_{c_{k+1}}_\cdots _{c_{2k}}}(-1)^{j-1}F_{dc_1\cdots c_{k}}\overset{\eqref{Appendix - equation - delta antisim}}{=}\\
&\overset{\phantom{\eqref{Appendix - equation - definicion hodge}}}{=}\frac{(2k)!}{(k+1)!}\sum_{j=1}^{k+1}F_{d[a_1\cdots a_{k}}\tensor{F}{^d_{a_{k+1}}_\cdots _{a_{2k}}_{]}}=\\
&\overset{\phantom{\eqref{Appendix - equation - definicion hodge}}}{=}\frac{(2k)!}{k!}\frac{1}{2}\Big(F_{d[a_1\cdots a_{k}}\tensor{F}{^d_{a_{k+1}}_\cdots _{a_{2k}}_{]}}+(-1)^kF_{d[a_{k+1}\cdots a_{2k}}\tensor{F}{^d_{a_{1}}_\cdots _{a_{k}}_{]}}\Big)=\\
&\overset{\phantom{\eqref{Appendix - equation - definicion hodge}}}{=}\frac{(2k)!}{k!}\frac{1+(-1)^k}{2}F_{d[a_1\cdots a_{k}}\tensor{F}{^d_{a_{k+1}}_\cdots _{a_{2k}}_{]}}=0
\end{align*}
because $k$ is odd. Besides, in the $\dagger$ equality we have used that $F$ is antisymmetric (then any contraction vanishes) and we have relabeled $c_j\to d$ and $c_{j+r}\to c_{j+r-1}$ for every $r\geq1$.
\end{proof}

    \subsection*{Hypersurfaces}\trassub

    \begin{lemma}\label{appendix - lemma - escalar de curvatura}\mbox{}\\
    	Let $\jmath:(\overline{M},\overline{g}=\jmath^*\!g)\to(M,g)$ be an embedding such that $\jmath(\overline{M})$ is a hypersurface, then
    	\begin{align*}
     R(g)=\overline{R}(\overline{g})+\varepsilon \mathrm{Tr}_{\overline{g}}(K)^2-2\varepsilon \langle K,K\rangle_{\overline{g}}-2\varepsilon\mathrm{div}\big(\vec{n}\,\mathrm{div}\,\vec{n}-\nabla_{\vec{n}}\vec{n}\big)
    	\end{align*}
    \end{lemma}
    \begin{proof}\mbox{}\\
    	The Gauss-Codazzi equation of lemma \ref{Mathematical background - theorem - Gauss R bar(R)} reads on abstract index notation
    \[\overline{g}_{ab}\tensor{\overline{Riem}}{^b_c_d_e}-\varepsilon \tensor{K}{_c_e}K_{da}+\varepsilon \tensor{K}{_c_a}K_{de}=g_{\alpha\beta}\tensor{Riem}{^\beta_\gamma_\delta_\sigma}(\jmath_*)^\gamma_{c}(\jmath_*)^\delta_d(\jmath_*)^\sigma_e(\jmath_*)^\alpha_a\]
    Taking the trace $a-d$ we get
    \begin{align*}
    \overline{\mathrm{Ric}}_{ce}&-\varepsilon \mathrm{Tr}_{\overline{g}}(K)K_{ce}+\varepsilon \tensor{K}{_c^a}K_{ae}=\textcolor{red}{\overline{g}^{ad}}g_{\alpha\beta}\tensor{Riem}{^\beta_\gamma_\delta_\sigma}(\jmath_*)^\gamma_{c}\textcolor{red}{(\jmath_*)^\delta_d}(\jmath_*)^\sigma_e\textcolor{red}{(\jmath_*)^\alpha_a}\overset{\eqref{Mathematical background - equation - eta=gamma^{-1}X_*X_*}}{=}\\
    &=\textcolor{red}{\zeta^{\alpha\delta}}\tensor{Riem}{_\alpha_\gamma_\delta_\sigma}(\jmath_*)^\gamma_{c}(\jmath_*)^\sigma_e\overset{\eqref{Mathematical background - equation - inverse metric n}}{=}\textcolor{red}{(g^{\alpha\delta}-\varepsilon n^\alpha n^\delta)}\tensor{Riem}{_\alpha_\gamma_\delta_\sigma}(\jmath_*)^\gamma_{c}(\jmath_*)^\sigma_e=\\
    &=\big(\mathrm{Ric}_{\gamma\sigma}-\varepsilon n^\alpha n^\delta \tensor{Riem}{_\alpha_\gamma_\delta_\sigma}\big)(\jmath_*)^\gamma_{c}(\jmath_*)^\sigma_e
    \end{align*}
    Taking now the trace $c-e$ we obtain
    \begin{align*}
    \overline{R}&({\overline{g}})-\varepsilon \mathrm{Tr}_{\overline{g}}(K)^2+2\varepsilon \langle K,K\rangle_{\overline{g}}=\big(\mathrm{Ric}_{\gamma\sigma}-\varepsilon n^\alpha n^\delta \tensor{Riem}{_\alpha_\gamma_\delta_\sigma}\big)\textcolor{brown}{(\jmath_*)^\gamma_{c}(\jmath_*)^\sigma_e\overline{g}^{ce}}\updown{\eqref{Mathematical background - equation - eta=gamma^{-1}X_*X_*}}{\eqref{Mathematical background - equation - inverse metric n}}{=}\\
    &=\big(\mathrm{Ric}_{\gamma\sigma}-\varepsilon n^\alpha n^\delta \tensor{Riem}{_\alpha_\gamma_\delta_\sigma}\big)\textcolor{brown}{\big(g^{\gamma\sigma}-\varepsilon n^\gamma n^\sigma\big)}=\\
    &=R(g)-\varepsilon n^\gamma n^\sigma\mathrm{Ric}_{\gamma\sigma}-\varepsilon n^\alpha n^\delta \tensor{Riem}{_\alpha_\gamma_\delta_\sigma}g^{\gamma\sigma}+n^\alpha n^\delta \tensor{Riem}{_\alpha_\gamma_\delta_\sigma}n^\gamma n^\sigma\overset{\eqref{Appendix - property - simetrias Riemann}}{=}\\
    &=R(g)-\varepsilon n^\gamma n^\sigma\mathrm{Ric}_{\gamma\sigma}-\varepsilon n^\alpha n^\delta \tensor{\mathrm{Ric}}{_\alpha_\delta}+0=R(g)-2\varepsilon  n^\sigma\tensor{Riem}{^\alpha_\gamma_\alpha_\sigma}n^\gamma\overset{\eqref{Appendix - definition - Riemann}}{=}\\
    &=R(g)-2\varepsilon n^\sigma\Big[\nabla_\alpha\nabla_\sigma n^\alpha-\nabla_\sigma\nabla_\alpha n^\alpha\Big]=\\
    &=R(g)-2\varepsilon \nabla_\alpha(n^\sigma\nabla_\sigma n^\alpha)+2\varepsilon \textcolor{teal}{(\nabla_\alpha n^\sigma)(\nabla_\sigma n^\alpha)}+2\varepsilon \nabla_\sigma(n^\sigma\nabla_\alpha n^\alpha)-2\varepsilon \textcolor{red}{(\nabla_\sigma n^\sigma)(\nabla_\alpha n^\alpha)}=\\
    &=R(g)-2\varepsilon(\textcolor{red}{\mathrm{Tr}_{\overline{g}}(K)^2}-\textcolor{teal}{K_{ab}K^{ab}})+2\varepsilon \nabla_\sigma(n^\sigma\nabla_\alpha n^\alpha-n^\alpha\nabla_\alpha n^\sigma)
    \end{align*}\newpage
    
    where in the last line we have used that
    \begin{align*}
     \nabla_\alpha &n^\beta=\delta_\alpha^\gamma\nabla_\gamma n^\beta\overset{\eqref{Appendix - equation - e.tau=delta-nn}}{=}\big[(e_\jmath)^c_\alpha(\tau_\jmath)_c^\gamma+\varepsilon n_\alpha n^\gamma\big]\nabla_\gamma n^\beta\updown{\eqref{Appendix - equation - definition nabla_b}}{\eqref{Appendix - equation - nabla n=K}}{=}-e^c_\alpha\tau^\beta_d K_c^d+\varepsilon n_\alpha n^\gamma\nabla_\gamma n^\beta
    \end{align*}
    Thus
    \begin{align*}
      &\nabla_\alpha n^\alpha=-e^c_\alpha\tau^\alpha_d K_c^d+\varepsilon n^\gamma n_\alpha \nabla_\gamma n^\alpha\overset{\eqref{Appendix - equation - e.tau=delta}}{=}-\delta^d_cK_c^d+\varepsilon n^\gamma\frac{1}{2}\nabla_\gamma(n_\alpha n^\alpha)=-\mathrm{Tr}_{\overline{g}}(K)+0\\
      &(\nabla_\alpha n^\beta)(\nabla_\beta n^\alpha )=\big(-\textcolor{red}{e^c_\alpha}\textcolor{teal}{\tau^\beta_d} K_c^d+\varepsilon \textcolor{red}{n_\alpha}n^\gamma\nabla_\gamma n^\beta\big)\big(-\textcolor{teal}{e^a_\beta}\textcolor{red}{\tau^\alpha_b} K_a^b+\varepsilon \textcolor{teal}{n_\beta}n^\sigma\nabla_\sigma n^\alpha \big)\overset{\eqref{Appendix - equation - e.tau=delta}}{=}\\
      \qquad&=\textcolor{red}{\delta^c_b}\textcolor{teal}{\delta_d^a}K_c^dK_a^b-\textcolor{teal}{0}-\textcolor{red}{0}+n^\gamma n^\sigma\frac{1}{4}\nabla_\sigma(\textcolor{red}{n_\alpha}n^\alpha)\nabla_\gamma(\textcolor{teal}{n_\beta}n^\beta)=K_{ca}K^{ca}
    \end{align*}
    \mbox{}\vspace*{-7ex}
    
\end{proof}
    
    \subsection*{Embeddings}\label{Appendix - section - embeddings}\trassub
    
       We recall that $\big(\peqsub{M}{\,\mathbb{Y}}\big)_{\!\raisemath{.7pt}{a}}^{\!\phantom{a}\raisemath{-1.5pt}{b}}:=\nabla_a Y^b-Y^{\scriptscriptstyle\perp} K^b_a$ and $\big(\peqsub{m}{\,\mathbb{Y}}\big)_{\!\raisemath{.7pt}{a}}:=K_{ab} Y^b+\varepsilon (\mathrm{d}Y^{\scriptscriptstyle\perp})_a$
        
        \begin{lemma}\mbox{}\\
        Let $\peqsub{f}{X}=F\smallcirc X$ and $\peqsub{v}{X}=V\smallcirc X$ be defined by composition with $F\in\Cinf{M}$ and $V\in\mathfrak{X}(M)$ fixed. Then their variations are given by \[\peqsub{D}{\left(X,\mathbb{Y}_X\!\right)}\peqsub{f}{X}=\nabla_{\mathbb{Y}_X}F\qquad\qquad\qquad\peqsub{D}{\left(X,\mathbb{Y}_X\!\right)}\peqsub{v}{X}^b=\nabla_{\mathbb{Y}_X}V^b\]
        \end{lemma}
    	\begin{proof}\mbox{}\\
    		$\peqsub{D}{\left(X,\mathbb{Y}_X\!\right)}\peqsub{f}{X}\overset{\eqref{Mathematical background - equation - definition variation}}{=}\left.\nabla_{\!\raisemath{-2pt}{\!\scriptscriptstyle\partial_t}}\right|_0(F\smallcirc X_t)=\mathrm{d}F(\nabla_{\partial_t}X_t)=\mathrm{d}F(\peqsub{\mathbb{Y}}{X})=\nabla_{\mathbb{Y}_X}F$\\
    		$\peqsub{D}{\left(X,\mathbb{Y}_X\!\right)}\peqsub{v}{X}^b\overset{\eqref{Mathematical background - equation - definition variation}}{=}\left.\nabla_{\!\raisemath{-2pt}{\!\scriptscriptstyle\partial_t}}\right|_0(V\smallcirc X_t)^b=\nabla_aV^b(\nabla_{\partial_t}X_t)^a=\nabla_{\mathbb{Y}_X}V^b$
    	\end{proof}
    
    	\begin{lemma}
    		\begin{align*}
    		&\left(\peqsub{D}{\left(X,\mathbb{Y}_X\!\right)}\,\tau\right){}_{\!b}^{\!\alpha}=\varepsilon n^\alpha \big(\peqsub{m}{\,\mathbb{Y}_{\!X}}\big)_{\!\raisemath{.7pt}{b}}+\tau^\alpha_c\big(\peqsub{M}{\,\mathbb{Y}_{\!X}}\big)_{\!\raisemath{.7pt}{a}}^{\!\phantom{b}\raisemath{-1.5pt}{c}}&&\left(\peqsub{D}{\left(X,\mathbb{Y}_X\!\right)}e\right){}^{\!b}_{\!\alpha}=\varepsilon n_\alpha (\peqsub{m}{\,\mathbb{Y}_{\!X}})^b-e^c_\alpha \big(\peqsub{M}{\,\mathbb{Y}_{\!X}}\big)_{\!\raisemath{.7pt}{c}}^{\!\phantom{c}\raisemath{-1.5pt}{b}}\\
    		&\left(\peqsub{D}{\left(X,\mathbb{Y}_X\!\right)}\,n\right){}^{\!\alpha}=-\tau^\alpha_b(\peqsub{m}{\,\mathbb{Y}_{\!X}})^b&&\left(\peqsub{D}{\left(X,\mathbb{Y}_X\!\right)}\,n\right){}_{\!\alpha}=-e^b_\alpha \big(\peqsub{m}{\,\mathbb{Y}_{\!X}}\big)_{\!\raisemath{.7pt}{b}}\\
    		&\left(\peqsub{D}{\left(X,\mathbb{Y}_X\!\right)}\,\gamma\right){}_{\!bc}=\big(\peqsub{M}{\,\mathbb{Y}_{\!X}}\big)_{\!\raisemath{.7pt}{bc}}+\big(\peqsub{M}{\,\mathbb{Y}_{\!X}}\big)_{\!\raisemath{.7pt}{cb}}&&\left(D\gamma^{-1}\right)^{dc}=-(D\gamma)^{dc}
    		\end{align*}
    	\end{lemma}
    	\begin{proof}\mbox{}\\
    		Given an embedding $X\in\mathrm{Emb}(\Sigma,M)$ and $Y,Z\in\mathfrak{X}(\Sigma)$, as the covariant derivative \eqref{Mathematical background - equation - nabla_a=tau nabla_alpha} is torsion-free we have
    		\[\peqsubfino{\nabla}{\!Y}{-0.2ex}(X_*Z)-\peqsubfino{\nabla}{\!Z}{-0.2ex}(X_*Y)-X_*[Y,Z]=0\]
    		Consider now a curve of embeddings, that we denote also $X$, and recall that we can consider it as a map $X:\R\times\Sigma\to M$ thanks to \eqref{Mathematical background - equation - C(R,C(M,N)) cong C(R x M,N)}. Taking now the vector field $(\partial_t,0)$ and $(0,Y)$ for $Y\in\mathfrak{X}(\Sigma)$, we obtain the important result
    		\begin{equation}\label{Appendix - equation - swapping covariant derivatives} \nabla_{(\partial_t,0)}X_*(0,Y)-\nabla_{(0,Y)}X_*(\partial_t,0)=0\qquad\quad\equiv\quad\qquad\nabla_{\partial_t}X_*Y-\nabla_YX_*\partial_t=0
    		\end{equation}
    		where $X_*\partial_t$ is, when evaluated at $t=0$, the velocity vector field of the variation at $X$ i.e.\ $\peqsub{\mathbb{Y}}{X}:=X_*\partial_t\in \peqsub{T}{X}\mathrm{Emb}(\Sigma,M)=\Gamma(X^*TM)$.
    		\begin{equation}
    		\left(\peqsub{D}{\left(X,\mathbb{Y}_X\!\right)}\peqsub{\tau}{X}\right)\!.Z\overset{\eqref{Mathematical background - equation - definition variation}}{=}\left.\nabla_{\!\raisemath{-2pt}{\!\scriptscriptstyle\partial_t}}\right|_0(\peqsub{\tau}{X}.Z)\overset{\eqref{Appendix - equation - swapping covariant derivatives}}{=}\peqsubfino{\nabla}{\!Z}{-0.2ex}(\left.X_*\partial_t\right|_{0})=\peqsubfino{\nabla}{\!Z}{-0.2ex}(\peqsub{\mathbb{Y}}{X})
    		\end{equation}

            $\blacktriangleright$ Variation of $\tau$\separprevia
            
            From the previous reasoning we have
    		\begin{align*}
    		\left(\peqsub{D}{\mathbb{Y}}\,\tau\right){}_{\!b}^{\!\alpha}&= \nabla_a\mathbb{Y}^\beta=\nabla_a\big(\tau^\beta_b Y^b+Y^{\scriptscriptstyle\perp} n^\beta\big)\overset{\ref{Mathematical background - theorem - Gauss lemma}}{=}\\
    		&=\tau^\beta_b\nabla_a Y^b+\varepsilon n^\beta K_{ab}Y^b+Y^{\scriptscriptstyle\perp} \nabla_an^\beta+n^\beta\nabla_a Y^{\scriptscriptstyle\perp}\overset{\eqref{Mathematical background - equation - Weingarten}}{=}\\
    		&=\tau^\beta_b\big(\nabla_a Y^b-Y^{\scriptscriptstyle\perp} K^b_a\big)+ \varepsilon n^\beta\big( K_{ab}Y^b+\varepsilon\nabla_a Y^{\scriptscriptstyle\perp}\big)=\tau^\beta_b\big(\peqsub{M}{\,\mathbb{Y}}\big)_{\!\raisemath{.7pt}{a}}^{\!\phantom{a}\raisemath{-1.5pt}{b}}+\varepsilon n^\beta \big(\peqsub{m}{\,\mathbb{Y}}\big)_{\!\raisemath{.7pt}{a}}
    		\end{align*}
    		
    		$\blacktriangleright$ Variation of $\vec{n}$\separprevia
    		
    		First notice that as $g(\vec{n},\vec{n})=\varepsilon$ we have
    		\[0=\peqsub{D}{\mathbb{Y}}\big(g_{\alpha\beta}n^\alpha n^\beta\big)=(\peqsub{D}{\mathbb{Y}}\,g)_{\alpha\beta}n^\alpha n^\beta+2g_{\alpha\beta}(\peqsub{D}{\mathbb{Y}}\,n)^\alpha n^\beta=0+2n_\alpha(\peqsub{D}{\mathbb{Y}}\,n)^\alpha\]
    		This implies that $(\peqsub{D}{\mathbb{Y}}\,n)^\alpha$ has no perpendicular part and, thus, is of the form $\peqsub{D}{\mathbb{Y}}\,\vec{n}=\tau.(\peqsub{D}{\mathbb{Y}}\,\vec{n})^{\scriptscriptstyle\top}$ for some $(\peqsub{D}{\mathbb{Y}}\,\vec{n})^{\scriptscriptstyle\top}\in\mathfrak{X}(\Sigma)$ to be determined. Now, for an arbitrary $Z\in\mathfrak{X}(\Sigma)$ we have
    		\begin{align*}
    		\peqsubfino{\gamma}{X}{-0.2ex}\Big((\peqsub{D}{\left(X,\mathbb{Y}_X\!\right)}&\peqsub{\vec{n}}{X})^{\scriptscriptstyle\top},Z\Big)=g\Big(X_*(\peqsub{D}{\left(X,\mathbb{Y}_X\!\right)}\peqsub{\vec{n}}{X})^{\scriptscriptstyle\top},X_*Z\Big)=g\big(\peqsub{D}{\left(X,\mathbb{Y}_X\!\right)}\peqsub{\vec{n}}{X},X_*Z\big)\overset{\eqref{Mathematical background - equation - definition variation}}{=}\\
    		&=g\Big(\left.\nabla_{\!\raisemath{-2pt}{\!\scriptscriptstyle\partial_t}}\right|_0\peqsub{\vec{n}}{X},X_*Z\Big)=\left.\nabla_{\!\raisemath{-2pt}{\!\scriptscriptstyle\partial_t}}\right|_0\Big[g\big(\peqsub{\vec{n}}{X},X_*Z\big)\Big]-g\Big(\peqsub{\vec{n}}{X},\left.\nabla_{\!\raisemath{-2pt}{\!\scriptscriptstyle\partial_t}}\right|_0X_*Z\Big)\updown{\peqsub{\vec{n}}{X}\perp X_*}{\eqref{Mathematical background - equation - definition variation}}{=}\\
    		&=-g\Big(\peqsub{\vec{n}}{X},\left(\peqsub{D}{\left(X,\mathbb{Y}_X\!\right)}\peqsub{\tau}{X}\right).Z\Big)=-g\Big(\peqsub{\vec{n}}{X},\left(\peqsub{\tau}{X}.\peqsub{M}{\,\mathbb{Y}}+\varepsilon \peqsub{\vec{n}}{X}\,\peqsub{m}{\,\mathbb{Y}}\right).Z\Big)=\\
    		&=0-\peqsub{m}{\,\mathbb{Y}}(Z)=-(\peqsub{m}{\,\mathbb{Y}})_aZ^a
    		\end{align*}
    		Thus we obtain that $\peqsub{D}{\mathbb{Y}}\,\vec{n}=\tau.(\peqsub{D}{\mathbb{Y}}\,\vec{n})^{\scriptscriptstyle\top}=-\tau.\peqsub{m}{\,\mathbb{Y}}$.\separpost
    		
    		$\blacktriangleright$ Variation of $n$ and $e$\separprevia
    		
    		Taking variations in \eqref{Appendix - equation - e.tau=delta-nn} for $r=1$, we obtain
    		\begin{align*}
    		 0_\alpha^\beta&=\peqsub{D}{\mathbb{Y}}\Big(e_\alpha^c\tau^\beta_c+\varepsilon n_\alpha n^\beta\Big)=\peqsub{D}{\mathbb{Y}}\big(e_\alpha^c\big)\tau^\beta_c+e_\alpha^c\peqsub{D}{\mathbb{Y}}\big(\tau^\beta_c\big)+\varepsilon\peqsub{D}{\mathbb{Y}}\big( n_\alpha\big) n^\beta+\varepsilon n_\alpha\peqsub{D}{\mathbb{Y}}\big(  n^\beta\big)=\\
    		 &=\peqsub{D}{\mathbb{Y}}\big(e_\alpha^c\big)\tau^\beta_c+e_\alpha^c\Big(\varepsilon n^\beta \big(\peqsub{m}{\,\mathbb{Y}_{\!X}}\big)_{\!\raisemath{.7pt}{c}}+\tau^\beta_d\big(\peqsub{M}{\,\mathbb{Y}_{\!X}}\big)_{\!\raisemath{.7pt}{c}}^{\!\phantom{c}\raisemath{-1.5pt}{d}}\Big)+\varepsilon\peqsub{D}{\mathbb{Y}}\big( n_\alpha\big) n^\beta-\varepsilon n_\alpha\tau^\beta_b(\peqsub{m}{\,\mathbb{Y}_{\!X}})^b=\\
    		 &=\Big[\peqsub{D}{\mathbb{Y}}\big(e_\alpha^d\big)+e_\alpha^c\big(\peqsub{M}{\,\mathbb{Y}_{\!X}}\big)_{\!\raisemath{.7pt}{c}}^{\!\phantom{c}\raisemath{-1.5pt}{d}}-\varepsilon n_\alpha(\peqsub{m}{\,\mathbb{Y}_{\!X}})^d\Big]\tau^\beta_d+\varepsilon n^\beta \Big[e_\alpha^c \big(\peqsub{m}{\,\mathbb{Y}_{\!X}}\big)_{\!\raisemath{.7pt}{c}}+\peqsub{D}{\mathbb{Y}}\big( n_\alpha\big)\Big]
    		\end{align*}
    		From where we read the variations of $e$ (the tangent part of the previous expression) and $n$ (the normal part).\separpost
    		
    		$\blacktriangleright$ Variation of $\peqsubfino{\gamma}{X}{-0.2ex}$\separprevia
    		
    		By definition of the pullback metric we have
    		\begin{align*}
    		  (\peqsub{D}{\mathbb{Y}}\,\gamma)(Y,Z)&=\peqsub{D}{\mathbb{Y}}\Big(g(\tau.Y,\tau.Z)\Big)=g\Big((\peqsub{D}{\mathbb{Y}}\tau).Y,\tau.Z\Big)+g\Big(\tau.Y,(\peqsub{D}{\mathbb{Y}}\tau).Z\Big)=\\
    		  &=g\Big((\tau.\peqsub{M}{\,\mathbb{Y}}+\varepsilon \vec{n}\,\peqsub{m}{\,\mathbb{Y}}).Y,\tau.Z\Big)+g\Big(\tau.Y,(\tau.\peqsub{M}{\,\mathbb{Y}}+\varepsilon \vec{n}\,\peqsub{m}{\,\mathbb{Y}}).Z\Big)\overset{\vec{n}\perp\tau}{=}\\
    		  &=g\Big(\tau.\peqsub{M}{\,\mathbb{Y}}.Y,\tau.Z\Big)+g\Big(\tau.Y,\tau.\peqsub{M}{\,\mathbb{Y}}.Z\Big)=\gamma\Big(\peqsub{M}{\,\mathbb{Y}}.Y,Z\Big)+\gamma\Big(Y,\peqsub{M}{\,\mathbb{Y}}.Z\Big)
    		\end{align*}
    		Thus $(\peqsub{D}{\mathbb{Y}}\,\gamma)_{ab}=\gamma_{cb}\big(\peqsub{M}{\,\mathbb{Y}}\big)_{\!\raisemath{.7pt}{a}}^{\!\phantom{a}\raisemath{-1.5pt}{c}}+\gamma_{ac}\big(\peqsub{M}{\,\mathbb{Y}}\big)_{\!\raisemath{.7pt}{b}}^{\!\phantom{b}\raisemath{-1.5pt}{c}}=(\peqsub{M}{\,\mathbb{Y}})_{ab}+(\peqsub{M}{\,\mathbb{Y}})_{ba}$.\separpost
    		
    		$\blacktriangleright$ Variation of $\peqsubfino{\gamma}{X}{-0.2ex}^{-1}$\separprevia
    		
    	Finally, taking the variation of $\gamma_{ab}\gamma^{bc}=\delta_a^c$ we have
    	\[0_a^c=(D\gamma)_{ab}\gamma^{bc}+\gamma_{ab}(D\gamma^{-1})^{bc}\qquad\longrightarrow\qquad(D\gamma^{-1})^{dc}=-\gamma^{ad}(D\gamma)_{ab}\gamma^{bc}=-(D\gamma)^{dc}\]
    	\end{proof}

        \begin{lemma}\mbox{}\\[1ex]
        	$\displaystyle\tensor{\left(D\nabla\right)}{^c_a_b}=\frac{\gamma^{cd}}{2}\Big(\nabla_a\tensor{(D\gamma)}{_d_b}+\nabla_b\tensor{(D\gamma)}{_a_d}-\nabla_d(D\gamma)_{ab}\Big)$
        \end{lemma}
    \begin{proof}\mbox{}\\
    	Let $\nabla$ be the Levi-Civita connection associated with the metric $\gamma$. It obviously depends on the metric $\gamma$ and is uniquely determined by the Koszul formula
    	\begin{align*}
    	2\gamma(\peqsub{\nabla}{V}W, U)&=V\big(\gamma(W,U)\big) +W\big(\gamma(V,U)\big)-U\big(\gamma(V,W)\big)+{}\\
    	&\quad+\gamma\big([V,W],U\big) -\gamma\big([V,U],W\big) - \gamma\big([W,U],V\big)
    	\end{align*}
    	Taking the variation we obtain
    	\begin{align*}
    	2(D&\gamma)\big(\peqsub{\nabla}{V}W, U\big)+2\gamma\big((D\nabla)(V,W), U\big)=V\big((D\gamma)(W,U)\big) +W\big((D\gamma)(V,U)\big)-U\big((D\gamma)(V,W)\big)+{}\\
    	&\quad+(D\gamma)\big([V,W],U\big) -(D\gamma)\big([V,U],W\big) -(D\gamma)\big([W,U],V\big)=\\
    	&=\big(\peqsub{\nabla}{V}D\gamma\big)(W,U) +\textcolor{red}{(D\gamma)\big(\peqsub{\nabla}{V}W,U\big)}+\textcolor{teal}{(D\gamma)\big(W,\peqsub{\nabla}{V}U\big) }+\big(\peqsub{\nabla}{W}D\gamma\big)(V,U)+\textcolor{red}{(D\gamma)\big(\peqsub{\nabla}{W}V,U\big)}+{}\\
    	&\quad+\textcolor{brown}{(D\gamma)\big(V,\peqsub{\nabla}{W}U\big)}-\big(\peqsub{\nabla}{U}D\gamma\big)(V,W)-\textcolor{teal}{(D\gamma)\big(\peqsub{\nabla}{U}V,W\big)}-\textcolor{brown}{(D\gamma)\big(V,\peqsub{\nabla}{U}W\big)}+{}\\
    	&\quad+\textcolor{red}{(D\gamma)\big([V,W],U\big)}-\textcolor{teal}{(D\gamma)\big([V,U],W\big)} -\textcolor{brown}{(D\gamma)\big([W,U],V\big)}\overset{\eqref{Appendix - equation - Lie bracket}}{=}\\
    	&=\big(\peqsub{\nabla}{V}D\gamma\big)(W,U) +\textcolor{red}{2(D\gamma)\big(\peqsub{\nabla}{V}W,U\big)}+\big(\peqsub{\nabla}{W}D\gamma\big)(V,U)-\big(\peqsub{\nabla}{U}D\gamma\big)(V,W)
    	\end{align*}
    	which leads, using abstract index notation, to
    	\begin{align*}
    	  2\gamma_{cd}\tensor{(D\nabla)}{^c_a_b}V^aW^b U^d=V^a\nabla_a(D\gamma)_{bd}W^bU^d W^b\nabla_b(\peqsub{\nabla}{W}D\gamma)_{ad}V^aU^d-U^d\nabla_d(\peqsub{\nabla}{U}D\gamma)_{ab}V^aW^b
    	\end{align*}
    	As it is true for every $V,W,U\in\mathfrak{X}(\Sigma)$ we obtain the formula that we wanted to prove.
    	\end{proof}
        
        \begin{lemma}\mbox{}\\[1ex]
        	$\displaystyle\peqsub{D}{Z}(\peqsub{{\star_g}}{\!Z})_k=\left([\peqsub{D}{Z}\peqsubfino{g}{Z}{-0.2ex}]_{n-k}-\frac{\mathrm{Tr}(\peqsub{D}{Z}\peqsubfino{g}{Z}{-0.2ex})}{2}\mathrm{Id}\right)(\peqsub{{\star_g}}{\!Z})_k$
        	where $[\peqsub{D}{Z}\peqsubfino{g}{Z}{-0.2ex}]_k:\Omega^k(M)\to \Omega^k(M)$ is given by
        	\begin{equation}\label{Appendix - equation - (Dg)_k}
        	\big[(\peqsub{D}{Z}\peqsubfino{g}{Z}{-0.2ex})_k\beta\big]_{b_1\cdots b_k}=\sum_{i=1}^k(\peqsub{D}{Z}\peqsubfino{g}{Z}{-0.2ex})_{b_i}^{\phantom{b_i}c}\beta_{b_1\cdots \underset{j)}{c}\cdots b_k}
        	\end{equation}
        	and $[\peqsub{D}{Z}\peqsubfino{g}{Z}{-0.2ex}]_0=0$
        \end{lemma}
        \begin{proof}\mbox{}\\
        	Let $k\geq1$, then
        	\begin{align*}
        	\alpha&\wedge\peqsub{D}{Z}(\peqsub{{\star_g}}{\!Z})_k\beta=\peqsub{D}{Z}\left[\alpha\wedge(\peqsub{{\star_g}}{\!Z})_k\beta\right]=\\
        	&=\peqsub{D}{Z}\left[\frac{1}{k!}\prod_{i=1}^k\peqsubfino{g}{Z}{-0.2ex}^{a_ib_i}\alpha_{a_1\cdots a_k}\beta_{b_1\cdots b_k}\peqsub{{\mathrm{vol}_g}}{\!Z}\right]=\peqsub{D}{Z}\left[\peqsubfino{{\langle\alpha,\beta\rangle_g}}{\!Z}{-0.2ex}\,\peqsub{{\mathrm{vol}_g}}{\!Z}\right]=\\
        	&=\frac{1}{k!}\sum_{j=1}^j(\peqsub{D}{Z}\peqsubfino{g}{Z}{-0.2ex}^{-1})^{a_jb_j}\left(\prod_{i\neq j}^k\peqsubfino{g}{Z}{-0.2ex}^{a_ib_i}\right)\alpha_{a_1\cdots a_k}\beta_{b_1\cdots b_k}\peqsub{{\mathrm{vol}_g}}{\!Z}+\frac{1}{k!}\peqsubfino{g}{Z}{-0.2ex}^{a_1b_1}\cdots\peqsubfino{g}{Z}{-0.2ex}^{a_kb_k}\alpha_{a_1\cdots a_k}\beta_{b_1\cdots b_k}(\peqsub{D}{Z}\peqsub{{\mathrm{vol}_g}}{\!Z})=\\
        	&=\frac{1}{k!}\sum_{j=1}^j\peqsubfino{g}{Z}{-0.2ex}^{a_jb_j}\tensor{(\peqsub{D}{Z}\peqsubfino{g}{Z}{-0.2ex}^{-1})}{_{b_j}^c}\left(\prod_{i\neq j}^k\peqsubfino{g}{Z}{-0.2ex}^{a_ib_i}\right)\alpha_{a_1\cdots a_k}\beta_{b_1\cdots \underset{j)}{c}\cdots b_k}\peqsub{{\mathrm{vol}_g}}{\!Z}+\peqsubfino{{\langle\alpha,\beta\rangle_g}}{\!Z}{-0.2ex}\,(\peqsub{D}{Z}\peqsub{{\mathrm{vol}_g}}{\!Z})\overset{\eqref{Appendix - equation - variacion inversa metrica}}{=}\\
        	&=\frac{1}{k!}\left(\prod_{i=1}^k\peqsubfino{g}{Z}{-0.2ex}^{a_ib_i}\right)\alpha_{a_1\cdots a_k}\big[-(\peqsub{D}{Z}\peqsubfino{g}{Z}{-0.2ex})_k\beta\big]_{b_1\cdots b_k}+\frac{\mathrm{Tr}(\peqsub{D}{Z}\peqsubfino{g}{Z}{-0.2ex})}{2}\peqsubfino{{\langle\alpha,\beta\rangle_g}}{\!Z}{-0.2ex}\,=\\
        	&=-\peqsubfino{{\langle\alpha,(\peqsub{D}{Z}\peqsubfino{g}{Z}{-0.2ex})_k\beta\big\rangle_g}}{\!Z}{-0.2ex}\,+\frac{\mathrm{Tr}(\peqsub{D}{Z}\peqsubfino{g}{Z}{-0.2ex})}{2}\peqsubfino{{\langle\alpha,\beta\rangle_g}}{\!Z}{-0.2ex}\,=\\
        	&=\alpha\wedge (\peqsub{{\star_g}}{\!Z})_k\left[- [\peqsub{D}{Z}\peqsubfino{g}{Z}{-0.2ex}]_k+\frac{\mathrm{Tr}(\peqsub{D}{Z}\peqsubfino{g}{Z}{-0.2ex})}{2}\right]\beta
        	\end{align*}
        	As $[\peqsub{D}{Z}\peqsubfino{g}{Z}{-0.2ex}]_0=0$ the $k=0$ case holds. We denote $[\peqsub{D}{Z}\peqsubfino{g}{Z}{-0.2ex}]_1=\peqsub{D}{Z}\peqsubfino{g}{Z}{-0.2ex}$, 
        	As the previous equation holds for every $\alpha\in\Omega^k(M)$ we have that $\peqsub{D}{Z}(\peqsub{{\star_g}}{\!Z})_k=(\peqsub{{\star_g}}{\!Z})_k\left(-[\peqsub{D}{Z}\peqsubfino{g}{Z}{-0.2ex}]_k+\frac{\mathrm{Tr}(\peqsub{D}{Z}\peqsubfino{g}{Z}{-0.2ex})}{2}\mathrm{Id}\right)$.\vspace*{2ex}
        	
        	Now as $(\peqsub{{\star_g}}{\!Z})_{n-k}(\peqsub{{\star_g}}{\!Z})_k=\mathrm{Id}$ if we compute its variation we get
        	\begin{align*}
        	0&=\peqsub{D}{Z}\left[(\peqsub{{\star_g}}{\!Z})_{n-k}(\peqsub{{\star_g}}{\!Z})_k\right]=\peqsub{D}{Z}\left[(\peqsub{{\star_g}}{\!Z})_{n-k}\right](\peqsub{{\star_g}}{\!Z})_k+(\peqsub{{\star_g}}{\!Z})_{n-k}\peqsub{D}{Z}\left[(\peqsub{{\star_g}}{\!Z})_k\right]=\\
        	&=(\peqsub{{\star_g}}{\!Z})_{n-k}\left(-[\peqsub{D}{Z}\peqsubfino{g}{Z}{-0.2ex}]_{n-k}+\frac{\mathrm{Tr}(\peqsub{D}{Z}\peqsubfino{g}{Z}{-0.2ex})}{2}\right)(\peqsub{{\star_g}}{\!Z})_k+(\peqsub{{\star_g}}{\!Z})_{n-k}(\peqsub{{\star_g}}{\!Z})_k\left(-[\peqsub{D}{Z}\peqsubfino{g}{Z}{-0.2ex}]_k+\frac{\mathrm{Tr}(\peqsub{D}{Z}\peqsubfino{g}{Z}{-0.2ex})}{2}\right)=\\
        	&=(\peqsub{{\star_g}}{\!Z})_{n-k}\left\{-[\peqsub{D}{Z}\peqsubfino{g}{Z}{-0.2ex}]_{n-k}(\peqsub{{\star_g}}{\!Z})_k+\frac{\mathrm{Tr}(\peqsub{D}{Z}\peqsubfino{g}{Z}{-0.2ex})}{2}(\peqsub{{\star_g}}{\!Z})_k-(\peqsub{{\star_g}}{\!Z})_k[\peqsub{D}{Z}\peqsubfino{g}{Z}{-0.2ex}]_k+(\peqsub{{\star_g}}{\!Z})_k\frac{\mathrm{Tr}(\peqsub{D}{Z}\peqsubfino{g}{Z}{-0.2ex})}{2}\right\}=\\
        	&=(\peqsub{{\star_g}}{\!Z})_{n-k}\left\{-(\peqsub{{\star_g}}{\!Z})_k[\peqsub{D}{Z}\peqsubfino{g}{Z}{-0.2ex}]_k-[\peqsub{D}{Z}\peqsubfino{g}{Z}{-0.2ex}]_{n-k}(\peqsub{{\star_g}}{\!Z})_k+\mathrm{Tr}(\peqsub{D}{Z}\peqsubfino{g}{Z}{-0.2ex})(\peqsub{{\star_g}}{\!Z})_k\right\}
        	\end{align*}
        	so $\peqsub{D}{Z}(\peqsub{{\star_g}}{\!Z})_k=\left([\peqsub{D}{Z}\peqsubfino{g}{Z}{-0.2ex}]_{n-k}-\frac{\mathrm{Tr}(\peqsub{D}{Z}\peqsubfino{g}{Z}{-0.2ex})}{2}\mathrm{Id}\right)(\peqsub{{\star_g}}{\!Z})_k$ which finishes the proof.
        \end{proof}

    \begin{lemma}\mbox{}\\[1ex]
    	$D\peqsub{\mathrm{vol}}{\gamma}=\dfrac{\mathrm{Tr}(D\gamma)}{2}\peqsub{\mathrm{vol}}{\gamma}\qquad\qquad\qquad
    	D\sqrt{\gamma}=\dfrac{\mathrm{Tr}(D\gamma)}{2}\sqrt{\gamma}$
    \end{lemma}
	\begin{proof}\mbox{}\\
	  In the proof of the previous lemma we obtained $D(\peqsub{\star}{\gamma})_k=(\peqsub{\star}{\gamma})_k\left(-[D\gamma]_k+\frac{\mathrm{Tr}(D\gamma)}{2}\mathrm{Id}\right)$. Applying it to $\peqsub{{\mathrm{vol}_g}}{\!Z}=(\peqsub{\star}{\gamma})_01$ and taking into account that $[D\gamma]_0=0$, we get the desired result. The second equation is immediate using that $\peqsub{\mathrm{vol}}{\gamma}=\sqrt{\gamma}\,\peqsub{\mathrm{vol}}{\Sigma}$.
	\end{proof}

\begin{lemma}\label{Appendix - lemma - divergencia densidad}\mbox{}\\
	Given $(M,g)$ a Lorentzian manifold and $X:\Sigma\hookrightarrow M$ an embedding such that $X(\Sigma)$ is a space-like hypersurface. Denote $\peqsub{\delta}{X}$ the codifferential over $\Sigma$ induced by the metric $\peqsubfino{\gamma}{X}{-0.2ex}:=X^*g$ and consider $\Lambda$ an antisymmetric $(k+1,0)$-density of weight $w=-1$, then $D(\delta\Lambda)=\delta(D\Lambda)$.
\end{lemma}
\begin{proof}\mbox{}\\
	We will now work for a moment in coordinates to express the covariant derivative, given by equation \eqref{Appendix - equation - covariant derivative} with weight $w=-1$, as
	\begin{align*}
	(\delta\Lambda)^{a_1\cdots a_k}&=-\nabla_{a_0}\Lambda^{a_0a_1\cdots a_k}=-\partial_{a_0}\Lambda^{a_0a_1\cdots a_k}-\sum_{i=0}^k\tensor*{\Lambda}{^{a_0}^{\cdots}^{\overset{\left.i\right)}{\rule{0ex}{1.5ex}d}}^{\cdots}^{a_k}}\tensor{\Gamma}{^{a_i}_{a_0}_d}+\Lambda^{a_0a_1\cdots a_k}\tensor{\Gamma}{^d_{a_0}_d}
	\end{align*}
	Notice that $\partial_c$ does not depend on the metric so we have
	\begin{align*}
	&\big(D\delta\Lambda\big)^{a_1\cdots a_k}=D\left(-\partial_{a_0}\Lambda^{a_0a_1\cdots a_k}-\sum_{i=0}^k\tensor*{\Lambda}{^{a_0}^{\cdots}^{\overset{\left.i\right)}{\rule{0ex}{1.5ex}d}}^{\cdots}^{a_k}}\tensor{\Gamma}{^{a_i}_{a_0}_d}+\Lambda^{a_0a_1\cdots a_k}\tensor{\Gamma}{^d_{a_0}_d}\right)=\\
	&\ =\textcolor{red}{-\partial_{a_0}(D\Lambda)^{a_0a_1\cdots a_k}-\sum_{i=0}^k\tensor*{(D\Lambda)}{^{a_0}^{\cdots}^{\overset{\left.i\right)}{\rule{0ex}{1.5ex}d}}^{\cdots}^{a_k}}\tensor{\Gamma}{^{a_i}_{a_0}_d}+(D\Lambda)^{a_0a_1\cdots a_k}\tensor{\Gamma}{^d_{a_0}_d}}-{}\\
	&\ \ \phantom{=}{}-\sum_{i=0}^k\tensor*{\Lambda}{^{a_0}^{\cdots}^{\overset{\left.i\right)}{\rule{0ex}{1.5ex}d}}^{\cdots}^{a_k}}\tensor{(D\Gamma)}{^{a_i}_{a_0}_d}+\Lambda^{a_0a_1\cdots a_k}\tensor{(D\Gamma)}{^d_{a_0}_d}\updown{\eqref{Appendix - equation - covariant derivative}}{\eqref{Appendix - equation - variacion nabla}}{=}\\
	&\ =\textcolor{red}{-\nabla_{a_0}(D\Lambda)^{a_0a_1\cdots a_k}}-\textcolor{teal}{\tensor*{\Lambda}{^d^{a_1}^{\cdots}^{a_k}}\tensor{(D\Gamma)}{^{a_0}_{a_0}_d}}
	-\sum_{i=1}^k\tensor*{\Lambda}{^{a_0}^{a_1}^{\cdots}^{\overset{\left.i\right)}{\rule{0ex}{1.5ex}d}}^{\cdots}^{a_k}}\tensor{(D\Gamma)}{^{a_i}_{a_0}_d}+\textcolor{teal}{\Lambda^{a_0a_1\cdots a_k}\tensor{(D\Gamma)}{^d_{a_0}_d}}=\\
	&\ =\delta(D\Lambda)^{a_1\cdots a_k}+\textcolor{teal}{0}-0=\delta(D\Lambda)
	\end{align*}
	where in the last equality we have used, apart from the definition of the $\delta$, that $\Lambda$ is antisymmetric in $a_0d$ while $(D\Gamma)$ is symmetric.	
\end{proof}
    
    \subsection*{Hypersurface deformation algebra}\trassub

    \begin{lemma}\label{appendix - lemma - hypersurface deformation algebra}
    	\begin{align}
    	\begin{split}
    	[\mathbb{V},\mathbb{W}]&=\Big(\peqsub{D}{\mathbb{V}}\ \!\!W^{\scriptscriptstyle\perp}-\peqsub{D}{\mathbb{W}}\ \!\!V^{\scriptscriptstyle\perp}+\mathrm{d}V^{\scriptscriptstyle\perp}(\vec{W}^{\scriptscriptstyle\top})-\mathrm{d}W^{\scriptscriptstyle\perp}(\vec{V}^{\scriptscriptstyle\top})\Big)\mathbbm{n}+{}\\
    	&\phantom{=}+\tau.\Big(\peqsub{D}{\mathbb{V}}\ \!\!\vec{W}^{\scriptscriptstyle\top}-\peqsub{D}{\mathbb{W}}\ \!\!\vec{V}^{\scriptscriptstyle\top}+\varepsilon\big(V^{\scriptscriptstyle\perp}\nabla^{\gamma_X}W^{\scriptscriptstyle\perp}-W^{\scriptscriptstyle\perp}\nabla^{\gamma_X}V^{\scriptscriptstyle\perp}\big)-[\vec{V}^{\scriptscriptstyle\top},\vec{W}^{\scriptscriptstyle\top}]\Big)
    	\end{split}
    	\end{align}
    \end{lemma}
\begin{proof}
    \begin{align*}
     	&\big[\mathbb{V},\mathbb{W}\big]= \big[V^{\scriptscriptstyle\perp}\mathbbm{n}+\tau.\vec{V}^{\scriptscriptstyle\top},W^{\scriptscriptstyle\perp}\mathbbm{n}+\tau.\vec{W}^{\scriptscriptstyle\top}\big]=\\
     	&=\peqsub{D}{\mathbb{V}}\Big(W^{\scriptscriptstyle\perp}\mathbbm{n}+\tau.\vec{W}^{\scriptscriptstyle\top}\Big)-\peqsub{D}{\mathbb{W}}\Big(V^{\scriptscriptstyle\perp}\mathbbm{n}+\tau.\vec{V}^{\scriptscriptstyle\top}\Big)=\\
     	&=\Big[\peqsub{D}{\mathbb{V}}W^{\scriptscriptstyle\perp}\!-\!\peqsub{D}{\mathbb{W}}V^{\scriptscriptstyle\perp}\Big]\mathbbm{n}+W^{\scriptscriptstyle\perp}\peqsub{D}{\mathbb{V}}\,\mathbbm{n}-V^{\scriptscriptstyle\perp}\peqsub{D}{\mathbb{W}}\,\mathbbm{n}+(\peqsub{D}{\mathbb{V}}\,\tau).\vec{W}^{\scriptscriptstyle\top}-(\peqsub{D}{\mathbb{W}}\,\tau).\vec{V}^{\scriptscriptstyle\top}+\tau.\Big[\peqsub{D}{\mathbb{V}}\vec{W}^{\scriptscriptstyle\top}\!-\!\peqsub{D}{\mathbb{W}}\vec{V}^{\scriptscriptstyle\top}\Big]\updown{\eqref{Appendix - equation - Dtau}}{\eqref{Appendix - equation - Dvec(n)}}{=}\\
      &=\Big[\peqsub{D}{\mathbb{V}}W^{\scriptscriptstyle\perp}\!-\!\peqsub{D}{\mathbb{W}}V^{\scriptscriptstyle\perp}\Big]\mathbbm{n}-\tau.\Big[W^{\scriptscriptstyle\perp}\peqsub{\vec{m}}{\mathbb{V}}-V^{\scriptscriptstyle\perp}\peqsub{\vec{m}}{\mathbb{W}}\Big]+\varepsilon\mathbbm{n}\Big[\vec{W}^{\scriptscriptstyle\top}\peqsub{\vec{m}}{\mathbb{V}}-\vec{V}^{\scriptscriptstyle\top}\peqsub{\vec{m}}{\mathbb{W}}\Big]+{}\\
      &\phantom{=}+\tau.\Big[\peqsub{M}{\mathbb{V}}.\vec{W}^{\scriptscriptstyle\top}\!-\!\peqsub{M}{\mathbb{W}}.\vec{V}^{\scriptscriptstyle\top}\Big]+\tau.\Big[\peqsub{D}{\mathbb{V}}\vec{W}^{\scriptscriptstyle\top}\!-\!\peqsub{D}{\mathbb{W}}\vec{V}^{\scriptscriptstyle\top}\Big]\updown{\eqref{Appendix - equation - perp M-perp M}}{\eqref{Appendix - equation - top m-top m}\eqref{Appendix - equation - top M+perp m}}{=}\\
      &=\Big[\peqsub{D}{\mathbb{V}}W^{\scriptscriptstyle\perp}\!-\!\peqsub{D}{\mathbb{W}}V^{\scriptscriptstyle\perp}\Big]\mathbbm{n}+\mathbbm{n}\Big[\mathrm{d}V^{\scriptscriptstyle\perp}(\vec{W}^{\scriptscriptstyle\top}\!)-\mathrm{d}W^{\scriptscriptstyle\perp}(\vec{V}^{\scriptscriptstyle\top}\!)\Big]+{}\\
      &\phantom{=}+\tau.\Big[[W^\top,V^\top]+\varepsilon(V^\perp\nabla W^\perp-W^\perp\nabla V^\perp)\Big]+\tau.\Big[\peqsub{D}{\mathbb{V}}\vec{W}^{\scriptscriptstyle\top}\!-\!\peqsub{D}{\mathbb{W}}\vec{V}^{\scriptscriptstyle\top}\Big]
      \end{align*}
      \mbox{}\vspace*{-7ex}
      
   \end{proof}
    
    \subsection*{Symplectic geometry}\trassub
    
    Let $S=\{\varphi:M\to\R\ /\ \square\varphi=0\}$ be the space of solutions to the Klein-Gordon equation with Dirichlet or Robin boundary bondition over a globally hyperbolic space-time $(M,g)$. We define the map $F:\mathrm{Emb}^\partial_{g\textrm{-sl}}(\Sigma,M)\times S\times S\to\R$ given by 
    \[F(X,\varphi_1,\varphi_2)=\int_{\Sigma}\peqsubfino{{\mathrm{vol}_\gamma}}{\!X}{-0.3ex}\left[\varphi_2\peqsubfino{{\mathcal{L}_{\vec{n}}}}{\!X}{-0.3ex}\varphi_1-\varphi_1\peqsubfino{{\mathcal{L}_{\vec{n}}}}{\!X}{-0.3ex}\varphi_2\right]\smallcirc X\]
    where $\peqsub{\vec{n}}{X}$ is the vector field along $X$ which is future directed and orthonormal to $X(\Sigma)$. 
    \begin{lemma}\label{Appendix - lemma - Omega(sol,sol) embedding}\mbox{}\\
    	$F$ does not depend on $X$.
    \end{lemma}
    \begin{proof}\mbox{}\\
      Although we could obtain this result using the variations we have already computed, it is much more direct with the following approach. Take two embeddings $X_1,X_2\in\mathrm{Emb}^\partial_{g\textrm{-sl}}(\Sigma,M)$ that do not intersect and consider the submanifold $N\subset M$ between $X_1(\Sigma)$, $X_2(\Sigma)$ and $\peqsub{\partial}{\Sigma}M$. We consider the $(n-1)$-form
      \[\beta=\varphi_1\peqsub{\star}{g}\mathrm{d}\varphi_2-\varphi_2\peqsub{\star}{g}\mathrm{d}\varphi_1\]
      and compute its differential
      \begin{align*}
      \mathrm{d}\beta&=\mathrm{d}\varphi_1\wedge\peqsub{\star}{g}\mathrm{d}\varphi_2+\varphi_1\mathrm{d}\peqsub{\star}{g}\mathrm{d}\varphi_2-\mathrm{d}\varphi_2\wedge\peqsub{\star}{g}\mathrm{d}\varphi_1-\varphi_2\mathrm{d}\peqsub{\star}{g}\mathrm{d}\varphi_1\updown{\eqref{Appendix - equation - definition Hodge (a,b)vol=a wedge*b}}{\eqref{appendix property star delta=d star}}{=}\\
      &=\peqsubfino{\left\langle\mathrm{d}\varphi_1\coma\mathrm{d}\varphi_2\right\rangle}{\!g}{-0.2ex}\peqsub{\mathrm{vol}}{g}-\varphi_1\peqsub{\star}{g}\delta\mathrm{d}\varphi_2-\peqsubfino{\left\langle\mathrm{d}\varphi_2\coma\mathrm{d}\varphi_1\right\rangle}{\!g}{-0.2ex}\peqsub{\mathrm{vol}}{g}-\varphi_2\mathrm{d}\peqsub{\star}{g}\mathrm{d}\varphi_1=0
      \end{align*}
      It is indeed zero because $\delta\mathrm{d}=0$ is precisely the Klein-Gordon equation and $\varphi_1,\varphi_2\in S$. Now we integrate $\mathrm{d}\beta\in\Omega^n(M)$ over $N$ and apply the Stoke's theorem \ref{Mathematical background - theorem - stokes}.
      \begin{align*}
        0&=\int_N\mathrm{d}\beta=\int_{\partial N}\peqsub{\imath^*}{\partial}\beta=\int_{\partial N}\peqsub{\imath^*}{\partial}\Big(\varphi_1\wedge\peqsub{\star}{g}\mathrm{d}\varphi_2-\varphi_2\wedge\peqsub{\star}{g}\mathrm{d}\varphi_1\Big)\overset{\ref{Appendix - lemma - j*(a wedge star b)=(a,i_nu b)vol}}{=}\\
        &=\int_{\partial N}\Big(\peqsubfino{\left\langle\varphi_1\coma\imath_{\vec{\mathcal{V}}}\mathrm{d}\varphi_2\right\rangle}{\!\!g}{-0.2ex}-\peqsubfino{\left\langle\varphi_2\coma\imath_{\vec{\mathcal{V}}}\mathrm{d}\varphi_1\right\rangle}{\!\!g}{-0.2ex}\!\Big)\peqsub{{\mathrm{vol}_g}}{\!\partial}\overset{\dagger}{=}\\
        &=\int_{X_2(\Sigma)}\!\Big(\!\peqsubfino{\left\langle\varphi_1\coma\imath_{-}\mathrm{d}\varphi_2\right\rangle}{\!\!g}{-0.2ex}-\peqsubfino{\left\langle\varphi_2\coma\imath_{-}\mathrm{d}\varphi_1\right\rangle}{\!\!g}{-0.2ex}\!\Big)\!\big(\peqsub{\vec{n}}{X_2}\big)\peqsub{{\mathrm{vol}_g}}{\!\partial}+\int_{\peqsub{\partial}{\Sigma}N}\!\Big(\!\peqsubfino{\left\langle\varphi_1\coma\imath_{-}\mathrm{d}\varphi_2\right\rangle}{\!\!g}{-0.2ex}-\peqsubfino{\left\langle\varphi_2\coma\imath_{-}\mathrm{d}\varphi_1\right\rangle}{\!\!g}{-0.2ex}\!\Big)\!(\vec{\nu})\peqsub{{\mathrm{vol}_g}}{\!\partial}+\\
        &\phantom{=}+\int_{X_1(\Sigma)}\!\Big(\!\peqsubfino{\left\langle\varphi_1\coma\imath_{-}\mathrm{d}\varphi_2\right\rangle}{\!\!g}{-0.2ex}-\peqsubfino{\left\langle\varphi_2\coma\imath_{-}\mathrm{d}\varphi_1\right\rangle}{\!\!g}{-0.2ex}\!\Big)\!\big(-\peqsub{\vec{n}}{X_1}\big)\peqsub{{\mathrm{vol}_g}}{\!\partial}\overset{\dagger}{=}\\
        &=\int_{X_2(\Sigma)}\!\Big(\varphi_1\peqsubfino{{\mathcal{L}_{\vec{n}}}}{\!X_2}{-0.3ex}\varphi_2-\varphi_2\peqsubfino{{\mathcal{L}_{\vec{n}}}}{\!X_2}{-0.3ex}\varphi_1\Big)\peqsub{{\mathrm{vol}_g}}{\!\partial}+\int_{\peqsub{\partial}{\Sigma}N}\!\Big(\varphi_1\imath_{\vec{\nu}}\mathrm{d}\varphi_2-\varphi_2\imath_{\vec{\nu}}\mathrm{d}\varphi_1\Big)\peqsub{{\mathrm{vol}_g}}{\!\partial}+\\
        &\phantom{=}+\int_{X_1(\Sigma)}\!\Big(-\varphi_1\peqsubfino{{\mathcal{L}_{\vec{n}}}}{\!X_1}{-0.3ex}\varphi_2+\varphi_2\peqsubfino{{\mathcal{L}_{\vec{n}}}}{\!X_1}{-0.3ex}\varphi_1\Big)\peqsub{{\mathrm{vol}_g}}{\!\partial}
      \end{align*}
      where in the $\dagger$ equality we have used that $\peqsubfino{g}{\partial}{-0.2ex}$ is the induced metric over the boundary and $\vec{\mathcal{V}}$ is the outer $g$-normal vector field of unit $1$ of $\partial N$. Besides, as the boundary of $N$ has three pieces $\partial N=X_1(\Sigma)\cup\peqsub{\partial}{\Sigma}M\cup X_2(\Sigma)$ we have
      \[\vec{\mathcal{V}}=\begin{cases}
      \peqsub{\vec{n}}{X_2}&X_2(\Sigma)\\
      \vec{\nu}&\peqsub{\partial}{\Sigma}N\\
      -\peqsub{\vec{n}}{X_1}&X_1(\Sigma)
      \end{cases}\]  
      Now using the Dirichlet or Robin boundary condition we get that $\varphi_1\imath_{\vec{\nu}}\mathrm{d}\varphi_2-\varphi_2\imath_{\vec{\nu}}\mathrm{d}\varphi_1$ vanishes over the boundary and, hence, the second integral is zero leading to
      \begin{equation}\int_{X_1(\Sigma)}\!\Big(\varphi_2\peqsubfino{{\mathcal{L}_{\vec{n}}}}{\!X_1}{-0.3ex}\varphi_1-\varphi_1\peqsubfino{{\mathcal{L}_{\vec{n}}}}{\!X_1}{-0.3ex}\varphi_2\Big)\peqsub{{\mathrm{vol}_g}}{\!\partial}=\int_{X_2(\Sigma)}\!\Big(\varphi_2\peqsubfino{{\mathcal{L}_{\vec{n}}}}{\!X_2}{-0.3ex}\varphi_1-\varphi_1\peqsubfino{{\mathcal{L}_{\vec{n}}}}{\!X_2}{-0.3ex}\varphi_2\Big)\peqsub{{\mathrm{vol}_g}}{\!\partial}
      \end{equation}
      Now noticing that the pullback of the metric through $X_i$ is $\peqsubfino{\gamma}{X_i}{-0.2ex}$, using lemma \ref{Appendix - lemma - volz^*g=z^*vol_g and star_z^*g=z^*star_g} and pulling back each integral with the corresponding $X_i$ we get that
      \[F(X_1,\varphi_1,\varphi_2)=F(X_2,\varphi_1,\varphi_2)\]
      We can obviously generalize this result to two intersecting embeddings by using an auxiliary one $X_3$ that do not intersect any of them.
  \end{proof}

    \subsection*{Parametrized theories}\trassub
    
   \begin{lemma}\label{appendix lemma variacion embedding energia cinetica}\mbox{}\\
   	Let $B\in\Cinf{M}$ (and let us denote $\peqsub{b}{Z}=Z^*B$), $F\in\Omega^k(M)$ and $Z\in\mathrm{Diff}(M)$. Given the action
   	\[S(F,Z)=\frac{1}{2}\int_M \peqsub{b}{Z}\,\peqsubfino{{\big\langle F,F\big\rangle_{\!g}}}{\!Z}{-0.2ex}\peqsub{{\mathrm{vol}_g}}{\!Z}\qquad\quad\text{then}\quad\qquad\peqsub{D}{(Z,\mathbb{V}_{\!Z})}S(F,Z)=-\int_M \peqsubfino{{\big\langle\hspace*{0.1ex}\peqsubfino{{\mathcal{L}_{\vec{\mathbb{V}}}}}{\!Z}{-0.2ex}F\hspace*{0.1ex},\hspace*{0.1ex}\peqsub{b}{Z} F\hspace*{0.1ex}\big\rangle_{\!g}}}{\!Z}{-0.2ex}\peqsub{{\mathrm{vol}_g}}{\!Z}\]
   \end{lemma}
   \begin{proof}\mbox{}\\
   	First notice that
   	\begin{align*}
   	  \mathcal{L}_{\vec{V}}\peqsubfino{{\big\langle F,F\big\rangle_{\!g}}}{\!Z}{-0.2ex}&=\frac{1}{k!}\mathcal{L}_{\vec{V}}\left(\prod_{i=1}^k\peqsubfino{g}{\!Z}{-0.2ex}^{a_ib_i}F_{a_1\cdots a_k}F_{b_1\cdots b_k}\right)=\\
   	  &=\frac{1}{k!}\mathcal{L}_{\vec{V}}\left(\prod_{i=1}^k\peqsubfino{g}{\!Z}{-0.2ex}^{a_ib_i}\right)F_{a_1\cdots a_k}F_{b_1\cdots b_k}+\frac{2}{k!}\prod_{i=1}^k\peqsubfino{g}{\!Z}{-0.2ex}^{a_ib_i}(\mathcal{L}_{\vec{V}}F)_{a_1\cdots a_k}F_{b_1\cdots b_k}=\\
   	  &=\frac{1}{k!}\mathcal{L}_{\vec{V}}\left(\prod_{i=1}^k\peqsubfino{g}{\!Z}{-0.2ex}^{a_ib_i}\right)F_{a_1\cdots a_k}F_{b_1\cdots b_k}+2\peqsubfino{{\big\langle \mathcal{L}_{\vec{V}}F,F\big\rangle_{\!g}}}{\!Z}{-0.2ex}
   \end{align*}
   Then we have
   \begin{align*}
	   &\peqsub{D}{(Z,\mathbb{V}_{\!Z})}S(F,Z)=\frac{1}{2k!}\int_M \peqsub{D}{(Z,\mathbb{V}_{\!Z})}\Big[\peqsub{b}{Z}\,\peqsubfino{{\big\langle F,F\big\rangle_{\!g}}}{\!Z}{-0.2ex}\peqsub{{\mathrm{vol}_g}}{\!Z}\Big]\updown{\eqref{Appendix - equation - variacion funcion}}{\eqref{Appendix - equation - variacion funcion}}{=}\\
	   &\ =\frac{1}{2}\int_M \left[\textcolor{red}{\left(\peqsub{\nabla}{\mathbb{V}_{\!Z}}\peqsub{b}{Z}\right)\peqsubfino{{\big\langle F,F\big\rangle_{\!g}}}{\!Z}{-0.2ex}}+\frac{\peqsub{b}{Z}}{k!}\peqsub{D}{(Z,\mathbb{V}_{\!Z})}\left(\prod_{i=1}^k\peqsubfino{g}{\!Z}{-0.2ex}^{a_ib_i}\right)F_{a_1\cdots a_k}F_{b_1\cdots b_k}+\textcolor{red}{\peqsub{b}{Z}\,\peqsubfino{{\big\langle F,F\big\rangle_{\!g}}}{\!Z}{-0.2ex}\peqsubfino{{\mathrm{div}_g}}{\!Z}{-0.2ex}\peqsub{\vec{\mathbb{V}}}{\!Z}}\right]\peqsub{{\mathrm{vol}_g}}{\!Z}\overset{\eqref{Appendix - equation - variacion inversa metrica}}{=}\\
	   &\ =\frac{1}{2}\int_M \left[\peqsub{b}{Z}\,\mathcal{L}_{\vec{V}}\peqsubfino{{\big\langle F,F\big\rangle_{\!g}}}{\!Z}{-0.2ex}-2\peqsub{b}{Z}\peqsubfino{{\big\langle\hspace*{0.1ex} \peqsubfino{{\mathcal{L}_{\vec{\mathbb{V}}}}}{\!Z}{-0.2ex}F\hspace*{0.1ex},\hspace*{0.1ex}F\hspace*{0.1ex}\big\rangle_{\!g}}}{\!Z}{-0.2ex}+\textcolor{red}{\peqsubfino{{\big\langle F,F\big\rangle_{\!g}}}{\!Z}{-0.2ex}\peqsubfino{{\mathrm{div}_g}}{\!Z}{-0.2ex}\big(\peqsub{b}{Z}\peqsub{\vec{\mathbb{V}}}{\!Z}\big)}\right]\peqsub{{\mathrm{vol}_g}}{\!Z}\overset{\eqref{Appendix - equation - derivada de lie funcion}}{=}\\
	   &\ =\frac{1}{2}\int_M \left[-2\peqsub{b}{Z}\peqsubfino{{\big\langle\hspace*{0.1ex} \peqsubfino{{\mathcal{L}_{\vec{\mathbb{V}}}}}{\!Z}{-0.2ex}F\hspace*{0.1ex},\hspace*{0.1ex}F\hspace*{0.1ex}\big\rangle_{\!g}}}{\!Z}{-0.2ex}+\peqsubfino{{\mathrm{div}_g}}{\!Z}{-0.2ex}\big(\peqsub{b}{Z}\,\peqsubfino{{\big\langle F,F\big\rangle_{\!g}}}{\!Z}{-0.2ex}\peqsub{\vec{\mathbb{V}}}{\!Z}\big)\right]\peqsub{{\mathrm{vol}_g}}{\!Z}\overset{(\refconchap{appendix equation integracion por partes})}{=}\\
	   &\ =-\!\int_M\!\peqsubfino{{\big\langle\hspace*{0.1ex}\peqsubfino{{\mathcal{L}_{\vec{\mathbb{V}}}}}{\!Z}{-0.2ex}F\hspace*{0.1ex},\hspace*{0.1ex}\peqsub{b}{Z} F\big\rangle_{\!g}}}{\!Z}{-0.2ex}\peqsub{{\mathrm{vol}_g}}{\!Z}+\frac{1}{2}\int_{\partial M} \imath_\partial^*\left(\peqsub{b}{Z}\,\peqsubfino{{\big\langle F,F\big\rangle_{\!g}}}{\!Z}{-0.2ex}\right)\peqsub{\nu}{\!Z}(\peqsub{\vec{\mathbb{V}}}{\!Z})\peqsubfino{{\mathrm{vol}_{g^{\partial}}}}{\!\!\!\!Z}{-0.2ex}\overset{\nu\perp\mathbb{V}}{=}-\!\int_M\! \peqsubfino{{\big\langle\hspace*{0.1ex}\peqsubfino{{\mathcal{L}_{\vec{\mathbb{V}}}}}{\!Z}{-0.2ex}F\hspace*{0.1ex},\hspace*{0.1ex}\peqsub{b}{Z} F\hspace*{0.1ex}\big\rangle_{\!g}}}{\!Z}{-0.2ex}\peqsub{{\mathrm{vol}_g}}{\!Z}
  \end{align*}
   \mbox{}\vspace*{-7ex}
   
\end{proof}

  \begin{lemma}\label{appendix - lemma - descomposicion dA}\mbox{}\\
  	Given a foliation over a globally hyperbolic manifold (with a transversal vector field $\partial_t$ and their associated lapse\index{Lapse} $\vec{N}$ and shift\index{Shift} $\mathbf{N}$), we have that for a given $A\in\Omega^k(M)$ its exterior derivative $\mathrm{d}A\in\Omega^{k+1}(M)$ can be decomposed as follows
  	\begin{align*}
  	&\peqsub{(\mathrm{d}A)}{\perp}=\varepsilon\frac{\mathcal{L}_{\partial_t}A^{\scriptscriptstyle\top}-\mathcal{L}_{\vec{N}}A^{\scriptscriptstyle\top}}{\mathbf{N}}-\frac{\mathrm{d}^{\scriptscriptstyle\top}(\mathbf{N}\peqsub{A}{\perp})}{\mathbf{N}}\\
  	&(\mathrm{d}A)^{\scriptscriptstyle\top}=\mathrm{d}^{\scriptscriptstyle\top}\!A^{\scriptscriptstyle\top}
  	\end{align*}  
  \end{lemma}
\begin{proof}\mbox{}\\
First notice that $n=\varepsilon\mathbf{N}\mathrm{d}t$, thus
\begin{align*}
\mathrm{d}(n\wedge\peqsub{A}{\perp})&=\varepsilon\mathrm{d}\mathbf{N}\wedge\mathrm{d}t\wedge\peqsub{A}{\perp}+(-1)^{|n|}n\wedge\mathrm{d}\peqsub{A}{\perp}=\\
&=-n\wedge\frac{\mathrm{d}\mathbf{N}}{\mathbf{N}}\wedge\peqsub{A}{\perp}-n\wedge\mathrm{d}\peqsub{A}{\perp}=-n\wedge\left(\frac{\mathrm{d}\mathbf{N}}{\mathbf{N}}\wedge\peqsub{A}{\perp}+\mathrm{d}\peqsub{A}{\perp}\right)
\end{align*}
Now using the decomposition $A=n\wedge\peqsub{A}{\perp}+A^{\scriptscriptstyle\top}$  we have
\begin{align*}
(\mathrm{d}A&)^{\scriptscriptstyle\top}=\mathrm{d}A-\varepsilon n\wedge\imath_{\vec{n}}\mathrm{d}A=\mathrm{d}(n\wedge\peqsub{A}{\perp}+A^{\scriptscriptstyle\top})-\varepsilon n\wedge\imath_{\vec{n}}\mathrm{d}(n\wedge\peqsub{A}{\perp}+A^{\scriptscriptstyle\top})=\\
&=-n\wedge\left(\frac{\mathrm{d}\mathbf{N}}{\mathbf{N}}\wedge\peqsub{A}{\perp}+\mathrm{d}\peqsub{A}{\perp}\right)+\textcolor{red}{\mathrm{d}(A^{\scriptscriptstyle\top})}-\varepsilon n\wedge\imath_{\vec{n}}\left[-n\wedge\left(\frac{\mathrm{d}\mathbf{N}}{\mathbf{N}}\wedge\peqsub{A}{\perp}+\mathrm{d}\peqsub{A}{\perp}\right)\right]-\textcolor{red}{\varepsilon n\wedge\imath_{\vec{n}}\mathrm{d}(A^{\scriptscriptstyle\top})}=\\
&=\textcolor{red}{\mathrm{d}^{\scriptscriptstyle\top}\!A^{\scriptscriptstyle\top}}-n\wedge\left(\frac{\mathrm{d}\mathbf{N}}{\mathbf{N}}\wedge\peqsub{A}{\perp}+\mathrm{d}\peqsub{A}{\perp}\right)+\varepsilon n\wedge\varepsilon\left(\frac{\mathrm{d}\mathbf{N}}{\mathbf{N}}\wedge\peqsub{A}{\perp}+\mathrm{d}\peqsub{A}{\perp}\right)-0=\mathrm{d}^{\scriptscriptstyle\top}\!A^{\scriptscriptstyle\top}
\end{align*}
\begin{align*}
&\peqsub{(\mathrm{d}A)}{\perp}+\frac{\mathrm{d}^{\scriptscriptstyle\top}(\mathbf{N}\peqsub{A}{\perp})}{\mathbf{N}}=\varepsilon\imath_{\vec{n}}\mathrm{d}A+\frac{\mathrm{d}(\mathbf{N}\peqsub{A}{\perp})-\varepsilon n \wedge\imath_{\vec{n}}\mathrm{d}(\mathbf{N}\peqsub{A}{\perp})}{\mathbf{N}}=\\
&=\varepsilon\imath_{\vec{n}}\mathrm{d}(n\wedge\peqsub{A}{\perp}+A^{\scriptscriptstyle\top})+\frac{1}{\mathbf{N}}\Big[\mathrm{d}\mathbf{N}\wedge\peqsub{A}{\perp}+\mathbf{N}\mathrm{d}\peqsub{A}{\perp}-\varepsilon n \wedge\imath_{\vec{n}}\big(\mathrm{d}\mathbf{N}\wedge\peqsub{A}{\perp}+\mathbf{N}\mathrm{d}\peqsub{A}{\perp}\big)\Big]=\\[1ex]
&=\varepsilon\imath_{\vec{n}}\left[-n\wedge\left(\frac{\mathrm{d}\mathbf{N}}{\mathbf{N}}\wedge\peqsub{A}{\perp}+\mathrm{d}\peqsub{A}{\perp}\right)\right]+\varepsilon\imath_{\vec{n}}\mathrm{d}A^{\scriptscriptstyle\top}+\frac{\mathrm{d}\mathbf{N}}{\mathbf{N}}\wedge\peqsub{A}{\perp}+\mathrm{d}\peqsub{A}{\perp}-\varepsilon n \wedge\imath_{\vec{n}}\left(\frac{\mathrm{d}\mathbf{N}}{\mathbf{N}}\wedge\peqsub{A}{\perp}+\mathrm{d}\peqsub{A}{\perp}\right)\updown{\eqref{Appendix - equation - regla de Leibniz producto interior}}{\eqref{appendix - formula - L=di+id}}{=}\\
&=-\varepsilon^2\left(\frac{\mathrm{d}\mathbf{N}}{\mathbf{N}}\wedge\peqsub{A}{\perp}+\mathrm{d}\peqsub{A}{\perp}\right)+\varepsilon(\mathcal{L}_{\vec{n}}-\mathrm{d}\imath_{\vec{n}})A^{\scriptscriptstyle\top}+\left(\frac{\mathrm{d}\mathbf{N}}{\mathbf{N}}\wedge\peqsub{A}{\perp}+\mathrm{d}\peqsub{A}{\perp}\right)=\\
&=\varepsilon\mathcal{L}_{\frac{1}{\mathbf{N}}(\partial_t-\vec{N})}A^{\scriptscriptstyle\top}-0\overset{\eqref{Appendix - equation - derivada de lie L_(fV)}}{=}\\
&=\frac{\varepsilon}{\mathbf{N}}\mathcal{L}_{\partial_t-\vec{N}}A^{\scriptscriptstyle\top}+\varepsilon\mathrm{d}\left(\frac{1}{\mathbf{N}}\right)\wedge\mathbf{N}\imath_{\vec{n}}A^{\scriptscriptstyle\top}=\\
&=\varepsilon\frac{\mathcal{L}_{\partial_t}A^{\scriptscriptstyle\top}-\mathcal{L}_{\vec{N}}A^{\scriptscriptstyle\top}}{\mathbf{N}}+0=\varepsilon\frac{\mathcal{L}_{\partial_t}A^{\scriptscriptstyle\top}-\mathcal{L}_{\vec{N}}A^{\scriptscriptstyle\top}}{\mathbf{N}}
\end{align*}
\mbox{}\vspace*{-7ex}

\end{proof}

\begin{lemma}\label{Appendix - equation - FL(w3)=Pi}\mbox{}\\
	Following the definitions of \eqref{Parametrized theories - equation - Lagrangian} and \eqref{Parametrized theories - definitions - H's}, we have
	\begin{align*}
		FL&(\mathbf{v}_{(\peqsubfino{q}{\perp}{-0.2ex},q,X)})\left(\mathbf{w}^3_{(\peqsubfino{q}{\perp}{-0.2ex},q,X)}\right)=-\int_\Sigma \mathbb{W}^\alpha\Big((\peqsubfino{e}{X}{-0.2ex})^b_\alpha\mathcal{H}_b+\varepsilon(\peqsub{n}{X})_\alpha\peqsubfino{\mathcal{H}}{\!\perp}{-0.1ex}\Big)\peqsub{\mathrm{vol}}{\Sigma}\,-{}\\
	&\hspace*{30ex}-\int_{\partial\Sigma}\mathbb{W}^\alpha\Big(\varepsilon\theta_\alpha\peqsubfino{\mathcal{H}}{\!\perp}{-0.1ex}^{\scriptscriptstyle\partial\,B}+\varepsilon n_\alpha\peqsubfino{\mathcal{H}}{\!\perp}{-0.1ex}^{\scriptscriptstyle\partial}+\mathcal{H}^{\scriptscriptstyle\partial}_a e^a_\alpha\Big)\peqsub{{\mathrm{vol}}}{\partial\Sigma}
	\end{align*}
\end{lemma}
\begin{proof}\mbox{}\\
	The Lagrangian \eqref{Parametrized theories - equation - Lagrangian} can be rewritten as
\begin{align*}
\begin{split}
	L(\mathbf{v}_{(\peqsubfino{q}{\perp}{-0.2ex},q,X)})&=\frac{1}{2}\left\llangle v-\mathcal{L}_{\peqsub{\vec{v}}{X}^{\scriptscriptstyle\top}}q-\mathrm{d}(\peqsub{n}{X}(\peqsub{\mathbb{V}}{X})\peqsubfino{q}{\perp}{-0.2ex}),\frac{v-\mathcal{L}_{\peqsub{\vec{v}}{X}^{\scriptscriptstyle\top}}q-\mathrm{d}(\peqsub{n}{X}(\peqsub{\mathbb{V}}{X})\peqsubfino{q}{\perp}{-0.2ex})}{\varepsilon\peqsub{n}{X}(\peqsub{\mathbb{V}}{X})}\right\rrangle_{\!\!\peqsubfino{\gamma}{\!X}{-0.4ex}}\!+\frac{1}{2}{\big\llangle\peqsub{n}{X}(\peqsub{\mathbb{V}}{X})}\mathrm{d}q,\mathrm{d}q\big\rrangle_{\!\peqsubfino{\gamma}{\!X}{-0.4ex}}\!+{}\\
	&\phantom{=}-\frac{1}{2}\big\llangle \peqsub{\theta}{X}(\peqsub{\mathbb{V}}{X})\peqsub{b}{X}^2\,\peqsubfino{q^{\scriptscriptstyle\partial}}{\perp}{-0.2ex},\,\peqsubfino{q^{\scriptscriptstyle\partial}}{\perp}{-0.2ex}\big\rrangle_{\!\peqsubfino{\gamma^{\scriptscriptstyle\partial}}{\!X}{-0.4ex}}\!-\frac{\varepsilon}{2}\big\llangle \peqsub{\theta}{X}(\peqsub{\mathbb{V}}{X})\peqsub{b}{X}^2\, \peqsubfino{q}{\partial}{-0.2ex}, \peqsubfino{q}{\partial}{-0.2ex}\big\rrangle_{\!\peqsubfino{\gamma^{\scriptscriptstyle\partial}}{\!X}{-0.4ex}}
\end{split}
\end{align*}
	\begin{align*}
&F\!L(\mathbf{v}_{(\peqsubfino{q}{\perp}{-0.2ex},q,X)})\left(\mathbf{w}^3_{(\peqsubfino{q}{\perp}{-0.2ex},q,X)}\right)=\left.\frac{\mathrm{d}}{\mathrm{d}\lambda}\right|_0 L(\mathbf{v}_{(\peqsubfino{q}{\perp}{-0.2ex},q,X)}+\lambda \mathbf{w}^3_{(\peqsubfino{q}{\perp}{-0.2ex},q,X)})=\\
&=\frac{1}{2}\left.\frac{\mathrm{d}}{\mathrm{d}\lambda}\right|_0\Bigg\{\left\llangle v-\mathcal{L}_{\peqsub{\vec{v}}{X}^{\scriptscriptstyle\top}+\lambda\peqsub{\vec{w}}{X}^{\scriptscriptstyle\top}}q-\mathrm{d}\Big(\peqsub{n}{X}(\peqsub{\mathbb{V}}{X}+\lambda\peqsub{\mathbb{W}}{X})\peqsubfino{q}{\perp}{-0.2ex}\Big)\hspace*{0.1ex},\hspace*{0.1ex}\frac{v-\mathcal{L}_{\peqsub{\vec{v}}{X}^{\scriptscriptstyle\top}+\lambda\peqsub{\vec{w}}{X}^{\scriptscriptstyle\top}}q-\mathrm{d}\Big(\peqsub{n}{X}(\peqsub{\mathbb{V}}{X}+\lambda\peqsub{\mathbb{W}}{X})\peqsubfino{q}{\perp}{-0.2ex}\Big)}{\varepsilon\peqsub{n}{X}(\peqsub{\mathbb{V}}{X}+\lambda\peqsub{\mathbb{W}}{X})}\right\rrangle_{\!\!\peqsubfino{\gamma}{\!X}{-0.4ex}}\!+{}\\
&+\frac{1}{2}{\big\llangle\peqsub{n}{X}(\peqsub{\mathbb{V}}{X}+\lambda\peqsub{\mathbb{W}}{X})}\mathrm{d}q,\mathrm{d}q\big\rrangle_{\!\peqsubfino{\gamma}{\!X}{-0.4ex}}\!-\frac{1}{2}\big\llangle \peqsub{\theta}{X}(\peqsub{\mathbb{V}}{X}+\lambda\peqsub{\mathbb{W}}{X})\peqsub{b}{X}^2\,\peqsubfino{q^{\scriptscriptstyle\partial}}{\perp}{-0.2ex},\,\peqsubfino{q^{\scriptscriptstyle\partial}}{\perp}{-0.2ex}\big\rrangle_{\!\peqsubfino{\gamma^{\scriptscriptstyle\partial}}{\!X}{-0.4ex}}\!-\frac{\varepsilon}{2}\big\llangle \peqsub{\theta}{X}(\peqsub{\mathbb{V}}{X}+\lambda\peqsub{\mathbb{W}}{X})\peqsub{b}{X}^2\, \peqsubfino{q}{\partial}{-0.2ex},\, \peqsubfino{q}{\partial}{-0.2ex}\big\rrangle_{\!\peqsubfino{\gamma^{\scriptscriptstyle\partial}}{\!X}{-0.4ex}}\!\Bigg\}=\\
&=\left\llangle -\mathcal{L}_{\peqsub{\vec{w}}{X}^{\scriptscriptstyle\top}}q-\mathrm{d}\Big(\peqsub{n}{X}(\peqsub{\mathbb{W}}{X})\peqsubfino{q}{\perp}{-0.2ex}\Big)\hspace*{0.1ex},\hspace*{0.1ex}\frac{v-\mathcal{L}_{\peqsub{\vec{w}}{X}^{\scriptscriptstyle\top}}q-\mathrm{d}\Big(\peqsub{n}{X}(\peqsub{\mathbb{V}}{X})\peqsubfino{q}{\perp}{-0.2ex}\Big)}{\varepsilon\peqsub{n}{X}(\peqsub{\mathbb{V}}{X})}\right\rrangle_{\!\!\peqsubfino{\gamma}{\!X}{-0.4ex}}\!-{}\\
&-\frac{1}{2}\left\llangle v-\mathcal{L}_{\peqsub{\vec{v}}{X}^{\scriptscriptstyle\top}}q-\mathrm{d}\Big(\peqsub{n}{X}(\peqsub{\mathbb{V}}{X})\peqsubfino{q}{\perp}{-0.2ex}\Big)\hspace*{0.1ex},\hspace*{0.1ex}\frac{v-\mathcal{L}_{\peqsub{\vec{v}}{X}^{\scriptscriptstyle\top}}q-\mathrm{d}\Big(\peqsub{n}{X}(\peqsub{\mathbb{V}}{X})\peqsubfino{q}{\perp}{-0.2ex}\Big)}{\varepsilon\peqsub{n}{X}(\peqsub{\mathbb{V}}{X})}\frac{\varepsilon\peqsub{n}{X}(\peqsub{\mathbb{W}}{X})}{\varepsilon\peqsub{n}{X}(\peqsub{\mathbb{V}}{X})}\right\rrangle_{\!\!\peqsubfino{\gamma}{\!X}{-0.4ex}}\!+{}\\
&+\frac{1}{2}{\big\llangle\peqsub{n}{X}(\peqsub{\mathbb{W}}{X})}\mathrm{d}q,\mathrm{d}q\big\rrangle_{\!\peqsubfino{\gamma}{\!X}{-0.4ex}}\!-\frac{1}{2}\big\llangle \peqsub{\theta}{X}(\peqsub{\mathbb{W}}{X})\peqsub{b}{X}^2\,\peqsubfino{q^{\scriptscriptstyle\partial}}{\perp}{-0.2ex},\,\peqsubfino{q^{\scriptscriptstyle\partial}}{\perp}{-0.2ex}\big\rrangle_{\!\peqsubfino{\gamma^{\scriptscriptstyle\partial}}{\!X}{-0.4ex}}\!-\frac{\varepsilon}{2}\big\llangle \peqsub{\theta}{X}(\peqsub{\mathbb{W}}{X})\peqsub{b}{X}^2\, \peqsubfino{q}{\partial}{-0.2ex},\, \peqsubfino{q}{\partial}{-0.2ex}\big\rrangle_{\!\peqsubfino{\gamma^{\scriptscriptstyle\partial}}{\!X}{-0.4ex}}\updown{\star}{\eqref{appendix - formula - L=di+id}}{=}\\
&=\left\llangle -\mathcal{\imath}_{\peqsub{\vec{w}}{X}^{\scriptscriptstyle\top}}\mathrm{d}q-\mathrm{d}\mathcal{\imath}_{\peqsub{\vec{w}}{X}^{\scriptscriptstyle\top}}q-\mathrm{d}\Big(\peqsub{n}{X}(\peqsub{\mathbb{W}}{X})\peqsubfino{q}{\perp}{-0.2ex}\Big)\hspace*{0.1ex},\hspace*{0.1ex}\frac{\underline{p}}{\sqrt{\peqsubfino{\gamma}{X}{-0.2ex}}}\right\rrangle_{\!\!\peqsubfino{\gamma}{\!X}{-0.4ex}}\!-\frac{1}{2}\left\llangle \frac{\underline{p}}{\sqrt{\peqsubfino{\gamma}{X}{-0.2ex}}}\hspace*{0.1ex},\hspace*{0.1ex}\frac{\underline{p}}{\sqrt{\peqsubfino{\gamma}{X}{-0.2ex}}}\varepsilon\peqsub{n}{X}(\peqsub{\mathbb{W}}{X})\right\rrangle_{\!\!\peqsubfino{\gamma}{\!X}{-0.4ex}}\!+{}\\
&+\frac{1}{2}{\big\llangle\peqsub{n}{X}(\peqsub{\mathbb{W}}{X})}\mathrm{d}q,\mathrm{d}q\big\rrangle_{\!\peqsubfino{\gamma}{\!X}{-0.4ex}}\!-\frac{1}{2}\big\llangle \peqsub{\theta}{X}(\peqsub{\mathbb{W}}{X})\peqsub{b}{X}^2\,\peqsubfino{q^{\scriptscriptstyle\partial}}{\perp}{-0.2ex},\,\peqsubfino{q^{\scriptscriptstyle\partial}}{\perp}{-0.2ex}\big\rrangle_{\!\peqsubfino{\gamma^{\scriptscriptstyle\partial}}{\!X}{-0.4ex}}\!-\frac{\varepsilon}{2}\big\llangle \peqsub{\theta}{X}(\peqsub{\mathbb{W}}{X})\peqsub{b}{X}^2\, \peqsubfino{q}{\partial}{-0.2ex},\, \peqsubfino{q}{\partial}{-0.2ex}\big\rrangle_{\!\peqsubfino{\gamma^{\scriptscriptstyle\partial}}{\!X}{-0.4ex}}\hspace*{-3ex}\updown{\eqref{appendix equation integracion por partes}}{\qquad\vec{\nu}^{\,\scriptscriptstyle\top}/|\vec{\nu}^{\,\scriptscriptstyle\top}|}{\hspace*{-2ex}=}\\
&=-\left\llangle \mathcal{\imath}_{\peqsub{\vec{w}}{X}^{\scriptscriptstyle\top}}\mathrm{d}q\hspace*{0.1ex},\hspace*{0.1ex}\frac{\underline{p}}{\sqrt{\peqsubfino{\gamma}{X}{-0.2ex}}}\right\rrangle_{\!\!\peqsubfino{\gamma}{\!X}{-0.4ex}}\!-\left\llangle \mathcal{\imath}_{\peqsub{\vec{w}}{X}^{\scriptscriptstyle\top}}q+\peqsub{n}{X}(\peqsub{\mathbb{W}}{X})\peqsubfino{q}{\perp}{-0.2ex}\hspace*{0.1ex},\hspace*{0.1ex}\frac{\delta\underline{p}}{\sqrt{\peqsubfino{\gamma}{X}{-0.2ex}}}\right\rrangle_{\!\!\peqsubfino{\gamma}{\!X}{-0.4ex}}\!-\left\llangle \mathcal{\imath}_{\peqsub{\vec{w}}{X}^{\scriptscriptstyle\top}}q+\peqsub{n}{X}(\peqsub{\mathbb{W}}{X})\peqsubfino{q}{\perp}{-0.2ex}\hspace*{0.1ex},\hspace*{0.1ex}\frac{\imath_{\vec{\nu}}\underline{p}}{|\vec{\nu}^{\,\scriptscriptstyle\top}|\sqrt{\peqsubfino{\gamma}{X}{-0.2ex}}}\right\rrangle_{\!\!\peqsubfino{\gamma^{\scriptscriptstyle\partial}}{\!X}{-0.4ex}}-{}\\
&-\frac{\varepsilon}{2}\left\llangle \frac{\underline{p}}{\sqrt{\peqsubfino{\gamma}{X}{-0.2ex}}}\hspace*{0.1ex},\hspace*{0.1ex}\frac{\underline{p}}{\sqrt{\peqsubfino{\gamma}{X}{-0.2ex}}}\peqsub{n}{X}(\peqsub{\mathbb{W}}{X})\right\rrangle_{\!\!\peqsubfino{\gamma}{\!X}{-0.4ex}}\!\!+\frac{1}{2}{\big\llangle\peqsub{n}{X}(\peqsub{\mathbb{W}}{X})}\mathrm{d}q,\mathrm{d}q\big\rrangle_{\!\peqsubfino{\gamma}{\!X}{-0.4ex}}\!-\frac{1}{2}\big\llangle \peqsub{\theta}{X}(\peqsub{\mathbb{W}}{X})\peqsub{b}{X}^2\,\peqsubfino{q^{\scriptscriptstyle\partial}}{\perp}{-0.2ex},\,\peqsubfino{q^{\scriptscriptstyle\partial}}{\perp}{-0.2ex}\big\rrangle_{\!\peqsubfino{\gamma^{\scriptscriptstyle\partial}}{\!X}{-0.4ex}}\!-\frac{\varepsilon}{2}\big\llangle \peqsub{\theta}{X}(\peqsub{\mathbb{W}}{X})\peqsub{b}{X}^2\, \peqsubfino{q}{\partial}{-0.2ex},\, \peqsubfino{q}{\partial}{-0.2ex}\big\rrangle_{\!\peqsubfino{\gamma^{\scriptscriptstyle\partial}}{\!X}{-0.4ex}}\!=\\
&=-\int_\Sigma\Big[\langle\imath_{-}\mathrm{d}q,\underline{p}\rangle_{\raisemath{-0.4ex}{\!\peqsubfino{\gamma}{\!X}{-0.4ex}}}+\langle\imath_{-}q,\delta\underline{p}\rangle_{\raisemath{-0.4ex}{\!\peqsubfino{\gamma}{\!X}{-0.4ex}}}\Big]\peqsub{\vec{w}}{X}^{\scriptscriptstyle\top}\peqsub{\mathrm{vol}}{\Sigma}-{}\\
&-\int_\Sigma\peqsub{W}{X}^{\scriptscriptstyle\perp}\left[\frac{1}{2}\left\langle \frac{\underline{p}}{\sqrt{\peqsubfino{\gamma}{X}{-0.2ex}}}\hspace*{0.1ex},\hspace*{0.1ex}\frac{\underline{p}}{\sqrt{\peqsubfino{\gamma}{X}{-0.2ex}}}\right\rangle_{\raisemath{0.2ex}{\!\!\!\peqsubfino{\gamma}{\!X}{-0.4ex}}}-\frac{\varepsilon}{2}\big\langle\mathrm{d}q,\mathrm{d}q\big\rangle_{\!\peqsubfino{\gamma}{\!X}{-0.4ex}}+\varepsilon\left\langle \peqsubfino{q}{\perp}{-0.2ex}\hspace*{0.1ex},\hspace*{0.1ex}\frac{\delta\underline{p}}{\sqrt{\peqsubfino{\gamma}{X}{-0.2ex}}}\right\rangle_{\raisemath{0.2ex}{\!\!\!\peqsubfino{\gamma}{\!X}{-0.4ex}}}\right]\sqrt{\peqsubfino{\gamma}{\!X}{-0.2ex}}\,\peqsub{\mathrm{vol}}{\Sigma}\,-{}\\
&-\int_{\partial \Sigma}\left\{\left\langle \mathcal{\imath}_{\peqsub{\vec{w}}{X}^{\scriptscriptstyle\top}}q+\varepsilon\peqsubfino{q}{\perp}{-0.2ex}\peqsub{W}{X}^{\scriptscriptstyle\perp}\hspace*{0.1ex},\hspace*{0.1ex}\frac{\sqrt{\peqsubfino{\gamma^{\scriptscriptstyle\partial}}{\!X}{-0.3ex}}}{\sqrt{\peqsubfino{\gamma}{\!X}{-0.3ex}}}\frac{\imath_{\vec{\nu}}\underline{p}}{|\vec{\nu}^{\,\scriptscriptstyle\top}|}\right\rangle_{\!\!\!\peqsubfino{\gamma^{\scriptscriptstyle\partial}}{\!X}{-0.4ex}}+\varepsilon\peqsub{\theta}{X}(\peqsub{\mathbb{W}}{X})\sqrt{\peqsubfino{\gamma^{\scriptscriptstyle\partial}}{\!X}{-0.3ex}}\,\frac{\peqsub{b}{X}^2}{2}\Big[\varepsilon\big\langle\peqsubfino{q^{\scriptscriptstyle\partial}}{\perp}{-0.2ex},\peqsubfino{q^{\scriptscriptstyle\partial}}{\perp}{-0.2ex}\big\rangle_{\!\peqsubfino{\gamma^{\scriptscriptstyle\partial}}{\!X}{-0.4ex}}\!+\big\langle \peqsubfino{q}{\partial}{-0.2ex},\peqsubfino{q}{\partial}{-0.2ex}\big\rangle_{\!\peqsubfino{\gamma^{\scriptscriptstyle\partial}}{\!X}{-0.4ex}}\,\Big]\right\}\peqsub{\mathrm{vol}}{\partial\Sigma}
\end{align*}
where in the $\star$ equality we have defined $\underline{p}=\peqsubfino{{\flat_\gamma}}{\!X}{-0.2ex}p\in\Omega^k(\Sigma)$.
\end{proof}

We now proceed to compute\label{Appendix - computation - omega}
\[(\peqsub{\jmath}{\mathrm{FC}})_*Y=\Big(\vecc[\scalebox{0.4}{$\perp$}]{Y}{q},\vecc{Y}{q},\vecc{Y}{X},\vecc{Y}{p},-\peqsub{D}{Y}\big(\varepsilon n \peqsubfino{\mathcal{H}}{\!\perp}{-0.1ex}+e.\mathcal{H}\big),-\peqsub{D}{Y}\big(\varepsilon \theta \peqsubfino{\mathcal{H}}{\!\perp}{-0.1ex}^{\scriptscriptstyle\partial\,B}+\varepsilon n\peqsubfino{\mathcal{H}}{\!\perp}{-0.1ex}^{\scriptscriptstyle\partial}+e.\mathcal{H}^{\scriptscriptstyle\partial}\big)\Big)\in\mathfrak{X}(\widetilde{\mathcal{P}})\]	
For that, we need the variations of the objects involved
\begin{align*}
&\mathcal{H}(q,p)=\Big(\imath_{-}\mathrm{d}q,p\Big)+\Big(\imath_{-}q,\delta p\Big)\in\Omega^1(\Sigma)\\
&\peqsubfino{\mathcal{H}}{\!\perp}{-0.1ex}(\peqsubfino{q}{\perp}{-0.2ex},q,X,p)=\frac{1}{2\sqrt{\peqsubfino{\gamma}{X}{-0.2ex}}}\Big(\underline{p}\hspace*{0.1ex},\hspace*{0.1ex}p\Big)-\frac{\varepsilon\sqrt{\peqsubfino{\gamma}{X}{-0.2ex}}}{2}\big\langle\mathrm{d}q,\mathrm{d}q\big\rangle_{\!\peqsubfino{\gamma}{\!X}{-0.4ex}}+\varepsilon\Big(\peqsubfino{q}{\perp}{-0.2ex}\hspace*{0.1ex},\hspace*{0.1ex}\delta p\Big)\in\Cinf{\Sigma}\\
&\mathcal{H}^{\scriptscriptstyle\partial}(q,p)=\left(\frac{\sqrt{\peqsubfino{\gamma^{\scriptscriptstyle\partial}}{\!X}{-0.3ex}}}{\sqrt{\peqsubfino{\gamma}{\!X}{-0.3ex}}}\frac{\nu}{|\vec{\nu}^{\,\scriptscriptstyle\top}|}\wedge\imath_{-}q,p\right)\in\Omega^1(\partial\Sigma)\\
&\peqsubfino{\mathcal{H}}{\!\perp}{-0.1ex}^{\scriptscriptstyle\partial}(\peqsubfino{q}{\perp}{-0.2ex},p)=\varepsilon\,\left(\frac{\sqrt{\peqsubfino{\gamma^{\scriptscriptstyle\partial}}{\!X}{-0.3ex}}}{\sqrt{\peqsubfino{\gamma}{\!X}{-0.3ex}}}\frac{\nu}{|\vec{\nu}^{\,\scriptscriptstyle\top}|}\wedge\peqsubfino{q}{\perp}{-0.2ex},p\right)\in\Cinf{\partial\Sigma}\\
&\peqsubfino{\mathcal{H}}{\!\perp}{-0.1ex}^{\scriptscriptstyle\partial\,B}(\peqsubfino{q}{\perp}{-0.2ex},q,X)=\frac{\sqrt{\peqsubfino{\gamma^{\scriptscriptstyle\partial}}{\!X}{-0.3ex}}}{2}\peqsub{b}{X}^2\Big[ \varepsilon\big\langle\peqsubfino{q^{\scriptscriptstyle\partial}}{\perp}{-0.2ex},\peqsubfino{q^{\scriptscriptstyle\partial}}{\perp}{-0.2ex}\big\rangle_{\!\peqsubfino{\gamma^{\scriptscriptstyle\partial}}{\!X}{-0.4ex}}\!+\big\langle \peqsubfino{q}{\partial}{-0.2ex},\peqsubfino{q}{\partial}{-0.2ex}\big\rangle_{\!\peqsubfino{\gamma^{\scriptscriptstyle\partial}}{\!X}{-0.4ex}}\Big]\in\Cinf{\partial\Sigma}
\end{align*}

\begin{lemma}\mbox{}\\
	$\displaystyle\frac{1}{2} \peqsub{D}{Y}\left(\peqsubfino{\langle\alpha,\alpha\rangle}{\!\gamma}{-0.2ex}\sqrt{\gamma}\right)=\peqsubfino{\left\langle D\alpha+\frac{(D\gamma)_k}{2}\alpha-\frac{\mathrm{Tr}(D\gamma)}{4}\alpha,\alpha\right\rangle}{\!\!\!\gamma}{0.2ex}\sqrt{\gamma}$
\end{lemma}
\begin{proof}
\begin{align*}
  D\Big(&\peqsubfino{\langle\alpha,\beta\rangle}{\!\gamma}{-0.2ex}\sqrt{\gamma}\Big)\peqsub{\mathrm{vol}}{\Sigma}=D\Big(\peqsubfino{\langle\alpha,\beta\rangle}{\!\gamma}{-0.2ex}\sqrt{\gamma}\,\peqsub{\mathrm{vol}}{\Sigma}\Big)=D\Big(\peqsubfino{\langle\alpha,\beta\rangle}{\!\gamma}{-0.2ex}\peqsub{\mathrm{vol}}{\gamma}\Big)\overset{\eqref{Appendix - equation - definition Hodge (a,b)vol=a wedge*b}}{=}D\big(\alpha\wedge\star_\gamma\beta\big)=\\
  &=(D\alpha)\wedge\star\beta+\alpha\wedge (D\star)\beta+\alpha\wedge\star(D\beta)\updown{\eqref{Appendix - equation - definition Hodge (a,b)vol=a wedge*b}}{\eqref{Appendix - equation - variacion Hodge}}{=}\\
  &=\peqsubfino{\big\langle D\alpha,\beta\big\rangle}{\!\!\gamma}{-0.2ex}\peqsub{\mathrm{vol}}{\gamma}+\alpha\wedge \star\left[D\gamma-\frac{\mathrm{Tr}(D\gamma)}{2}\mathrm{Id}\right]\beta+\peqsubfino{\big\langle\alpha,D\beta \big\rangle}{\!\!\gamma}{-0.2ex}\peqsub{\mathrm{vol}}{\gamma}=\\
  &=\left(\peqsubfino{\big\langle D\alpha,\beta\big\rangle}{\!\!\gamma}{-0.2ex}+\peqsubfino{\big\langle\alpha,D\gamma(\beta)\big\rangle}{\!\!\gamma}{-0.2ex}-\frac{\mathrm{Tr}(D\gamma)}{2}\peqsubfino{\big\langle\alpha,\beta\big\rangle}{\!\!\gamma}{-0.2ex}+\peqsubfino{\big\langle\alpha,D\beta \big\rangle}{\!\!\gamma}{-0.2ex}\right)\sqrt{\gamma}\,\peqsub{\mathrm{vol}}{\Sigma}
\end{align*}
\mbox{}\vspace*{-7ex}

\end{proof}

In order to simplify the computation, from now on we will often use at our convenience the metric scalar product $\peqsubfino{\langle\,,\rangle}{\!\gamma}{-0.2ex}$ or the usual pairing $(\,,)$. They are of course related by $\peqsubfino{\gamma}{X}{-0.2ex}$. 
\begin{align*}
&\peqsub{D}{Y}\mathcal{H}=\Big(\imath_{-}\mathrm{d}\vecc{Y}{q},p\Big)+\Big(\imath_{-}\mathrm{d}q,\vecc{Y}{p}\Big)+\Big(\imath_{-}\vecc{Y}{q},\delta p\Big)+\Big(\imath_{-}q,\delta \vecc{Y}{p}\Big)\\
&\peqsub{D}{Y}\peqsubfino{\mathcal{H}}{\!\perp}{-0.1ex}=-\frac{\mathrm{Tr}(\peqsub{D}{Y}\gamma)}{2\sqrt{\peqsubfino{\gamma}{X}{-0.2ex}}}\Big(\underline{p}\hspace*{0.1ex},\hspace*{0.1ex}p\Big)+\frac{1}{2\sqrt{\peqsubfino{\gamma}{X}{-0.2ex}}}\Big((\peqsub{D}{Y}\gamma)_k\underline{p}\hspace*{0.1ex},\hspace*{0.1ex}p\Big)+\frac{1}{\sqrt{\peqsubfino{\gamma}{X}{-0.2ex}}}\Big(\underline{p}\hspace*{0.1ex},\hspace*{0.1ex}\vecc{Y}{p}\Big)-{}\\
&\hspace*{8ex}-\varepsilon\sqrt{\peqsubfino{\gamma}{X}{-0.2ex}}\peqsubfino{\left\langle \mathrm{d}\vecc{Y}{q}+\frac{(\peqsub{D}{Y}\gamma)_k}{2}\mathrm{d}q-\frac{\mathrm{Tr}(\peqsub{D}{Y}\gamma)}{4}\mathrm{d}q,\mathrm{d}q\right\rangle}{\!\!\!\gamma}{0.2ex}+\varepsilon\Big(\vecc[\perp]{Y}{q}\hspace*{0.1ex},\hspace*{0.1ex}\delta p\Big)+\varepsilon\Big(\peqsubfino{q}{\perp}{-0.2ex}\hspace*{0.1ex},\hspace*{0.1ex}\delta \vecc{Y}{p}\Big)\overset{\star}{=}\\
&\phantom{\peqsub{D}{Y}\peqsubfino{\mathcal{H}}{\!\perp}{-0.1ex}}=\frac{1}{\sqrt{\peqsubfino{\gamma}{X}{-0.2ex}}}\Big(\underline{p}\hspace*{0.1ex},\hspace*{0.1ex}\vecc{Y}{p}\Big)-\varepsilon\sqrt{\peqsubfino{\gamma}{X}{-0.2ex}}\peqsubfino{\left\langle \mathrm{d}\vecc{Y}{q},\mathrm{d}q\right\rangle}{\!\gamma}{0.2ex}+\varepsilon\Big(\vecc[\perp]{Y}{q}\hspace*{0.1ex},\hspace*{0.1ex}\delta p\Big)+\varepsilon\Big(\peqsubfino{q}{\perp}{-0.2ex}\hspace*{0.1ex},\hspace*{0.1ex}\delta \vecc{Y}{p}\Big)+\peqsubfino{\chi^{\scriptscriptstyle Y}}{2}{-0.2ex}\!\!\left(\frac{\underline{p}}{\sqrt{\peqsubfino{\gamma}{X}{-0.2ex}}}\right)-\varepsilon\peqsubfino{\chi^{\scriptscriptstyle Y}}{4}{-0.2ex}\!(\mathrm{d}q)\\
&\peqsub{D}{Y}\mathcal{H}^{\scriptscriptstyle\partial}=\left(\frac{\sqrt{\peqsubfino{\gamma^{\scriptscriptstyle\partial}}{\!X}{-0.3ex}}}{\sqrt{\peqsubfino{\gamma}{\!X}{-0.3ex}}}\frac{\nu}{|\vec{\nu}^{\,\scriptscriptstyle\top}|}\wedge\imath_{-}\vecc{Y}{q},p\right)+\left(\frac{\sqrt{\peqsubfino{\gamma^{\scriptscriptstyle\partial}}{\!X}{-0.3ex}}}{\sqrt{\peqsubfino{\gamma}{\!X}{-0.3ex}}}\frac{\nu}{|\vec{\nu}^{\,\scriptscriptstyle\top}|}\wedge\imath_{-}q,\vecc{Y}{p}\right)\\
&\peqsub{D}{Y}\peqsubfino{\mathcal{H}}{\!\perp}{-0.1ex}^{\scriptscriptstyle\partial}=\varepsilon\left(\frac{\sqrt{\peqsubfino{\gamma^{\scriptscriptstyle\partial}}{\!X}{-0.3ex}}}{\sqrt{\peqsubfino{\gamma}{\!X}{-0.3ex}}}\frac{\nu}{|\vec{\nu}^{\,\scriptscriptstyle\top}|}\wedge\vecc[\perp]{Y}{q},p\right)+\varepsilon\left(\frac{\sqrt{\peqsubfino{\gamma^{\scriptscriptstyle\partial}}{\!X}{-0.3ex}}}{\sqrt{\peqsubfino{\gamma}{\!X}{-0.3ex}}}\frac{\nu}{|\vec{\nu}^{\,\scriptscriptstyle\top}|}\wedge\peqsubfino{q}{\perp}{-0.2ex},\vecc{Y}{p}\right)\\
&\peqsub{D}{Y}\peqsubfino{\mathcal{H}}{\!\perp}{-0.1ex}^{\scriptscriptstyle\partial\,B}\!=\sqrt{\peqsubfino{\gamma^{\scriptscriptstyle\partial}}{\!X}{-0.3ex}}\peqsub{b}{X}^2\left[ \varepsilon\peqsubfino{{\left\langle  \vecc[\perp]{Y}{q}^{\scriptscriptstyle\partial}\!+\!\frac{(\peqsub{D}{Y}\gamma)_k}{2}\peqsubfino{q^{\scriptscriptstyle\partial}}{\perp}{-0.2ex}\!-\!\frac{\mathrm{Tr}(\peqsub{D}{Y}\gamma)}{4}\peqsubfino{q^{\scriptscriptstyle\partial}}{\perp}{-0.2ex},\peqsubfino{q^{\scriptscriptstyle\partial}}{\perp}{-0.2ex}\right\rangle_{\!\!\!\gamma}}}{\!\partial}{-.3ex}\!+\peqsubfino{{\left\langle  \vecc{Y}{q}^{\scriptscriptstyle\partial}\!+\!\frac{(\peqsub{D}{Y}\gamma)_k}{2}\peqsubfino{q}{\partial}{-0.2ex}\!-\!\frac{\mathrm{Tr}(\peqsub{D}{Y}\gamma)}{4}\peqsubfino{q}{\partial}{-0.2ex},\peqsubfino{q}{\partial}{-0.2ex}\right\rangle_{\!\!\!\gamma}}}{\!\partial}{-.3ex}\,\right]+{}\\
&\hspace*{10ex}+\frac{\peqsub{D}{Y}\sqrt{\peqsubfino{\gamma^{\scriptscriptstyle\partial}}{\!X}{-0.3ex}}}{2}\big[ \varepsilon\peqsubfino{\left\langle\peqsubfino{q^{\scriptscriptstyle\partial}}{\perp}{-0.2ex},\peqsubfino{q^{\scriptscriptstyle\partial}}{\perp}{-0.2ex}\right\rangle}{\!\gamma^\partial}{0.2ex}\!+\peqsubfino{\left\langle \peqsubfino{q}{\partial}{-0.2ex},\peqsubfino{q}{\partial}{-0.2ex}\right\rangle}{\!\gamma^\partial}{0.2ex}\big]\peqsub{b}{X}^2+\frac{\sqrt{\peqsubfino{\gamma^{\scriptscriptstyle\partial}}{\!X}{-0.3ex}}}{2}\big[ \varepsilon\peqsubfino{\left\langle\peqsubfino{q^{\scriptscriptstyle\partial}}{\perp}{-0.2ex},\peqsubfino{q^{\scriptscriptstyle\partial}}{\perp}{-0.2ex}\right\rangle}{\!\gamma^\partial}{0.2ex}\!+\peqsubfino{\left\langle \peqsubfino{q}{\partial}{-0.2ex},\peqsubfino{q}{\partial}{-0.2ex}\right\rangle}{\!\gamma^\partial}{0.2ex}\big]\peqsub{D}{Y}\peqsub{b}{X}^2=\\
&\phantom{\peqsub{D}{Y}\peqsubfino{\mathcal{H}}{\!\perp}{-0.1ex}^{\scriptscriptstyle\partial\,B}}=\peqsub{b}{X}^2\sqrt{\peqsubfino{\gamma^{\scriptscriptstyle\partial}}{\!X}{-0.3ex}}\left[ \varepsilon\peqsubfino{\left\langle\vecc[\perp]{Y}{q}^{\scriptscriptstyle\partial},\peqsubfino{q^{\scriptscriptstyle\partial}}{\perp}{-0.2ex}\right\rangle}{\!\gamma^\partial}{0.2ex}\!+\peqsubfino{\left\langle\vecc{Y}{q}^{\scriptscriptstyle\partial},\peqsubfino{q}{\partial}{-0.2ex}\right\rangle}{\!\gamma^\partial}{0.2ex}\right]+\peqsub{b}{X}^2\left[ \peqsubfino{\chi}{4}{-0.2ex}^{\scriptscriptstyle\partial Y}\!(\peqsubfino{q^{\scriptscriptstyle\partial}}{\perp}{-0.2ex})+\peqsubfino{\chi}{4}{-0.2ex}^{\scriptscriptstyle\partial Y}\!(\peqsubfino{q}{\partial}{-0.2ex})\right]+\left[\frac{\peqsub{D}{Y}\sqrt{\peqsubfino{\gamma^{\scriptscriptstyle\partial}}{\!X}{-0.3ex}}}{\sqrt{\peqsubfino{\gamma^{\scriptscriptstyle\partial}}{\!X}{-0.3ex}}}+\frac{\peqsub{D}{Y}\peqsub{b}{X}^2}{\peqsub{b}{X}^2}\right]\peqsubfino{\mathcal{H}}{\!\perp}{-0.1ex}^{\scriptscriptstyle\partial\,B}
\end{align*}
In the $\star$ equality we have defined for a given $\alpha\in\Omega^k(\Sigma)$ (where $\overline{\alpha}=\#_\gamma\alpha$) the map
\[\peqsubfino{\chi^{\scriptscriptstyle Y}}{k}{-0.2ex}\!(\alpha)=\frac{\sqrt{\peqsubfino{\gamma}{X}{-0.2ex}}}{2}\peqsubfino{\big\langle(\peqsub{D}{Y}\gamma)\alpha,\alpha\big\rangle}{\!\!\gamma}{-0.2ex}-\frac{\mathrm{Tr}(\peqsub{D}{Y}\gamma)}{k}\sqrt{\peqsubfino{\gamma}{X}{-0.2ex}}\peqsubfino{\big\langle\alpha,\alpha\big\rangle}{\!\!\gamma}{-0.2ex}=\frac{\sqrt{\peqsubfino{\gamma}{X}{-0.2ex}}}{2}\Big((\peqsub{D}{Y}\gamma)\alpha,\overline{\alpha}\Big)-\frac{\mathrm{Tr}(\peqsub{D}{Y}\gamma)}{k}\sqrt{\peqsubfino{\gamma}{X}{-0.2ex}}\Big(\alpha,\overline{\alpha}\Big)\]
Now applying the previous computations together with \eqref{Appendix - equation - Dn} and \eqref{Appendix - equation - De} --omitting, for the moment, the subindexes for $M$ and $m$-- and taking into account that $\peqsub{\mathbb{Z}}{X}=\peqsub{Z}{X}^{{\raisemath{0.2ex}{\!\scriptscriptstyle\perp}}}\peqsub{\vec{n}}{X}+\peqsub{\tau}{X}.\peqsub{\vec{z}}{X}^{\,\scriptscriptstyle\top}$ we have
\begin{align*}
&\bullet\ \peqsub{\mathbb{Z}}{X}^\beta\peqsub{D}{Y}\Big((\peqsub{e}{X})^b_\beta\mathcal{H}_b+\varepsilon(\peqsub{n}{X})_\beta\peqsub{\mathcal{H}}{\perp}\Big)=\\
&=\peqsub{\mathbb{Z}}{X}^\beta(\peqsub{D}{Y}\peqsub{e}{X})^b_\beta\mathcal{H}_b+\peqsub{\mathbb{Z}}{X}^\beta(\peqsub{e}{X})^b_\beta(\peqsub{D}{Y}\mathcal{H})_b+\peqsub{\mathbb{Z}}{X}^\beta\varepsilon(\peqsub{D}{Y}\peqsub{n}{X})_\beta\peqsub{\mathcal{H}}{\perp}+\peqsub{\mathbb{Z}}{X}^\beta\varepsilon(\peqsub{n}{X})_\beta \peqsub{D}{Y}\peqsub{\mathcal{H}}{\perp}=\\
&=(Z^{{\raisemath{0.2ex}{\scriptscriptstyle\perp\!}}} m^b-Z^cM^{\ b}_c)\mathcal{H}_b+\left[\Big(\imath_{-}\mathrm{d}\vecc{Y}{q},p\Big)+\Big(\imath_{-}\mathrm{d}q,\vecc{Y}{p}\Big)+\Big(\imath_{-}\vecc{Y}{q},\delta p\Big)+\Big(\imath_{-}q,\delta \vecc{Y}{p}\Big)\right](\vec{z}^{\,\scriptscriptstyle\top})-\varepsilon Z^c m_c\peqsub{\mathcal{H}}{\perp}+{}\\
&\phantom{=}+Z^{{\raisemath{0.2ex}{\scriptscriptstyle\perp\!}}}\left[\frac{1}{\sqrt{\peqsubfino{\gamma}{X}{-0.2ex}}}\Big(\underline{p}\hspace*{0.1ex},\hspace*{0.1ex}\vecc{Y}{p}\Big)-\varepsilon\sqrt{\peqsubfino{\gamma}{X}{-0.2ex}}\peqsubfino{\left\langle \mathrm{d}\vecc{Y}{q},\mathrm{d}q\right\rangle}{\!\gamma}{0.2ex}+\varepsilon\Big(\vecc[\perp]{Y}{q}\hspace*{0.1ex},\hspace*{0.1ex}\delta p\Big)+\varepsilon\Big(\peqsubfino{q}{\perp}{-0.2ex}\hspace*{0.1ex},\hspace*{0.1ex}\delta \vecc{Y}{p}\Big)+\peqsubfino{\chi^{\scriptscriptstyle Y}}{2}{-0.2ex}\!\!\left(\frac{\underline{p}}{\sqrt{\peqsubfino{\gamma}{X}{-0.2ex}}}\right)-\varepsilon\peqsubfino{\chi^{\scriptscriptstyle Y}}{4}{-0.2ex}\!(\mathrm{d}q)\right]=\\
&=\peqsub{C}{\mathbb{Y}}(\peqsub{\mathbb{Z}}{X})+\varepsilon Z^{{\raisemath{0.2ex}{\scriptscriptstyle\perp\!}}}\Big(\vecc[\perp]{Y}{q}\hspace*{0.1ex},\hspace*{0.1ex}\delta p\Big)+\Big(\imath_{\vec{z}^{\scriptscriptstyle\top}}\vecc{Y}{q},\delta p\Big)+\Big(\imath_{\vec{z}^{\scriptscriptstyle\top}}\mathrm{d}\vecc{Y}{q},p\Big)-\varepsilon\sqrt{\peqsubfino{\gamma}{X}{-0.2ex}}Z^{{\raisemath{0.2ex}{\scriptscriptstyle\perp\!}}}\peqsubfino{\left\langle \mathrm{d}\vecc{Y}{q},\mathrm{d}q\right\rangle}{\!\gamma}{0.2ex}+{}\\
&\phantom{=}+\left(\imath_{\vec{z}^{\scriptscriptstyle\top}}\mathrm{d}q+Z^{{\raisemath{0.2ex}{\scriptscriptstyle\perp\!}}}\frac{\underline{p}}{\sqrt{\peqsubfino{\gamma}{X}{-0.2ex}}},\vecc{Y}{p}\right)+\Big(\varepsilon Z^{{\raisemath{0.2ex}{\scriptscriptstyle\perp\!}}}\peqsubfino{q}{\perp}{-0.2ex}+\imath_{\vec{z}^{\scriptscriptstyle\top}}q,\delta \vecc{Y}{p}\Big)\\
&\bullet\ \peqsub{C}{\mathbb{Y}}(\peqsub{\mathbb{Z}}{X}):=(Z^{{\raisemath{0.2ex}{\scriptscriptstyle\perp\!}}}\peqsub{m}{\mathbb{Y}}^b-Z^c(\peqsub{M}{\mathbb{Y}})^{\ b}_c)\mathcal{H}_b-\varepsilon Z^c (\peqsub{m}{\mathbb{Y}})_c\peqsub{\mathcal{H}}{\perp}+Z^{{\raisemath{0.2ex}{\scriptscriptstyle\perp\!}}}\left[\peqsubfino{\chi^{\scriptscriptstyle Y}}{2}{-0.2ex}\!\!\left(\frac{\underline{p}}{\sqrt{\peqsubfino{\gamma}{X}{-0.2ex}}}\right)-\varepsilon\peqsubfino{\chi^{\scriptscriptstyle Y}}{4}{-0.2ex}\!(\mathrm{d}q)\right]\\[1ex]
&\bullet\ \peqsub{\mathbb{Z}}{X}^\beta \peqsub{D}{Y}\Big((\peqsub{e}{X})^b_\beta\mathcal{H}^{\scriptscriptstyle\partial}_b+\varepsilon(\peqsub{n}{X})_\beta\peqsub{\mathcal{H}}{\perp}^{\scriptscriptstyle\partial}+\varepsilon(\peqsub{\theta}{X})_\beta\peqsub{\mathcal{H}}{\perp}^{\scriptscriptstyle\partial\,B}\Big)=\\
&=\peqsub{\mathbb{Z}}{X}^\beta(\peqsub{D}{Y}e)^b_\beta\mathcal{H}^{\scriptscriptstyle\partial}_b+(\peqsub{D}{Y}\mathcal{H}^{\scriptscriptstyle\partial})_b\vec{z}^{\,b}+\varepsilon\peqsub{\mathbb{Z}}{X}^\beta(\peqsub{D}{Y}n)_\beta\peqsub{\mathcal{H}^{\scriptscriptstyle\partial}}{\perp}+ Z^{{\raisemath{0.2ex}{\scriptscriptstyle\perp\!}}}\peqsub{D}{Y}\peqsub{\mathcal{H}^{\scriptscriptstyle\partial}}{\perp}+\varepsilon\peqsub{\mathbb{Z}}{X}^\beta(\peqsub{D}{Y}\theta)_\beta\peqsub{\mathcal{H}}{\perp}^{\scriptscriptstyle\partial\,B}+\peqsubfino{Z^{{\raisemath{0.2ex}{\scriptscriptstyle\perp\!}}}}{\partial}{-0.2ex}\peqsub{D}{Y}\peqsub{\mathcal{H}}{\perp}^{\scriptscriptstyle\partial\,B}=\\
&=(Z^{{\raisemath{0.2ex}{\scriptscriptstyle\perp\!}}} m^b-Z^cM^{\ b}_c)\mathcal{H}^{\scriptscriptstyle\partial}_b+\left(\frac{\sqrt{\peqsubfino{\gamma^{\scriptscriptstyle\partial}}{\!X}{-0.3ex}}}{\sqrt{\peqsubfino{\gamma}{\!X}{-0.3ex}}}\frac{\nu}{|\vec{\nu}^{\,\scriptscriptstyle\top}|}\wedge\imath_{-}\vecc{Y}{q},p\right)\!(\vec{z})+\left(\frac{\sqrt{\peqsubfino{\gamma^{\scriptscriptstyle\partial}}{\!X}{-0.3ex}}}{\sqrt{\peqsubfino{\gamma}{\!X}{-0.3ex}}}\frac{\nu}{|\vec{\nu}^{\,\scriptscriptstyle\top}|}\wedge\imath_{-}q,\vecc{Y}{p}\right)\!(\vec{z})-\varepsilon Z^c m_c\peqsub{\mathcal{H}}{\perp}^{\scriptscriptstyle\partial}+{}\\
&+\varepsilon Z^{{\raisemath{0.2ex}{\scriptscriptstyle\perp\!}}}\left(\frac{\sqrt{\peqsubfino{\gamma^{\scriptscriptstyle\partial}}{\!X}{-0.3ex}}}{\sqrt{\peqsubfino{\gamma}{\!X}{-0.3ex}}}\frac{\nu}{|\vec{\nu}^{\,\scriptscriptstyle\top}|}\wedge\vecc[\perp]{Y}{q},p\right)+\varepsilon Z^{{\raisemath{0.2ex}{\scriptscriptstyle\perp\!}}}\left(\frac{\sqrt{\peqsubfino{\gamma^{\scriptscriptstyle\partial}}{\!X}{-0.3ex}}}{\sqrt{\peqsubfino{\gamma}{\!X}{-0.3ex}}}\frac{\nu}{|\vec{\nu}^{\,\scriptscriptstyle\top}|}\wedge\peqsubfino{q}{\perp}{-0.2ex},\vecc{Y}{p}\right)-\varepsilon \peqsub{\vec{z}^{\,b}}{\partial}m^{\scriptscriptstyle\partial}_b\peqsub{\mathcal{H}}{\perp}^{\scriptscriptstyle\partial\,B}+{}\\
&+\peqsubfino{Z^{{\raisemath{0.2ex}{\scriptscriptstyle\perp\!}}}}{\partial}{-0.2ex}\peqsub{b}{X}^2\sqrt{\peqsubfino{\gamma^{\scriptscriptstyle\partial}}{\!X}{-0.3ex}}\left[ \varepsilon\peqsubfino{\left\langle\vecc[\perp]{Y}{q}^{\scriptscriptstyle\partial},\peqsubfino{q^{\scriptscriptstyle\partial}}{\perp}{-0.2ex}\right\rangle}{\!\gamma^\partial}{0.2ex}\!+\peqsubfino{\left\langle\vecc{Y}{q}^{\scriptscriptstyle\partial},\peqsubfino{q}{\partial}{-0.2ex}\right\rangle}{\!\gamma^\partial}{0.2ex}\right]+\peqsubfino{Z^{{\raisemath{0.2ex}{\scriptscriptstyle\perp\!}}}}{\partial}{-0.2ex}\peqsub{b}{X}^2\left[ \peqsubfino{\chi}{4}{-0.2ex}^{\scriptscriptstyle\partial Y}\!(\peqsubfino{q^{\scriptscriptstyle\partial}}{\perp}{-0.2ex})+\peqsubfino{\chi}{4}{-0.2ex}^{\scriptscriptstyle\partial Y}\!(\peqsubfino{q}{\partial}{-0.2ex})\right]+\peqsubfino{Z^{{\raisemath{0.2ex}{\scriptscriptstyle\perp\!}}}}{\partial}{-0.2ex}\left[\frac{\peqsub{D}{Y}\sqrt{\peqsubfino{\gamma^{\scriptscriptstyle\partial}}{\!X}{-0.3ex}}}{\sqrt{\peqsubfino{\gamma^{\scriptscriptstyle\partial}}{\!X}{-0.3ex}}}+\frac{\peqsub{D}{Y}\peqsub{b}{X}^2}{\peqsub{b}{X}^2}\right]\peqsubfino{\mathcal{H}}{\!\perp}{-0.1ex}^{\scriptscriptstyle\partial\,B}\overset{\star}{=}\\
&=\peqsub{C^{\scriptscriptstyle\partial}}{\mathbb{Y}}(\peqsub{\mathbb{Z}}{X})+\peqsub{C^{\scriptscriptstyle\partial\,B}}{\mathbb{Y}}(\peqsub{\mathbb{Z}}{X})+\varepsilon Z^{{\raisemath{0.2ex}{\scriptscriptstyle\perp\!}}}\left(\frac{\sqrt{\peqsubfino{\gamma^{\scriptscriptstyle\partial}}{\!X}{-0.3ex}}}{\sqrt{\peqsubfino{\gamma}{\!X}{-0.3ex}}}\frac{\nu}{|\vec{\nu}^{\,\scriptscriptstyle\top}|}\wedge\vecc[\perp]{Y}{q},p\right)+\frac{Z^{{\raisemath{0.2ex}{\scriptscriptstyle\perp\!}}}}{|\vec{\nu}^{\,\scriptscriptstyle\top}|}\peqsub{b}{X}^2\sqrt{\peqsubfino{\gamma^{\scriptscriptstyle\partial}}{\!X}{-0.3ex}}\left[ \varepsilon\peqsubfino{\left\langle\vecc[\perp]{Y}{q}^{\scriptscriptstyle\partial},\peqsubfino{q^{\scriptscriptstyle\partial}}{\perp}{-0.2ex}\right\rangle}{\!\gamma^\partial}{0.2ex}\!+\peqsubfino{\left\langle\vecc{Y}{q}^{\scriptscriptstyle\partial},\peqsubfino{q}{\partial}{-0.2ex}\right\rangle}{\!\gamma^\partial}{0.2ex}\right]+{}\\
&+\left(\frac{\sqrt{\peqsubfino{\gamma^{\scriptscriptstyle\partial}}{\!X}{-0.3ex}}}{\sqrt{\peqsubfino{\gamma}{\!X}{-0.3ex}}}\frac{\nu}{|\vec{\nu}^{\,\scriptscriptstyle\top}|}\wedge\imath_{\vec{z}}\vecc{Y}{q},p\right)+\frac{\sqrt{\peqsubfino{\gamma^{\scriptscriptstyle\partial}}{\!X}{-0.3ex}}}{\sqrt{\peqsubfino{\gamma}{\!X}{-0.3ex}}}\left(\frac{\nu}{|\vec{\nu}^{\,\scriptscriptstyle\top}|}\wedge\big(\imath_{\vec{z}}q+\varepsilon Z^{{\raisemath{0.2ex}{\scriptscriptstyle\perp\!}}}\peqsubfino{q}{\perp}{-0.2ex}\big),\vecc{Y}{p}\right)\\
&\bullet\ \peqsub{C^{\scriptscriptstyle\partial}}{\mathbb{Y}}(\peqsub{\mathbb{Z}}{X})=(Z^{{\raisemath{0.2ex}{\scriptscriptstyle\perp\!}}}\peqsub{m}{\mathbb{Y}}^b-Z^c(\peqsub{M}{\mathbb{Y}})^{\ b}_c)\mathcal{H}^{\scriptscriptstyle\partial}_a-\varepsilon Z^c (\peqsub{m}{\mathbb{Y}})_c\peqsub{\mathcal{H}}{\perp}^{\scriptscriptstyle\partial}\\
&\bullet\ \peqsub{C^{\scriptscriptstyle\partial\,B}}{\mathbb{Y}}(\peqsub{\mathbb{Z}}{X})=-\varepsilon \peqsub{\vec{z}^{\,b}}{\partial} (\peqsub{m}{Y}^{\scriptscriptstyle\partial})_c\peqsub{\mathcal{H}}{\perp}^{\scriptscriptstyle\partial\,B}+\frac{Z^{{\raisemath{0.2ex}{\scriptscriptstyle\perp\!}}}}{|\vec{\nu}^{\,\scriptscriptstyle\top}|}\peqsub{b}{X}^2\left[ \peqsubfino{\chi}{4}{-0.2ex}^{\scriptscriptstyle\partial Y}\!(\peqsubfino{q^{\scriptscriptstyle\partial}}{\perp}{-0.2ex})+\peqsubfino{\chi}{4}{-0.2ex}^{\scriptscriptstyle\partial Y}\!(\peqsubfino{q}{\partial}{-0.2ex})\right]+\frac{Z^{{\raisemath{0.2ex}{\scriptscriptstyle\perp\!}}}}{|\vec{\nu}^{\,\scriptscriptstyle\top}|}\left[\frac{\peqsub{D}{Y}\sqrt{\peqsubfino{\gamma^{\scriptscriptstyle\partial}}{\!X}{-0.3ex}}}{\sqrt{\peqsubfino{\gamma^{\scriptscriptstyle\partial}}{\!X}{-0.3ex}}}+\frac{\peqsub{D}{Y}\peqsub{b}{X}^2}{\peqsub{b}{X}^2}\right]\peqsubfino{\mathcal{H}}{\!\perp}{-0.1ex}^{\scriptscriptstyle\partial\,B}
\end{align*}
where in the $\star$ equality we have used that $\peqsubfino{Z^{{\raisemath{0.2ex}{\scriptscriptstyle\perp\!}}}}{\partial}{-0.2ex}|\vec{\nu}^{\,\scriptscriptstyle\top}|=Z^{{\raisemath{0.2ex}{\scriptscriptstyle\perp\!}}}$. Notice that in the last expression if $b$ vanishes, then its variation and $\peqsub{\mathcal{H}}{\perp}^{\scriptscriptstyle\partial\,B}$ also vanish and, therefore, $\peqsub{C^{\scriptscriptstyle\partial\,B}}{\mathbb{Y}}(\peqsub{\mathbb{Z}}{X})=0$. Now we proceed to plug these expressions into \eqref{Parametrized theories - equation - omega final} taking into account that $(\alpha,\vec{v})=\peqsubfino{\langle\alpha,\underline{v}\rangle}{\gamma}{-0.2ex}$.
\begin{align*}
&\omega_{(\peqsubfino{q}{\perp}{-0.2ex},q,X,p)}\Big((\vecc[\scalebox{0.4}{$\perp$}]{Y}{q},\vecc{Y}{q},\vecc{Y}{X},\vecc{Y}{p}),(\vecc[\scalebox{0.4}{$\perp$}]{Z}{q},\vecc{Z}{q},\vecc{Z}{X},\vecc{Z}{p})\Big)
=\\ &=\int_\Sigma\left[\Big(\vecc{Y}{q},\vecc{Z}{p}\Big)+\peqsub{D}{Y}(\varepsilon\peqsubfino{\mathcal{H}}{\!\perp}{-0.1ex}n+e.\mathcal{H})_\alpha\vecc{Z}{X}^\alpha\right]\peqsub{\mathrm{vol}}{\Sigma}+\int_{\partial\Sigma}\peqsub{D}{Y}(\varepsilon\peqsubfino{\mathcal{H}}{\!\perp}{-0.1ex}^{\scriptscriptstyle\partial}n+e.\mathcal{H}^{\scriptscriptstyle\partial})_\alpha\vecc{Z}{X}^\alpha\peqsub{\mathrm{vol}}{\partial\Sigma}-(Y\leftrightarrow Z)=\\ &=\int_\Sigma\bigg[\Big(\vecc{Y}{q},\vecc{Z}{p}\Big)+\peqsub{C}{\mathbb{Y}}(\mathbb{Z})+\varepsilon Z^{{\raisemath{0.2ex}{\scriptscriptstyle\perp\!}}}\Big(\vecc[\perp]{Y}{q}\hspace*{0.1ex},\hspace*{0.1ex}\delta p\Big)+\peqsubfino{\langle\imath_{\vec{z}^{\scriptscriptstyle\top}}\vecc{Y}{q},\delta \underline{p}\rangle}{\!\gamma}{-0.4ex}+\peqsubfino{\langle\imath_{\vec{z}^{\scriptscriptstyle\top}}\mathrm{d}\vecc{Y}{q},\underline{p}\rangle}{\!\gamma}{-0.4ex}-\textcolor{red}{\varepsilon\sqrt{\gamma}\peqsubfino{\left\langle \mathrm{d}\vecc{Y}{q},Z^{{\raisemath{0.2ex}{\scriptscriptstyle\perp\!}}}\mathrm{d}q\right\rangle}{\!\gamma}{0.2ex}}\,+{}\\
&+\left(\imath_{\vec{z}^{\scriptscriptstyle\top}}\mathrm{d}q+Z^{{\raisemath{0.2ex}{\scriptscriptstyle\perp\!}}}\frac{\underline{p}}{\sqrt{\gamma}},\vecc{Y}{p}\right)+\textcolor{teal}{\peqsubfino{\langle\varepsilon Z^{{\raisemath{0.2ex}{\scriptscriptstyle\perp\!}}}\peqsubfino{q}{\perp}{-0.2ex}+\imath_{\vec{z}^{\scriptscriptstyle\top}}q,\delta \underline{\vecc{Y}{p}}\rangle}{\!\gamma}{-0.2ex}}\bigg]\peqsub{\mathrm{vol}}{\Sigma} +\int_{\partial\Sigma}\frac{\sqrt{\gamma^{\scriptscriptstyle\partial}}}{\sqrt{\gamma}}\bigg[\frac{\sqrt{\gamma}}{\sqrt{\gamma^{\scriptscriptstyle\partial}}}\Big(\peqsub{C^{\scriptscriptstyle\partial}}{\mathbb{Y}}(\mathbb{Z})+\peqsub{C^{\scriptscriptstyle\partial\,B}}{\mathbb{Y}}(\mathbb{Z})\Big)+{}\\
&+\varepsilon Z^{{\raisemath{0.2ex}{\scriptscriptstyle\perp\!}}}\peqsubfino{\left\langle\frac{\nu}{|\vec{\nu}^{\,\scriptscriptstyle\top}|}\wedge\vecc[\perp]{Y}{q},\underline{p}\right\rangle}{\!\!\!\gamma}{0.2ex}+\frac{Z^{{\raisemath{0.2ex}{\scriptscriptstyle\perp\!}}}}{|\vec{\nu}^{\,\scriptscriptstyle\top}|}b^2\sqrt{\gamma}\left[ \varepsilon\peqsubfino{\left\langle\vecc[\perp]{Y}{q}^{\scriptscriptstyle\partial},\peqsubfino{q^{\scriptscriptstyle\partial}}{\perp}{-0.2ex}\right\rangle}{\!\gamma^\partial}{0.2ex}\!+\peqsubfino{\left\langle\vecc{Y}{q}^{\scriptscriptstyle\partial},\peqsubfino{q}{\partial}{-0.2ex}\right\rangle}{\!\gamma^\partial}{0.2ex}\right]+\peqsubfino{\left\langle\frac{\nu}{|\vec{\nu}^{\,\scriptscriptstyle\top}|}\wedge\imath_{\vec{z}}\vecc{Y}{q},\underline{p}\right\rangle}{\!\!\!\gamma}{0.2ex}+{}\\
&+\peqsubfino{\left\langle\frac{\nu}{|\vec{\nu}^{\,\scriptscriptstyle\top}|}\wedge\big(\imath_{\vec{z}}q+\varepsilon Z^{{\raisemath{0.2ex}{\scriptscriptstyle\perp\!}}}\peqsubfino{q}{\perp}{-0.2ex}\big),\underline{\vecc{Y}{p}}\right\rangle}{\!\!\!\gamma}{0.2ex}\Bigg]\peqsub{\mathrm{vol}}{\partial\Sigma}-(Y\leftrightarrow Z)\updown{\eqref{Appendix - equation - (v wedge beta,omega)=(beta,i_v omega)}}{\eqref{appendix equation integracion por partes}}{=}\\
&=\int_\Sigma\Big[\peqsub{C}{\mathbb{Y}}(\mathbb{Z})+\varepsilon \Big(\vecc[\perp]{Y}{q}\hspace*{0.1ex},\hspace*{0.1ex}Z^{{\raisemath{0.2ex}{\scriptscriptstyle\perp\!}}}\delta p\Big)+\peqsubfino{\langle\vecc{Y}{q},\underline{\vecc{Z}{p}}\rangle}{\!\gamma}{-0.3ex}+\peqsubfino{\langle \vecc{Y}{q},\underline{z^{\scriptscriptstyle\top}}\!\!\wedge\delta \underline{p}\rangle}{\!\gamma}{-0.3ex}+\textcolor{brown}{\peqsubfino{\langle \mathrm{d}\vecc{Y}{q},\underline{z^{\scriptscriptstyle\top}}\!\!\wedge\underline{p}\rangle}{\!\gamma}{-0.3ex}}-\textcolor{red}{\varepsilon\sqrt{\gamma}\peqsubfino{\left\langle \vecc{Y}{q},\delta(Z^{{\raisemath{0.2ex}{\scriptscriptstyle\perp\!}}}\mathrm{d}q)\right\rangle}{\!\gamma}{0.2ex}}\,+{}\\
&+\left(\imath_{\vec{z}^{\scriptscriptstyle\top}}\mathrm{d}q+Z^{{\raisemath{0.2ex}{\scriptscriptstyle\perp\!}}}\frac{\underline{p}}{\sqrt{\gamma}},\vecc{Y}{p}\right)+\textcolor{teal}{\peqsubfino{\big\langle\varepsilon\mathrm{d}( Z^{{\raisemath{0.2ex}{\scriptscriptstyle\perp\!}}}\peqsubfino{q}{\perp}{-0.2ex})+\mathrm{d}\imath_{\vec{z}^{\scriptscriptstyle\top}}q,\underline{\vecc{Y}{p}}\big\rangle}{\!\!\gamma}{-0.1ex}}\Big]\peqsub{\mathrm{vol}}{\Sigma}+\int_{\partial\Sigma}\Big(\peqsub{C^{\scriptscriptstyle\partial}}{\mathbb{Y}}(\mathbb{Z})+\peqsub{C^{\scriptscriptstyle\partial\,B}}{\mathbb{Y}}(\mathbb{Z})\Big)\peqsub{\mathrm{vol}}{\partial\Sigma}+{}\\
&+\int_{\partial\Sigma}\frac{\sqrt{\gamma^{\scriptscriptstyle\partial}}}{\sqrt{\gamma}}\bigg[\varepsilon \frac{Z^{{\raisemath{0.2ex}{\scriptscriptstyle\perp\!}}}}{|\vec{\nu}^{\,\scriptscriptstyle\top}|}\peqsubfino{\left\langle\vecc[\perp]{Y}{q},\imath_{\vec{\nu}}\underline{p}\right\rangle}{\!\gamma}{0.2ex}+\frac{Z^{{\raisemath{0.2ex}{\scriptscriptstyle\perp\!}}}}{|\vec{\nu}^{\,\scriptscriptstyle\top}|}b^2\sqrt{\gamma}\left[ \varepsilon\peqsubfino{\left\langle\vecc[\perp]{Y}{q}^{\scriptscriptstyle\partial},\peqsubfino{q^{\scriptscriptstyle\partial}}{\perp}{-0.2ex}\right\rangle}{\!\gamma^\partial}{0.2ex}\!+\peqsubfino{\left\langle\vecc{Y}{q}^{\scriptscriptstyle\partial},\peqsubfino{q}{\partial}{-0.2ex}\right\rangle}{\!\gamma^\partial}{0.2ex}\right]+\peqsubfino{\left\langle\imath_{\vec{z}}\vecc{Y}{q},\frac{\imath_{\vec{\nu}}\underline{p}}{|\vec{\nu}^{\,\scriptscriptstyle\top}|}\right\rangle}{\!\!\!\gamma}{0.2ex}+{}\\
&+\peqsubfino{\left\langle\imath_{\vec{z}}q+\varepsilon Z^{{\raisemath{0.2ex}{\scriptscriptstyle\perp\!}}}\peqsubfino{q}{\perp}{-0.2ex},\frac{\imath_{\vec{\nu}}\underline{\vecc{Y}{p}}}{|\vec{\nu}^{\,\scriptscriptstyle\top}|}\right\rangle}{\!\!\!\gamma}{0.2ex}-\textcolor{red}{\varepsilon\sqrt{\gamma}\peqsubfino{\left\langle \vecc{Y}{q},\frac{Z^{{\raisemath{0.2ex}{\scriptscriptstyle\perp\!}}}}{|\vec{\nu}^{\,\scriptscriptstyle\top}|}\imath_{\vec{\nu}}\mathrm{d}q\right\rangle}{\!\!\!\gamma}{0.2ex}}-\textcolor{teal}{\peqsubfino{\left\langle\varepsilon Z^{{\raisemath{0.2ex}{\scriptscriptstyle\perp\!}}}\peqsubfino{q}{\perp}{-0.2ex}+\imath_{\vec{z}^{\scriptscriptstyle\top}}q,\frac{\imath_{\vec{\nu}}\underline{\vecc{Y}{p}}}{|\vec{\nu}^{\,\scriptscriptstyle\top}|}\right\rangle}{\!\!\!\gamma}{-0.2ex}}\Bigg]\peqsub{\mathrm{vol}}{\partial\Sigma}-(Y\leftrightarrow Z)\updown{\eqref{appendix - formula - L=di+id}}{\eqref{appendix equation integracion por partes}}{=}\\
&=\int_\Sigma\Big[\peqsub{C}{\mathbb{Y}}(\mathbb{Z})+\varepsilon \Big(\vecc[\perp]{Y}{q}\hspace*{0.1ex},\hspace*{0.1ex}Z^{{\raisemath{0.2ex}{\scriptscriptstyle\perp\!}}}\delta p\Big)+\peqsubfino{\Big\langle \vecc{Y}{q},\underline{\vecc{Z}{p}}+\underline{z^{\scriptscriptstyle\top}}\!\!\wedge\delta \underline{p}+\textcolor{brown}{\delta(\underline{z^{\scriptscriptstyle\top}}\!\!\wedge\underline{p})}-\varepsilon\sqrt{\gamma}\delta(Z^{{\raisemath{0.2ex}{\scriptscriptstyle\perp\!}}}\mathrm{d}q)\Big\rangle}{\!\!\gamma}{0.2ex}\,+{}\\
&+\left(\mathcal{L}_{\vec{z}^{\scriptscriptstyle\top}}q+Z^{{\raisemath{0.2ex}{\scriptscriptstyle\perp\!}}}\frac{\underline{p}}{\sqrt{\gamma}}+\varepsilon\mathrm{d}( Z^{{\raisemath{0.2ex}{\scriptscriptstyle\perp\!}}}\peqsubfino{q}{\perp}{-0.2ex}),\vecc{Y}{p}\right)\Big]\peqsub{\mathrm{vol}}{\Sigma}+\int_{\partial\Sigma}\Big(\peqsub{C^{\scriptscriptstyle\partial}}{\mathbb{Y}}(\mathbb{Z})+\peqsub{C^{\scriptscriptstyle\partial\,B}}{\mathbb{Y}}(\mathbb{Z})\Big)\peqsub{\mathrm{vol}}{\partial\Sigma}+{}\\
&+\int_{\partial\Sigma}\frac{\sqrt{\gamma^{\scriptscriptstyle\partial}}}{|\vec{\nu}^{\,\scriptscriptstyle\top}|\sqrt{\gamma}}\bigg[\peqsubfino{\left\langle\varepsilon Z^{{\raisemath{0.2ex}{\scriptscriptstyle\perp\!}}}\vecc[\perp]{Y}{q}+\imath_{\vec{z}}\vecc{Y}{q},\imath_{\vec{\nu}}\underline{p}\right\rangle}{\!\gamma}{0.2ex}+Z^{{\raisemath{0.2ex}{\scriptscriptstyle\perp\!}}}b^2\sqrt{\gamma}\left[ \varepsilon\peqsubfino{\left\langle\vecc[\perp]{Y}{q}^{\scriptscriptstyle\partial},\peqsubfino{q^{\scriptscriptstyle\partial}}{\perp}{-0.2ex}\right\rangle}{\!\gamma^\partial}{0.2ex}\!+\peqsubfino{\left\langle\vecc{Y}{q}^{\scriptscriptstyle\partial},\peqsubfino{q}{\partial}{-0.2ex}\right\rangle}{\!\gamma^\partial}{0.2ex}\right]-{}\\
&-\varepsilon\sqrt{\gamma}\peqsubfino{\left\langle \vecc{Y}{q},Z^{{\raisemath{0.2ex}{\scriptscriptstyle\perp\!}}}\imath_{\vec{\nu}}\mathrm{d}q\right\rangle}{\!\gamma}{0.2ex}+\textcolor{brown}{\peqsubfino{\Big\langle \vecc{Y}{q},\imath_{\vec{\nu}}(\underline{z^{\scriptscriptstyle\top}}\!\!\wedge\underline{p})\Big\rangle}{\!\!\gamma}{0.2ex}}\Bigg]\peqsub{\mathrm{vol}}{\partial\Sigma}-(Y\leftrightarrow Z)\updown{\star}{\eqref{Appendix - equation - regla de Leibniz producto interior}}{=}\\
&=\int_\Sigma\left[\peqsub{C}{\mathbb{Y}}(\mathbb{Z})+\varepsilon \Big(\vecc[\perp]{Y}{q}\hspace*{0.1ex},\hspace*{0.1ex}Z^{{\raisemath{0.2ex}{\scriptscriptstyle\perp\!}}}\delta p\Big)\!+\!\peqsubfino{\Big\langle \vecc{Y}{q},\underline{\vecc{Z}{p}}-\underline{\mathcal{L}_{\vec{z}}p}-\varepsilon\sqrt{\gamma}\delta(Z^{{\raisemath{0.2ex}{\scriptscriptstyle\perp\!}}}\mathrm{d}q)\Big\rangle}{\!\!\gamma}{0.2ex}\!\,-\!\left(\mathcal{L}_{\vec{y}^{\scriptscriptstyle\top}}q+Y^{{\raisemath{0.2ex}{\scriptscriptstyle\perp\!}}}\frac{\underline{p}}{\sqrt{\gamma}}+\varepsilon\mathrm{d}( Y^{{\raisemath{0.2ex}{\scriptscriptstyle\perp\!}}}\peqsubfino{q}{\perp}{-0.2ex}),\vecc{Z}{p}\right)\!\right]\peqsub{\mathrm{vol}}{\Sigma}+{}\\
&+\int_{\partial\Sigma}\Big(\peqsub{C^{\scriptscriptstyle\partial}}{\mathbb{Y}}(\mathbb{Z})+\peqsub{C^{\scriptscriptstyle\partial\,B}}{\mathbb{Y}}(\mathbb{Z})\Big)\peqsub{\mathrm{vol}}{\partial\Sigma}+\int_{\partial\Sigma}\frac{\varepsilon \sqrt{\gamma^{\scriptscriptstyle\partial}}}{|\vec{\nu}^{\,\scriptscriptstyle\top}|\sqrt{\gamma}}Z^{{\raisemath{0.2ex}{\scriptscriptstyle\perp\!}}}\bigg[\peqsubfino{\left\langle\vecc[\perp]{Y}{q},\imath_{\vec{\nu}}\underline{p}\right\rangle}{\!\gamma}{0.2ex}+b^2\sqrt{\gamma}\left[ \peqsubfino{\left\langle\vecc[\perp]{Y}{q}^{\scriptscriptstyle\partial},\peqsubfino{q^{\scriptscriptstyle\partial}}{\perp}{-0.2ex}\right\rangle}{\!\gamma^\partial}{0.2ex}\!+\varepsilon\peqsubfino{\left\langle\vecc{Y}{q}^{\scriptscriptstyle\partial},\peqsubfino{q}{\partial}{-0.2ex}\right\rangle}{\!\gamma^\partial}{0.2ex}\right]-{}\\
&-\sqrt{\gamma}\peqsubfino{\left\langle \vecc{Y}{q},\imath_{\vec{\nu}}\mathrm{d}q\right\rangle}{\!\gamma}{0.2ex}-\peqsubfino{\Big\langle \vecc{Y}{q},\peqsub{\nu}{\perp}\underline{p})\Big\rangle}{\!\!\gamma}{0.2ex}\Bigg]\peqsub{\mathrm{vol}}{\partial\Sigma}-(Y\leftrightarrow Z)\updown{\dagger}{\eqref{Appendix - lemma - g=gamma+nu nu}\eqref{appendix - formula - i_x^2=0}}{=}\\
&=\int_\Sigma\varepsilon Z^{{\raisemath{0.2ex}{\scriptscriptstyle\perp\!}}} \Big(\vecc[\perp]{Y}{q}\hspace*{0.1ex}-\mathcal{L}_{\vec{y}}\peqsubfino{q}{\perp}{-0.2ex}+\peqsub{\imath}{\mathrm{d}Y^{\!\scriptscriptstyle\perp}}q,\hspace*{0.1ex}\delta p\Big)\peqsub{\mathrm{vol}}{\Sigma}+\int_{\partial\Sigma}\varepsilon Z^{{\raisemath{0.2ex}{\scriptscriptstyle\perp\!}}} \peqsubfino{\left\langle\peqsub{\jmath}{\partial}^*\big(\vecc[\perp]{Y}{q}\hspace*{0.1ex}-\mathcal{L}_{\vec{y}}\peqsubfino{q}{\perp}{-0.2ex}+\peqsub{\imath}{\mathrm{d}Y^{\!\scriptscriptstyle\perp}}q\big),\peqsub{\jmath}{\partial}^*\left(\frac{\imath_{\vec{\nu}}\underline{p}}{\sqrt{\gamma}}\right)+b^2\peqsubfino{q^{\scriptscriptstyle\partial}}{\perp}{-0.2ex}\right\rangle}{\!\!\!\gamma^\partial}{0.2ex}\frac{\peqsubfino{{\mathrm{vol}_\gamma}}{\!\partial\!X}{-0.4ex}}{|\vec{\nu}^{\,\scriptscriptstyle\top}|}+{}\\
&+\int_\Sigma\left(\vecc{Y}{q}-\mathcal{L}_{\vec{y}^{\scriptscriptstyle\top}}q-Y^{{\raisemath{0.2ex}{\scriptscriptstyle\perp\!}}}\frac{\underline{p}}{\sqrt{\gamma}}-\varepsilon\mathrm{d}( Y^{{\raisemath{0.2ex}{\scriptscriptstyle\perp\!}}}\peqsubfino{q}{\perp}{-0.2ex}),\vecc{Z}{p}-\mathcal{L}_{\vec{z}}p-\varepsilon\sqrt{\gamma}\#_\gamma\delta(Z^{{\raisemath{0.2ex}{\scriptscriptstyle\perp\!}}}\mathrm{d}q)\right)\peqsub{\mathrm{vol}}{\Sigma}+{}\\
&-\int_{\partial\Sigma}\varepsilon Z^{{\raisemath{0.2ex}{\scriptscriptstyle\perp\!}}}\peqsubfino{\left\langle\jmath^*\big(\vecc{Y}{q}-\mathcal{L}_{\vec{y}^{\scriptscriptstyle\top}}q-Y^{{\raisemath{0.2ex}{\scriptscriptstyle\perp\!}}}\frac{\underline{p}}{\sqrt{\gamma}}-\varepsilon\mathrm{d}( Y^{{\raisemath{0.2ex}{\scriptscriptstyle\perp\!}}}\peqsubfino{q}{\perp}{-0.2ex})\big),\jmath^*(\imath_{\vec{\nu}}\mathrm{d}q)+\peqsub{\nu}{\perp}\frac{\underline{p}^{\scriptscriptstyle\partial}}{\sqrt{\gamma}}-\varepsilon b^2\peqsubfino{q}{\partial}{-0.2ex}\right\rangle}{\!\!\!\gamma^\partial}{0.2ex}\frac{\peqsubfino{{\mathrm{vol}_\gamma}}{\!\partial\!X}{-0.4ex}}{|\vec{\nu}^{\,\scriptscriptstyle\top}|}+{}\\
&+\int_\Sigma\left[\peqsub{C}{\mathbb{Y}}(\mathbb{Z})+\varepsilon Z^{{\raisemath{0.2ex}{\scriptscriptstyle\perp\!}}} \Big(\mathcal{L}_{\vec{y}}\peqsubfino{q}{\perp}{-0.2ex}-\peqsub{\imath}{\mathrm{d}Y^{\!\scriptscriptstyle\perp}}q,\hspace*{0.1ex}\delta p\Big)-\left(\mathcal{L}_{\vec{y}^{\scriptscriptstyle\top}}q+Y^{{\raisemath{0.2ex}{\scriptscriptstyle\perp\!}}}\frac{\underline{p}}{\sqrt{\gamma}}+\varepsilon\mathrm{d}( Y^{{\raisemath{0.2ex}{\scriptscriptstyle\perp\!}}}\peqsubfino{q}{\perp}{-0.2ex}),\mathcal{L}_{\vec{z}}p+\varepsilon\sqrt{\gamma}\#_\gamma\delta(Z^{{\raisemath{0.2ex}{\scriptscriptstyle\perp\!}}}\mathrm{d}q)\right)\right]\peqsub{\mathrm{vol}}{\Sigma}+{}\\
&+\int_{\partial\Sigma}\left[\frac{\peqsub{C^{\scriptscriptstyle\partial}}{\mathbb{Y}}(\mathbb{Z})}{\sqrt{\gamma^{\scriptscriptstyle\partial}}}+\frac{\peqsub{C^{\scriptscriptstyle\partial\,B}}{\mathbb{Y}}(\mathbb{Z})}{\sqrt{\gamma^{\scriptscriptstyle\partial}}}+\frac{\varepsilon Z^{{\raisemath{0.2ex}{\scriptscriptstyle\perp\!}}} }{|\vec{\nu}^{\,\scriptscriptstyle\top}|}\peqsubfino{\left\langle\peqsub{\jmath}{\partial}^*\big(\mathcal{L}_{\vec{y}}\peqsubfino{q}{\perp}{-0.2ex}-\peqsub{\imath}{\mathrm{d}Y^{\!\scriptscriptstyle\perp}}q\big),\peqsub{\jmath}{\partial}^*\left(\frac{\imath_{\vec{\nu}}\underline{p}}{\sqrt{\gamma}}\right)+b^2\peqsubfino{q^{\scriptscriptstyle\partial}}{\perp}{-0.2ex}\right\rangle}{\!\!\!\gamma^\partial}{0.2ex}-\right.\\
&\qquad\qquad\left.-\frac{\varepsilon Z^{{\raisemath{0.2ex}{\scriptscriptstyle\perp\!}}} }{|\vec{\nu}^{\,\scriptscriptstyle\top}|}\peqsubfino{\left\langle\jmath^*\big(\mathcal{L}_{\vec{y}^{\scriptscriptstyle\top}}q+Y^{{\raisemath{0.2ex}{\scriptscriptstyle\perp\!}}}\frac{\underline{p}}{\sqrt{\gamma}}+\varepsilon\mathrm{d}( Y^{{\raisemath{0.2ex}{\scriptscriptstyle\perp\!}}}\peqsubfino{q}{\perp}{-0.2ex})\big),\jmath^*(\imath_{\vec{\nu}}\mathrm{d}q)+\peqsub{\nu}{\perp}\frac{\underline{p}^{\scriptscriptstyle\partial}}{\sqrt{\gamma}}-\varepsilon b^2\peqsubfino{q}{\partial}{-0.2ex}\right\rangle}{\!\!\!\gamma^\partial}{0.2ex}\right]\peqsubfino{{\mathrm{vol}_\gamma}}{\!\partial\!X}{-0.4ex}-(Y\leftrightarrow Z)
\end{align*}
In the $\star$ equality, the Lie derivative of $\underline{p}$ appears using the definitions of the wedge \eqref{Appendix - definition - wedge} and the codifferential \eqref{Appendix - definition - delta=-nabla} together with the Leibniz rule for the covariant derivative and properties \eqref{Appendix - label - descomposicion delta} and \eqref{Appendix - equation - delta antisim}. Besides we have used that $0=\nu_\alpha Z^\alpha=\varepsilon\peqsub{\nu}{\perp}Z^{\scriptscriptstyle\perp}+\nu_az^a$. Meanwhile, on the $\dagger$ equality we have added and subtracted terms anticipating the right answer. Now it is a very long (and quite uninteresting) computation to prove that the last three lines vanish which leads to the final result \eqref{Parametrized theories - equation - omega final explicita}.

\subsection*{General relativity}\trassub

\begin{lemma}\label{Appendix - lemma - DR}\mbox{}\\
 The variation of the scalar curvature $R$ in the direction $h\in T_g\mathrm{Met}(M)$ is given by
 \[\peqsub{D}{(g,h)}R(g)=-h^{ab}\mathrm{Ric}(g)_{ab}+\nabla^d\Big(\nabla^bh_{bd}-\nabla_d\peqsub{\mathrm{Tr}}{g}h\Big)\]
\end{lemma}
\begin{proof}\mbox{}\\
Let us first compute the variation of the Ricci curvature at $p\in M$ using a local Lorentz frame $\{x_k\}$ (in particular, $\Gamma(p)=0$).
\begin{align*}
\peqsub{D}{(g,h)}\mathrm{Ric}(g)_{ab}&=\peqsub{D}{(g,h)}(\peqsub{R}{g})^c_{\ acb}=\\
&=\peqsub{D}{(g,h)}\left[\partial_c\Gamma(g)^c_{\ ab}-\partial_b\Gamma(g)^c_{\ ac}+\Gamma(g)^c_{\ cd}\Gamma(g)^d_{\ ab}-\Gamma(g)^c_{\ bd}\Gamma(g)^d_{\ ac}\right]\overset{\Gamma(p)=0}{=}\\
&=\partial_c\peqsub{D}{(g,h)}\Gamma(g)^c_{\ ab}-\partial_b\peqsub{D}{(g,h)}\Gamma(g)^c_{\ ac}+0\overset{\Gamma(p)=0}{=}\\[1.2ex]
&=\nabla_c\peqsub{D}{(g,h)}\Gamma(g)^c_{\ ab}-\nabla_b\peqsub{D}{(g,h)}\Gamma(g)^c_{\ ac}
\end{align*}
The last expression, being tensorial, is valid for any coordinate system. Besides, we recall that  $\peqsub{D}{(g,h)}\Gamma=\peqsub{D}{(g,h)}\nabla$ is given by equation \eqref{Appendix - equation - variacion nabla} with $\peqsub{D}{(g,h)}g=h$, which can be rewritten as
\[\tensor{(D\nabla)}{^c_a_b}=\frac{1}{2}\Big(\nabla_a\tensor{h}{^c_b}+\nabla_b\tensor{h}{_a^c}-\nabla^ch_{ab}\Big)\]
Thus the variation of the scalar curvature is given by
\begin{align*}
\peqsub{D}{(g,h)}&R(g)=\peqsub{D}{(g,h)}\Big(g^{ab}\mathrm{Ric}(g)_{ab}\Big)=\mathrm{Ric}(g)_{ab}(\peqsub{D}{(g,h)}g^{ab})+g^{ab}\peqsub{D}{(g,h)}\mathrm{Ric}(g)_{ab}\overset{\eqref{Appendix - equation - variacion inversa metrica}}{=}\\
&=-\mathrm{Ric}(g)_{ab}g^{ac}[\peqsub{D}{(g,h)}g_{cd}]g^{db}+g^{ab}\nabla_c\peqsub{D}{(g,h)}\tensor{\Gamma(g)}{^c_a_b}-g^{ab}\nabla_b\tensor{\peqsub{D}{(g,h)}\Gamma(g)}{^c_a_c}=\\
&=-h^{ab}\mathrm{Ric}(g)_{ab}+\frac{g^{ab}}{2}\nabla_c\Big(\textcolor{red}{\nabla_a\tensor{h}{_b^c}+\nabla_b\tensor{h}{_a^c}}-\textcolor{teal}{\nabla_dh_{ab}}\Big)-\frac{g^{ab}}{2}\nabla_b\Big(\textcolor{teal}{\nabla_a\tensor{h}{_c^c}}+\textcolor{brown}{\nabla^dh_{ad}-\nabla^ch_{ac}}\Big)=\\
&=-h^{ab}\mathrm{Ric}(g)_{ab}+\nabla^d\Big(\textcolor{red}{\nabla^bh_{bd}}-\textcolor{teal}{\nabla_d\peqsub{\mathrm{Tr}}{g}h}\Big)
\end{align*}
\mbox{}\vspace*{-7ex}

\end{proof}

\begin{lemma}\label{Appendix - lemma - dH}\mbox{}\\
	Given the Hamiltonian \eqref{GR - equation - Hamiltonian} its differential is given by
	  \begin{align*}
	&\mathrm{d}_{(\mathbf{N},N,\gamma;\pi)}H\Big( \peqsub{Z}{\boldsymbol{\mathbf{N}}},\peqsub{Z}{\boldsymbol{\vec{N}}},\peqsub{Z}{\boldsymbol{\gamma}},\peqsub{Z}{\boldsymbol{\pi}}\Big)=\int_\Sigma\peqsub{\mathrm{vol}}{\Sigma}\,\Bigg\{\peqsub{Z}{\boldsymbol{\vec{N}}}^a\mathcal{H}_a+\peqsub{Z}{\boldsymbol{\mathbf{N}}}\peqsub{\mathcal{H}}{\perp}+(\peqsub{Z}{\boldsymbol{\pi}})^{ab}\Big[(\mathcal{L}_{\vec{N}}\gamma)_{ab}+2\mathbf{N}K_{ab}\Big]-{}\\
	&\hspace*{30ex}-(\peqsub{Z}{\boldsymbol{\gamma}})_{ab}\Big[(\mathcal{L}_{\vec{N}}\pi)^{ab}-\mathbf{N}\beta^{ab}+\sqrt{\gamma}\Big(\nabla^a\nabla^b-\gamma^{ab}\nabla_d\nabla^d\Big)\mathbf{N}\Big]\Bigg\}
	\end{align*}
	where $\mathcal{H}$ and $\peqsubfino{\mathcal{H}}{\perp}{-0.2ex}$ are defined in \eqref{GR - equation - H_tangente} and \eqref{GR - equation - H_perpendicular}.
\end{lemma}
\begin{proof}\mbox{}\\
	First notice that $\nabla_c\pi^{cb}$ depends on $\pi$ but also on the metric through the covariant derivative and also through $\pi$ because it is a $(0,2)$-density tensor field. We will now work for a moment in coordinates to express the covariant derivative, given by equation \eqref{Appendix - equation - covariant derivative} with weight $w=-1$, as
	\[\nabla_c\pi^{cb}=\partial_c\pi^{cb}+\tensor{\Gamma}{^c_c_d}\pi^{db}+\tensor{\Gamma}{^b_c_d}\pi^{cd}-\tensor{\Gamma}{^d_d_c}\pi^{cb}\]
	Notice that $\partial_c$ does not depend on the metric and, by definition $D\pi=\peqsub{Y}{\boldsymbol{\pi}}$. Therefore we have
	\begin{align*}
	D&\Big(\nabla_c\pi^{cb}\Big)=D\Big(\partial_c\pi^{cb}+\tensor{\Gamma}{^c_c_d}\pi^{db}+\tensor{\Gamma}{^b_c_d}\pi^{cd}-\tensor{\Gamma}{^d_d_c}\pi^{cb}\Big)=\\
	&\ =\partial_c\peqsub{Y}{\boldsymbol{\pi}}^{cb}+\tensor{\Gamma}{^c_c_d}\peqsub{Y}{\boldsymbol{\pi}}^{db}+\tensor{\Gamma}{^b_c_d}\peqsub{Y}{\boldsymbol{\pi}}^{cd}-\tensor{\Gamma}{^d_d_c}\peqsub{Y}{\boldsymbol{\pi}}^{cb}+\textcolor{gray}{(D\Gamma)^c_{\ cd}\pi^{db}}+(D\Gamma)^b_{\ cd}\pi^{cd}-\textcolor{gray}{(D\Gamma)^b_{\ bc}\pi^{cb}}\updown{\eqref{Appendix - equation - covariant derivative}}{\eqref{Appendix - equation - variacion nabla}}{=}\\
	&\ =\nabla_c\peqsub{Y}{\boldsymbol{\pi}}^{cb}+\frac{\gamma^{be}}{2}\Big(\nabla_c(\peqsub{Y}{\boldsymbol{\gamma}})_{de}+\nabla_d(\peqsub{Y}{\boldsymbol{\gamma}})_{ce}-\nabla_e(\peqsub{Y}{\boldsymbol{\gamma}})_{cd}\Big)\pi^{cd}
	\end{align*}
	which is valid for any coordinate system (globally). Thus
\begin{align*}
(D\mathcal{H})_{a}&=-2D\Big(\gamma_{ab}\nabla_c\pi^{cb}\Big)=-2(D\gamma)_{ab}\nabla_c\pi^{cb}-2\gamma_{ab}D\Big(\nabla_c\pi^{cb}\Big)=\\
&=-2(\peqsub{Y}{\boldsymbol{\gamma}})_{ab}\nabla_c\pi^{cb}-2\gamma_{ab}\nabla_c\peqsub{Y}{\boldsymbol{\pi}}^{cb}-2\gamma_{ab}\frac{\gamma^{be}}{2}\Big(\nabla_c(\peqsub{Y}{\boldsymbol{\gamma}})_{de}+\nabla_d(\peqsub{Y}{\boldsymbol{\gamma}})_{ce}-\nabla_e(\peqsub{Y}{\boldsymbol{\gamma}})_{cd}\Big)\pi^{cd}=\\
&=-\textcolor{teal}{2(\peqsub{Y}{\boldsymbol{\gamma}})_{ad}\nabla_c\pi^{cd}}-2\nabla_c(\peqsub{Y}{\boldsymbol{\pi}})^{c}_{\ a}-\pi^{cd}\Big(\textcolor{teal}{\nabla_c(\peqsub{Y}{\boldsymbol{\gamma}})_{da}+\nabla_d(\peqsub{Y}{\boldsymbol{\gamma}})_{ca}}-\nabla_a(\peqsub{Y}{\boldsymbol{\gamma}})_{cd}\Big)\updown{\pi}{\mathrm{sym.}}{=}\\
&=-\textcolor{teal}{2\nabla_c\left[(\peqsub{Y}{\boldsymbol{\gamma}})_{ad}\pi^{cd}\right]}-2\nabla_c(\peqsub{Y}{\boldsymbol{\pi}})^{c}_{\ a}+\pi^{cd}\nabla_a(\peqsub{Y}{\boldsymbol{\gamma}})_{cd}
\end{align*}
On the other hand we have
\begin{align*}
D&\peqsub{\mathcal{H}}{\perp}=D\left(-\frac{\varepsilon}{\sqrt{\gamma}}\pi^{ab}\pi^{cd}\left[\gamma_{ac}\gamma_{bd}-\frac{1}{n-1}\gamma_{ab}\gamma_{cd}\right]-\sqrt{\gamma}\Big(R^{(3)}_\gamma-2\Lambda\Big)\right)=\\
&=\left[\frac{\textcolor{red}{\varepsilon }D\sqrt{\gamma}}{\sqrt{\gamma}\textcolor{red}{\sqrt{\gamma}}}\pi^{ab}\textcolor{red}{\pi^{cd}}-\frac{2\textcolor{red}{\varepsilon}}{\textcolor{red}{\sqrt{\gamma}}}(D\pi)^{ab}\textcolor{red}{\pi^{cd}}\right]\textcolor{red}{\left[\gamma_{ac}\gamma_{bd}-\frac{1}{n-1}\gamma_{ab}\gamma_{cd}\right]}-{}\\
&\qquad-\frac{\varepsilon}{\sqrt{\gamma}}\pi^{ab}\textcolor{brown}{\pi^{cd}}\left[2(D\gamma)_{ac}\textcolor{brown}{\gamma_{bd}}-\frac{2}{n-1}(D\gamma)_{ab}\textcolor{brown}{\gamma_{cd}}\right]-(D\sqrt{\gamma})\Big(R^{(3)}_\gamma-2\Lambda\Big)-\sqrt{\gamma}(DR^{(3)}_\gamma)\updown{\eqref{Appendix - equation - variacion determinante}}{\eqref{Appendix - lemma - DR}}{=}\\
&=\left[\frac{ Tr(\peqsub{Y}{\boldsymbol{\gamma}})}{2}\pi^{ab}-2(\peqsub{Y}{\boldsymbol{\pi}})^{ab}\right]\textcolor{red}{\frac{\varepsilon}{\sqrt{\gamma}}\left[\pi_{ab}-\frac{1}{n-1}\gamma_{ab}\pi\right]}-2\pi^{ab}\frac{\varepsilon}{\sqrt{\gamma}}\left[(\peqsub{Y}{\boldsymbol{\gamma}})_{ac}\textcolor{brown}{\pi^{c}_{\ b}}-\frac{1}{n-1}(\peqsub{Y}{\boldsymbol{\gamma}})_{ac}\gamma^c_b\textcolor{brown}{\pi}\right]-{}\\
&\qquad -\frac{Tr(\peqsub{Y}{\boldsymbol{\gamma}})}{2}\sqrt{\gamma}\Big(R^{(3)}_\gamma-2\Lambda\Big)-\sqrt{\gamma}\Big[-\peqsub{Y}{\boldsymbol{\gamma}}^{ac}\mathrm{Ric}(\gamma)_{ac}+\nabla^d\Big(\nabla^b(\peqsub{Y}{\boldsymbol{\gamma}})_{bd}-\nabla_d(\peqsub{Y}{\boldsymbol{\gamma}})^a_{\ a}\Big)\Big]\overset{\eqref{GR - equation - pi=k... K=pi...}}{=}\\
&\ =\left[\textcolor{teal}{\frac{ Tr(\peqsub{Y}{\boldsymbol{\gamma}})}{2}\pi^{ab}}-2(\peqsub{Y}{\boldsymbol{\pi}})^{ab}\right]\textcolor{red}{(-K_{ab})}-\textcolor{teal}{2\pi^{ab}(\peqsub{Y}{\boldsymbol{\gamma}})_{ac}}\textcolor{brown}{(-K^{c}_{\ b})}-\textcolor{teal}{\frac{Tr(\peqsub{Y}{\boldsymbol{\gamma}})}{2}\sqrt{\gamma}\Big(R^{(3)}_\gamma-2\Lambda\Big)}+{}\\
&\qquad +\sqrt{\gamma}\Big[\textcolor{teal}{\peqsub{Y}{\boldsymbol{\gamma}}^{ac}\mathrm{Ric}(\gamma)_{ac}}-\nabla^d\Big(\nabla^b(\peqsub{Y}{\boldsymbol{\gamma}})_{bd}-\nabla_d(\peqsub{Y}{\boldsymbol{\gamma}})^a_{\ a}\Big)\Big]=\\
&\ =2(\peqsub{Y}{\boldsymbol{\pi}})^{ab}\textcolor{red}{K_{ab}}+\textcolor{teal}{(\peqsub{Y}{\boldsymbol{\gamma}})_{ac}\left(2\pi^{ab}\textcolor{brown}{K^{c}_{\ b}}+ \sqrt{\gamma}\mathrm{Ric}(\gamma)^{ac}-\frac{\gamma^{ac}}{2}\left[\pi^{bd}\textcolor{red}{K_{bd}}+\sqrt{\gamma}(R^{(3)}_\gamma-2\Lambda)\right]\right)}-{}\\
&\qquad-\sqrt{\gamma}\nabla^d\Big(\nabla^b(\peqsub{Y}{\boldsymbol{\gamma}})_{bd}-\nabla_d(\peqsub{Y}{\boldsymbol{\gamma}})^a_{\ a}\Big)=\\
&\ =2(\peqsub{Y}{\boldsymbol{\pi}})^{ab}K_{ab}+(\peqsub{Y}{\boldsymbol{\gamma}})_{ac}\beta^{ac}-\sqrt{\gamma}\nabla^d\Big(\nabla^b(\peqsub{Y}{\boldsymbol{\gamma}})_{bd}-\nabla_d(\peqsub{Y}{\boldsymbol{\gamma}})^a_{\ a}\Big)
\end{align*}
where we have defined
\begin{align*}
\beta^{ac}&=2\pi^{ab}K^{c}_{\ b}+ \sqrt{\gamma}\mathrm{Ric}(\gamma)^{ac}-\frac{\gamma^{ac}}{2}\left[\pi^{bd}\textcolor{red}{K_{bd}}+\sqrt{\gamma}(R^{(3)}_\gamma-2\Lambda)\right]=\\
&=2\pi^{ab}K^{c}_{\ b}+ \sqrt{\gamma}\left[\mathrm{Ric}(\gamma)^{ac}-\gamma^{ac}(R^{(3)}_\gamma-2\Lambda)\right]-\frac{\gamma^{ac}}{2}\peqsub{\mathcal{H}}{\perp}=\\
&=\sqrt{\gamma}R^{ac}+2\pi^{ab}K^{c}_{\ b}-\gamma^{ac}\pi^{bd}K_{bd}+\frac{\gamma^{ac}}{2}\peqsub{\mathcal{H}}{\perp}=\\
&=\sqrt{\gamma}R^{ac}-\frac{2\varepsilon}{\sqrt{\gamma}}\left(\pi^{ad}\pi^c_{\ d}-\frac{1}{n-1}\pi^{ac}\pi\right)+\frac{\varepsilon}{\sqrt{\gamma}}\left(\pi_{bd}\pi^{bd}-\frac{1}{n-1}\pi^2\right)\gamma^{ac}+\frac{\gamma^{ac}}{2}\peqsub{\mathcal{H}}{\perp}
\end{align*}
So finally we have
\begin{align*}
&\mathrm{d}_{(\mathbf{N},N,\gamma;\pi)}H\Big( \peqsub{Z}{\boldsymbol{\mathbf{N}}},\peqsub{Z}{\boldsymbol{\vec{N}}},\peqsub{Z}{\boldsymbol{\gamma}},\peqsub{Z}{\boldsymbol{\pi}}\Big)=\int_\Sigma\peqsub{\mathrm{vol}}{\Sigma}\Big[(\peqsub{D}{Z}N)^a\mathcal{H}_a+N^a(\peqsub{D}{Z}\mathcal{H})_a+(\peqsub{D}{Z}\mathbf{N})\peqsub{\mathcal{H}}{\perp}+\mathbf{N}(\peqsub{D}{Z}\peqsub{\mathcal{H}}{\perp})\Big]=\\
&=\int_\Sigma\peqsub{\mathrm{vol}}{\Sigma}\Bigg\{\peqsub{Z}{\boldsymbol{\vec{N}}}^a\mathcal{H}_a-N^a\Big(2\nabla_c\left[(\peqsub{Z}{\boldsymbol{\gamma}})_{ab}\pi^{cb}\right]+2\nabla_c(\peqsub{Z}{\boldsymbol{\pi}})^{c}_{\ a}-\pi^{cd}\nabla_a(\peqsub{Z}{\boldsymbol{\gamma}})_{cd}\Big)+\peqsub{Z}{\boldsymbol{\mathbf{N}}}\peqsub{\mathcal{H}}{\perp}+{}\\
&\qquad\qquad+\mathbf{N}\Big(2(\peqsub{Z}{\boldsymbol{\pi}})^{ab}K_{ab}+(\peqsub{Z}{\boldsymbol{\gamma}})_{ac}\beta^{ac}-\sqrt{\gamma}\nabla^d\big\{\nabla^b(\peqsub{Z}{\boldsymbol{\gamma}})_{bd}-\nabla_d(\peqsub{Z}{\boldsymbol{\gamma}})^a_{\ a}\big\}\Big)\Bigg\}\overset{\eqref{appendix equation integracion por partes}}{=}\\
&=\int_\Sigma\peqsub{\mathrm{vol}}{\Sigma}\Bigg\{\peqsub{Z}{\boldsymbol{\vec{N}}}^a\mathcal{H}_a+2\textcolor{red}{(\peqsub{Z}{\boldsymbol{\gamma}})_{ab}}\pi^{cb}\nabla_cN^a+2\textcolor{brown}{(\peqsub{Z}{\boldsymbol{\pi}})^{c}_{\ a}}\nabla_cN^a-\textcolor{red}{(\peqsub{Z}{\boldsymbol{\gamma}})_{ab}}\nabla_c(\pi^{ab}N^c)+{}\\
&\quad+\peqsub{Z}{\boldsymbol{\mathbf{N}}}\peqsub{\mathcal{H}}{\perp}+2\mathbf{N}\textcolor{brown}{(\peqsub{Z}{\boldsymbol{\pi}})^{ab}}K_{ab}+\textcolor{red}{(\peqsub{Z}{\boldsymbol{\gamma}})_{ac}}\mathbf{N}\beta^{ac}-\sqrt{\gamma}\textcolor{red}{(\peqsub{Z}{\boldsymbol{\gamma}})_{bd}}\nabla^b\nabla^d\mathbf{N}+\sqrt{\gamma}\textcolor{red}{(\peqsub{Z}{\boldsymbol{\gamma}})^a_{\ a}}\nabla_d\nabla^d\mathbf{N}\Bigg\}=\\
&=\int_\Sigma\peqsub{\mathrm{vol}}{\Sigma}\Bigg\{\peqsub{Z}{\boldsymbol{\vec{N}}}^a\mathcal{H}_a+\peqsub{Z}{\boldsymbol{\mathbf{N}}}\peqsub{\mathcal{H}}{\perp}+\textcolor{brown}{(\peqsub{Z}{\boldsymbol{\pi}})^{ab}}\Big[(\mathcal{L}_{\vec{N}}\gamma)_{ab}+2\mathbf{N}K_{ab}\Big]+{}\\
&\quad-\textcolor{red}{(\peqsub{Z}{\boldsymbol{\gamma}})_{ab}}\Big[-\textcolor{teal}{\pi^{cb}\nabla_cN^a-\pi^{ca}\nabla_cN^b+\nabla_c(\pi^{ab}N^c)}-\mathbf{N}\beta^{ab}+\sqrt{\gamma}\Big(\nabla^a\nabla^b-\gamma^{ab}\nabla_d\nabla^d\Big)\mathbf{N}\Big]\Bigg\}\overset{\eqref{Appendix - equation - derivada de lie abstract index}}{=}\\
&=\int_\Sigma\peqsub{\mathrm{vol}}{\Sigma}\Bigg\{\peqsub{Z}{\boldsymbol{\vec{N}}}^a\mathcal{H}_a+\peqsub{Z}{\boldsymbol{\mathbf{N}}}\peqsub{\mathcal{H}}{\perp}+(\peqsub{Z}{\boldsymbol{\pi}})^{ab}\Big[(\mathcal{L}_{\vec{N}}\gamma)_{ab}+2\mathbf{N}K_{ab}\Big]-{}\\
&\qquad\qquad-(\peqsub{Z}{\boldsymbol{\gamma}})_{ab}\Big[\textcolor{teal}{(\mathcal{L}_{\vec{N}}\pi)^{ab}}-\mathbf{N}\beta^{ab}+\sqrt{\gamma}\Big(\nabla^a\nabla^b-\gamma^{ab}\nabla_d\nabla^d\Big)\mathbf{N}\Big]\Bigg\}
\end{align*}
\mbox{}\vspace*{-7ex}

\end{proof}

\begin{lemma}\label{Appendix - lemma - Y_K}\mbox{}\\
	Assuming the dynamic equations \eqref{GR - equation - Hamiltonian equations pi} the variation of
	\[K_{ab}=\frac{-\varepsilon}{\sqrt{\gamma}}\left[\pi_{ab}-\frac{\pi}{n-1}\gamma_{ab}\right]\]
	is given by
	\[\vecc{Y}{K}^{ab}=(\mathcal{L}_{\vec{N}}K)^{ab}-\varepsilon\nabla_a\nabla_b\mathbf{N}+\mathbf{N}\Bigg(\varepsilon R_{ab}-KK_{ab}+2K_{ac}\tensor{K}{^d_b}-\frac{\varepsilon\gamma_{ab}}{n-1}\left[2\Lambda-\frac{\peqsub{\mathcal{H}}{\perp}}{2\sqrt{\gamma}}\right]\Bigg)\]
\end{lemma}
\begin{proof}\mbox{}\\
	For this computation we consider the operator $D-\mathcal{L}_{\vec{N}}$ and take advantage of the fact that the Lie derivative is a variation with respect to the embedding with zero lapse.
	\begin{align*}
	&(\vecc{Y}{K})_{ab}-(\mathcal{L}_{\vec{N}}K)_{ab}\overset{\eqref{GR - equation - pi=k... K=pi...}}{=}(D-\mathcal{L}_{\vec{N}})\left(\frac{-\varepsilon}{\sqrt{\gamma}}\left[\gamma_{ac}\gamma_{bd}-\frac{\gamma_{ab}\gamma_{cd}}{n-1}\right]\pi^{cd}\right)=\\
	&=-\varepsilon\frac{-(D-\mathcal{L}_{\vec{N}})\sqrt{\gamma}}{\sqrt{\gamma}\sqrt{\gamma}}\left[\pi_{ab}-\frac{\pi}{n-1}\gamma_{ab}\right]-\frac{\varepsilon}{\sqrt{\gamma}}\Bigg[\tensor{\pi}{^c_b}(D-\mathcal{L}_{\vec{N}})\gamma_{ac}+\tensor{\pi}{_a^d}(D-\mathcal{L}_{\vec{N}})\gamma_{bd}-{}\\
	&-\frac{(D-\mathcal{L}_{\vec{N}})\gamma_{ab}}{n-1}\pi-\gamma_{ab}\frac{(D-\mathcal{L}_{\vec{N}})\gamma_{cd}}{n-1}\pi^{cd}+\left(\gamma_{ac}\gamma_{bd}-\frac{\gamma_{ab}\gamma_{cd}}{n-1}\right)(D-\mathcal{L}_{\vec{N}})\pi^{cd}\Bigg]\updown{\eqref{Appendix - equation - variacion determinante}}{\eqref{GR - equation - Hamiltonian equations pi}}{=}\\
	&=\frac{-\mathrm{Tr}(2\mathbf{N}K)}{2}K_{ab}-\frac{\varepsilon}{\sqrt{\gamma}}\Bigg[\textcolor{red}{\tensor{\pi}{^d_b}}2\mathbf{N}K_{ac}+\tensor{\pi}{_a^d}2\mathbf{N}K_{bd}-\frac{2\mathbf{N}K_{ab}}{n-1}\textcolor{red}{\pi}-\gamma_{ab}\frac{2\mathbf{N}K_{cd}}{n-1}\pi^{cd}+{}\\
	&+\left(\gamma_{ac}\gamma_{bd}-\frac{\gamma_{ab}\gamma_{cd}}{n-1}\right)\Big(-\mathbf{N}\beta^{cd}+\sqrt{\gamma}\Big(\nabla^c\nabla^d-\gamma^{cd}\nabla_e\nabla^e\big)\mathbf{N}\Big)\Bigg]\overset{\eqref{GR - equation - pi=k... K=pi...}}{=}\\
	&=\frac{-\mathrm{Tr}(2\mathbf{N}K)}{2}K_{ab}+2\mathbf{N}\Bigg[K_{ac}\textcolor{red}{\tensor{K}{^d_b}}+(\tensor{K}{_a^d}-\gamma^d_aK)K_{bd}-\gamma_{ab}\frac{K_{cd}}{n-1}(K^{cd}-\gamma^{cd}K)\Bigg]-{}\\
	&-\frac{\varepsilon}{\sqrt{\gamma}}\left(-\mathbf{N}\beta_{ab}+\sqrt{\gamma}\big(\textcolor{brown}{\nabla_a\nabla_b}-\textcolor{teal}{\gamma_{ab}\nabla_e\nabla^e}\big)\mathbf{N}-\gamma_{ab}\frac{-\mathbf{N}\beta+\sqrt{\gamma}\textcolor{teal}{\big(\nabla_e\nabla^e-n\nabla_e\nabla^e\big)\mathbf{N}}}{n-1}\right)=\\
	&=-\varepsilon\textcolor{brown}{\nabla_a\nabla_b}\mathbf{N}+\mathbf{N}\Bigg[-KK_{ab}+2K_{ac}\tensor{K}{^d_b}+2\tensor{K}{_a^d}K_{bd}-2KK_{ab}-\frac{2\gamma_{ab}}{n-1}(K_{cd}K^{cd}-K^2)\Bigg]+{}\\
	&+\frac{\varepsilon\mathbf{N}}{\sqrt{\gamma}}\left(\beta_{ab}-\gamma_{ab}\frac{\beta}{n-1}\right)\overset{\star}{=}\\
	&=-\varepsilon\nabla_a\nabla_b\mathbf{N}+\mathbf{N}\Bigg[-\textcolor{magenta}{3KK_{ab}}+2K_{ac}\tensor{K}{^d_b}+2\textcolor{red}{\tensor{K}{_a^d}K_{bd}}-\frac{2\varepsilon\gamma_{ab}}{n-1}\textcolor{teal}{\varepsilon(K_{cd}K^{cd}-K^2)}\Bigg]+{}\\
	&+\mathbf{N}\left(\varepsilon R_{ab}-2(\textcolor{red}{\tensor{K}{_a^d}K_{bd}}-\textcolor{magenta}{KK_{ab}})-\frac{\varepsilon\gamma_{ab}}{n-1}\left[\frac{\peqsub{\mathcal{H}}{\perp}}{2\sqrt{\gamma}}-\textcolor{teal}{\varepsilon(K^{cd}K_{cd}-K^2)}+R\right]\right)=\\
	&=-\varepsilon\nabla_a\nabla_b\mathbf{N}+\mathbf{N}\Bigg(\varepsilon R_{ab}-\textcolor{magenta}{KK_{ab}}+2K_{ac}\tensor{K}{^d_b}-\frac{\varepsilon\gamma_{ab}}{n-1}\left[\frac{\peqsub{\mathcal{H}}{\perp}}{2\sqrt{\gamma}}+\textcolor{teal}{\varepsilon(K^{cd}K_{cd}-K^2)}+R\right]\Bigg)\overset{\eqref{GR - equation - H_perpendicular}}{=}\\
	&=-\varepsilon\nabla_a\nabla_b\mathbf{N}+\mathbf{N}\Bigg(\varepsilon R_{ab}-KK_{ab}+2K_{ac}\tensor{K}{^d_b}-\frac{\varepsilon\gamma_{ab}}{n-1}\left[2\Lambda-\frac{\peqsub{\mathcal{H}}{\perp}}{2\sqrt{\gamma}}\right]\Bigg)
	\end{align*}
	In the $\star$ equality we have used the definition of $\beta$ (see proof of lemma \ref{Appendix - lemma - dH}) to obtain
	\begin{align*}
	  &\frac{\beta_{ab}}{\sqrt{\gamma}}-\gamma_{ab}\frac{\beta/\sqrt{\gamma}}{n-1}=R_{ab}+\textcolor{red}{\varepsilon(K_{cd}K^{cd}-K^2)\gamma_{ab}}-2\varepsilon(K_{ad}\tensor{K}{_b^d}-KK_{ab})+\textcolor{teal}{\frac{\peqsub{\mathcal{H}}{\perp}}{2\sqrt{\gamma}}\gamma_{ab}}-{}\\
	  &-\frac{\gamma_{ab}}{n-1}\left[R+\textcolor{red}{\varepsilon(K_{cd}K^{cd}-K^2)n-2\varepsilon(K_{cd}\tensor{K}{^c^d}-K^2)}+\textcolor{teal}{\frac{\peqsub{\mathcal{H}}{\perp}}{2\sqrt{\gamma}}n}\right]=\\
	  &=R_{ab}-2\varepsilon(K_{ad}\tensor{K}{_b^d}-KK_{ab})+\frac{\gamma_{ab}}{n-1}\left[-\textcolor{teal}{\frac{\peqsub{\mathcal{H}}{\perp}}{2\sqrt{\gamma}}}+\textcolor{red}{\varepsilon(K_{cd}K^{cd}-K^2)}-R\right]
	\end{align*}	
	\mbox{}\vspace*{-6ex}
	
\end{proof}

\begin{lemma}\label{Appendix - lemma - DH and DH_perp}\mbox{}\\
	Let $\peqsub{\mathcal{H}}{\perp}$ and $\mathcal{H}_a$ be defined by \eqref{GR - equation - H_tangente}-\eqref{GR - equation - H_perpendicular}, then the variations with respect to the dynamics \eqref{GR - equation - Hamiltonian equations pi} are given by
	\begin{align*}
	&(D\mathcal{H})_a=(\mathcal{L}_{\vec{N}}\mathcal{H})_a+\peqsubfino{\mathcal{H}}{\perp}{-0.2ex}\nabla_a\mathbf{N}\\[1ex]
	&(D\peqsub{\mathcal{H}}{\perp})=\varepsilon\mathbf{N}\nabla^a\mathcal{H}_a+2\varepsilon\mathcal{H}_a\nabla^a\mathbf{N}
	\end{align*}
\end{lemma}
\begin{proof}\mbox{}\\
	For these computations we consider again the operator $D-\mathcal{L}_{\vec{N}}$ and use the computations we obtained in the proof of lemma \ref{Appendix - lemma - dH}.
	\begin{align*}
	\big[(&D-\mathcal{L}_{\vec{N}})\mathcal{H}\big]_a=-2\nabla_c\left[\pi^{cd}(D\!-\!\mathcal{L}_{\vec{N}})\gamma_{ad}\right]-2\nabla_c[(D\!-\!\mathcal{L}_{\vec{N}})\pi^{c}_{\ a}]+\pi^{cd}\nabla_a[(D\!-\!\mathcal{L}_{\vec{N}})\gamma_{cd}]\overset{\eqref{GR - equation - Hamiltonian equations pi}}{=}\\
	&=-2\nabla_c\left[\textcolor{brown}{\pi^{cd}2\mathbf{N}K_{ad}}\right]-2\nabla_c[\textcolor{brown}{-\mathbf{N}\beta^{c}_{\ a}}+\textcolor{red}{\sqrt{\gamma}(\nabla^c\nabla_a-\gamma^c_a\nabla^d\nabla_d)\mathbf{N}}]+\textcolor{magenta}{\pi^{cd}\nabla_a[2\mathbf{N}K_{cd}]}\overset{\star}{=}\\
	&=2\nabla_c\left[\textcolor{brown}{\mathbf{N}\Big\{\sqrt{\gamma}R^c_{\ a}-\textcolor{cyan}{K_{bd}\pi^{bd}\gamma^c_a}+\frac{\gamma^a_c}{2}\peqsub{\mathcal{H}}{\perp}\Big\}}\right]-2\textcolor{red}{\sqrt{\gamma}(\nabla_c\nabla_a\nabla^c-\nabla_a\nabla_c\nabla^c)\mathbf{N}}+{}\\
	&\qquad+\textcolor{magenta}{\nabla_a[\textcolor{cyan}{2\mathbf{N}K_{cd}\pi^{cd}}]-2\mathbf{N}K_{cd}\nabla_a\pi^{cd}}\overset{\eqref{Appendix - definition - Riemann}}{=}\\
	&=2\sqrt{\gamma}\mathbf{N}\textcolor{teal}{\nabla_cR^c_{\ a}}+2\sqrt{\gamma}R^c_{\ a}\nabla_c\mathbf{N}+\textcolor{cyan}{0}+\peqsub{\mathcal{H}}{\perp}\nabla_a\mathbf{N}+\mathbf{N}\nabla_a\peqsub{\mathcal{H}}{\perp}-2\textcolor{red}{\sqrt{\gamma}\tensor{R}{^c_f_c_a}\nabla^f\mathbf{N}}+{}\\
	&\qquad+\textcolor{magenta}{2\varepsilon\sqrt{\gamma}\mathbf{N}K_{cd}\nabla_a(K^{cd}-\gamma^{cd}K)}\overset{\eqref{Appendix - equation - div Ric=grad R}}{=}\\[-0.8ex]
	&=\peqsub{\mathcal{H}}{\perp}\nabla_a\mathbf{N}+2\sqrt{\gamma}R_{ca}\nabla^c\mathbf{N}-2\textcolor{red}{\sqrt{\gamma}R_{ba}\nabla^b\mathbf{N}}+\sqrt{\gamma}\mathbf{N}\nabla_a\Bigg\{\textcolor{teal}{R}+\frac{\peqsub{\mathcal{H}}{\perp}}{\sqrt{\gamma}}+\textcolor{magenta}{\Big[\varepsilon(K_{cd}K^{cd}-K^2)\Big]}\Bigg\}\overset{\eqref{GR - equation - H_perpendicular}}{=}\\
	&=\peqsub{\mathcal{H}}{\perp}\nabla_a\mathbf{N}+0+\sqrt{\gamma}\mathbf{N}2\nabla_a\Lambda=\\
	&=\peqsub{\mathcal{H}}{\perp}\nabla_a\mathbf{N}
	\end{align*}
	where in the $\star$ equality we have used that the covariant derivatives commute over functions and that $\beta$ can be rewritten, using \eqref{GR - equation - pi=k... K=pi...}, as
	\[\beta^{ac}=\sqrt{\gamma}R^{ac}+2\tensor{\pi}{^c_d}K^{ad}-K_{bd}\pi^{bd}\gamma^{ac}+\frac{\gamma^{ac}}{2}\peqsub{\mathcal{H}}{\perp}\]
	Analogously for $\peqsub{\mathcal{H}}{\perp}$ we have
	\begin{align*}
	&(D\!-\!\mathcal{L}_{\vec{N}})\peqsub{\mathcal{H}}{\perp}=2K_{ab}(D\!-\!\mathcal{L}_{\vec{N}})\pi^{ab}+\beta^{ab}(D\!-\!\mathcal{L}_{\vec{N}})\gamma_{ab}-\sqrt{\gamma}\Big(\nabla^a\nabla^b-\gamma^{ab}\nabla^d\nabla_d\Big)(D\!-\!\mathcal{L}_{\vec{N}})\gamma_{ab}\overset{\eqref{GR - equation - Hamiltonian equations pi}}{=}\\  
	&=2K_{ab}\Big[\!-\textcolor{brown}{\mathbf{N}\beta^{ab}}+\sqrt{\gamma}(\nabla^a\nabla^b-\gamma^{ab}\nabla_d\nabla^d)\mathbf{N}\Big]+\textcolor{brown}{\beta^{ab}2\mathbf{N}K_{ab}}-\textcolor{red}{\sqrt{\gamma}\Big(\nabla^a\nabla^b-\gamma^{ab}\nabla^d\nabla_d\Big)[2\mathbf{N}K_{ab}]}=\\ 
	&=2\sqrt{\gamma}K_{ab}(\nabla^a\nabla^b-\gamma^{ab}\nabla_d\nabla^d)\mathbf{N}-\textcolor{red}{2\sqrt{\gamma}\mathbf{N}\Big(\nabla^a\nabla^b-\gamma^{ab}\nabla^d\nabla_d\Big)K_{ab}}-{}\\
	&\qquad-\textcolor{red}{2\sqrt{\gamma}K_{ab}\Big(\nabla^a\nabla^b-\gamma^{ab}\nabla^d\nabla_d\Big)\mathbf{N}-2\sqrt{\gamma}\Big(2\nabla^a\mathbf{N}\cdot{}\nabla^bK_{ab}-2\nabla^d\mathbf{N}\cdot{}\nabla_dK\Big)}=\\
	&=-2\sqrt{\gamma}\mathbf{N}\nabla^a\Big(\nabla^bK_{ab}-\nabla_aK\Big)-4\sqrt{\gamma}\nabla^a\mathbf{N}\cdot{}\Big(\nabla^bK_{ab}-\nabla_aK\Big)\updown{\eqref{GR - equation - H_tangente}}{\eqref{GR - equation - pi=k... K=pi...}}{=}\\
	&=\varepsilon\mathbf{N}\nabla^a\mathcal{H}_a+2\varepsilon\mathcal{H}_a\nabla^a\mathbf{N}
	\end{align*}
	\mbox{}\vspace*{-7ex}
	
\end{proof}

\begin{lemma}\label{Appendix - lemma - DH DH_perp Dlambda uni}\mbox{}\\
	Let $\peqsub{\mathcal{H}}{\perp}$, $\mathcal{H}_a$ and $\peqsub{\mathcal{H}}{\lambda}$ be defined by \eqref{GR - equation - H_tangente uni}-\eqref{GR - equation - H_lambda uni}, then the variations with respect to the dynamics \eqref{GR - equation - Hamiltonian equations pi uni} are given by
	\begin{align*}
	&(D\mathcal{H})_a=\big(\peqsub{\mathcal{H}}{\perp}+2\lambda\sqrt{\gamma}\big)\nabla_a\mathbf{N}+\peqsub{\mathcal{H}}{\lambda}\nabla_a\lambda+2\epsilon\nabla_a\lambda\\[1ex]
	&D\Big[\peqsub{\mathcal{H}}{\perp}+2\lambda\sqrt{\gamma}\Big]=\mathcal{L}_{\vec{N}}\Big[\peqsub{\mathcal{H}}{\perp}+2\lambda\sqrt{\gamma}\Big]+\varepsilon\mathbf{N}\nabla^a\mathcal{H}_a+2\varepsilon\mathcal{H}_a\nabla^a\mathbf{N}+2\sqrt{\gamma}\Big(\peqsub{Y}{\boldsymbol{\lambda}}-\mathcal{L}_{\vec{N}}\lambda\Big)\\
	&D\peqsub{\mathcal{H}}{\lambda}=\mathcal{L}_{\vec{N}}\peqsub{\mathcal{H}}{\lambda}+2\sqrt{\gamma}\left[\peqsub{Y}{\boldsymbol{\mathbf{N}}}-\mathcal{L}_{\vec{N}}\mathbf{N}+\frac{\varepsilon\pi}{(n-1)\sqrt{\gamma}}\mathbf{N}^2\right]
	\end{align*}
\end{lemma}
\begin{proof}\mbox{}\\  
  	To prove this lemma we only have to adapt the proof of lemma \ref{Appendix - lemma - DH and DH_perp} taking into account that $\peqsub{\beta}{\mathrm{uni}}=\beta+\lambda\sqrt{\gamma}\,\gamma$ in the solution of $\peqsub{Y}{\boldsymbol{\gamma}}$.
	\begin{align*}
	\big[(D\!-\!\mathcal{L}_{\vec{N}})\mathcal{H}\big]_a&\overset{\ref{Appendix - lemma - DH DH_perp Dlambda uni}}{=}(\peqsub{\mathcal{H}}{\perp}+2\lambda\sqrt{\gamma})\nabla_a\mathbf{N}+2\mathbf{N}\sqrt{\gamma}\nabla_c\lambda\overset{\eqref{GR - equation - H_lambda uni}}{=}\\
	&\overset{\phantom{\ref{Appendix - lemma - DH DH_perp Dlambda uni}}}{=}(\peqsub{\mathcal{H}}{\perp}+2\lambda\sqrt{\gamma})\nabla_a\mathbf{N}+\big(\peqsub{\mathcal{H}}{\lambda}+2\epsilon\big)\nabla_c\lambda=\\
	&\overset{\phantom{\ref{Appendix - lemma - DH DH_perp Dlambda uni}}}{=}\big(\peqsub{\mathcal{H}}{\perp}+2\lambda\sqrt{\gamma}\big)\nabla_a\mathbf{N}+\peqsub{\mathcal{H}}{\lambda}\nabla_a\lambda+2\epsilon\nabla_a\lambda
	\end{align*}
	From the proof of \eqref{Appendix - lemma - dH} we have that
	\begin{align*}
	D&\peqsub{\mathcal{H}}{\perp}=2(\peqsub{Y}{\boldsymbol{\pi}})^{ab}K_{ab}+(\peqsub{Y}{\boldsymbol{\gamma}})_{ac}\beta^{ac}-\sqrt{\gamma}\nabla^d\Big(\nabla^b(\peqsub{Y}{\boldsymbol{\gamma}})_{bd}-\nabla_d(\peqsub{Y}{\boldsymbol{\gamma}})^a_{\ a}\Big)
	\end{align*}
	Now we can follow the proof of lemma \ref{Appendix - lemma - DH and DH_perp} but taking into account that when we plug the equation for $\peqsub{Y}{\boldsymbol{\gamma}}$ a $\peqsub{\beta}{\mathrm{uni}}$ will appear.
	\begin{align*}
	&(D\!-\!\mathcal{L}_{\vec{N}})\Big[\peqsub{\mathcal{H}}{\perp}+2\lambda\sqrt{\gamma}\Big]=\\
	&\qquad=\varepsilon\mathbf{N}\nabla^a\mathcal{H}_a+2\varepsilon\mathcal{H}_a\nabla^a\mathbf{N}-2K_{ab}\textcolor{brown}{\mathbf{N}\lambda\sqrt{\gamma}\gamma^{ab}}+2\sqrt{\gamma}(D\!-\!\mathcal{L}_{\vec{N}})\lambda+2\lambda(D\!-\!\mathcal{L}_{\vec{N}})\sqrt{\gamma}\updown{\eqref{GR - equation - Hamiltonian equations pi uni}}{\eqref{Appendix - equation - variacion determinante}}{=}\\
	&\qquad=\varepsilon\mathbf{N}\nabla^a\mathcal{H}_a+2\varepsilon\mathcal{H}_a\nabla^a\mathbf{N}-\textcolor{red}{2\mathbf{N}\lambda\sqrt{\gamma}K}+2\sqrt{\gamma}\Big(\peqsub{Y}{\boldsymbol{\lambda}}-\mathcal{L}_{\vec{N}}\lambda\Big)+\textcolor{red}{2\lambda\frac{\varepsilon\mathbf{N}\pi}{n-1}}\overset{\eqref{GR - equation - pi=k... K=pi... uni}}{=}\\
	&\qquad=\varepsilon\mathbf{N}\nabla^a\mathcal{H}_a+2\varepsilon\mathcal{H}_a\nabla^a\mathbf{N}+2\sqrt{\gamma}\Big(\peqsub{Y}{\boldsymbol{\lambda}}-\mathcal{L}_{\vec{N}}\lambda\Big)
	\end{align*}
	Finally we compute the variation of $\peqsub{\mathcal{H}}{\lambda}$
	\begin{align*}
	  (D\!-\!\mathcal{L}_{\vec{N}})\peqsub{\mathcal{H}}{\lambda}&= 2(D\!-\!\mathcal{L}_{\vec{N}})\Big(\mathbf{N}\sqrt{\gamma}-\epsilon\Big)=2\sqrt{\gamma}(D\!-\!\mathcal{L}_{\vec{N}})\mathbf{N}\,+\mathbf{N}(D\!-\!\mathcal{L}_{\vec{N}})\sqrt{\gamma}=\\
	  &=2\sqrt{\gamma}\Big(\peqsub{Y}{\boldsymbol{\mathbf{N}}}-\mathcal{L}_{\vec{N}}\mathbf{N}\Big)+\mathbf{N}\frac{\varepsilon\mathbf{N}\pi}{n-1}=2\sqrt{\gamma}\left[\peqsub{Y}{\boldsymbol{\mathbf{N}}}-\mathcal{L}_{\vec{N}}\mathbf{N}+\frac{\varepsilon\pi}{(n-1)\sqrt{\gamma}}\mathbf{N}^2\right]
	\end{align*}
	
	\mbox{}\vspace*{-6ex}
	
\end{proof}

\end{appendix}
\cleardoublepage